\documentclass[a4paper, 12pt]{article}
\usepackage[top=3cm, bottom=2cm, left=3cm, right=2cm]{geometry}
\usepackage{booktabs}
\usepackage{tkz-euclide}
\usepackage{tocloft}

\usepackage[utf8]{inputenc}
\usepackage{amsmath, amsfonts, amssymb,amsthm,amsbsy}
\numberwithin{equation}{section}
\usepackage[subentrycounter,seeautonumberlist,nonumberlist=true, acronym]{glossaries} 
\usepackage{float}
\usepackage{graphicx}
\usepackage{pdfpages}
\usepackage{pdflscape}
\usepackage{multicol}
\usepackage{comment}
\usepackage{colortbl}
\usepackage{epstopdf}
\AppendGraphicsExtensions{.gif}
\usepackage[nottoc,notlot,notlof]{tocbibind}
 \usepackage[citecounter=true,  
 backend=biber,
 maxnames=4, 
 style=abnt-numeric,
 giveninits=true, 
 repeatfields,
 sortcites = false,
]{biblatex}
\DeclareCiteCommand{\cite}[\mkbibbrackets]
  {\usebibmacro{cite:init}%
   \usebibmacro{prenote}}%
  {\usebibmacro{citeindex}%
   \usebibmacro{cite:comp}}%
  {\iftoggle{comp}{}{\multicitedelim}}%
  {\usebibmacro{cite:dump}%
   \usebibmacro{postnote}}%

\DeclareCiteCommand{\parencite}[\mkbibbrackets]%
  {\usebibmacro{cite:init}%
   \usebibmacro{prenote}}%
  {\usebibmacro{citeindex}%
   \usebibmacro{cite:comp}}%
  {\iftoggle{comp}{}{\multicitedelim}}%
  {\usebibmacro{cite:dump}%
\usebibmacro{postnote}}%

\makeatletter
  \renewbibmacro*{textcite}{
    \iftoggle{comp}{%
      \iffieldequals{namehash}{\cbx@lasthash}%
        {\usebibmacro{cite:comp}}%
        {\usebibmacro{cite:dump}%
        \ifbool{cbx:parens}%
          {\printtext{\bibclosebracket}\global\boolfalse{cbx:parens}}%
          {}%
        \iffirstcitekey%
          {}%
          {\textcitedelim}%
        \usebibmacro{cite:init}%
        \printtext[bibhyperref]{%
          \ifnameundef{labelname}%
            {\printfield[citetitle]{labeltitle}}%
            {\printnames{labelname}}%
        }
        \setunit*{\printdelim{namelabeldelim}}%
        \printtext{\bibopenbracket}\global\booltrue{cbx:parens}%
        \ifnumequal{\value{citecount}}{1}%
          {\usebibmacro{prenote}}%
          {}%
        \usebibmacro{cite:comp}%
        \stepcounter{textcitecount}%
        \savefield{namehash}{\cbx@lasthash}}%
    }{%
      \iffieldequals{namehash}{\cbx@lasthash}%
            {\setunit{\multicitedelim}}%
            {\printtext[bibhyperref]{%
              \ifnameundef{labelname}%
                {\printfield[citetitle]{labeltitle}}%
                {\printnames{labelname}
             }}%
            \setunit*{\printdelim{namelabeldelim}}%
            \printtext{\bibopenbracket}\global\booltrue{cbx:parens}%
            \stepcounter{textcitecount}%
            \savefield{namehash}{\cbx@lasthash}}%
          \ifnumequal{\value{citecount}}{1}%
            {\usebibmacro{prenote}}%
            {}%
          \usebibmacro{cite}%
          \setunit{%
            \ifbool{cbx:parens}%
              {\bibclosebracket\global\boolfalse{cbx:parens}}%
              {}%
            \textcitedelim}%
    }%
}%

  \renewbibmacro*{textcite:postnote}{%
    \usebibmacro{postnote}%
    \ifthenelse{\value{multicitecount}=\value{multicitetotal}}%
      {\setunit{}%
      \printtext{%
        \ifbool{cbx:parens}%
          {\bibclosebracket\global\boolfalse{cbx:parens}}%
          {}}}%
      {\setunit{%
        \ifbool{cbx:parens}%
          {\bibclosebracket\global\boolfalse{cbx:parens}}%
          {}%
        \textcitedelim}}}%

  \DeclareCiteCommand{\cbx@textcite}%
    {\iftoggle{comp}{\usebibmacro{cite:init}}{\usebibmacro{textcite:init}}}%
    {\usebibmacro{citeindex}%
    \usebibmacro{textcite}}%
    {}%
    {\iftoggle{comp}{%
      \usebibmacro{cite:dump}%
          \usebibmacro{postnote}%
          \ifbool{cbx:parens}%
            {\bibclosebracket\global\boolfalse{cbx:parens}}%
            {}%
    }{%
      \usebibmacro{textcite:postnote}%
}}%
\makeatother

\DeclareFieldFormat{labelnumberwidth}{\mkbibbrackets{#1}}

\usepackage[brazil]{babel}
\usepackage{csquotes}
\usepackage{notoccite}
\usepackage{tcolorbox} 
\usepackage[perpage]{footmisc}
\usepackage{bm}
\usepackage{caption}
\usepackage{subcaption}
\usepackage{textcomp}
\usepackage{epigraph}
\usepackage{cmap}
\usepackage[T1]{fontenc}
\usepackage{xcolor}
\usepackage{graphicx}
\usepackage{ragged2e} 
\usepackage[normalem]{ulem}
\usepackage{boxedminipage}

\usepackage{esvect}
\newcommand*{\putunder}[2]{%
  {\mathop{#1}_{\textstyle #2}}%
}
\usepackage{siunitx}
\usepackage[titletoc]{appendix}
\usepackage[inline]{enumitem}
\usepackage{setspace}
\setlist{itemsep=0pt,parsep=0pt} 

\usepackage{xurl}
\usepackage[
  pdfencoding=auto, 
  psdextra, 
]{hyperref}
\usepackage[open, openlevel=2]{bookmark}
\hypersetup{
  colorlinks   = true, 
  urlcolor     = blue, 
  linkcolor    = blue, 
  citecolor   = blue 
}
\emergencystretch=\maxdimen
\usepackage{rotating}
\usepackage{titlesec}
\titleformat*{\section}{\large\bfseries}
\usepackage{orcidlink}

\definecolor{Gray}{gray}{0.9}
\definecolor{darkgreen}{cmyk}{1,0.4,0.8,0}
\definecolor{lightgreen}{cmyk}{0,0.16,1,0}
\definecolor{lightyellow}{rgb}{0.95, 0.95, 0.8}

\DeclareMathOperator{\Tr}{Tr}
\DeclareMathOperator{\sgn}{sgn}

\begin{document}
\pagestyle{empty}
\hypersetup{pageanchor=false}
\pagenumbering{gobble}
\thispagestyle{empty}
\vspace{2cm}
\begin{figure}[ht]
\centering
\includegraphics[scale=0.25]{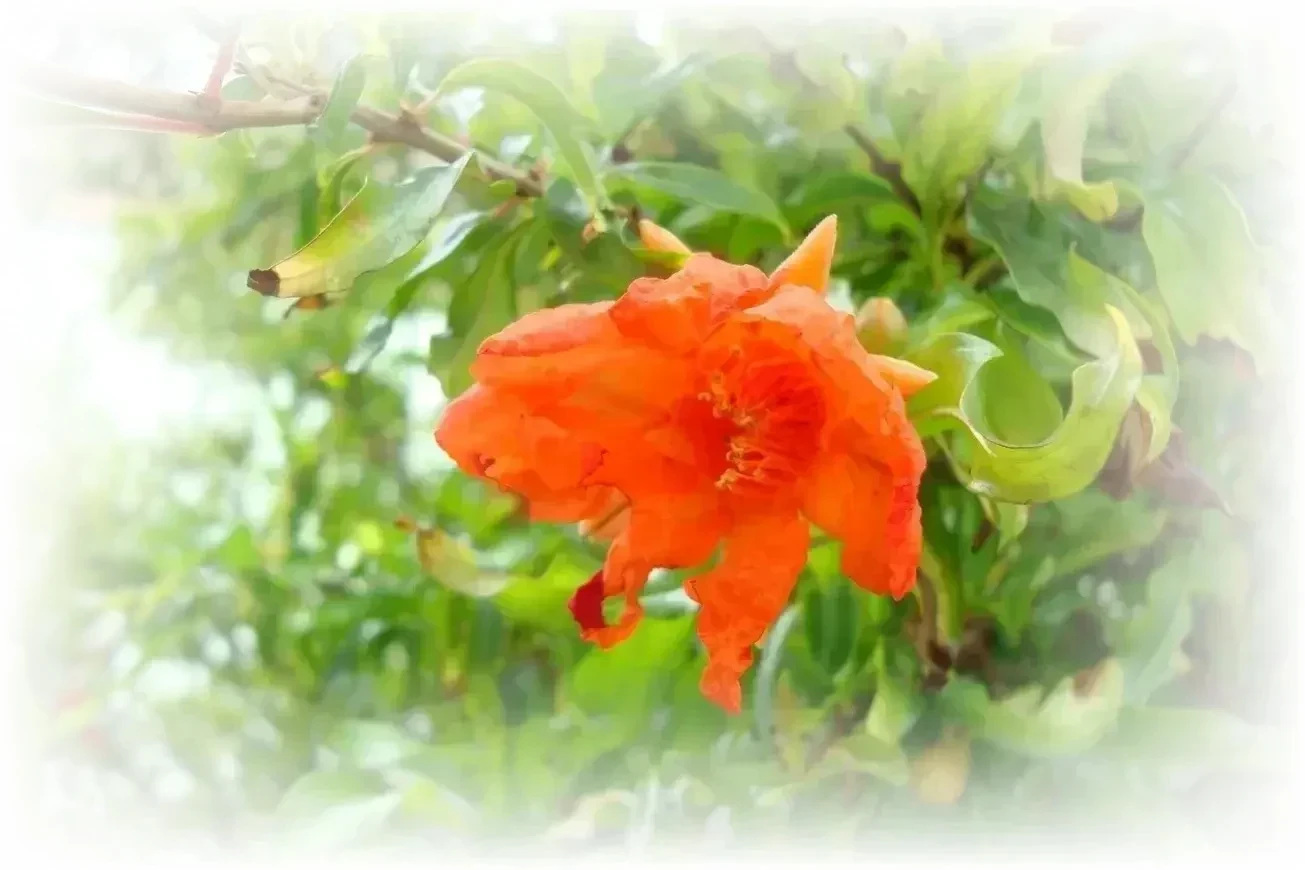}
\label{fig:logo}
\end{figure}
\begin{center}

\vspace{1cm}
	\textbf{\Large REGULARIZAÇÃO, APRENDIZAGEM PROFUNDA E INTERDISCIPLINARIDADE EM PROBLEMAS INVERSOS MAL-POSTOS}
	
\vspace{1cm}

\textbf{{\large Perguntas e Respostas}}
	
\vspace{3cm}
	
\begin{center}
\begin{tabular}{ c l }
Roberto Gutierrez Beraldo \orcidlink{0000-0001-6986-3435} & Universidade Federal do ABC  \\ 
 Ricardo Suyama \orcidlink{0000-0002-8398-5268} & Universidade Federal do ABC  \\  
\end{tabular}
\end{center}

\vspace{2cm}
	
{\large \textbf{Book Preprint}}
	
\vspace{\fill}

SANTO ANDRÉ (SÃO PAULO) - BRASIL \\
2025

\end{center}

\newpage
\justify
\thispagestyle{empty}

\section*{\centering PREPRINT DISCLAIMER}

\begin{itemize}
\item \textbf{Sugestões, colaborações, correções e reclamações} são muito bem-vindas e podem ser enviadas para \href{mailto:roberto.gutierrez@ufabc.edu.br}{roberto.gutierrez@ufabc.edu.br} ou para \href{mailto:ricardo.suyama@ufabc.edu.br}{ricardo.suyama@ufabc.edu.br}.
\item Esta é a primeira versão do livro, que ainda não passou por um processo de revisão por pares e editoração formal. Assim, de tempos em tempos ele será atualizado. 
\item A forma escolhida para o livro foi de perguntas e respostas. A maioria das seções foi definida desta maneira, quando adequado. O leitor pode acompanhar o livro do começo ao fim ou ir rapidamente para a pergunta de interesse; 
\item Considerando que há muitas perguntas, a profundidade das respostas varia. O objetivo não era esgotar o assunto, mas sim agrupar elementos e referências relevantes dentro de seu contexto;
\item Existem muitas expressões usualmente utilizadas na língua inglesa. Ao longo do livro, a maioria delas foi traduzida para o português. No entanto, os acrônimos e siglas foram mantidos em inglês, além de incluir uma lista de traduções específicas no material pré-textual. O objetivo é facilitar a busca por tais palavras-chave; 
\item Nas citações, há indicações de páginas, capítulos, apêndices e demais partes da referência, para facilitar a localização do conceito de interesse. No entanto, entre edições ou repositórios, essas indicações podem ser diferentes. No caso de ambiguidades, a página indicada pode ser desconsiderada;
\item As URLs foram verificadas durante a escrita, mas não há como garantir que elas funcionem em momentos posteriores. 
\end{itemize}

\vfill

\newpage
\thispagestyle{empty}
\section*{\centering PREFÁCIO DA PRIMEIRA EDIÇÃO}
A área de pesquisa de problemas inversos tem como objetivo, principalmente, estudar seu comportamento e propor soluções para eles. Um exemplo é a obtenção de informações de um fenômeno a partir apenas de medidas indiretas disponíveis. O desafio é que eles são usualmente mal-postos: podem não ter solução, não ter solução única ou cuja solução é instável a pequenas flutuações nos dados. Com origem na área de matemática aplicada, resolver tais problemas encontra aplicações também nas áreas da física, engenharias e da computação. Nesse contexto, a teoria da inversão de Tikhonov \cite{tikhonov1977solutions} foi um marco, pois trouxe bases teóricas importantes para entendê-los e também desenvolveu o método de regularização como proposta prática de solução. 

Na literatura produzida no Brasil, existem livros sobre problemas inversos e de regularização de modo mais geral \cite{Neto2005}, ou focando em algoritmos específicos para sua solução \cite{Neto2016} e aplicações específicas \cite{2016menin}. Há também diversas dissertações e teses sobre o assunto e livros do IMPA \cite{baumeister2005topics, bleyer2015novel}, mas em língua inglesa. Sendo assim, por que haver mais um livro sobre o assunto?

Existem artigos e livros que são muito completos, mas voltados à matemática aplicada, como \cite{Benning2018, engl1996regularization}, exigindo mais conhecimentos prévios sobre o assunto. Existem também livros que descrevem problemas inversos discretos, partindo principalmente de conceitos de álgebra linear \cite{aster2019parameter, hansen2010discrete, Mueller2012}. Assim, a nossa primeira motivação foi escrever um material sobre problemas inversos discretos que fosse razoavelmente acessível, sob o ponto de vista da língua ou mesmo da notação matemática.

Nos últimos anos, soluções de ponta a ponta de aprendizagem profunda vêm sendo propostas para solução de problemas inversos, contribuindo em várias aplicações \cite{Adler2021, Bai2020, Belthangady2019, Koh2021, Ongie2020, Su2022}. Também estão sendo propostos métodos que unem os paradigmas dos métodos de regularização e de aprendizagem profunda na solução de problemas mal-postos, utilizando as informações do modelo e dos dados em conjunto \cite{Arridge2019}.  Partindo do pressuposto que as informações obtidas nos modelos e nos dados são complementares, a nossa segunda motivação para escrita deste livro foi explorar e introduzir tais métodos integrados para novos leitores, apresentando diversas possibilidades.  

Finalmente, regularização pode ter diferentes significados dependendo do seu contexto  \cite{Chen2002} e interpretações desenvolvidas na aprendizagem profunda \cite{goodfellow2016deep} podem ficar cada vez mais distantes da definição original de Tikhonov. Considerando regularização um conceito polissêmico, a nossa última motivação foi discutir essas interpretações sob o ponto de vista da multidisciplinaridade e da interdisciplinaridade. O livro discute exemplos de ensino de problemas inversos, mostrando algumas iniciativas que colaboram para divulgação da área de forma acessível, discutindo as vantagens que uma abordagem interdisciplinar pode trazer.

\newpage
\thispagestyle{empty}
\section*{\centering PREFACE TO THE FIRST EDITION}
The research area of inverse problems aims to study their behavior and propose solutions to them. An example is obtaining information about a phenomenon from indirect measurements available only. The challenge is that these problems are usually ill-posed: they may have no solution, no unique solution or their solution is unstable due to small fluctuations in the data. Originating in the field of applied mathematics, solving such problems also finds applications in the fields of physics, engineering, and computing. In this context, Tikhonov's inversion theory \cite{tikhonov1977solutions} was a milestone, as it provided important theoretical basis for understanding them and also developed the regularization method as a practical solution proposal.

In the literature produced in Brazil, there are books on inverse and regularization problems in a more general way \cite{Neto2005}, or focusing on specific algorithms for their solution \cite{Neto2016} and specific applications \cite{2016menin}. There are also several dissertations and theses on the subject and IMPA books \cite{baumeister2005topics, bleyer2015novel}, but written in English. So why should there be another book on the subject?

Some articles and books are very complete, but focused on applied mathematics, such as \cite{Benning2018, engl1996regularization}, requiring more prior knowledge on the subject. Other books describe discrete inverse problems, mainly based on linear algebra concepts \cite{aster2019parameter, hansen2010discrete, Mueller2012}. Our first motivation was to write material about discrete inverse problems that was reasonably accessible, from the point of view of language or even mathematical notation.

In recent years, end-to-end deep learning solutions have been proposed for solving inverse problems, contributing to various applications \cite{Adler2021, Bai2020, Belthangady2019, Koh2021, Ongie2020, Su2022}. Methods are also being proposed that combine the paradigms of regularization and deep learning methods to solve ill-posed problems, using information from the model and the data together \cite{Arridge2019}.  Based on the assumption that the information obtained from models and data is complementary, our second motivation for writing this book was to explore and introduce integrated methods to new readers, presenting several possibilities.

Finally, regularization can have different meanings depending on its context \cite{Chen2002}, and interpretations developed in deep learning \cite{goodfellow2016deep} can become increasingly distant from Tikhonov's original definition. Considering regularization a polysemic concept, our last motivation was to discuss these interpretations from the point of view of multidisciplinarity and interdisciplinarity. The book discusses examples of teaching inverse problems, showing some initiatives that contribute to making the area accessible and discussing the advantages that results from an interdisciplinary approach.

\newpage
\section*{\centering ORGANIZAÇÃO DO TEXTO}
Os capítulos são descritos a seguir de acordo com a pergunta que eles
visam responder:
\begin{itemize}
\item \textbf{Capítulo \ref{sec:illposed}}: \textit{Quais são os problemas que se quer resolver?} O principal tema do livro é problemas inversos mal-postos. Este capítulo diferencia problemas diretos, inversos, bem-postos e mal-postos;
\item \textbf{Capítulo \ref{sec:deblur_forward}}: \textit{Aplicação: Como ver deblurring como um problema inverso?} Ilustração dos conceitos do Capítulo \ref{sec:illposed} no problema de \textit{deblurring};
\item \textbf{Capítulo \ref{sec:variational}}: \textit{Qual é o método utilizado para tentar resolvê-los?} Este capítulo discute uma forma de resolver problemas mal-postos, o método de regularização de Tikhonov. Sobre ele, são discutidas algumas das principais formas de termos de regularização;
\item \textbf{Capítulo \ref{sec:de_nonblind}}: \textit{Aplicação: Como obter imagens mais nítidas com deblurring?} Ilustração dos conceitos do Capítulo \ref{sec:variational} no problema de \textit{deblurring};
 \item \textbf{Capítulo \ref{sec:integrated}}: \textit{Existem formas de unir abordagens baseadas em modelos e em dados?}  Este capítulo pressupõe que o leitor conheça a visão geral de aprendizado de máquina e aprendizagem profunda. Após uma breve revisão de conceitos, regularização será rediscutida nesse novo contexto, já que ela adquiriu novos significados. Na sequência, o capítulo discute propostas encontradas na literatura que unem o método de regularização e aprendizagem profunda. Ao final, algumas limitações sobre aprendizagem profunda em problemas inversos são revistas, incluindo questões de interpretabilidade e reprodutibilidade;
\item \textbf{Capítulo \ref{sec:polysemy}}: \textit{Quais são algumas das diferentes interpretações de regularização?}  O capítulo descreve dez possíveis interpretações dos autores sobre o que regularização é ou como ela atua na solução. Também foi verificada a ocorrência desses significados em livros técnicos de problemas inversos e aprendizagem de máquina;
\item \textbf{Capítulo \ref{sec:interdisciplinarity}}: \textit{Quais são algumas das implicações de tantas áreas de pesquisa diferentes explorarem o conceito de regularização?} Após trazer diversos exemplos de conceitos abordados por diferentes áreas de pesquisa, discute-se a polissemia de regularização e possíveis caminhos interdisciplinares da área de problemas inversos, destacando algumas implicações dessas características para o ensino. 
\end{itemize}

\newpage
\thispagestyle{empty}
\section*{\centering AGRADECIMENTOS}

\vfill

Agradecemos o Prof. Dr. Fernando S. de Moura, Prof. Dr. Marcelo Zanotello e Dr. Leonardo A. Ferreira, que deram contribuições valiosas durante a escrita deste livro.  

\vfill

\newpage
 \section*{\centering EPÍGRAFE}\label{part0epi}

\vfill

\epigraph{«\textit{The rapidly increasing use of computational technology requires the development of computational algorithms for solving broad classes of problems. But just what do we mean by the ``solution'' of a problem? What requirements must the algorithms for finding a ``solution'' satisfy?}»}{Tikhonov e Arsenin \cite[Pág. 1]{tikhonov1977solutions}}

\vspace{2cm}

\epigraph{«\textit{Sometimes a computing machine does do something rather weird that we hadn’t expected. In principle one could have predicted it, but in practice it’s usually too much trouble. Obviously if one were to predict everything a computer was going to do one might just as well do without it.}»}{Alan Turing \cite{Turing2004}}

\vfill

\newpage
\renewcommand{\listfigurename}{LISTA DE FIGURAS}
\listoffigures
\renewcommand{\listtablename}{LISTA DE TABELAS}
\listoftables
 
\newpage
 \section*{\centering LISTA DE ACRÔNIMOS E SIGLAS}
\begin{flushleft}
\begin{tabular}{ l l }
\textbf{ADAM}&Adaptive Moment Estimation\\                                           \textbf{ADMM}&Alternating Direction Method of Multipliers\\                       \textbf{ANN}&Artificial Neural Network\\                                                     
\textbf{CGLS} & Conjugate Gradient Least Squares \\                                  
\textbf{CNN}&Convolutional Neural Network\\                                             
\textbf{CPU}&Central Processing Unit\\                                                       
\textbf{CS}&Compressed Sensing\\                                                             
\textbf{CT}&Computed Tomography\\                                                         
\textbf{DCT}&Discrete Cosine Transform\\                                                
\textbf{DFT}&Discrete Fourier Transform\\                                            
\textbf{DIP}&Deep Image Prior\\                                                                 
\textbf{DNN}&Deep Neural Network\\                                                         
\textbf{EIT}&Electrical Impedance Tomography\\                                        
\textbf{ERM}&Empirical Risk Minimization\\                                                 
\textbf{FBP}&Filtered Back Projection\\                                                                                                                      
\textbf{FFT}&Fast Fourier Transform\\                                                                                                                      
\textbf{FISTA}&Fast Iterative Soft Thresholding Algorithm\\                          
\textbf{GCV} &  Generalized Cross-Validation \\                                            
\textbf{GMRES} &  Generalized Minimal Residual Method \\   
\textbf{GPU}&Graphics Processing Unit\\                                                       
\textbf{GSVD}& Generalized Singular Value Decomposition\\                         
\textbf{HQS}&Half-Quadratic Splitting\\                                                        
\textbf{IRLS}&Iterative Re-weighted Least Squares\\                                    
\textbf{ISTA}&Iterative Soft Thresholding Algorithm\\                                   
\textbf{LASSO}&Least Absolute Shrinkage and Selection Operator\\              
\textbf{MAP}&Maximum A Posteriori\\                                            
\textbf{MaxEnt}&Maximum Entropy\\                                           
\textbf{MLE}&Maximum Likelihood Estimation\\                                        
\end{tabular}

\begin{tabular}{ l l }
\textbf{MSE}&Mean Squared Error\\                                                           
\textbf{NSR}&Noise-to-Signal Ratio\\
\textbf{OLS}&Ordinary Least Squares\\                                                     
$P^3$&Plug-and-Play Prior\\                                                                      
\textbf{PDE}& Partial Differential Equation\\                                              
\textbf{PSF}&Point Spread Function\\                                                        
\textbf{RAM}&Random Access Memory\\                                                   
\textbf{RED}&Regularization by Denoising\\                                                   
 \textbf{RIP}&Restricted Isometry Property\\                                              
\textbf{SGD}&Stochastic Gradient Descent\\                                              
\textbf{SNR}&Signal-to-Noise Ratio\\     
\textbf{SRM}&Structural Risk Minimization\\                                              
\textbf{STEM} &  Science, Technology, Engineering, and Mathematics\\
\textbf{SVD}&Singular Value Decomposition\\                                            
\textbf{SVM}&Support Vector Machine\\                                                 
\textbf{TSVD}&Truncated Singular Value Decomposition\\                         
\textbf{TV}&Total Variation\\                                                                     
\textbf{UFABC}&Federal University of ABC\\                                                                     
\textbf{VC}& Vapnik-Chervonenkis \\
\end{tabular}
\end{flushleft}

\newpage
 \section*{\centering LISTA DE TERMOS TRADUZIDOS}
\begin{center}
\textbf{Capítulo \ref{sec:illposed}}
\end{center}

\begin{table}[H]
\begin{center}
\begin{tabular}{ l l }
Cenário matemático & \textit{Mathematical setting}\\
Declaração do problema & \textit{Problem statement}\\
Estimação de parâmetros & \textit{Parameter estimation} \\
Fotobranqueamento & \textit{Photobleaching} \\
Identificação do sistema & \textit{System identification}\\
 & \quad ou \textit{model identification} \\
Matriz inversa à direita & \textit{Right inverse} \\
Matriz inversa à esquerda & \textit{Left inverse} \\
Problemas bem-postos & \textit{Well-posed problems}\\
 & \quad ou  \textit{Well-defined problems}\\
Problemas diretos & \textit{Forward problems} \\
 & \quad ou  \textit{direct problems} \cite{Bertero2021}\\
Problemas inversos & \textit{Inverse problems} \\
Problemas mal-postos & \textit{Ill-posed problems} \\
 & \quad ou \textit{Improperly posed problems} \\
Solução de norma mínima & \textit{Minimum norm solution} \\
Solução ingênua & \textit{Naive solution} \\
Subamostragem & \textit{Downsampling} \\
Tomografia computadorizada & \textit{Computed tomography} \\
Tomografia por impedância elétrica & \textit{Electrical impedance tomography} \\
\end{tabular}
\end{center}
\end{table}

\begin{center}
\textbf{Capítulo \ref{sec:deblur_forward}}
\end{center}

\begin{table}[H]
\begin{center}
\begin{tabular}{ l l }
Borrar & \textit{Blur} (verbo) \\
Borrão & \textit{Blur} (substântivo) \\
Condição de contorno replicada & \textit{Replicate boundary condition} \\
Deconvolução cega & \textit{Blind deblurring}\\
Deconvolução míope & \textit{Myopic deconvolution}\\
 & \quad ou \textit{Semi-blind deconvolution} \\
Filtragem espacial linear & \textit{Linear spatial filtering}\\
Função de espalhamento pontual & \textit{Point spread function}\\
Melhoramento de imagem & \textit{Image enhancement}\\
Nítida & \textit{Sharp}\\
\end{tabular}
\end{center}
\end{table}

\newpage
\begin{center}
\textbf{Capítulo \ref{sec:variational}}
\end{center}

\begin{table}[H]
\begin{center}
\begin{tabular}{ l l }
Amostragem comprimida & \textit{Compressed sensing} \\
Contínuas por partes & \textit{Piecewise continuous} \\
Diferenças para trás & \textit{Backward difference} \\
Função de perda & \textit{Loss function} \\
Imagens de treinamento & \textit{Training images} \\
Irregulares, ásperas, rugosas & \textit{Rough} \\
Irregularidade, aspereza, rugosidade & \textit{Roughness} \\
Máxima entropia & \textit{Maximum entropy} \\
Máximo \textit{a posteriori} & \textit{Maximum a posteriori} \\
Mínimos quadrados total & \textit{Total least-squares} \\
Otimização em dois níveis & \textit{Bilevel optimization} \\
Planas & \textit{Flat} \\
\textit{Priors} baseados em amostras & \textit{Sample-based priors} \\
Problema de Tikhonov na forma geral & \textit{General-form Tikhonov problem} \\
Problema de Tikhonov na forma padrão & \textit{Standard-form Tikhonov problem} \\
Qualidade do ajuste & \textit{Goodness of fit} \\
Quase monotônico & \textit{Quasimonotonic} \\
Rede elástica & \textit{Elastic net}\\
Regularização dupla & \textit{Double regularization} \\
Regularização que promove a esparsidade & \textit{Sparsity-promoting regularisation}$^*$\\
Restrição das propriedades físicas & \textit{Physics constrains} \\
Restrição suave & \textit{Soft constraint} \\
Suave & \textit{Smooth} \\
Termo de fidelidade & \textit{Data fidelity}  \\
 & \quad ou \textit{Data misfit} \\
Transformação reversa & \textit{Back-transformation} \\
Validação cruzada generalizada & \textit{Generalized cross-validation} \\
Variação total & \textit{Total variation} \\
\end{tabular}
\end{center}
\end{table}

\vspace{-4mm}

$^*$ Outras expressões incluem: \textit{sparse regularization}, \textit{sparsity-enforcing regularisation}, \textit{sparsity-encouraging penalty terms}, \textit{sparsity regularization} e \textit{sparse recovery algorithms}.

\newpage
\begin{center}
\textbf{Capítulo \ref{sec:de_nonblind}}
\end{center}

\begin{table}[H]
\begin{center}
\begin{tabular}{ l l }
Ajuste fino & \textit{Fine tuning} \\
Decomposição em valores singulares & \textit{Generalized singular value} \\
\quad generalizada & \quad \textit{decomposition} \\
Expansão em valores singulares & \textit{Singular value expansion} \\
Gradientes conjugados para  & \textit{Conjugate gradient} \\
\quad mínimos quadrados & \quad \textit{least squares} \\
Método do residual mínimo & \textit{Generalized minimal residual} \\
\quad generalizado & \quad \textit{method} \\
Posto numérico & \textit{Rank} \\
Problemas com deficiência de posto & \textit{Rank-deficient problems} \\
Razão ruído-sinal & \textit{Noise-to-signal ratio} \\
Razão sinal-ruido & \textit{Signal-to-noise ratio} \\
SVD amortecida & \textit{Damped} SVD \\
\end{tabular}
\end{center}
\end{table}

\begin{center}
\textbf{Capítulo \ref{sec:integrated}}
\end{center}

\begin{table}[H]
\begin{center}
\begin{tabular}{ l l }
Arquiteturas guiadas pela física & \textit{Physics-guided architectures}\\
Arquiteturas informadas pela física & \textit{Physics-informed architectures}\\
Caixa-preta &  \textit{Black box}\\
\textit{Design} estruturado de rede & \textit{Structured network design}\\
Camadas totalmente conectadas & \textit{Fully conected layers} \\ 
Diferenciação automática & \textit{Automatic differentiation}\\
Dissipação do gradiente & \textit{Vanishing gradient}\\
Extração de características & \textit{Feature extraction}\\
Função de ativação & \textit{Activation function} \\
Microscopia sem lente & \textit{Lens-free microscopy}\\
Minimização do risco empírico & \textit{Empirical risk minimization} \\
\textit{Perceptron} multicamadas & \textit{Multilayer perceptron}\\
Rasa &  \textit{Shallow} \\
Retropropagação & \textit{Backpropagation}\\
Separação de variáveis & \textit{Variable splitting}\\
Sobreparametrizada  & \textit{Overparameterized} \\
Teorema do ``não há almoço grátis'' & \textit{No free lunch theorem})\\
Unidades & \textit{Units} \\
\end{tabular}
\end{center}
\end{table}

\newpage

\begin{center}
\textbf{Capítulo \ref{sec:polysemy}}
\end{center}

\begin{table}[H]
\begin{center}
\begin{tabular}{ l l }
Compacidade & \textit{Compactness}\\
Informação \textit{a priori} fraca & \textit{Weak a priori information}\\
Limitantes & \textit{Bounds}\\
Matrizes circulantes de blocos & \textit{Block circulant matrices} \\
\quad com blocos circulantes & \quad  \textit{with circulant blocks} \\
Método dos residuais & \textit{Residual method}\\
Minimização do risco estrutural & \textit{Structural risk minimization}\\
Quase-soluções & \textit{Quasi-solutions}\\
Pesos dos filtros & \textit{Filter weights} \\
Problemas bem-postos vizinhos &  \textit{Neighboring well-posed problems} \\
Regularização variacional & \textit{Variational regularization} \\
Restrição rígida & \textit{Hard constraint} \\
Restrição suave & \textit{Soft constraint} \\
Suavização dos dados & \textit{Data smoothing}\\
\end{tabular}
\end{center}
\end{table}

\vfill

Observação: Não há expressões específicas no Capítulo \ref{sec:interdisciplinarity}. 

\newpage
\begin{center}
\textbf{Apêndices}
\end{center}

\begin{table}[H]
\begin{center}
\begin{tabular}{ l l }
Absolutamente contínuas & \textit{Absolutely continuous}\\
Aprendizagem da transformada & \textit{Transform learning}\\
Aprendizagem de dicionário esparsa & \textit{Sparse dictionary learning}\\
Ajuste de hiperparâmetros & \textit{Hyperparameter tuning} \\
Colocação &  \textit{Collocation}\\
Densidade \textit{a priori} & \textit{Prior density} \\
Decomposição em valores singulares& \textit{Truncated singular value}\\ 
\quad truncada & \quad \textit{decomposition}\\ 
Equalização de canais & \textit{Channel equalization}\\
Erro médio quadrático & \textit{Mean squared error}\\
Espaço gerado &  \textit{span}\\
Estimativa de máxima verossimilhança & \textit{Maximum likelihood estimation}\\
Estimativa de máximo a posteriori &  \textit{Maximum a posteriori estimation}\\
Memória de acesso aleatório & \textit{Random access memory}\\
Métodos diretos & \textit{Direct methods}\\
Mínimos quadrados ordinário & \textit{Ordinary least squares}\\
Mínimos quadrados regularizado & \textit{Regularized output least squares} \\
Não-suave & \textit{Non-smooth}\\
Posto linha completo & \textit{Full row rank} \\
Produto interno & \textit{Inner product} ou \textit{Dot product} \\
Propriedade de isometria restrita & \textit{Restricted isometry property}\\
Rede elástica & \textit{Elastic net}\\
Relaxação convexa & \textit{Convex relaxation}\\
Suave & \textit{Smooth} \\
Transformação por codificação & \textit{Transform coding}\\
Transformada discreta de Fourier & \textit{Discrete Fourier transform}\\
Transformada rápida de Fourier &  \textit{Fast Fourier transform}\\ 
\end{tabular}
\end{center}
\end{table}
 
\newpage
\section*{\centering LISTA DE SÍMBOLOS}

A lista a seguir traz os principais símbolos presentes nos capítulos, exceto apêndices. A notação básica é:
\begin{itemize}
\item Matrizes: $\mathbf{A}$ ou $\bm{\Sigma}$, possuem dimensão $[m \times  n]$, $[m \times  m]$ ou $[n \times  n]$ 
\item Vetores: $\mathbf{a}$ ou $\bm{\sigma}$ possuem dimensão $[m  \times  1]$ ou $[n \times 1]$ 
\item Escalares: $a$ ou $\sigma$  possuem dimensão $[1  \times  1]$ 
\end{itemize}

\begin{multicols}{2}

\subsection*{Capítulo \ref{sec:illposed}}
\noindent\begin{tabular}{ m{1.3cm} m{6.2cm}}
$\mathcal{A}$ & Operador direto \\
$ f $ &Entrada (parâmetros) \\
$F$  & Espaço dos parâmetros\\
$g$ &Saída (dados)  \\
$G$  & Espaço dos dados\\
$\forall$ & Para todo\\
$\in$ & 	É membro de\\
$\Delta g$ & Variação de $g$  \\
$\mathcal{A}^{-1}$ & Operador inverso \\
$k(t,s)$ &\textit{Kernel} de Hilbert–Schmidt  \\
$h(t-s)$ & Resposta ao impulso  \\ 
$\mathcal{L}\{\cdot\}(s) $ & Transformada de Laplace  \\
$\mathbf{y}$ &Medidas  \\
$\mathbf{x}$ &Parâmetros  \\
$\mathbf{A}$ &Operador direto \\
$i$, $j$ & Índices das matrizes  \\
$A_{i,j}$ & $(i,j)$-ésimo valor de $\mathbf{A}$ \\
$m$ & Nº de linhas da matriz  \\
 $n$ & Nº de colunas da matriz  \\
$\mathbf{a}_n$ &Enésima coluna de $\mathbf{A}$  \\
$x_i$ &i-ésimo valor de $\mathbf{x}$  \\
\end{tabular}

\noindent\begin{tabular}{ m{1.3cm} m{6.2cm}}
$\approx$ & Aproximadamente  \\
$\vert\vert \cdot \vert \vert_F$ & Norma de Frobenius \\
$\vert\vert \cdot \vert \vert_p$ & Norma $\ell_p$ \\
$p$ & Índice da norma $\ell_p$\\
$\bm{\epsilon}$ &Erros do modelo \\
$\bm{\delta}$ &Erros dos dados \\
$\mathbf{y}_{\delta}$ &Medidas com erros  \\
$\hat{\mathbf{x}}$ &Grandeza estimada \\
$\mathbf{A}^{-1}$ & Matriz inversa  \\
$\mathbf{I}$ &Matriz identidade   \\
$\mathbf{A}^T$ & Matriz transposta  \\
$\mathbf{A}^+_d$ &Matriz pseudoinversa à direita  \\
$\mathbf{A}^+_e$ &Matriz pseudoinversa à esquerda  \\
s.t. &Sujeito a  \\
$\sigma$ &Valores singulares  \\
$\mathbf{u}$ &Vetores singulares à direita \\
$\mathbf{v}$ &Vetores singulares à esquerda  \\ 
$\mathbf{U}, \mathbf{V} $ & Matrizes ortogonais da SVD  \\
$ \mathbf{\Sigma}$& Matriz diagonal de $\sigma$   \\
$\mathbf{A}^*$ & Conjugado transposto  \\
$diag(\cdot)$ & Matriz diagonal \\
$cond(\mathbf{A})$ & Número de condição de $\mathbf{A}$ \\
$\mathcal{O}$ & Grande-O \\
\end{tabular}

\newpage

\subsection*{Capítulo \ref{sec:deblur_forward}}
\noindent\begin{tabular}{ m{1.3cm} m{6.2cm}}
$\mathbf{Y}$ &Matriz de imagem degradada\\
$\mathbf{X}$ &Matriz de imagem nítida\\
$\mathbf{H}$ &Matriz do \textit{kernel}\\
$\mathbf{N}$ &Matriz de ruído\\
$\mathbf{X} * \mathbf{Y} $ &  Convolução de $\mathbf{X}$ com $\mathbf{Y}$\\
$\mathbf{h}$ &\textit{Kernel} $\mathbf{H}$ vetorizado  \\
$\mu$ & Média \\
$s$ & Desvio padrão\\ 
$s^2$ & Variância\\ 
$\mathbf{\Gamma}$ & Matriz de covariância \\
$\mathcal{N}(\mu,\mathbf{\Gamma})$ &Distribuição gaussiana \\
$\sim$ & É distribuído como  \\
$\mathbf{A}_{sub}$& Matriz de subamostragem\\
\end{tabular}

\vfill

\subsection*{Capítulo \ref{sec:variational}}
\noindent\begin{tabular}{ m{1.3cm} m{6.2cm}}
$\hat{\mathbf{x}}_{\lambda}$ &Regularização de Tikhonov   \\
$\mathcal{L}(\cdot)$ &Função de perda  \\
$\Omega(\cdot)$ &Termo de regularização  \\
$\lambda$ &Parâmetro de regularização  \\
$\mathcal{M}(\cdot)$ &Funcional  \\
$ \mathbf{A}^+_{\lambda}$ &Matriz de reconstrução \\
$\mathbf{0}$ & Vetor nulo\\
$G(\lambda)$ &GCV: Função\\
$c$& GCV: número de pontos \\
$Tr(\cdot)$ &Traço  \\
$\mathbf{L}$ &Matriz de regularização  \\
$N(\mathbf{A})$ & Núcleo de $\mathbf{A}$ \\ 
$\cap$& Intersecção entre conjuntos \\ 
$\emptyset$& Conjunto vazio \\ 
$\mathbf{L_{1 \hspace{1mm} d1}}$ & $\mathbf{L}$ de primeira derivada \\
$\mathbf{L_{2 \hspace{1mm} d1}}$ & Outra $\mathbf{L}$ de primeira derivada \\
$\mathbf{L}_{d2}$ & Matriz de segunda derivada \\
$\mathbf{H_{d2}}$ & Kernel laplaciano em 2D\\
$\mathbf{X}$ &Matriz de uma imagem\\
$\mathbf{H} * \mathbf{X}$ & Convolução entre $\mathbf{H}$ e $\mathbf{X}$ \\
$\mathbf{x}^*$ &Valor de referência fixo \\
$\mathbf{x}^*_k$ &Valor de referência variável \\
$k$ & Índice da iteração \\
$\mathbf{I}_N$ & Matriz $\mathbf{I}$  $n \times n$ \\
$\mathbf{I}_P$ & Matriz $\mathbf{I}$  $p \times p$\\
$\mathbf{A}^*$ & Hermitiano de $\mathbf{A}$\\
$\vert\vert \mathbf{x} \vert\vert_{p,q}$ & Norma mista \\
$\nabla \mathbf{x}$ &  Gradiente de $\mathbf{x}$\\
\end{tabular}

\noindent\begin{tabular}{ m{1.3cm} m{6.2cm}}
$\omega_i$ & Máxima entropia: pesos \\ 
$L_{aug}$ &Lagrangiano aumentado \\ 
$\rho$ &ADMM: escalar  \\
$\mathbf{z}$ &ADMM: variável auxiliar  \\
$\mathbf{u}$ &ADMM: variável dual  \\
$ f(\cdot)$    & \textit{Denoiser}\\
$\mathbf{D}$ & Dicionário \\
$\mathbf{s}_D$ & Representação de $\bm{x}$ em $\mathbf{D}$ \\
$\overline{\mathbf{x}}$ & $\mathbf{x}$ na forma padrão\\
$\mathbf{P}$ & Precondicionador \\
$\mathbf{s}_P$ & Representação de $\bm{x}$ em $\mathbf{P}$ \\
$\mathbf{\Gamma}$ & Matriz de covariância \\
$a$ & Atlas anatômico: escalar \\
$\Omega_{\bm{\theta}}(\mathbf{x})$ & $\Omega(\cdot)$ parametrizado por $\bm{\theta}$\\
$\bm{\theta}$ & Parâmetros\\
$\mathbf{x}_t$ & Modelos de treinamento\\
$\hat{\mathbf{x}}_{\bm{\theta}}$ & $\mathbf{x}$ estimada com  $\bm{\theta}$ \\
\end{tabular}

\vfill\null
\columnbreak

\subsection*{Capítulo \ref{sec:integrated}}
\noindent\begin{tabular}{ m{1.3cm} m{6.2cm}}
  $h_{\bm{\theta}}( \cdot)$ & Estrutura candidata \\
  $\bm{\theta}$ & Parâmetros de $h$\\
 ($\mathbf{x}_t^{i}, \mathbf{y}_t^{i}$) & i-ésimo par de treinamento\\
 $N$ &Número de amostras \\
 $\Sigma_{i=1}^N$ &Somatório de 1 até N  \\
 $\Psi_{\bm{\theta}}(\cdot)$  & Rede neural como uma função\\
 $\mathbf{b}$ & \textit{Bias} \\
$\mathbf{w}$ & Pesos da rede\\
$ f(\cdot)$    & Função de ativação\\
\end{tabular}

\subsection*{Capítulo \ref{sec:polysemy}}

\noindent\begin{tabular}{ m{1.3cm} m{6.2cm}}
$ \mathcal{R}_{\lambda}(\cdot)$ &Operador de regularização  \\
$M$ & Ivanov: subconjunto de $F$ \\
$c$,$c_0$  & Ivanov: limitante superior \\
$\bm{\delta}_1$, $\bm{\delta}_2$ & Limitantes supeiores\\
$\pi(\mathbf{x})$ & Densidade de probabilidade de $\mathbf{x}$ \\ 
$\pi(\mathbf{x})_{priori}$ & Densidade \textit{a priori} de $\mathbf{x}$ \\ 
$\pi(\mathbf{x})_{post}$ & Densidade \textit{a posteriori} de $\mathbf{x}$ \\ 
$\pi(\mathbf{x} | \mathbf{y})$ & Densidade condicional \\ 
$\beta$ & Coeficientes de regressão linear\\
$S$ & SRM: Subconjunto\\
$\subset$ & Pertencente a \\
\end{tabular}

\end{multicols}

\newpage
\renewcommand\contentsname{SUMÁRIO}
\setcounter{tocdepth}{3} 
\tableofcontents

\newpage
\pagestyle{plain}
\hypersetup{pageanchor=true}
\pagenumbering{arabic} 
\setcounter{page}{1}
\justify
\section{PROBLEMAS INVERSOS MAL-POSTOS: O QUE SÃO OS PROBLEMAS QUE QUEREMOS RESOLVER?}\label{sec:illposed}

\subsection{Introdução}

Três problemas diferentes são descritos a seguir e deve-se dizer qual que é a semelhança entre eles. Suponha que:
\begin{itemize}
\item Um dispositivo não possa ser aberto, como um sistema do tipo caixa preta. No seu circuito elétrico, há dois terminais de entrada, dois de saída e vários resistores. Deseja-se obter a resistência de cada um deles, individualmente \cite{ufabcDC1, ufabcDC2};
\item O tamanho de um estádio da antiguidade tenha sido medido por Hércules, contando seus próprios passos. Deseja-se calcular a altura do Hércules a partir do tamanho desse estádio \cite{ufabcDC6}; 
\item Você quer resolver um quebra-cabeça lógico chamado de nonograma. Em uma malha quadriculada, você deve descobrir qual é o desenho oculto apenas sabendo quantos quadrados coloridos há em cada linha e coluna dessa malha \cite{ufabcDC4}.
\end{itemize}
A semelhança entre eles é que as informações disponíveis para resolvê-los são outras, ou seja, indiretas. Não se tem acesso aos resistores, ao Hércules e nem ao desenho.

Uma das grandes áreas de pesquisa da ciência e da engenharia busca construir modelos numéricos de sistemas para relacionar seus parâmetros físicos com os dados provenientes de observações \cite[pág. 1]{aster2019parameter}. Em muitos casos, a validação dos modelos e sua comparação com dados reais é um grande desafio, pois as observações possíveis e disponíveis são apenas medidas indiretas da grandeza de interesse. 

Problemas que exigem a interpretação de medidas indiretas ou incompletas dão origem aos chamados problemas inversos \cite[pág. xi]{Mueller2012}. Em \cite{Kabanikhin2008}, discute-se alguns critérios utilizados para definir tais problemas e listas de exemplos são encontradas em \cite{aster2019parameter} e \cite[Capítulo 1]{baumeister2005topics}. Ao longo do presente capítulo, os conceitos principais dos problemas inversos serão apresentados. A seguir, eles são exemplificados a partir da tomografia computadorizada, tomografia por impedância elétrica e da microscopia de fluorescência. 

\subsection{$1^o$ Exemplo de problema inverso: Como reconstruir imagens a partir de sombras de raio-X?}

Técnicas tomográficas buscam a visualização de dentro de objetos a partir de sinais coletados que sejam externos ao objeto. Elas trazem como principais características serem não-invasivas e não-destrutivas. Há diversas técnicas tomográficas e uma das mais conhecidas e associadas ao termo é a Tomografia Computadorizada (CT) \cite{Natterer2001}, \cite[págs. 21-32]{Mueller2012}. Nela, uma fonte emite raios-X sobre o objeto de interesse, que percorrem caminhos diferentes dentro dele e sendo atenuados (absorvidos) de modos diferentes de acordo com os materiais que atravessam, sendo então captados por um detector de lado oposto à fonte de raios-X.  

Essa técnica é muito útil para visualizar o interior do corpo humano. Na região de interesse, o  coeficiente de absorção de um mesmo tecido (subdomínio) é aproximadamente constante, de modo que tecidos musculares e outros tecidos moles aparecem transparentes (ou quase) nas imagens, se comparado à informação dos diferentes tipos de ossos \cite[pág. 237]{kaipio2005statistical}. Essas diferentes atenuações  permitem que o diagnóstico seja feito. 

Assim, o problema fundamental da CT é de reconstruir um objeto a partir de suas projeções, ou sombras \cite{buzug2008computed}, também conhecido como sinograma, ficando caracterizado como um problema de medidas indiretas \cite[pag. 5]{Mueller2012}. Com interpretação clara do significado físico dos valores de atenuação de raios-X, do desenvolvimento de novos sensores e métodos matemáticos e computacionais para reconstrução, a técnica trouxe avanços e novas possibilidades para a medicina diagnóstica no campo da radiografia \cite{buzug2008computed}, como maior contraste do que a radiografia convencional e a visão tridimensional, com cortes em diferentes sentidos. 

\begin{figure}[H]
        \centering
        \includegraphics[width = \textwidth]{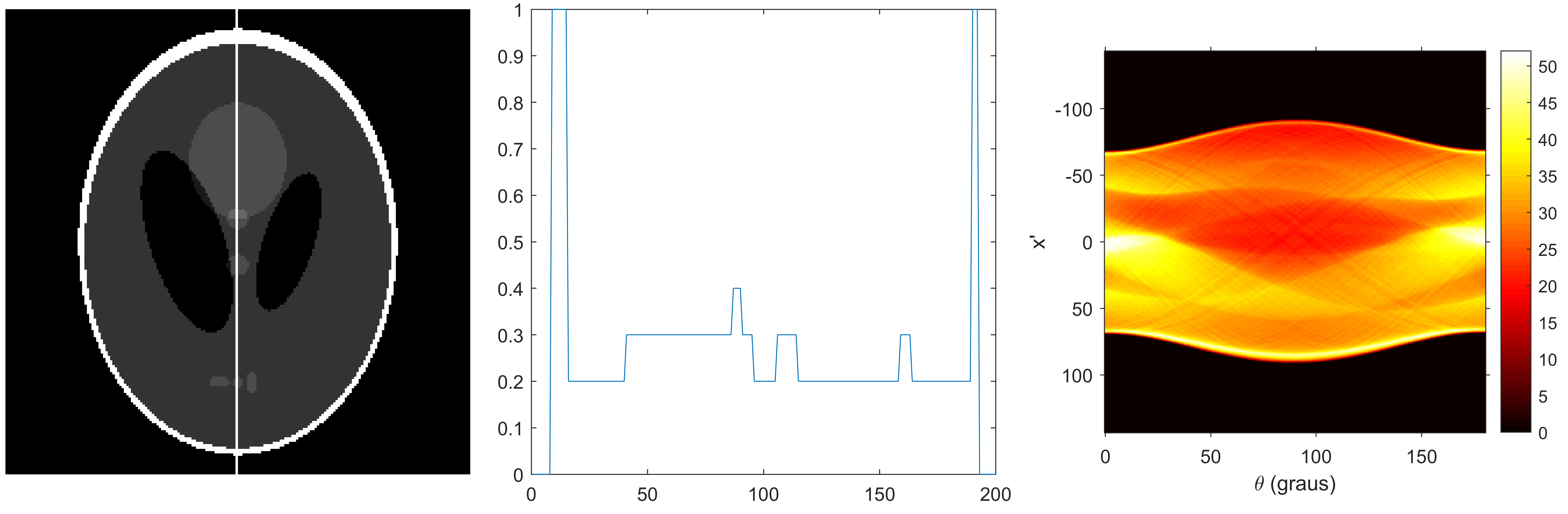}
        \caption[Imagem Tomográfica, valores do corte indicado e seu sinograma.]{Imagem Tomográfica, valores do corte indicado e seu sinograma. Fonte: Próprio autor.}
        \label{fig:CT2}
    \end{figure}

Para comparação, uma chapa convencional de raios-X é obtida a partir de uma única projeção, mas a imagem de CT é reconstruída a partir de muitos ângulos de projeção (de $0^{\circ}$ até $179^{\circ}$) em torno do objeto. O conjunto dessas projeções organizadas na forma de uma imagem é chamado de sinograma. Na Figura \ref{fig:CT2} é mostrado uma imagem tomográfica, seu sinograma completo com todas as projeções e os valores da imagem tomográfica ao longo da reta indicada nessa região de interesse.

\subsection{$2^o$ Exemplo de problema inverso: Como reconstruir imagens a partir de tensões elétricas?}

A Tomografia por Impedância Elétrica (EIT) é uma técnica para obtenção do mapa de resistividades elétricas dentro de um objeto de interesse. De modo não invasivo, eletrodos posicionados na superfície da região impõe corrente elétrica e os potenciais elétricos resultantes, obtidos nessa mesma interface, são utilizados para reconstrução das imagens tomográficas \cite[pág. 159]{Mueller2012}. Com sua alta resolução temporal, a EIT permite uma monitoração contínua do paciente. Além disso, a EIT não apresenta efeitos adversos causados por radiação ionizante e seu uso pode ser por tempo prolongado, desde que as normas de segurança elétrica sejam respeitadas.  

Partindo-se das Equações de Maxwell, busca-se calcular as medidas de potencial elétrico em um modelo computacional \cite{Holder}. Esse modelo deve representar a região de interesse das medidas. As medidas calculadas são comparadas com os dados experimentais dos eletrodos e o modelo é então atualizado. Assim, estima-se as resistividades dessa região tais como elas são, ou como elas variam ao longo do tempo.

\begin{figure}[H]
        \centering
                \includegraphics[width = 0.25\textwidth]{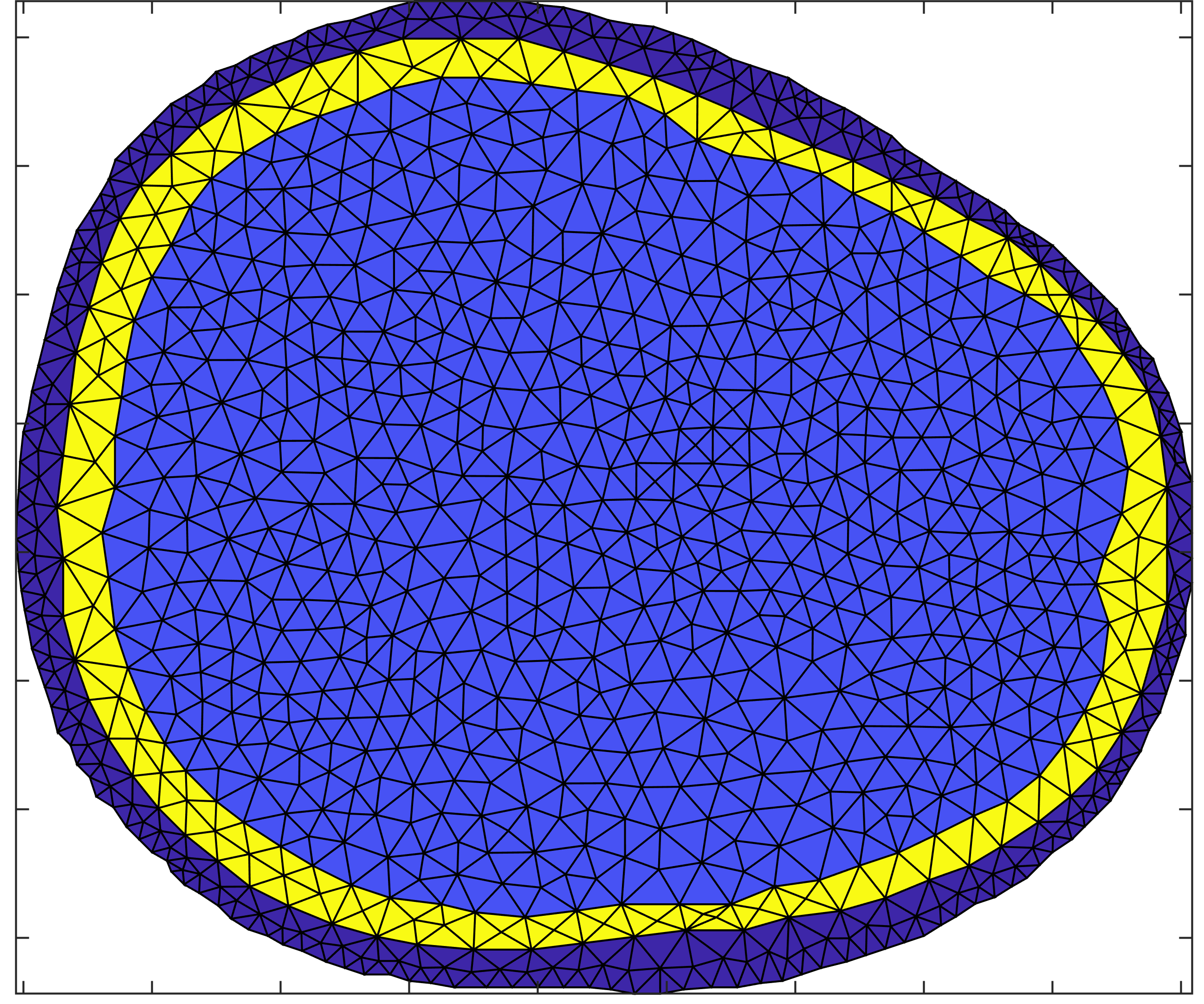}
        \includegraphics[width =1\textwidth]{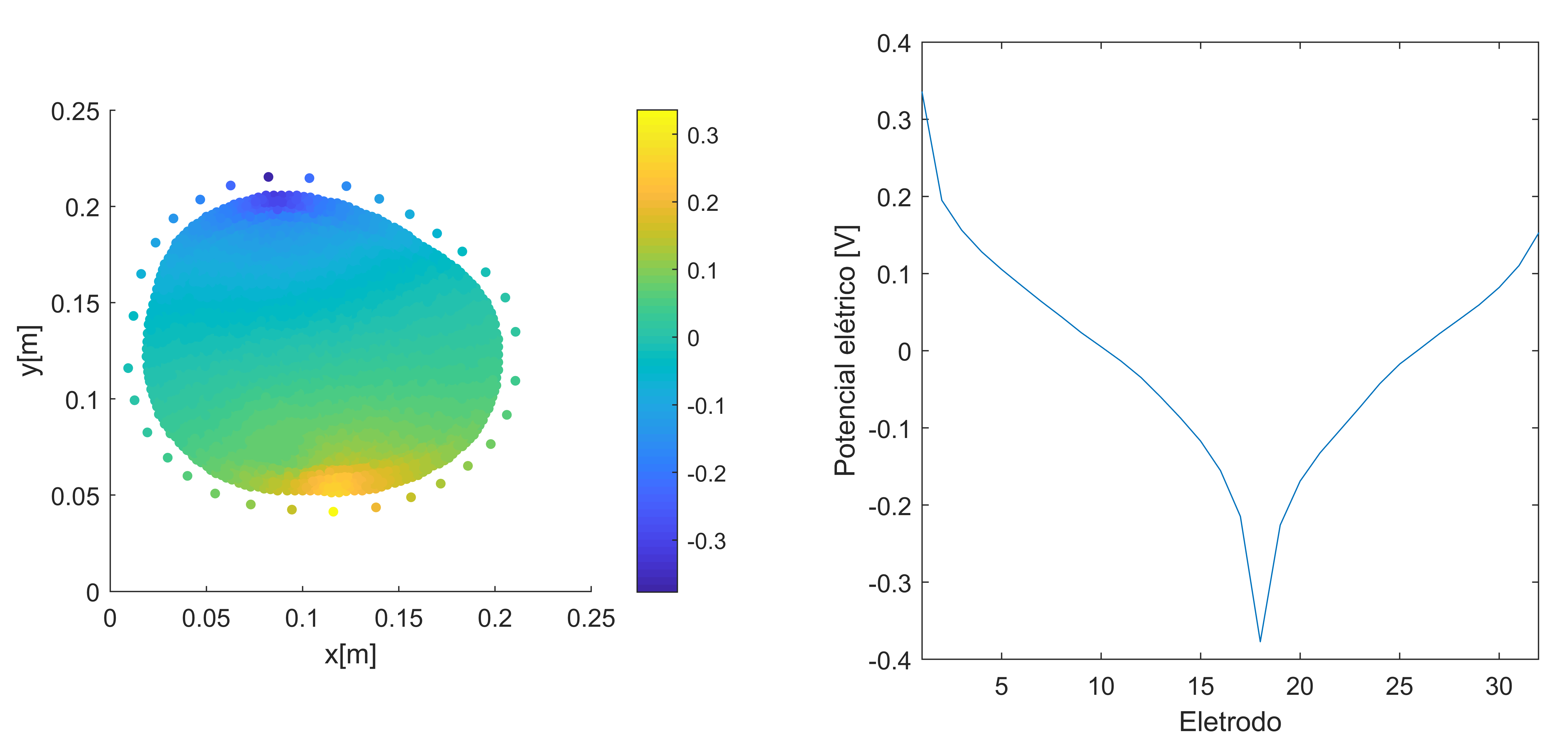}
        \caption[Tensões calculadas em um modelo de cabeça humana]{Tensões calculadas em um modelo de cabeça humana. Fonte: Próprio autor.}
        \label{fig:TIE2}
    \end{figure}

A Figura \ref{fig:TIE2} mostra o problema direto de EIT, o resultado de tensão elétrica em uma malha de elementos finitos após injeção de corrente elétrica pelos eletrodos. Na imagem de cima é mostrada uma malha de elementos finitos para representar a região de interesse (cabeça humana). Abaixo, são mostrados os potenciais elétricos dos nós da malha (esquerda) e nós dos eletrodos (direita). 

\subsection{$3^o$ Exemplo de problema inverso: Como visualizar estruturas biológicas em imagens borradas?}

\begin{figure}[H]
     \centering
   \begin{subfigure}[b]{0.49\textwidth}
         \centering
         \includegraphics[width=0.6\textwidth]{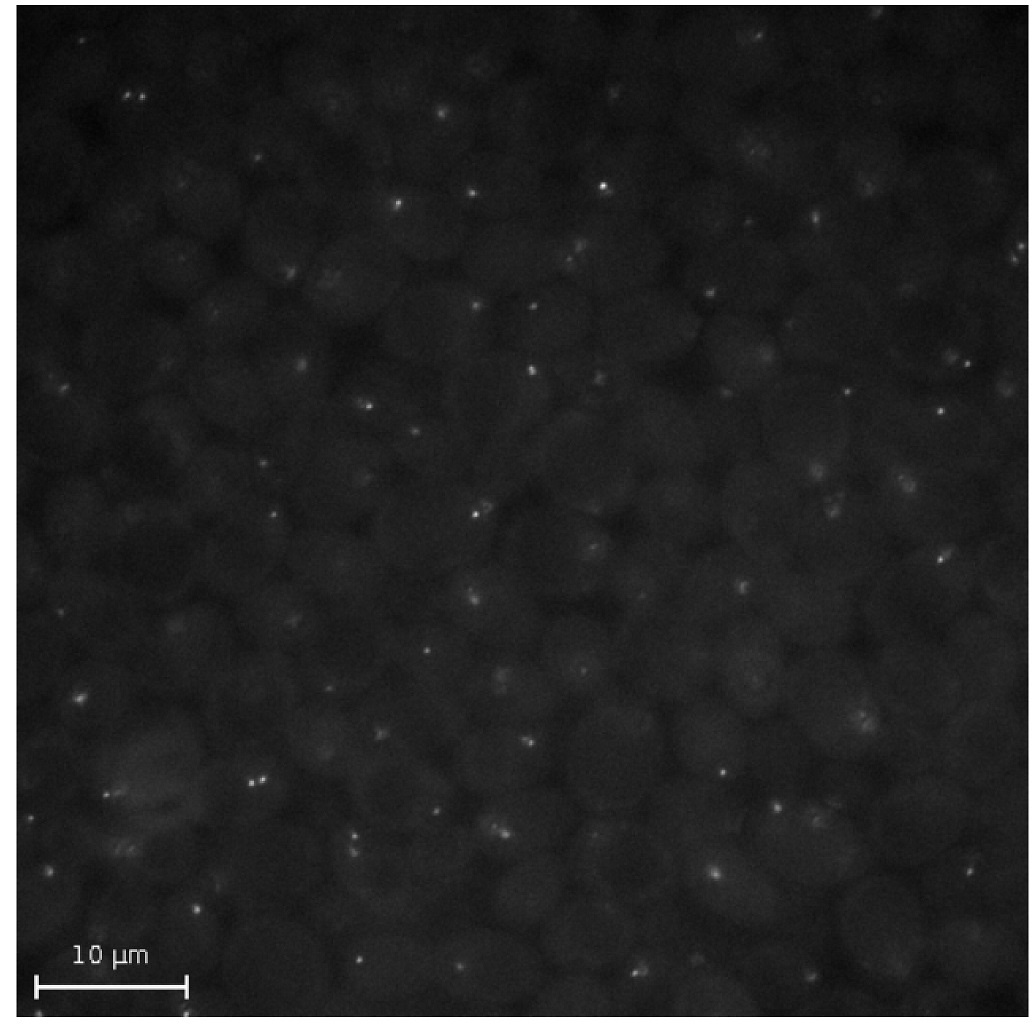}
         \caption{Fluorescência}
                  \label{fig:mf1a}
      \end{subfigure}
     \hfill
     \begin{subfigure}[b]{0.49\textwidth}
         \centering
                  \includegraphics[width=0.6\textwidth]{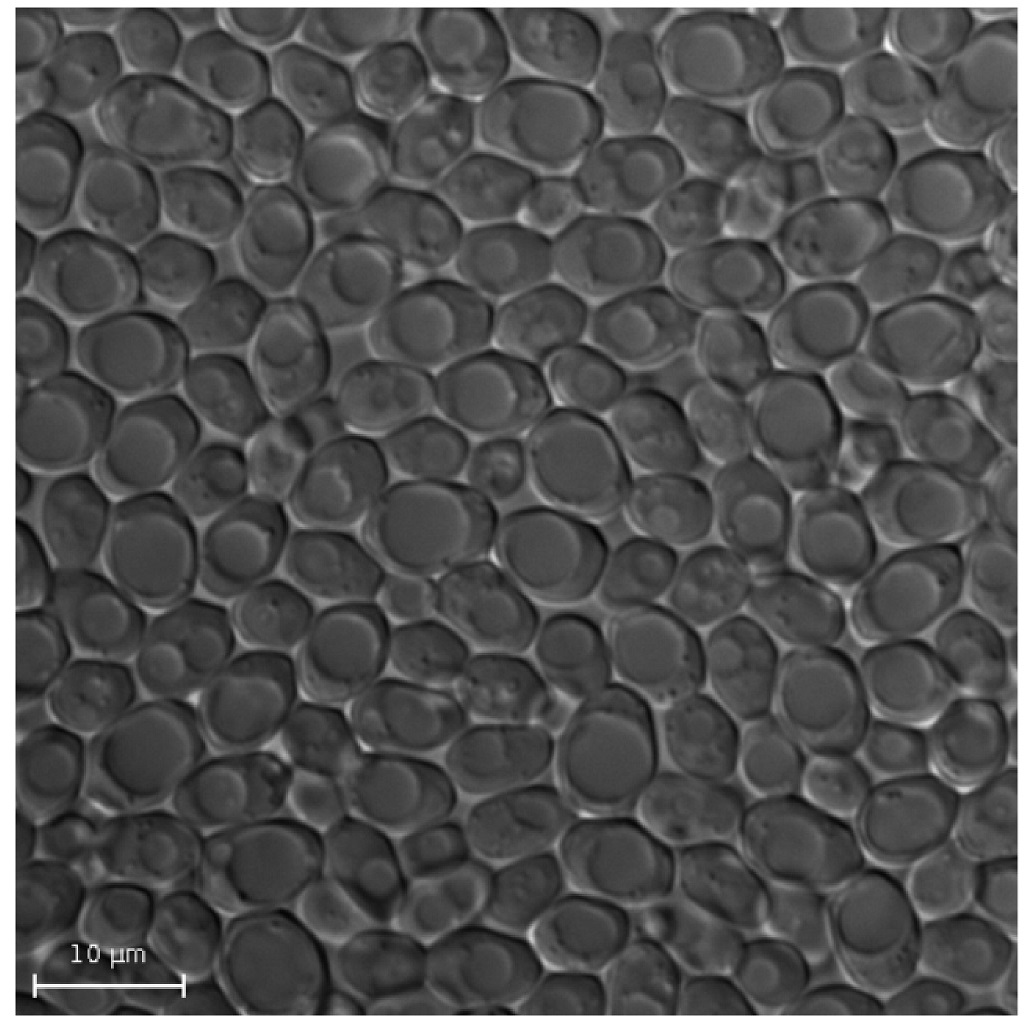}
         \caption{Contraste de interferência diferencial}
                  \label{fig:mf2b}
       \end{subfigure}
     \caption[Imagens microscópicas do repositório público de imagem YRC.]{Imagens microscópicas do repositório público de imagem YRC. Fonte: Retirado de \cite{Riffle2010}.}
     \label{fig:fluorescence}
 \end{figure}

A microscopia de fluorescência é atualmente uma ferramenta essencial na biologia molecular e celular, pois é uma importante fonte de informações de características celulares como distribuição, estrutura e função, e também uma ferramenta poderosa para aquisição de dados para análise de processos fisiológicos ao longo do tempo.  A Figura \ref{fig:mf1a} apresenta a imagem de fluorescência de \textit{Saccharomyces cerevisiae} com filtro ciano, onde são vistas proteínas SPC29 \cite{Riffle2010}. Ela traz a informação da localização das proteínas, mas não apresenta informações estruturais, que podem ser obtidas por microscopia de contraste de interferência diferencial conforme visto na Figura \ref{fig:mf2b}. Por tanto, a união dessas duas informações é de interesse\footnote{Imagens retiradas de \url{https://images.yeastrc.org/imagerepo/viewExperiment.do?id=844}.}.

Essa técnica é baseada no uso de fluoróforos, substâncias que se ligam a moléculas ou estruturas celulares específicas e emitem radiação fluorescente quando excitadas com fontes de luz apropriadas. Centenas de fluoróforos são conhecidos e eles permitem que apenas os alvos de interesse emitam luz e, portanto, sejam imageados (partindo do pressuposto de que a fluorescência de fundo é desprezível), o que permite a geração de imagens com alta especificidade e seletividade. A técnica também permite o estudo de células vivas intactas, moléculas individuais de interesse e interações intermoleculares ou intramoleculares. 

A qualidade de imagens pode ser deteriorada por motivos como reações fotoquímicas do fluoróforo, que incluem o fotobranqueamento, em que o fluoróforo se degrada e perde sua capacidade de emitir radiação fluorescente, esmaecendo a imagem resultante; e a fototoxicidade, em que a reação fotoquímica resulta em uma substância que danifica a amostra do biomaterial em análise \cite{kubitscheck2017fluorescence}. A qualidade da imagem também é limitada pelo equipamento utilizado, pelas leis da óptica e pelas características da amostra em análise e pelos fatores de degradação, que necessitam atenção especial, pois influenciam a interpretação correta da imagem por especialistas ou para posterior uso eficaz de outras ferramentas de análise, que incluem métodos de registro de imagem, segmentação e detecção de objetos. 

Um dos principais fatores de degradação que limitam a identificação de estruturas pequenas é que as imagens obtidas são borradas. A fonte pontual da luz não é detectada apenas em um pixel, ela se espalha em diversos pixels e em diversas camadas do objeto (que pode ser um conjunto de células, por exemplo). A falta de foco presente se relaciona com a baixa resolução espacial na microscopia padrão de fluorescência de campo amplo. Ela pode ser causada pelo fato de o objeto ser tridimensional, então informações de outros planos focais influenciam aquele que está sendo observado, além a presença de aberrações esféricas da lente, ou quando a interface entre a amostra e o meio de imersão apresenta índices de refração que se desviam dos valores nominais. 

O ruído é outro fator de degradação relevante. Limites de intensidade de luz e tempo de exposição, para não danificar a amostra, resultam em menor número de fótons adquiridos, diminuindo a relação sinal-ruído, potencializando artefatos causados por ruído de disparo \cite {Zhang2019}. Outras fontes de ruído incluem o ruído de corrente escura, ruído de leitura, ruído de fundo \cite{kubitscheck2017fluorescence} e ruído térmico, que também prejudicar a visualização. 

Em suma, na microscopia de fluorescência é necessário superar a falta de foco e a presença de ruído de Poisson \cite{Bertero2009}, principalmente quando não se utilizam equipamentos mais caros e especializados. Caso eles não estejam disponíveis, a alternativa se torna utilizar técnicas de processamento de imagem com tal objetivo \cite{Belthangady2019}.

\subsection{Quais são as diferenças entre problemas bem-postos e mal-postos?}\label{sec:illposedprob}

Em áreas como na engenharia ou física, um modelo é obtido quando se quer representar o funcionamento de um sistema físico em termos matemáticos, usualmente incluindo parâmetros e constantes físicas. Uma vez que o modelo é obtido, é necessário também resolvê-lo. Considerando que a entrada do sistema seja conhecida e que os parâmetros do sistema são conhecidos, um objetivo pode ser calcular a saída do sistema. Há também o caminho inverso, obtendo os parâmetros ou as entradas a partir dos dados de saída. Entre as duas direções, será que alguma é mais difícil de resolver? Para responder, é necessário distinguir problemas bem-postos e mal-postos.

A origem da área de problemas inversos se deu na matemática, de modo que é necessário rigor em cada definição que descreve o problema. O foco do presente capítulo não é definir esses espaços para uma aplicação específica, mas sim o de descrever o problema geral da área de problemas inversos mal-postos. No entanto, é necessário dizer que na área de análise funcional, ou mesmo da matemática aplicada, a descrição em termos de espaços \cite{Shima2016} e operadores \cite[Seção 6.24]{tarantola2005inverse} é necessária para formalizar a declaração do problema e o cenário matemático.

No início do século XX \cite{1902hadamard}, Hadamard estudou equações de operadores conforme
\begin{equation}
\mathcal{A} f = g,
\label{eq:canonical}
\end{equation}
onde $\mathcal{A}$ é um operador contínuo e injetivo que mapeia elementos $f$ do espaço $F$ para elementos $g$ do espaço $G$. 

Ele descobriu que existiam equações desse tipo em que o operador inverso $\mathcal{A}^{-1}$ era descontínuo e pequenas variações  $g + \Delta g$ no lado direito da equação poderiam causar uma grande variação na solução $f$ \cite[Subseção 1.3.1]{Vapnik2006}. Em outras palavras, a operação inversa era instável e não era possível resolver através de $f = \mathcal{A}^{-1} g$ \cite{Bell1978}. 

Então, Hadamard definiu problemas bem-postos \cite{1902hadamard}, que em uma interpretação atual obedece às seguintes condições \cite[pág. 2]{hansen2010discrete}, \cite[pág. 1]{Lavrentiev1967}, \cite[págs. 7-8]{tikhonov1977solutions}:
\begin{itemize}
\item \textbf{Existência}: A solução da Equação \eqref{eq:canonical} existe, $ \forall g \in G$ os dados admissíveis; 
\item \textbf{Unicidade}: A solução da Equação \eqref{eq:canonical} é única em $F$, $ \forall g \in G$ os dados admissíveis; 
\item \textbf{Estabilidade}: A solução depende continuamente dos dados $g$. 
\end{itemize}

Em relação aos espaços $F$ e $G$, sua definição depende do problema. Em problemas inversos, eles incluem espaços métricos \cite[págs. 27-8]{tikhonov1977solutions}, espaços de Hilbert separáveis \cite[pág. 7]{kaipio2005statistical}, espaços de Banach \cite[pág. 27]{bleyer2015novel} ou espaços de funções \cite[pág. 1]{Nair2009}.

Assim, para que o problema seja bem-posto, além de $\mathcal{A}$ ser bijetivo, o operador inverso $\mathcal{A}^{-1}$ deve ser contínuo \cite[Definição 1]{Chen2002}, \cite[Subseção 3.3]{Mueller2012}. Se, pelo menos, uma das condições de Hadamard não é respeitada, o problema é mal-posto. Considerando que $f$ seja a causa e $g$ a consequência, a característica mal-posta pode ser entendida devido à não-causalidade da inversão, já que isso aumentaria a entropia, havendo perda de informação no sistema (não-reversível), pela segunda lei da termodinâmica \cite[pág. 2]{Calvetti2018a}.

Hadamard considerou que esses problemas não corresponderiam à realidade física \cite[pág. 2]{hansen2010discrete}, visão que persistiu por décadas na matemática \cite[Subseção 3.6.2]{Courant1989}. Mais do que resolver um problema matemático abstrato, hoje se sabe que há diversas aplicações práticas possíveis a partir da solução de problemas mal-postos. Em \cite{Kabanikhin2008}, discute-se diversos exemplos de problemas mal-postos em cálculo, geometria, equações diferenciais e integrais, para citar alguns.
 
Mesmo para problemas bem-postos e mal-postos existem outras definições. Em \cite[Capítulos 5 e 6]{Serovaiskii2003} e em \cite[Capítulos 1 e 2]{Dontchev1993}, os autores diferenciam o que é um problema bem-posto no sentido de Hadamard de um problema bem-posto no sentido de Tikhonov. Já em \cite[Subseção 1.5]{Bunge2019}, são descritas regras de formação para que um problema seja bem definido. Via de regra, é importante que as definições utilizadas estejam claras para o leitor, evitando assim ambiguidades.

\subsection{Quais são as diferenças entre problemas diretos e inversos?}\label{sec:forwardp}

Na subseção anterior, o objetivo era, de modo simplificado, definir um problema que seria possível de ser resolvido. Para o caso de um problema específico em que há um modelo disponível, é também importante definir a direção do problema que se quer resolver. Assim, é necessário diferenciar problemas diretos de problemas inversos\footnote{Essa discussão é ampliada na Subseção \ref{sec:poincare}.}. 

Retomando a Equação \eqref{eq:canonical}, $g$ representa dados medidos, $f$ são os parâmetros do modelo e $\mathcal{A}$ é um modelo físico-matemático que relaciona essas duas grandezas. Problemas diretos são baseados na modelagem matemática de um operador direto $\mathcal{A}$ \cite[Seção 1]{Arridge2019}, que busca prever o comportamento do sistema tanto nos casos em que há dados experimentais (verificação dos modelos) quanto nos casos em que ainda não há. Logo, o problema direto seria a obtenção de $g$ quando se tem $\mathcal{A}$ e $f$ disponíveis \cite[pág. 5]{Hansen1998}. É razoável requerer que as mesmas causas produzam as mesmas consequências. Logo, muitos problemas diretos da física matemática são bem-postos. Para ilustrar, esse é o caso quando se deseja estimar um estado futuro de um sistema, a partir do conhecimento do estado atual dele e das leis físicas que regem o problema \cite[pág. 3]{engl1996regularization}. 

Os problemas inversos são aqueles na qual se busca obter $f$ (causa) quando se tem $\mathcal{A}$ (modelo) e $g$ (consequência) disponíveis. Logo, sempre deve existir um problema direto correspondente a um problema inverso. Assim, resolver problemas inversos pode significar:
\begin{itemize}
\item Obter a quantidade de interesse $f$ quando só estão disponíveis medidas indiretas $g$  \cite[pág. 1]{kaipio2005statistical}, o que inclui obter parâmetros do interior de um sistema a partir de medidas obtidas em seu contorno;
\item Tornar  o invisível visivel \cite{Uhlmann2014};
\item Resolver um problema inverso como buscar a causa de um dado efeito \cite{baumeister2005topics};
\item Buscar a resposta para uma pergunta que não se conhece \cite{Keller1976};
\item Obter estados anteriores de um sistema a partir do estado atual, o que ressalta a sua característica de não-causalidade \cite[págs. 1-2]{kaipio2005statistical}

\end{itemize}
É importante ainda destacar que nem todo problema inverso é mal-posto, e vice-versa \cite[pág. 4]{Mueller2012}, \cite[Subseção 1.3]{tikhonov1977solutions}.

\subsection{Quais problemas inversos podem ser descritos pela equação integral de Fredholm de primeiro tipo?}\label{sec:fredh}

Diversos problemas práticos podem ser tratados como problemas inversos mal-postos e uma lista pode ser encontrada em \cite[págs. 1-6]{bleyer2015novel}. Vários deles são dados por equações integrais de Fredholm de primeiro tipo \cite[pág. 7]{hansen2010discrete}, \cite[pág. 41]{Mueller2012}. Seja a equação integral no intervalo arbitrário $a\leq t \leq b$. Assim, eles podem ser descritos conforme 
\begin{equation}
\int_a^b k(t,s) f(s) ds = g(t),
\label{eq:fred}
\end{equation}
onde $f(s)$ e $g(t)$ são funções e $k(t,s)$ é um \textit{kernel} que descreve a relação linear entre elas. 

Este tipo de equação integral pode descrever diversos modelos lineares de engenharia:
\begin{itemize}
\item A influência do instrumento na medição \cite[pág. 75]{aster2019parameter}, representada pela convolução da resposta ao impulso do sistema $h(t-s)$ com o sinal $f(s)$, isto é, 
\begin{equation}
\int_{-\infty}^{\infty} h(t-s) f(s) ds = g(t).
\label{eq:deco}
\end{equation}
De fato, é possível mostrar que sistemas lineares e invariantes no tempo podem  ser expressos como convoluções \cite[págs. 211-2]{aster2019parameter};

\item A transformada de Laplace de um sinal \cite[pág. 133]{Mueller2012};
\begin{equation}
\mathcal{L}\{f(t)\}(s)=\int_{0}^{\infty}f(t)e^{-st} dt;
\label{eq:laplace}
\end{equation}
\item Prospecções geofísicas, como levantamentos gravimétricos, onde a massa de uma região é relacionada com a aceleração gravitacional daquele local \cite[pág. 5]{hansen2010discrete}.
\end{itemize}
A obtenção e desenvolvimento de $k(t,s)$ é a modelagem matemática em si do fenômeno físico. A avaliação da integral da Equação \eqref{eq:fred}, estimar a saída $g(t)$ conhecendo-se tanto $k(t,s)$ e $f(s)$ é chamado de problema direto e respeita a relação de que as causas precedem os efeitos, a direção da causalidade \cite[pág. 2]{Calvetti2018a}, \cite[pág. 1]{kaipio2005statistical}. 

Na Equação \eqref{eq:fred}, o problema direto pode ser escrito em termos de um operador direto $\mathcal{A}$, que realiza um mapeamento entre espaços de funções \cite[pág. 8]{Montegranario2014},
\begin{equation}
\mathcal{A} f = \int_a^b k(t,s) f(s) ds, 
\label{eq:fred2}
\end{equation}
onde $\mathcal{A}$ define um operador linear \cite[pág. 7]{hansen2010discrete}, \cite[pág. 41]{Mueller2012}, o que deixa clara a relação das Equações \eqref{eq:canonical} e \eqref{eq:fred2}. Retomando a Equação \eqref{eq:fred}, tanto o problema direto quanto o inverso partem dessa mesma equação, a diferença é o que se quer calcular: 
\begin{itemize}

\item Realizar a deconvolução significa reconstruir $f(s)$, sem a influência do instrumento de medição, a partir de $g(t)$ e de $k(t,s) = h(t-s)$;
\item Reconstruir o sinal original $f(s)$ a partir de amostras da sua transformada de Laplace $g(t)$, ou seja, calcular a transformada inversa de Laplace \cite[pág. 133]{Mueller2012};
\item No caso de levantamentos gravimétricos, deseja-se reconstruir a densidade $f(s)$ de uma distribuição de massas de uma região, da qual não se tem acesso diretamente, a partir da aceleração da gravidade associada $g(t)$, que pode ser medida.

\item A propagação de calor por um material ao longo do tempo é modelada pela equação do calor, uma equação diferencial parcial (PDE) \cite[Subseção 2.2]{Mueller2012}. No problema conhecido como \textit{backward heat equation}, de calcular a distribuição de temperaturas em um instante de tempo anterior ao atual, sua solução é dada por uma equação de Fredholm de primeiro tipo \cite[págs. 36-7]{Mueller2012}; 

\end{itemize}

Problemas inversos dados por equações integrais de Fredholm do primeiro tipo, com \textit{kernel} de Hilbert-Schmidt são mal-postos \cite[págs. 40-1]{Mueller2012}. O lema de Riemann–Lebesgue explica a instabilidade de problemas inversos originados pela Equação \eqref{eq:fred}: Componentes de alta frequência de $f$ são amortecidas no problema direto, de modo que $g$ é mais suave que $f$, implicando que no problema inverso haverá a amplificação da perturbação se ela tiver componente de alta frequência \cite[pág. 8]{hansen2010discrete}.

\subsection{É possível resolver problemas inversos com computadores?}
Problemas inversos podem apresentar operadores diretos com dimensão infinita, bem como a necessidade de estimação de infinitos parâmetros desse modelo \cite[págs. 49-50]{Neto2005}. O problema direto pode ser modelado, por exemplo, por uma PDE ou por equações integrais. Para alguns modelos é possível deduzir expressões analíticas que descrevem a resposta do sistema, mas, quando isso não é possível, são necessários métodos numéricos e ferramentas computacionais com tal objetivo. 

Para resolver tanto o problema direto quanto o problema inverso com o uso de computadores, é necessário discretizar $\mathcal{A}$ e $f$, considerando que as amostras das medidas $g$ já sejam discretas também, passando a possuir dimensão finita \cite[pág. 23]{hansen2010discrete}. Dessa forma, a PDE é substituída por uma equação algébrica. Caso a PDE seja linear, em geral a discretização resulta em um sistema de equações algébricas lineares, e caso a PDE seja não-linear, o sistema é não-linear \cite[pág. 310]{pinchover2005introduction}. 

Computacionalmente, o resultado da discretização desses modelos pode ser representado por uma matriz $\mathbf{A}$ com dimensões finitas. Seja também $\mathbf{y}$ o vetor das medidas e $\mathbf{x}$ o vetor de parâmetros que se deseja reconstruir, ambos com dimensões finitas. Problemas inversos não-lineares apresentam a forma geral \cite[pág. 257]{aster2019parameter} 
\begin{equation}
\mathbf{y} = \mathbf{A}( \mathbf{x} ).
\label{eq:eq111}
\end{equation}
Alguns exemplos de aplicações a partir da Equação \eqref{eq:eq111} incluem a discretização de uma PDE, como a equação de Laplace em  tomografia por impedância elétrica \cite[pág. 165]{Mueller2012}, alguns efeitos não-lineares da tomografia computadorizada (CT) \cite[Seção 6.1.2]{Woo2012} e problemas sísmicos \cite{Adler2021}. A solução de problemas inversos não-lineares não apresenta um \textit{framework} único, sendo necessárias ferramentas específicas para cada aplicação na sua solução \cite[pág. xi]{Mueller2012}.

\begin{figure}[H]
\begin{centering}
\includegraphics[width = 0.6\textwidth]{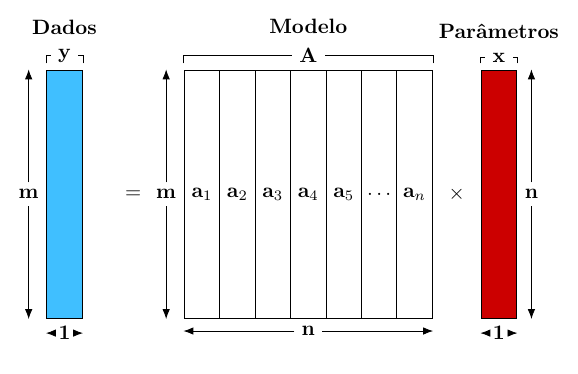}
\caption[Ilustração de um problema de medidas indiretas.]{Ilustração de um problema de medidas indiretas. Fonte: Próprio autor.}
\label{fig:my_label}
\end{centering}
\end{figure}

Neste livro, também serão discutidos problemas inversos lineares e discretos, dados por sistemas de equações lineares na forma matricial conforme
\begin{equation}
\underset{m \times 1}{\mathbf{y}} \quad = \quad \underset{m \times n}{\mathbf{A}} \quad \underset{n \times 1}{\mathbf{x}},
\label{eq:eq1}
\end{equation}
onde $m$ e $n$ são dimensões (que definem o tamanho dos vetores e das matrizes), $\mathbf{x}$ é um vetor de parâmetros, $\mathbf{y}$ é um vetor de medidas e a matriz $\mathbf{A}$ representa a transformação linear do espaço do modelo para o espaço dos dados, um modelo derivado de leis físicas que relaciona as grandezas. Esse sistema é ilustrado na Figura \ref{fig:my_label}. 

Em alguns exemplos de aplicações, quando o operador direto é discretizado, $\mathbf{A}$ pode representar: a convolução de um sinal com a resposta ao impulso de um sistema \cite[Subseção 2.1]{Mueller2012}; em microscopia de fluorescência, a convolução com a função de espalhamento pontual relativa às lentes do microscópio; a discretização da transformada de Radon \cite[Subseção 2.3]{Mueller2012} utilizada em CT; discretização da transformada de Laplace \cite[Subseção 10.2]{Mueller2012}; e assim por diante.

Seja no caso linear ou no caso não-linear, a obtenção de $\mathbf{x}$ a partir do conhecimento de  $\mathbf{A}$ e $\mathbf{y}$ é chamada de estimação de parâmetros. Já a obtenção de $\mathbf{A}$ a partir do conhecimento de  $\mathbf{x}$ e $\mathbf{y}$ é chamada de identificação do sistema \cite[pág. 2]{aster2019parameter}.

\subsection{Quais são os fatores de discrepâncias entre dados e o modelo discretizado?}

Antes de discutir soluções para problemas mal-postos discretos, é necessário avaliar o modelo computacional que será utilizado. Quando possível, busca-se garantir que o problema direto seja tão acurado quanto possível, tendo claras quais são as condições de contorno e as hipóteses simplificadoras utilizadas. Por outro lado, em muitos casos práticos, tanto $\mathbf{A}$ quanto $\mathbf{y}$ são conhecidos apenas aproximadamente. 

\subsubsection{A aproximação do modelo computacional é suficiente?}
Idealmente, qualquer vetor de saída $\mathbf{y}$ calculado a partir da Equação \eqref{eq:eq1} deve ser uma combinação linear das colunas $\mathbf{a}_i$, $i = 1\cdots n$, da matriz $\mathbf{A}$
\begin{equation}
\begin{aligned}
\mathbf{y}_{calculado} & = \hspace{1mm}\mathbf{A}_{calculada} \mathbf{x} \\
& =  \begin{pmatrix}
\vert & \vert & & \vert\\
\mathbf{a}_1 & \mathbf{a}_2 & \cdots & \mathbf{a}_n\\
\vert & \vert & & \vert\\
\end{pmatrix} 
\begin{pmatrix}
x_1 \\
x_2 \\
\vdots\\
x_n \\
\end{pmatrix} \\
& =  \hspace{1mm} x_1 \mathbf{a}_1 + x_2 \mathbf{a}_2 + \cdots + x_n \mathbf{a}_n.
\label{eq:eq2}
\end{aligned}
\end{equation}
Estes são os possíveis valores de saída que o modelo é capaz de representar ao resolver o problema direto. Em termos da álgebra linear, o espaço coluna de $\mathbf{A}$ é o conjunto de todos os vetores de saída $\mathbf{y}$ de tal forma que $\mathbf{A} \mathbf{x} = \mathbf{y}$ possui solução \cite[pág. 64-5]{golub2013matrix}, mas não há garantias de que medidas $\mathbf{y}_{medido}$ estejam no espaço coluna de $\mathbf{A}$, pois isso seria admitir que o modelo matemático é perfeito. Para Tikhonov \cite[pág. 6]{tikhonov1977solutions}, essas situações poderiam ser traduzidas como uma diferença finita entre o modelo e a realidade como
\begin{equation}
\vert \vert \mathbf{A}_{exata} - \mathbf{A}_{calculada} \vert \vert^2_F \leq \bm{\epsilon},
\label{eq:sistema2}
\end{equation}
onde $\bm{\epsilon}$ representa erros associados ao modelo. A norma matricial de Frobenius foi utilizada por ser uma operação com matrizes \cite[pág. 71]{golub2013matrix}. Caso o valor de $\bm{\epsilon}$ fosse conhecido, essa informação poderia ser incluída na solução, mas sua estimação não é simples.

Consequentemente, o resultado $\mathbf{y}_{calculado}$ é apenas uma aproximação do dado medido
\begin{equation}
\mathbf{y}_{calculado} \approx \mathbf{y}_{medido}.
\label{eq:eq6}
\end{equation}
Apesar da diferença entre modelo e realidade ilustrada por $\bm{\epsilon}$, muitas vezes se parte da hipótese do conhecimento exato do modelo, sem nenhuma atualização de  $\mathbf{A}$ na solução. 

\subsubsection{Qual é o efeito que ruídos exercem no modelo aproximado?}
Outros fatores podem tirar o vetor $\mathbf{y}_{medido}$ do espaço coluna do sistema $\mathbf{A}$. Mesmo que a dedução do modelo seja adequada, há questões relativas à sua representação em um computador. Dada a necessidade de discretização do modelo contínuo, há erros associados ao arredondamento e truncamento dos valores dentro da representação numérica com um número limitado de dígitos. Além disso, o teste decisivo para o modelo é a comparação dos seus resultados com medidas experimentais, onde há a presença de ruídos, sistemáticos ou aleatórios. Tudo isso implica no conhecimento aproximado de $\mathbf{y}$ \cite[pág. 6]{tikhonov1977solutions}, 
\begin{equation}
\label{eq:sistema3}
\vert \vert  \mathbf{y}_{exato} - \mathbf{y}_{medido} \vert \vert^2_2 \leq \bm{\delta},
\end{equation}
onde $\bm{\delta}$ representa ruídos\footnote{Termo também encontrado como perturbações ou erros.}  associados às medidas\footnote{Neste capítulo, $\bm{\delta}$ não se refere a nenhum tipo de ruído específico, exceto se indicado.}. A matriz $\mathbf{A}_{calculada}$ será daqui para frente chamada apenas de $\mathbf{A}$ e o vetor $\mathbf{y}_{medido}$ de $\mathbf{y}$, exceto quando necessário.

Nenhum instrumento de medição possui resolução infinita, há incertezas inerentes ao processo de medição, da conversão analógico-digital e da precisão numérica finita. O resultado é traduzido em uma medida ruidosa $\mathbf{y}_{\delta}$, conforme
\begin{equation}
\mathbf{A} \mathbf{x} \approx \mathbf{y}_{\delta},
\label{eq:eq7x}
\end{equation}
onde $\bm{\delta}$ é um ruído intrínseco à medida, que pode ser considerado uma realização fixa de um processo estocástico \cite[pág. 5]{Mueller2012}. No caso de um ruído aditivo\footnote{Dependendo da referência, o ruído aditivo $\bm{\delta}$ é escrito do lado esquerdo do sistema, ao invés do lado direito como na Equação \eqref{eq:eq7}. Uma vantagem em escrever do lado direito da equação é que isso facilita na hora de ilustrar os efeitos de $\bm{\delta}$ na solução ingênua, conforme Equação \eqref{eq:caso1_sol3}.  }, essa relação se torna
\begin{equation}
\mathbf{A} \mathbf{x} \approx \mathbf{y}_{exato} + \bm{\delta},
\label{eq:eq7}
\end{equation}
onde $\mathbf{y}_{\delta} = \mathbf{y}_{exato} + \bm{\delta}$. Em algumas aplicações, é de conhecimento que $\bm{\delta}$ pode ser considerado branco, aditivo e gaussiano, mas também há casos quando ele apresenta outras características, por exemplo quando ele possui outra cor \cite[pág. 43]{hansen2010discrete} ou quando ele é dependente do sinal, como o ruído de Poisson \cite[págs. 44-5]{hansen2010discrete}. 

Em dados simulados, o ruído é definido e incluído pelo pesquisador, mas em dados experimentais nem sempre ele pode ser aproximado por um ruído branco e gaussiano. Inclusive, ele pode ter natureza desconhecida.  Assim, é importante buscar informações sobre a instrumentação utilizada para captura do sinal medido e incluir a informação do nível e do tipo do ruído, tornando-o mais próximo da realidade dos dados disponíveis \cite[pág. 163]{hansen2010discrete}, o que pode impactar nos resultados obtidos a partir de tais dados.

\subsubsection{A simples inversão de $\mathbf{A}$ é suficiente para obter soluções adequadas?}
O sistema $\mathbf{A} \mathbf{x} = \mathbf{y}$ terá solução única se, e somente se, $\mathbf{A}^{-1}$ existir, o que ocorre quando a matriz $\mathbf{A}$ não é singular. A primeira abordagem seria a solução ingênua de inverter diretamente o operador $\mathbf{A}$ \cite[pág. 30]{hansen2010discrete}. Considerando que $\mathbf{A}$ seja quadrada e que o ruído seja aditivo, a solução ingênua é dada por
\begin{equation}
\begin{aligned}
\hat{\mathbf{x}} & \approx  \mathbf{A}^{-1}\left(\mathbf{y}+\bm{\delta}\right) \\
 &  \approx \mathbf{A}^{-1}\mathbf{y} + \mathbf{A}^{-1} \bm{\delta},
\label{eq:caso1_sol3}
\end{aligned}
\end{equation} 
onde $\mathbf{A}^{-1}$ é a matriz inversa\footnote{Conhecida como \textit{two sided inverse}, pois se ela existir terá a propriedade que $\mathbf{A}^{-1} \mathbf{A} = \mathbf{I} = \mathbf{A} \mathbf{A}^{-1} $} de $\mathbf{A}$. Doravante o subscrito em $\mathbf{y}_{\delta}$ será omitido, exceto quando seja necessário destacá-lo. 

Quando um problema inverso mal-posto é discretizado, o resultado é um sistema mal condicionado. Isso é mostrado, por exemplo, para o caso de uma equação de Fredholm de primeiro tipo \cite[Seção 3.3]{hansen2010discrete}. Logo, a Equação \eqref{eq:caso1_sol3} mostra a instabilidade na solução de problemas inversos mal-postos, pois $\mathbf{A}$ é mal condicionada, ou quase singular, e haverá a amplificação dos ruídos. Caso $\mathbf{A}$ fosse retangular, esse também seria o caso utilizando-se a inversa generalizada de Moore-Penrose. Neste caso, a Equação \eqref{eq:eq1} não possui uma solução clássica e é necessário definir uma solução aproximada, desde que ela seja estável para pequenas variações nos dados iniciais  $(\mathbf{A}, \mathbf{y})$.

\subsubsection{Em problemas simulados, o que são crimes de inversão?}

Suponha que os dados $\mathbf{y}$ sejam obtidos utilizando  simulação computacional,  através da Equação \eqref{eq:eq1}, tendo conhecimento tanto de $\mathbf{A}$ quanto de $\mathbf{x}$. Na ausência de ruídos nas medidas, a reconstrução é perfeita mesmo com a solução ingênua da Equação \eqref{eq:caso1_sol3}, sem regularização. Isso só é possível pela perfeita concordância entre o modelo e os dados. 

Nesse contexto, deve-se ter o cuidado para evitar os crimes de inversão, que acontecem quando se tenta resolver o problema inverso utilizando o mesmo modelo que gerou os dados simulados \cite[pág. 139]{hansen2010discrete}, \cite[pág. 5]{kaipio2005statistical}. O crime de inversão ajuda artificialmente na solução, podendo indicar resultados melhores do que o são. 

Em um primeiro momento, é vantajoso testar o algoritmo proposto utilizando o mesmo modelo $\mathbf{A}$ nos problemas direto e inverso, apenas avaliando os efeitos dos ruídos $\bm{\delta}$ no problema inverso mal-posto. Se o algoritmo não funcionar assim, ele não funcionará em nenhuma outra circunstância. Em seguida, deve-se resolver o problema sem cometer crime de inversão, para verificar se o algoritmo é robusto \cite[pág. 163]{hansen2010discrete}. Nesse sentido, não há crime de inversão quando se utilizam dados experimentais, os dados de interesse de fato.

\subsubsection{Durante a solução, a quantidade de dados disponíveis é adequada, incompleta ou em excesso?}\label{sec:excess}
Não há garantia de que o número de medidas disponíveis (que resulta no número de equações do modelo) seja igual ao número de incógnitas (que a matriz $\mathbf{A}$ seja quadrada):
\begin{itemize}
\item Sistemas subdeterminados, com mais parâmetros do que medidas (mais incógnitas do que equações), gerando uma matriz $\mathbf{A}$ com mais colunas do que linhas;
\item Sistemas sobredeterminados, com mais medidas do que parâmetros (mais equações do que incógnitas), gerando uma matriz $\mathbf{A}$ com mais linhas do que colunas.
\end{itemize} 
Isso pode ser entendido tanto do ponto de vista da disponibilidade de dados quanto da discretização do modelo. Uma discretização pequena pode resultar em problemas sobredeterminados. Uma discretização em excesso pode até ser beneficial na solução do problema direto, mas pode também levar a um sistema muito subdeterminado, pois a quantidade de dados disponível seria a mesma. 

Em todo caso, a inversão de matriz é definida apenas para matrizes quadradas. Em vez de se buscar a relação de igualdade entre os lados de um sistema $\mathbf{A} \mathbf{x} = \mathbf{y}$ inconsistente (sem solução), como nos casos de sistemas sobredeterminados \cite[pág. 46]{goodfellow2016deep}, pode-se buscar  a solução que minimize a soma do quadrado entre o termo da medida e do modelo, o método dos mínimos quadrados \cite[pág. 260]{golub2013matrix}, 
\begin{equation}
\hat{\mathbf{x}} = \arg\min\limits_{\mathbf{x}}  \frac{1}{2} \vert \vert \mathbf{A}\mathbf{x} - \mathbf{y} \vert \vert^2_2 ,
\label{eq:otimizacao1}
\end{equation}
onde $\hat{\mathbf{x}}$ é a solução calculada. Ressalta-se que uma breve revisão sobre normas de vetores e a notação utilizada neste trabalho é encontrada no Apêndice \ref{Ap:normas1}.

Pode-se mostrar (Apêndice \ref{Ap:sobredet}) que a solução da Equação \eqref{eq:otimizacao1} é dada por 
\begin{equation}
\hat{\mathbf{x}} = \left(\mathbf{A}^T \mathbf{A} \right)^{-1} \mathbf{A}^T \mathbf{y},
\label{eq:normalequation}
\end{equation}
conforme \cite[pág. 260]{golub2013matrix}. Sua forma compacta é $\hat{\mathbf{x}} =\mathbf{A}^+_e \mathbf{y}$, onde $\mathbf{A}^+_e = \left(\mathbf{A}^T \mathbf{A} \right)^{-1} \mathbf{A}^T$ é uma matriz inversa à esquerda, pois $\mathbf{A}^{+}_e \mathbf{A}=\mathbf{I}$, uma matriz pseudoinversa. Um exemplo de interpolação que utiliza a Equação \eqref{eq:normalequation} é discutida no Apêndice \ref{Ap:pseudo}.

Uma forma para lidar com um sistema que tem infinitas soluções (como em sistemas subdeterminados, que não são necessariamente inconsistentes) é a de buscar a solução de norma mínima \cite[pág. 62]{calvetti2007introduction}, \cite[pág. 46]{goodfellow2016deep}, uma otimização com restrição dado por
\begin{equation}
\hat{\mathbf{x}} = \arg\min\limits_{\mathbf{x}}\vert \vert \mathbf{x} \vert \vert_2^2 \hspace{3mm} \text{s.t.}\hspace{3mm} \mathbf{A} \mathbf{x}= \mathbf{y},
\label{eq:casosub_sol}
\end{equation}
cuja solução é dada por \cite[Seção 2.3]{Monticelli2012}
\begin{equation}
\hat{\mathbf{x}} = \mathbf{A}^T \left(\mathbf{A} \mathbf{A}^T\right)^{-1} \mathbf{y} 
\label{eq:normalequation3}
\end{equation}
cuja forma compacta é $\hat{\mathbf{x}} = \mathbf{A}^+_d \mathbf{y}$, onde $\mathbf{A}^+_d = \mathbf{A}^T \left(\mathbf{A} \mathbf{A}^T\right)^{-1}$ é a matriz inversa à direita, pois $\mathbf{A}\mathbf{A}^{+}_d=\mathbf{I}$, outra matriz pseudoinversa. No Apêndice \ref{Ap:sobredet} há a derivação de \eqref{eq:normalequation3}.

Quando $\mathbf{A}$ é mal condicionada, $\mathbf{A}^{-1}$ pode ser obtida numericamente, mas $\mathbf{A}^{-1}\mathbf{y}$ amplifica  ruídos. Quando $\mathbf{A}$ é singular, tanto $\mathbf{A}^T \mathbf{A}$ quanto $\mathbf{A} \mathbf{A}^T$ também são singulares e a utilização direta das matrizes pseudoinversas também não é adequada.  

\subsection{O que é a decomposição em valores singulares e qual sua importância para problemas inversos lineares?}\label{app-svd}

Outra forma de analisar problemas inversos é através da decomposição em valores singulares (SVD) de seu operador direto $\mathbf{A}$. Problemas inversos lineares são similares, mesmo em aplicações diferentes, no sentido que eles são completamente descritos pela expansão em valores singulares de $\mathcal{A}$ \cite[pág. xi] {Mueller2012}, retomando a notação da Equação \eqref{eq:canonical}, e compartilham métodos de solução \cite[pág. 10]{aster2019parameter}. Com a discretização de um problema linear, é realizada a decomposição em valores singulares\footnote{Que é o equivalente discreto da expansão em valores singulares.} da matriz $\mathbf{A}$, pois os valores singulares obtidos pela SVD são aproximações adequadas e podem ser utilizados na sua análise \cite[Subseção 3.3]{hansen2010discrete}. 

A SVD é uma técnica de fatoração de matrizes reais ou complexas, uma generalização da decomposição em autovalores e autovetores, que pode ser utilizada tanto para matrizes quadradas quanto para matrizes retangulares. Um valor singular e os correspondentes vetores singulares de uma matriz $\mathbf{A}_{m \times m}$ são, respectivamente, um escalar não-negativo $\sigma$ e um par de vetores unitários ${\mathbf{u}}$ e ${\mathbf{v}} $ se
\begin{equation}
\mathbf{A} {\mathbf{v}}=\sigma {\mathbf{u}}  \quad \text{e} \quad \mathbf{A} ^{*}{\mathbf{u}}=\sigma {\mathbf{v}},
\label{eq:SVD0}
\end{equation}
onde $\mathbf{u}$ e ${\mathbf{v}}$ são os vetores singulares à esquerda e vetores singulares à direita, respectivamente, e $\mathbf{A}^*$ denota a matriz conjugada transposta \cite{Strang1993}, \cite[corolário 2.4.2]{golub2013matrix}. 

Seja $\mathbf{\Sigma} = diag\left(\sigma_1, \dots, \sigma_m \right)$ uma matriz diagonal com os valores singulares em ordem decrescente; $\mathbf{U} = (\mathbf{u}_1,\dots, \mathbf{u}_m)$ uma matriz cujas colunas são formadas pelos vetores à esquerda e $\mathbf{V} = (\mathbf{v}_1,\dots, \mathbf{v}_m)$ uma matriz cujas colunas são formadas pelos vetores à direita. As relações da Equação \eqref{eq:SVD0} são reescritas conforme
\begin{equation}
\mathbf{A} \mathbf{V} =\mathbf{U} \mathbf{\Sigma} 
\label{eq:dvs0}
\end{equation}
\begin{equation}
\mathbf{A} ^{*}\mathbf{U}= \mathbf{V} \mathbf{\Sigma}. \label{eq:dvs1}
\end{equation}
Multiplicando-se ambos os lados da Equação \eqref{eq:dvs0} por $\mathbf{V}^*$, e considerando que $\mathbf{V} \mathbf{V}^* = \mathbf{I}$, é possível escrever a SVD da matriz como o produto de três matrizes, conforme
\begin{equation}
\begin{aligned}
\mathbf{A} \mathbf{V} \mathbf{V}^* & = \mathbf{U} \mathbf{\Sigma} \mathbf{V}^* \\
\putunder{\mathbf{A}}{m \times m}  & = \putunder{\begin{pmatrix}
\vert & & \vert\\
\mathbf{u}_1 & \cdots & \mathbf{u}_n\\
\vert & & \vert\\
\end{pmatrix}}{m \times m}
\putunder{\begin{pmatrix}
\sigma_1 & & 0\\
& \ddots & \\
0 & & \sigma_m\\
\end{pmatrix}}{m \times m}
\putunder{\begin{pmatrix}
\vert & & \vert\\
\mathbf{v}_1 & \cdots & \mathbf{v}_n\\
\vert & & \vert\\
\end{pmatrix}^{*}}{m \times m}.
\end{aligned}
\label{eq:dvs35}
\end{equation}
Para uma matriz $\mathbf{A}$ quadrada $m \times m$ cujos valores são reais, $\mathbf{U}$ e $ \mathbf{V}$ são matrizes ortonormais e valem as relações $ \mathbf{V^{*}} = \mathbf{V^{T}}$, $\mathbf{U}^T \mathbf{U} = \mathbf{V}^T \mathbf{V} = \mathbf{I} $, como utilizado na Equação \eqref{eq:dvs35}. Assim, para $\mathbf{A}$ com apenas valores reais, a SVD é dada por  
\begin{equation}
\mathbf{A} = \mathbf{U} \mathbf{\Sigma} \mathbf{V}^T,
\label{eq:dsv3texto}
\end{equation}
onde $\mathbf{U} = (\mathbf{u}_1,\dots, \mathbf{u}_m)$ e $\mathbf{V} = (\mathbf{v}_1,\dots, \mathbf{v}_m)$ são matrizes ortonormais, $\mathbf{U}^T \mathbf{U} = \mathbf{U} \mathbf{U}^T = \mathbf{I} $, $\mathbf{V}^T \mathbf{V} = \mathbf{V} \mathbf{V}^T = \mathbf{I}$ e $ \mathbf{\Sigma}$ é uma matriz diagonal, cujos valores singulares $\sigma_i$, para $i = 1, ..., m$, são números reais não-negativos colocados em ordem decrescente por convenção. 

A operacionalização do cálculo da SVD é mais direta de ser entendida do que o seu significado. Por exemplo, é possível relacionar os valores singulares de $\mathbf{A}$ com os autovalores de $\mathbf{A}^T \mathbf{A}$ e $\mathbf{A} \mathbf{A}^T$ \cite[pág. 57]{aster2019parameter}. Para mais informações sobre a SVD e seu cálculo, ver \cite{golub2013matrix}.

Algumas propriedades da SVD que se relacionam com o discutido são:
\begin{itemize}
\item O maior valor singular $\sigma_1 $ é igual ao módulo de $\vert \vert \mathbf{A} \vert\vert_2$ \cite[corolário 2.4.3]{golub2013matrix};
\item O número de valores singulares não-nulos indica o posto de $\mathbf{A}$ \cite[corolário 2.4.6]{golub2013matrix};
\item É possível escrever a SVD como \cite[corolário 2.4.7]{golub2013matrix}, \cite[pág. 29]{hansen2010discrete}: 
\begin{equation}
\begin{aligned}
\putunder{\mathbf{A}}{m \times n} & = \quad \putunder{\mathbf{u}_1}{m \times 1} \sigma_1 \putunder{(\mathbf{v}_1)^T}{1 \times n} + \dots + \putunder{\mathbf{u}_m}{m \times 1} \sigma_m \putunder{(\mathbf{v}_m)^T}{1 \times n} \\
& = \quad \sum^k_{i=1} \mathbf{u}_i \sigma_i \mathbf{v}_i^T
\label{eq:dsv10}
\end{aligned}
\end{equation}
\item O número de condição de uma matriz pode ser escrito em termos da razão entre o seu maior e o seu menor (não-nulo) valor singular
\begin{equation}
cond(\mathbf{A}) = \frac{\sigma_1}{\sigma_n}. 
\label{eq:cond} 
\end{equation}
\end{itemize}

\subsubsection{Como avaliar a dificuldade de se resolver um determinado problema mal-posto discreto utilizando a decomposição em valores singulares?}
O desafio de resolver um problema inverso é quando a matriz $\mathbf{A}$ é mal condicionada. A matriz é singular se cond($\mathbf{A}$) = $\infty$ e a matriz é mal condicionada se $\sigma_1 \ggg \sigma_n$. Os seguintes problemas inversos apresentam $\mathbf{A}$ aproximadamente singulares: os problema mal-posto, cuja discretização resulta sistemas mal condicionados, e os problemas com deficiência de posto. Para diferenciá-los, é necessária a visualização da queda dos seus $\sigma_i$, que é informativa sobre sua instabilidade \cite[pág. 74]{aster2019parameter}, \cite{hansen2010discrete}, \cite[pág. 45]{Mueller2012}. 

Problemas com deficiência de posto apresentam uma queda suave dos  $\sigma_i$, seguida de uma queda brusca até aproximadamente zero. Problemas mal-postos apresentam $\sigma_i$ que decaem gradualmente para zero, sem uma distinção clara entre os valores singulares nulos e não-nulos \cite[Subseção 7.4]{hansen2010discrete}. Nesse sentido, o conceito de posto numérico acaba sendo mais associado a problemas com deficiência de posto do que a problemas mal-postos \cite[Pág. 2]{Hansen1998}. Algoritmos para cálculo do posto podem ser baseados na SVD, sendo necessário definir uma tolerância, por exemplo associada à precisão numérica da representação, para que sejam considerados apenas os valores singulares maiores que ela. 

A queda de  $\sigma_i$ pode ser mais acentuada ou menos, sendo um indicativo de quanta informação poderá ser recuperada. De acordo com \cite[Definição 3.3.2]{Mueller2012}, se existir um número real $\alpha$ tal que $\sigma_n = \mathcal{O}(n^{-\alpha})$, então:
\begin{itemize}

    \item Se $0<\alpha\leq1$, o problema é levemente mal-posto;
    \item Se $\alpha>1$, o problema é moderadamente mal-posto;
    \item  Se $\sigma_n = \mathcal{O}(e^{-\alpha n})$, isto é, se os valores singulares decaírem mais rapidamente do que uma função exponencial qualquer, o problema é severamente mal-posto.
\end{itemize} 
A SVD permite outra visualização para análise do problema inverso, o gráfico dos coeficientes da SVD $\mathbf{u}_i^T \mathbf{y}$ e da solução $\frac{\mathbf{u}_i^T \mathbf{y}}{\sigma_i}$ em conjunto com os valores singulares. Essa visualização é chamada de gráfico de Picard e é importante para verificar a condição discreta de Picard, na qual os valores de $\mathbf{u}_i^T \mathbf{y}_{exato}$ devem decair mais rapidamente do que $\sigma_i$ \cite[pág. 37]{hansen2010discrete}. Quando ela não é respeitada, pode não haver a existência de soluções adequadas \cite[pág. 47]{hansen2010discrete}. Na presença de ruídos, os coeficientes $\mathbf{u}_i^T \mathbf{y}_{\bm{\delta}}$ decaem até chegarem ao nível do ruído $\bm{\delta}$. Nesse nível, os coeficientes param de cair, ficando em um \textit{plateau} no nível do ruído, e a condição de Picard deixa de ser satisfeita \cite[pág. 12]{hansen2010discrete}, mostrando de outra forma que o ruído limita a reconstrução.

\newpage
\section{É POSSÍVEL VER O \textit{DEBLURRING} COMO UM PROBLEMA INVERSO?}\label{sec:deblur_forward}

\subsection{Introdução}
Ao tirar fotografias, dependendo das condições de aquisição, a imagem resultante pode sair borrada. Borrar uma imagem é um exemplo da influência do instrumento de medidas (câmera) na aquisição desses sinais 2D (fotografias). Assim, resolver o problema direto significa calcular a imagem borrada a partir de uma imagem nítida, idealmente da forma mais realista possível. 

Um modelo de \textit{blur} é a convolução entre o sinal e a resposta ao impulso do sistema que é, no exemplo, a função de espalhamento pontual (PSF) \cite{hansen2006deblurring}. No caso de imagens 2D, a convolução é bidimensional. Sejam as funções de duas variáveis $f(\cdot)$, $g(\cdot)$  e $h(\cdot)$, uma imagem nítida, uma imagem borrada e a PSF, respectivamente. O modelo \cite[págs. 167]{aster2019parameter} que relaciona essas grandezas é dado por
\begin{equation}
f(t,z)*h(t,z) = g(t,z)
\label{eq:convdisc1}
\end{equation}
\begin{equation}
f(t,z) * h(t,z) = \int_{- \infty}^\infty \int_{- \infty}^\infty h(t-s_1, z-s_2) f(s_1, s_2) ds_1 ds_2.
\label{eq:blur}
\end{equation}
Nesse caso, o problema direto é calcular $g(t,z)$ a partir de $f(t,z)$ e $h(t,z)$. 

\begin{figure}[H]
     \centering
     \begin{subfigure}[b]{0.32\textwidth}
         \centering
         \includegraphics[width=\textwidth]{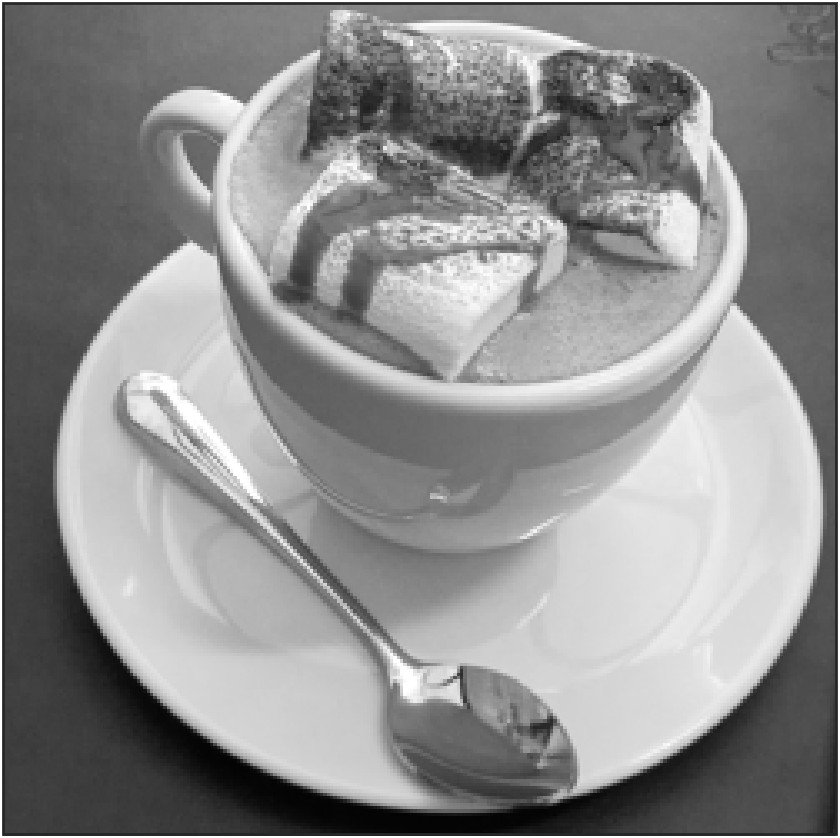}
         \caption{Imagem original}
         \label{fig:01_002a}
     \end{subfigure}

     \begin{subfigure}[b]{0.32\textwidth}
         \centering
                  \includegraphics[width=1\textwidth]{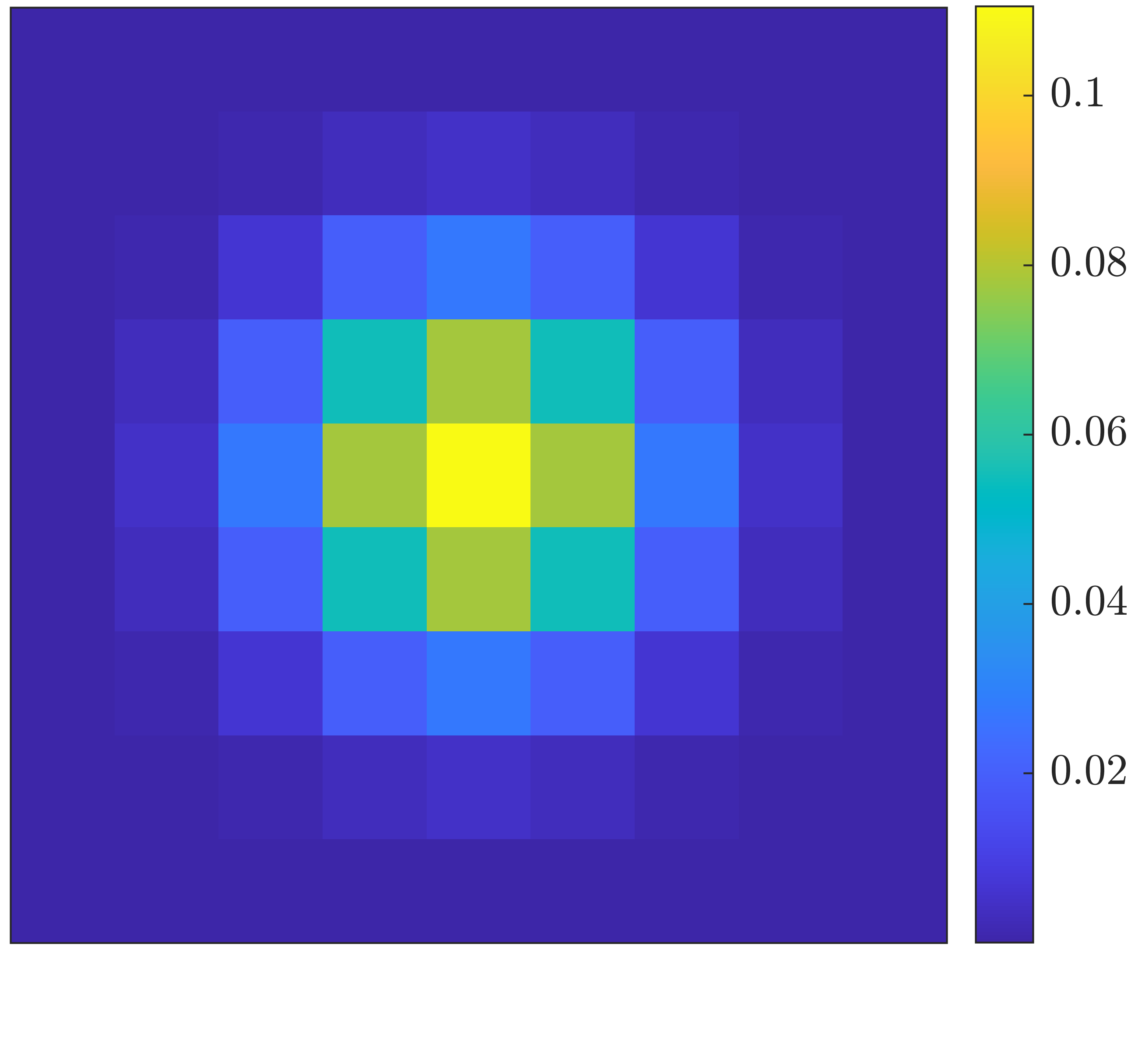}
         \caption{PSF gaussiana isotrópica}
         \label{fig:01_002b}
     \end{subfigure}     
     \hfill
     \begin{subfigure}[b]{0.32\textwidth}
         \centering
                  \includegraphics[width=\textwidth]{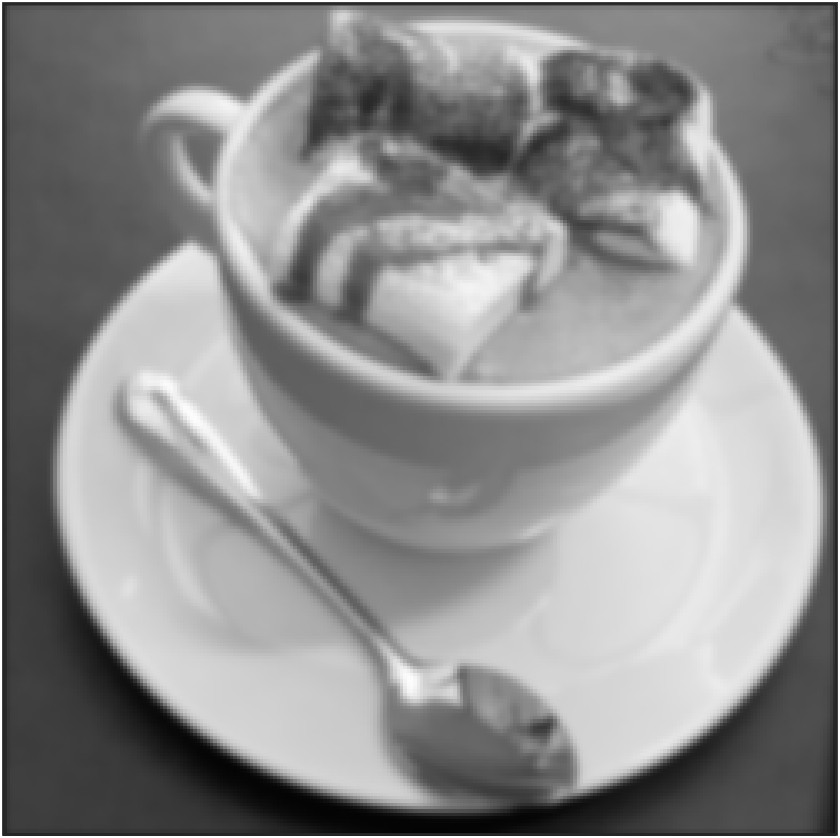}
         \caption{Imagem borrada com b)}
         \label{fig:01_002c}
     \end{subfigure}
     \hfill
          \begin{subfigure}[b]{0.32\textwidth}
         \centering
                  \includegraphics[width=\textwidth]{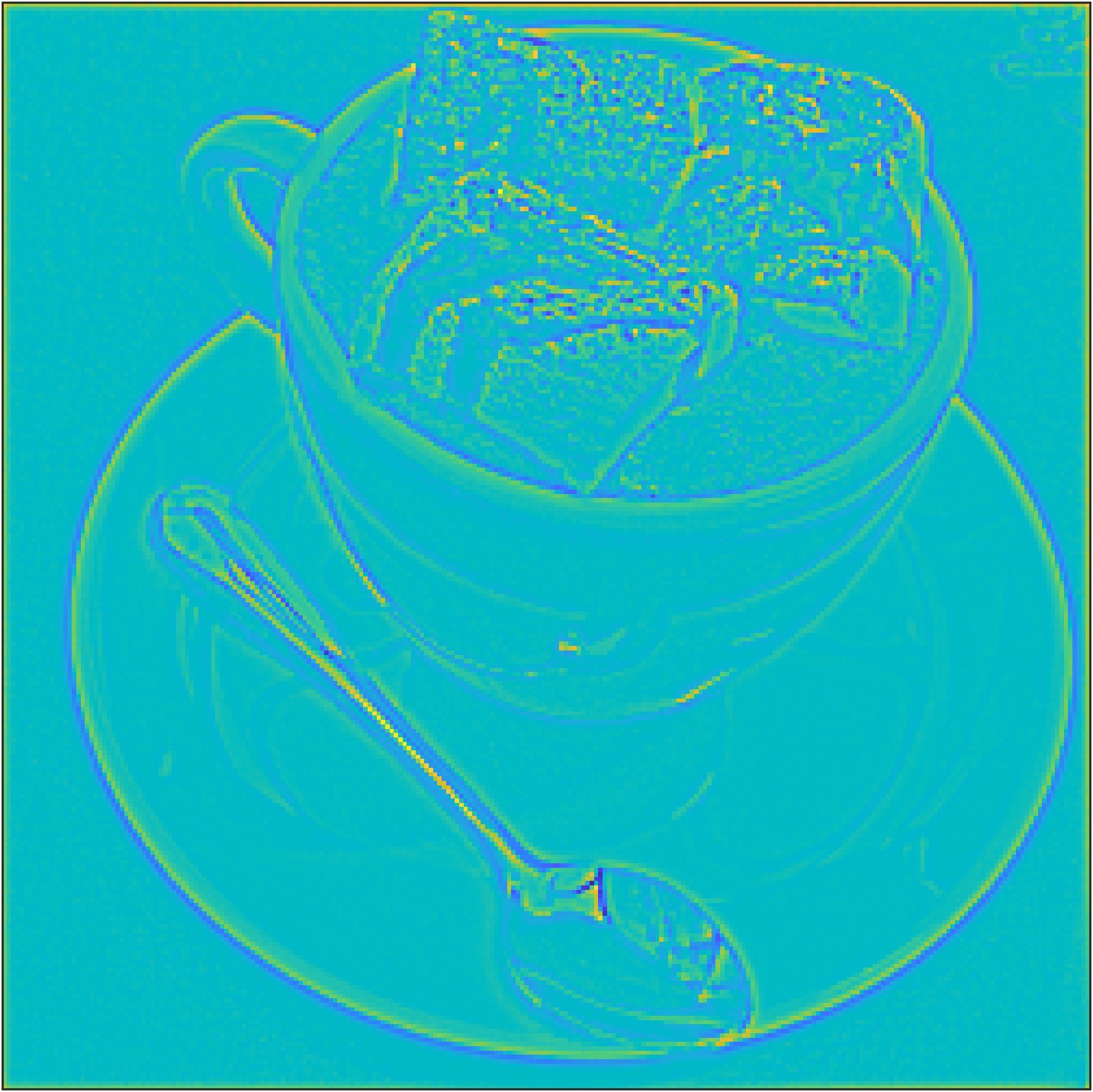}
         \caption{Diferença entre a) e c)}
         \label{fig:01_002d}
     \end{subfigure}

     \begin{subfigure}[b]{0.32\textwidth}
         \centering
         \includegraphics[width=1\textwidth]{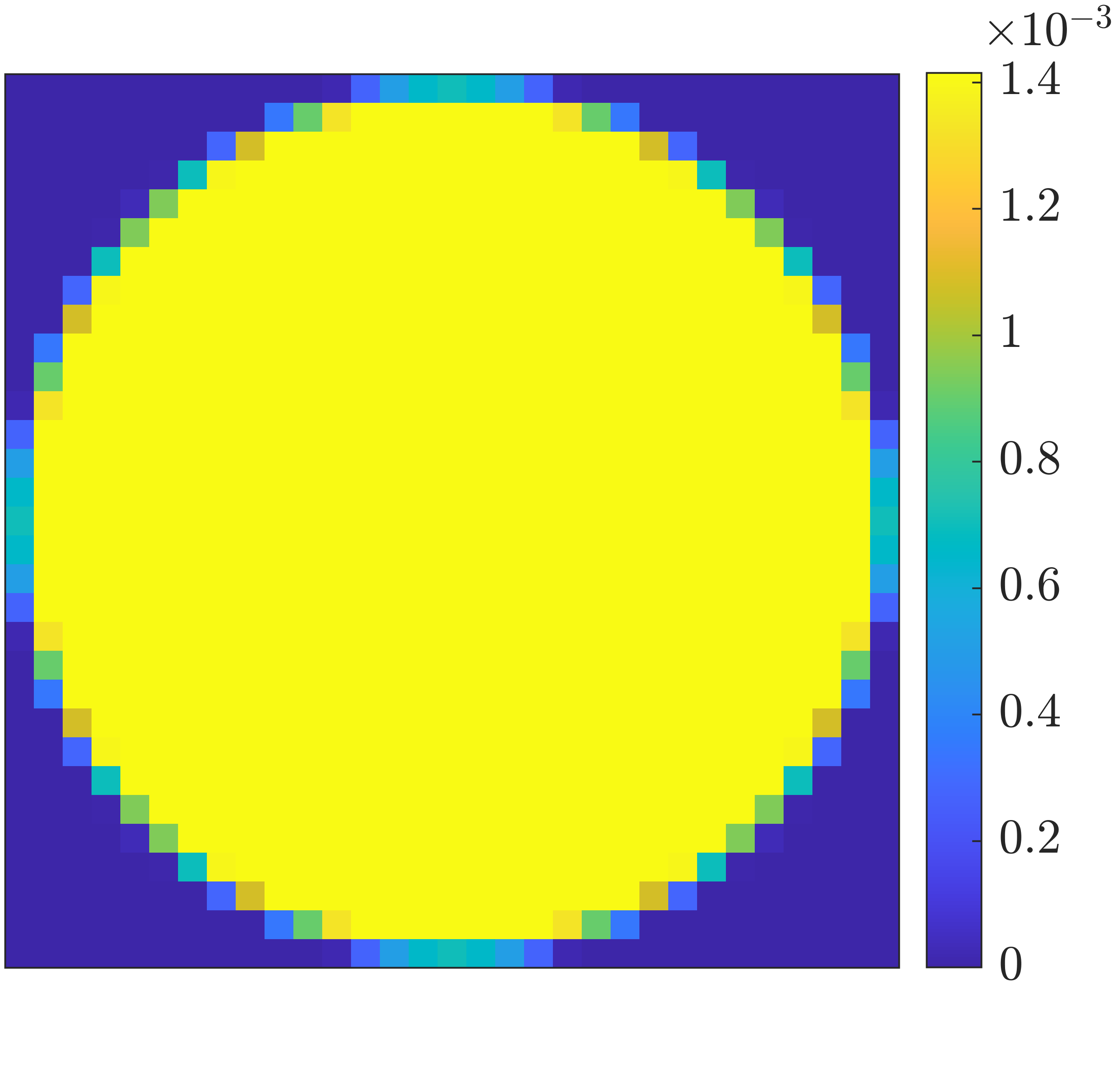}
         \caption{PSF gaussiana anisotrópica}
                  \label{fig:01_002e}
      \end{subfigure}
     \hfill
     \begin{subfigure}[b]{0.32\textwidth}
         \centering
                  \includegraphics[width=\textwidth]{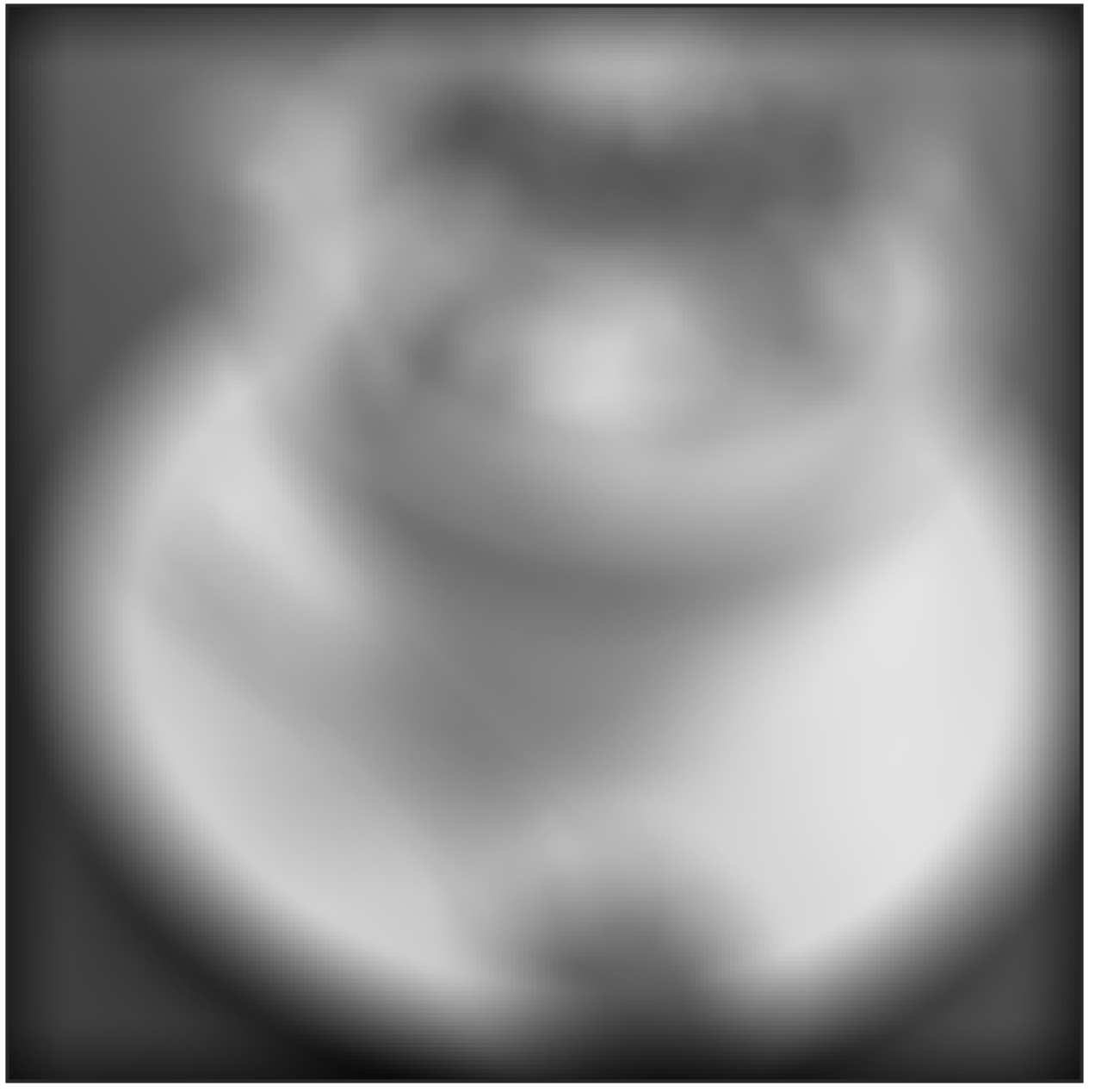}
         \caption{Imagem borrada com e)}
                  \label{fig:01_002f}
       \end{subfigure}
     \hfill
          \begin{subfigure}[b]{0.32\textwidth}
         \centering
                  \includegraphics[width=\textwidth]{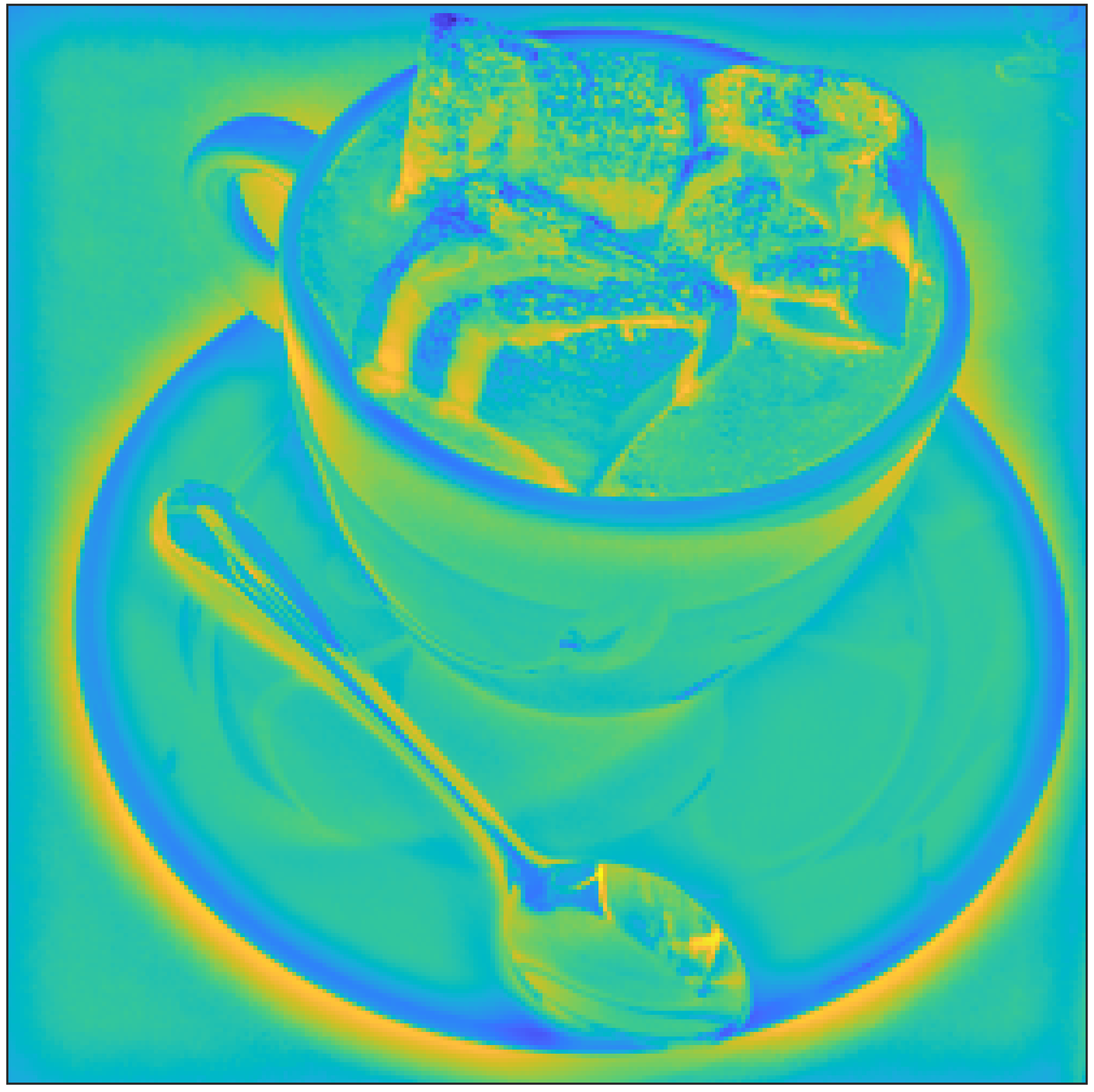}
         \caption{Diferença entre a) e f)}
                  \label{fig:01_002g}
     \end{subfigure}

     \begin{subfigure}[b]{0.32\textwidth}
         \centering
         \includegraphics[width=1\textwidth]{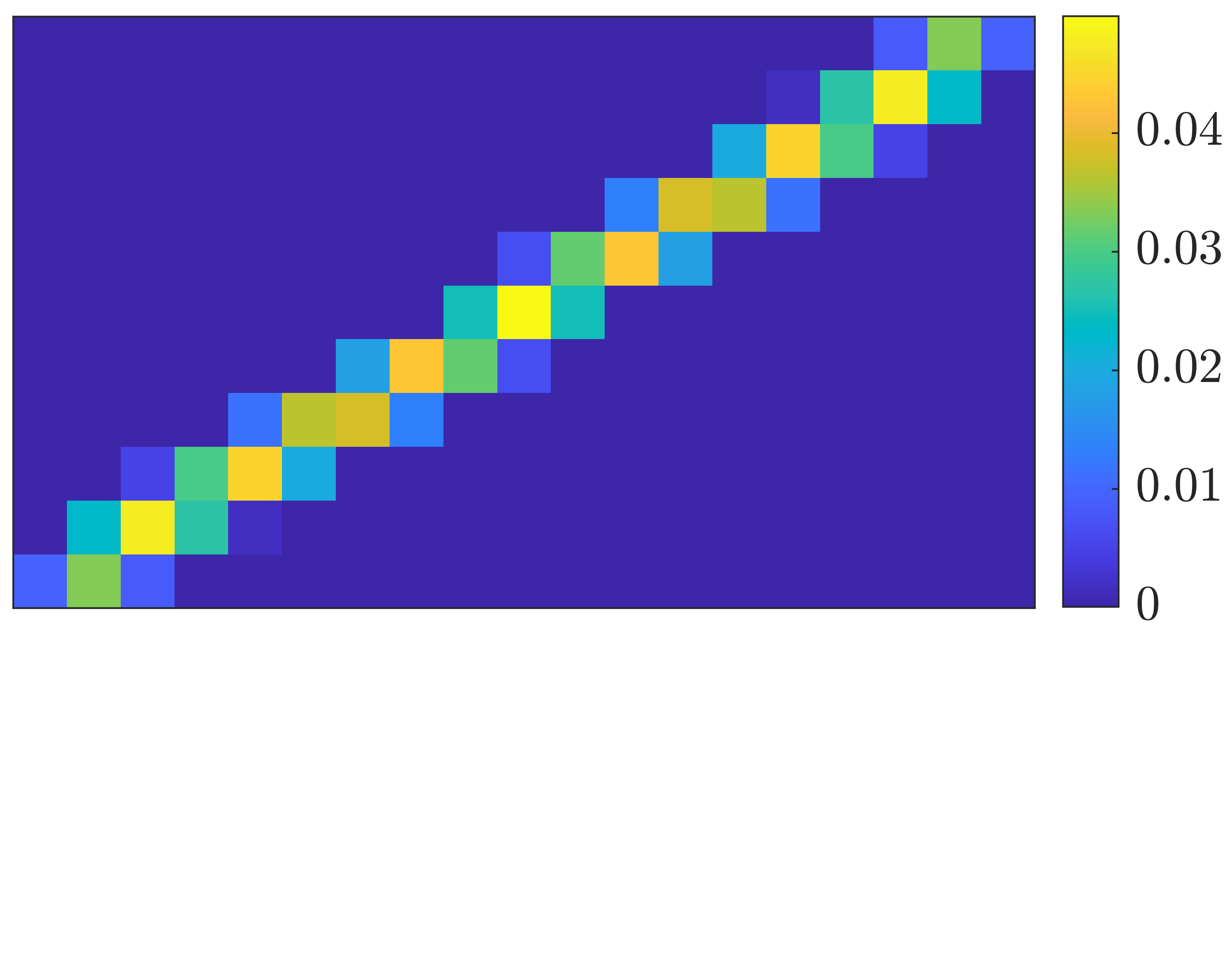}
         \caption{PSF de movimento}
                  \label{fig:01_002h}
     \end{subfigure}
     \hfill
     \begin{subfigure}[b]{0.32\textwidth}
         \centering
                  \includegraphics[width=\textwidth]{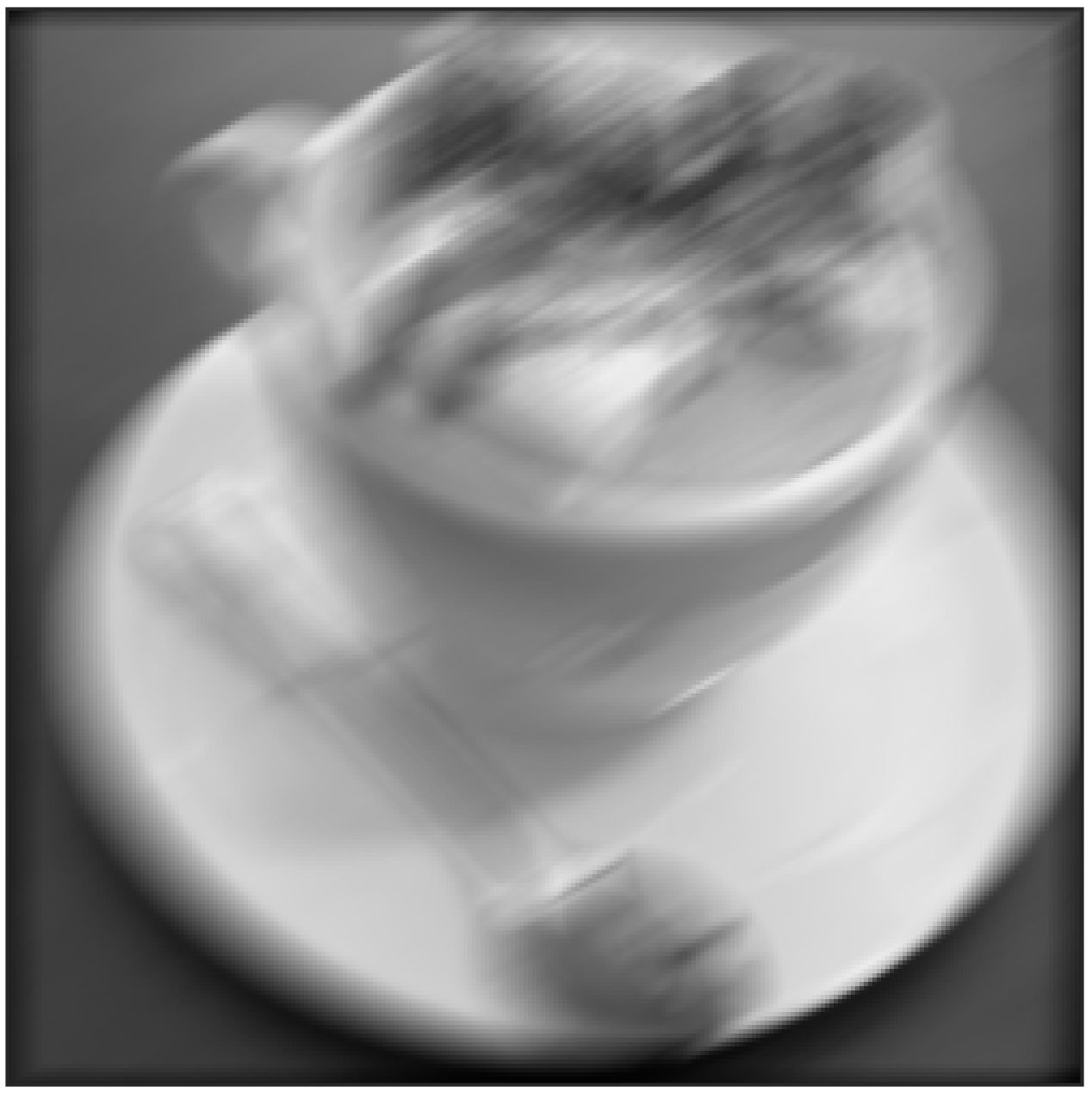}
         \caption{Imagem borrada com h)}
                  \label{fig:01_002i}
     \end{subfigure}
     \hfill
          \begin{subfigure}[b]{0.32\textwidth}
         \centering
                  \includegraphics[width=\textwidth]{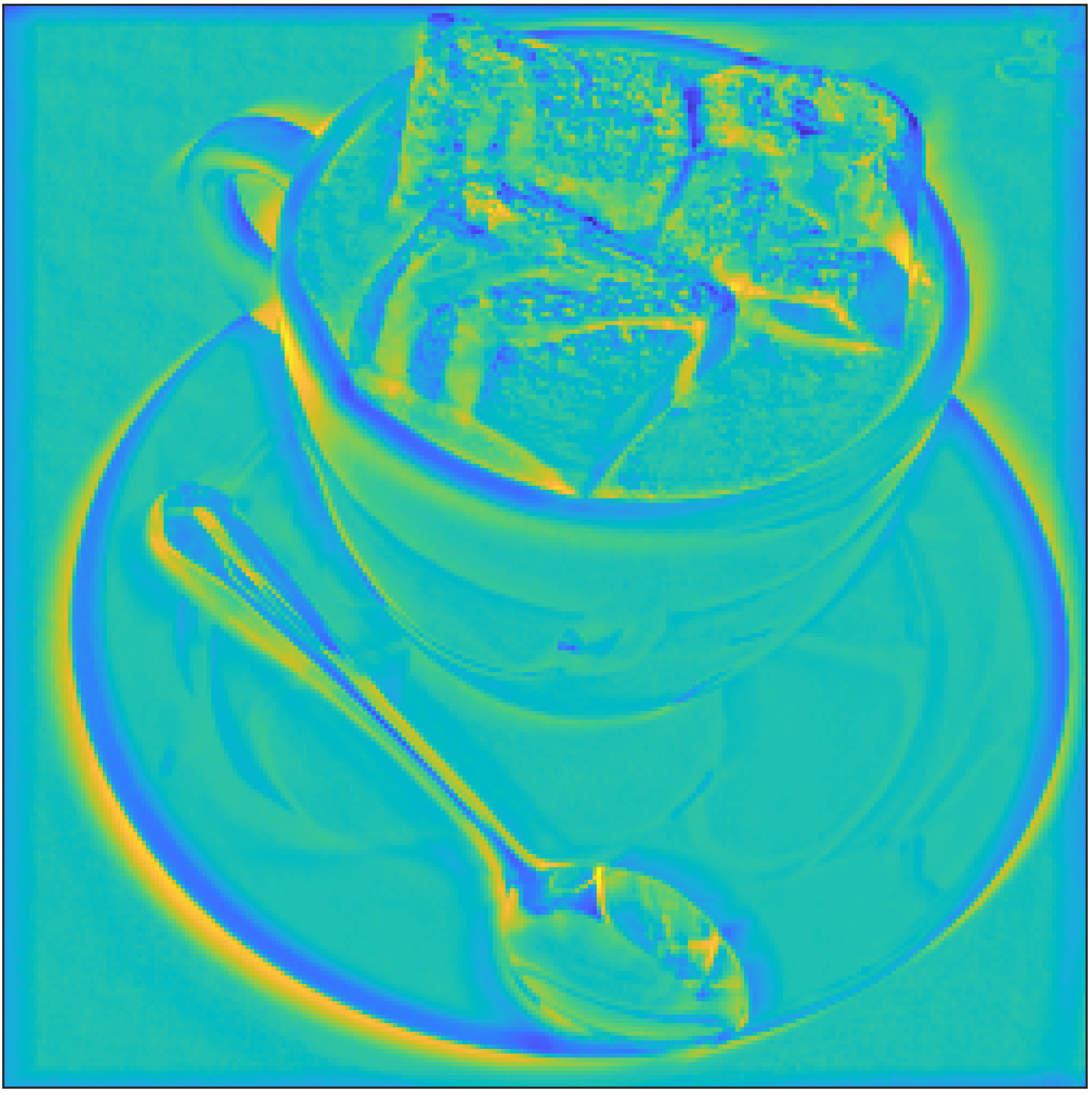}
         \caption{Diferença entre a) e i)}
                  \label{fig:01_002j}
      \end{subfigure}

\caption[Exemplos de convoluções com diferentes PSFs.]{Exemplos de convoluções com diferentes PSFs. Fonte: Próprio autor.}
\label{fig:01_002}
\end{figure}

Há diferentes tipos de PSFs e cada uma pode resultar em efeitos característicos após a convolução. Seja a Figura \ref{fig:01_002a}, na qual há o exemplo de uma imagem nítida. Assim:
\begin{itemize}
\item A falta de foco pode ser obtida com PSF gaussiana. Se o desvio padrão é o mesmo ao longo das duas dimensões, a PSF gaussiana é isotrópica; se o desvio padrão é diferente entre os sentidos vertical e horizontal, ela é anisotrópica. Na Figura \ref{fig:01_002b} há a PSF gaussiana isotrópica e na Figura  \ref{fig:01_002c} há a simulação de \textit{blur} pequeno com esse filtro. Na Figura \ref{fig:01_002d}, a diferença entre as duas mostra que se perde a nitidez das bordas, informações de alta frequência;

\item A PSF da falta de foco também pode ser modelada como um disco \cite[págs. 25-6]{hansen2006deblurring}, na qual a região dentro do disco apresenta um valor constante e a região fora do disco apresenta valores nulos. Nas Figuras \ref{fig:01_002e}, \ref{fig:01_002f} e \ref{fig:01_002g} há um \textit{blur} grande com disco de PSF (quanto maior o raio, maior é o \textit{blur});

\item Há também o borrão por movimento, quando o objeto ou a câmera está em movimento em relação ao outro, que apresenta direcionalidade. 
Se a reconstrução é realizada a partir de uma única imagem, chama-se de \textit{single-image deblurring}. Existem também algoritmos que utilizam mais de um \textit{frame} na reconstrução, visando melhor caracterização do movimento. Nas Figuras \ref{fig:01_002h}, \ref{fig:01_002i} e \ref{fig:01_002j} é apresentado o caso de uma PSF de movimento com ângulo de 30$^o$. 
\end{itemize}

Quando se considera o \textit{blur} uniforme ao longo de toda a imagem (invariante no espaço), não há grandes problemas para a simulação da imagem borrada, mas é difícil que se consiga um resultado realista através deste procedimento \cite{Koh2021}. Para simular um \textit{blur} que não é uniforme, é necessário segmentar a imagem para definir o \textit{blur} de cada \textit{pixel}, mas isso pode não ser trivial \cite{Koh2021}. Desenvolver o modelo de um \textit{blur} cada vez mais realista pode ser o próprio problema de pesquisa. 

\begin{figure}[!ht]
\centering
\includegraphics[width = 0.6\textwidth]{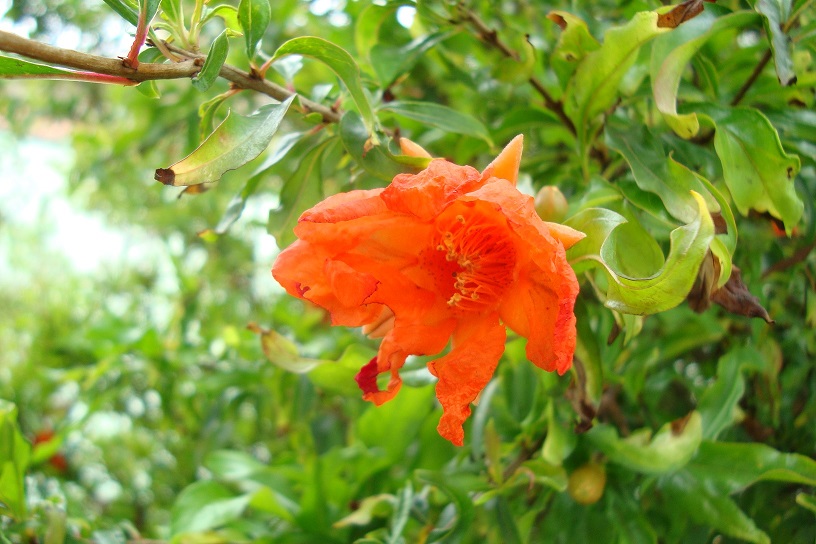} \caption[Fotografia de \textit{blur} não-uniforme.]{Fotografia de \textit{blur} não-uniforme. Fonte: Próprio autor.} 
\label{fig:01_0002}
\end{figure}

Na Figura \ref{fig:01_0002} é mostrado um \textit{blur} não-simulado. Ele é não-uniforme, variante no espaço. A flor aparece nítida no centro da imagem, mas quanto mais ao fundo, mais borrada a imagem fica. O desafio é que, diferente da Figura \ref{fig:01_002} com \textit{blur} simulado, não há um gabarito da Figura \ref{fig:01_0002} onde todas as regiões são nítidas. 

\subsection{Como que é definida a convolução discreta?}
\begin{figure}[H]
\centering
\includegraphics[width = 0.7\textwidth]{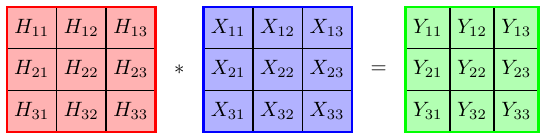}
\caption[Operação de convolução.]{Operação de convolução. Fonte: Próprio autor.}
\label{deblur0}
\end{figure}

A Equação \eqref{eq:blur} pode ser discretizada e, pela propriedade da comutatividade da convolução, reescrita como
\begin{equation}
\begin{aligned}
\mathbf{Y}  & = \mathbf{H} * \mathbf{X} \\
Y(i,j) = & \sum_{k_1=-\infty}^{\infty} \sum_{k_2=-\infty}^{\infty} H(k_1,k_2) X(i-k_1,j-k_2).
\label{eq:convdisc3}
\end{aligned}
\end{equation}
Essa operação é mostrada na Figura \ref{deblur0}, onde os elementos $Y(i,j)$ são denotados $Y_{ij}$ (da mesma forma será utilizado $H_{ij}$ e $X_{ij}$). Ela substitui o valor de um \textit{pixel} por outro valor, que depende do próprio \textit{pixel} e dos \textit{pixels} vizinhos. Em relação à nomenclatura, a matriz $\mathbf{H}$ é usualmente denotada \textit{kernel}, mas nessa aplicação em específico também pode ser entendida como a PSF do sistema. Já o processo como um todo da convolução com um \textit{kernel} é uma filtragem espacial linear \cite[Subseção 3.4]{gonzalez2018}. 

\begin{figure}[H]
\centering
\includegraphics[width = 0.7\textwidth]{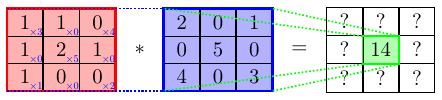}
\caption[Ilustração do cálculo da convolução.]{Ilustração do cálculo da convolução. Fonte: Próprio autor.}
\label{fig:deblur1}
\end{figure}

 Para ilustrar, o cálculo do elemento $Y_{22}$ pela Equação \eqref{eq:convdisc3} é dado por
\begin{equation}
Y_{22} = H_{33} X_{11} + H_{32} X_{12} + H_{31} X_{13} + H_{23} X_{21} + H_{22} X_{22} + H_{21} X_{23} + H_{13} X_{31} + H_{12} X_{32} + H_{11} X_{33}.
\end{equation} 
Um exemplo numérico é visto na Figura \ref{fig:deblur1}, onde $Y_{22} = 14$ é calculado por
\begin{equation}
Y_{22} = 3 \times 1 + 0 \times1 + 4\times 0 + 0 \times1 + 5 \times2 + 0\times 1 + 1 \times1 + 0 \times0 + 2 \times0.
\end{equation}

\subsection{Qual é a influência das condições de contorno na convolução?}
\begin{figure}[H]
\centering
\includegraphics[width = 0.75\textwidth]{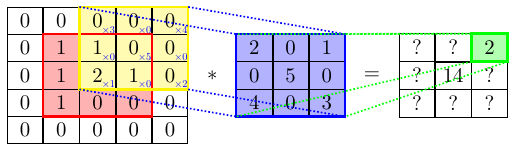}
\caption[Condição de contorno nula na convolução.]{Condição de contorno nula na convolução. Fonte: Próprio autor.}
\label{fig:deblur2}
\end{figure}

As condições de contorno definem como os \textit{pixels} fora do campo de visão influenciam os \textit{pixels} dentro das bordas \cite{hansen2006deblurring}. Na Figura \ref{fig:deblur2} é ilustrada a condição de contorno nula, quando todos os \textit{pixels} externos são considerados nulos. 

O valor numérico de $Y_{13} = 2$ pode ser calculado por
\begin{equation} 
Y_{13} = 3 \times 0 + 0 \times 0 + 4\times 0 + 0 \times 1 + 5 \times 0 + 0\times 0 + 1 \times 2 + 0 \times 1 + 2 \times 0.
\end{equation}

\begin{figure}[H]
     \centering
     \begin{subfigure}[b]{0.37\textwidth}
         \centering
         \includegraphics[width=1\textwidth]{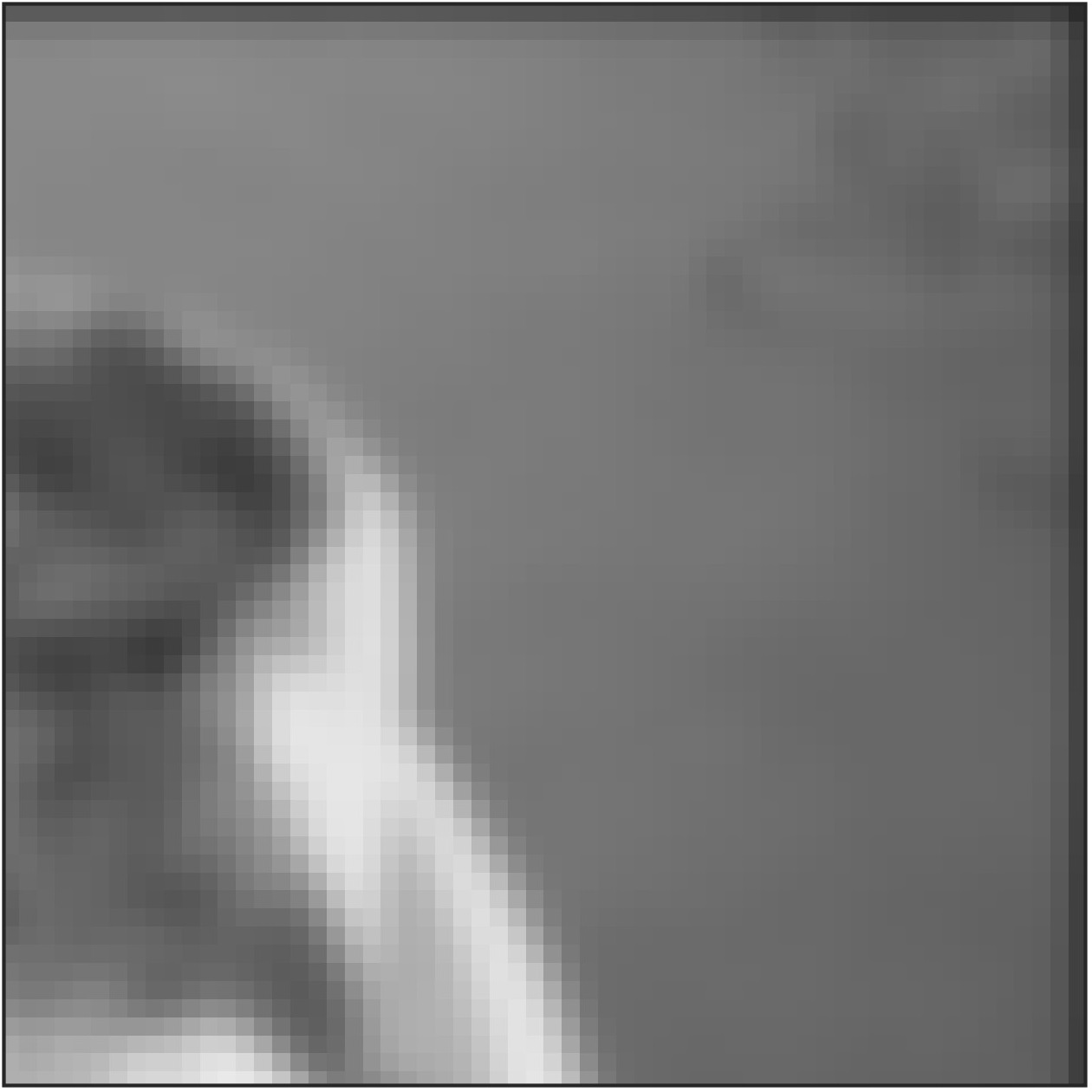}
         \caption{Condição de contorno nula}
         \label{fig:01_003a}
     \end{subfigure}     
     \begin{subfigure}[b]{0.37\textwidth}
         \centering
                  \includegraphics[width=1\textwidth]{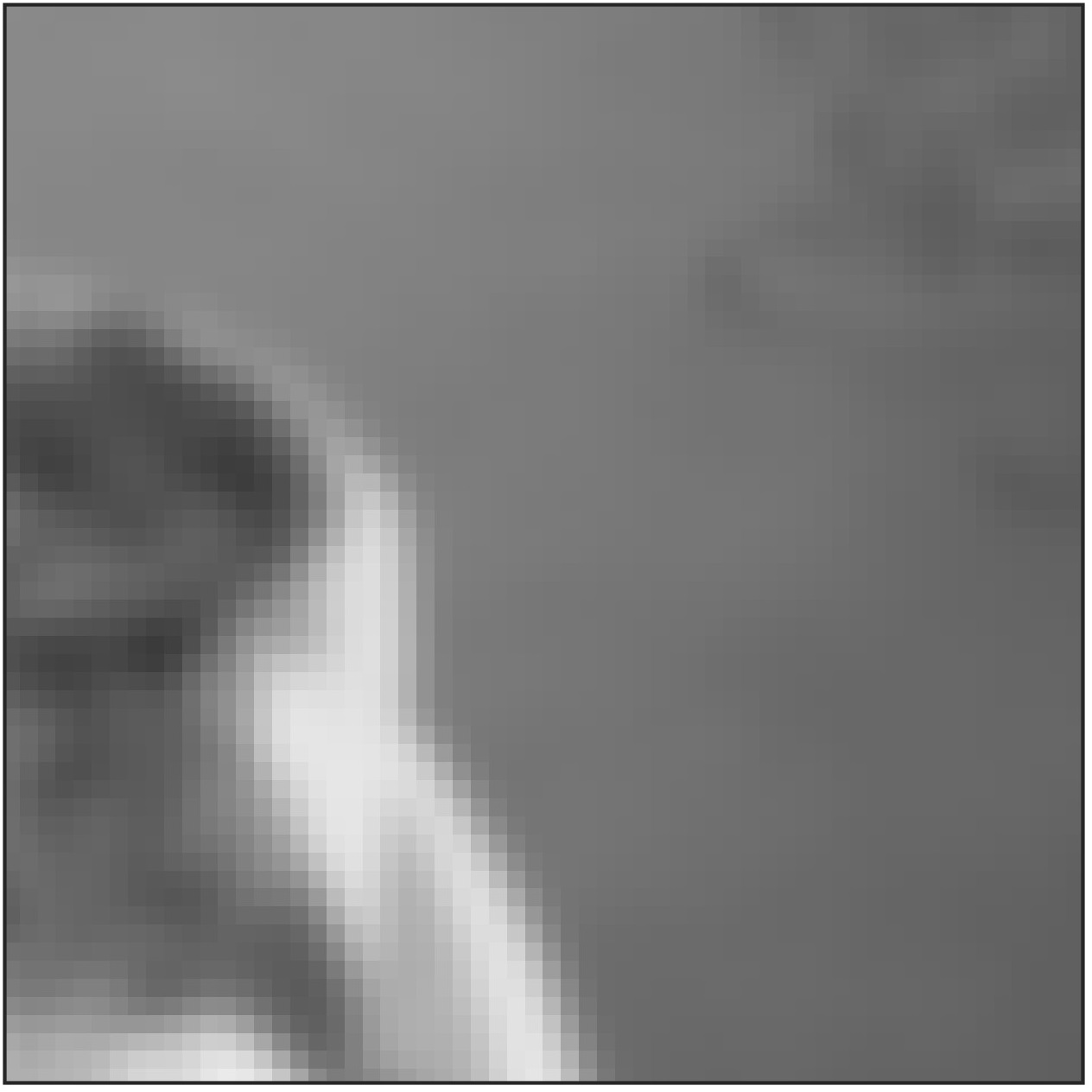}
         \caption{Condição de contorno replicada}
         \label{fig:01_003b}
     \end{subfigure}
\caption[Comparação entre duas condições de contorno.]{Comparação entre duas condições de contorno. Fonte: Próprio autor}
\label{fig:01_003}
\end{figure}

A Figura \ref{fig:01_002b} mostra o resultado da Equação \eqref{eq:convdisc3} com a PSF gaussiana e condições de contorno nulas, que podem resultar em um contorno escuro na borda, conforme visto na Figura \ref{fig:01_003a}, a aproximação do seu canto direito superior direito. Com a condição de contorno replicada, o valor do \textit{pixel} fora da imagem é dado pelo valor do \textit{pixel} da borda mais próximo. Assim, a borda escura não é mais observada na Figura \ref{fig:01_003b}. 

\subsection{Como definir a convolução como um sistema linear de equações?}

Seja $\mathbf{x}$ a concatenação das colunas da imagem nítida $\mathbf{X}$; $\mathbf{y}$ a concatenação das colunas da imagem borrada $\mathbf{Y}$; e $\mathbf{h}$ a concatenação das colunas do \textit{kernel} $\mathbf{H}$. Assim, a Equação \eqref{eq:convdisc3} pode ser escrita na forma da multiplicação matriz-vetor $\mathbf{A}\mathbf{x} = \mathbf{y}$ \cite[Cap. 4]{hansen2006deblurring}, onde a matriz $\mathbf{A}$ representa a operação de convolução com $\mathbf{h}$. Nesse caso, a vetorização de $\mathbf{h}$ não traz benefícios, pois a sua distribuição em $\mathbf{A}$, embora estruturada, depende apenas de seus componentes $h_{ij}$.

Para tal sistema, pode-se considerar condições de contorno nulas \cite[pág. 37]{hansen2006deblurring}. Já em \cite[pág. 38]{hansen2006deblurring} é discutido como a condição de contorno periódica resulta em uma matriz $\mathbf{A}$ circulante de blocos com blocos circulantes e como no caso de condições de contorno reflexivas $\mathbf{A}$ pode ser escrita como a soma de matrizes de bloco Hankel e Toeplitz. 

\begin{figure}[H]
     \centering
     \begin{subfigure}[b]{0.37\textwidth}
         \centering
         \includegraphics[width=1\textwidth]{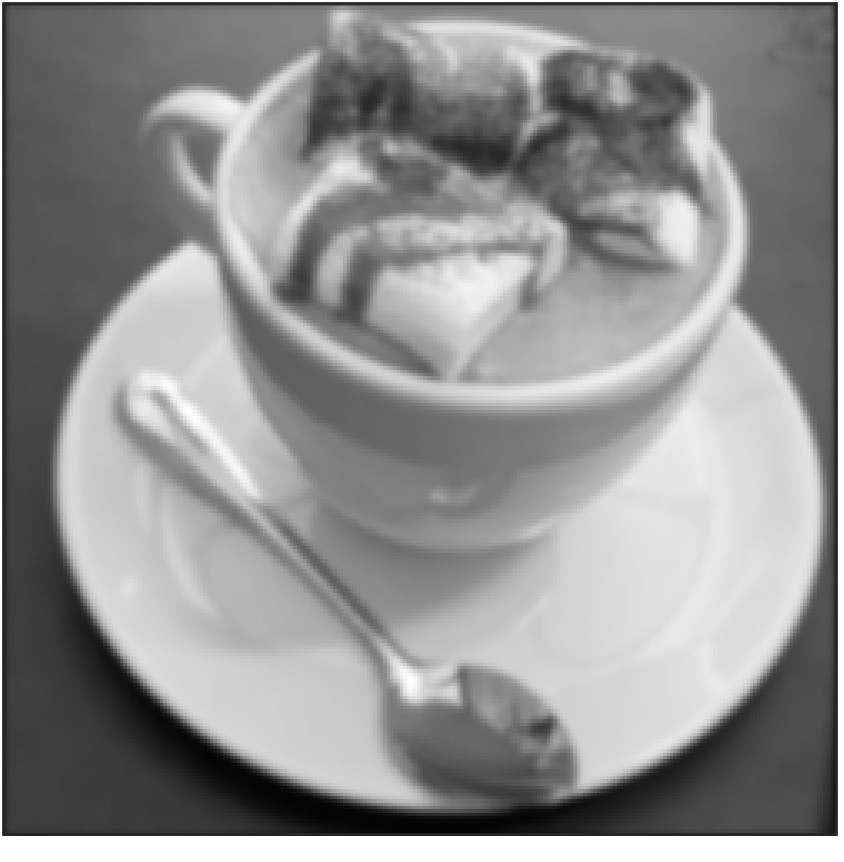}
         \caption{Multiplicação $\mathbf{A}\mathbf{x}$}
         \label{fig:bludifa}
     \end{subfigure}     
     \begin{subfigure}[b]{0.37\textwidth}
         \centering
                  \includegraphics[width=1.32\textwidth]{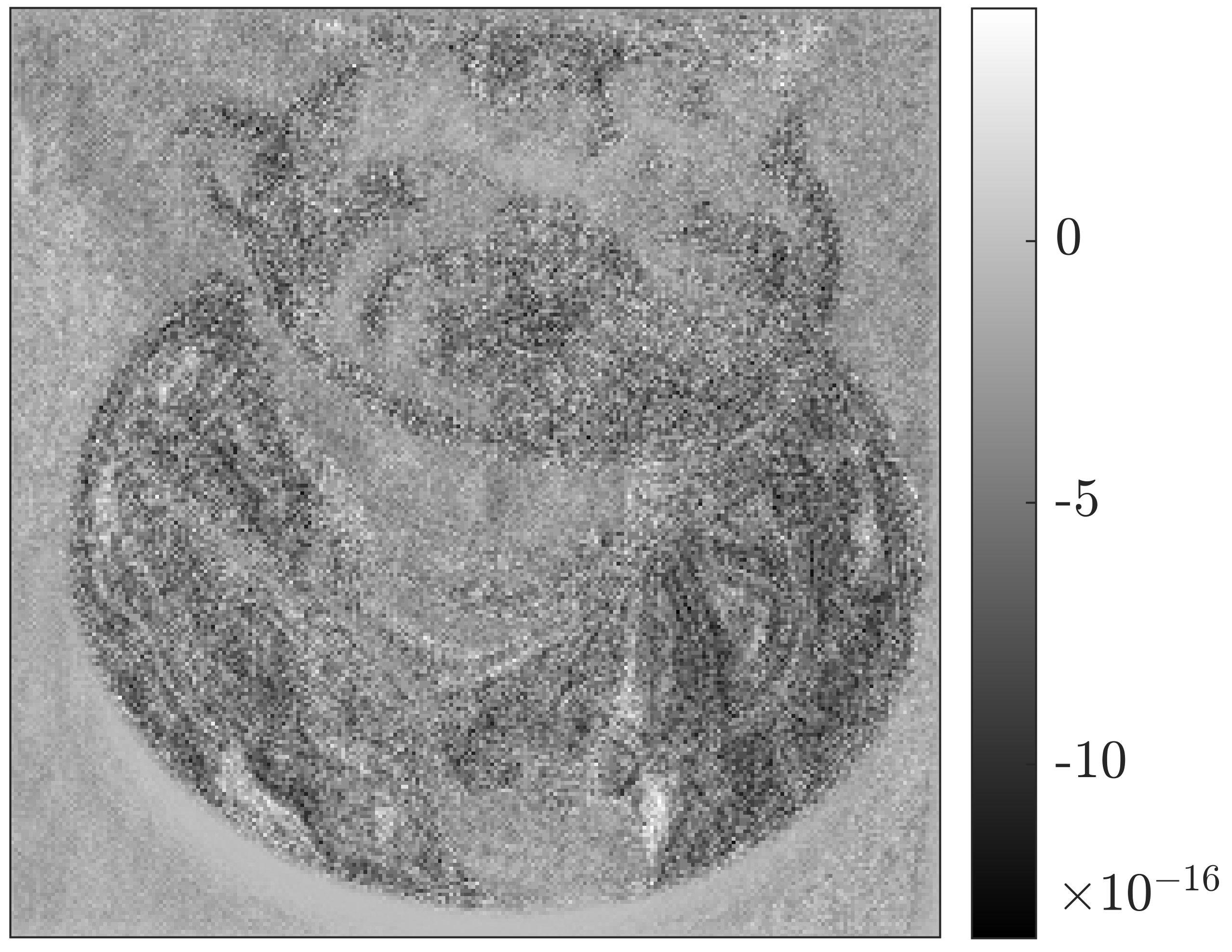}
         \caption{Diferença para convolução.}
         \label{fig:bludifb}
     \end{subfigure}
\caption[a) Convolução como sistema linear. b) Diferença entre as Figuras \ref{fig:01_002b} e \ref{fig:bludifa}.]{a) Convolução como sistema linear. b) Diferença entre as Figuras \ref{fig:01_002b} e \ref{fig:bludifa}. Fonte: Próprio autor.}
\label{fig:bludif}
\end{figure}

A Figura \ref{fig:01_008} mostra o resultado da convolução como um sistema linear de equações.  Na Figura \ref{fig:bludifa} é mostrada a convolução com a mesma PSF da Figura \ref{fig:01_002}, mas calculado na forma $\mathbf{A}\mathbf{x} = \mathbf{y}$, ao invés da convolução convencional. A Figura \ref{fig:bludifb} mostra que a diferença entre os resultados está na ordem da precisão numérica, sendo, portanto, equivalentes na prática. A escolha de utilizar uma ou outra forma pode ser decorrente do algoritmo que será utilizado para processamento dos dados, alguns exigindo a disponibilidade explícita de $\mathbf{A}$, enquanto outros não.

\subsection{Qual informação que a visualização dos valores singulares   traz?}

\begin{figure}[H]
     \centering
     \begin{subfigure}[b]{0.3\textwidth}
         \centering
         \includegraphics[width=0.92\textwidth]{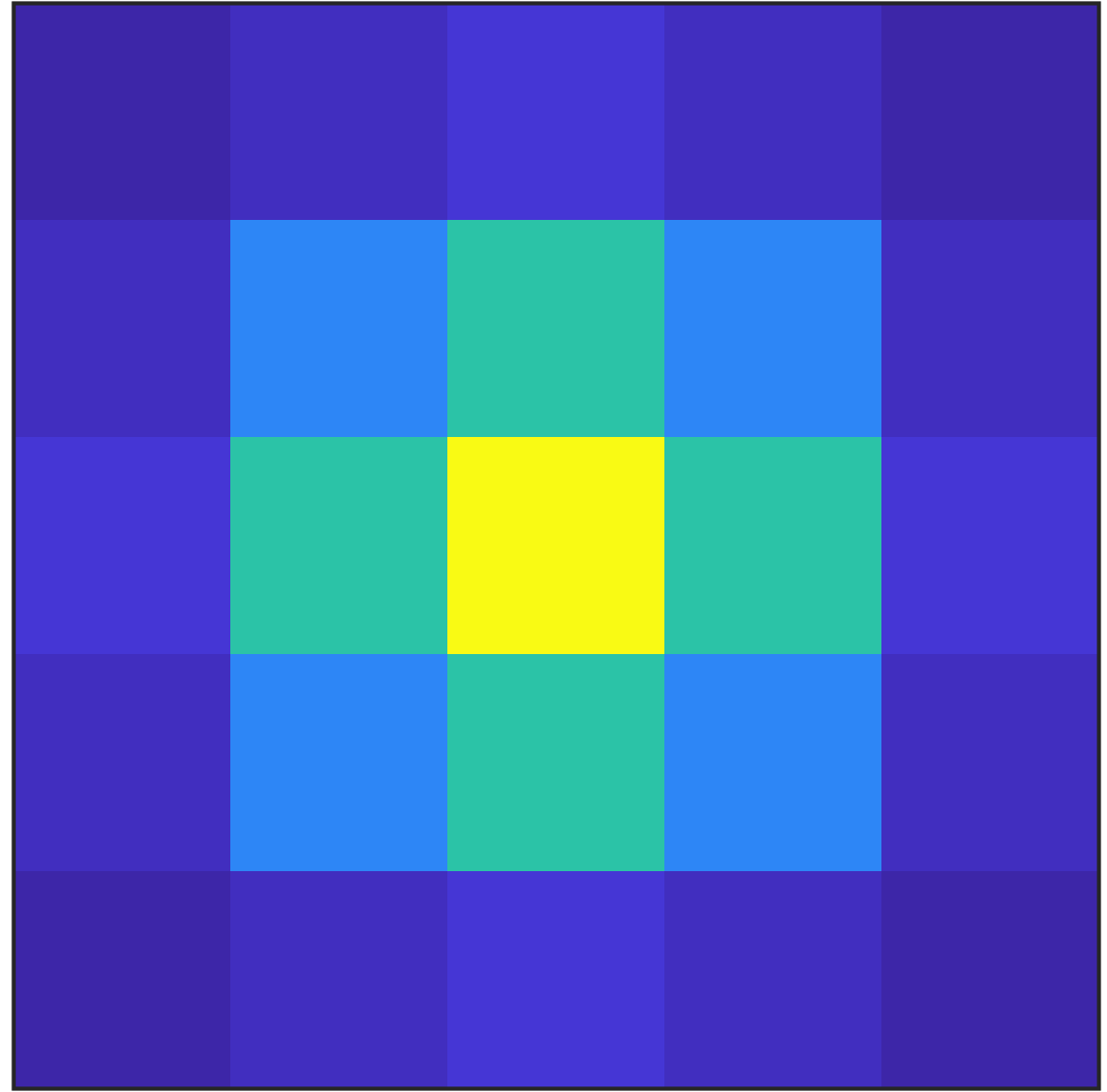}
         \caption{\textit{Blur} menor}
         \label{fig:01_008a}
     \end{subfigure}
     \hfill
     \begin{subfigure}[b]{0.3\textwidth}
         \centering
                  \includegraphics[width=0.92\textwidth]{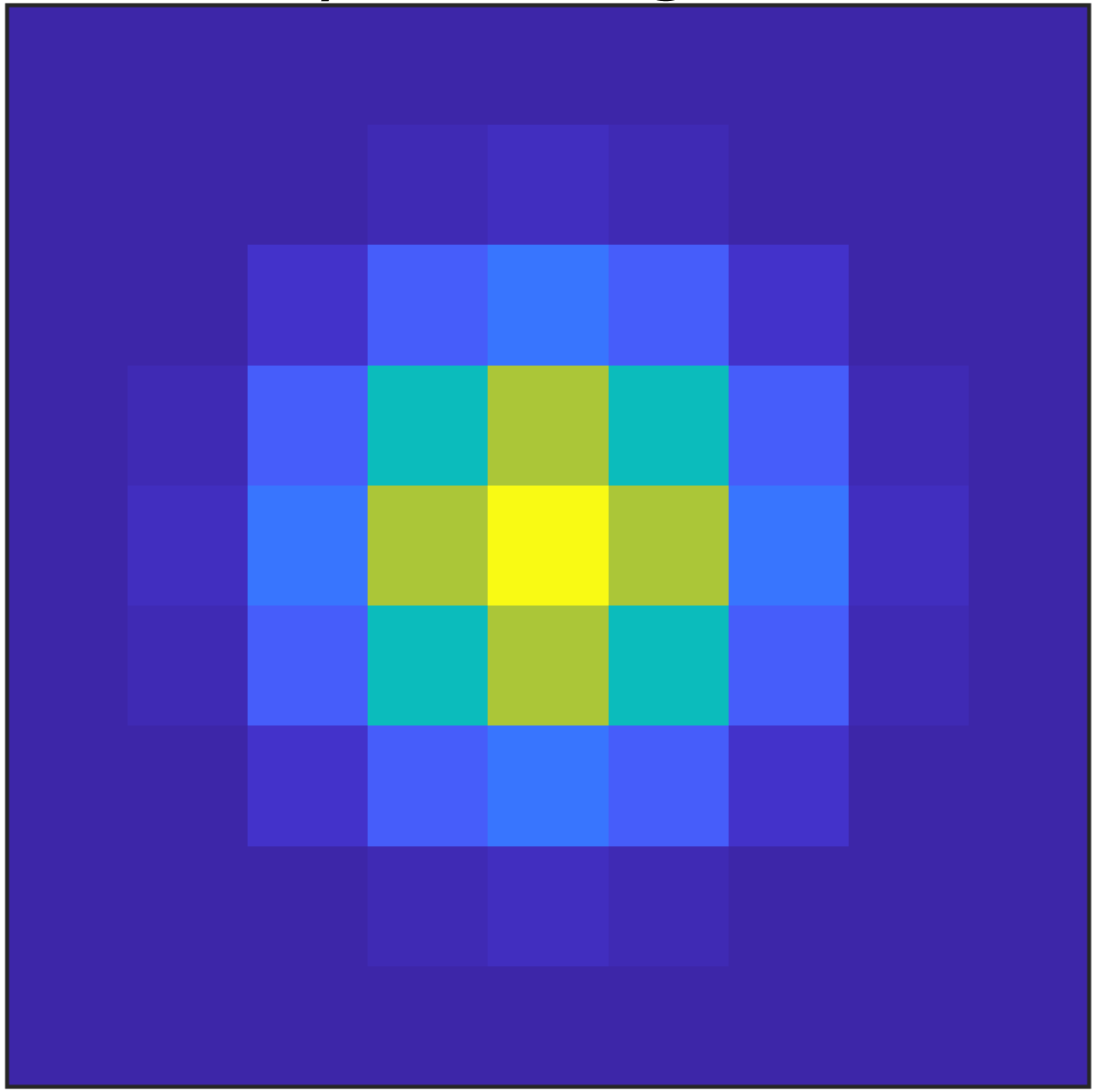}
         \caption{\textit{Blur} original}
         \label{fig:01_008b}
     \end{subfigure}
     \hfill
          \begin{subfigure}[b]{0.3\textwidth}
         \centering
                  \includegraphics[width=0.92\textwidth]{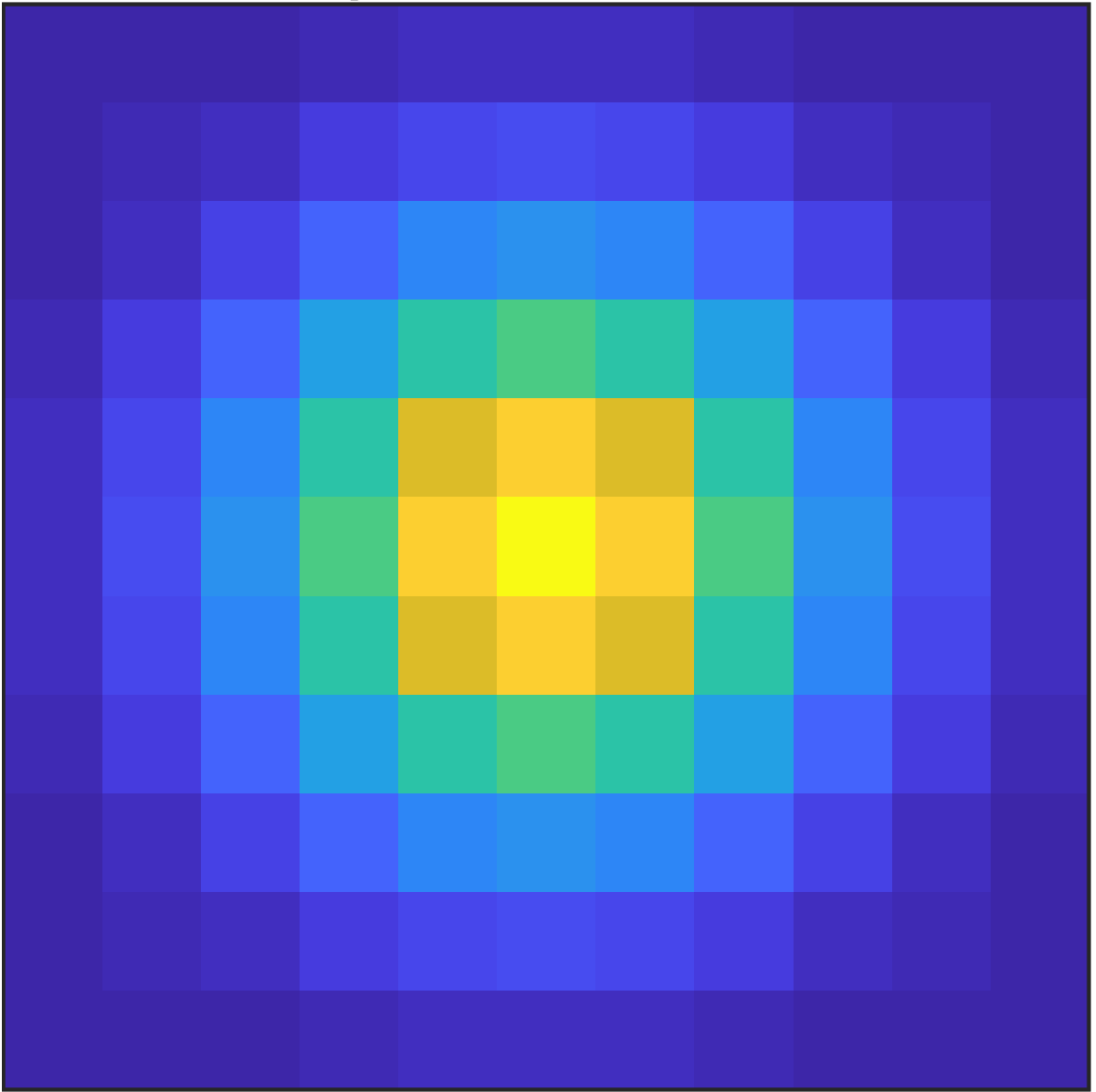}
         \caption{\textit{Blur} maior}
         \label{fig:01_008c}
     \end{subfigure}
\caption[PSFs gaussianas com diferentes tamanhos.]{PSFs gaussianas com diferentes tamanhos. Fonte: Próprio autor.}
\label{fig:01_008}

\end{figure}

Considerando a decomposição em valores singulares discutida na Subseção \ref{app-svd}, cada $\mathbf{A}$ deve ser analisada individualmente. No \textit{deblurring}, a Condição de Picard é satisfeita quando se consideram medidas sem ruído $\mathbf{y}_{exato}$ \cite[pág. 69]{hansen2006deblurring}. Agora, sejam também as PSF gaussianas das Figura \ref{fig:01_008}, onde o \textit{blur} original é relativo à Figura \ref{fig:01_002b}. Para cada $\mathbf{A}$ montada a partir dessas PSF, foram calculados os números de condição, obtendo-se $cond(\mathbf{A})_{a)} \approx 617$,  $cond(\mathbf{A})_{b)} \approx 3.26 \times 10^6$ e $cond(\mathbf{A})_{c)} \approx 7.56 \times 10^{11}$. Ou seja, quanto maior o \textit{blur}, pior é o condicionamento de $\mathbf{A}$.

\begin{figure}[H]
\centering
\includegraphics[width=1\textwidth]{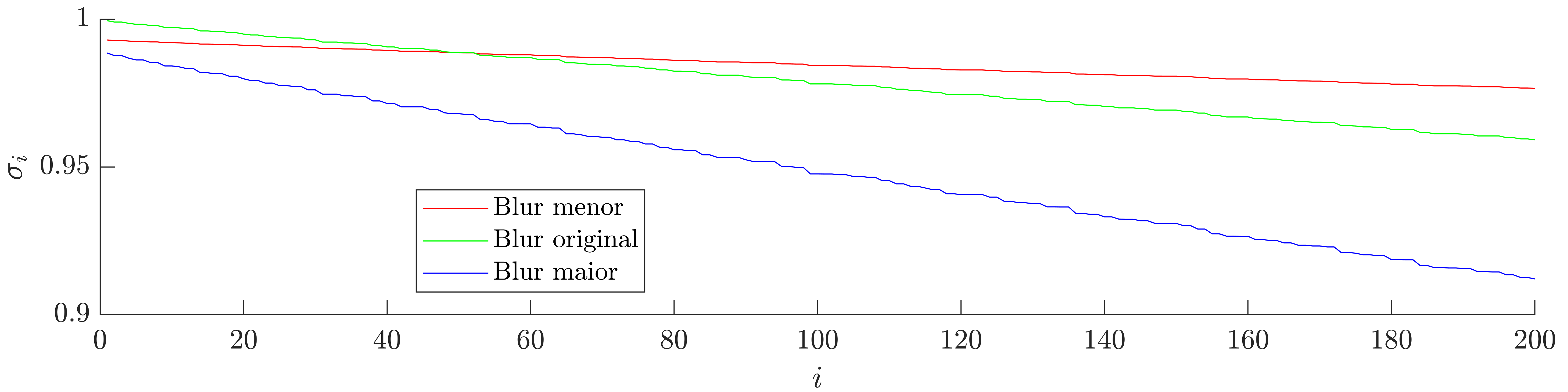} 
\caption[Os 200 maiores $\sigma_i$ para diferentes PSFs.]{Os 200 maiores $\sigma_i$ para diferentes PSFs. Fonte: Próprio autor.}
\label{fig:01_009}
\end{figure}

Para as PSFs dadas na Figura \ref{fig:01_008}a-c), a Figura \ref{fig:01_009} mostra os 200 maiores $\sigma_i$ de cada $\mathbf{A}$. Observa-se que, quanto maior é a PSF, mais rápida é a queda de $\sigma_i$ \cite[págs. 58-9]{hansen2006deblurring}, mesmo que o primeiro $\sigma_i$ do \textit{blur} original seja maior. 

\subsection{Qual informação que a visualização dos vetores singulares traz?}

\begin{figure}[H]
     \centering
     \begin{subfigure}[b]{0.3\textwidth}
         \centering
         \includegraphics[width=1\textwidth]{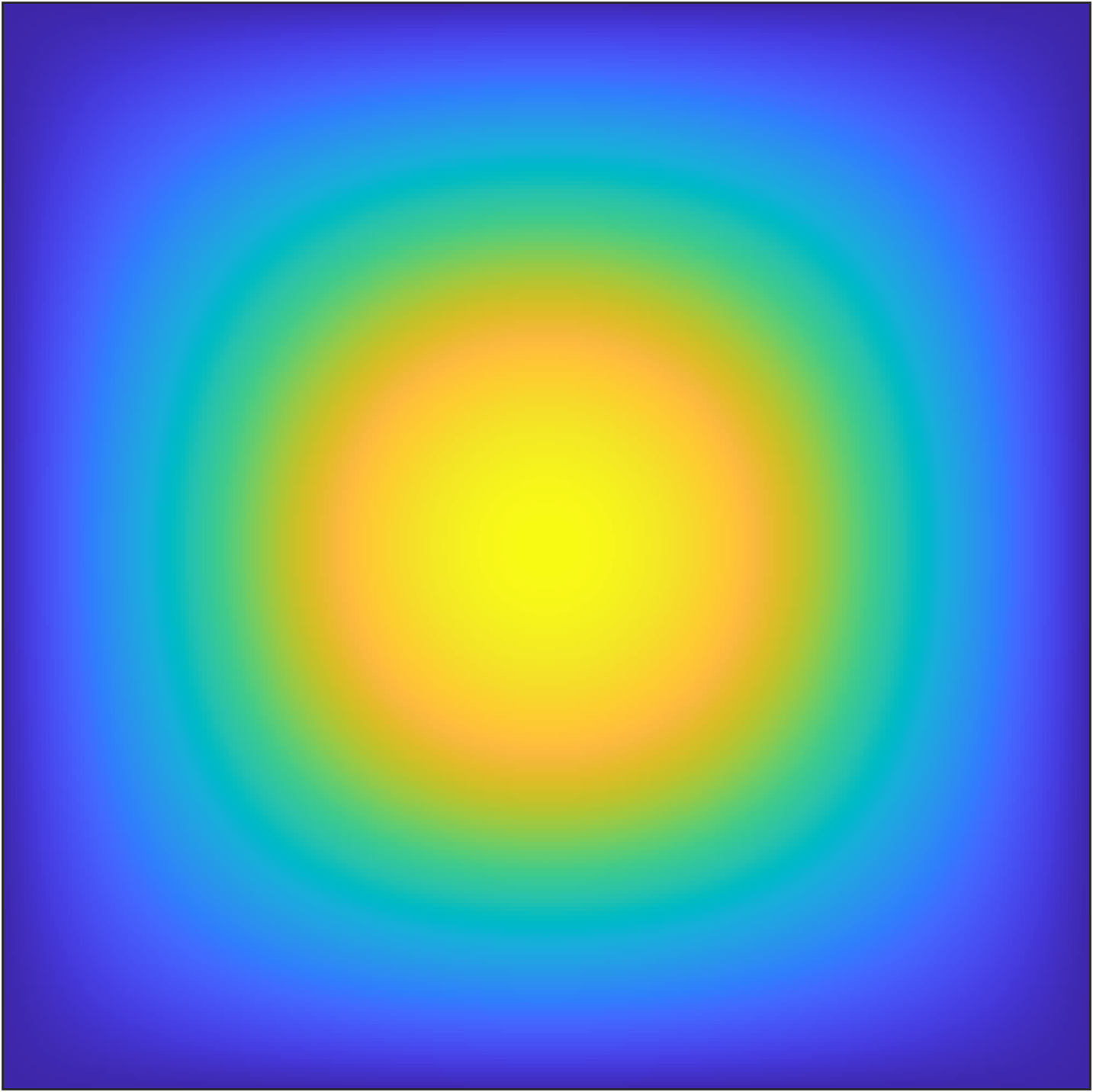}
         \caption{Coluna 1}
         \label{fig:01_006a}
     \end{subfigure}
     \hfill
     \begin{subfigure}[b]{0.3\textwidth}
         \centering
                  \includegraphics[width=1\textwidth]{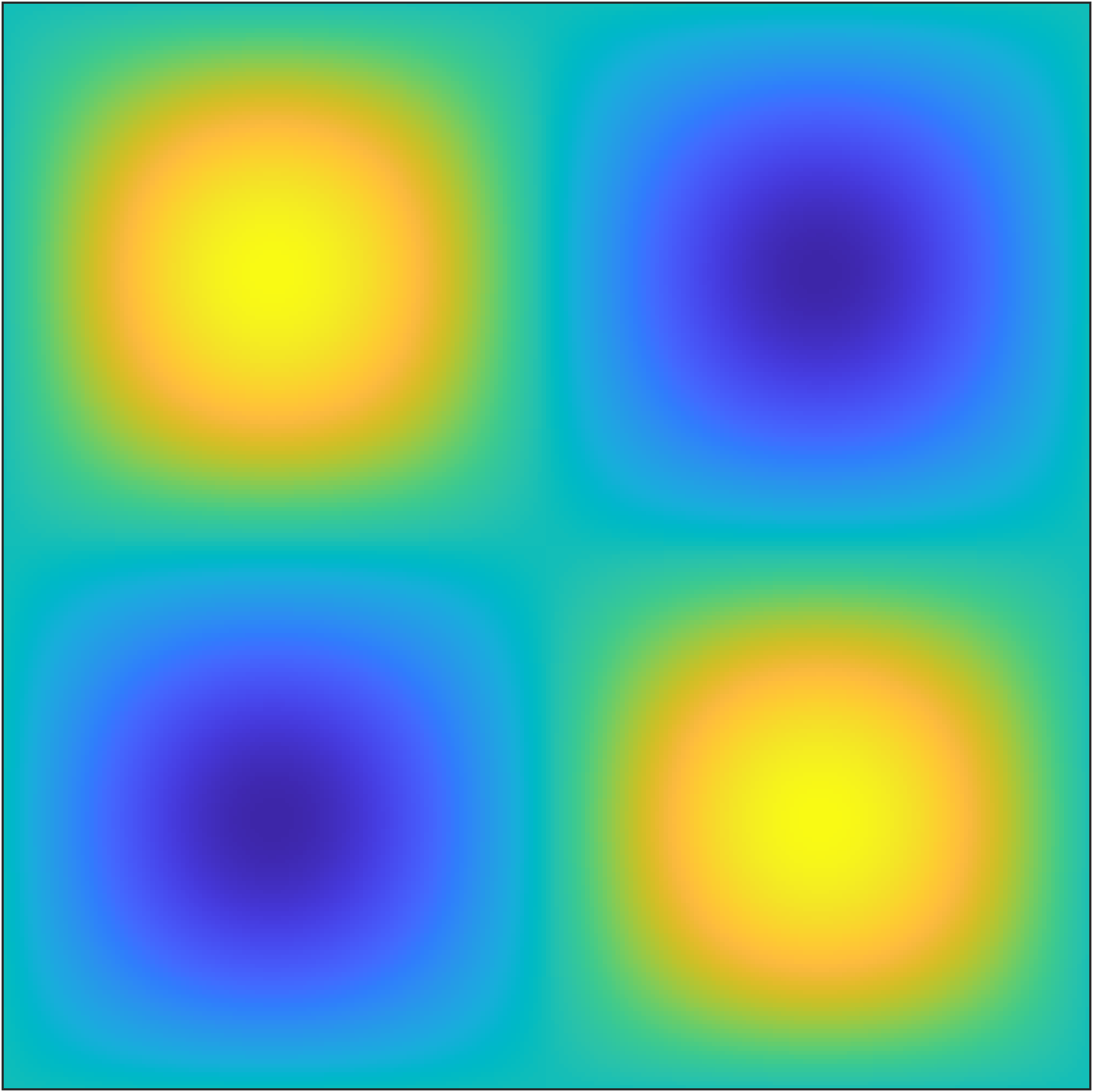}
         \caption{Coluna 4}
         \label{fig:01_006b}
     \end{subfigure}
     \hfill
          \begin{subfigure}[b]{0.3\textwidth}
         \centering
                  \includegraphics[width=1\textwidth]{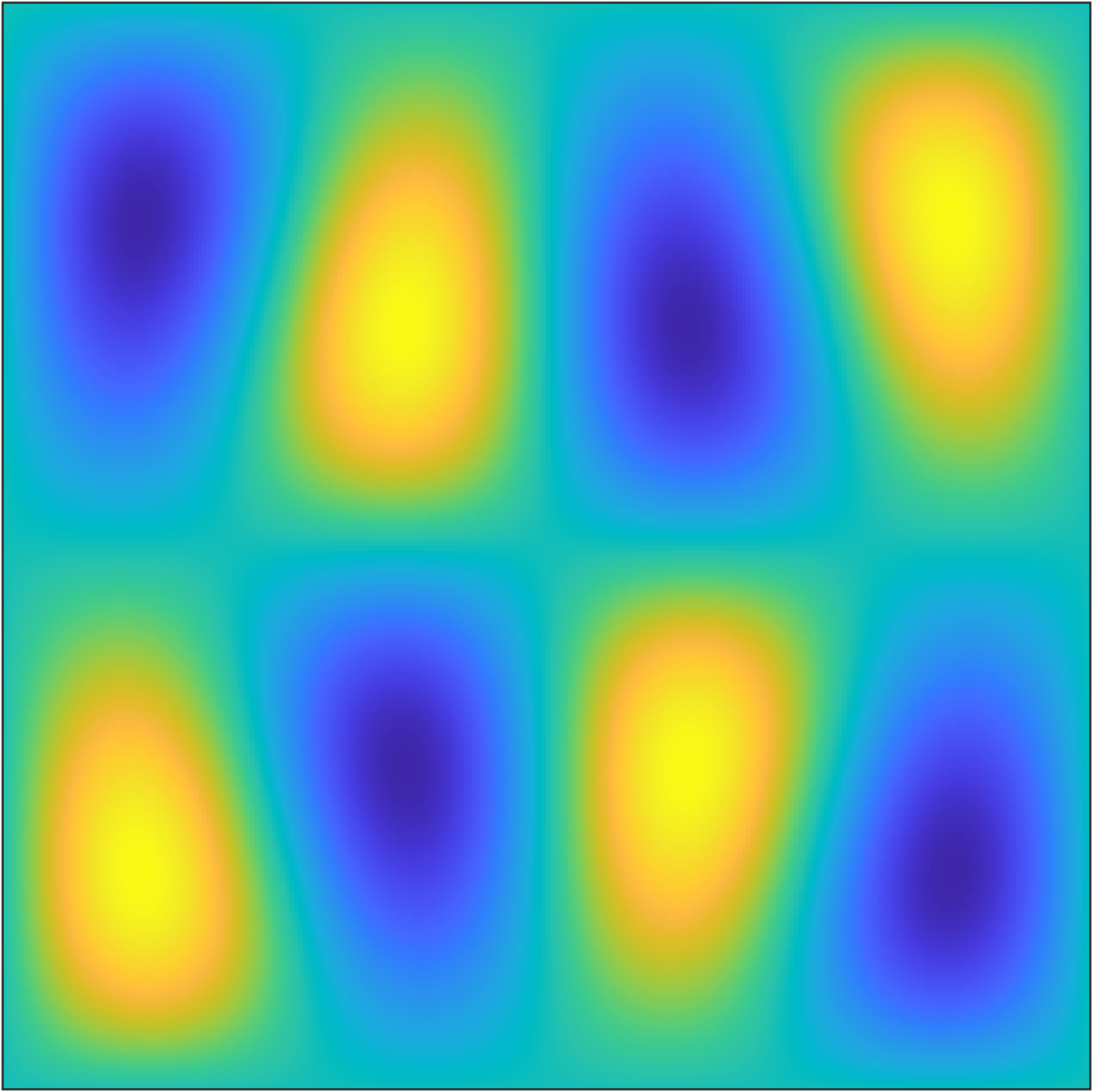}
         \caption{Coluna 12}
         \label{fig:01_006c}
     \end{subfigure}
          \begin{subfigure}[b]{0.3\textwidth}
         \centering
         \includegraphics[width=1\textwidth]{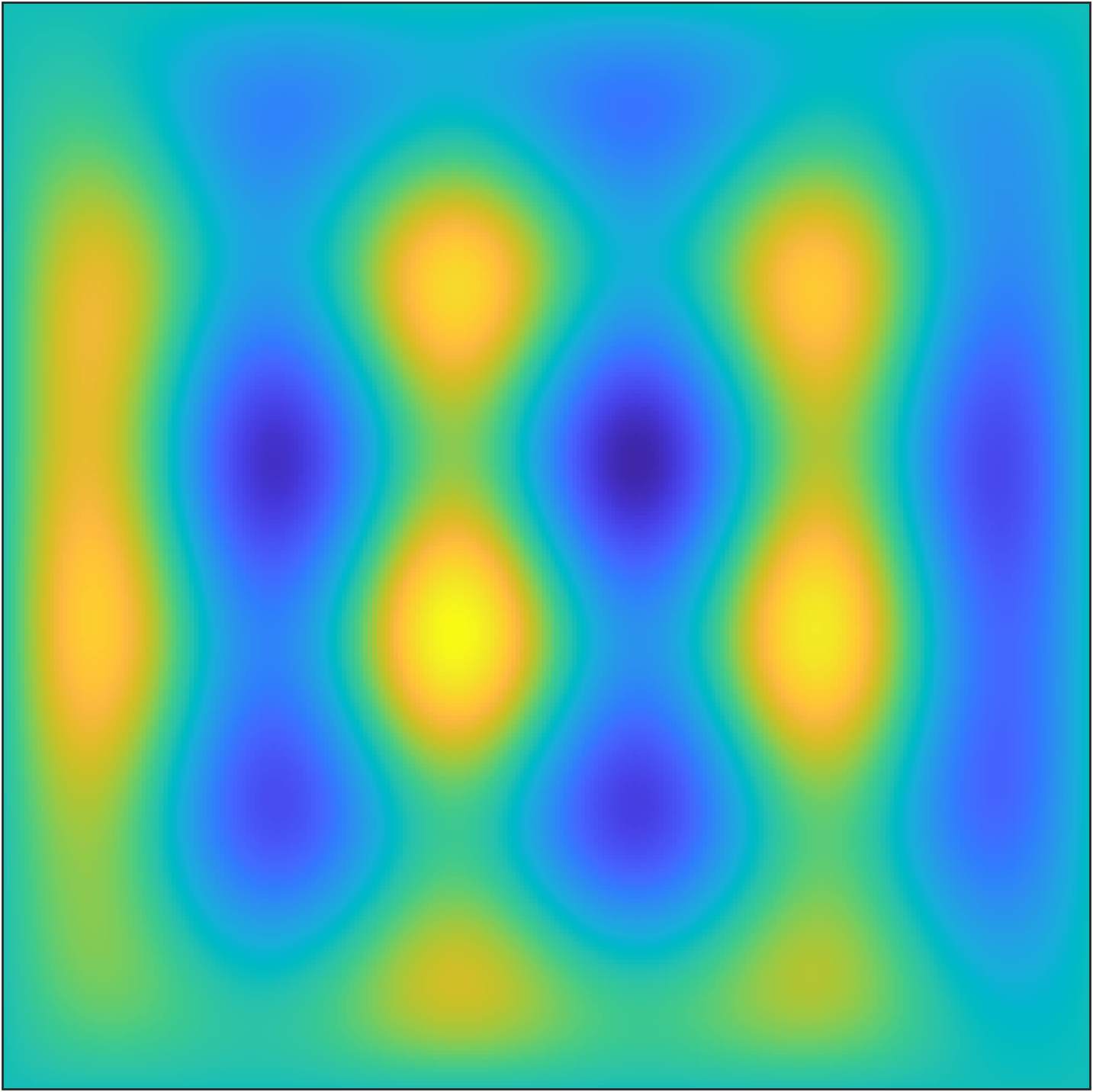}
         \caption{Coluna 24}
         \label{fig:01_006d}
     \end{subfigure}
     \hfill
     \begin{subfigure}[b]{0.3\textwidth}
         \centering
                  \includegraphics[width=1\textwidth]{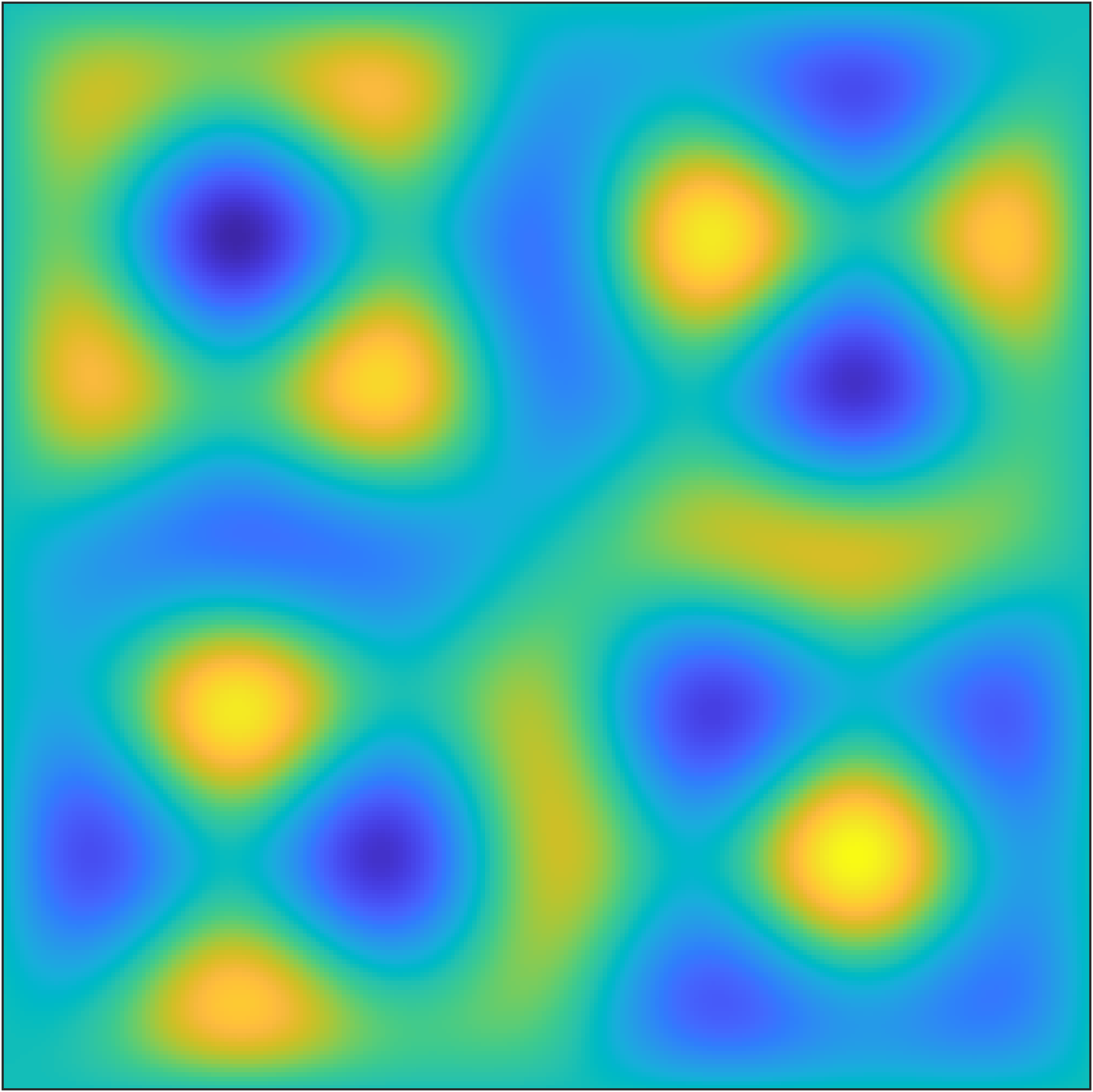}
         \caption{Coluna 36}
         \label{fig:01_006e}
     \end{subfigure}
     \hfill
          \begin{subfigure}[b]{0.3\textwidth}
         \centering
                  \includegraphics[width=1\textwidth]{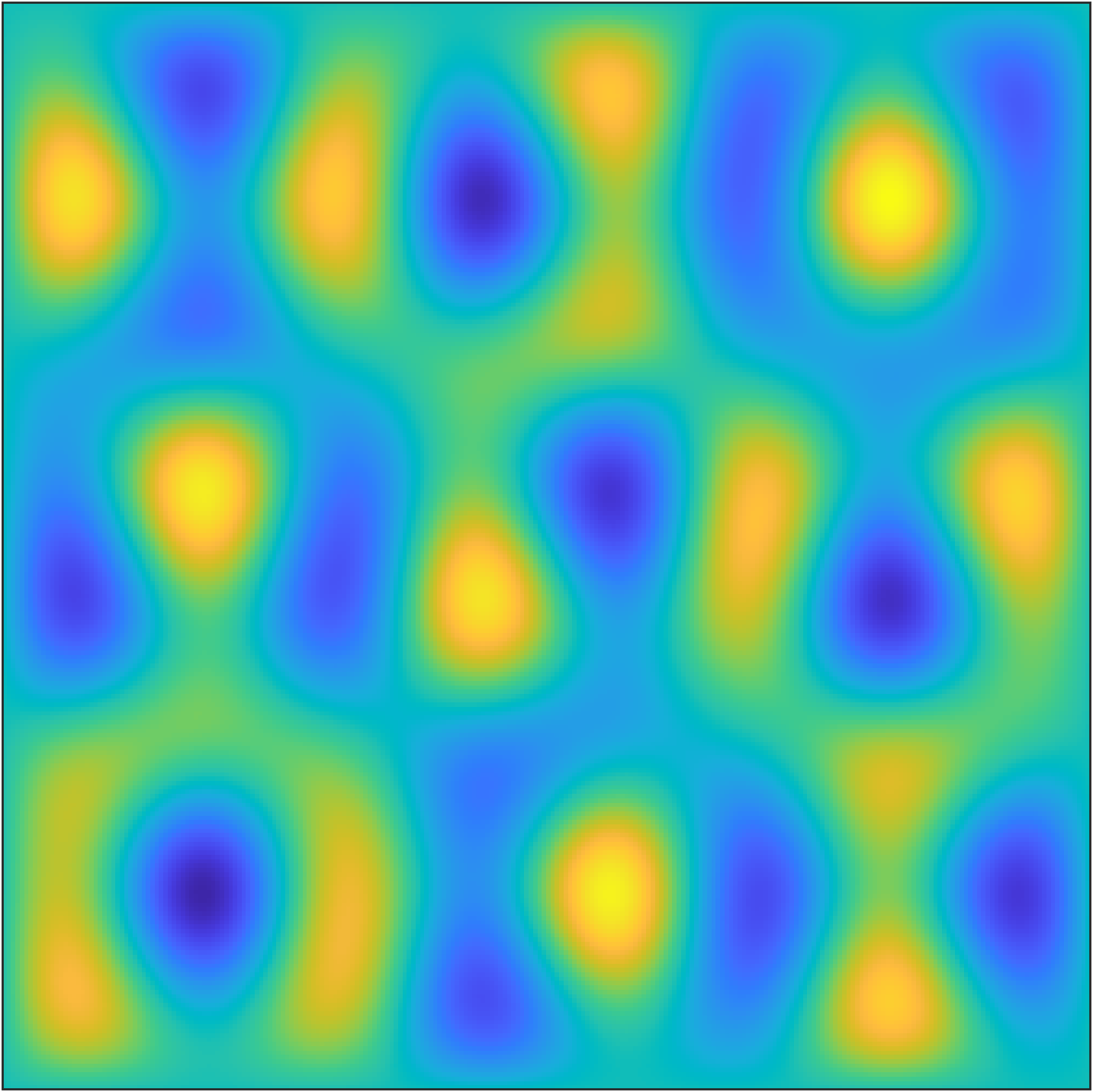}
         \caption{Coluna 50}
         \label{fig:01_006f}
     \end{subfigure}
\caption[Vetores $\mathbf{v}_i$ como imagens associadas aos $\sigma_i$.]{Vetores $\mathbf{v}_i$ como imagens associadas aos $\sigma_i$. Fonte: Próprio autor.}
\label{fig:01_006}
\end{figure}

Quando o problema é 2D, os vetores singulares à direita também podem ser expressos como imagens \cite[pág. 62]{hansen2006deblurring}. Seja a SVD da matriz $\mathbf{A}$ relativo ao \textit{blur} da Figura \ref{fig:01_002b}. Na Figura \ref{fig:01_006} são vistos vetores singulares a direita na forma de imagens (matrizes), relacionados com valores singulares $\sigma_i$. Conforme o valor singular $\sigma_i$ diminui, as frequências de $\mathbf{v}_i$ aumentam. Ou seja, a SVD é outra base de representação para $\mathbf{A}$, mas o significado físico dos vetores e valores singulares já não são tão claros quanto as colunas de $\mathbf{A}$ ou $\mathbf{x}$ originais. 

\subsection{Quais são exemplos de ruídos presentes em imagens?}

\begin{figure}[H]
     \centering
     \begin{subfigure}[t]{0.4\textwidth}
         \centering
         \includegraphics[width=0.8\textwidth]{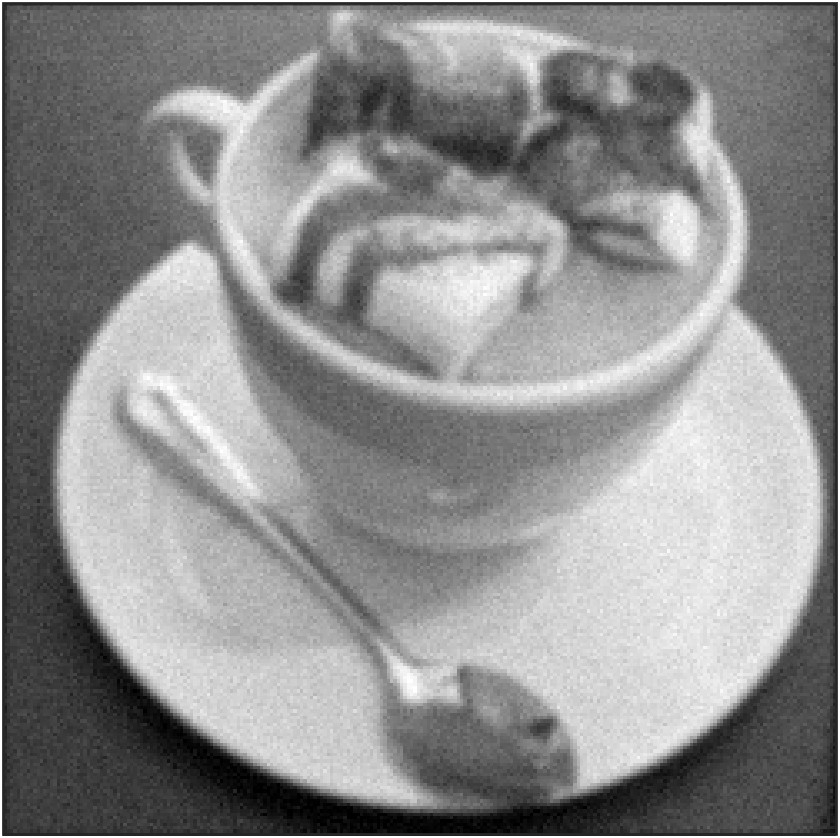}
         \caption{Imagem com ruído gaussiano}
         \label{fig:01_004a}
     \end{subfigure}
     \begin{subfigure}[t]{0.4\textwidth}
         \centering
                  \includegraphics[width=0.8\textwidth]{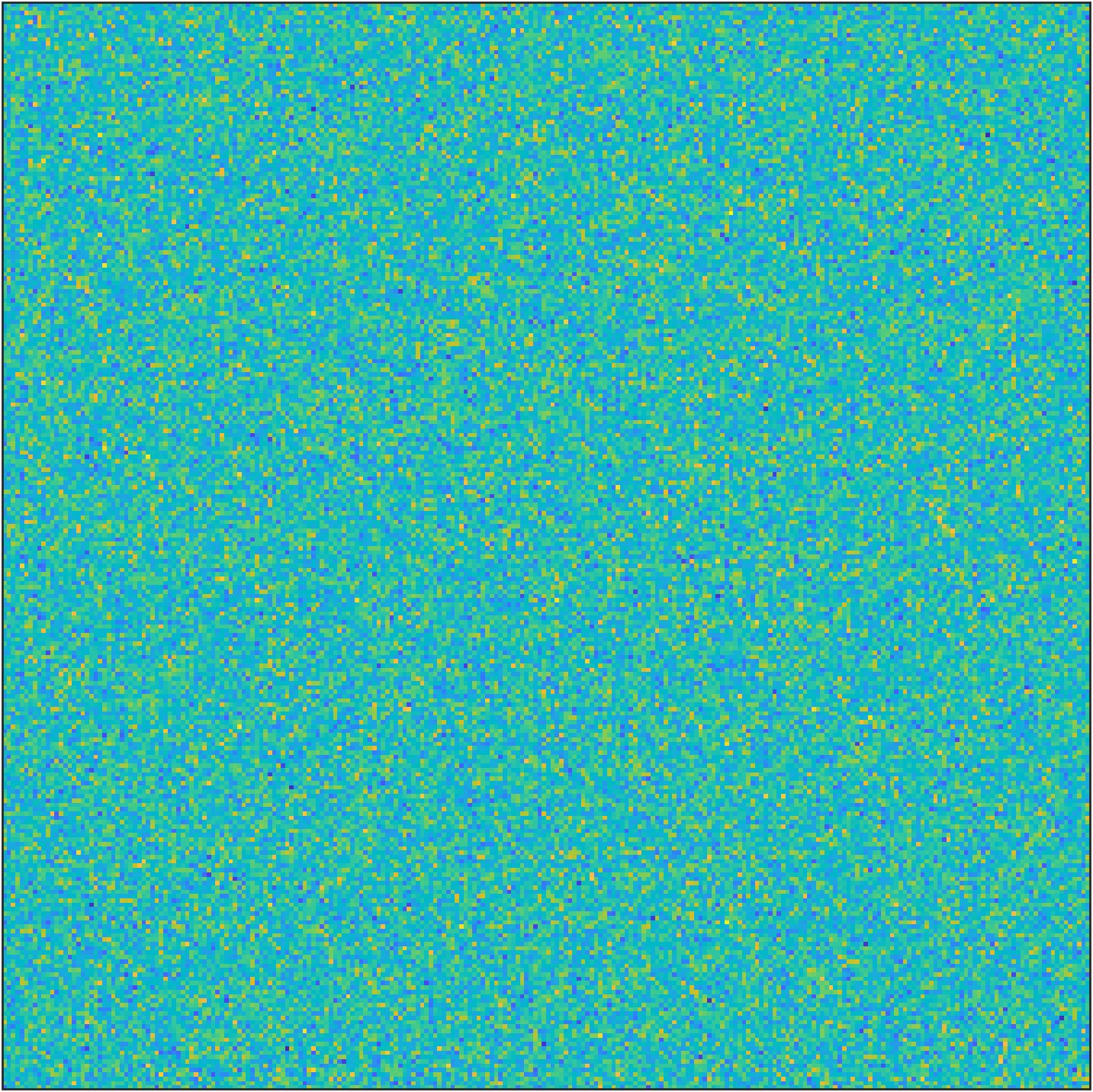}
         \caption{Ruído gaussiano}
         \label{fig:01_004b}
     \end{subfigure}     
     
          \begin{subfigure}[t]{0.4\textwidth}
         \centering
         \includegraphics[width=0.8\textwidth]{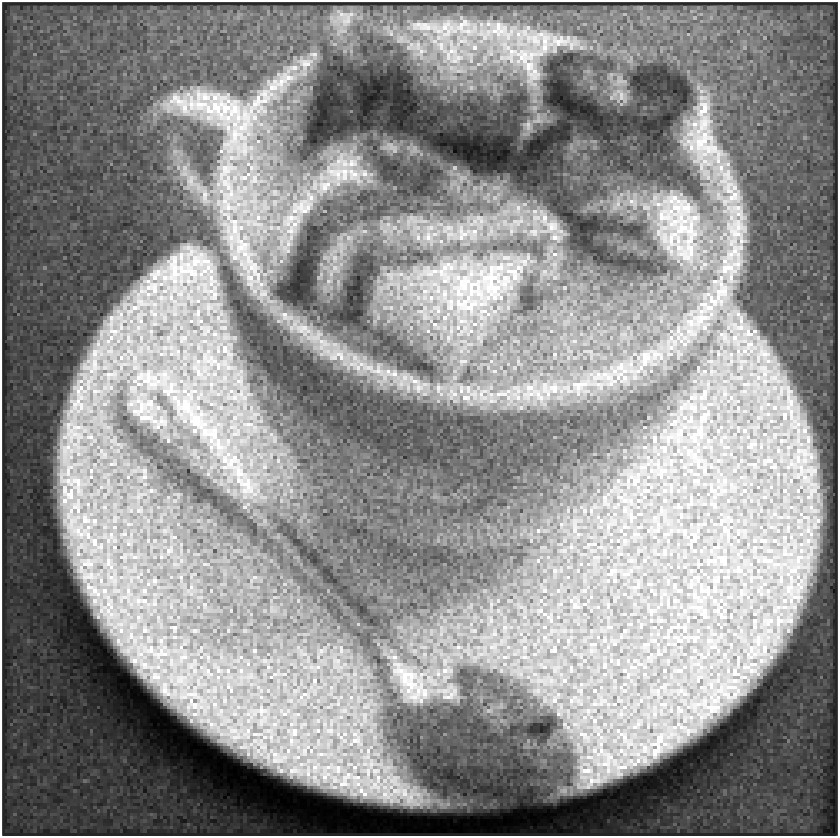}
         \caption{Imagem com ruído de Poisson}
         \label{fig:poissona}
     \end{subfigure}
     \begin{subfigure}[t]{0.4\textwidth}
         \centering
                  \includegraphics[width=0.8\textwidth]{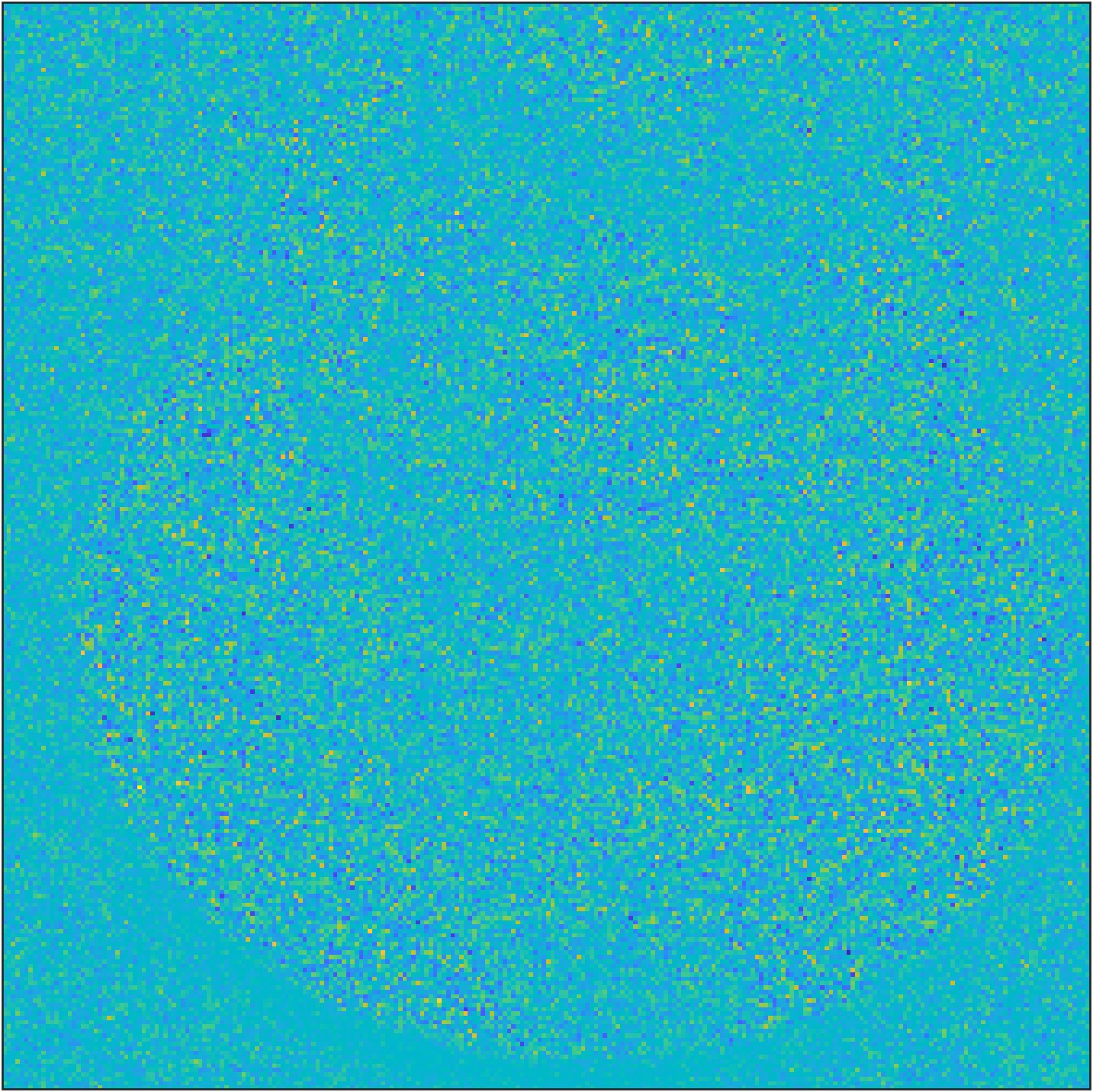}
         \caption{Ruído de Poisson}
         \label{fig:poissonb}
     \end{subfigure}

\caption[Ruídos gaussiano e de Poisson.]{Ruídos gaussiano e de Poisson. Fonte: Próprio autor}
\label{fig:01_004}
\end{figure}

Além do modelo de \textit{blur}, é necessário considerar que os dados podem apresentar ruído. Em dados experimentais, o ruído pode ser desconhecido, o que exige sua estimação para cada caso ou a sua aproximação para um modelo de ruído conhecido. Na Figura \ref{fig:01_004} são mostrados o ruído gaussiano e o ruído de Poisson. Ao invés da forma vetorizada $\bm{\delta}$, nesse caso o ruído pode ser redimensionado para uma matriz e visualizado como uma imagem. 

\begin{itemize}
\item  Um modelo de ruído $\bm{\delta}$ convencional é o ruído aditivo, branco e gaussiano, com média $\mu = 0$ e covariância $ \mathbf{\Gamma} = s^2 \mathbf{I}$, podendo ser escrito como $\bm{\delta} \sim \mathcal{N}(0,s^{2} \mathbf{I})$. Considera-se que não há correlação entre o vetor de medidas $\mathbf{y}$ e $\bm{\delta}$, seguindo a notação vetorizada,  e que não há correlação entre realizações diferentes de $\bm{\delta}$, sendo sua matriz de covariância  uma matriz identidade escalada pela sua variância $s^2$ \cite[seção 3.5.1]{hansen2010discrete}. As Figuras \ref{fig:01_004a} e  \ref{fig:01_004b} mostram a adição de ruído gaussiano na imagem, que se distribui da mesma forma em toda a imagem. 
 
  \item  Em outras aplicações, o ruído é caracterizado por uma distribuição de Poisson \cite{Bertero2009} sendo proporcional ao sinal (e não mais aditivo). Um exemplo é mostrado nas Figuras \ref{fig:poissona} e \ref{fig:poissonb}, na qual se observa que ele acompanha o formato do objeto.
  
  \end{itemize}

\subsection{Quais são diferentes problemas inversos que envolvem deconvolução?}

O objetivo do presente estudo é abordar o problema inverso, recuperar imagens nítidas a partir das imagens borradas e ruidosas. Em dados reais, é provável que haja ruído nas medidas. Mesmo assim, há trabalhos que chamam este problema apenas de \textit{deblurring}, enquanto outros o caracterizam como \textit{deblurring} e \textit{denoising} simultâneos. Considerando a notação da Equação \eqref{eq:convdisc3}, o \textit{deblurring} é um problema de deconvolução e diferentes problemas inversos são relacionados:
\begin{itemize}
\item \textbf{Estimação de parâmetros}\footnote{Estimação de parâmetros é um termo mais geral na área de problemas inversos. Nesse caso, é usual chamar de \textit{non-blind deblurring}, pois o modelo é conhecido.}: Quando se deseja estimar $\mathbf{X}$ dos parâmetros a partir do conhecimento do sistema $\mathbf{H}$ e da saída $\mathbf{Y}$ obtidas \cite{aster2019parameter}. No \textit{deblurring}, é quando se busca a imagem nítida original a partir da imagem borrada e do modelo de \textit{blur}. O conhecimento da PSF pode ser exato ou não. Trabalhos como \cite{Bertero2009} chamam o problema de deconvolução míope quando a PSF é apenas aproximada.

\item \textbf{Identificação de sistemas}: É um termo mais geral na área de problemas inversos. Nesse caso, trata-se da estimação da PSF, quando se quer estimar $\mathbf{H}$ a partir da entrada $\mathbf{X}$ e da saída $\mathbf{Y}$ \cite{aster2019parameter, baumeister2005topics}. No \textit{deblurring}, é quando se busca a PSF a partir das imagens nítida e borrada disponíveis \cite[págs. 15, 24]{meresescu2018};
\item \textbf{Deconvolução cega}: Quando se deseja estimar tanto $\mathbf{X}$ quanto $\mathbf{H}$. No \textit{deblurring}, seria o caso de estimar tanto a imagem nítida quanto a PSF que a borrou, tendo disponível apenas da imagem borrada \cite[págs. 99-100]{hansen2006deblurring}.
\end{itemize}

De acordo com \cite{Yu2014}, a expressão \textit{reconstrução} indica que há um modelo matemático de degradação associado ao problema, seja na notação $\mathbf{H}$ e $\mathbf{N}$ ou $\mathbf{A}$ e $\bm{\delta}$. Logo, estimação de parâmetros, identificação de sistemas e deconvolução cega são reconstruções que exigem, para cada uma delas, estratégias específicas. Uma expressão mais geral também encontrada é melhoramento de imagem, que é utilizada, segundo o mesmo autor, quando não há necessidade de menção explicita a um modelo. 

\subsection{Como escrever o problema de super-resolução pela composição de mais de um operador direto?}
Em algumas aplicações é possível separar as contribuições do operador direto em matrizes distintas. Um exemplo é conhecido como o problema de super-resolução \cite[Subseções 3.9.6.3 e 3.13.7]{bovik2005handbook}. Na literatura, ele é encontrado de duas formas: uma que considera apenas a subamostragem do sinal original; e uma outra em que há dois fatores de degradação: ao mesmo tempo que o sinal é subamostrado, ele também é degradado, por exemplo borrado.

 O segundo caso pode ser separado em um operador $\mathbf{A}$ de degradação e um $\mathbf{A}_{sub}$ de subamostragem, conforme
\begin{equation}
\mathbf{A}_{sub} = \frac{1}{2}\begin{pmatrix}
1 & 1 & 0 & 0 & 0 & 0 & 0 & 0 &  \cdots\\
0 & 0 & 1 & 1 & 0 & 0 & 0 & 0 & \\
0 & 0 & 0 & 0 & 1 & 1 & 0 & 0 & \cdots\\
0 & 0 & 0 & 0 & 0 & 0 & 1 & 1 & \\
\vdots & & \vdots &  & \vdots &  & \vdots & & \ddots\\
\end{pmatrix} \in \mathbb{R}^{(n/2) \times n}.
 \label{eq:downs} 
\end{equation}
É importante notar que  nesse caso a ordem de multiplicação as matrizes afeta o resultado final. Considerando que primeiro o sinal é subamostrado para ter metade da(s) dimensão(ões) original(is) e depois ele é borrado, tem-se: 
\begin{itemize}
\item No caso 1D, seja $\mathbf{x}$ um sinal nítido e $\mathbf{y}$ o sinal subamostrado e suavizado. Assim,
\begin{equation}
\underset{\frac{n}{2} \times 1}{\mathbf{y}_{downsampling}} \quad = \quad \underset{\frac{n}{2} \times n}{\mathbf{A}_{sub}} \quad \underset{n \times n}{\mathbf{A}} \quad \underset{n \times 1}{\mathbf{x}}. 
\label{eq:super1}
\end{equation}
\item No caso 2D, é necessário subamostrar as direções horizontal e vertical. Seja $\mathbf{x}$ uma imagem nítida e de alta resolução que foi vetorizada, de modo que $\mathbf{A}\mathbf{x}$  uma imagem borrada. Converte-se suas dimensões de volta para uma imagem, denotada por $\mathbf{Y}_{blurred}$. Assim, a obtenção da baixa resolução é possível através de 
\begin{equation}
\underset{\frac{n}{2} \times \frac{n}{2}}{\mathbf{Y}_{downsampling}} \quad = \quad \underset{\frac{n}{2} \times n}{\mathbf{A}_{sub}} \quad \underset{n \times n}{\mathbf{Y}_{blurred}}\quad  \underset{n \times \frac{n}{2}}{\mathbf{A}_{sub}^T}. 
\label{eq:super2}
\end{equation}
\end{itemize}

A escrita do modelo na forma de um sistema linear, que envolve multiplicação matriz-vetor, pode não ser a mais eficiente computacionalmente, principalmente em problemas de grande escala. Conforme ainda será discutido, existem algoritmos que implementam essas operações sem montagem explícita das matrizes $\mathbf{A}$ e $\mathbf{A}_{sub}$, diminuindo a necessidade de armazenamento de grandes matrizes.

\newpage
\section{REGULARIZAÇÃO VARIACIONAL: QUAL É O MÉTODO UTILIZADO PARA TENTAR RESOLVÊ-LOS?}\label{sec:variational}

\subsection{Introdução}

Problemas diretos e problemas inversos dependem da definição de um mesmo operador direto $\mathcal{A}$, o que muda é o que se deseja calcular, se é a entrada, parâmetros do sistema e/ou a saída.  Quando $\mathcal{A}$ é discretizado em uma matriz $\mathbf{A}$, ele passa a ter dimensão finita, podendo ser utilizados métodos computacionais para solução dos dois problemas. O desafio é que muitos problemas inversos são mal-postos, os dados apresentam ruídos e a inversão ingênua através de $\mathbf{A}^{-1}$ não é suficiente para obter soluções adequadas.

A partir do trabalho de Tikhonov \cite{tikhonov1977solutions}, soluções para problemas mal-postos podem ser obtidas através do método de regularização \cite{Benning2018}, um método que na sua forma original é considerado bem entendido \cite{Gerth2021}, mas que continua permitindo o desenvolvimento de novas propostas e extensões \cite{Arridge2019}. Delimitando o escopo\footnote{No Capítulo \ref{sec:polysemy}, discute-se regularização de modo mais geral.}, o presente capítulo visa discutir a regularização variacional \cite[pág. 8]{Benning2018}, cuja forma é dada a partir do funcional\footnote{De acordo com \cite[pág. 8]{Montegranario2014}, funcional denota um mapeamento de um conjunto de funções para um conjunto de números.}
\begin{equation}
\hat{\mathbf{x}}_{\lambda} = \arg\min\limits_{\mathbf{x}} \left[ \mathcal{L} \left(\mathbf{A}, \mathbf{x}, \mathbf{y} \right) + \lambda^2 \Omega(\mathbf{x}) \right],
 \label{eq:tikhonovgeral}
\end{equation}
onde  $\Omega(\mathbf{x})$ é o termo de regularização (ou funcional de regularização, ou regularizador \cite[pág. 1129]{Caselle2011}) e mede a regularidade da solução \cite[págs. 61, 172]{hansen2010discrete}; $\mathcal{L}$ é a função de perda que mede a diferença dos dados para a saída do modelo; $\lambda$ é um parâmetro de regularização que define quanto peso será dado para cada termo. 

Logo, na regularização variacional é necessário definir $\mathcal{L}(\cdot)$, $\Omega(\mathbf{x})$ e a forma como $\lambda$ será escolhido, mas a sua ideia geral continua sendo a de transformar um problema mal-posto em um problema bem-posto. 

O termo $\Omega(\mathbf{x})$ permite a restrição do espaço de soluções a partir das informações suplementares sobre ela disponíveis \cite[ pág. 59]{tikhonov1977solutions}. Diferentes regularizadores podem refletir diferentes características esperadas das soluções, como a suavidade \cite[ pág. 70]{tikhonov1977solutions}, características físicas  \cite[pág. 8]{Bertero2021}, esparsidade \cite{Daubechies2004, Daubechies2016}, entre outros. Um requisito importante é que $\Omega(\mathbf{x})$ seja baseado nas informações \textit{a priori} que se tem da solução $\mathbf{x}$, sendo independentes das medidas $\mathbf{y}$. Disso o nome \textit{a priori} ou \textit{prior} \cite[pág. 2]{kaipio2005statistical}. 

Na proposta original de Tikhonov, o termo $\Omega(\mathbf{x})$ não era arbitrário. Enquanto a função de perda era quadrática, o termo de regularização $\Omega(\mathbf{x})$ era conhecido como \textit{stabilizing functional}, apresentando propriedades como ser contínuo, não-negativo \cite[pág. 50-1, 71]{tikhonov1977solutions} e quase monotônico \cite[pág. 56]{tikhonov1977solutions}, \cite[págs 420-1]{Vapnik2006}. Nesse contexto, o regularizador pode ser visto como um termo de penalização da literatura de otimização, o que permite interpretar ambos como uma restrição suave \cite[pág. 20]{andreasson2020an}, em que o algoritmo é desestimulado a seguir o determinado caminho de soluções indesejadas.  Para compor um método de regularização, ainda são necessários critérios para a escolha de $\lambda$. Estes serão discutidos brevemente no contexto da regularização clássica de Tikhonov, mas em \cite[Capítulo 7]{Hansen1998} é encontrada uma revisão maior sobre o assunto. 

Outra teoria desenvolvida para solução de problemas inversos é a inversão bayesiana \cite{tarantola2005inverse}, discutida no Apêndice \ref{Ap:estimador}. Nela, quando utilizado o estimador máximo \textit{a posteriori} (MAP), a escolha de $\mathcal{L}$ é guiada pelo modelo de ruído (usualmente aditivo), enquanto a escolha de $\Omega(\mathbf{x})$ é uma densidade de probabilidade \textit{a priori} de $\mathbf{x}$ \cite{kaipio2005statistical}. As combinações de $\Omega(\mathbf{x})$ e $\mathcal{L}(\cdot)$ resultam em diversos métodos encontrados na literatura \cite[pág. 111]{alvarez2017digital}. 

Neste capítulo, os exemplos consideram o termo de fidelidade quadrático, isto é, $\mathcal{L} \left(\mathbf{A}, \mathbf{x}, \mathbf{y} \right) = \vert \vert \mathbf{A} \mathbf{x} - \mathbf{y} \vert \vert^2_2$, quando o modelo de ruído é considerado branco e gaussiano \cite[Exemplo 5]{kaipio2005statistical}, mas sem explorar a inversão bayesiana de fato. 

Na sequência, são descritas diferentes formas para o regularizador $\Omega(\mathbf{x})$\footnote{Ver Capítulo \ref{sec:de_nonblind} para ilustração de alguns deles em aplicações de processamento de imagens.}. É importante ressaltar que uma implementação adequada de regularização depende do problema que se quer resolver \cite[pág. 192]{hansen2010discrete} e que não existe um método de regularização que seja superior a outro, pois cada um tem suas vantagens dependendo da aplicação desejada \cite[pág. 2]{Hansen1998}.

\subsection{Qual é a forma e significado da regularização clássica de Tikhonov?}\label{sec:tikhclas}

Possivelmente, a forma mais conhecida da Equação \eqref{eq:tikhonovgeral} é a regularização clássica de Tikhonov.  Ela apresenta tanto $\mathcal{L}$ quanto $\Omega(\mathbf{x})$ na forma de norma $\ell_2^2$. Em um certo sentido, esse método se assemelha à união das  Equações \eqref{eq:otimizacao1} e \eqref{eq:casosub_sol} para problemas subdeterminados e sobredeterminados, pois ao mesmo tempo em que se busca minimizar um termo quadrático, é colocada uma restrição na norma quadrática do vetor $\mathbf{x}$. 

O operador de regularização proposto por Tikhonov trata a inversão como um problema de otimização, sendo conhecida como formulação variacional. Seja o funcional $\mathcal{M}(\lambda, \mathbf{x}, \mathbf{y})$ definido considerando-se $\Omega(\mathbf{x}) = \vert \vert \mathbf{x} \vert \vert_2^2$, conforme
\begin{equation}
\mathcal{M}(\lambda, \mathbf{x}, \mathbf{y}) = \vert \vert \mathbf{A} \mathbf{x} - \mathbf{y} \vert \vert^2_2 + \lambda^2 \vert \vert \mathbf{x} \vert \vert_2^2.
\label{eq:functional}
\end{equation}
O problema desejado de otimização sem restrições é obtido pela minimização de $\mathcal{M}(\lambda, \mathbf{x}, \mathbf{y})$, obtendo-se  a solução regularizada $\hat{\mathbf{x}}_\lambda$ de acordo com
\begin{equation}
\hat{\mathbf{x}}_{\lambda} = \arg\min\limits_{\mathbf{x}} \left[\mathcal{M}(\lambda, \mathbf{x}, \mathbf{y}) \right].
\label{eq:functional2}
\end{equation}
Substituindo a Equação \eqref{eq:functional} na Equação \eqref{eq:functional2}, a sua função objetivo\footnote{Esse termo também é encontrado como  função custo ou função de energia} é dada por  
\begin{equation}
\hat{\mathbf{x}}_{\lambda} = \arg\min\limits_{\mathbf{x}} \left[ \vert \vert \mathbf{A} \mathbf{x} - \mathbf{y} \vert \vert^2_2 + \lambda^2 \vert \vert  \mathbf{x} \vert \vert_2^2 \right].
\label{eq:tikhonov1}
\end{equation}
Na Equação \eqref{eq:tikhonov1}, tanto o termo de fidelidade quanto o regularizador resultam em valores escalares. A norma nesse caso funciona como uma medida do comprimento do argumento $\mathbf{x}$, um vetor.  Isso também é válido para a Equação \eqref{eq:tikhonovgeral}, o que permite entender $\mathcal{L}$ como um critério/medida sobre $\mathbf{A}$, $\mathbf{x}$ e $\mathbf{y}$, enquanto $\Omega$ atua apenas sobre  $\mathbf{x}$. 

O Apêndice \ref{Ap:interp} traz no exemplo de interpolação de um sinal ruidoso uma forma de visualizar o papel da regularização no controle da norma da solução. 

\subsubsection{Quais são os três componentes básicos da regularização clássica de Tikhonov?}
   
Na Equação \eqref{eq:tikhonov1}, há dois termos e um parâmetro \cite[pág. 8]{Benning2018}. Sobre eles:
\begin{itemize}
\item O termo de fidelidade dos dados $\vert \vert \mathbf{A}\mathbf{x} - \mathbf{y}\vert \vert^2_2$ denota uma medida de performance do algoritmo \cite[pág. 308]{Bertero2021}, pois traz a diferença entre os valores produzidos pelo modelo $\mathbf{A}\mathbf{x}$ e os dados medidos $\mathbf{y}$. Ele pode ser visto como $\vert \vert \mathbf{y}_{calculado} - \mathbf{y}_{medido}\vert \vert^2_2$, um resíduo que representa o quão bem o modelo consegue prever as medidas ruidosas. Se este termo for muito grande, a solução não é boa, pois de fato o problema não foi resolvido. Por outro lado, esse residual não deve ser tão pequeno quando a ordem de valores dos ruídos, para que eles não sejam ajustados juntos na solução. Como esse termo inclui o operador direto $\mathbf{A}$, ele traz informações e restrições das propriedades físicas do problema;
\item O termo $ \vert \vert \mathbf{x}\vert \vert^2_2$ visa controlar a norma do vetor $\mathbf{x}$ de modo que ela seja pequena e que possa suprimir parte da amplificação dos ruídos decorrente da inversão. A norma deve ser mínima dentro das condições físicas impostas pelo modelo $\mathbf{A}$ e suas condições de contorno, que façam sentido físico para a respectiva aplicação. Como $\mathbf{x}$ representa uma quantidade física e o termo de fidelidade traz as restrições físicas do problema, não se espera que o resultado desse regularizador seja nulo, mas sim que ele restrinja o tamanho de $\mathbf{x}$. O resultado esperado é que $ \vert \vert \mathbf{x}\vert \vert^2_2$ tenha a menor norma possível considerando o modelo físico disponível;
\item O equilíbrio entre esses dois termos é dado pelo parâmetro de regularização $\lambda$ para poder estabilizar a inversão da matriz \cite{Prato2008}. Há um \textit{trade-off} entre a acurácia e a estabilidade \cite[pág. 2804]{Chen2002}: Quanto maior $\lambda$, maior peso à minimização de $ \vert \vert \mathbf{x}\vert \vert^2_2$, tornando-o cada vez mais regular; Quanto menor $\lambda$, maior peso para o ajuste dos dados ruidosos, resultando em soluções menos regulares. Com $\lambda = 0$ obtém-se a solução ingênua, que não é adequada, mas é natural pensar em $\lambda$ como um número pequeno, para não enviesar muito a solução. 

\end{itemize}

É importante notar que a escolha de cada componente também depende da dimensionalidade do problema que se quer resolver. O Apêndice \ref{sec:class} discute diferentes abordagens para a Equação \eqref{eq:tikhonov1} considerando o tamanho do sistema, enquanto o Apêndice \ref{sec:otimi1} discute otimizadores específicos e \textit{toolboxes} encontradas na literatura.

\subsubsection{Como é possível minimizar o funcional obtido?}
Reescreve-se a Equação \eqref{eq:tikhonov1} como um problema de mínimos quadrados \cite[pág. 61]{hansen2010discrete}, conforme
\begin{equation}
\hat{\mathbf{x}} = \arg\min\limits_{\mathbf{x}} 
\left|\left|
\begin{pmatrix}
\mathbf{A}\\ 
\lambda \textbf{I}\\ 
\end{pmatrix}\mathbf{x} - \begin{pmatrix}
\mathbf{y}\\ 
\mathbf{0}
\end{pmatrix} \right| \right|^2_2.
\label{eq:tikhonov18}
\end{equation} 
Calculando o gradiente em relação à $\mathbf{x}$ da Equação \eqref{eq:tikhonov18}, o argumento do minizador, que antes resultava em um escalar, vai resultar em um sistema linear de equações para obtenção de  $\mathbf{x}$, conforme desejado. Igualando o gradiente a zero, obtém-se a solução de mínimos quadrados dada por \cite[pág. 61]{hansen2010discrete}
\begin{equation}
\hat{\mathbf{x}}_{\lambda} =\left( \mathbf{A}^T \mathbf{A} + \lambda^2 \mathbf{I}^T \mathbf{I} \right)^{-1} \mathbf{A}^T \mathbf{y}, 
\label{eq:tiksol}
\end{equation}
conhecida como regularização clássica de Tikhonov, uma solução em um único passo. 

Observa-se que o termo $\vert \vert \mathbf{x} \vert \vert_2^2$ que controla a norma da solução na Equação \eqref{eq:tikhonov1} se reflete na adição de uma matriz identidade na Equação \eqref{eq:tiksol}, demonstrando a facilidade da sua implementação. Costuma-se escrever $\mathbf{I}^T \mathbf{I}$ apenas como $\mathbf{I}$, mas essa notação facilita o entendimento para quando é utilizada uma matriz $\mathbf{L}$ qualquer no termo de regularização, conforme será visto adiante.

Seguindo a notação de matriz pseudoinversa utilizada nas Equações \eqref{eq:normalequation} e \eqref{eq:normalequation3}, a Equação \eqref{eq:tiksol} pode ser reescrita de modo compacto conforme  
\begin{equation}
\hat{\mathbf{x}}_{\lambda} = \mathbf{A}^+_{\lambda} \mathbf{y},
\label{eq:tiksol2}
\end{equation}
onde $\mathbf{A}^+_{\lambda} = \left( \mathbf{A}^T \mathbf{A} + \lambda^2 \mathbf{I} \right)^{-1} \mathbf{A}^T $ é uma matriz de reconstrução \cite{Adler2019}. Logo, a inversão regularizada se torna uma multiplicação por uma matriz. 

No Apêndice \ref{App:svd2}, a regularização clássica de Tikhonov é obtida a partir da decomposição em valores singulares, trazendo um outro ponto de vista sobre como essa regularização atua na solução dos problemas mal-postos.

\subsubsection{Quais são as principais regras de escolha de $\lambda$?}

O parâmetro de regularização $\lambda$ é componente da regularização variacional e sua escolha é de extrema importância para a solução que será obtida. A escolha visual do parâmetro de regularização pode servir como primeiro teste \cite[pág. 199]{bovik2005handbook}, mas nada garante que em outras condições (de medidas, nível de ruído, modelo) o parâmetro para uma determinada melhor reconstrução seja o mesmo.  Assim, é importante que existam métodos de escolha objetivos de $\lambda$ para evitar o \textit{overfitting}, quando os parâmetros estão demasiadamente ajustados para um determinado conjunto de dados, podendo falhar no tratamento de outro conjunto de dados. 

As regras para a escolha de $\lambda$ podem ser \textit{a priori}, quando assumem o conhecimento (ou a possibilidade de estimação) dos ruídos $\bm{\delta}$ nos dados. Um exemplo dessa estratégia é o  princípio da discrepância de Morozov \cite[págs. 83-4]{engl1996regularization}:
\begin{itemize} 
\item Não é razoável esperar que $\vert \vert \mathbf{A} \mathbf{x}_{\lambda} - \mathbf{y} \vert \vert^2_2$ seja menor do que $\vert\vert\bm{\delta}\vert\vert_2$ da medida, pois isso poderia implicar que a inclusão de ruídos em seu ajuste. Caso $\bm{\delta}$ seja conhecido, o princípio de Discrepância de Morozov é uma regra para escolha \textit{a posteriori} de $\lambda$ que parte da diferença entre as medidas e a reconstrução $\vert \vert \mathbf{A} \mathbf{x}_{\lambda} - \mathbf{y} \vert \vert_2 \leq \vert\vert\bm{\delta}\vert\vert_2$. Dessa forma, o parâmetro $\lambda$ através do princípio da discrepância pode ser calculado considerando que $\vert \vert \mathbf{A} \mathbf{x}_{\lambda} - \mathbf{y} \vert \vert_2 = \vert\vert\bm{\delta}\vert\vert_2$ \cite[pág. 90]{hansen2010discrete}. Há também o método de discrepância generalizado que permite incluir também informação sobre erros no operador direto $\mathbf{A}$ \cite[pág. 179-81]{Hansen1998}.
\end{itemize} 

Quando essa informação não está disponível, são utilizadas regras \textit{a posteriori}, pois não usam a informação explícita de $\bm{\delta}$ e são baseadas em valores calculados a partir dos dados disponíveis \cite[pág. 76]{kern2016numerical}. Regras \textit{a posteriori} também são conhecidas como heurísticas \cite[pág. 177]{hansen2010discrete} ou \textit{error-free} \cite[pág. 108]{engl1996regularization}. Dois exemplos dessa estratégia são a Curva-L e a validação cruzada generalizada:

\begin{itemize} 
\item Obtém-se a Curva-L calculando-se a norma da solução $\vert \vert \mathbf{x}_{\lambda} \vert \vert_2$ e a norma do residual $\vert \vert \mathbf{A} \mathbf{x}_{\lambda} - \mathbf{y} \vert \vert_2$ para uma ampla faixa de parâmetros $\lambda$ e gerando um gráfico $log-log$ deles \cite[págs. 73-5]{hansen2010discrete}. Ela possui esse nome pelo formato característico que ela exibe, lembrando a letra L. A partir da obtenção da Curva-L busca-se a solução que está exatamente no canto da curva, ou seja, na região de maior variação da curva. Essa solução indicaria um equilíbrio entre os termos da norma da solução e da norma residual \cite[pág. 71]{hansen2010discrete}. Para uma formulação matemática desse critério, evitando escolha visual do canto, ver \cite[pág. 110]{engl1996regularization};

\item O parâmetro $\lambda$ também pode ser escolhido através da validação cruzada generalizada (GCV), a partir da minimização de uma função $G(\lambda)$ \cite[págs. 116-7]{aster2019parameter} \cite[págs. 95-6]{hansen2010discrete}. Seja $c$ o número de pontos de dados e $\hat{\mathbf{x}}_{\lambda}$ a solução para um determinado $\lambda$, então se busca pelo menor valor de 
\begin{equation}
G(\lambda) = \frac{c \vert \vert \mathbf{A} \hat{\mathbf{x}}_{\lambda} - \mathbf{y} \vert \vert_2^2 }{\Tr\left (\mathbf{I} - \mathbf{A} \mathbf{A}^+_{\lambda}\right)}
\end{equation} 
\begin{equation}
\lambda = \underset{\lambda}{\arg\min} [G(\lambda)], 
\end{equation} 
onde $\mathbf{A}^+_{\lambda}$ é a matriz de reconstrução da Equação \eqref{eq:tiksol2}. O gráfico de $G(\lambda)$ possui uma parte plana (constante) e uma íngreme (crescente) para localizar a posição de seu valor mínimo. Assim como a Curva-L, são necessárias várias reconstruções para obtê-la, o que pode ser custoso em problemas de grande dimensionalidade. 
\end{itemize} 

\subsection{O que é a regularização generalizada de Tikhonov e como ela se diferencia da regularização clássica?}\label{sec:tikhgene}

O funcional da Equação \eqref{eq:tikhonov1} e sua solução na Equação \eqref{eq:tiksol} são expressões muito conhecidas e encontradas há décadas na literatura \cite[pág. 103]{tikhonov1977solutions}.  Apesar disso, ela é apenas uma parte dos métodos de regularização que existem. Mesmo Tikhonov e Arsenin previam uma expressão mais geral \cite[pág. 57]{tikhonov1977solutions}, na qual o termo de fidelidade era quadrático, mas $\Omega(\mathbf{x})$ poderia ter outras formas. 

Outra forma é a chamada de regularização generalizada de Tikhonov, descrita por
\begin{equation}
\hat{\mathbf{x}} = \arg\min\limits_{\mathbf{x}} \left[ \vert \vert \mathbf{A} \mathbf{x} - \mathbf{y} \vert \vert^2_2 + \lambda^2 \vert \vert \mathbf{L} \mathbf{x} \vert \vert_2^2 \right],
\label{eq:tikhonov12}
\end{equation}
onde $ \mathbf{L} $ é chamada de matriz de regularização. Caso $ \mathbf{L} = \mathbf{I} $, tem-se a regularização clássica de Tikhonov, ou de ordem zero \cite[pág. 104]{aster2019parameter}. Para outras matrizes $\mathbf{L}$ denomina-se regularização generalizada de Tikhonov \cite[pág. 171]{hansen2010discrete}, ou \textit{weighted generalized inverse} \cite[pág. 197]{engl1996regularization}. Dessa forma, é esperada uma estrutura esperada para o vetor $\mathbf{x}$ e não apenas que ele possua a menor norma possível. 

Seja $\mathbf{0}$ um vetor com componentes nulos. Para obter a solução da Equação \eqref{eq:tikhonov12}, o procedimento é análogo ao caso da regularização clássica. Parte-se de
\begin{equation}
\hat{\mathbf{x}} = \arg\min\limits_{\mathbf{x}} 
\left|\left|
\begin{pmatrix}
\mathbf{A}\\ 
\lambda \textbf{L}\\ 
\end{pmatrix}\mathbf{x} - \begin{pmatrix}
\mathbf{y}\\ 
\mathbf{0}
\end{pmatrix} \right| \right|^2_2,
\label{eq:tikhonov23}
\end{equation} 
A sua solução também pode ser encontrada a partir da primeira derivada deste funcional igualada a zero, resultando em \cite[pág. 61]{hansen2010discrete} 
\begin{equation}
\hat{\mathbf{x}} =\left( \mathbf{A}^T \mathbf{A} + \lambda^2 \mathbf{L}^T \mathbf{L} \right)^{-1} \mathbf{A}^T \mathbf{y}.
\label{eq:tiksolgen}
\end{equation}
Para que a solução seja única, é necessário impor que os núcleos de $\mathbf{A}$ e $\mathbf{L}$ não possuam elementos em comum, isto é, que $N(\mathbf{A}) \cap N(\mathbf{L}) = \emptyset$, além de que $\mathbf{L}$ tenha posto-linha completo  \cite[pág. 104]{aster2019parameter}, \cite[Seção 2.1.2]{Hansen1998}.  

As matrizes $\mathbf{L}$ não serão  necessariamente bem-condicionadas e não é uma exigência que elas possuam inversas. Elas podem ser retangulares ou não apresentarem posto completo. É possível que tanto $\mathbf{A}$ quanto $\mathbf{L}$ sejam matrizes mal-condicionadas, de modo que não seja possível inverter $\left(\mathbf{A}^T \mathbf{A} \right)$ nem $\left( \mathbf{L}^T \mathbf{L} \right)$ individualmente, mas que quando somadas na Equação \eqref{eq:tiksol} o cálculo de $\left(\mathbf{A}^T \mathbf{A} + \lambda^2 \mathbf{L}^T \mathbf{L}\right)^{-1} $ torna possível obter a solução desejada.  Assim, deve-se avaliar as reconstruções para diferentes níveis de ruído $\vert \vert \bm{\delta} \vert \vert _2$ e parâmetros de regularização $\lambda$, pois se $\vert \vert \bm{\delta} \vert \vert _2$ é grande e $\lambda$ é pequeno, a própria solução regularizada pode ser sensível a ruídos. 

Outro ponto importante é que a forma de $\mathbf{L}$ depende tanto da informação a priori que
se tem quanto da forma que se espera para os dados x. Seja:
\begin{itemize}
\item Um sinal unidimensional (1D) regularmente discretizado no espaço, na forma de um vetor. Um exemplo em deconvolução é um sinal da forma de degrau unitário, que após aquisição por um instrumento de medida é suavizado, arredondado;

\item Um sinal bidimensional (2D) regularmente discretizado no espaço. Um exemplo é uma imagem nítida disposta em um \textit{grid} regular, com pixels igualmente espaçados. Nesse caso, pode-se buscar trabalhar diretamente com a imagem na forma de uma matriz, ou concatenar suas colunas para a forma de um vetor;

\item Um vetor que representa quantidades distribuídas irregularmente no espaço. Um exemplo é uma região de interesse que foi discretizada com o método dos elementos finitos em uma malha irregular.
\end{itemize}

Conforme ilustrado na Subseção \ref{sec:suave}, $\mathbf{L}$ será diferente em cada um desses casos.

 \subsubsection{Como utilizar regularização para promover soluções suaves?}\label{sec:suave}

Métodos de regularização buscam que a solução do problema inverso apresente a propriedade de estabilidade, o que pode ser entendido como a solução depender suavemente (ou continuamente) dos dados de entrada \cite[pág. 50]{Neto2005}. Suavidade, por sua vez, pode ser relacionada com a quantidade de derivadas que uma função possui  \cite[Apêndice A.2.1]{Choksi2022}. Em \cite[pág. 50]{Neto2005}, a terceira condição de Hadamard é descrita de modo que a solução deveria depender suavemente dos dados e que suavidade, por vezes, pode ser traduzida por continuidade. 

Assim, visando a regularidade da solução, pode-se partir da suposição de que as soluções sejam as mais suaves possíveis \cite[pág. 53]{hansen2010discrete}, \cite{Phillips1962}. Nesse caso, pode-se supor que os parâmetros $\mathbf{x}$ sejam valores discretos de uma função que é diferenciável (uma ou mais vezes) \cite[pág. 80]{kaipio2005statistical}.  Ainda na Equação \eqref{eq:tiksolgen}, é possível definir a matriz de regularização $\mathbf{L}$ para representar a discretização de operadores derivativos de primeira, segunda ou de maior ordem. De modo geral, essas escolhas correspondem ao conhecimento \textit{a priori} que as soluções são suaves, pois soluções que não sejam suaves (ou que sejam oscilatórias) poderiam levar a valores muito grandes das derivadas e do funcional como consequência, não sendo candidatas \cite[pág. 9]{Benning2018}. 

Essa ideia não é recente. Em 1962, Phillips desenvolveu um método numérico de solução para equações de Fredholm de primeiro tipo na qual se buscava a solução mais suave possível \cite{Phillips1962}, que apresentasse segundas derivadas que fossem contínuas por partes. Essa solução foi proposta pouco tempo antes da teoria de regularização de Tikhonov e ela também poderia ser escrita na forma da Equação \eqref{eq:tiksolgen}, o que faz com que esse método seja encontrado na literatura como sendo de Tikhonov-Phillips. 

Na prática, a forma explícita de $\mathbf{L}$ depende da discretização utilizada no problema que se quer resolver, bem como na sua dimensionalidade:
\begin{itemize}

\item \textbf{Matrizes $\mathbf{L}$ para sinais 1D}: Para sinais 1D, o operador discreto de primeira derivada $\mathbf{L_{1 \hspace{1mm} d1}}$, que caracteriza a regularização de Tikhonov de primeira ordem, favorece soluções que são relativamente planas \cite[pág. 103]{aster2019parameter}. Com condições de contorno nulas \cite[págs. 175-6]{hansen2010discrete},  ela apresenta uma linha a menos do que o número de colunas e é descrita por
\begin{equation}
\mathbf{L_{1 \hspace{1mm} d1}} = \begin{pmatrix}
1 & -1 & & \\
& \ddots & \ddots & \\
& & 1 & -1 \\ 
\end{pmatrix}.
\label{eq:ld1} 
\end{equation}
Há também forma alternativa invertível \cite[pág. 115]{calvetti2007introduction}, quando se escreve
\begin{equation}
\mathbf{L_{2 \hspace{1mm} d1}} = \begin{pmatrix}
1 & & \\
1 & -1 & & \\
& \ddots & \ddots & \\
& & 1 & -1 \\ 
\end{pmatrix}.
\label{eq:ld1alt} 
\end{equation}
Também para sinais 1D e também com condições de contorno nulas, o operador de segunda derivada $\mathbf{L_{d2}}$  (ou laplaciano), que define a regularização de Tikhonov de segunda ordem, penaliza soluções que sejam mais irregulares no sentido da segunda derivada ao quadrado  \cite[pág. 104]{aster2019parameter},
\begin{equation}
\mathbf{L_{d2}} = \begin{pmatrix}
1 & -2 & 1 & & \\
& \ddots & \ddots & \ddots & \\
& & 1 & -2 & 1 \\
\end{pmatrix}, 
\label{eq:ld2}
\end{equation}
com duas linhas a menos do que o número de colunas. 

Quando $\mathbf{L} = \mathbf{I}$, $\mathbf{L} = \mathbf{L_{d1}}$, ou $\mathbf{L} = \mathbf{L_{d2}}$, a mesma operação é aplicada a todo o domínio, uma operação invariante no espaço, mas isso não é obrigatório. É possível definir a matriz $\mathbf{L}$ considerando outras condições de contorno. A matriz $\mathbf{L}_{d2}$ com condições de contorno reflexivas \cite[págs. 175-6]{hansen2010discrete} é
\begin{equation}
 \begin{pmatrix}
- 1 & 1 & & &\\
1 & -2 & 1 & & \\
& \ddots & \ddots & \ddots & \\
& & 1 & -2 & 1 \\
& &  & -1 & 1 \\
\end{pmatrix},
\label{eq:ld2r}
\end{equation}
onde o número de linhas é igual ao número de colunas. 

\item \textbf{Matrizes $\mathbf{L}$ para imagens 2D}:  Considerando que os \textit{pixels} (parâmetros) de uma imagem estejam em um \textit{grid} regular, com espaçamento padronizado, é preciso adaptar as matrizes das Equações \eqref{eq:ld1} e \eqref{eq:ld2} para o caso 2D, considerando  as duas direções (vertical e horizontal) presentes na imagem \cite[pág. 92-5]{hansen2006deblurring}. Em caso de solução suave, pode-se utilizar o operador laplaciano, obtido a partir de matrizes de primeira derivada e suas transpostas, representando derivadas nas direções verticais e horizontais da imagem \cite[pág. 177]{hansen2010discrete}, \cite[pág. 82-3]{kaipio2005statistical}, também sendo possível definir diferentes condições de contorno para ela.
\end{itemize}

\begin{itemize}
\item \textbf{Matrizes $\mathbf{L}$ para malhas de elementos finitos}: Diferente de imagens, onde os \textit{pixels} são regularmente espaçados, os elementos de uma malha podem não ter essa característica, tornando necessárias outras formas de definir as matrizes de regularização  $\mathbf{L}$. Por exemplo, no caso de um operador de primeira derivada, cada linha da matriz é composta por vetores $[l , 0, ..., -l ]$ onde a posição de $l$ e $-l$ é dada pelos dois elementos que compartilham uma mesma aresta e o valor de $l$ é o comprimento dessa aresta \cite[pág. 3]{Borsic2010}. 
 
 Outro regularizador pode ser definido quando $\mathbf{L}$ representa um filtro passa-altas \cite{Adler1996}, obtido subtraindo um filtro passa-baixas gaussiano de uma matriz identidade. Primeiro é calculada uma matriz de distância entre os centroides de todos os elementos em relação a todos os elementos. A partir dessa matriz de distância, o filtro passa-baixas é calculado a partir de uma função gaussiana e o filtro passa-altas é obtido. A montagem dessa matriz é descrita em \cite[págs. 115-6]{2013moura}. Sua utilização também penaliza soluções abruptas, suavizando a reconstrução e um aspecto positivo é que filtros passa-baixas gaussiano não promovem \textit{ringing} \cite[pág. 277]{gonzalez2018}. 
 
É possível ainda utilizar essa matriz que representa o filtro passa-altas em conjunto  com a informação \textit{a priori} estrutural, em que a matriz $\mathbf{L}$ é ponderada por uma matriz diagonal para buscar representar mudanças abruptas das incógnitas (parâmetros) \cite[pág. 6]{Calvetti2018a}; bem como a regularização espacial regional, na qual a matriz de regularização apresenta valores não-nulos apenas dentro de certas regiões, evitando suavizar regiões de transições agudas \cite[pág. 167]{Horesh}. Essas duas formas podem ser utilizadas em imagens médicas, quando há um conhecimento prévio da distribuição dos tecidos biológicos que vão compor a imagem. 
\end{itemize}

Essas matrizes de regularização não são exclusivas à uma aplicação.  Existem diferentes nomes dentro da busca por soluções suavizadas, dependendo da área. No contexto de ajuste, interpolação e suavização de \textit{B-splines} \cite{Eilers1996, Ramsay1997} e de regressão de \textit{splines} penalizada \cite[Seção 3.5]{ruppert2003semiparametric}, o termo de regularização generalizada com matriz derivativa é chamado de \textit{roughness penalty} e a matriz $\mathbf{L}$ em si de \textit{roughening matrices} \cite[pág. 103]{aster2019parameter}. A palavra \textit{roughness} tem o significado contrário da palavra  \textit{smoothness}, ou seja, irregularidade, o contrário de suavidade. Nesse caso,  $\lambda$ controla o \textit{trade-off} entre  a qualidade do ajuste e a irregularidade da solução \cite[pág. 2797]{Chen2002}. No Apêndice \ref{Ap:missing}, ilustra-se os efeitos de $\mathbf{I}$, $\mathbf{L_{1 \hspace{1mm} d1}}$, $\mathbf{L_{d2}}$ como matrizes de regularização em problemas de \textit{missing data}.

\subsubsection{Como interpretar a matriz de regularização como convolução com um \textit{kernel}?}

No caso de sinais 1D, as matrizes da Equação \eqref{eq:ld1}  ou da Equação \eqref{eq:ld2} são circulantes e podem ser entendidas a partir da operação da convolução \cite[pág. 232]{aster2019parameter}. Especificamente no caso da Equação \eqref{eq:ld2}, o resultado da multiplicação matriz-vetor $\mathbf{L_{d2}} \mathbf{x}$ será equivalente à convolução do sinal com o vetor $[-1 , 2, -1]$, que representa o operador laplaciano discreto. Isto é, equivalente à $\mathbf{L_{d2}} \mathbf{x} = \mathbf{x} * [-1 , 2, -1]$, com alguma diferença possivelmente no início e no final do sinal resultante, por conta da escolha das condições de contorno.

 Para o caso 2D de imagens, essa relação também é possível. Seja o \textit{kernel} laplaciano em 2D $\mathbf{H_{d2}}$, conforme 
\begin{equation}
\mathbf{H_{d2}} = \begin{pmatrix}
0 & 1 & 0  \\
1 & -4 & 1\\
0 & 1 & 0 
\end{pmatrix}.
\label{eq:hd2}
\end{equation}
Seja $\mathbf{X}$ uma imagem e $\mathbf{x}$ essa mesma imagem só que vetorizada, isto é, com suas colunas concatenadas verticalmente. É possível escrever a operação de convolução com um \textit{kernel} como uma matriz de regularização circulante \cite[pág. 177]{hansen2010discrete}, obtém-se um vetor $\mathbf{L} \mathbf{x}$, que pode ser transformado de volta à forma de uma imagem. Essa imagem resultante será equivalente ao resultado de $\mathbf{X} *\mathbf{H_{d2}}$, desde que se utilizem condições de contorno apropriadas. 

Para alguns algoritmos iterativos não é necessário montar $\mathbf{L}$ explicitamente, mas sim utilizar alguma função, como a convolução, diminuindo o custo computacional para seu armazenamento. Além disso, entendendo $\mathbf{L}$ como a operação de convolução com um \textit{kernel} possibilita utilizar outros \textit{kernels} além do laplaciano.

\subsubsection{Qual é a vantagem de incluir um valor de referência fixo?}

Quando há informação \textit{a priori} disponível de que a solução $\mathbf{x}$ apresenta um valor próximo a um outro vetor $\mathbf{x}^*$ \cite[pág. 69]{Mueller2012}, é possível incluí-lo no funcional da Equação \eqref{eq:tikhonov12} e enviesar  a solução \cite[pág. 13]{Hansen1998}. O novo funcional é dado por \cite[pág. 33]{paivi}
\begin{equation}
\hat{\mathbf{x}} = \arg\min\limits_{\mathbf{x}} \left[ \vert \vert \mathbf{y}- \mathbf{A} \mathbf{x} \vert \vert^2_2 + \lambda^2 \vert \vert \mathbf{L} \left(\mathbf{x} - \mathbf{x}^*\right) \vert \vert^2_2 \right] 
\label{eq:eqmult33}
\end{equation}
 \begin{equation}
\hat{\mathbf{x}} =  \left( \mathbf{A}^T \mathbf{A} + \lambda^2 \mathbf{L}^T \mathbf{L} \right)^{-1} \left(\mathbf{A}^T \mathbf{y} + \lambda^2 \mathbf{L}^T \mathbf{L} \mathbf{x}^* \right),
\label{eq:eqmult3}
\end{equation}
englobando o caso da Equação \eqref{eq:tiksolgen} quando $\mathbf{x}^* = \mathbf{0}$ for um vetor nulo. 

Esse novo vetor é as vezes chamado de valor de referência  \cite[pág. 60]{Neto2005} ou mesmo como chute inicial da solução \cite[pág. 99]{Camargo2013}.  A Equação \eqref{eq:eqmult3} pode ser utilizada, por exemplo, no caso de um sistema dinâmico cuja solução é esperada para em um determinado instante de tempo. Este mesmo valor $\mathbf{x}^*$ também pode ser utilizado como valor inicial $\mathbf{x}_0$ em algoritmos iterativos, ao invés do vetor nulo usual.

Uma forma de interpretar a solução clássica de Tikhonov que inclui $\mathbf{x}^*$ é notar que no termo de penalização ele aparece como uma diferença de vetores dentro de uma norma euclidiana ao quadrado $\vert\vert\mathbf{x} - \mathbf{x}^*\vert\vert_2^2$. Ou seja, essa operação é o quadrado da distância euclidiana entre dois vetores. Dentro da minimização, isso significa que se busca uma solução que esteja próxima de $\mathbf{x}^*$, de modo que a expressão toda $\vert \vert \mathbf{x} - \mathbf{x}^*\vert \vert^2_2$ apresente valores pequenos. Quando $\mathbf{x}^* = \mathbf{0}$, é como se fosse uma distância até a origem. 

\subsubsection{Qual é a vantagem de incluir um valor de referência variável?}\label{sec:disappearing}

Quando o problema é linear e valor de referência é fixo, há soluções em um só passo conforme Equação \eqref{eq:eqmult3}. Por outro lado, o valor de referência não precisa ser fixo dentro de um algoritmo iterativo. Ele pode ser variável e escrito como $\mathbf{x}_k^*$. Um exemplo é quando $\mathbf{x}_k^* = \mathbf{x}_{k-1}$, o resultado de $\mathbf{x}$ na iteração anterior. Essa proposta é conhecida como \textit{disappearing Tikhonov regularization} \cite[pág. 143]{Parikh2014}, conforme:
\begin{equation}
\hat{\mathbf{x}} = \arg\min\limits_{\mathbf{x}} \left[ \vert \vert \mathbf{y}- \mathbf{A} \mathbf{x} \vert \vert^2_2 + \lambda^2 \vert \vert \mathbf{x} - \mathbf{x}_{k-1} \vert \vert^2_2 \right],
\label{eq:eqmult53}
\end{equation}
onde $\mathbf{L} = \mathbf{I}$ foi escolhido apenas por simplificação. Na medida que o algoritmo convergir, o termo $\vert \vert \mathbf{x} - \mathbf{x}_{k-1} \vert \vert^2_2$ se tornará cada vez menor, de modo que a contribuição desse regularizador quadrático irá sumindo, o que define seu nome \cite[pág. 143]{Parikh2014}.

Seja $n$ o número total de iterações. Seja também a sequência de soluções $\mathbf{x}_k$, para $k = 1, \dots, n$, de soluções de um algoritmo iterativo. Para observar a convergência do algoritmo, a cada iteração $k$ podem ser calculadas as grandezas $\vert \vert \mathbf{A} \mathbf{x}_k - \mathbf{y}\vert \vert_2^2$ e $\vert \vert \mathbf{x}_k - \mathbf{x}_{k-1} \vert \vert^2_2$, bem como a soma das duas, para acompanhar se de fato ela está sendo minimizada. Nesse caso, o regularizador penalizaria $\mathbf{x}_k - \mathbf{x}_{k-1}$, grandes diferenças entre a solução da iteração atual e a solução da iteração anterior, ou seja, penalizaria maiores diferenças para trás. 

\subsubsection{É possível incluir mais de um termo de regularização no funcional?}
A regularização generalizada de Tikhonov não é limitada a uma matriz de regularização. Outra possibilidade é a chamada regularização multiparâmetros, quando se utiliza mais de uma matriz de regularização simultaneamente, ou seja, quando se concatenam matrizes de regularização formando $\mathbf{L} = \left[ \lambda_1 \mathbf{L}_{1}, \lambda_2 \mathbf{L}_{2}, \cdots, \lambda_i \mathbf{L}_{i}\right]^T$, onde $i$ é o número de matrizes de regularização. A regularização multiparâmetros com dois termos é definida através de
\begin{equation}
\hat{\mathbf{x}} = \arg\min\limits_{\mathbf{x}} \left[ \vert \vert \mathbf{y} - \mathbf{A} \mathbf{x} \vert \vert^2_2 
+  \lambda^2_1 \vert \vert \mathbf{L}_{1} \left(\mathbf{x} - \mathbf{x}^*_1\right) \vert \vert^2_2
+  \lambda^2_2 \vert \vert \mathbf{L}_{2} \left(\mathbf{x} - \mathbf{x}^*_2\right) \vert \vert^2_2 \right],
\label{eq:eqmult1}
\end{equation}
onde $\mathbf{L}_1$ e $\mathbf{L}_2$ são matrizes de regularização e $\lambda_1$ e $\lambda_2$ seus parâmetros. 

Conforme Subseção \ref{Ap:multi}, a solução da Equação \eqref{eq:eqmult1} é dada por
\begin{equation}
\hat{\mathbf{x}} =\left( \mathbf{A}^T \mathbf{A} + \lambda_1^2 \mathbf{L}_1^T \mathbf{L}_1 + \lambda_2^2 \mathbf{L}_2^T \mathbf{L}_2 \right)^{-1} \left(\mathbf{A}^T \mathbf{y} + \lambda_1^2 \mathbf{L}_1^T \mathbf{L}_1 \mathbf{x}^*_1 + \lambda_2^2 \mathbf{L}_2^T \mathbf{L}_2 \mathbf{x}^*_2\right).
\label{eq:eqmult2}
\end{equation}
Deve-se destacar que é possível utilizar mais do que dois regularizadores também. Cada matriz de regularização pode ter seu próprio parâmetro $\lambda_i$, eles não precisam ter o mesmo peso na otimização. Ainda que seja custosa, é possível a escolha dos $\lambda_i$ através de análogos multidimensionais da curva-L, chamadas de hipersuperfícies-L \cite{Belge_2002}.

\subsubsection{Existe formulação alternativa da regularização de Tikhonov?}

A Equação \eqref{eq:tiksol} é uma forma muito encontrada para a regularização clássica de Tikhonov, semelhante a uma matriz inversa à esquerda, mas é possível obter uma forma que se assemelha à matriz inversa à direita \cite[pág. 223]{engl1996regularization}, \cite[pág. 79]{kaipio2005statistical}, conforme
\begin{equation}
\hat{\mathbf{x}}_{\lambda} =\mathbf{A}^T \left( \mathbf{A} \mathbf{A}^T + \lambda^2 \mathbf{I}^T \mathbf{I} \right)^{-1} \mathbf{y},  
\label{eq:tiksolx}
\end{equation}
ou de modo compacto como $\hat{\mathbf{x}}_{\lambda} = \mathbf{A}^+_{\lambda} \mathbf{y}$, onde nesse caso $\mathbf{A}^+_{\lambda} =\mathbf{A}^T \left( \mathbf{A}\mathbf{A}^T + \lambda^2 \mathbf{I}^T \mathbf{I} \right)^{-1}$. 

Seja $\mathbf{A}$ uma matriz $n \times p$ composta por valores reais, $\mathbf{I}_N$ uma matriz identidade $n \times n$ e $\mathbf{I}_P$ outra matriz identidade, mas $p \times p$. Mostra-se que as Equações \eqref{eq:tiksol} e \eqref{eq:tiksolx} são equivalentes a partir da relação de igualdade
\begin{equation}
\underset{p \times n}{\mathbf{A}^T}  \hspace{3mm} \underset{n \times p}{\mathbf{A}} \hspace{3mm} \underset{p \times n}{\mathbf{A}^T} \hspace{3mm} + \hspace{3mm} \underset{p \times n}{\mathbf{A}^T}\hspace{3mm}  (\lambda \underset{n \times n}{\mathbf{I}_N}) \hspace{3mm} = \hspace{3mm} \underset{p \times n}{\mathbf{A}^T}  \hspace{3mm} \underset{n \times p}{\mathbf{A}} \hspace{3mm} \underset{p \times n}{\mathbf{A}^T} \hspace{3mm} + \hspace{3mm} (\lambda \underset{p \times p}{\mathbf{I}_P})  \hspace{3mm} \underset{p \times n}{\mathbf{A}^T},
\label{eq:apendiceD}
\end{equation}
Colocando $\mathbf{A}^T$ em evidência:
\begin{equation}
\underset{p \times n}{\mathbf{A}^T}  \left(\underset{n \times p}{\mathbf{A}} \hspace{3mm} \underset{p \times n}{\mathbf{A}^T} +  \lambda \underset{n \times n}{\mathbf{I}_N}\right) = \left(\underset{p \times n}{\mathbf{A}^T}  \hspace{3mm} \underset{n \times p}{\mathbf{A}} + \lambda \underset{p \times p}{\mathbf{I}_P}\right)  \underset{p \times n}{\mathbf{A}^T}.
\end{equation}
Multiplicando pela esquerda os dois lados da equação por $\left(\mathbf{A}^T   \mathbf{A} + \lambda \mathbf{I}_P\right)^{-1}$ e multiplicando pela direita os dois lados da equação por $\left(\mathbf{A}  \mathbf{A}^T +  \lambda \mathbf{I}_N\right)^{-1}$ obtém-se
\begin{equation}
\begin{aligned}
\underset{p \times p}{\underline{\left(\mathbf{A}^T  \mathbf{A} + \lambda \mathbf{I}_P\right)^{-1}}} \hspace{2mm} \underset{p \times n}{\underline{\mathbf{A}^T  \left(\mathbf{A} \mathbf{A}^T +  \lambda \mathbf{I}_N\right)}} \hspace{2mm}  \underset{n \times n}{ \underline{\left(\mathbf{A}  \mathbf{A}^T +  \lambda \mathbf{I}_N\right)^{-1}}} = \\  \underset{p \times p}{\underline{\left(\mathbf{A}^T   \mathbf{A} + \lambda \mathbf{I}_P\right)^{-1}}} \hspace{2mm} \underset{p \times n}{\underline{ \left(\mathbf{A}^T \mathbf{A} + \lambda \mathbf{I}_P\right) \mathbf{A}^T}}  \hspace{2mm}  \underset{n \times n}{ \underline{ \left(\mathbf{A} \mathbf{A}^T +  \lambda \mathbf{I}_N\right)^{-1}}},
\end{aligned}
\end{equation}
que pode ser simplifcada para a relação procurada, isto é, 
\begin{equation}
\left(\mathbf{A}^T  \mathbf{A} + \lambda \mathbf{I}_P\right)^{-1} \mathbf{A}^T  \mathbf{I}_N = \mathbf{I}_P \mathbf{A}^T \left(\mathbf{A} \mathbf{A}^T +  \lambda \mathbf{I}_N\right)^{-1}
\end{equation}
\begin{equation}
\underset{p \times n}{\underline{\left(\mathbf{A}^T  \mathbf{A} + \lambda \mathbf{I}_P\right)^{-1}\mathbf{A}^T}}  = \underset{p \times n}{\underline{\mathbf{A}^T \left(\mathbf{A} \mathbf{A}^T +  \lambda \mathbf{I}_N\right)^{-1}}}.
\end{equation}
Para considerar valores complexos em $\mathbf{A}$, seja $\mathbf{A}^*$ o seu hermitiano.  Conforme \cite[Apêndice 2]{Grech2008}, essa relação passa a ser
\begin{equation}
\left( \mathbf{A}^* \mathbf{A} + \lambda \mathbf{I} \right)^{-1} \mathbf{A}^* = \mathbf{A}^* \left( \mathbf{A} \mathbf{A}^* + \lambda \mathbf{I} \right)^{-1}.
\label{eq:igualdade}
\end{equation}
Assim, as Equações \eqref{eq:tiksol} e \eqref{eq:tiksolx} podem ser equivalentes sob determinadas condições. Para escolher entre as duas, faz sentido utilizar a Equação \eqref{eq:tiksol} se o sistema for sobredeterminado, com $m \gg n$, e a Equação \eqref{eq:tiksolx} para sistemas subdeterminados, com $n \gg m$ \cite[pág. 150]{calvetti2007introduction}, para poupar tempo computacional e melhorar a acurácia e estabilidade da inversão. 

Uma alternativa à regularização generalizada de Tikhonov da Equação \eqref{eq:tiksolgen} é 
\begin{equation}
\hat{\mathbf{x}} =\mathbf{A}^T \left( \mathbf{A} \mathbf{A}^T + \lambda^2 \mathbf{L}^T \mathbf{L} \right)^{-1} \mathbf{y},
\label{eq:tiksolx2}
\end{equation}
mas dessa vez os casos em que ela não é equivalente à Equação \eqref{eq:tiksolgen} são mais claros.

 Relembrando a Equação \eqref{eq:eq1}, os dados $\mathbf{y}$ são $[m \times 1]$, os parâmetros $\mathbf{x}$ são $[n \times 1]$ e a matriz $\mathbf{A}$ que os relaciona tem tamanho [$m \times n$].  Quando $\mathbf{A}$ for retangular, a decisão de se usar a primeira ou a segunda forma depende também da forma como a matriz $\mathbf{L}$ é montada. A partir da Equação \eqref{eq:tiksolgen}, tanto $\mathbf{A}^T \mathbf{A} $ quanto $\mathbf{L}^T \mathbf{L}$ apresentam tamanho $n \times n$, ou seja, tem relação com o vetor de parâmetros. Por outro lado, utilizando a Equação \eqref{eq:tiksolx2}, tanto $\mathbf{A} \mathbf{A}^T$ quanto $ \mathbf{L}^T \mathbf{L}$ devem ter tamanho $m \times m$, ou seja, há relação com o vetor de dados. Assim, a matriz de reconstrução da Equação \eqref{eq:tiksolx2} estaria na \textit{data form} segundo autores como \cite{Adler2009}.
 
 Na regularização generalizada de Tikhonov, pode haver  matrizes de regularização $\mathbf{L}_N$ de tamanho $n \times n$ ou $\mathbf{L}_P$ de tamanho $p \times p$. A dificuldade é obter a igualdade da Equação \eqref{eq:igualdade} incluindo $\mathbf{L}_N$ e $\mathbf{L}_P$. Considerando que $\lambda$ seja o mesmo nos dois lados da Equação \eqref{eq:apendiceD}, a relação de igualdade seria
\begin{equation}
 \underset{p \times n}{\mathbf{A}^T}\hspace{3mm}   \underset{n \times n}{\mathbf{L}_N} \hspace{3mm} = \hspace{3mm} \underset{p \times p}{\mathbf{L}_P}  \hspace{3mm} \underset{p \times n}{\mathbf{A}^T}. 
\end{equation}
No entanto, além da matriz identidade, não é trivial indicar matrizes de regularização que possuam tal propriedade, pois a multiplicação de matrizes não apresenta a propriedade da comutatividade, ainda mais quando  $\mathbf{L}_N$ e $\mathbf{L}_P$ são retangulares. 

Obter tais matrizes não é necessário, só foi uma forma de discutir um caso mais geral para a Equação \eqref{eq:igualdade}. É possível utilizar as matrizes $\mathbf{L}$ discutidas até agora, como operadores de primeira ou de segunda derivada, na Equação \eqref{eq:tiksolx2}, desde que elas sejam corretamente adaptadas para o respectivo espaço e discretização. A única consequência é que o resultado não será igual ao da Equação \ref{eq:tiksolgen} para um mesmo valor de $\lambda$.

\subsubsection{Se o modelo é imperfeito, é possível regularizá-lo também?}

Tikhonov reconhecia que haveria também um erro $\bm{\epsilon}$ associado à $\mathbf{A}$ do modelo \cite[pág. 6]{tikhonov1977solutions}, conforme Equação \eqref{eq:sistema2}, mas $\mathbf{A}$ não é atualizada na Equação \eqref{eq:tiksol}. Mesmo que o valor de $\bm{\epsilon}$ não seja conhecido, pode ser de conhecimento que o modelo é imperfeito e deve ser atualizado. Além de regularizadores sobre $\mathbf{x}$, é possível também incluir regularizadores sobre $\mathbf{A}$ para sua atualização \cite{Bleyer2013}.  Métodos que levam em conta tanto $\bm{\epsilon}$ quanto $\bm{\delta}$ são chamados de métodos de regularização dupla \cite[Capítulo 4]{bleyer2015novel}.

No contexto da regularização multiparâmetros, um algoritmo que permite incluir esse erro no funcional é o algoritmo dos mínimos quadrados total \cite[pág. 106]{Arridge2019}, \cite{Golub1999}, conforme

\begin{equation}
\left(\hat{\mathbf{x}}, \Delta\hat{\mathbf{A}} \right) = \arg\min\limits_{x, \Delta\mathbf{A}} \left[ \vert \vert \mathbf{y} - (\mathbf{A}+\Delta\mathbf{A}) \mathbf{x} \vert \vert^2_2 +  \lambda^2_1 \vert \vert \mathbf{L}\mathbf{x} \vert \vert^2_2 +  \lambda^2_2 \vert \vert \Delta\mathbf{A} \vert \vert^2_F \right],
\label{eq:eqmult7}
\end{equation}
onde $\vert\vert \cdot \vert\vert_F$ denota a norma de Frobenius e $\Delta\mathbf{A}$ é a correção desejada de $\mathbf{A}$.

\subsection{Como utilizar regularização para promover soluções esparsas?}
As propostas até aqui revisadas são de regularizações lineares (em relação à $\mathbf{x}$) para problemas inversos mal-postos \cite[pág. 51]{engl1996regularization}. Nas últimas décadas o interesse mudou para métodos de regularização não-lineares, inclusive para modelos lineares, motivado pelo sucesso de técnicas como a variação total ou técnicas que buscam soluções esparsas \cite{Benning2018}, que obtém soluções que não são necessariamente suaves. 

É uma característica marcante na regularização de clássica de Tikhonov a presença de termos quadráticos, tanto no termo do resíduo quanto no termo de regularização, buscando por soluções suaves. Entretanto, em alguns casos o interesse está em soluções que não sejam suaves para manter o contorno do objeto de interesse, incluindo soluções descontínuas e soluções esparsas \cite{Benning2018}. Uma das formas de obtê-las não é apenas mudando a matriz de regularização $\mathbf{L}$, mas sim utilizando outras normas, especialmente escrevendo-se uma norma $\ell_p$ no termo de regularização, 
\begin{equation}
\hat{\mathbf{x}} = \arg\min\limits_{\mathbf{x}} \left[ \vert \vert \mathbf{A} \mathbf{x} - \mathbf{y} \vert \vert^2_2 + \lambda^2 \vert \vert \mathbf{L} \mathbf{x} \vert \vert_p^p \right],
\label{eq:normap}
\end{equation}
cuja solução depende da definição de $p$, discutido no Apêndice \ref{Ap:normas1}.

Um problema de otimização deve ser avaliado em termos da convexidade e da diferenciabilidade dos termos do funcional. Na regularização de Tikhonov, um dos casos de maior interesse,  $p = 2$ e o problema de otimização é convexo, então qualquer mínimo local é também um mínimo global, além de ser diferenciável em todos os pontos. Ainda que existam propostas para  $p > 2$ \cite{Adil2019}, um grande interesse está nos casos em que $ p = 1$ e $p=0$, pois a solução nestes casos tende a ser esparsa. 

\subsubsection{Qual é a importância da norma $\ell_1$ na obtenção de soluções esparsas?}
A utilização de $p < 2$ favorece a reconstrução de objetos que são esparsos na base de representação escolhida \cite{Daubechies2004, Daubechies2016}. Esparsidade, nesse caso, significa que há parâmetros cujos valores ótimos são nulos \cite[pág. 236]{goodfellow2016deep}. Esse tipo de regularização é chamada de regularização que promove a esparsidade. No caso em que $1 \leq p \leq 2$, o regularizador penaliza menos os grandes valores no argumento do que a norma $\ell_2$ e ainda são funções convexas do argumento \cite[pág. 194]{bovik2005handbook}. 

O uso de norma $\ell_1$ é bastante investigada \cite[pág. 181]{aster2019parameter}, já que é o menor valor de $p$ em que ainda apresenta convexidade em relação a $\mathbf{x}$, mas não são diferenciáveis qualquer que seja $\mathbf{x}$, o que pode trazer dificuldades na solução \cite[pág. 46]{aster2019parameter}. Neste caso, o funcional tem a forma 
\begin{equation}
\hat{\mathbf{x}} = \arg\min\limits_{\mathbf{x}} \left[ \vert \vert \mathbf{A} \mathbf{x} - \mathbf{y} \vert \vert^2_2 + \lambda^2 \vert \vert \mathbf{L} \mathbf{x} \vert \vert_1 \right].
\label{eq:norma1}
\end{equation}
Não há uma fórmula fechada para sua solução, mas existem algoritmos iterativos com tal finalidade, como o \textit{iterative shrinkage-thresholding algorithm} (ISTA), o \textit{fast-iterative shrinkage-thresholding algorithm} (FISTA)   \cite{Daubechies2016}, o \textit{alternating direction method of multipliers} (ADMM) \cite[págs. 196-201]{aster2019parameter} e o \textit{iteratively reweighted least squares} (IRLS).

Alguns deles são especializados para o caso em que a matriz de regularização é a matriz identidade $\mathbf{I}$. Outros algoritmos possibilitam a utilização de outras matrizes de regularização $\mathbf{L}$. Para ilustrar, o Apêndice \ref{Ap:IRLS} descreve a visão geral do IRLS, que pode ser utilizado para aproximar normas $\ell_1$ por normas $\ell_2$ iterativamente tanto no regularizador quanto no termo de fidelidade (ou em ambos). 

Retomando a regularização multiparâmetros, outra opção, conhecida na área da estatística como rede elástica \cite[pág. 213]{majumdar2019compressed}, \cite{zou2005}, é utilizar um termo de regularização com norma $\ell_1$ e outro com norma $\ell_2$, conforme 
\begin{equation}
\hat{\mathbf{x}} = \underset{\mathbf{x}}{\arg\min}\left[ \vert \vert \mathbf{y} - \mathbf{A}\mathbf{x} \vert \vert^2_2+\lambda_{1}^2 \vert\vert \mathbf{x} \vert\vert^{2}_2+\lambda _{2}^2 \vert \vert \mathbf{x} \vert\vert_{1} \right].
\label{eq:elastic}
\end{equation} 
Fica clara a diversidade possível de combinações no contexto da regularização multiparâmetros. Outro exemplo para recuperação esparsa, encontrado em \cite{Ding2020}, inclui regularização não-convexa conforme  
\begin{equation}
\hat{\mathbf{x}} = \underset{\mathbf{x}}{\arg\min}\left[ \vert \vert \mathbf{y} - \mathbf{A}\mathbf{x} \vert \vert^2_2+\lambda_{1}^2 \vert\vert \mathbf{x} \vert\vert_{1} - \lambda _{2}^2 \vert \vert \mathbf{x} \vert\vert_{2} \right],
\label{eq:sparsl21}
\end{equation} 
onde, neste caso, $\lambda_{1}\geq\lambda_{2}\geq0$. 

\subsubsection{E no caso da não-convexidade da norma $\ell_p$ para $p<1$? }

Até agora as normas verificadas resultavam em funcionais convexos, mas também existem soluções para normas $\ell_p$ que introduzem a não-convexidade ao problema. Na área da estatística, a generalização de uma norma $\ell_p$ qualquer é conhecida como regressão de Bridge \cite{Park2011}. Novamente buscando soluções esparsas, busca-se utilizar $0 < p < 1$. O desafio é que o problema deixa de ser convexo, correndo o risco de obter um mínimo local e não um mínimo global \cite{majumdar2019compressed}, além de não ser diferenciável em todos os pontos.

O caso limite é $p=0$, em que $\ell_0$ não define uma norma propriamente dita. Sua interpretação é de que $ \vert \vert \mathbf{x} \vert \vert_{0}$ seria equivalente a contagem do número de valores não-nulos do vetor \cite{Donoho2006} (para isso é necessário considerar que $0^0 = 0$). Se a quantidade e a posição dos valores não-nulos de $\mathbf{x}$ fossem conhecidos, chamada de solução do oráculo \cite[pág. 9]{majumdar2019compressed}, isso poderia ser utilizado como informação \textit{a priori}, mas dificilmente ela estará disponível na prática. Portanto, busca-se a solução a partir de um problema de otimização \cite{baraniuk2007}, conforme 
\begin{equation}
\hat{\mathbf{x}} = \arg\min\limits_{\mathbf{x}}\vert \vert \mathbf{x} \vert \vert_0 \hspace{3mm} \text{s.t.} \hspace{3mm}  \mathbf{A} \mathbf{x} = \mathbf{y}.
\label{eq:norma0}
\end{equation}
Esse é um problema NP-difícil \cite[pág. 10]{majumdar2019compressed} e não há um algoritmo melhor do que um de força bruta para resolvê-lo, mas várias propostas são encontradas na literatura \cite{majumdar2019compressed}. A utilização de uma norma $\ell_0$ pode parecer apenas um procedimento teórico, mas vem intimamente ligado à aplicação chamada de amostragem comprimida (CS) \cite{schulz2009compressive}, o que estimulou a sua pesquisa e trouxe importantes resultados práticos \cite{Donoho2006}. Uma breve apresentação sobre CS é encontrada no Apêndice \ref{Ap:CS}. Uma justificativa no seu interesse é que ela possibilita a aquisição exata de sinais através de um número de amostras menor do que o necessário através da amostragem convencional \cite{majumdar2019compressed}, mesmo que o teorema de amostragem de Nyquist não seja obedecido \cite{baraniuk2007}. 

Vale notar que há ainda regularizadores para promover esparsidade que partem da norma $\ell_p^p$. Em \cite{Xu2010}, por exemplo, os autores exploram um regularizador não-convexo considerando $p = 1/2$, conforme 
\begin{equation}
\Omega(\mathbf{x}) = \sum_{i=1}^{m} \vert x_i \vert^{\frac{1}{2}},
\label{eq:norma12meio}
\end{equation}
onde $m$ é o tamanho do vetor. Em relação à notação utilizada na Equação \eqref{eq:lpnormp}, esse regularizador poderia ser identificado como $\vert \vert \mathbf{x} \vert\vert_{1/2}^{1/2}$.

\subsection{Regularização por variação total: Como obter soluções com transições abruptas?}
Na regularização de Tikhonov, descontinuidades do modelo são suavizadas, pois modelos suaves são menos penalizados com a norma $\ell_2$ do que transições abruptas. Na regularização por variação total (TV) \cite[pág. 195]{aster2019parameter},  \cite{Rudin1992, Gilboa2018, Wu2010}, o regularizador representa a discretização do gradiente $ \nabla\mathbf{x}$ junto da norma $\ell_1$ ou $\ell_2$, buscando por soluções com gradiente esparso dos parâmetros \cite{Benning2018}. Ele é apropriado quando se espera ter descontinuidades e funções \textit{piecewise-constant} \cite[Seção 7.4]{aster2019parameter} \cite[Seção 6.4]{Mueller2012}. A sua forma pode ser reescrita de acordo com a aplicação:
\begin{itemize}
\item Para aplicações unidimensionais \cite[pág. 195]{aster2019parameter}, a regularização por TV anisotrópica retoma a matriz da Equação \eqref{eq:ld1} como 
\begin{equation}
\hat{\mathbf{x}} = \arg\min\limits_{\mathbf{x}} \left[ \vert \vert \mathbf{A} \mathbf{x} - \mathbf{y} \vert \vert^2_2 + \lambda^2 \vert \vert \mathbf{L}_{d1} \mathbf{x} \vert \vert_1 \right].
\label{eq:TV11d1}
\end{equation}
\item Para aplicações em imagens \cite[págs. 195-6]{aster2019parameter}, utiliza-se a norma $\ell_1$ do operador gradiente, que inclui a diferença de valores dos \textit{pixels} adjacentes. Assim, a regularização por TV anisotrópica em duas dimensões é dada por 
\begin{equation}
\hat{\mathbf{x}} = \arg\min\limits_{\mathbf{x}} \left[ \vert \vert \mathbf{A} \mathbf{x} - \mathbf{y} \vert \vert^2_2 + \lambda^2 \vert \vert \nabla \mathbf{x} \vert \vert_1 \right],
\label{eq:TV11d3}
\end{equation}
onde $ \vert \vert \nabla \mathbf{x} \vert \vert_1 = \sum _{i,j} \sqrt{\vert x_{i+1,j}-x_{i,j}\vert^{2}} + \sum_{i,j} \sqrt{\vert x_{i,j+1}-x_{i,j}\vert^{2}}$.

\item Na proposta original do método, os autores de \cite{Rudin1992} definiram regularização por Variação Total isotrópica que utiliza a norma $\ell_2$ do operador gradiente, que em duas dimensões é dada por 
\begin{equation}
\hat{\mathbf{x}} = \arg\min\limits_{\mathbf{x}} \left[ \vert \vert \mathbf{A} \mathbf{x} - \mathbf{y} \vert \vert^2_2 + \lambda^2 \vert \vert \nabla \mathbf{x} \vert \vert_2 \right],
\label{eq:TV11d4}
\end{equation}
onde $ \vert \vert \nabla \mathbf{x} \vert \vert_2 = \sum _{i,j}{\sqrt {\vert x_{i+1,j}-x_{i,j}\vert^{2}+\vert x_{i,j+1}-x_{i,j}\vert^{2}}}$, conforme \cite[pág. 195]{aster2019parameter}. 

\item No caso de uma malha de elementos finitos bidimensional, novamente essas diferenças são calculadas entre elementos que compartilham arestas \cite{Borsic2010}, enquanto no caso tridimensional as diferenças são calculadas em relação aos elementos que compartilham faces. Vale notar que esses elementos não estão, necessariamente, em posições adjacentes no vetor de parâmetros, o que muda a forma de $\mathbf{L}$. 
\end{itemize}

Por conta da norma $\ell_1$, observa-se que o termo de regularização da variação total não é linear \cite{Rudin1992}, não é diferenciável em todos os pontos, não é quadrático, mas é convexo \cite[pág. 268]{2010Chambolle}, o que permite utilização de técnicas de otimização convexa. Assim, não há soluções em um único passo para se resolver o problema inverso com Variação Total e são necessários algoritmos especializados para isso. Uma lista é encontrada em \cite{Dahl2009}. 

A formulação variacional permite novas propostas, como regularização multiparâmetros onde um termo é a regularização de Tikhonov e o outro é de TV \cite{Gholami2013}, ou incluir direções de gradiente preferenciais para sua realização \cite{Kongskov2019}. Uma revisão de TV para diferentes modelos de ruídos é encontrada em \cite{Rodriguez2013}. Há ainda a regularização por TV generalizada  \cite{Bredies2010}, que utiliza ordens maiores das derivadas.

\subsection{Regularização por máxima entropia: Quais são os desafios de termos de regularização não-lineares?}
Do ponto de vista matemático, um método de regularização linear é definido em \cite[Definição 4.2]{Benning2018}, mas é possível também definir métodos de regularização para englobar a regularização não-linear em relação aos parâmetros $\mathbf{x}$ que se quer estimar. Isso independe do modelo $\mathbf{A}$ ser linear ou não em relação à $\mathbf{x}$. O foco se volta para a forma de $\Omega(\mathbf{x})$. Em \cite[Seção 4]{Benning2018} são discutidos os fundamentos teóricos para a regularização não-linear, enquanto uma definição formal é mostrada em \cite[Definição 4.7]{Benning2018}. 

Do ponto de vista computacional, pode-se pensar nos métodos de regularização que resultam em sistemas de equações não-lineares. Não é necessário se restringir à formas em que a matriz de regularização seja linear com os parâmetros $\mathbf{x}$, como no caso em que  $\vert \vert \mathbf{L} \mathbf{x} \vert \vert_2^2$ quando $\mathbf{L}$ não depende de $\mathbf{x}$ . A utilização de termos de regularização quadráticos em relação aos parâmetros traz a facilidade de que as condições necessárias de primeira ordem ainda sejam equações lineares \cite[Seção 5.3]{engl1996regularization}, \cite[pág. 180]{luenberger2015linear}. 

Isso não vale para regularizadores que partem de medidas da informação baseadas na entropia de Shannon. Um dos regularizadores não-quadráticos desse tipo é a regularização por máxima entropia (MaxEnt) \cite{Smith1985, Amato1991}, \cite[pág. 192]{bovik2005handbook}, \cite{Silva2016}, conforme 
\begin{equation}
\Omega(\mathbf{x}) = \sum_{i=1}^{m} x_i \log(\omega_i x_i), 
\label{eq:shannon}
\end{equation}
onde $m$ é o tamanho do vetor $\mathbf{x}$, $x_i$ são valores positivos de $\mathbf{x}$ e $\omega_i$ são pesos relativos a esses valores \cite{Hansen2007}. Esse termo de regularização possui uma relação não-linear com os parâmetros. Como consequência, a função custo  é não-linear conforme 
\begin{equation}
\hat{\mathbf{x}} = \underset{\mathbf{x}}{\arg\min} \left[ \vert \vert \mathbf{A}\mathbf{x} - \mathbf{y} \vert \vert^2_2 + \lambda^2 \sum_{i=1}^{m} x_i \log(\omega_i x_i) \right].
\label{eq:shannon2}
\end{equation}
Essa medida pode ser interpretada como a mais objetiva em relação à informação indisponível (ou corrompida por ruídos) do vetor de medidas \cite{Hansen2007}, \cite[pág. 192]{bovik2005handbook}. Por conta da presença de uma função logaritmo, a solução encontrada é necessariamente positiva. Nota-se ainda que $-\sum_{i=1}^{m} x_i \log(\omega_i x_i)$ é a medida da entropia de $\mathbf{x}$, o que define o nome dessa regularização \cite[pág. 28]{Hansen2007}. Há trabalhos que tratam a máxima entropia não apenas como um termo de regularização, mas como um princípio qual a teoria de regularização pode ser estabelecida \cite[págs. 2793-4]{Chen2002}. 

O regularizador é convexo e diferenciável e algoritmos iterativos podem ser utilizados para sua a solução \cite[pág. 138]{engl1996regularization}, mas a complexidade desses algoritmos aumenta. Não se considerava fácil comparar os resultados de regularizações de Tikhonov e a de máxima entropia, pois as duas poderiam obter bons resultados \cite[pág. 140]{engl1996regularization}.

\subsubsection{Existem outras medidas de entropia que podem ser utilizadas como regularizadores?}

Recentemente, \cite{Menin} partiu tanto de generalizações da entropia, como a Entropia de Tsallis, quanto da função logaritmo generalizada para obter uma forma unificada das duas regularizações em um termo só, trazendo \textit{insights} sobre essa relação entre elas. Em \cite{Velho2006}, os autores também utilizaram a entropia não-extensiva de Tsallis como regularizador. 

Outra relação poderia considerar a entropia de Renyi \cite{Silva2016}, que contém a entropia de Shannon como caso especial. Nota-se que há trabalhos que citam que a entropia de Renyi poderia ser utilizada como termo de regularização \cite[pág. 13]{Bercher2008}, \cite[pág. 2822]{Chen2002},  \cite[pág. 8]{Koenker2006}. 

\subsection{Regularização por \textit{denoising}: Como buscar por soluções com o menor ruído possível?}\label{sec:red}  

Não é sempre que se sabe qual é a característica esperada para a solução \cite{Romano2017} e em alguns casos admite-se que não há regularizadores adequados para determinadas tarefas \cite{Reehorst2019},  o que torna difícil traduzir características esperadas da reconstrução na forma de regularizadores. Assim, uma outra forma de regularização encontrada na literatura busca privilegiar soluções com o menor ruído possível ao utilizar um \textit{denoiser} como regularizador para problemas que não sejam necessariamente o \textit{denoising} \cite{Romano2017}. 

Entre eles está a proposta do \textit{Regularization by Denoising} (RED)  \cite{Romano2017}, na qual um regularizador explícito é construído a partir de
\begin{equation}
\Omega(\mathbf{x}) = \frac{1}{2} \mathbf{x}^T \left(\mathbf{x} - f(\mathbf{x})\right), 
\label{eq:RED}
\end{equation}
onde $f(\mathbf{x})$ é um \textit{denoiser}.  O funcional resultante é dado por
\begin{equation}
\hat{\mathbf{x}} = \underset{\mathbf{x}}{\arg\min} \left[ \vert \vert \mathbf{A}\mathbf{x} - \mathbf{y} \vert \vert^2_2 + \lambda^2 \frac{1}{2} \mathbf{x}^T \left(\mathbf{x} - f(\mathbf{x})\right) \right],
\label{eq:RED2}
\end{equation}
 e cuja otimização pode ser feita com métodos como gradiente descendente, ponto fixo ou o próprio ADMM, trazendo flexibilidade na sua implementação \cite[págs. 61-2]{Arridge2019}.
 
Diversos trabalhos incluem etapas de \textit{denoising} na solução do problema inverso. Alguns exemplos ilustrativos são:
\begin{itemize}
\item Em \cite{Reehorst2019}, os autores reinterpretaram o RED considerando o regularizador em termos quadráticos e desenvolveram algoritmos eficientes para sua implementação;
\item Em \cite{VenkaPlugplay2013}, os autores propuseram o \textit{plug-and-play prior}, na qual eles incluíram o \textit{denoiser} como uma das etapas de um algoritmo iterativo baseado em separação de variáveis. Especificamente, eles substituíram um subproblema de otimização do ADMM pela aplicação direta do \textit{denoiser}; 
\item Em \cite{zhang2021plugandplay}, os autores utilizaram o algoritmo \textit{half-quadratic splitting} (HQS)  para separação de variáveis e também substituíram um subproblema de otimização dele pela aplicação direta do \textit{denoiser}. Uma diferença relevante é que os autores utilizaram redes neurais convolucionais profundas como \textit{denoiser}.
\end{itemize}

\subsection{Como unir de métodos de projeção e de regularização?}

Diversos métodos iterativos que apresentam o fenômeno da semiconvergência (ver Apêndice \ref{sec:class}) podem ser entendidos dentro do contexto dos métodos de projeção (descritos no Apêndice \ref{Ap:projection}). Observa-se na Equação  \eqref{eq:proj2} que não há um termo de regularização e, mesmo assim, eles podem levar a efeitos de regularização \cite{Santos1996}, \cite[pág. 117]{hansen2010discrete}. Porém, a união de métodos de projeção com métodos regularização pode tornar os algoritmos iterativos mais robustos \cite[págs. 126, 221]{engl1996regularization}, \cite[págs. 127, 131]{hansen2010discrete}. 

Métodos que unem projeção e regularização são chamados de métodos híbridos e trazem as vantagens tanto da parte da projeção quanto da parte da regularização \cite[pág. 129]{hansen2010discrete}. Partindo da Equação \eqref{eq:proj2} e sendo o regularizador quadrático, um método híbrido \cite[pág. 127]{hansen2010discrete} apresenta a forma geral conforme 
\begin{equation}
\bm{\hat{s}}_D = \arg\min\limits_{\mathbf{x}} \left[ \vert \vert \mathbf{A} \mathbf{D} \mathbf{s}_D - \mathbf{y} \vert \vert^2_2 + \lambda^2 \vert \vert \mathbf{L}\mathbf{s}_D \vert \vert_2^2 \right],
\label{eq:norma2proj}
\end{equation}
onde $\mathbf{x}$ pode ser recuperado também pela Equação \eqref{eq:CS1}. 

Duas observações são necessárias.  Em termos de implementação, é possível tanto regularizar primeiro e projetar depois, quanto  projetar primeiro ou regularizar depois \cite[Figura 6.10]{hansen2010discrete}. Além disso, nesse caso é necessário tanto uma regra para escolha de  $\lambda$ quanto um critério de para do número de iterações do algoritmo, o que para problemas de grande escala pode ser desafiador, já que nem sempre é viável refazer muitas vezes uma mesma simulação para adequação dos parâmetros.  

\subsubsection{Como utilizar métodos híbridos para promoverem esparsidade?}
Para utilizar algoritmos de reconstrução com norma $\ell_1$ ou $\ell_0$ efetivamente, com melhor performance, é necessário admitir que o sinal seja esparso \cite{Daubechies2016}, mas a maioria dos sinais de aplicações práticas, como sinais biomédicos ou imagens, não o são em sua base original \cite[pág. 57]{majumdar2019compressed}. Caso ele não seja, pode-se buscar mudar a base de representação dos sinais para uma em que eles sejam esparsos em algum dicionário e então utilizar regularizadores que promovam esparsidade de modo mais eficiente, obtendo melhores resultados na reconstrução. 

Caso seja utilizada a norma $\ell_1$ no termo de regularização buscando soluções esparsas, o funcional resultante é conhecido como \textit{synthesis prior formulation} \cite[pág. 57]{majumdar2019compressed},
\begin{equation}
\mathbf{\hat{s}}_D = \arg\min\limits_{s} \left[ \vert \vert \mathbf{A} \mathbf{D} \mathbf{s}_D - \mathbf{y} \vert \vert^2_2 + \lambda^2 \vert \vert \mathbf{s}_D \vert \vert_1 \right],
\label{eq:norma1v}
\end{equation}
de modo a incluir restrições de esparsidade no vetor de representação $\mathbf{s}$. A solução da Equação \eqref{eq:norma1v} permite recuperar o sinal original através da equação de síntese \eqref{eq:CS1}, mas $\mathbf{D}$ deve ser possuir inversa, sendo ortogonal ou \textit{tight-frame} \cite[pág. 58]{majumdar2019compressed}.

Na \textit{co-sparse analysis prior formulation}\footnote{Co-esparsa significa esparsa no domínio de transformação \cite[pág. 58]{majumdar2019compressed}}, resolve-se para $\mathbf{x}$ conforme
\begin{equation}
\hat{\mathbf{x}} = \arg\min\limits_{\mathbf{x}} \left[ \vert \vert \mathbf{A} \mathbf{x} - \mathbf{y} \vert \vert^2_2 + \lambda^2 \vert \vert \mathbf{D}^{-1} \mathbf{x} \vert \vert_1 \right].
\label{eq:norma111}
\end{equation}
Para solução da Equação \eqref{eq:norma111}, pode-se utilizar algoritmos como \textit{majorization minimization}, \textit{split bregman} e \textit{greedy analysis pursuit} \cite[págs. 58, 60, 62]{majumdar2019compressed}.

\subsection{Como modificar a regularização de Tikhonov para melhor eficiência de algoritmos iterativos?}
A Equação \eqref{eq:norma2proj} define a união de métodos de projeção e de regularização com termos quadráticos. A Equação \eqref{eq:norma1v} busca a representação esparsa do sinal para poder utilizar algoritmos com norma $\ell_1$ de modo eficiente. Além dessas duas, existem outras propostas encontradas na literatura que possuem funcionais semelhantes, mas apresentam objetivos diferentes. Uma delas é a transformação para a forma padrão que visa a utilização de alguns algoritmos iterativos na solução e a outra é a utilização de precondicionadores que visam melhorar a condição do problema.

\subsubsection{Quais são as vantagens de se utilizar uma transformação para forma padrão?}

Seja o problema de Tikhonov na forma padrão, aquela da regularização clássica de Tikhonov (Subseção \ref{sec:tikhclas}). Seja também o problema de Tikhonov na forma geral aquela da regularização generalizada de Tikhonov (Subseção  \ref{sec:tikhgene}). Dependendo do algoritmo iterativo utilizado, pode ser necessário reescrever da forma geral para a forma padrão  para poder utilizá-lo \cite[págs. 172, 181]{hansen2010discrete}, \cite[pág. 221]{engl1996regularization}.

A partir da Equação \eqref{eq:tikhonov12}, quando $\mathbf{L} \neq \mathbf{I}$ e que $\mathbf{L}$ possua inversa, a ideia da transformação para forma padrão é a de se resolver o funcional equivalente dado por 
\begin{equation}
\mathbf{\overline{x}} = \arg\min\limits_{\overline{x}} \left[ \vert \vert \mathbf{A} \mathbf{L}^{-1} \mathbf{\overline{x}} - \mathbf{y} \vert \vert^2_2 + \lambda^2 \vert \vert \mathbf{\overline{x}} \vert \vert_2^2 \right].
\label{eq:tikhstandard}
\end{equation}
Uma vez obtido $\mathbf{\overline{x}}$, a solução desejada é calculada pela transformação reversa conforme
\begin{equation}
\hat{\mathbf{x}} = \mathbf{L}^{-1} \mathbf{\overline{x}}.
\label{eq:tikhstandard2}
\end{equation}
O desafio é que $ \mathbf{L}$ pode ser singular, por não ter posto completo ou por não ser quadrada, de modo que não se resume a pensar em invertê-la. Os casos em que $ \mathbf{L}$ possuem mais linhas do que colunas, ou vice-versa, são discutidos em \cite[pág. 181]{hansen2010discrete} e outras formas de realizar essa transformação que sejam mais adequadas em \cite[Seção 8.5]{hansen2010discrete}.

\subsubsection{Quais são as vantagens de se utilizar precondicionadores e \textit{priorconditioners}?}

É possível estabelecer uma relação entre abordagens bayesianas descritas no Apêndice \ref{Ap:estimador} e algoritmos iterativos truncados. A performance de alguns algoritmos iterativos pode diminuir quando o sistema de equações é mal-condicionado. Uma possibilidade de melhorar a solução é através de seu precondicionamento. 

Seja a matriz $\mathbf{P}$ não-singular de modo que $\mathbf{P}\mathbf{x} = \mathbf{s}_{P}$. Retomando a Equação \eqref{eq:eq1}, resolver esse sistema linear será equivalente a resolver \cite[pág. 82]{Mueller2012} 
\begin{equation}
\mathbf{A}\mathbf{P}^{-1} \mathbf{s}_{P} = \mathbf{y},
\label{eq:precond2}
\end{equation}
onde $\mathbf{s}_P$ são os parâmetros que estão relacionados com um precondicionador $\mathbf{P}$.

Observa-se que a posição de $\mathbf{P}$ na Equação \eqref{eq:precond2} está a direita de $\mathbf{A}$. Portanto, $\mathbf{P}$ é um precondicionador a direita, mas existe também aquele pela esquerda ou por ambos os lados \cite[págs. 107-8]{calvetti2007introduction}. Deseja-se que esse novo sistema resulte tanto em um aumento de velocidade de convergência quanto em um sistema de equações melhor condicionado que o original \cite[pág. 179]{aster2019parameter}, \cite[págs. 81-2]{Mueller2012} e a escolha do precondicionador $\mathbf{P}$ pode determinante para o sucesso do algoritmo. No entanto, sabe-se que $\mathbf{A}$ nem sempre é invertível e há a presença de ruídos em $\mathbf{y}$, sendo necessária regularização adicional. 

A partir da Equação \eqref{eq:precond2} pode-se reescrever a Equação \eqref{eq:tikhonov12} para incluir o precondicionador $\mathbf{P}$ como \cite[pág. 110]{calvetti2007introduction} 
\begin{equation}
\bm{\hat{s}}_P = \arg\min\limits_{s} \left[ \vert \vert \mathbf{A} \mathbf{P}^{-1} \mathbf{s}_{P} - \mathbf{y} \vert \vert^2_2 + \lambda^2 \vert \vert \mathbf{s}_{P} \vert \vert_2^2 \right].
\label{eq:precond3}
\end{equation}
Em algumas situações pode-se utilizar métodos iterativos que são regularizados pelo truncamento e baseados na semiconvergência, conforme discutido, de modo que não há a necessidade de acelerar a convergência  \cite{Calvetti2005}. 

Outra possibilidade é utilizar essa formulação não com o objetivo de melhorar o condicionamento do sistema, mas de incluir informações estatísticas \textit{a priori} sobre a solução desejada. Nesse caso, os precondicionadores são chamados de \textit{priorconditioners} e fazem essa ponte de algoritmos iterativos com métodos bayesianos de inversão \cite[Capítulo 6]{calvetti2007introduction},  \cite{Calvetti2005},  \cite[pág. 8]{Gazzola2018}. 

Nesses trabalhos, os autores discutem a construção de \textit{priorconditioners} a partir de amostras representativas da solução, onde são calculadas a matriz de covariância e o vetor médio dessas amostras. Além de \textit{priorconditioners}, esses termos também podem ser utilizados, respectivamente, como a matriz de regularização $\mathbf{L}$ e o valor de referência $\mathbf{x}^*$ na regularização de Tikhonov.

\subsection{Existem formas para obter componentes de regularizadores a partir de dados?}\label{sec:dados}

Os regularizadores discutidos podem ser caracterizados sob alguns aspectos:
\begin{itemize}
\item Todos os termos de regularização descritos até agora podem ser chamados de \textit{handcrafted} \cite[págs. 26, 38]{Arridge2019}, aquele cuja forma é definida de antemão pelo pesquisador. Muitas vezes, esse é o caso dos \textit{priors} explícitos discutidos;
\item Ao mesmo tempo, há opções de $\Omega(\mathbf{x})$ que são de propósito geral, cuja escolha deve ser feita de acordo com o problema que se quer resolver, mas elas não são personalizadas para uma aplicação em específico. 
\end{itemize}
Logo, deve-se discutir a possibilidade de desenvolver regularizadores que não são \textit{handcrafted} e que sejam específicos para uma aplicação. 

De fato, uma outra forma de obter regularizadores é através de obter seus componentes a partir de dados, seja com estatística convencional ou com aprendizado de máquina, em diferentes etapas da inversão. Três abordagens serão discutidas: \textit{priors} baseados em amostras, a obtenção de uma base de representação para o sinal com aprendizado de dicionário e aprendizagem de parâmetros do regularizador com otimização em dois níveis.

\subsubsection{É possível incluir informação anatômica como \textit{prior}?}\label{sec:atlas}

É possível obter \textit{priors} a partir de amostras \cite{Calvetti2018a}, \cite[Seção 3.3.5]{kaipio2005statistical}. Um exemplo é quando se tem acesso a um conjunto de soluções típicas que apresentam as estruturas de interesse, como no caso de imagens médicas, pois estruturas são esperadas em cada região do corpo humano, mas há uma variação anatômica entre as pessoas e é importante que essas incertezas estejam contempladas na reconstrução. 

No contexto da inversão bayesiana (Apêndice \ref{Ap:estimador}), esta é uma forma a obter uma densidade de probabilidade \textit{a priori} da solução desejada. Os chamados atlas anatômicos \cite[pág. 70]{kaipio2005statistical} permitem que, para aquela população, sejam estimados informações que não são triviais de se expressar em termos quantitativos feitos à mão \cite{Calvetti2018a}. Para cada tipo de imagem médica, a grandeza representada é diferente. Um exemplo de aplicação é para tomografia por impedância elétrica, cuja média e covariância estão relacionadas com valores de resistividade \cite{Moura2021}.  Neste caso, calcula-se o valor médio da grandeza de interesse como $\mathbf{x}^*$, além da sua matriz de covariância $\bm{\Gamma}$ entre os sujeitos, cuja inversa é utilizada na solução regularizada. Especificamente, o atlas anatômico apresenta a forma 
 \begin{equation}
\Omega(\mathbf{x}) = \left( \mathbf{x} - \mathbf{x}^*\right)^T  \bm{\Gamma}^{-1} \left( \mathbf{x} - \mathbf{x}^*\right).
\end{equation}
Isso significa que o atlas anatômico é uma forma de calcular o termo de regularização a partir dos dados, mas não está exatamente na forma $\Omega(\mathbf{x}) = \vert \vert \mathbf{L} \left(\mathbf{x} - \mathbf{x}^*\right)\vert \vert^2_2$. Caso isso fosse desejado, seria necessário fatorar a matriz $\bm{\Gamma}^{-1} = \mathbf{L}^T \mathbf{L}$, por exemplo com a decomposição Cholesky \cite[pág. 112]{calvetti2007introduction}. A matriz $\bm{\Gamma}$ pode não apresentar inversa \cite[págs. 70-1]{kaipio2005statistical} e uma possibilidade é a adição de uma matriz identidade 
\begin{equation}
\Omega(\mathbf{x}) = \left( \mathbf{x} - \mathbf{x}^*\right)^T \left( \bm{\Gamma} + a \mathbf{I} \right)^{-1} \left( \mathbf{x} - \mathbf{x}^*\right), 
\end{equation}
onde $k$ é um escalar \cite[pág. 50]{Camargo2013}. 

Dessa forma, o regularizador representa informações estatísticas dos parâmetros de uma população \cite{Calvetti2018a}, \cite[pág. 70]{kaipio2005statistical}. Através do conceito de \textit{subspace prior} \cite[pág. 71]{kaipio2005statistical}, ainda é possível reescrever essa expressão como um termo de regularização generalizada de Tikhonov $\Omega(\mathbf{x}) = \vert \vert \mathbf{L} \left( \mathbf{x} - \mathbf{x}^* \right) \vert \vert^2_2 $, evitando a inversão explícita de $\bm{\Gamma}$. 

O \textit{prior} deve ser independente dos dados sobre o qual se fará a estimação (a reconstrução), de modo que os dados do paciente específico (com uma possível patologia) não são incluídos na construção do atlas. Mas as imagens médicas devem permitir a detecção de anomalias e se o atlas foi desenvolvido apenas com pacientes saudáveis (excluindo casos patológicos), ou mesmo com poucas amostras de casos patológicos, a solução pode ficar muito enviesada, sendo incapaz da solução reproduzir esses \textit{outliers}, que era para ser o objetivo inicial \cite{calvetti2007introduction}. Nestes casos, pode-se fazer uma avaliação da medida de informação de uma patologia em um atlas anatômico \cite{Nakanishi2020} e adicionar amostras com a anomalia esperada \cite{calvetti2007introduction}, \cite[pág. 223]{kaipio2005statistical}.

\subsubsection{É possível obter representações com aprendizado de dicionário?}

A Equação \eqref{eq:norma2proj} descreveu um método híbrido entre regularização e projeção. No lugar de utilizar uma base fixa desde o início, é possível obter dicionários $\mathbf{D}$ a partir de dados, conforme Apêndice \ref{sec:baseslearned}. Alguns exemplos incluem:

\begin{itemize}
\item Em \cite{Soltani2017, Soltani2016}, realizou-se a aprendizado de dicionário com o objetivo de incorporar informações sobre a textura do objeto a partir de imagens de treinamento. É importante que eles os dados utilizados para o cálculo do dicionário sejam independentes dos dados que serão utilizados na reconstrução. Nesse caso, os autores dividem os problemas em primeiro construir o dicionário a partir de imagens de treinamento e depois utilizar o dicionário no problema de reconstrução em uma equação similar à Equação \eqref{eq:norma2proj};

\item Outra forma para manter os dados independentes é a aprendizado de dicionário multimodal, quando o dicionário de uma modalidade (por exemplo de imagens médicas) é utilizada na reconstrução de outra modalidade \cite{TONG2016153}; 

\item Há também aprendizado de dicionário simultânea à reconstrução \cite[pág. 54]{Arridge2019}, \cite{LewisD2019}. Em \cite{Soltani2017, Soltani2016}, os autores argumentam que isso viola o princípio de que a informação \textit{a priori} deve ser independente dos dados dos quais se fará a reconstrução. Mesmo assim, os resultados experimentais podem ser obtidos e podem ser adequados empiricamente.
 
\end{itemize}

\subsubsection{É possível obter parâmetros do regularizador a partir de dados?}

Utilizando dados, é possível definir regularizadores com estruturas conhecidas, mas parametrizados por $\bm{\theta}$ e denotados por $\Omega_{\bm{\theta}}(\mathbf{x})$. Os valores ótimos de $\bm{\theta}$ são obtidos em dados separados daqueles que se vai utilizar na reconstrução a partir de otimização em dois níveis, na qual os termos de regularização são parametrizados por $\bm{\theta}$ e apresentam formas pré-definidas diversas \cite{Haber2003}, como a variação total generalizada \cite{DelosReyes2016}. Os parâmetros $\bm{\theta}$ são obtidos diretamente a partir de dados da tarefa \cite[pág. 45]{Arridge2019}, \cite{Haber2003}. Com aprendizagem supervisionada, essa formulação permite realizar tanto a reconstrução $\hat{\mathbf{x}}_{\bm{\theta}}$ quanto estimar os parâmetros $\hat{\bm{\theta}}$ do regularizador ótimo \cite[pág. 45-7]{Arridge2019}, \cite{Haber2003}, \cite{DelosReyes2016},  conforme
\begin{equation}
\begin{aligned}
\hat{\bm{\theta}}  & = \arg\min_{\bm{\theta}} \left[\mathcal{L}_1\left( \hat{\mathbf{x}}_{\bm{\theta}}, \mathbf{x}_t \right) \right]\\
& s.t. \quad \hat{\mathbf{x}}_{\bm{\theta}} = \arg\min_{\mathbf{x}} \left[ \mathcal{L}_2(\mathbf{A}(\mathbf{x}),\mathbf{y}) + \Omega_{\bm{\theta}}(\mathbf{x}) \right],
\end{aligned}
\label{eq:learning2}
\end{equation}
onde $ \mathcal{L}_1$ e $ \mathcal{L}_2$ são funções de perda e $\mathbf{x}_t$ são modelos \textit{ground truth} de treinamento. 

É interessante notar que autores em \cite{Haber2003}, \cite{DelosReyes2016}, partem da otimização em dois níveis não apenas para obter $\Omega(\mathbf{x})$ diferentes, mas sim argumentando que eles obtém aqueles que são ótimos para a tarefa. A busca por regularizadores ótimos ainda é objeto de pesquisa \cite{alberti2021learning, DelosReyes2016}, mas sempre se deve atentar na definição do critério de otimalidade em cada caso. De qualquer forma, a otimização em dois níveis tem como desafio a sua solução numérica \cite{Afkham2021}, pois ela é custosa e se torna impraticável computacionalmente quando $\Omega_{\bm{\theta}}$ se encontra em um espaço com dimensão muito grande \cite{Lunz2018}.

\newpage

\section{QUAIS SÃO OS PRINCIPAIS ALGORITMOS PELO MÉTODO DE REGULARIZAÇÃO PARA \textit{NON-BLIND DEBLURRING}?}\label{sec:de_nonblind}

\subsection{Introdução}

\begin{figure}[H]
     \centering
     \begin{subfigure}[b]{0.34\textwidth}
         \centering
         \includegraphics[width=1\textwidth]{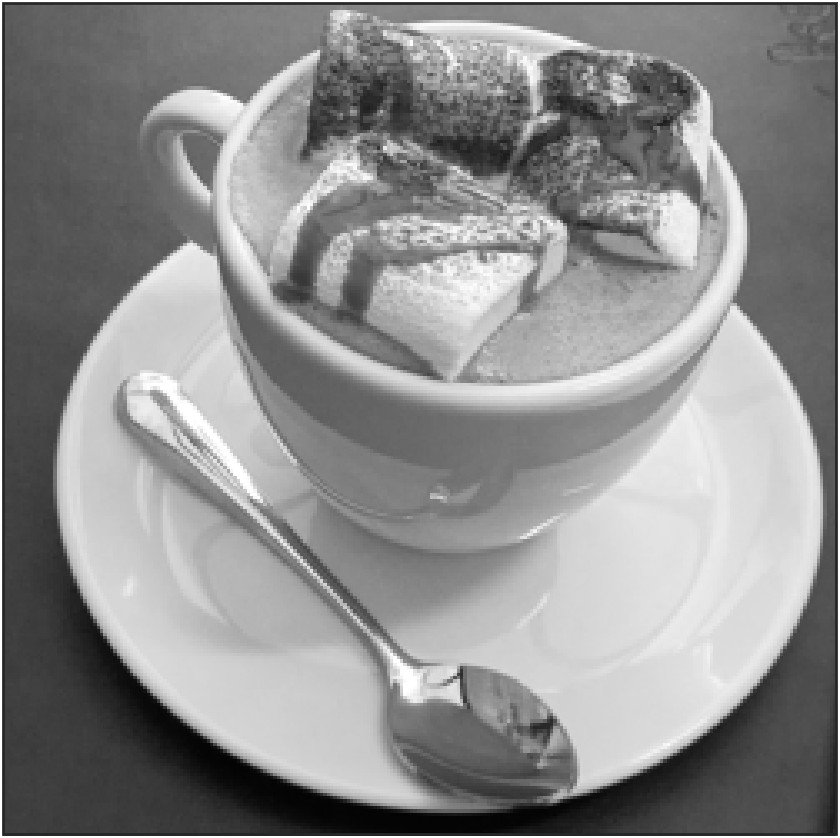}
         \caption{Mesmo operador e sem ruído}
         \label{fig:01_005a}
     \end{subfigure}
     \begin{subfigure}[b]{0.34\textwidth}
         \centering
                  \includegraphics[width=1\textwidth]{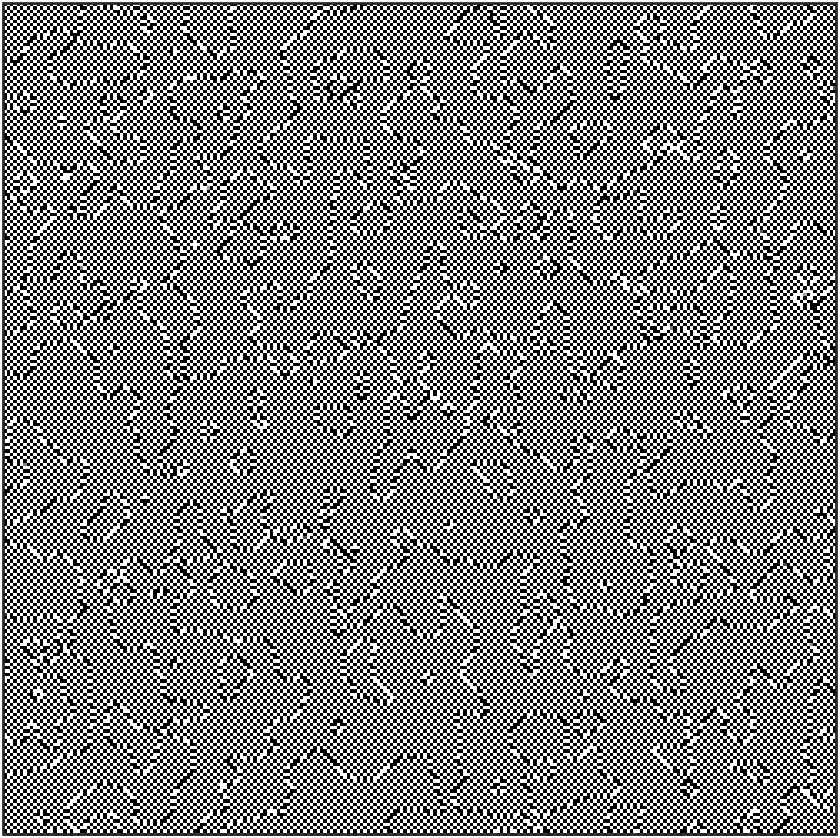}
         \caption{Mesmo operador e com ruído}
         \label{fig:01_005b}
     \end{subfigure}
\caption[Solução ingênua com crime de inversão.]{Solução ingênua com crime de inversão. Fonte: Próprio autor.}
\label{fig:01_005}
\end{figure}

 A Figura \ref{fig:01_005} mostra a inversão ingênua da Equação \eqref{eq:caso1_sol3}, sem regularização, para o \textit{blur} gaussiano isotrópico da Figura \ref{fig:01_002}. Quando o operador direto é conhecido e não há ruído aditivo, a reconstrução é perfeita, conforme Figura \ref{fig:01_005a}. No entanto, quando se adiciona ruído gaussiano, a Figura \ref{fig:01_005b} mostra que solução ingênua não é adequada \cite[pág. 5]{hansen2006deblurring}, ilustrando a necessidade de regularização.

Todas as reconstruções neste capítulo foram realizadas com o conhecimento de $\mathbf{A}$ ou $\mathbf{h}$, ou seja, com crime de inversão. Especificamente, utilizou-se o \textit{blur} gaussiano isotrópico da Figura \ref{fig:01_002}, o mesmo da solução ingênua anteriormente. Mesmo com o conhecimento exato de $\mathbf{A}$, o mal-condicionamento da matriz $\mathbf{A}$ e a presença de ruído nas medidas não permite a sua solução perfeitamente. Isso ilustra a dificuldade de se resolver problemas inversos mal-postos, que é ainda mais difícil quando se utilizam dados experimentais. 

Existem diversos algoritmos para realizar a reconstrução da imagem nítida a partir da imagem borrada quando se tem conhecimento da PSF. Retomando a Equação \eqref{eq:tikhonovgeral}, ressalta-se que alguns algoritmos seguem a formulação com matrizes conforme
\begin{equation}
\hat{\mathbf{X}}_{\lambda} = \arg\min\limits_{\mathbf{X}} \left[ \mathcal{L} \left(\mathbf{H}*\mathbf{X}, \mathbf{Y} \right) + \lambda \Omega(\mathbf{X}) \right],
 \label{eq:deblurringgeral}
\end{equation}
enquanto outros seguem uma formulação com vetores 
\begin{equation}
\hat{\mathbf{x}}_{\lambda} = \arg\min\limits_{\mathbf{x}} \left[ \mathcal{L} \left(\mathbf{A}\mathbf{x}, \mathbf{y} \right) + \lambda \Omega(\mathbf{x}) \right].
 \label{eq:deblurringgeral2}
\end{equation}
Considerando a função de perda $ \mathcal{L}$ quadrática, é possível avaliar o efeito da escolha de diferentes regularizadores $\Omega(\mathbf{x})$ no \textit{deblurring} utilizando as Equações \eqref{eq:deblurringgeral} e \eqref{eq:deblurringgeral2}. Em relação à implementação dos algoritmos de otimização:

\begin{itemize}
\item Métodos em um passo, como a regularização generalizada de Tikhonov, podem depender da montagem explícita das matrizes $\mathbf{A}$ e $\mathbf{L}$, conforme Equação \eqref{eq:tiksolgen}. 
\item Métodos iterativos, como o de gradientes conjugados, ou métodos híbridos podem depender apenas das multiplicações $\mathbf{A}\mathbf{x}$, $\mathbf{A}^T\mathbf{x}$, $\mathbf{L}\mathbf{x}$ e $\mathbf{L}^T\mathbf{x}$, sem precisar montar a matriz $\mathbf{A}$ durante a reconstrução. Isso permite utilizar algoritmos eficientes para solução do problema inverso
\item Em alguns casos, é possível trabalhar diretamente com as imagens e o \textit{kernel} sem vetorizá-las, isto é,  na forma de matrizes $\mathbf{H}$ e $\mathbf{X}$, respectivamente. Logo, o cálculo da imagem borrada é realizado através de $\mathbf{H}*\mathbf{X}$, na qual há implementações eficientes a operação de convolução, evitando também a montagem explícita de $\mathbf{A}$. 
\end{itemize}

As Figuras \ref{fig:01_012}-\ref{fig:01hybrid} ilustram o uso de diferentes regularizadores para realizar \textit{deblurring}. A Tabela \ref{Table:especificacoes_estudo} mostra as \textit{toolboxes} utilizadas.

{\centering 
\begin{longtable}{|| c  c  c || }
\caption{Toolboxes e códigos externos utilizados.}
\label{Table:especificacoes_estudo}  \\ \hline
 \rowcolor{lightyellow} \textbf{Figura} & \textbf{Tarefa} & \textbf{Fonte} \\ \hline\hline
\ref{fig:bludifa} & Montagem de $\mathbf{A}$ &  \cite[Exemplo 6.2]{aster2019parameter}\footnote{\url{https://github.com/brianborchers/PEIP}} \\
\ref{fig:laplacian} & Operador laplaciano $\mathbf{L}$ & IR Tools \cite{Gazzola2018}\footnote{\url{https://github.com/jnagy1/IRtools}} \\
\ref{fig:l1} & Norma $\ell_1$& \cite[Algoritmos 7.2-4]{aster2019parameter}\\
\ref{fig:deconvtv} & Variação total & deconvtv \cite{Chan2011}\footnote{\url{https://www.mathworks.com/matlabcentral/fileexchange/43600-deconvtv-fast-algorithm-for-total-variation-deconvolution}}e  \cite{Mueller2012}\footnote{\url{https://wiki.helsinki.fi/xwiki/bin/view/mathstatHenkilokunta/Henkil\%C3\%B6t/Siltanen\%2C\%20Samuli/Inverse\%20Problems\%20Book\%20Page/X-ray\%20tomography\%20with\%20matrices/Matrix-free\%20X-ray\%20tomography/Matrix-free\%20X-ray\%20tomography\%20with\%20sparse\%20data/2D\%20deconvolution/}} \\
\ref{fig:maxent} & Máxima entropia & Regtools \cite{Hansen2007}\footnote{\url{https://www.imm.dtu.dk/~pcha/Regutools/}} \\ 
\ref{fig:medianpnp} &  \textit{Plug-and-play} ADMM &  \cite{Chan2016}\footnote{\url{https://www.mathworks.com/matlabcentral/fileexchange/60641-plug-and-play-admm-for-image-restoration}} \\
\ref{fig:red} & Regularização por \textit{denoising} &  \cite{Romano2017}\footnote{\url{https://github.com/google/RED}} \\
\ref{fig:01hybrid} & Métodos iterativos truncados & IR Tools \cite{Gazzola2018}\\ \hline
\end{longtable}
}

\subsection{Regularização clássica de Tikhonov}

\begin{figure}[H]
     \centering
     \begin{subfigure}[b]{0.32\textwidth}
         \centering
         \includegraphics[width=\textwidth]{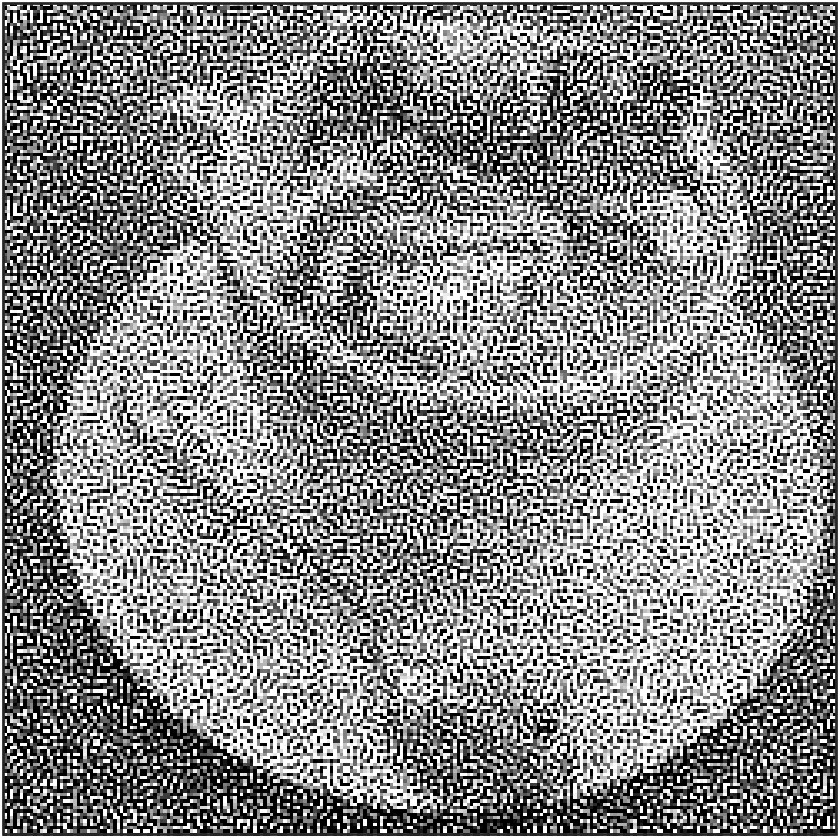}
         \caption{$\lambda = 8e-05$}
         \label{fig:01_012a}
     \end{subfigure}
     \hfill
     \begin{subfigure}[b]{0.32\textwidth}
         \centering
                  \includegraphics[width=\textwidth]{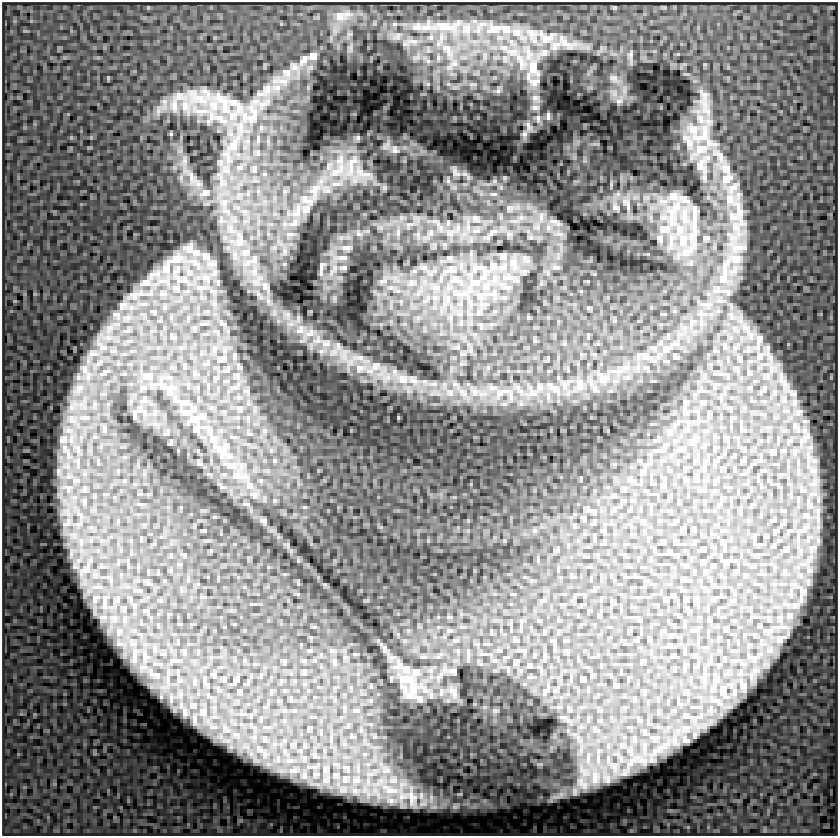}
         \caption{$\lambda = 0.0025$}
         \label{fig:01_012b}
     \end{subfigure}
          \hfill
     \begin{subfigure}[b]{0.32\textwidth}
         \centering
                  \includegraphics[width=\textwidth]{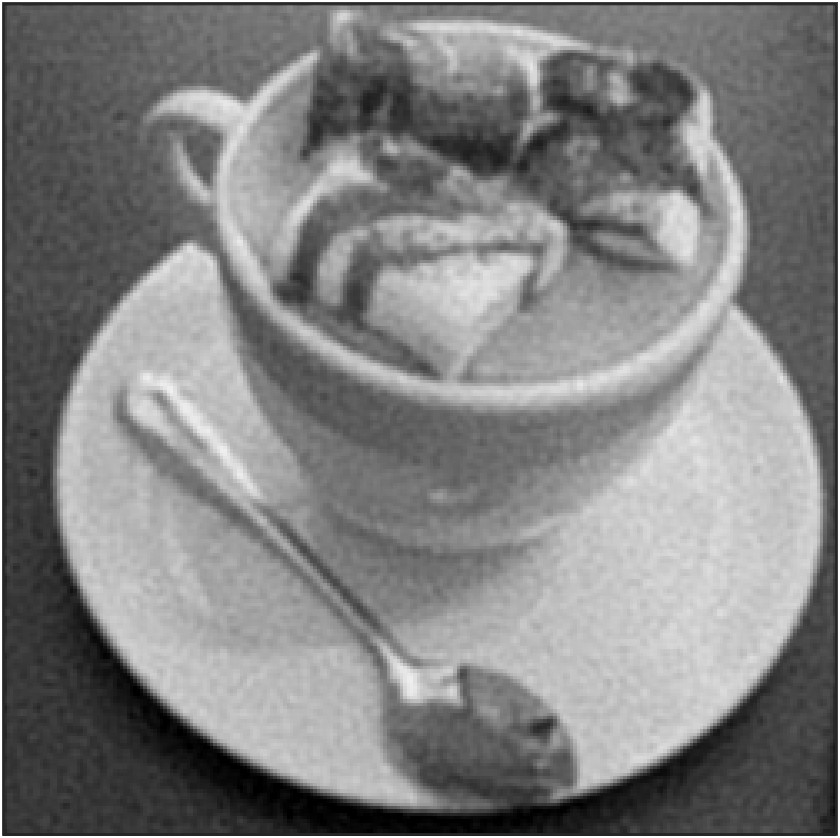}
         \caption{$\lambda = 0.1$}
         \label{fig:01_012c}
     \end{subfigure}
\caption[Regularização clássica de Tikhonov para diferentes valores de $\lambda$.]{Regularização clássica de Tikhonov para diferentes $\lambda$. Fonte: Próprio autor.}
\label{fig:01_012}
\end{figure}

A Figura \ref{fig:01_012} mostra o resultado da regularização clássica de Tikhonov. Observa-se a influência de $\lambda$ na solução. Um valor pequeno de $\lambda$ não permite filtrar os ruídos de modo adequado, enquanto um valor grande de $\lambda$ suaviza a solução além do necessário. 

\begin{figure}[H]
\centering
\includegraphics[width=1\textwidth]{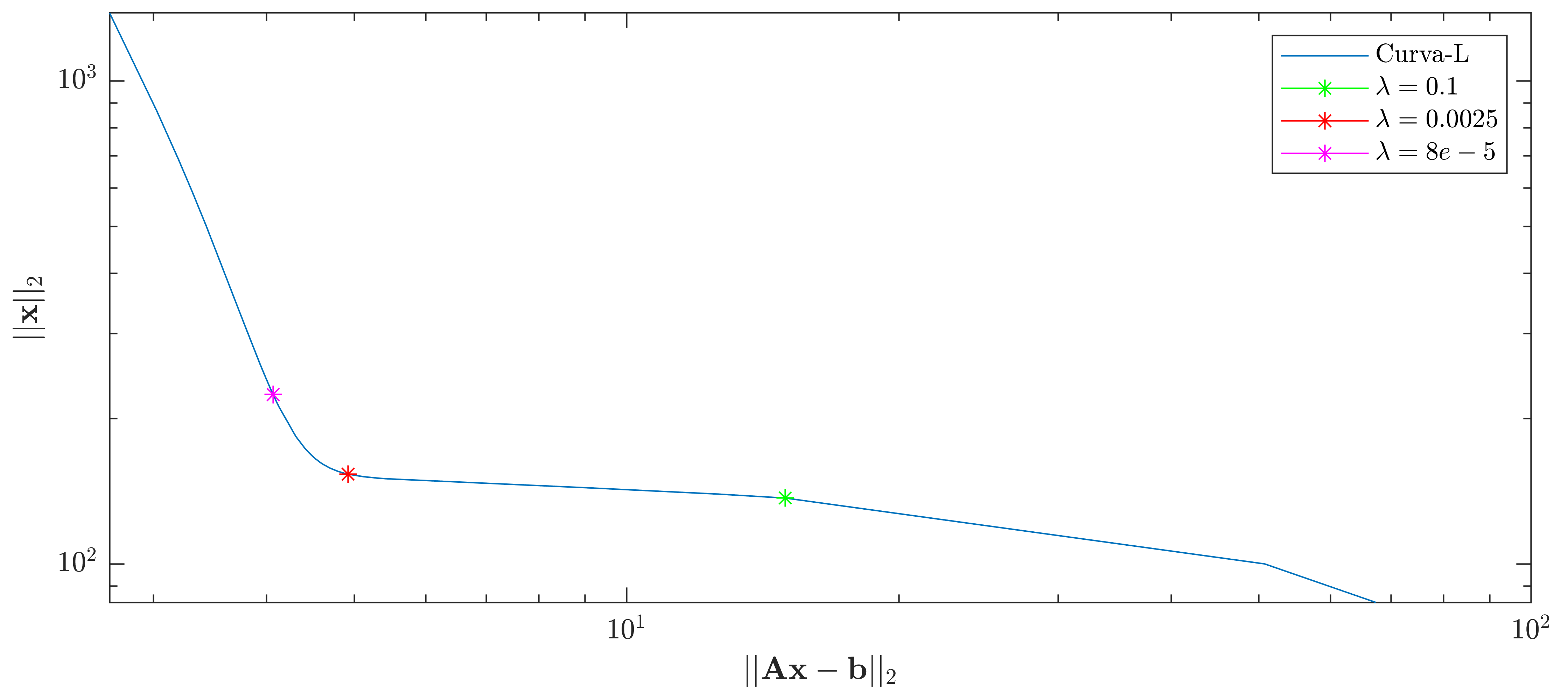} 
\caption[Curva-L para as soluções regularizadas.]{Curva-L para as soluções regularizadas. Fonte: Próprio autor.}
\label{fig:01_016}
\end{figure}

Na Figura \ref{fig:01_016} é mostrada a Curva-L para $10^{-6} \leq \lambda \leq 0.8$, um critério de escolha de $\lambda$. Nela, há um segmento aproximadamente horizontal, um segmento aproximadamente vertical e um canto que separa essas duas regiões, gerando um gráfico que lembra a letra ``L'', origem de seu nome.  Os valores de $\lambda$ das soluções da Figura \ref{fig:01_012} são destacados. 

A Figura \ref{fig:01_016} também permite obter o valor de $\lambda$ segundo o princípio de discrepância de Morozov. Localizando no eixo das abcissas o valor de $\vert \vert \bm{\delta} \vert \vert_2$, que nesse exemplo vale $\vert \vert \bm{\delta} \vert \vert_2 \approx 6.22$, obtém-se um $ \lambda$ maior do que o $\lambda$ do canto da Curva-L.

\subsection{Regularização generalizada de Tikhonov} 

\begin{figure}[H]
     \centering
     \begin{subfigure}[b]{0.32\textwidth}
         \centering
         \includegraphics[width=\textwidth]{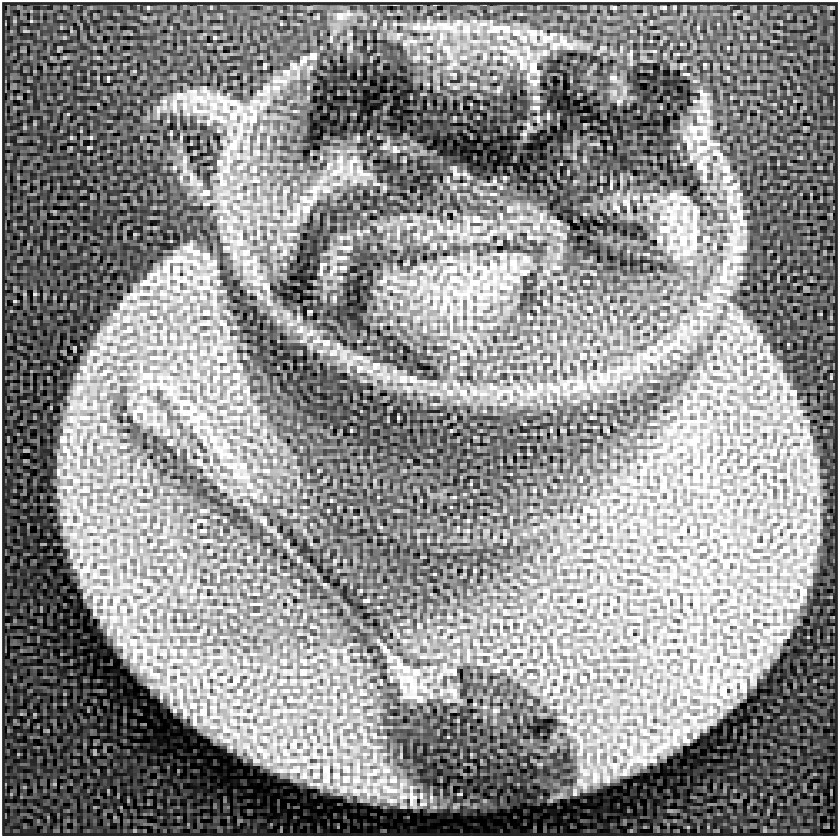}
         \caption{$\lambda = 8e-05$}
         \label{fig:laplaciana}
     \end{subfigure}
     \hfill
     \begin{subfigure}[b]{0.32\textwidth}
         \centering
                  \includegraphics[width=\textwidth]{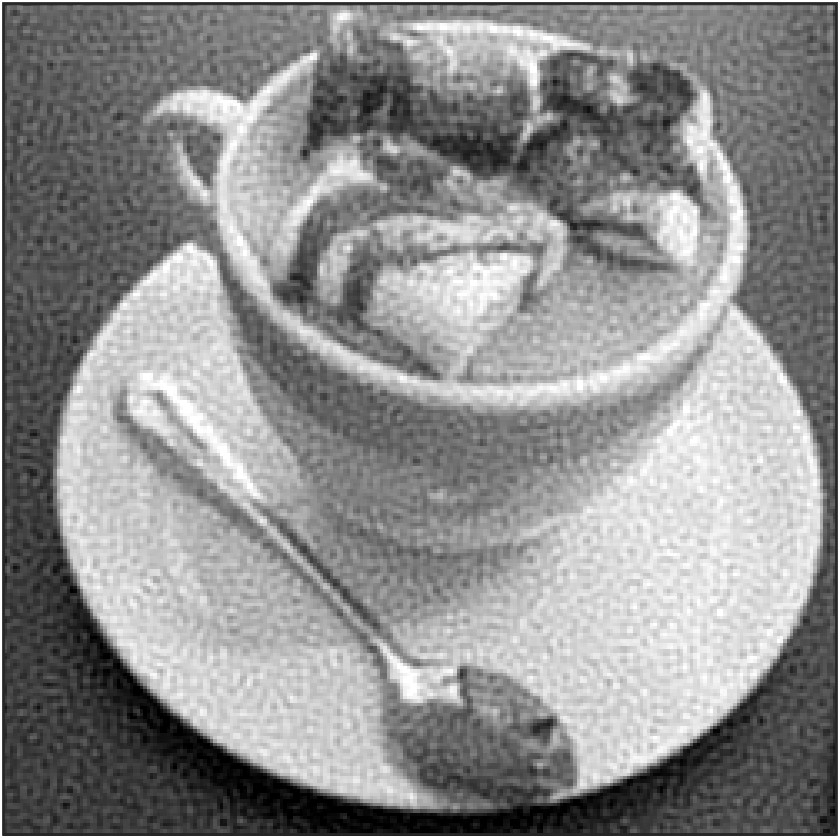}
         \caption{$\lambda = 0.0025$}
         \label{fig:laplacianb}
     \end{subfigure}
          \hfill
     \begin{subfigure}[b]{0.32\textwidth}
         \centering
                  \includegraphics[width=\textwidth]{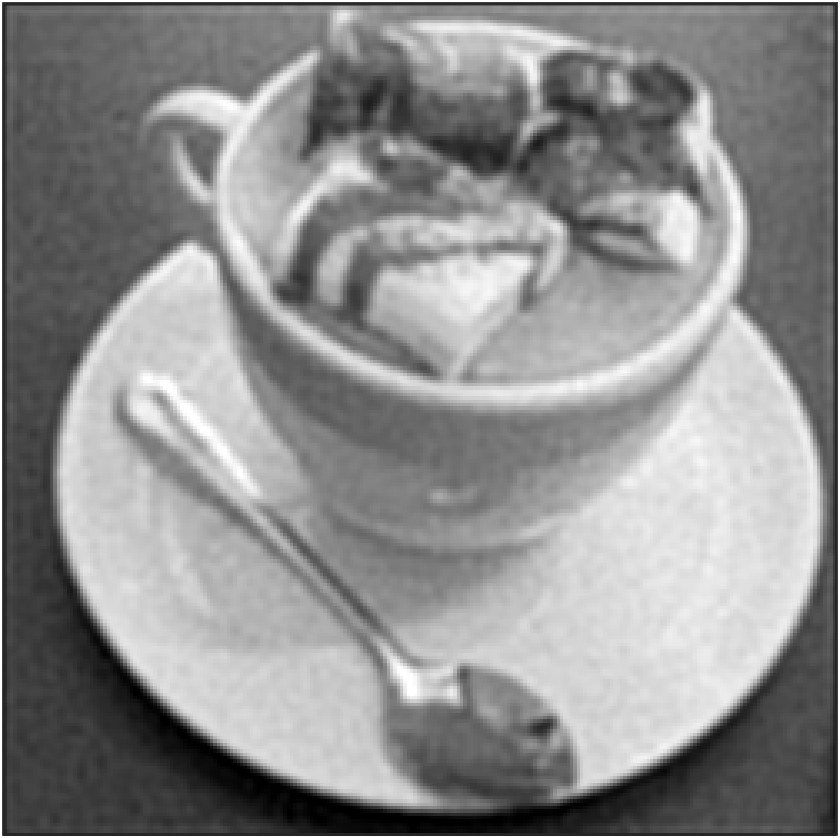}
         \caption{$\lambda = 0.1$}
         \label{fig:laplacianc}
     \end{subfigure}
\caption[\textit{Deblurring} com operador laplaciano.]{\textit{Deblurring} com operador laplaciano. Fonte: Próprio autor.}
\label{fig:laplacian}
\end{figure}

Na Figura \ref{fig:laplacian} são mostradas reconstruções para os mesmos $\lambda$ da Figura \ref{fig:01_012}, mas, utilizando o operador laplaciano como regularizador, os resultados são ainda mais suavizados. Ainda que a Figura \ref{fig:laplacianb} seja mais nítida do que a imagem degradada, uma regularização que obtenha soluções suavizadas pode não ser a escolha mais adequada no caso do \textit{deblurring}, já que o objetivo é recuperar altas frequências.

\subsection{Filtro de Wiener}

\begin{figure}[H]
     \centering
     \begin{subfigure}[b]{0.35\textwidth}
         \centering
         \includegraphics[width=0.90\textwidth]{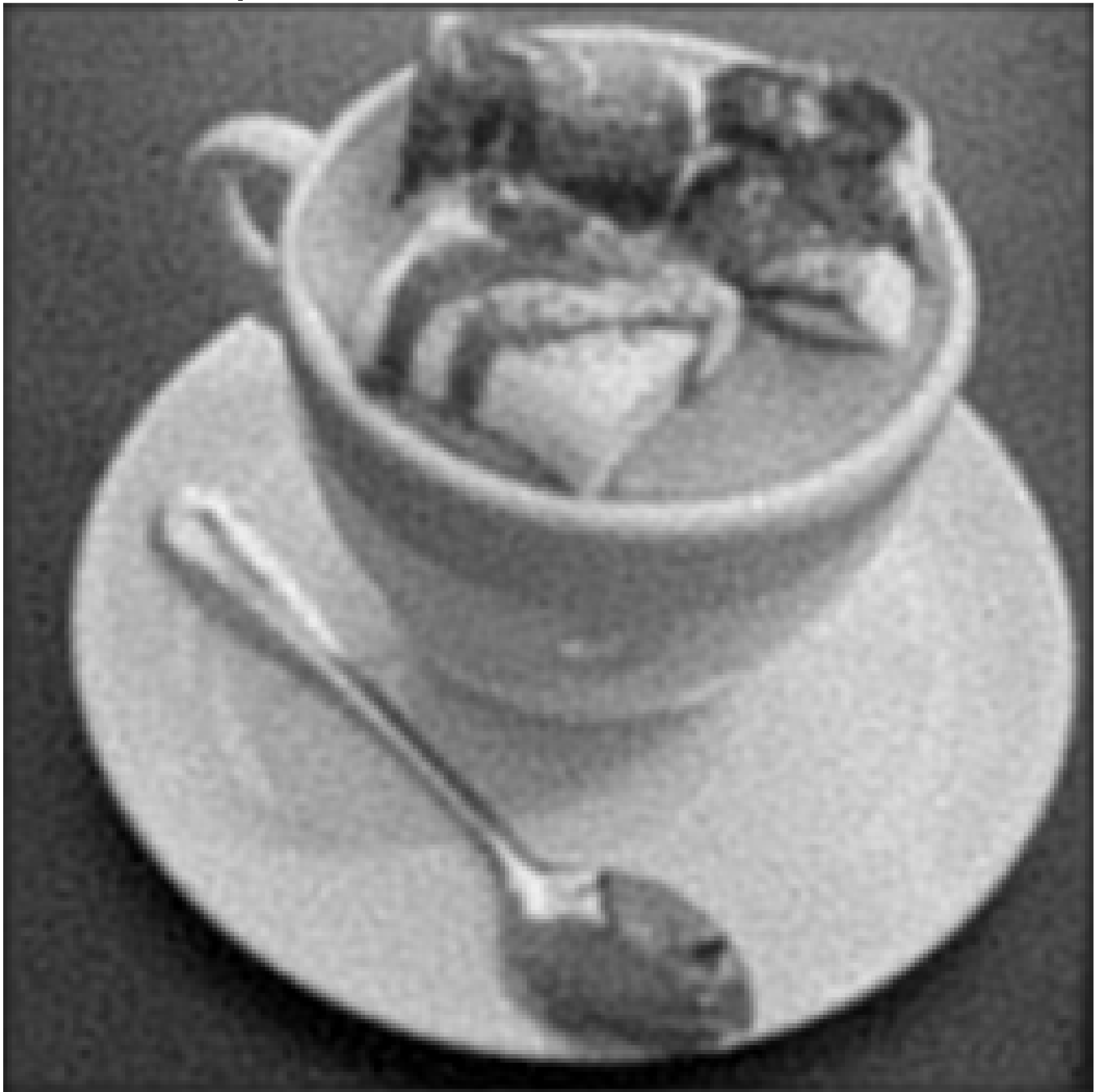}
         \caption{NSR = 0.01}
         \label{fig:fftwienera}
     \end{subfigure}
     \begin{subfigure}[b]{0.35\textwidth}
         \centering
         \includegraphics[width=0.90\textwidth]{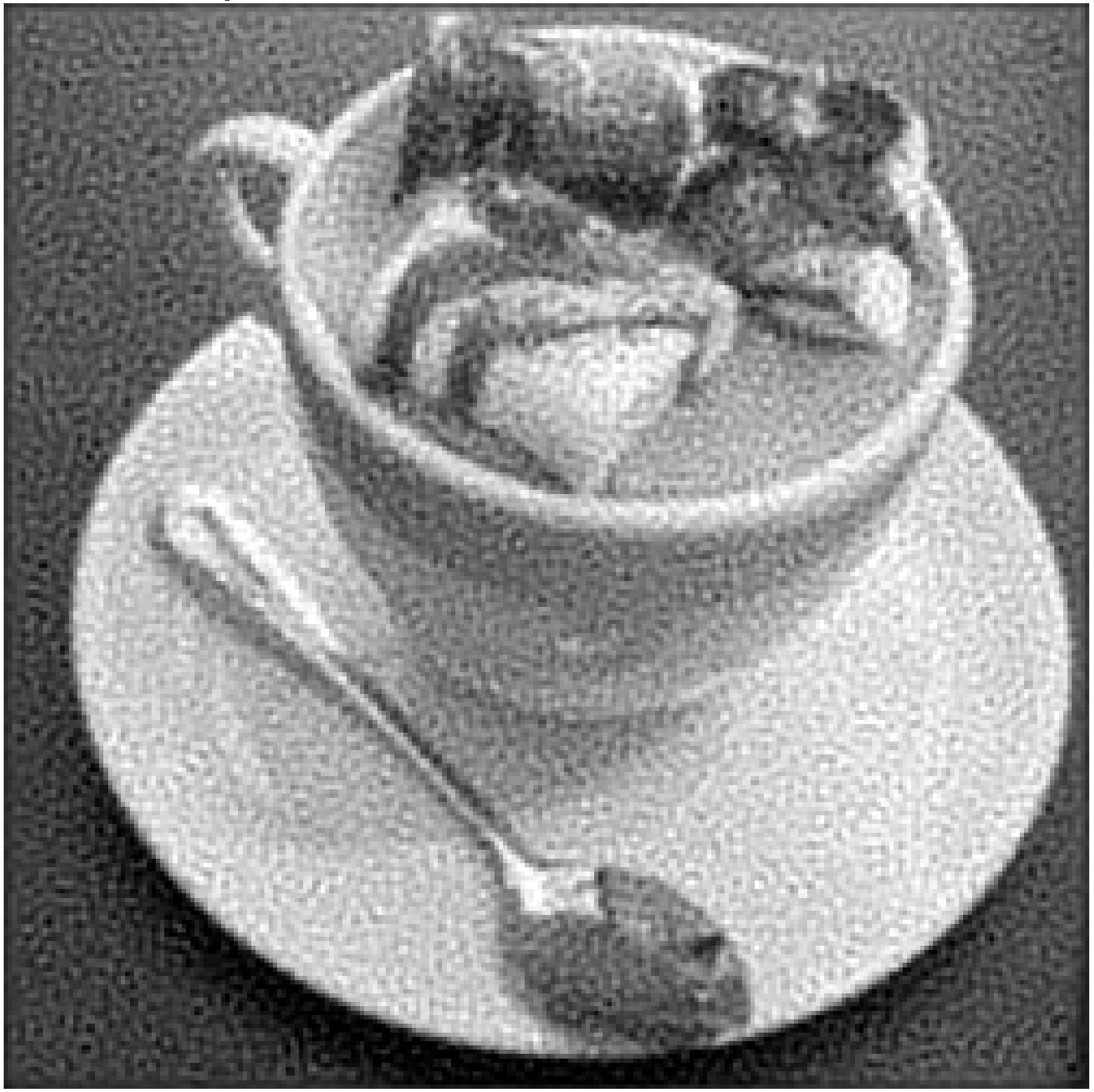}
         \caption{NSR = 0.1}
         \label{fig:fftwienerb}
     \end{subfigure}
\caption[Imagens reconstruídas com filtro de Wiener.]{Imagens reconstruídas com filtro de Wiener. Fonte: Próprio Autor. }
\label{fig:fftwiener}
\end{figure}

Seja a razão entre o ruído e o sinal (NSR), o  recíproco da razão sinal-ruido (SNR). Na Figura \ref{fig:fftwiener} são mostradas duas reconstruções com o filtro de Wiener: Para NSR, a imagem fica mais suavizada; Para NSR menor, a imagem fica com ruídos mais acentuados. Esse comportamento é semelhante ao da Figura \eqref{fig:01_012}.

\subsection{Reconstrução esparsa com norma $\ell_1$}

\begin{figure}[H]
     \centering
     \begin{subfigure}[b]{0.32\textwidth}
         \centering
         \includegraphics[width=1\textwidth]{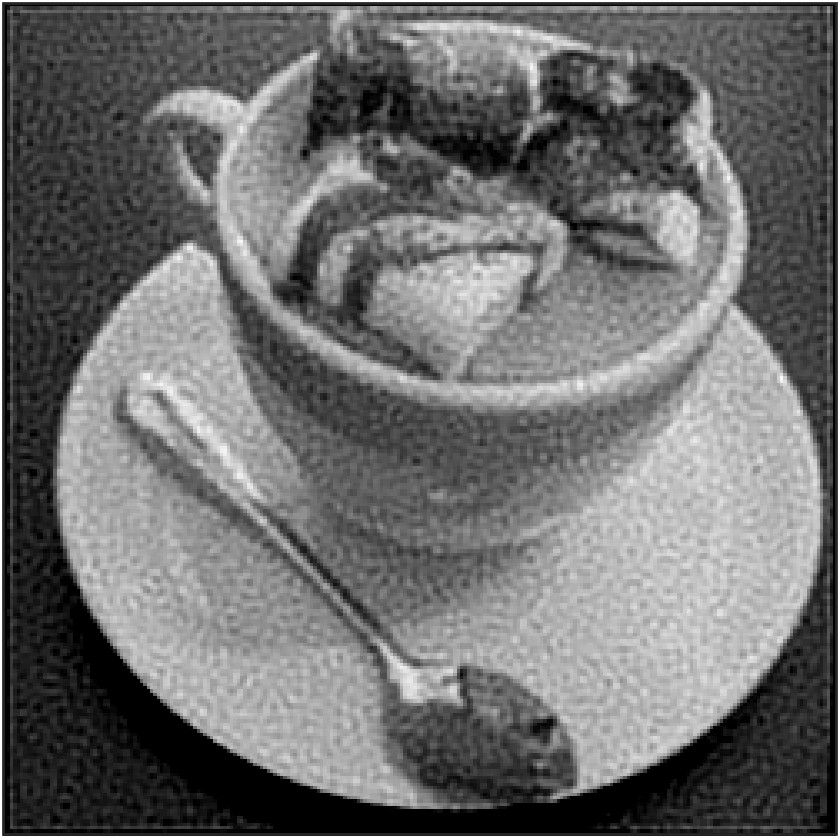}
         \caption{Utilizando FISTA}
         \label{fig:l1a}
     \end{subfigure}
               \hfill
     \begin{subfigure}[b]{0.32\textwidth}
         \centering
                  \includegraphics[width=1\textwidth]{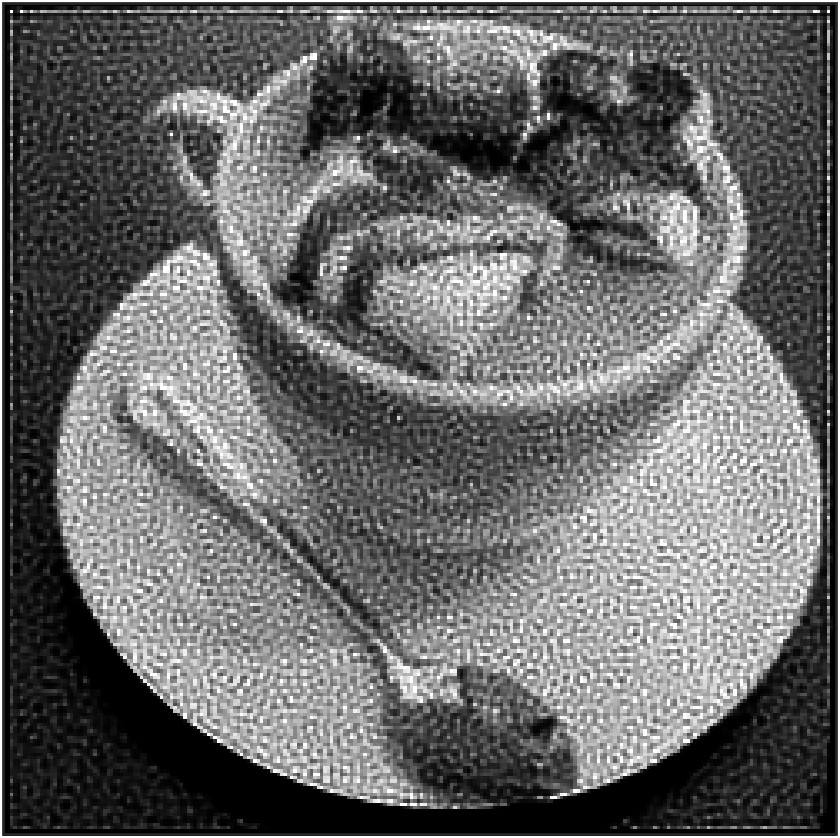}
         \caption{Utilizando IRLS}
         \label{fig:l1b}
     \end{subfigure}  
     \hfill
     \begin{subfigure}[b]{0.32\textwidth}
         \centering
                  \includegraphics[width=1\textwidth]{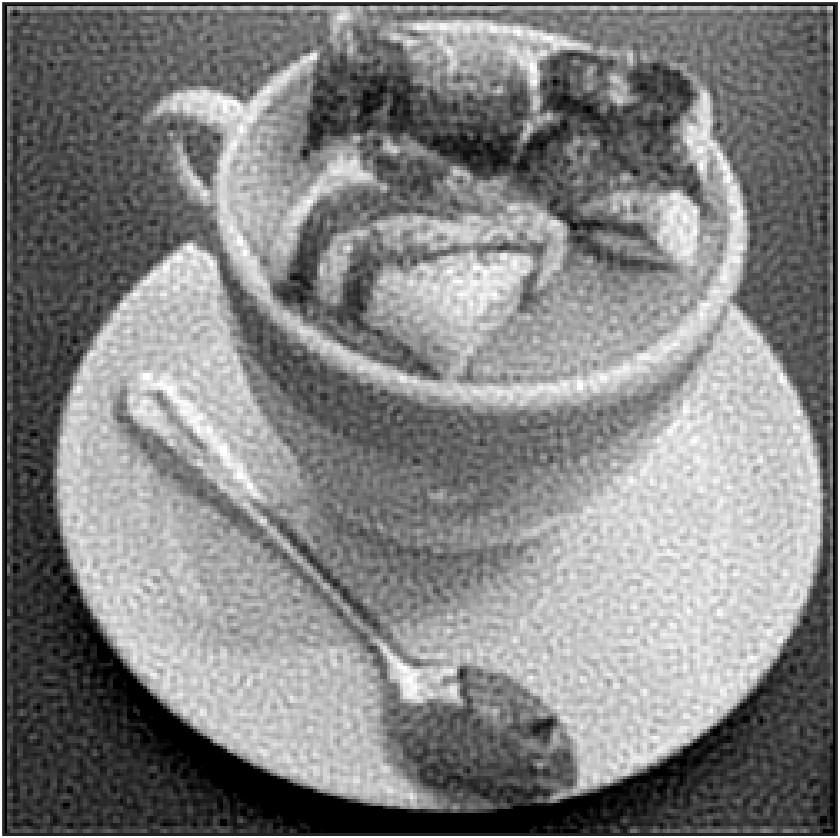}
         \caption{Utilizando ADMM}
         \label{fig:l1c}
     \end{subfigure}   
\caption[\textit{Deblurring} com norma $\ell_1$.]{\textit{Deblurring} com norma $\ell_1$. Fonte: Próprio autor.}
\label{fig:l1}
\end{figure}

Mesmo sabendo que a solução \textit{ground truth} não é esparsa, a Figura \ref{fig:l1} mostra a reconstrução com norma $\ell_1$ no termo de regularização e $\mathbf{L}= \mathbf{I}$, conforme Equação \eqref{eq:norma1}. Na Figura \ref{fig:l1a}, isso foi realizado através do algoritmo FISTA \cite[Algoritmo 7.2]{aster2019parameter};  Na Figura \ref{fig:l1b}, isso foi realizado através do algoritmo IRLS \cite[Algoritmo 7.3]{aster2019parameter}; e na Figura \ref{fig:l1c} foi com o ADMM \cite[Algoritmo 7.4]{aster2019parameter}. Nelas, não há uma suavização excessiva, mas, dependendo da escolha de $\lambda$ e demais  parâmetros do algoritmo, pode haver o escurecimento da imagem resultante.  

\subsection{Regularização de variação total}

\begin{figure}[H]
     \centering
     \begin{subfigure}[b]{0.32\textwidth}
         \centering
         \includegraphics[width=\textwidth]{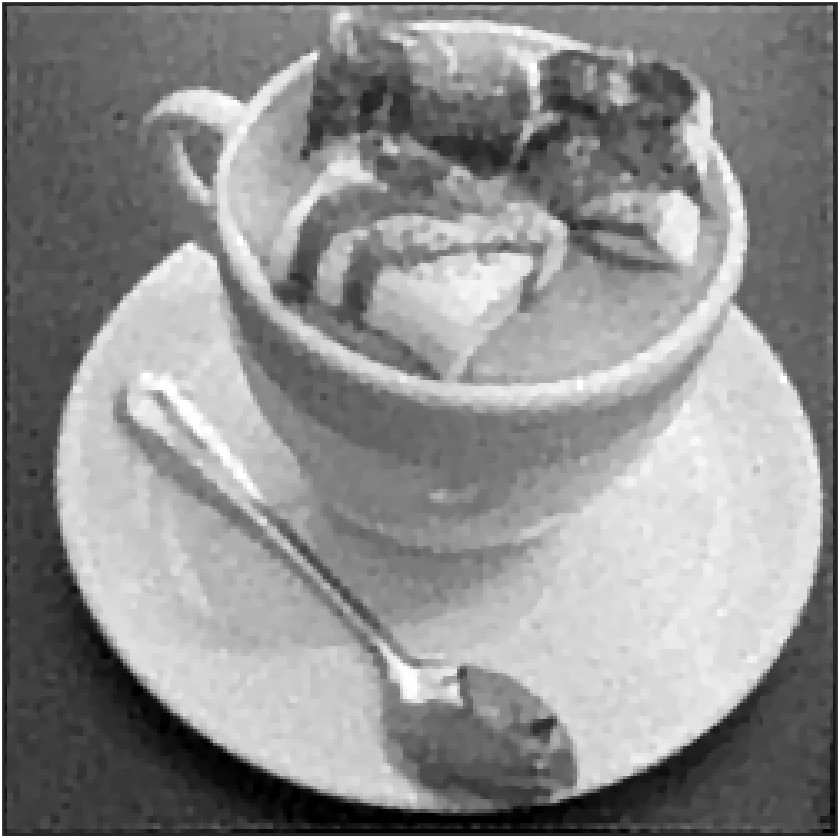}
         \caption{TV isotrópica nítida} 
         \label{fig:deconvtva}
     \end{subfigure}
     \hfill
     \begin{subfigure}[b]{0.32\textwidth}
         \centering
                  \includegraphics[width=\textwidth]{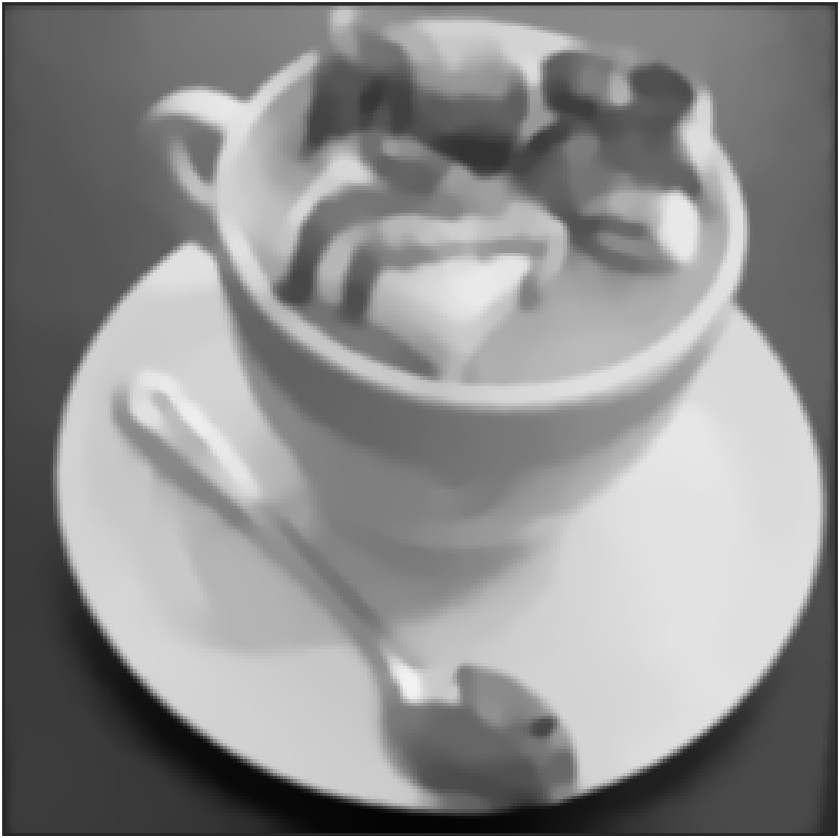}
         \caption{TV isotrópica suavizada} 
         \label{fig:deconvtvb}
     \end{subfigure}
          \hfill
     \begin{subfigure}[b]{0.32\textwidth}
         \centering
                  \includegraphics[width=\textwidth]{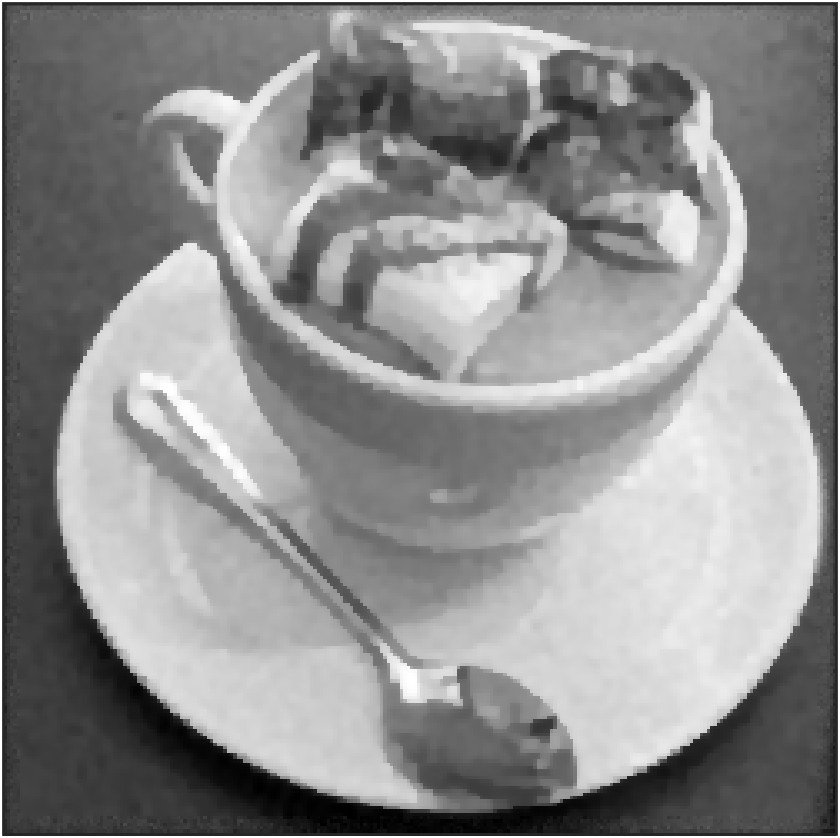}
         \caption{TV anisotrópica} 
         \label{fig:deconvtvc}
     \end{subfigure}
\caption[\textit{Deblurring} via Variação Total.]{\textit{Deblurring} via Variação Total. Fonte: Próprio autor.}
\label{fig:deconvtv}
\end{figure}

 As Figuras \ref{fig:deconvtva} e  \ref{fig:deconvtvb} mostram o \textit{Deblurring} via TV isotrópica, Equação \eqref{eq:TV11d4},  realizada de acordo com \cite{Chan2011}. A diferença entre elas é a escolha de parâmetros, que pode tornar os resultados mais nítidos ou bem suavizados, mas ambas com transições abruptas. Já a Figura \ref{fig:deconvtvc} mostra o \textit{Deblurring} via TV anisotrópica, Equação \eqref{eq:TV11d3}, obtida com iterações de Barzilai e Borwein \cite[págs. 90-2]{Mueller2012}. 

\subsection{Regularização de máxima entropia}

\begin{figure}[H]
     \centering
     \begin{subfigure}[b]{0.32\textwidth}
         \centering
         \includegraphics[width=\textwidth]{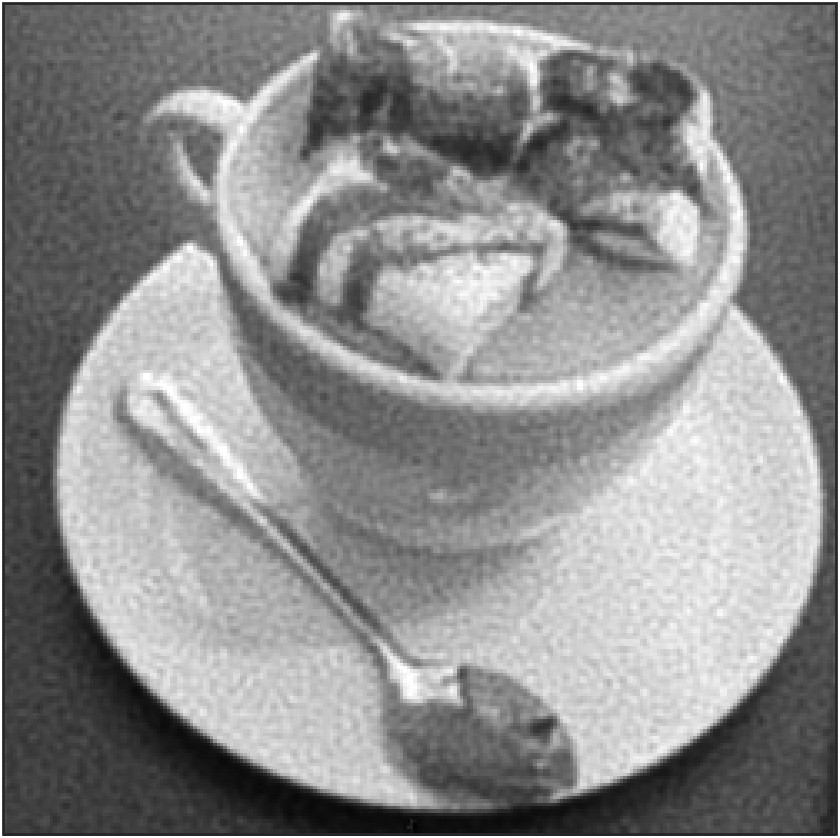}
         \caption{$\lambda = 0.2$}
         \label{fig:maxenta}
     \end{subfigure}
     \hfill
     \begin{subfigure}[b]{0.32\textwidth}
         \centering
                  \includegraphics[width=\textwidth]{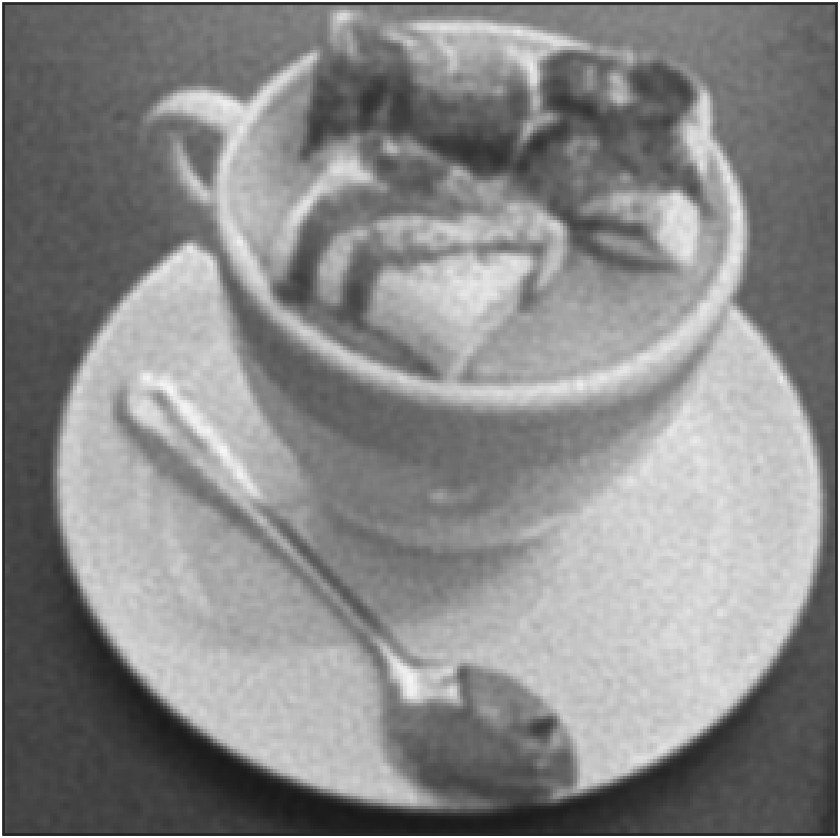}
         \caption{$\lambda = 0.4$}
         \label{fig:maxentb}
     \end{subfigure}
          \hfill
     \begin{subfigure}[b]{0.32\textwidth}
         \centering
                  \includegraphics[width=\textwidth]{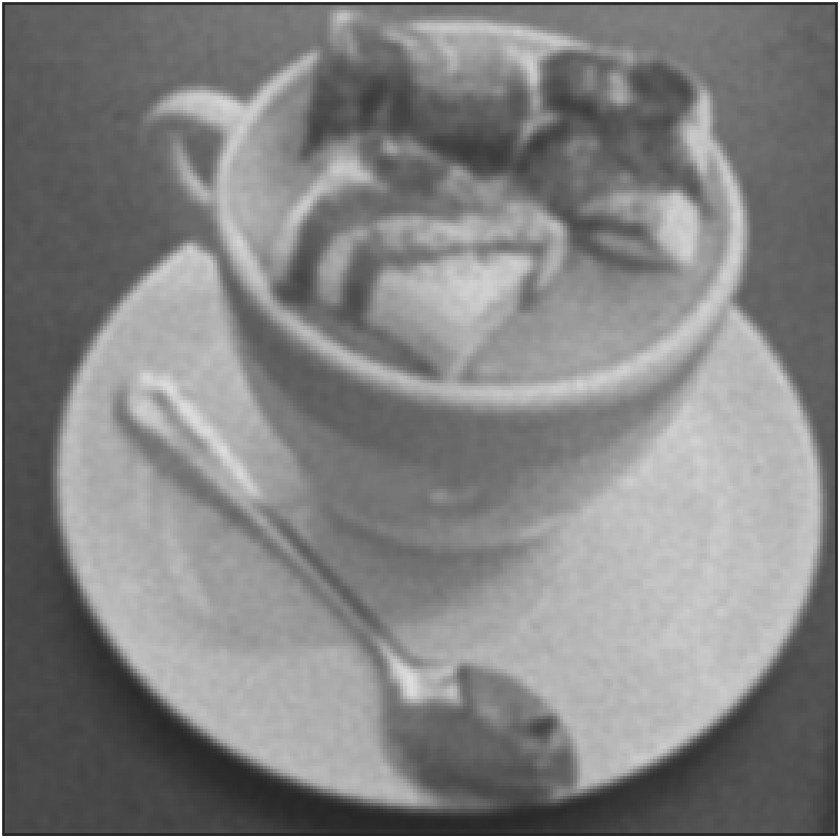}
         \caption{$\lambda = 0.6$}
         \label{fig:maxentc}
     \end{subfigure}
\caption[\textit{Deblurring} via Máxima Entropia.]{\textit{Deblurring} via Máxima Entropia. Fonte: Próprio autor.}
\label{fig:maxent}
\end{figure}

 Na Figura \ref{fig:maxent} é mostrada a regularização de máxima entropia, Equação \eqref{eq:shannon2}, obtida com o algoritmo gradientes conjugados não-linear disponibilizado em \cite{Hansen2007}. A reconstrução foi realizada para diferentes parâmetros de regularização $\lambda$ e, quanto maior seu valor, mais suavizada é a solução obtida.

\subsection{Plug-and-play prior e regularização por denoising}

\begin{figure}[H]
     \centering
     \begin{subfigure}[b]{0.32\textwidth}
         \centering
         \includegraphics[width=\textwidth]{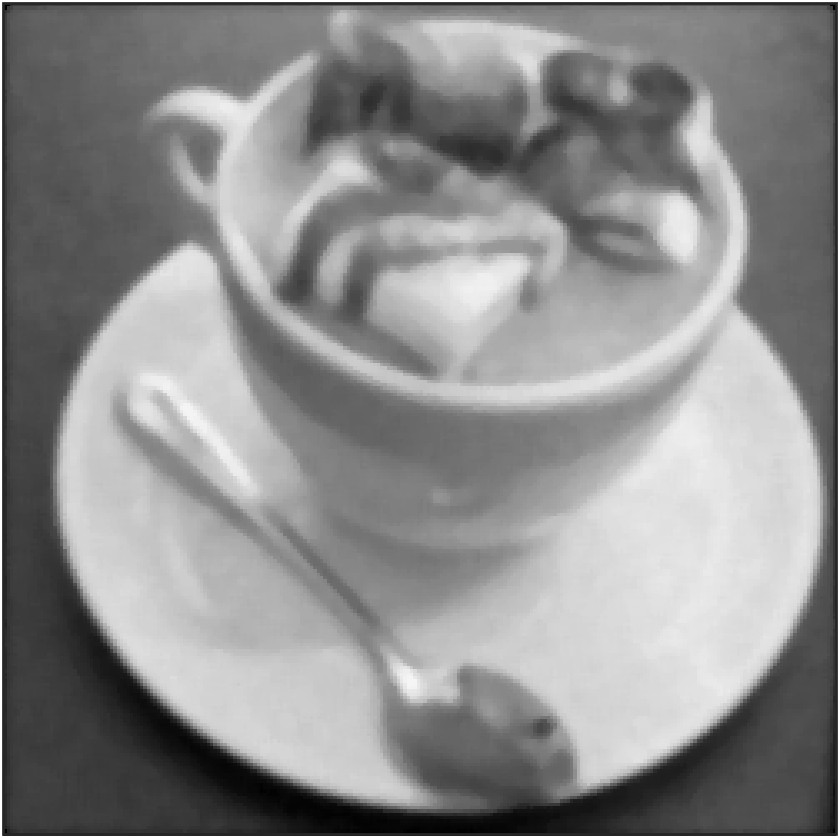}
         \caption{Filtro da mediana}
         \label{fig:medianpnpa}
     \end{subfigure}
     \hfill
     \begin{subfigure}[b]{0.32\textwidth}
         \centering
                  \includegraphics[width=\textwidth]{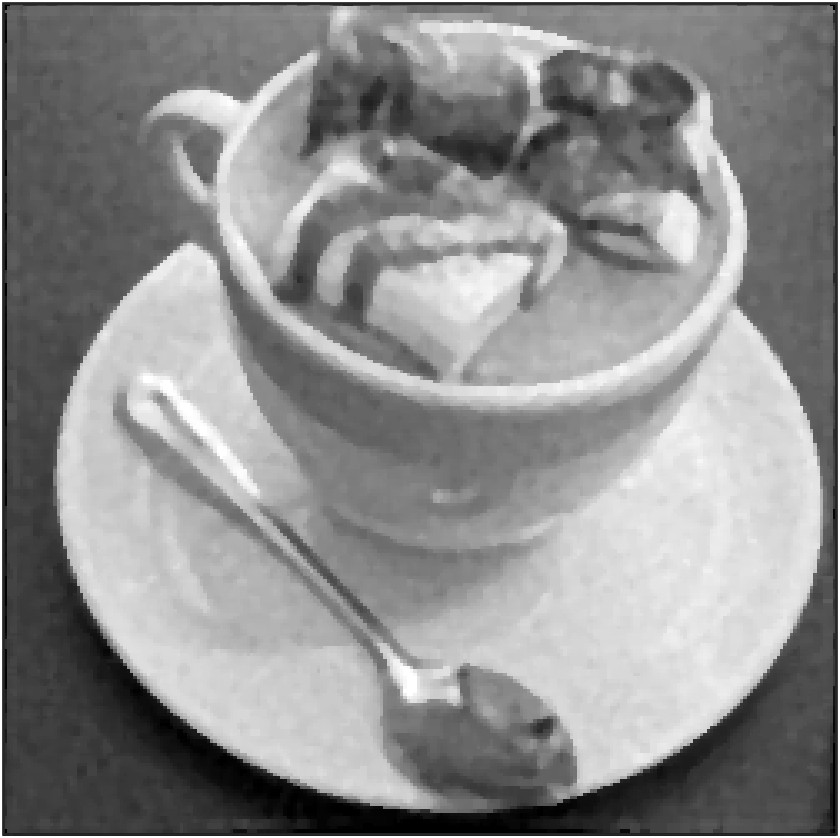}
         \caption{$\lambda = 1$}
         \label{fig:medianpnpb}
     \end{subfigure}
          \hfill
     \begin{subfigure}[b]{0.32\textwidth}
         \centering
                  \includegraphics[width=\textwidth]{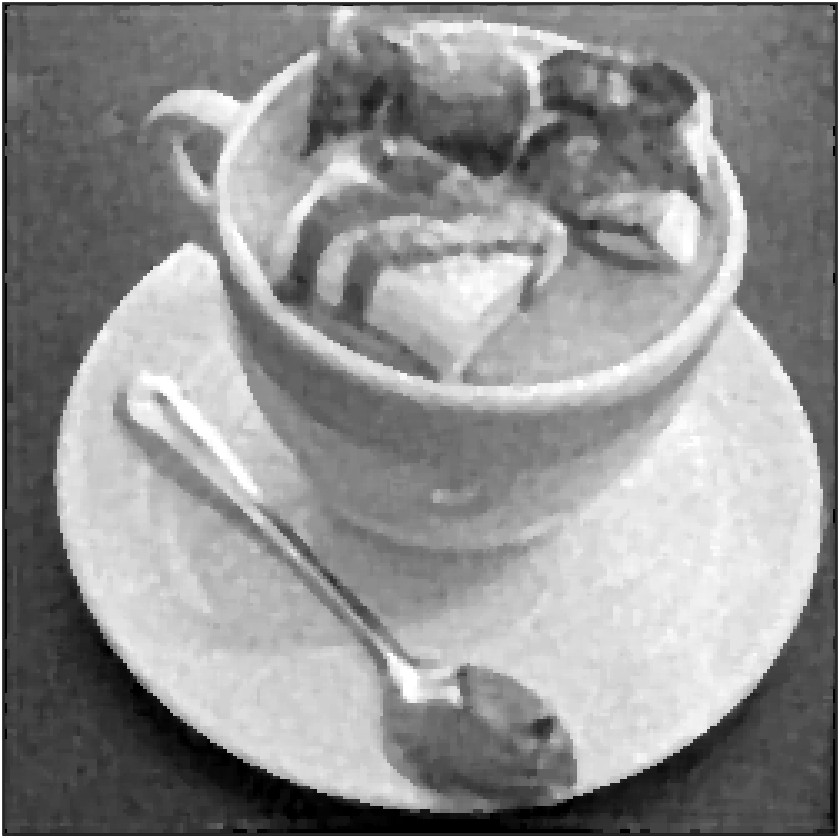}
         \caption{$\lambda = 2$}
         \label{fig:medianpnpc}
     \end{subfigure}
\caption[\textit{Deblurring} via \textit{plug-and-play prior}.]{\textit{Deblurring} via \textit{plug-and-play prior}. Fonte: Próprio autor.}
\label{fig:medianpnp}
\end{figure}

Na Figura \ref{fig:medianpnpa} é mostrada a aplicação de um filtro da mediana de tamanho [5,5]. Esse mesmo filtro  foi utilizado no \textit{framework} do $P^3$, um algoritmo baseado no ADMM, das Figuras \ref{fig:medianpnpb} e \ref{fig:medianpnpc}, variando-se o valor de $\lambda$. Observa-se que o resultado do $P^3$ é melhor do que a simples aplicação do filtro da mediana. 

\begin{figure}[H]
     \centering
     \begin{subfigure}[b]{0.32\textwidth}
         \centering
         \includegraphics[width=\textwidth]{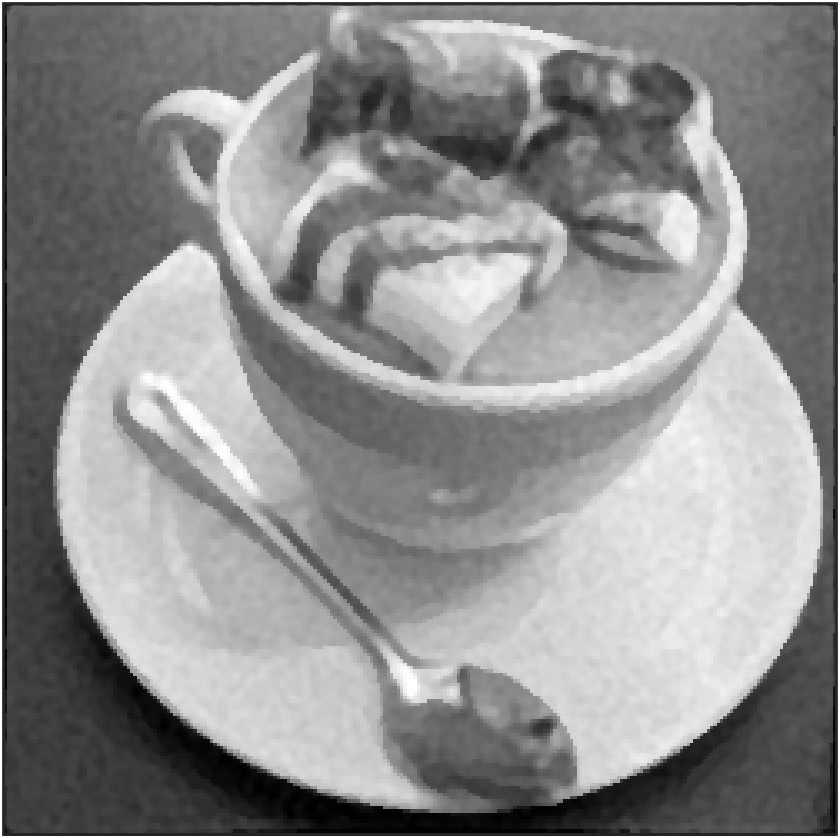}
         \caption{Ponto fixo}
         \label{fig:medianreda}
     \end{subfigure}
     \hfill
     \begin{subfigure}[b]{0.32\textwidth}
         \centering
                  \includegraphics[width=\textwidth]{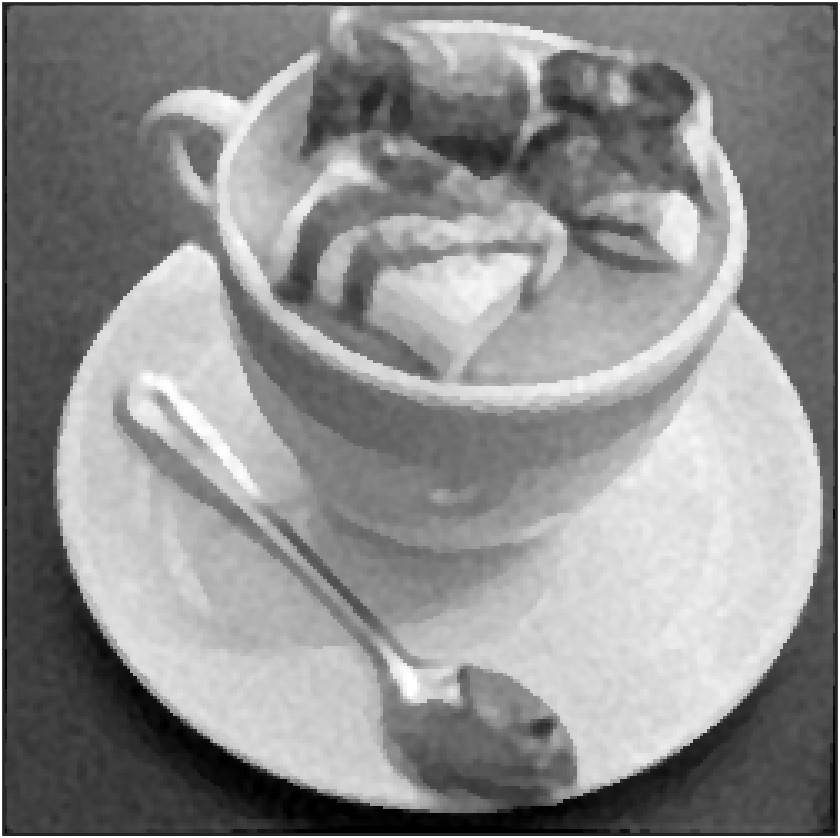}
         \caption{Máxima descida}
         \label{fig:medianredb}
     \end{subfigure}
          \hfill
     \begin{subfigure}[b]{0.32\textwidth}
         \centering
                  \includegraphics[width=\textwidth]{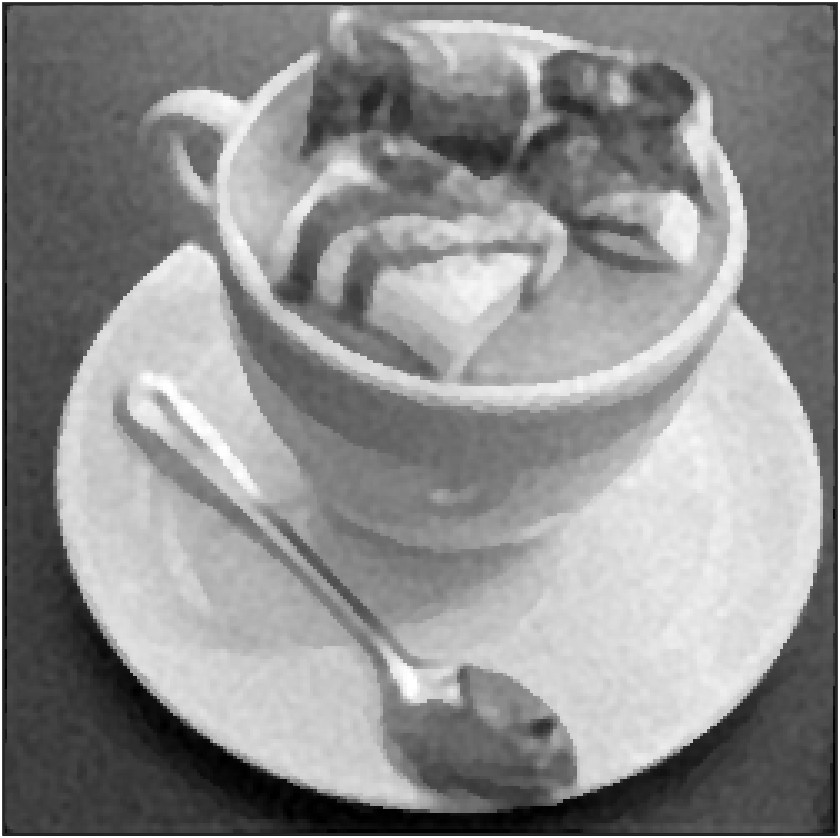}
         \caption{ADMM}
         \label{fig:medianredc}
     \end{subfigure}
\caption[\textit{Deblurring} via RED.]{\textit{Deblurring} via RED. Fonte: Próprio autor.}
\label{fig:red}
\end{figure}

Na Figura \ref{fig:red} são mostrados os resultados das três implementações do RED (ADMM, máxima descida e ponto fixo) quando se utiliza o mesmo filtro da mediana da Figura \ref{fig:medianpnpa} como seu \textit{denoiser}. 

Observa-se que os resultados dessas três implementações do RED são semelhantes entre si, mas isso é coerente com o fato das três minimizarem o mesmo funcional e também porque foi utilizado o mesmo \textit{denoiser}. 

Não se buscou que as soluções de $P^3$ e RED fossem idênticas. A aplicação do filtro da mediana em uma imagem borrada torna a imagem ainda mais borrada, mas as Figuras \ref{fig:medianpnp} e \ref{fig:red} ilustram que detalhes das bordas dos objetos foram recuperados. Isso mostra que a inclusão de \textit{denoisers} pode ser adequada em  problemas inversos mais gerais, como o \textit{deblurring}. 

Seja comparando $P^3$ e RED, ou as três implementações do RED entre si, ressalta-se que cada algoritmo iterativo possui seus próprios parâmetros do otimizador que devem ser ajustados. A escolha entre elas depende se uma determinada forma traz a flexibilidade necessária para a aplicação e se é possível resolver o mesmo problema com um algoritmo que utilize menos parâmetros, menos complexo. 

\subsection{Métodos iterativos truncados}

\begin{figure}[H]
     \centering
     \begin{subfigure}[b]{0.32\textwidth}
         \centering
                  \includegraphics[width=\textwidth]{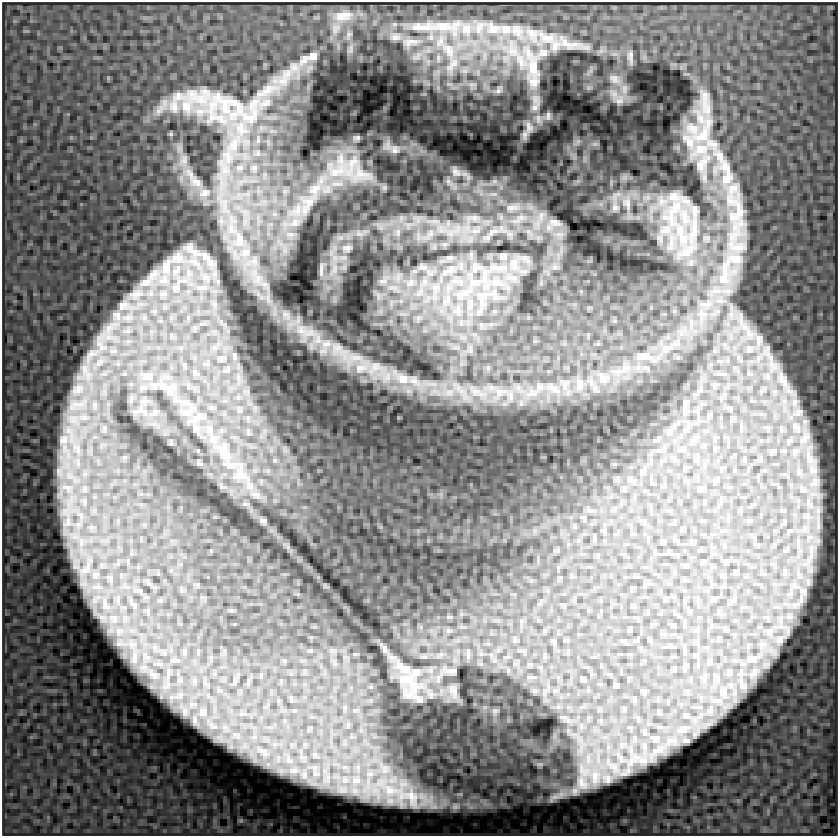}
         \caption{CGLS}
         \label{fig:01hybridb}
     \end{subfigure}
     \hfill
          \begin{subfigure}[b]{0.32\textwidth}
         \centering
         \includegraphics[width=\textwidth]{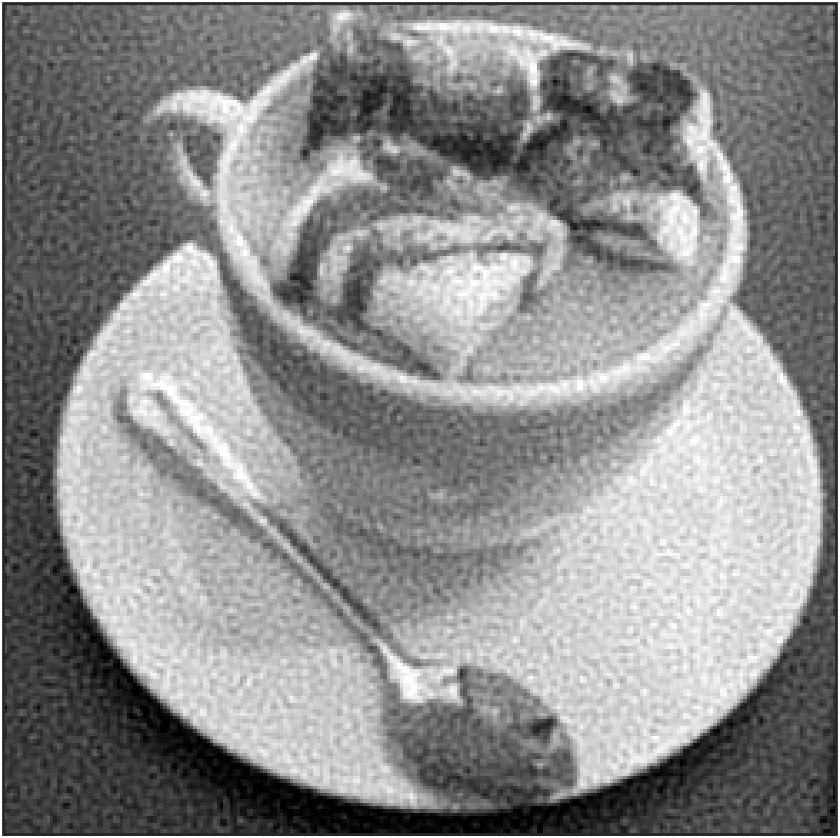}
         \caption{GMRES}
         \label{fig:01hybrida}
     \end{subfigure}
          \hfill
     \begin{subfigure}[b]{0.32\textwidth}
         \centering
                  \includegraphics[width=\textwidth]{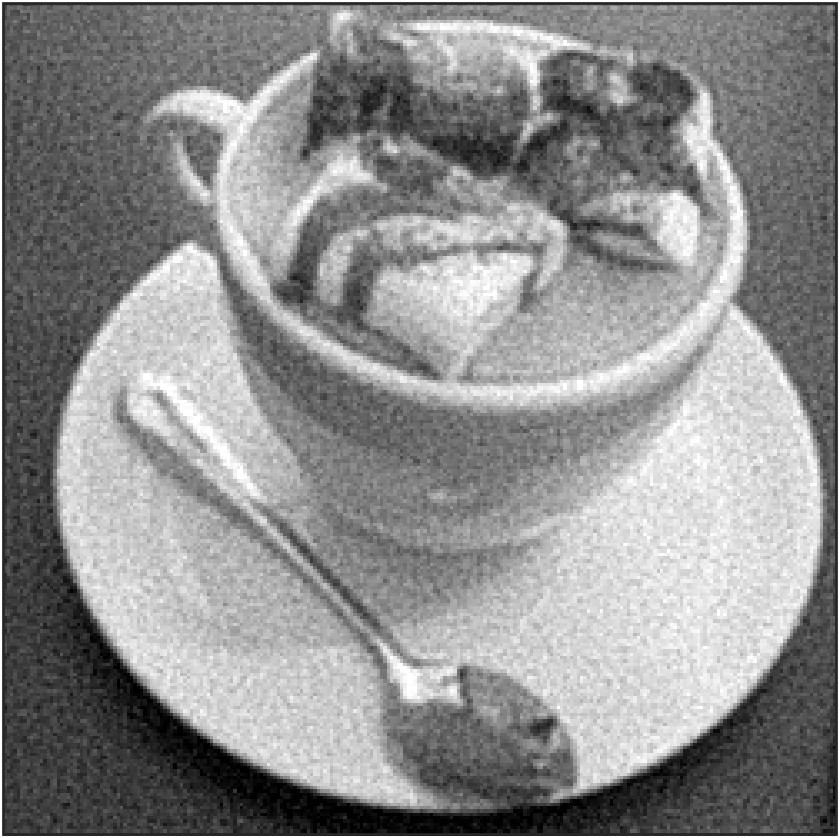}
         \caption{TV heurística}
         \label{fig:01hybridc}
     \end{subfigure}
\caption[\textit{Deblurring} com métodos iterativos.]{\textit{Deblurring} com métodos iterativos. Fonte: Próprio autor.}
\label{fig:01hybrid}
\end{figure}

 Seja o funcional da Equação \eqref{eq:tikhonov12}, quando $\mathbf{L} = \mathbf{I}$. Existem alternativas diferentes da Equação \eqref{eq:tiksolgen}. Pode-se utilizar um algoritmo iterativo \cite[pág. 82]{Mueller2012}.
\begin{itemize}
\item Na Figura \ref{fig:01hybridb}, foi utilizado o método dos gradientes conjugados para mínimos quadrados (CGLS) com 10 iterações, baseado em \cite{Gazzola2018}. Não há termo de regularização e a reconstrução é realizada pelo truncamento das iterações, baseado no fenômeno da semiconvergência. 

\item  Na Figura \ref{fig:01hybrida} é mostrada a reconstrução com iterações do método do residual mínimo generalizado (GMRES), baseado em \cite{Gazzola2018}. Ele é um método híbrido, o funcional da Equação \eqref{eq:tikhonov12}, combinando , que consiste projeções para subespaços de Krylov junto de regularização para estabilizar o comportamento de semiconvergência do problema mal-posto. O critério de parada utilizado foi o GCV e ele foi satisfeito com 8 iterações. 
\item Finalmente, na Figura \ref{fig:01hybridc}, foi utilizado um algoritmo heurístico para aproximar a regularização de variação total \cite{Gazzola2018}. 
\end{itemize}

Elas podem parecer mais ruidosas do que alguns resultados obtidos anteriormente. Por outro lado, elas trazem características desejadas, como automatização no critério de parada e grande flexibilidade de escolhas de seus componentes. É possível que reconstruções mais agradáveis ao olhar humano sejam obtidas, mas seria necessário um ajuste fino de seus parâmetros.

\newpage
\section{COMO UNIR ABORDAGENS BASEADAS EM DADOS COM O MÉTODO DE REGULARIZAÇÃO?}\label{sec:integrated}

\subsection{Introdução}\label{sec:deeplearning}

Atualmente, inteligência artificial, aprendizado de máquina e aprendizagem profunda são temas recorrentes tanto na ciência quanto na grande mídia. Expressões como \textit{ChatGPT}, \textit{DeepSeek}, \textit{Gemini}, \textit{Copilot}, cultura \textit{data-driven}, inteligência artificial generativa, \textit{deep fakes}, \textit{AlphaGo} e muitos outros já fazem parte do vocabulário da população. 

Diversos problemas são estudados em aprendizado de máquina. O ponto em comum entre eles é que tarefas são realizadas sem programação explícita para elas \cite{Bzdok2018} e a noção de aprendizagem, que os seus resultados tendem a ser melhores quanto maior o número de dados disponível \cite[pág. 2]{mitchell1997machine}. Existem métodos supervisionados para obter modelos preditivos e realizar classificação e regressão. Existem métodos não-supervisionados para reconhecimento de padrões e realizar clusterização, associação e redução de dimensionalidade. Existem ainda métodos de aprendizagem por reforço, em que o modelo aprende através de tentativa e erro, interagindo e modificando situações visando maximização de recompensas futuras. 

Este capítulo não traz uma revisão completa da teoria de aprendizagem, pois existem diferentes linhas de pesquisa nessa área \cite{goodfellow2016deep, mitchell1997machine, vapnik1998statistical}. Partindo do exemplo de um regressor definido por rede neural profunda, busca-se estabelecer relações entre aprendizagem, problemas inversos e o método de regularização. 

\subsection{Como relacionar aprendizado de máquina com problemas inversos?}\label{sec:general}

No Capítulo \ref{sec:illposed}, as Equações \eqref{eq:eq111} e \eqref{eq:eq1} apresentavam a relação linear e não-linear, respectivamente, entre o espaço do modelo com parâmetros $\mathbf{x}$ e o espaço dos dados $\mathbf{y}$, onde $\mathbf{A}$ denotava um modelo derivado de leis físicas que os relacionavam. A partir disso, o Capítulo \ref{sec:variational} descreveu o método de regularização de Tikhonov, que busca realizar a inversão aproximada do operador direto $\mathbf{A}$ de modo que a solução respeite as três condições de Hadamard, principalmente estabilidade. 

Mesmo com resultados concretos e robutos, existem algumas limitações nessa abordagem, como a indisponibilidade do operador direto em alguns casos, ou simplificações necessárias para resolvê-lo \cite[pág. 3]{Arridge2019}. Utilizar um modelo acurado, computacionalmente custoso, nem sempre é suficiente quando os dados são muito ruidosos \cite[pág. 105]{Arridge2019}. No caso de problemas lineares, em muitos casos é difícil definir qual é o melhor regularizador \cite{DelosReyes2016}, ao mesmo tempo que em problemas não-lineares não há garantias que os parâmetros ótimos do modelo possam ser obtidos \cite{Adler2021}. 

Na medida que os sistemas de aquisição, armazenamento, processamento e compartilhamento de dados ficaram mais sofisticados, o paradigma de aprendizado de máquina começou a ter destaque nas mesmas tarefas. Mesmo abdicando de modelagem explicativa e de  hipóteses fortes sobre a física do problema \cite{Kording2018}, elas conseguem resultados relevantes utilizando estruturas genéricas e utilizando grande quantidade de dados disponíveis para resolver as tarefas  \cite{goodfellow2016deep}. 

Muitos algoritmos de aprendizado de máquina são de propósito geral, no sentido de que eles podem realizar um mesmo tipo de tarefa, como classificação ou regressão, em diferentes aplicações, apenas variando-se a base de dados utilizada. Um dos grandes objetivos é realizar a predição, isto é, obter resultados em dados novos, nunca vistos pela máquina de aprendizado. Considerando que o erro da predição em novos dados possa ser calculado, a capacidade da máquina generalizar está associada à máquina de aprendizado obter o menor erro possível nesses novos dados  \cite[pág. 429]{vapnik1998statistical}.

\subsubsection{Para uma máquina de aprendizagem genérica, como a tarefa de regressão se relaciona com resolver problemas inversos?}

Seja o problema de regressão em aprendizado de máquina, na qual a entrada e a saída de um regressor podem ser escalares, matrizes ou tensores. Existem dois espaços (entrada e saída) e o regressor deve realizar o mapeamento entre eles. Nos capítulos anteriores, o operador direto $\mathbf{A}$ realizava o mapeamento entre o espaço dos parâmetros do modelo para o espaço dos dados (medidas). No aprendizado de máquina, busca-se realizar o mapeamento diretamente entre conjuntos de dados diferentes, sem necessidade da matriz $\mathbf{A}$ explícita. Com dados $\mathbf{x}$ e $\mathbf{y}$ disponíveis, pode-se tentar realizar o mapeamento nas duas direções, seja resolvendo o problema direto ($\mathbf{x}$ para $\mathbf{y}$) ou  inverso ($\mathbf{y}$ para $\mathbf{x}$). 

É necessário desenvolver um modelo capaz de representar os dados de interesse e que possa ser atualizado na medida em que houver mais dados disponíveis. Seja uma estrutura candidata $h_{\bm{\theta}}( \cdot) = h(\cdot, \bm{\theta})$, parametrizada por $\bm{\theta}$ \cite[Seção 8.2]{Deisenroth2020}. Se o objetivo é realizar o mapeamento de $\mathbf{x}$ para $\mathbf{y}$, a regressão teria a forma de 
\begin{equation}
\mathbf{y} = h_{\bm{\theta}}(\mathbf{x}),
\label{eq:NNinf}
\end{equation}
realizando a tarefa desejada. Neste caso, a solução de um problema direto. 

É possível também tentar realizar a tarefa na outra direção, de $\mathbf{y}$ para $\mathbf{x}$,  conforme 
\begin{equation}
\mathbf{x} = h_{\bm{\theta}}(\mathbf{y}),
\label{eq:NNinf2}
\end{equation}
que é o problema inverso, tarefa de interesse.  Exemplos desse tipo incluem recuperar sinais nítidos a partir de sinais degradadas ou realizar reconstruções de sinais a partir de medidas indiretas e de dados parciais.

\subsubsection{Como atualizar os parâmetros da máquina de aprendizagem?}

As $N$ amostras disponíveis para treinamento da máquina de aprendizagem devem ser representativas dos dados de entrada e de saída. No problema inverso, a i-ésima amostra de treinamento das entradas e das saídas são denotadas, respectivamente, por $\mathbf{y}_t^{i}$ e $\mathbf{x}_t^{i}$. Em outras palavras, é um caso de aprendizagem supervisionada. 

É mais convencional considerar $\mathbf{x}$ a entrada e  $\mathbf{y}$ a saída. No entanto, conforme Equação \eqref{eq:NNinf2},  $\mathbf{x}$ é a  saída da máquina de aprendizagem, pois são os parâmetros que devem ser estimados. A entrada da máquina de aprendizagem é $\mathbf{y}$, pois são os dados disponíveis. Dessa forma, a notação fica coerente com os capítulos anteriores. 

A etapa de atualizar $\bm{\theta}$ é chamada de treinamento e pode ser visto como um problema de otimização. A minimização do risco empírico (ERM) é dada por
\begin{equation}
\hat{\bm{\theta}} = \arg\min\limits_{\bm{\theta}} \left[ \frac{1}{N} \sum_{i=1}^{N} \mathcal{L} \left(h_{\bm{\theta}}(\mathbf{y}_t^{i}), \mathbf{x}_t^{i} \right) \right], 
\label{eq:ERM0}
\end{equation}
onde $\hat{\bm{\theta}}$ são os valores estimados de $\bm{\theta}$ após otimização. 

Supondo que uma função geradora de amostras (dados) fosse conhecida, seria possível calcular medidas de risco de um modelo preditivo. Como essa função geradora é deconhecida, é necessário trabalhar com as suas saídas, os dados. O nome empírico vem do fato de que existe um número $N$ finito de amostras que devem ser obtidas para que o treinamento seja realizado.

\subsubsection{É possível formular matematicamente conceitos como complexidade do modelo, generalização e consistência?}\label{sec:vapnik}

Na área do aprendizado de máquina, há a noção intuitiva de generalização, na qual um algoritmo consegue ter bons resultados mesmo para dados fora do conjunto de treinamento. O aprendizado estatístico, conhecida também como Teoria VC (Vapnik-Chervonenkis) \cite{Vapnik2019}, traz um \textit{framework} para aprendizado de máquina visando resolver problemas preditivos. Ela discute, entre outras coisas, as condições para uma máquina de aprendizagem ser capaz de generalizar, trazendo um formalismo matemático para cada etapa desse processo. 

Nessa teoria, os autores argumentam que apenas dois fatores são responsáveis por ela, o valor do risco empírico e a capacidade (caracterizada por sua dimensão VC \cite[pág. 147]{vapnik1998statistical}) do conjunto de funções admissíveis. Um método que seja capaz de controlar esses dois fatores é fortemente e universalmente consistente \cite[pág. 429]{Vapnik2006}. 

Essa é visão geral de consistência, mas são encontradas outras definições, como a consistência fraca, em que a convergência acontece para um número infinito de amostras, enquanto a consistência forte essa convergência aconteceria com probabilidade $P = 1$. Já a consistência universal descreve que há convergência forte quaisquer que sejam os pares dos dados \cite[pág. 91-2]{devroye1996a}. Em suma, consistência é relacionada à capacidade de generalização da máquina de aprendizagem e sem consistência, a aprendizagem não exerceria a sua função preditiva.

Para o aprendizado estatístico ser considerado completo, há dois problemas de seleção: o primeiro problema de seleção é a busca de um subconjunto de funções admissíveis que inclua aquelas funções que realizem uma boa aproximação, enquanto o segundo problema de seleção é buscar a melhor função de aproximação nesse subconjunto \cite{Vapnik2019}. É possível mostrar que a ERM da Equação \eqref{eq:ERM0} é consistente sob determinadas condições \cite[págs. 101-4] {cherkassky2007learning}.

\subsubsection{É necessário regularizar a etapa de aprendizagem?}\label{sec:general2}

A ERM é um problema mal-posto para um número fixo de dados de treinamento \cite[pág. 85]{Arridge2019}, \cite[pág. 25]{cherkassky2007learning}, \cite{Mukherjee2006}. A ERM também é propensa ao \textit{overfitting} \cite[pág. 276]{goodfellow2016deep}. Uma primeira ideia é acrescentar um regularizador como termo aditivo. 

Nesse contexto é proposto o risco empírico regularizado \cite[Equação 23]{Adler2021}, que deve ser minimizado conforme 
\begin{equation}
\hat{\bm{\theta}} = \arg\min\limits_{\bm{\theta}} \left[ \frac{1}{N} \sum_{i=1}^{N} \mathcal{L} \left(h_{\bm{\theta}}(\mathbf{y}_t^{i}), \mathbf{x}_t^{i} \right) + \lambda^2 \Omega(\bm{\theta}) \right], 
\label{eq:ERM4}
\end{equation}
onde $\hat{\bm{\theta}}$ são os valores estimados de $\bm{\theta}$ após otimização. Os demais termos $\mathcal{L} \left( \cdot, \cdot \right)$, $\lambda$ e  $\Omega(\cdot)$ têm papéis semelhantes na Equação \eqref{eq:tikhonovgeral}. 

É possível traçar um paralelo das Equações \eqref{eq:ERM4} e \eqref{eq:tikhonovgeral} do método de regularização, mesmo com origens e contextos diferentes. Na otimização, não se calcula $\mathbf{x}$ diretamente. O problema se torna obter os parâmetros $\bm{\theta}$ para que a estrutura candidata $h_{\bm{\theta}}$ pela Equação \eqref{eq:ERM4} seja capaz de realizar a tarefa adequadamente pela Equação \eqref{eq:NNinf2}. Essa diferença se dá que tanto na otimização quanto no regularizador, agora calculados sobre $\bm{\theta}$. 

A pergunta da presente subseção pode ser respondida a partir de outro ponto de vista. Em \cite{Scholkopf2012} se discute predição do efeito a partir da causa (causal) e predição da causa a partir do efeito (anticausal). Isso é realizado sem menção ao método de Tikhonov ou regularização, explicitamente. Ou seja, não é possível dizer que regularização é uma condição estritamente necessária para obter resultados relevantes em aprendizagem, mesmo que em certos casos possa ajudar. 
  
\subsubsection{Quais são os componentes básicos de um algoritmo de aprendizado de máquina?}\label{sec:general3}

Os ingredientes principais para se obter uma máquina de aprendizagem são os dados, a estrutura candidata, a função de perda, regularização, otimização e métrica de performance \cite{goodfellow2016deep}. Eles devem sempre ser escolhidos de modo adequado ao problema que se quer resolver. Alguns comentários sobre cada um deles:

\begin{itemize}
\item \textbf{Dados}: Quanto mais dados de treinamento, melhor. É claro que a regressão realizada não deve se limitar ao conjunto de dados de treinamento. O desejável é que o regressor seja capaz de lidar com novos dados, que não estavam disponíveis durante o treinamento. O conceito de generalização descreve quando o algoritmo é capaz de realizar boas predições em dados nunca vistos \cite{geron2019hands-on}, \cite[pág. 110]{goodfellow2016deep}. De fato, não haveria necessidade de predição se os dados de treinamento determinassem completamente a solução da tarefa \cite[pág. 73-4]{cherkassky2007learning}. Para avaliar a performance, deve-se dividir os dados pelo menos em dois conjuntos, treinamento e teste, evitando assim resolver um problema mais fácil do que realmente ele é. Supõe-se que há uma distribuição de probabilidade que define a geração dos dados. É importante imaginar que esses dados de treinamento foram gerados pela mesma distribuição dos dados de teste;

\item \textbf{Métrica de performance}: A performance da otimização é avaliada indiretamente, pois não é avaliada no treinamento da Equação \eqref{eq:ERM4} para dados de treinamento, mas sim na inferência daEquação \eqref{eq:NNinf2} para dados de teste.  Intuitivamente existe um paralelo entre realizar a inferência nos dados de treinamento e crimes de inversão, pois ambos facilitam artificialmente a obtenção de melhores resultados, sem garantir boa performance em novos dados. 

\item \textbf{Estrutura da máquina de aprendizagem}: Uma estrutura candidata $h_{\bm{\theta}}$ deve ser escolhida de acordo com o problema que se quer resolver, um problema de seleção. Busca-se um equilíbrio na flexibilidade do modelo de se ajustar nos dados disponíveis (complexidade) para o número finito de dados disponíveis \cite[pág. 73]{cherkassky2007learning}. Se o modelo for muito simples, ele não é capaz de representar os dados e há o \textit{underfitting}. Se o modelo for muito complexo, ele pode se ajustar e detectar padrões que não são dos dados, mas de ruídos e há o \textit{overfitting} \cite{geron2019hands-on}.  Nota-se ainda que há parâmetros $\bm{\theta}$, atualizáveis durante o treinamento e não são escolhidos manualmente, e hiperparâmetros, que são escolhidos manualmente e a otimização não os atualiza;

\item \textbf{Função de perda}: Medem a qualidade de aproximação produzida pela máquina de aprendizagem. Por convenção, $\mathcal{L}$ resulta em valores positivos, de modo que quanto maior o valor calculado, pior é a aproximação resultante \cite[pág. 25]{cherkassky2007learning}. Existem diversos tipos e, na regressão,  eles podem ser associados a modelos de ruídos aditivos \cite[pág. Tabela 2]{Chen2002}, \cite[pág. 109]{alvarez2017digital}. Outras $\mathcal{L}$ estão mais associadas a algum tipo de tarefa, sendo diferentes em problemas de classificação e regressão;

\item \textbf{Regularização}: Em aprendizado de máquina, a regularização não é só relacionada com a minimização da Equação \eqref{eq:ERM4}. A utilização de métodos de regularização em problemas de aprendizagem vista diminuir o \textit{overfitting} \cite{Prato2008}, obtendo melhores resultados para dados não-observados.  Isso será mais discutido a diante. Ressalta-se que, na etapa de treinamento, nem sempre há um termo de regularização aditivo calculado sobre $\bm{\theta}$. Isso depende do tipo do problema e se adicionar esse regularizador traz vantagens na solução do problema;

\item \textbf{Algoritmo de otimização}: Dependendo da estrutura que se escolhe, a Equação \eqref{eq:ERM4} pode se tornar um problema de otimização não-linear e não-convexo  \cite[pág. 177]{goodfellow2016deep}. Nesses casos, algoritmos iterativos baseados em gradientes podem diminuir consideravelmente o funcional regularizado, mas sem garantias de se obter mínimos globais. Nesse contexto, ferramentas de diferenciação automática são essenciais, pois nem sempre é possível obter soluções analíticas.

\end{itemize}
Esses componentes também são necessários para resolver problemas inversos. Logo, é possível imaginar que existam relações entre os dois paradigmas, com aproximações e afastamentos de acordo com suas características e métodos próprios.

\subsubsection{É possível incluir regularizadores na seleção de características?}\label{sec:feature}
Outra relação possível entre algoritmos de aprendizado de máquina e regularização é na literatura de extração de características, quando são discutidos \textit{Embedded Methods} dentro de métodos de seleção de características \cite{Guyon2006, Nilsson2007}. Nesse problema, deve-se achar qual que é o subconjunto de características que permita obter a maior generalização possível, ou menor risco. Discutindo a seleção de características como um problema de otimização, e adicionando um termo de regularização, algoritmos SVM com norma $\ell_1$ e LASSO são discutidos para seleção de variáveis.

\subsection{Quando que um problema de aprendizado de máquina se torna um problema de aprendizagem profunda?}

A Equação \eqref{eq:ERM4} pode ser utilizada na aprendizagem profunda quando a estrutura candidata tem uma forma específica, isto é, quando ela é dada por uma rede neural artificial (ANN) \cite[pág. 1]{Prince2023}. Simbolicamente, isso será denotado através de $h_{\bm{\theta}} = \Psi_{\bm{\theta}}( \cdot)$.

Nessa rede neural, a sua primeira camada é chamada de camada visível ou camada de entrada, pois ela contém as próprias variáveis observáveis, sejam elas vetores, matrizes ou tensores. A última camada é a camada de saída, que retorna o valor de interesse, o resultado. Há ainda camadas intermediárias entre elas, conhecidas como camadas ocultas, pois não há uma saída visível associada a elas.

Se a rede neural apresentar poucas camadas ocultas ela é chamada de rasa. Quando a ANN possui muitas camadas, ela é chamada de rede neural profunda (DNN) \cite{goodfellow2016deep}, mas essa definição não é rigorosa em relação ao número de camadas em si. Fato é que, com muitas camadas, DNNs muitas vezes são sobreparametrizadas, representando com alta dimensionalidade o sinal de interesse. Isso naturalmente pode levar ao \textit{overfitting} da rede sobre os dados, motivo pelo qual são necessárias estratégias para evitá-lo (ou, ao menos, diminuí-lo). 

Existem mais diferenças entre aprendizado de máquina e aprendizagem profunda. Para simplificar a questão, pode-se retomar a ideia de que se deseja obter um operador para realizar tarefas, seja de regressão ou outras, realizando mapeamento entre espaços. Os parâmetros e hiperparâmetros devem ser atualizados através de um problema de minimização e, uma vez obtidos e fixos, os resultados desse operador são avaliados com métricas de performance. A grande questão é que muitos resultados de alta performance estão sendo obtidos quando se utiliza DNNs como estrutura candidata, o que justifica o seu estudo \cite{goodfellow2016deep}. 

Existe uma grande variedade de arquiteturas e as mais conhecidas são apresentadas em revisões como \cite{Khamparia2019, Shrestha2019}. Outros exemplos são específicos para resolver problemas inversos de ponta a ponta, tais como utilizar redes neurais convolucionais (CNNs) em problemas de imagens \cite[Capítulo 9]{goodfellow2016deep}. Nelas, os filtros de cada camada são aprendidos, visando a extração de informações da imagem original a partir da convolução com esses filtros. Assim, uma característica é a invariância à translação, o que significa que as características extraídas da imagem pode estar em qualquer parte dela \cite[pág. 342]{goodfellow2016deep}. Aplicações incluem: \textit{deblurring} de movimento \cite{Koh2021}; \textit{denoising}, super-resolução ou conversão entre diferentes modalidades \cite{Kaji2019}; reconstrução de imagens de microscopia de fluorescência \cite{Belthangady2019} e de imagens médicas \cite{Ongie2020}; além de problemas inversos sísmicos \cite{Adler2021}.

\subsubsection{Para ilustrar uma rede neural, quais são os componentes básicos de um \textit{perceptron} multicamadas?}

\begin{figure}[htpb]
\includegraphics[width=0.8\textwidth]{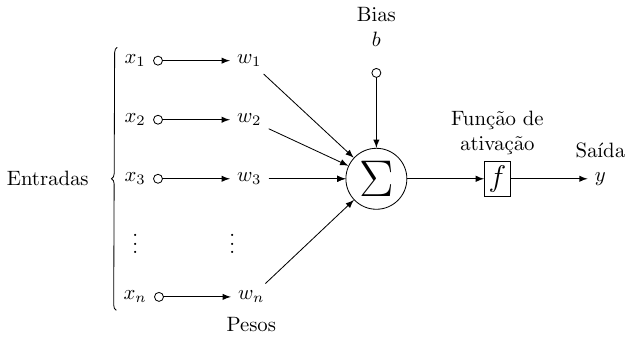}
\caption[Diagrama de uma unidade.]{Diagrama de uma unidade. Fonte: Próprio autor. }
\label{fig:my_neuron}
\end{figure}

A aprendizagem profunda se baseia na ideia de uma hierarquia de conceitos, onde sinais complexos, complicados, podem ser representados a partir de estruturas mais simples \cite[págs. 1,2]{goodfellow2016deep}. Para que isso seja feito, estruturas simples chamadas unidades, ou neurônios artificiais, são seus componentes básicos, conforme Figura \ref{fig:my_neuron}. Depois, elas são conectadas em camadas e essas camadas são conectadas em outras camadas, formando a rede neural artificial que será utilizada como $h(\cdot, \bm{\theta})$ na tarefa desejada.

No \textit{perceptron} multicamadas (MLP), uma unidade é caracterizada por pesos $\bm{w} = (w_1, w_2, \dots, w_n)$, por \textit{biases}\footnote{Na ausência de qualquer entrada, a saída da unidade é enviesada para o valor $\mathbf{b}$)} $\mathbf{b} = (b_1, b_2, \dots, b_n) $ e uma operação $f$, conhecida como função de ativação. É realizada sobre a soma ponderada de cada componente $x_1, x_2, \dots, x_n$ do vetor de entrada pelos pesos. Isso pode ser escrito como uma transformação afim seguida de uma não-linearidade 
\begin{equation}
\begin{aligned}
\mathbf{y} = & \quad f\left( (w_1 x_1 + b_1) + (w_2 x_2+ b_2) + \dots + (w_n x_n + b_n) \right) \\
= & \quad f\left( \bm{w}^T \mathbf{x} + \mathbf{b}\right),
\label{neuron}
\end{aligned}
\end{equation}
operações que são mostradas na Figura \eqref{fig:my_neuron}.

\begin{figure}[H]
\begin{center}
\includegraphics[width=0.85\textwidth]{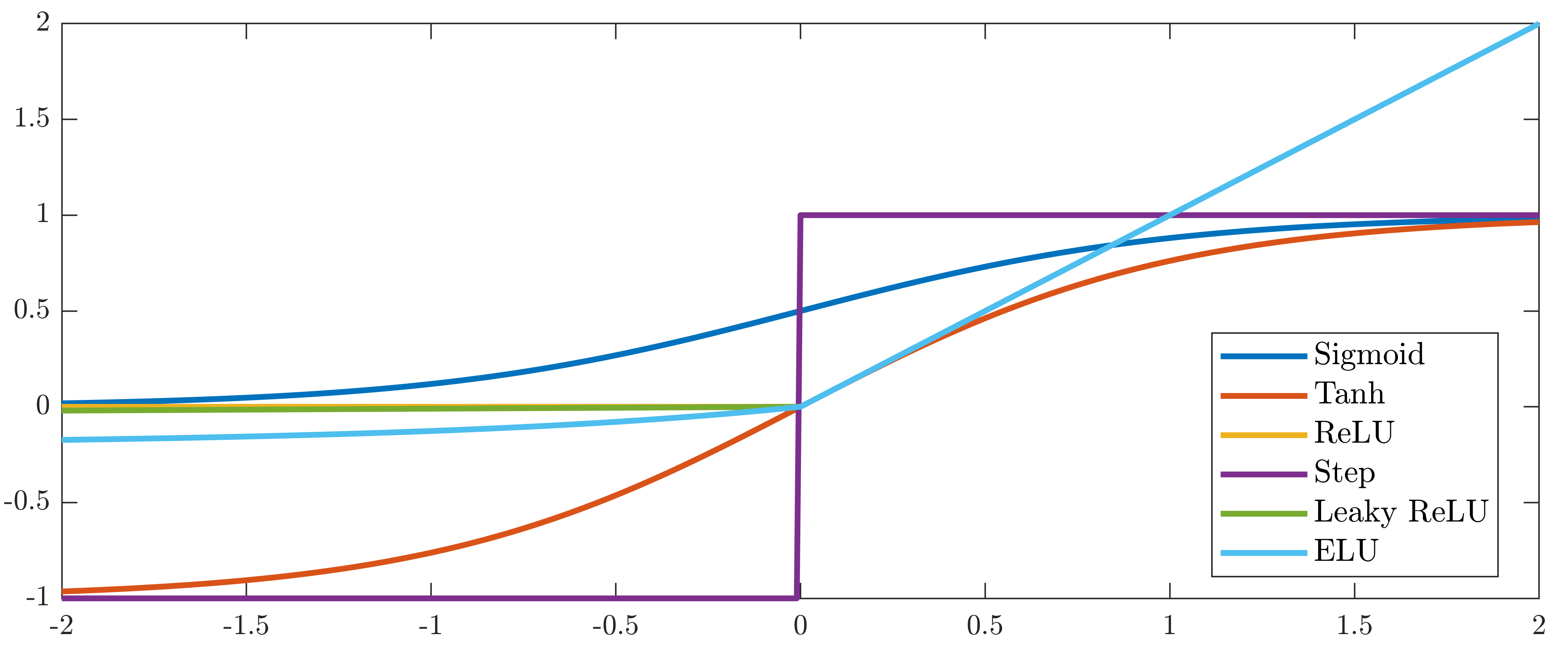}
\caption[Exemplos de funções de ativação.]{Exemplos de funções de ativação. Fonte: Próprio autor.}
\label{fig:activation}
\end{center}
\end{figure}

As funções de ativação $f$ da Equação \eqref{neuron} são responsáveis por incluir a não-linearidade nas redes neurais. Entre elas existem funções sigmoides, tangente hiperbólicas, ReLU (\textit{rectified linear unit}), \textit{softmax} \textit{softplus}, entre outras \cite[págs. 69, 175, 184-6, 195]{goodfellow2016deep} e algumas são mostradas na Figura \ref{fig:activation}.

Essa é apenas uma forma possível para as unidades de redes neurais e há outras classes de redes que partem de operações diferentes. Redes convolucionais, por exemplo, são pensadas principalmente para utilização com imagens \cite[Capítulo 9]{goodfellow2016deep}. Para isso, são aprendidos filtros diferentes em cada camada da rede, visando a extração de informações da imagem original a partir da convolução com esses filtros. Isso resulta por exemplo em invariância à translação da técnica, o que significa que a característica extraída pode estar em qualquer parte da imagem \cite[pág. 342]{goodfellow2016deep}.

\subsubsection{Na aprendizagem profunda, como a união de unidades simples formam a rede neural como um todo?}
\begin{figure}[H]
\begin{center}
\includegraphics[width=0.5\textwidth]{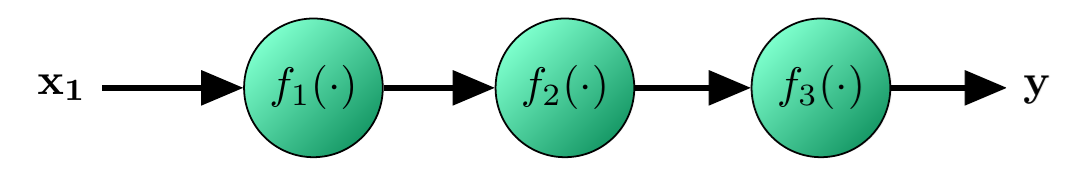}
\caption[Composição de funções.]{Composição de funções. Fonte: Próprio autor. }
\label{fig:composition}
\end{center}
\end{figure}

\begin{figure}[H]
\begin{center}
\includegraphics[width=0.65\textwidth]{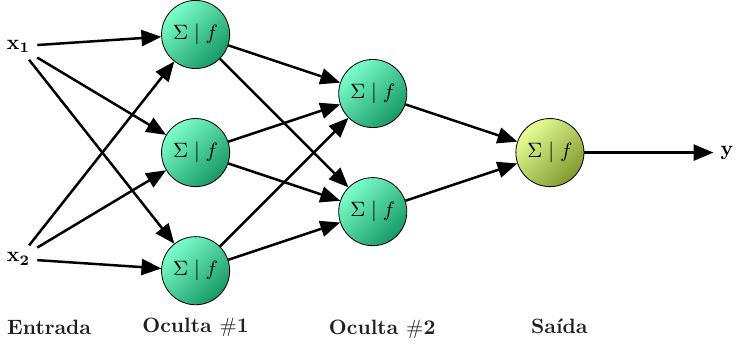}
\caption[Esquema de uma rede neural com várias camadas. ]{Esquema de uma rede neural com várias camadas. Fonte: Próprio autor. }
\label{fig:my_MLP}
\end{center}
\end{figure}

Suponha que a sequência de operações sobre um vetor de entrada seja definida da seguinte forma. Seja o sobrescrito $i$ relativo à camada de número $i$ da rede. Na Figura \ref{fig:my_MLP}, a camada para $i=3$ é a camada de saída. Logo, a saída em si da rede da  poderia ser reescrita como
\begin{equation} 
\mathbf{y} = f^{(3)} \left( f^{(2)} [f^{(1)} (\mathbf{x}, \bm{\theta}^{(1)}), \bm{\theta}^{(2)}], \bm{\theta}^{(3)} \right).
\label{eq:layers}
\end{equation}
De modo genérico, isso pode ser representado conforme Figura \ref{fig:composition}. 

Considerando as unidades conforme Equação \eqref{neuron}, os parâmetros da rede $\bm{\theta}$ são os pesos $\mathbf{w}^{(i)}$ do conjunto de unidades de uma determinada camada $i$ e os \textit{biases} $\mathbf{b}^{(i)}$. Assim, reescreve-se a Equação \eqref{eq:layers} como 
\begin{equation}
\mathbf{y} = f^{(3)} \left( \mathbf{w}^{(3)} \left[ f^{(2)} (\mathbf{w}^{(2)}\left[ f^{(1)}(\mathbf{w}^{(1)}\mathbf{x} + \mathbf{b}^{(1)}) \right] + \mathbf{b}^{(2)})\right] + \mathbf{b}^{(3)} \right)
\label{eq:inf1}
\end{equation}
Enquanto os parâmetros são $\mathbf{w}$ e $\mathbf{b}$, os hiperparâmetros definem a rede, incluindo o número de unidades, de camadas (a sua largura e profundidade), como essas camadas se conectam e a função de ativação \cite[pág. 197]{goodfellow2016deep}. 

No caso de uma MLP, uma rede neural \textit{feedfoward} (sem ligação de unidades com outros de camadas anteriores), há muito mais ligações acontecendo. Um exemplo é mostrado na Figura \ref{fig:my_MLP}. Na camada de entrada há duas \textit{features} e na camada de saída há apenas um valor. Entre elas, observa-se que todas as unidades da primeira camada oculta estão ligadas a todas as unidades da segunda camada oculta. Essas camadas são totalmente conectadas. O mapeamento de $\mathbf{x}$ para $\mathbf{y}$ é a composição das unidades. 

Para obter uma expressão de toda a operação da rede neural, a notação matemática se torna mais complexa e seria mais adequado reunir os pesos e operações na forma matricial. 

\subsection{Na aprendizagem profunda, existem outras definições para regularização?}\label{sec:new}

Nos últimos anos, diferentes definições de regularização estão sendo propostas na literatura de aprendizagem profunda, o que exige uma atenção a esse conceito em cada referência:

\begin{itemize}
\item Em \cite{Moradi2019}, os autores fazem uma revisão sobre diversas estratégias de regularização em aprendizagem profunda. Nela, os autores trazem regularização como incluir conhecimento de senso comum sobre os parâmetros do algoritmo de aprendizagem que não podem ser obtidos diretamente dos dados. Só que os autores dizem que isso é mais relevante quando os dados são insuficientes. Assim, regularização incluiria qualquer componente do processo de aprendizagem ou predição que compensaria a falta de dados para realizar melhor essas tarefas;
\item Em \cite[pág. 120]{goodfellow2016deep}, regularização é descrita como as modificações no algoritmo proposto que buscam reduzir o erro de generalização, mas sem necessariamente reduzir o erro de treinamento. Os autores discutem propósito e formas mais gerais para regularização em \cite[Capítulo 7]{goodfellow2016deep} e uma lista de estratégias de regularização em aprendizado de representação é encontrada em \cite[págs. 555-7]{goodfellow2016deep}. Os autores também entendem regularização como expressar explicitamente ou implicitamente uma preferência entre funções e a própria estrutura da DNN teria esse papel ao hierarquizar a representação \cite[págs. 120, 203, 505, 555]{goodfellow2016deep};
\item Em \cite{2017kukacka}, os autores consideraram que a definição anterior de \cite[pág. 120]{goodfellow2016deep} era restrita demais, pois métodos conhecidos de regularização como \textit{weight decay} podem também reduzir o erro de treinamento. Assim, os autores descrevem regularização como qualquer técnica que faça o algoritmo produzir melhores resultados no conjunto de teste, isto é, generalizar melhor, sem qualquer menção ao erro de treinamento. A partir dela, os autores descrevem formas de regularização em relação aos dados, arquitetura, função de erro, termo de regularização e método de otimização \cite{2017kukacka}. 
\end{itemize}

Nos livros em geral, a definição de regularização acaba sendo relacionada com a área de pesquisa em que ela está sendo utilizada. É possível que as novas definições de regularização em aprendizagem profunda se tornem mais distantes em relação ao significado original. Para que a definição original não se perca, seriam necessários novos trabalhos dedicados a mostrar que cada método de regularização no contexto de aprendizagem profunda pode (ou não) ser entendida como métodos de regularização no sentido de Tikhonov ou se partiram de outra definição de regularização. Unir evidências de resultados experimentais com o fundamento teórico das técnicas permite que se tenha mais clareza sobre o que de fato está sendo realizado e obtido.

\subsubsection{Como regularizar o treinamento de ANNs?}

A partir dessas definições, há diversas formas de regularização em aprendizagem profunda:
\begin{itemize}
\item \textit{Weight decay} (Subseção \ref{sec:weight}), em que há um termo aditivo e quadrático de penalização sobre os parâmetros da rede, de modo que alguns autores o consideram equivalente à regularização clássica de Tikhonov (de ordem zero) e à regressão de Ridge \cite{Chen2002};
\item \textit{Dropout}, em que a função custo não é modificada com termo adicional, mas sim alguns pesos das unidades da rede são zerados durante o treinamento, com uma certa probabilidade, de modo que é como treinar diferentes redes neurais, cujo resultado final será um efeito médio entre elas  \cite{nielsen2018, wei2020implicit}; 
\item \textit{Data augmentation}, quando se aumenta a quantidade de amostras na base de dados de treinamento a partir de transformações nos dados originais ou realizando simulações para complementar dados experimentais; 
\item \textit{Early stopping} (Subseção \ref{sec:early}) significa truncar as iterações da otimização, sem esperar convergência assintóptica; 
\item A escolha adequada da arquitetura e do otimizador podem ser consideradas formas de regularização implícita \cite[pág. 165]{Jakubovitz2019}. Um exemplo é no \textit{Deep Image Prior} (DIP) \cite{Ulyanov2020}, técnica que utiliza redes convolucionais generativas para obter os resultados. Ao considerar que o resultado deve ser gerado por uma CNN, a própria arquitetura teria o papel de regularizador \cite{Dittmer2019}, pois imagens naturais seriam mais facilmente representadas do que ruídos.  
\end{itemize}

No Apêndice \ref{subsec:implicit} foi discutida a diferença entre \textit{priors} explícitos e \textit{priors} implícitos. Métodos de regularização que englobam um termo aditivo no funcional de otimização, como o \textit{weight decay}, são classificados como explícitos.  Em alguns desses casos, a técnica pode ser relacionada com a inferência bayesiana com o estimador MAP, facilitando sua interpretação, mas isso nem sempre é possível.  

Em aprendizagem profunda, essa distinção é importante, pois muitos fatores, além do termo aditivo, podem ser responsáveis por mitigar o \textit{overfitting} em uma solução de problemas mal-postos com redes neurais e é de interesse que eles sejam conhecidos. Muitos desses regularizadores podem ser considerados implícitos, atuando mais indiretamente sobre a aprendizagem da função \cite[pág. 165]{Jakubovitz2019}. Por exemplo, o próprio otimizador iterativo, como o gradiente descendente estocástico (SGD), também apresentam efeitos de regularização \cite{2017kukacka} e há trabalhos que focam na regularização implícita deles \cite{Smith2021}. Mais importante do que essa classificação é a variedade de efeitos de regularização obtidos dentro das novas definições.

\subsubsection{Qual é o \textit{trade-off} em soluções regularizadas com termo aditivo?}

É possível analisar a relação entre \textit{bias} e variância da solução regularizada da Equação \eqref{eq:ERM_RLS}.  Para essa mesma discussão na regressão linear, ver Subseção \ref{sec:bias}. Se a regularização diminui o \textit{overfitting} do modelo aos dados, ela é uma forma de controlar o compromisso entre a acurácia de uma solução pela ERM e a complexidade do modelo obtido \cite[pág. 73-4]{cherkassky2007learning}, \cite[Seção 8.2.3]{Deisenroth2020}, \cite[pág. 188]{alvarez2017digital}. Especificamente:
\begin{itemize}
\item Quanto maior é $\lambda$, maior é o \textit{bias} e menor é a variância, de modo que o modelo se torna mais simples, com menos graus de liberdade nos parâmetros. Assim será mais difícil será para que resulte em \textit{overfitting} nos dados \cite{geron2019hands-on};
\item  Por outro lado, se $\lambda$ for muito grande, também pode resultar em \textit{underfitting} que, no contexto de aprendizagem, significa que mesmo com mais dados de treinamento, a performance do algoritmo não será melhor \cite{geron2019hands-on}.
\end{itemize}
Assim, $\lambda$ é responsável por controlar o peso que se dá para a complexidade da solução, ou seu grau de suavidade, e o erro no conjunto de treinamento \cite{Prato2008}. Há autores que relacionam $\lambda$ diretamente ao grau de generalização obtido \cite{Poggio1990}. Nesse contexto, $\lambda$ é um hiperparâmetro do algoritmo, não um parâmetro do modelo. Após a sua escolha \textit{a priori}, ele não é  atualizado durante a etapa de treinamento \cite{geron2019hands-on}.

\subsubsection{Qual é a relação entre regularização com norma $\ell_2$ e \textit{weight decay}?}\label{sec:weight}

Na Equação \eqref{eq:ERM4}, quando $\Omega(\bm{\theta}) = \vert \vert \bm{\theta} \vert \vert^2_2$, a técnica é conhecida como \textit{weight decay} \cite[pág. 70]{cherkassky2007learning}, \cite{Krogh1991}, conforme
\begin{equation}
\hat{\bm{\theta}} = \arg\min\limits_{\bm{\theta}} \left[ \frac{1}{N} \sum_{i=1}^{N} \vert\vert h_{\bm{\theta}}(\mathbf{y}_t^{i}) - \mathbf{x}_t^{i} \vert\vert_2^2 + \lambda^2 \vert \vert \bm{\theta} \vert \vert^2_2 \right].
\label{eq:ERM_RLS}
\end{equation}
Aqui, a definição de \textit{weight decay} é baseada em \cite[Subseção 12.2]{Hinton1989}, \cite[Equação 1]{Krogh1991} e \cite[Subseção 8.2]{Vapnik1992}. 

É possível dizer que ela é análoga à regularização clássica de Tikhonov \cite{Chen2002}. O \textit{weight decay} também pode ser entendido como uma estimação MAP, onde o termo de regularização corresponde a uma distribuição de probabilidades \textit{a priori} gaussiana sobre os parâmetros $\theta$ do modelo, incorporando preferências sobre os valores de $\bm{\theta}$ \cite[pág. 505]{goodfellow2016deep}. Nesse sentido, outros termos poderiam ser utilizados, como $\vert \vert \bm{\theta} \vert \vert_1$.

Por outro lado, partindo da definição de \textit{weight decay} encontrada em \cite[Equação 3]{Hanson1988}, os autores de \cite{Loshchilov2018} descreveram as diferenças desta para a  regularização com norma $\ell_2^2$ da Equação \eqref{eq:ERM_RLS}. Eles argumentam que, em alguns casos, elas são equivalentes, como quando se utiliza o algoritmo de otimização SGD. No entanto, quando se utiliza métodos de gradientes adaptativos como a estimação adaptativa de momento (ADAM), essa equivalência não é mais verificada.

 \subsubsection{Como visualizar os efeitos da regularização?}\label{sec:visualreg}
 
Suponha que um determinado modelo seja definido por uma função $f$ desconhecida (modelo  caixa-preta), que permite obter um valor de saída a partir de dois parâmetros $\theta_1$ e $\theta_2$. O objetivo é obter $\theta_1$ e $\theta_2$ tais que a saída seja minima. 

\begin{figure}[H]
     \centering
   \begin{subfigure}[b]{0.49\textwidth}
         \centering
         \begin{tikzpicture}
  \node (img)  {\includegraphics[scale=0.5]{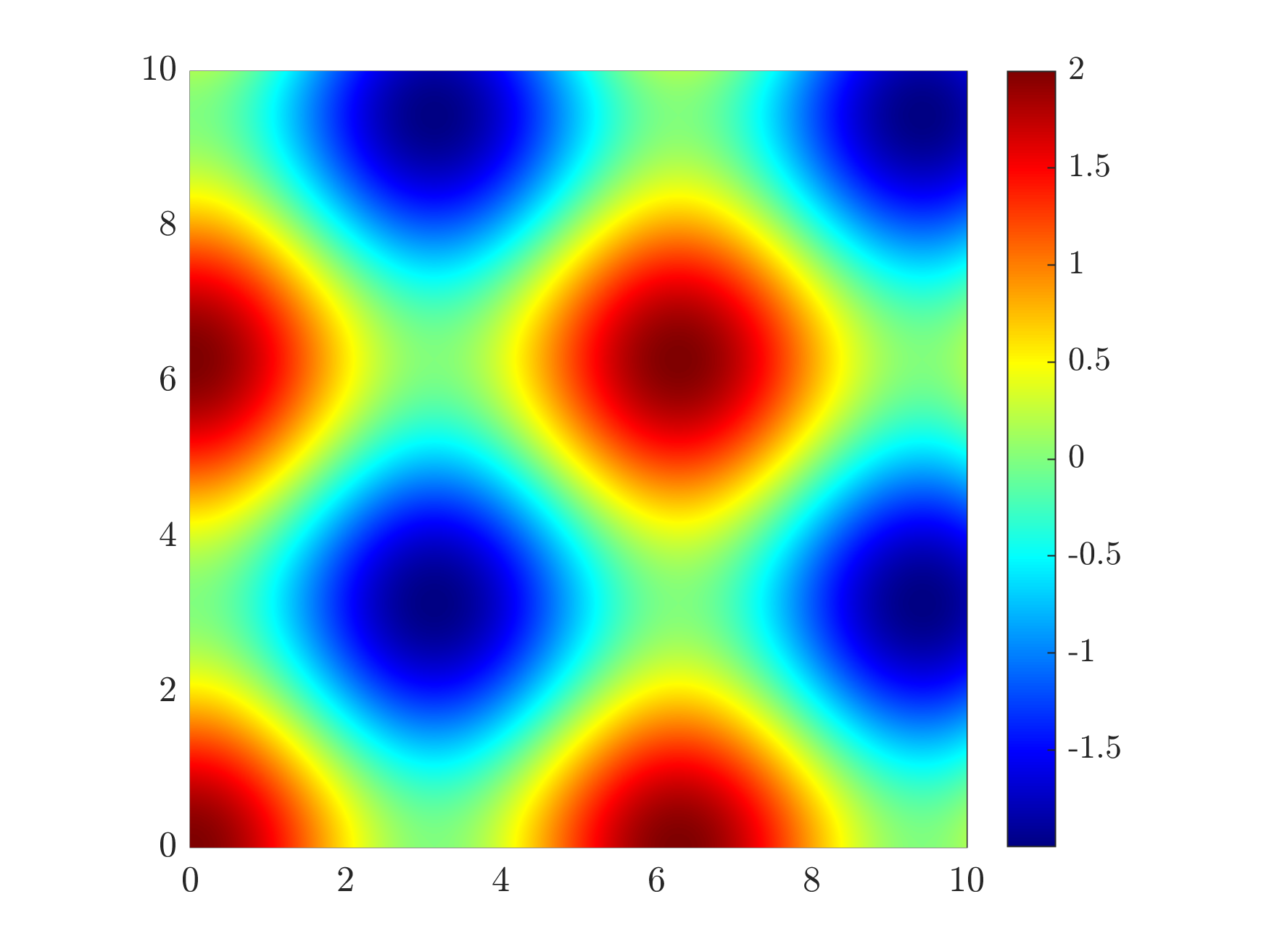}};
  \node[below=of img, node distance=0cm, yshift=1cm] {$\theta_2$};
  \node[left=of img, node distance=0cm, rotate=90, anchor=center,yshift=-0.7cm] {$\theta_1$};
 \end{tikzpicture}
         \caption{$\cos(\theta_1) + \cos(\theta_2)$}
                  \label{fig:visualreg1a}
      \end{subfigure}
     \hfill
     \begin{subfigure}[b]{0.49\textwidth}
         \centering
             \begin{tikzpicture}
  \node (img)  {\includegraphics[scale=0.5]{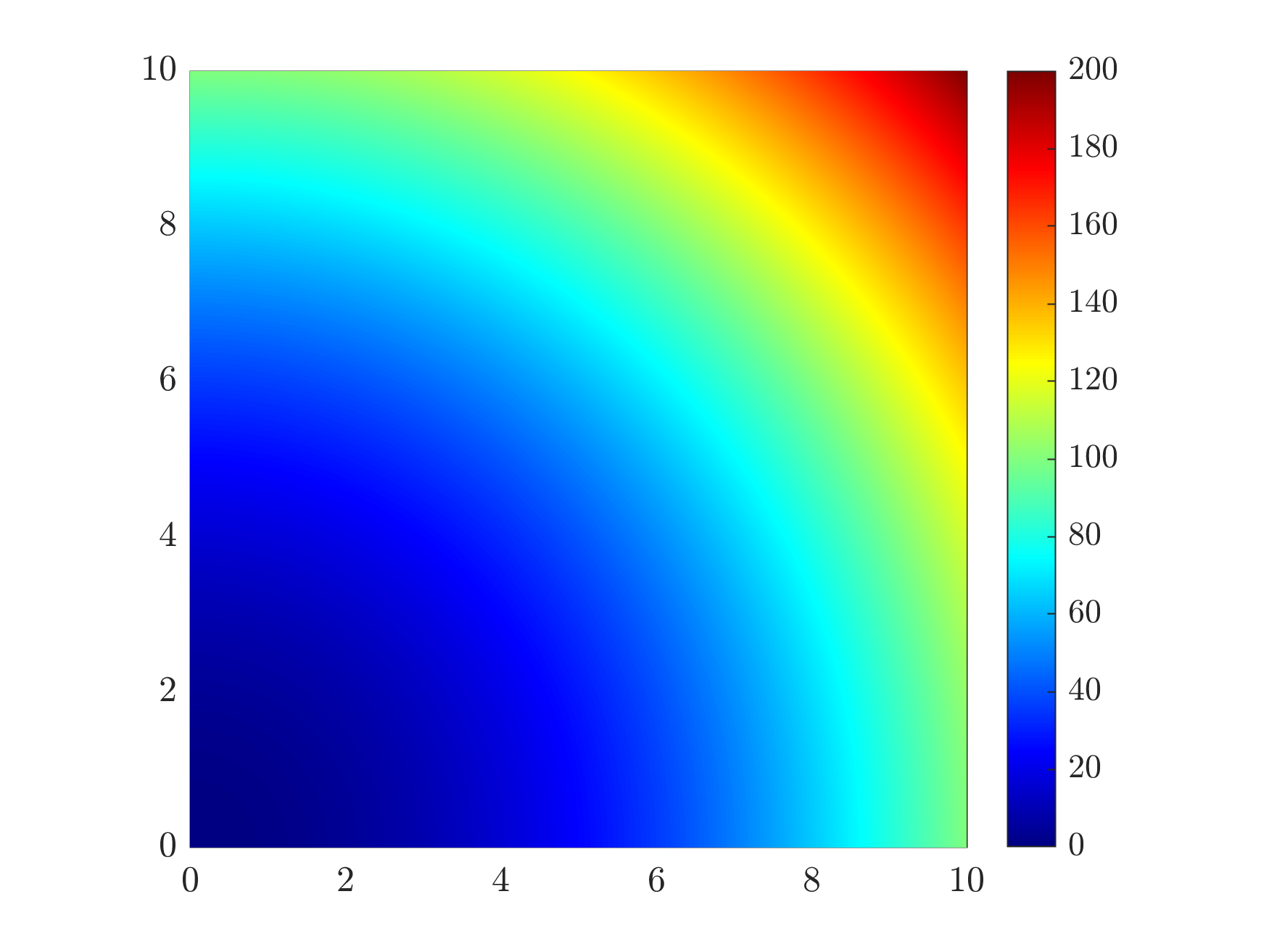}};
  \node[below=of img, node distance=0cm, yshift=1cm] {$\theta_2$};
  \node[left=of img, node distance=0cm, rotate=90, anchor=center,yshift=-0.7cm] {$\theta_1$};
\end{tikzpicture}
         \caption{$\theta_1^2 + \theta_2^2$}
                  \label{fig:visualreg1b}
       \end{subfigure}
     \caption[Visualização das saídas do modelo desconhecido (esquerda) e do termo de regularização (direita).]{Visualização das saídas do modelo desconhecido (esquerda) e do termo de regularização (direita). Fonte: Próprio autor.}
     \label{fig:visualreg1}
 \end{figure}
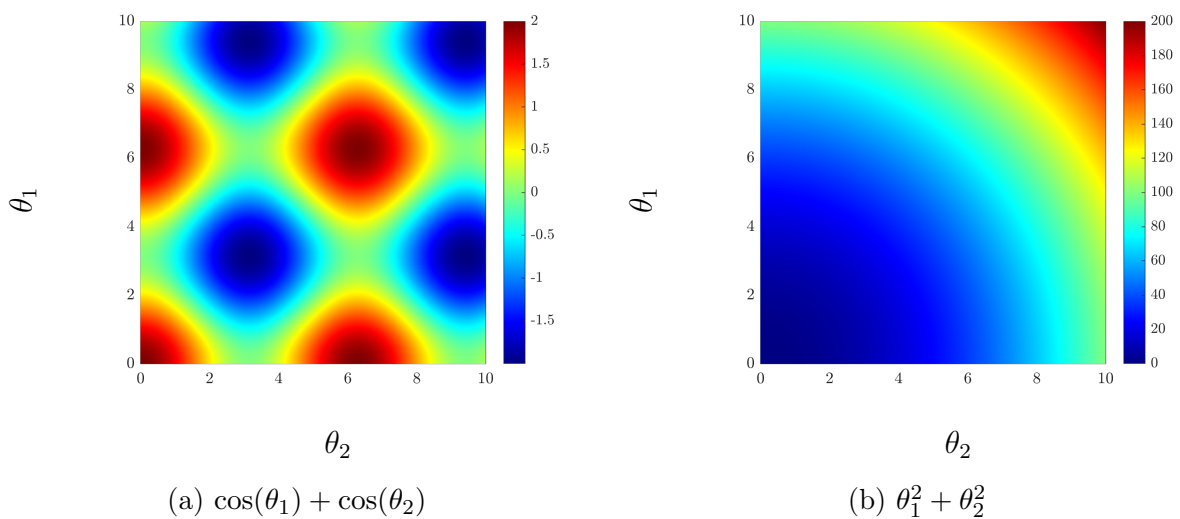

Neste exemplo específico, seja $f(\theta_1, \theta_2) = \cos(\theta_1) + \cos(\theta_2)$. Ela não apresenta um mínimo global, apenas mínimos locais. A Figura \ref{fig:visualreg1a} mostra a visualização dos resultados em um intervalo de valores positivos. Observa-se que existe mais de um par que resulta em $f(\theta_1, \theta_2) = 0$, as regiões de cor azul escuro. Em outras palavras, existem múltiplos, possívelmente infinitos pares possíveis. Qual par escolher?

Neste caso, a regularização se torna um critério para essa escolha. A Figura \ref{fig:visualreg1b} mostra a visualização apenas do termo de regularização quadrático $\theta_1^2 + \theta_2^2$. Sendo uma função convexa, apresenta mínimo global. Logo, ela pode ser utilizado como critério, privilegiando soluções com normas $\ell_2^2$ menores.  

Quando se adiciona a saída do modelo desconhecido e o termo de regularização, o mapa dessa superfície é modificado. Se isso for feito de maneira adequada, espera-se que agora possa existir um mínimo global. Agora, ao invés de minimizar apenas $\cos(\theta_1) + \cos(\theta_2)$, o objetivo se torna minimizar $\cos(\theta_1) + \cos(\theta_2) + \lambda^2(\theta_1^2 + \theta_2^2)$. 

\begin{figure}[H]
     \centering
   \begin{subfigure}[b]{0.49\textwidth}
         \centering
         \begin{tikzpicture}
  \node (img)  {\includegraphics[scale=0.5]{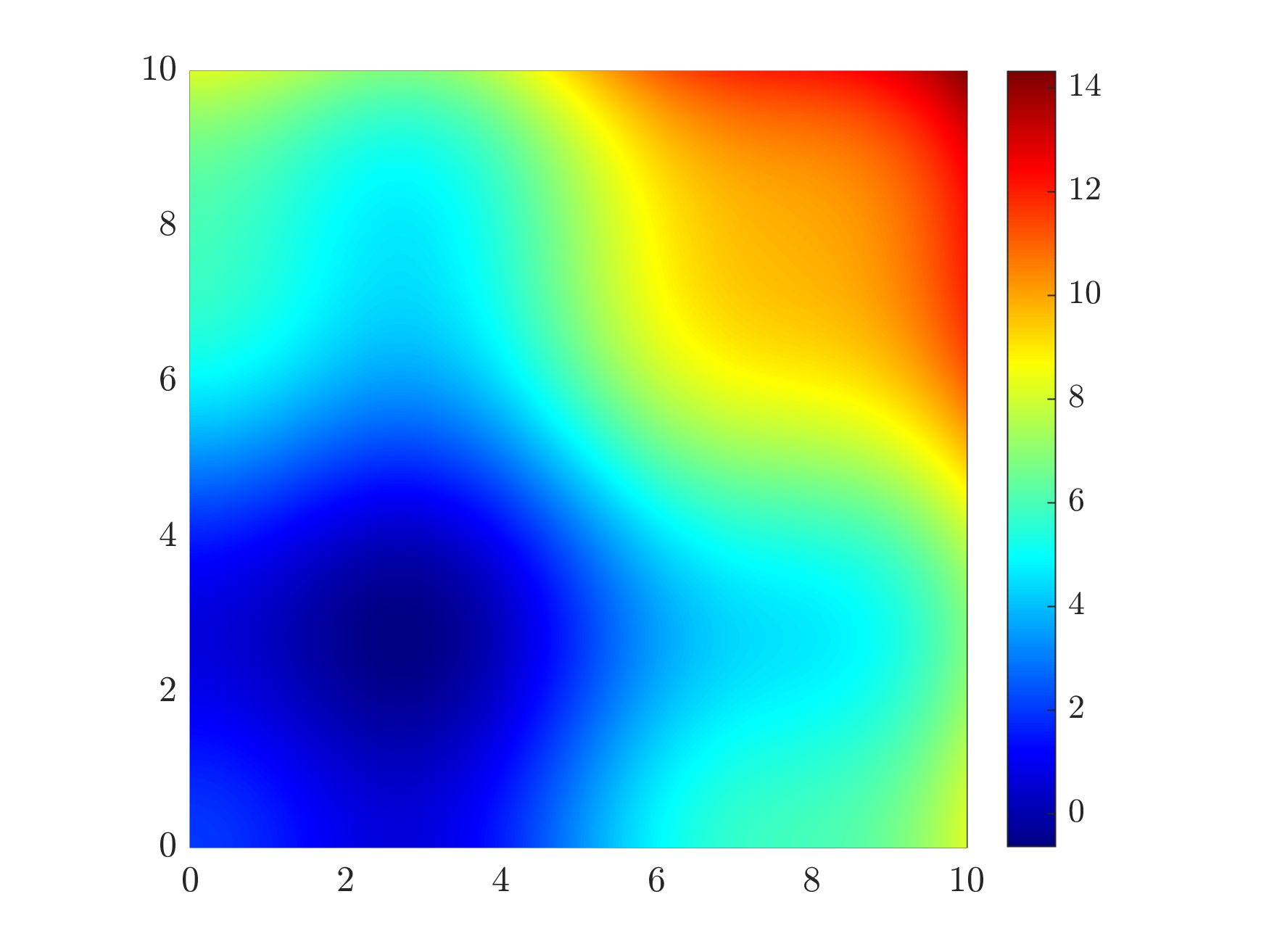}};
  \node[below=of img, node distance=0cm, yshift=1cm] {$\theta_2$};
  \node[left=of img, node distance=0cm, rotate=90, anchor=center,yshift=-0.7cm] {$\theta_1$};
 \end{tikzpicture}
         \caption{$\cos(\theta_1) + \cos(\theta_2) + 0.08(\theta_1^2 + \theta_2^2)$}
                  \label{fig:visualreg2a}
      \end{subfigure}
     \hfill
     \begin{subfigure}[b]{0.49\textwidth}
         \centering
                  \begin{tikzpicture}
  \node (img)  {\includegraphics[scale=0.5]{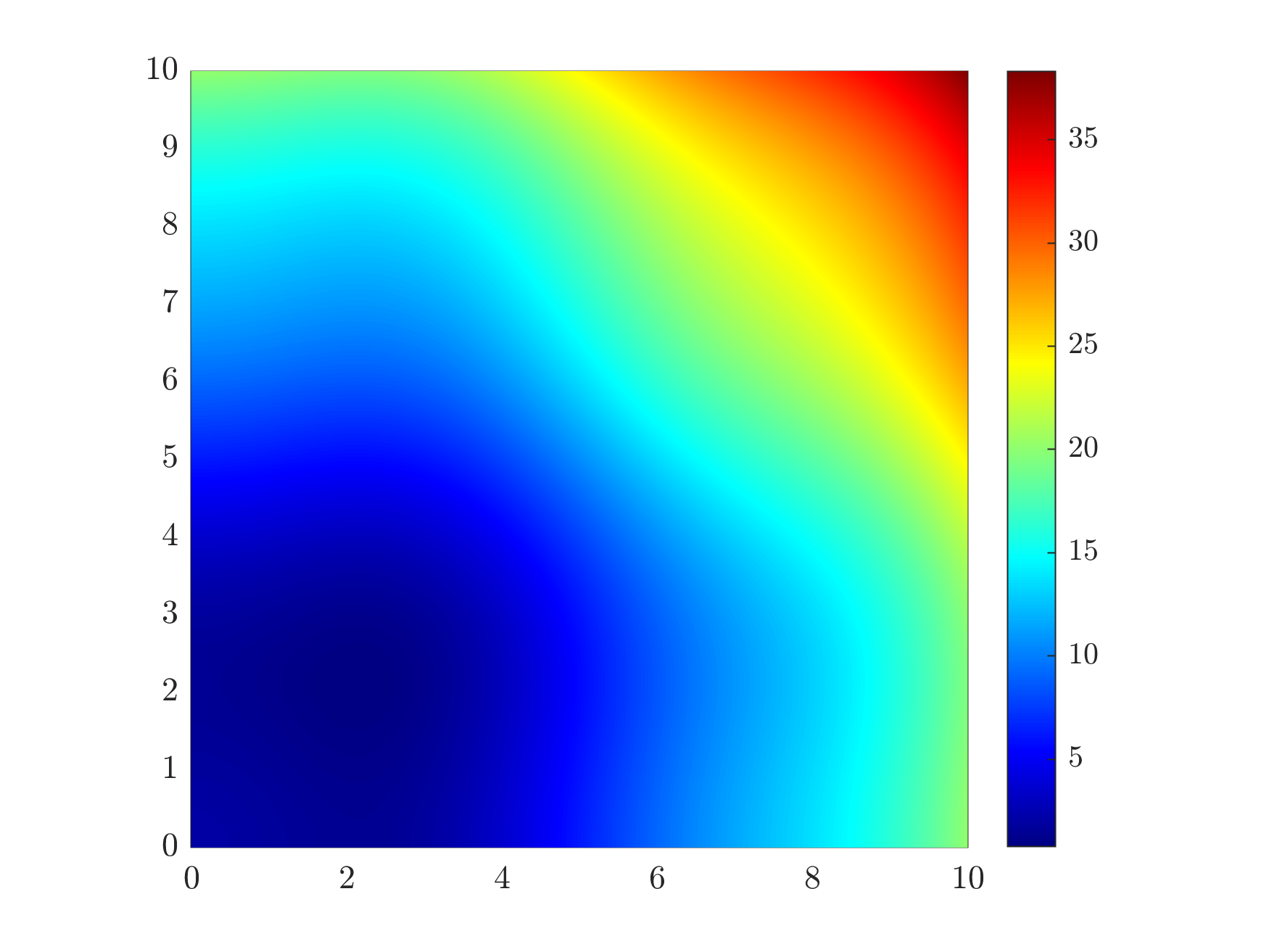}};
  \node[below=of img, node distance=0cm, yshift=1cm] {$\theta_2$};
  \node[left=of img, node distance=0cm, rotate=90, anchor=center,yshift=-0.7cm] {$\theta_1$};
\end{tikzpicture}
         \caption{$\cos(\theta_1) + \cos(\theta_2) + 0.2(\theta_1^2 + \theta_2^2)$}
                  \label{fig:visualreg2b}
       \end{subfigure}
     \caption[Visualização das soluções regularizadas.]{Visualização das soluções regularizadas. Fonte: Próprio autor.}
     \label{fig:visualreg2}
 \end{figure}
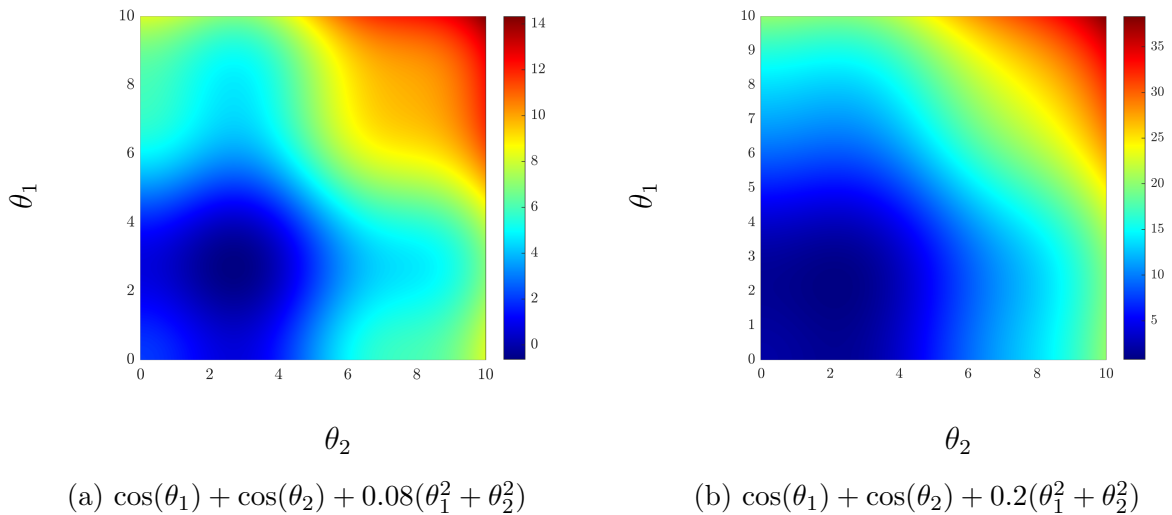

Conforme Figura \ref{fig:visualreg2}, o resultado depende da escolha de um valor fixo de $\lambda$.  A Figura \ref{fig:visualreg2a} mostra os resultados quando $\lambda^2 = 0.08$ e já é possível observar o mínimo global. Já a Figura \ref{fig:visualreg2b} apresenta $\lambda^2 = 0.2$, um valor maior. Quanto maior é $\lambda$, o termo de regularização domina a solução final e o mapa se torna cada vez mais parecido com aquele da Figura \ref{fig:visualreg1b}. Se $\lambda$ for grande em excesso, a solução regularizada pode acabar perdendo o propósito inicial do problema. Com valores adequados de $\lambda$, essa mistura dos dois termos permite achar o par  $\theta_1^2 + \theta_2^2$ da função original desconhecida, mas que apresente a menor norma como critério de escolha. 

\subsubsection{Qual é a analogia entre \textit{early stopping} e semiconvergência?}\label{sec:early}

É necessário um critério de parada das iterações do algoritmo iterativo, o que pode variar bastante entre as arquiteturas utilizadas e o poder computacional disponível. Um exemplo é conhecido como \textit{early stopping}, na qual o erro do funcional no conjunto de treinamento é acompanhado do erro em um conjunto de validação. Assim, o treinamento da ANN é encerrado antes do total de iterações, assim que o erro no conjunto de validação começa a aumentar, visando obter uma melhor performance de generalização \cite[págs. 343-4]{Bishop1995book}.  Se o \textit{early stopping} diminui o erro de generalização, ele pode ser considerado uma forma de regularização, nesse caso, implícita \cite[pág. 2829]{Chen2002}. 

Alguns autores interpretam que ANNs primeiro aprendem padrões mais simples presentes em todas as amostras de dados para depois se ajustar aos exemplos específicos e ao ruído \cite{Krueger2017DeepND, 2017kukacka}. Nesse caso, parar iterações antes permite que a ANN não aprenda o ruído. Apesar de motivação diferente, pode ser feito um paralelo com os algoritmos iterativos baseados no fenômeno da semiconvergência, utilizado em alguns métodos iterativos para problemas inversos mal-postos \cite[pág. 110]{hansen2010discrete}. Neles, não se esperava que o algoritmo convergisse para um grande número de iterações e era definido um critério de parada antes que ele começasse a divergir. 

\subsection{Existindo novas interpretações e definições para regularização, quais são as possíveis consequências?}

Um dos grandes desafios da área de aprendizagem profunda é tentar entender a habilidade das ANNs de generalizar. Em \cite[pág. 229]{goodfellow2016deep} os autores argumentam que o controle de capacidade de um modelo não deve ser a busca pelo modelo com número ideal de parâmetros, mas sim pode ser um modelo grande, como em aprendizagem profunda, que foi regularizado de maneira adequada. Logo, em \cite[pág. 120]{goodfellow2016deep}, os autores consideram que regularização é um assunto central tão importante quanto a otimização em si.
 
A partir do teorema sem almoço grátis, os autores argumentam que não há uma melhor forma geral de regularização, de modo que se deve buscar a mais adequada para a tarefa em particular que se deseja resolver \cite[pág. 120]{goodfellow2016deep}. Essa visão é compatível com as novas definições apresentadas e a escolha de qual utilizar pode acabar sendo guiada inicialmente pelos bons resultados experimentais. Mesmo os autores de \cite{2017kukacka} reconhecem que a definição que eles propõe está mais alinhada com a literatura de aprendizado de máquina do que com a literatura de problemas inversos, cuja definição é mais restritiva. 

 Uma ANN é capaz de generalizar mesmo sem regularização de um termo aditivo. Essa situação é diferente de determinados problemas inversos mal-postos, cuja ausência de termo de regularização impossibilita sua solução. Em aprendizado de máquina, se há diversos fatores que apresentam efeitos implícitos de regularização, não é trivial dizer a responsabilidade na melhoria da generalização. Cada componente do algoritmo tem sua influência e deve-se avaliar caso a caso. Um exemplo é  a regularização implícita do próprio otimizador utilizado quando ele é baseado em gradientes \cite{Barrett2021, Razin2020, Razin2021}. 
 
 \subsubsection{Quais são as relações entre a regularização de Tikhonov e as novas definições?}

Existem trabalhos que reinterpretam estratégias de aprendizado de máquina no contexto de métodos de regularização:
\begin{itemize}
\item Em \cite{Bishop1995} os autores discutem como adicionar ruído nos dados durante o treinamento pode melhorar a sua generalização;
\item Em \cite[pág. 345]{Bishop1995book}, \cite{Collobert2004}, os autores discutem que, no caso de $\mathcal{L}$ quadrática, o \textit{early stopping} seria equivalente ao \textit{weight decay};
\item Em \cite{Poggio1990} a aprendizagem de um mapeamento suave entre entradas e saídas a partir de redes neurais foi interpretada como um problema de reconstrução de hipersuperfícies a partir de dados esparsos. Nessa analogia, aprendizagem significa coletar exemplos e estimar a altura da hipersuperfície para aqueles pontos e generalização significa obter os valores da hipersuperfície onde não há exemplos. Isso requer interpolação (no caso sem ruído), ou de modo mais geral, a aproximação da superfície entre os pontos dos dados. A informação dos dados não é suficiente para reconstruir unicamente o mapeamento nas regiões onde os dados não estão disponíveis e, assim, o problema de aproximação é mal-posto \cite{Burger2003, Poggio1990}. Como solução, foi utilizada a regularização de Tikhonov, que também foi relacionada com a capacidade de generalização do algoritmo resultante \cite{Krkov2004, Krkov2005}.
\end{itemize}

Mais do que indicar analogias ou semelhanças dos problemas de otimização, diversos trabalhos reescreveram o problema da aprendizagem com base nos conceitos de problemas inversos lineares e regularização \cite{Burger2000, girosi1995,  Krkov2012, Lu2013, devito2005}, de modo que os algoritmos para a solução de problemas inversos lineares e instáveis pudessem ser adaptados para o \textit{framework} de aprendizagem \cite{Prato2008}. Em \cite[Tabela 1]{devito2005} é possível ver um resumo das relações obtidas, que depende de fatores como a definição do operador direto e de entender o papel do ruído $\bm{\delta}$ na aprendizagem \cite{devito2005}. 

É importante que solução de tarefas de aprendizagem através da ERM sejam consistentes, para que elas sejam preditivas, ao mesmo tempo que elas sejam estáveis e robustas \cite{Mukherjee2006}. Essa discussão permite que a consistência seja relacionada com a propriedade da convergência da estabilidade (no sentido de Hadamard) \cite{devito2005}. Em \cite{Mukherjee2006} os autores definem a característica de bem-posta de modo que ela suficiente para generalização, além de ser necessária e suficiente para a consistência da ERM. Ou seja, que certas formas de definir bem-posto e consistente são equivalentes. 

Estabelecer relações formais entre problemas inversos no contexto de métodos de regularização e aprendizado de máquina ainda é investigado. Um trabalho recente nessa linha é  \cite{Burger2021}, onde o autor reinterpreta o problema de aprendizado de máquina sob o ponto de vista da teoria da regularização e reinterpreta métodos variacionais para problemas inversos sob o ponto de vista da minimização do risco.

\subsection{O que são métodos integrados?}

Para resolver problemas inversos mal-postos, a escolha sobre utilizar um paradigma baseado em modelos ou baseado em dados pode ser feita a partir da quantidade dos dados disponíveis e do grau de conhecimento do operador direto \cite[Figura 1.5]{inman2005damage}, \cite{Ongie2020}. Ao mesmo tempo, a existência de diferentes interpretações sobre regularização (estabilizar problemas mal-postos e prevenir o \textit{overfitting} em problemas de aprendizagem) sugere que a utilização conjunta entre os dois paradigmas pode trazer benefícios e gerar novas formas de solução \cite{Adler2021}. 

Assim, o presente capítulo discute métodos que integram aprendizagem profunda e modelos físico-matemáticos na solução de problemas inversos mal-postos. Uma extensa revisão sobre métodos integrados é encontrada em \cite{Arridge2019}. É importante ressaltar que essa abordagem é uma via de mão dupla, conforme será discutido adiante \cite{Arridge2019, Kording2018}. Existem componentes comuns entre essas propostas que devem ser observados e discutidos: operador direto, regularização, otimização e a rede neural profunda. 
 
Ainda relacionado com esse capítulo, o Apêndice \ref{ap:checklist} discute questões de reprodutibilidade de métodos integrados e traz também um \textit{checklist} sobre reprodutibilidade para ajudar na documentação dessas propostas.

\subsubsection{Como modelos podem ajudar a aprendizagem profunda?}

Quando se conhece um modelo da aplicação, mas a aprendizagem é realizada de modo totalmente agnóstico, é necessário (re)aprender a estrutura do modelo físico \cite{Lunz2018}. Como esse conhecimento de domínio (quando disponível) pode ser incorporado em algoritmos de aprendizagem profunda? Visando a utilização eficiente do conhecimento do operador direto nas ANNs, algumas propostas encontradas na literatura são:
\begin{itemize}

\item \textbf{\textit{Deep unrolling} (ou \textit{deep unfolding}) de algoritmos iterativos de reconstrução:} Algoritmos iterativos convencionais para reconstrução de problemas inversos podem ser desdobrados ou desenrolados (traduções literais de \textit{unfolding} e \textit{unrolling}) em uma ANN, de modo que seja possível de aprender seus parâmetros ao invés de predeterminá-los \cite{Bai2020, Monga2021}. Em outras palavras, esse método é um esquema iterativo aprendido \cite[Subseção 5.1.4]{Arridge2019}, na qual cada iteração no processo é interpretado como uma camada da ANN \cite{Arridge1}. O \textit{deep unrolling} pode ser visto como um \textit{design} estruturado de rede que busca classes de funções para melhor representar o mapeamento desejado \cite[Seção 4.2]{Bai2020}. Esse tipo de proposta é versátil e já foi utilizado, por exemplo, com o algoritmo ISTA \cite{Zhang2018}; na solução de \textit{deblurring} \cite{Li2019unfolding}; com o algoritmo de Gauss-Newton \cite{Yang2020}; na solução de problemas inversos não-lineares com valores complexos \cite{Takabe2020}; e para problemas gerais de restauração de imagens \cite{Mou2022};

\item \textbf{Arquiteturas guiadas pela física}: Arquiteturas guiadas pela física são ANNs que combinam ao operador direto como parte do \textit{loop} de treinamento. Realizando a retropropagação através do operador direto em conjunto da ANN, busca-se a aprendizagem de parâmetros desconhecidos ou incertos do operador direto de maneira não-supervisionada ou semi-supervisionada \cite{Adler2021}. Propostas desse tipo se assemelham a estruturas de \textit{autoencoder}, onde o \textit{encoder} são redes neurais especializadas enquanto o \textit{decoder} é o operador direto \cite[págs. 112-3]{Adler2021}. Exemplos em problemas sísmicos são encontradas em \cite{Alfarraj2019, Biswas2019};

\item \textbf{\textit{Deep Image Prior} para restauração de imagens}: Redes neurais convolucionais generativas podem ser utilizadas para gerar os parâmetros $\mathbf{x}$ de um modelo linear $\mathbf{A}\mathbf{x}$. No caso do \textit{Deep Image Prior} (DIP) \cite{Ulyanov2020}, o treinamento é realizado sem dados de treinamento, mas com apenas o próprio vetor de medidas ruidoso $\mathbf{y}_{\delta}$ e o conhecimento de $\mathbf{A}$. Na proposta original do DIP, os autores inclusive consideraram que o regularizador não seria necessário, sendo substituído pelo \textit{prior} implícito relativo à estrutura da rede neural utilizada. Ao mesmo tempo, eles interpretam que o DIP pode ser visto como um regularizador, que penaliza infinitamente tudo o que não pode ser gerado pela CNN \cite{Ulyanov2020}. Assim, alguns autores entenderam essa técnica como uma forma de \textit{regularização pela arquitetura}, pois é a definição da CNN que ajudaria na solução de problemas mal-postos \cite{Dittmer2019}. Nesse contexto, os autores também consideram que ele é \textit{handcrafted} \cite{Ulyanov2020}, já que a escolha dessa estrutura é feita pelo pesquisador. Empiricamente o DIP trouxe resultados positivos em diversas tarefas de reconstrução de imagens, mas ainda são necessários estudos teóricos para embasar seus resultados \cite{Belthangady2019, Feng2020, Liu2019}. Um exemplo de aplicação em imagens médicas é para a tomografia computadorizada de baixa dose ou de ângulos limitados \cite{Baguer2020, Barutcu2021}. Outra abordagem é a de incluir o DIP na substituição de $\mathbf{x}$ de algoritmos iterativos diferentes. Um  exemplo está em \cite{Cascarano2021}, na qual os autores utilizaram um esquema de ADMM com regularização de variação total junto do DIP;

\item \textbf{\textit{Data augmentation}}: Quando não se tem dados de treinamento em quantidades suficientes, é possível que aconteça o \textit{overfitting} da rede neural sob os dados. Ao invés de modificar a arquitetura ara diminuir o número de parâmetros, ou modificar a função custo adicionando regularização, uma forma de tentar resolver o problema da falta de dados é através das técnicas de \textit{data augmentation}. Elas buscam aumentar o número de amostras sem nova coleta de dados, apenas com processamento de sinais. As formas são variadas e incluem transformações lineares, adição de ruído, a mistura de imagens, aumento no espaço das cores, filtros com \textit{kernels}, apagar partes aleatórias dos dados. Se há modelo físico disponível, pode-se simular novos dados para complementar os dados disponíveis, ou mesmo em casos em que não há nenhuma medida experimental \cite[pág. 6]{inman2005damage}. Há também trabalhos que buscam o \textit{data augmentation} com técnicas de aprendizagem profunda, por exemplo utilizando redes neurais adversariais generativas para gerar os novos dados \cite{Liu2020, Shorten2019}.  

\item \textbf{\textit{Combinação entre propostas}}: Quando possível, as propostas podem ser utilizadas em conjunto com outras técnicas. Um exemplo é o DeepRED \cite{Mataev_2019_ICCV}, quando o DIP é utilizado em conjunto com regularização por \textit{denoising} (RED). Os autores argumentam que essa união pode ser feita de modo altamente efetivo, resultando em um algoritmo de reconstrução não-supervisionado sem a exigência de diferenciabilidade do \textit{denoiser} e obtendo imagens com melhor reconstrução \cite{Mataev_2019_ICCV}. Sua minimização é realizada através do otimizador ADMM, sendo que no passo relativo aos parâmetros da rede neural  é utilizado o otimizador ADAM.  No DeepRED, os autores indicam que há potencial na utilização de \textit{denoisers} obtidos por aprendizagem profunda para se obter melhores resultados \cite{Mataev_2019_ICCV}. Sem trazerem garantia de convergência, eles argumentam que os algoritmos apresentaram convergência empírica em seus resultados numéricos e a principal vantagem é que os resultados obtidos pelo DeepRED são superiores ao do RED e ao DIP individualmente.  Variações desta proposta incluem outros termos de regularização no \textit{framework} do DeepRED, como a norma $\ell_0$ sobre a imagem latente \cite{Feng2020}.

\end{itemize}

\subsubsection{Como a aprendizagem profunda pode ajudar abordagens guiadas por modelos, como o método de regularização?}\label{sec:deepreg}

É possível utilizar ANNs $\Psi_{\bm{\theta}}$ para ajudar na solução de problemas mal-postos junto do método de regularização, mas o conhecimento aprendido fica codificado nos seus parâmetros $\bm{\theta}$. Isso significa que não é relacionada à uma densidade de probabilidades explícita, como na inversão bayesiana, mas que ainda pode ser utilizada para realizar operações que ajudem na solução. 
\begin{itemize}

\item \textbf{Aprendizagem de parâmetros que definem o modelo}: O operador direto $\mathbf{A}$ pode ser dependente de parâmetros ou grandezas, de modo que a ANN pode ser utilizada para estimá-los (ou atualizá-los) antes da reconstrução. Seja o exemplo de \textit{deblurring}, onde $\mathbf{A}$ depende do conhecimento da PSF do sistema de aquisição. Pode-se utilizar ANNs para estimar essa PSF e então resolver como \textit{non-blind deblurring} (ou \textit{semi-blind}, quando há o conhecimento apenas aproximado da PSF) \cite[Figura 4]{Koh2021}. Exemplos em microscopia de deconvolução são encontrados em \cite{He2020, Shajkofci2020}, onde os autores estimaram uma PSF variante no espaço e na profundidade, respectivamente;

\item \textbf{Aprendizagem da física}: Pode-se realizar a aprendizagem dos operadores diretos \cite[pág. 105]{Arridge2019}. Existem propostas de CNNs na qual os filtros convolucionais representam derivadas parciais de até uma determinada ordem. Após treinamento, seria possível obter a PDE dos sistemas examinados \cite{Long2018}. Há também arquiteturas informadas pela física que resolvem tarefas de aprendizagem supervisionadas ao mesmo tempo que respeitam as leis físicas descritas por PDEs, ou seja, utilizam PDEs como restrição na aprendizagem \cite{Raissi2019}; 

\item \textbf{Aprendizagem de pré ou pós-processamento}: Em algumas aplicações, métodos de regularização resultam em reconstruções com muitos artefatos. As ANNs podem corrigir esses artefatos \cite[pág. 94]{Arridge2019}, seja realizando pré-processamento dos dados ou pós-processamento dos dados. Um exemplo de pós-processamento em tomografia por impedância elétrica é encontrado em \cite{Martin2017} e um exemplo em microscopia sem lente em \cite{Herv2020}. Neste último, não era possível remover um tipo de artefato em um único problema de otimização. Logo, os autores alternaram métodos de otimização para reconstrução e utilização de uma DNN para correção dos artefatos; 

\item \textbf{Aprendizagem de parâmetro de regularização}: Existirem diversas regras para escolha de $\lambda$, mas há autores que dizem que não é clara a escolha ótima para uma determinada aplicação \cite{Afkham2021}. Nesse caso, ANNs podem ser utilizadas para aprendizagem do valor ótimo de $\lambda$ em problemas inversos \cite{Afkham2021} e em aprendizado estatístico \cite{deVito2020};

\item \textbf{Aprendizagem de regularizadores}:

Alguns trabalhos desenvolvidos nos últimos anos utilizaram ANNs como funcionais de regularização não-lineares $\Omega_{\bm{\theta}}(\mathbf{x}) = \Psi_{\bm{\theta}}(\mathbf{x})$, que são muitas vezes sobreparametrizados e escaláveis para espaços de alta dimensão \cite[Subseção 4.7]{Arridge2019}, \cite{Liu2020, Lunz2018, Ongie2020}. Ou seja, os termos de regularização podem ser aprendidos para codificar informações \textit{a priori} diretamente dos dados e penalizarem soluções indesejadas \cite[págs. 62-6]{Arridge2019}. Uma forma de realizar tais propostas é primeiro treinar $\Psi_{\bm{\theta}}$ e depois, com os parâmetros $\bm{\theta}$ fixos, incluí-la como regularizador, ou como componente dele, na etapa de reconstrução. Trabalhos deste tipo incluem \cite{Bobin2021, Fang2020, Kobler2020, Li2019, Li_2020, Lunz2018,  Obmann_2021, Wu2017, Wu2021};

\item \textbf{Aprendizagem de \textit{priors}}: Há trabalhos em que parte dos componentes do regularizador é dada por ANN. Em \cite{Li2019}, na aplicação de \textit{blind deblurring}, os autores utilizaram um termo de regularização quadrático, e convencional, sobre o \textit{kernel}, e um segundo regularizador dado por um classificador binário, com o objetivo de indicar se a imagem estava borrada (valor $1$) ou nítida (valor $0$), direcionando a solução para imagens nítidas.   Em  \cite{Fang2020}, uma ANN é utilizada para detectar bordas nas imagens e essa informação é utilizada no termo de regularização como um valor de referência. Mesmo quando não se conhece explicitamente a densidade de probabilidade subjacente aos dados em que houve aprendizagem, as inferências da ANN podem ser úteis. Essa informação fica codificada na ANN e ela acaba funcionando como um \textit{prior} implícito \cite{kadkhodaie2021solving}; 

\item \textbf{Aprendizagem de atualizações e correções para operadores diretos imperfeitos}: Quando o operador direto é imperfeito, alguns algoritmos iterativos podem ser adaptados para o atualizarem através de métodos de aprendizagem, se houver dados de treinamento disponíveis. Dois exemplos são o \textit{total least-squares} e o \textit{learned Landweber} \cite[págs. 106-9]{Arridge2019};

\item \textbf{Aprendizagem de \textit{denoisers}}: O método de \textit{plug-and-play priors} ($P^3$) \cite{VenkaPlugplay2013} utiliza algoritmos com separação de variáveis, como o ADMM ou HQS, para dividir um funcional de otimização em subproblemas mais fáceis de resolver, individualmente, do que o problema original. Assim, o método reconhece um dos subproblemas como uma etapa de \textit{denoising}, mas não o resolve como um problema de otimização, esta etapa é substituída por um operador de \textit{denoiser} em si. Em trabalhos recentes, o subproblema do \textit{denoising} está sendo resolvido utilizando-se ANNs treinadas em dados representativos, funcionando como um regularizador implícito \cite{Siavash2020, Zhang2019plug}. O regularizador não apresenta uma forma explícita, o que limita a interpretação bayesiana como uma distribuição de probabilidade \textit{a priori} quando se usa o estimador MAP \cite[pág. 61]{Arridge2019}. Esses \textit{black-box denoisers} \cite{Arridge2019} apresentam diferentes nomes como \textit{deep plug-and-play denoising} \cite{Fu2021},  \textit{deep denoiser prior} \cite{zhang2021plugandplay} ou \textit{deep plug-and-play prior} \cite{ZHAO2020137}, apesar de motivação semelhante (se não idêntica). As aplicações incluem detecção de anomalias \cite{Fu2021}, completar matrizes de baixo posto  \cite{ZHAO2020137} ou problemas inversos lineares de modo mais geral \cite{Guo2019, kadkhodaie2021solving}. Em \cite{Liu2020}, por exemplo, os autores separaram o \textit{deblurring} do \textit{denoising} da microscopia de deconvolução utilizando ADMM, resolvendo-os separadamente como subproblemas. Enquanto para o \textit{deblurring} é utilizada uma forma de regularização de Tikhonov na frequência, o \textit{denoising} é resolvido com CNNs. No caso do RED \cite{Romano2017}, também é possível utilizar CNNs como o \textit{denoiser} \cite{Arridge2019};

\item \textbf{Aprendizagem de um operador proximal}: Outra abordagem com ADMM é utilizar uma ANN para aprender um operador proximal que busca restringir as soluções para serem imagens naturais. Em \cite{Chang_2017_ICCV}, a ANN é um operador de quase-projeção para o conjunto  de imagens naturais, sendo possível aplicá-la em problemas inversos diferentes, como amostragem comprimida, \textit{inpainting}, \textit{denoising} e super-resolução, sem a necessidade de ser retreinada. Similar à ideia de redes adversariais generativas, os autores buscam uma arquitetura que seja capaz de gerar novas amostras da distribuição de probabilidade desconhecida, de imagens no domínio da reconstrução (imagens naturais, no caso), sem estimar essa distribuição explicitamente. Ou seja, a rede aprende uma distribuição e busca aprender um projetor para ela, o que é entendido como um \textit{prior} implícito do problema \cite{Kaji2019}. No entanto, a proposta depende que haja imagens de referência (de alta resolução ou as respectivas \textit{ground-truth}) para treinamento, o que nem sempre está disponível \cite{Kaji2019}. Um exemplo é o caso de reconstruir imagens de CT de baixa dose tendo disponível bases de dados de CT de dose completa e alta resolução;

\item \textbf{Aprendizagem de um projetor para um subespaço conhecido}: Existem problemas de otimização com restrição em que essa restrição é que a solução pertença à um subconjunto conhecido. Uma forma de resolvê-los é através de algoritmos iterativos, como gradiente descendente ou iterações de Landweber, que dão um passo na direção de otimizar o funcional sem restrições, mas na sequência há a projeção do resultado para o conjunto desejado. Em \cite{Gupta2018}, por exemplo, os autores utilizaram o algoritmo do gradiente descendente projetado, onde eles utilizaram ANNs para a etapa de projeção da solução em um subespaço convexo;

\item \textbf{Aprendizagem de representação esparsa dos sinais}: ANNs podem ser utilizadas como operadores não-lineares de síntese com o objetivo de tornar o sinal esparso nessa nova base para que algoritmos eficientes baseados em norma $\ell_1$ possam ser utilizados \cite[Equação 6]{Obmann2020}. Outro trabalho que explora a aprendizagem de \textit{priors} relativos à esparsidade é \cite{Wu2021}, na qual os autores obtém \textit{priors} a partir de coleções de imagens de referências (externos) e a partir da própria imagem degradada (internos). Os autores utilizam um regularizador com norma $\ell_2^2$ e outro com norma $\ell_1$, incluindo os termos obtidos com aprendizagem.

\end{itemize}

\subsection{Quais são as dificuldades e cuidados na utilização de ANNs?}

\subsubsection{Quais são as limitações de ANNs em problemas inversos?}

A solução de problemas inversos mal-postos utilizando ANNs de ponta a ponta também apresenta limitações, como:
\begin{itemize}
\item Instabilidade e falta de garantias teóricas em problemas mal-postos \cite{Arridge1, Gottschling2020, LewisD2019};

\item Problemas inversos específicos podem não ter dados disponíveis em quantidade suficiente \cite{LewisD2019} ou não haver dados \textit{ground truth} \cite{Benning2018};
\item Grande especificidade \cite{Chang_2017_ICCV}, isto é, a necessidade de retreinar a rede quando o problema muda, como níveis variados de ruídos ou \textit{blur} \cite{LewisD2019}, quando muda o equipamento de aquisição dos dados \cite{Arridge2019}.
\end{itemize}

Além destas limitações, em \cite{Ongie2020}, os autores apresentam desafios desses métodos de ponta a ponta em problemas inversos do mundo real, como em imagens médicas:
 
\begin{itemize}
\item A utilização de um operador direto diferente nas etapas de treinamento e de teste pode levar ao surgimento de artefatos nas reconstruções. Um exemplo é quando se modela o problema direto para um equipamento, mas a reconstrução é feita para dados de outro equipamento, com marca e modelo diferente. É necessário avaliar a robustez do algoritmo de reconstrução, se haverá a necessidade de retreinar a rede, ou mesmo remodelar o operador direto para cada caso, o que pode não ser prático;
\item Por hipótese, os dados de treinamento são representativos da aplicação desejada, ou seja, que as características aprendidas no conjunto de treinamento são suficientes para inferência e reconstrução a partir dos dados de teste. Em imagens médicas, os autores discutem que esse nem sempre pode ser o caso, pois pacientes podem apresentar geometrias que não são usuais, bem como patologias como tumores que não estavam inclusas nos dados de treinamento;
\item A dificuldade de analisar e  interpretar os resultados de modelos de aprendizagem profunda, principalmente no que se refere ao entendimento teórico de seus resultados e garantias. Como exemplo, eles trazem o caso do \textit{Deep Image Prior} \cite{Ulyanov2020}, uma técnica de aprendizagem profunda cujo entendimento teórico ainda é escasso;
\item A criação de artefatos, principalmente no caso de modelos generativos. Os autores dizem que um modelo de aprendizagem profunda que mapeie diretamente das medidas para imagens pode gerar imagens realísticas com a ausência ou presença de características diferentes daquelas que seriam esperadas, como no caso da presença ou ausência de tumores em imagens médicas;
\item A identificação de modos de falha são difíceis de serem realizadas. Quando o operador direto é muito subdeterminado ou que os dados apresentam \textit{outliers}, pode-se obter imagens de alta qualidade, mesmo que elas estejam erradas. Mais uma vez seria importante ter modelos de aprendizagem profunda robustos à ruídos, \textit{outliers} ou  capazes de caracterizar as incertezas dos seus resultados. 
 \end{itemize}
Ao utilizar métodos integrados, nem todas essas limitações são resolvidas, já que são (por enquanto) inerentes ao se utilizar redes neurais profundas. É necessário ter clareza das vantagens que a proposta traz, bem como limitações e modos de falha. 

\subsubsection{Por que o \textit{design ad hoc} da arquitetura de ANNs é criticado?}\label{sec:vapnik2}

Ao longo dos últimos anos, o tamanho das redes neurais vem aumentando consideravelmente. Ainda em 2016, o número de amostras das bases de dados, o número de conexões de uma unidade e número total de unidades das redes neurais em si, já alcançavam números tão grandes quanto $10^9$ amostras, quase $10^4$ conexões por unidade e quase $10^7$ em uma única rede \cite[Figuras 1.8 e 1.10 e 1.11]{goodfellow2016deep}. Esse tamanho considerável só foi possível graças à disponibilidade de unidades centrais de processamento (CPU) mais rápidas, a unidades de processamento gráfico (GPU) de propósito geral, que podem realizar cálculos tipicamente conduzidos por CPUs, melhor conexão de \textit{internet} e melhores \textit{softwares} para sistemas de processamento distribuído \cite[pág. 20]{goodfellow2016deep}. Considerando que já é possível obter ANNs tão grandes, a escolha da quantidade de unidades e suas conexões deixa de ser pelas limitações de \textit{hardware} e passam a ser próprias do \textit{design} da rede \cite[pág. 24]{goodfellow2016deep}. 

Uma rede neural artificial apresenta muitas operações e não-linearidades. Calcular a sua saída é relativamente fácil. Difícil é responder se é possível escolher a arquitetura de maneira ótima, se e como ela depende de características do sinal de interesse. O teorema da aproximação universal argumenta que uma rede neural \textit{feedforward} com uma camada oculta pelo menos, com número suficiente de unidades e funções de ativação adequadas será capaz de representar uma ampla gama de sinais se os pesos da rede forem apropriados, mas não garante que o algoritmo de treinamento da rede será capaz de obter os parâmetros ótimos \cite[pág. 198]{goodfellow2016deep}. No sentido desse teorema, não há uma estrutura da rede precisa para representar o sinal, mas sim uma direção que se espera ter uma boa performance \cite[pág. 170]{goodfellow2016deep}. Mesmo entre pesquisadores de aprendizagem profunda, entende-se que não é possível prever qual arquitetura terá melhor performance \cite[pág. 192]{goodfellow2016deep}, apesar dessa etapa impor premissas sobre o mapeamento entrada-saída esperado \cite{2017kukacka}. 

O teorema ``não há almoço grátis'' traz o argumento que, na média para todas as distribuições geradoras de dados possíveis, todos os algoritmos de aprendizado de máquina apresentam a mesma taxa de erro quando classificam pontos que não foram observados na fase do treinamento. Autores como \cite[pág. 120]{goodfellow2016deep} interpretam esse teorema dizendo que nenhum algoritmo de aprendizado de máquina é universalmente melhor do que o outro, o que leva à necessidade de buscar um algoritmo que seja mais adequado para uma determinada tarefa específica. Na prática, há um \textit{design} manual da rede, por vezes baseado em tentativa e erro, guiado pelo conjunto de validação \cite[pág. 198, 295]{goodfellow2016deep}, ou mesmo pelos resultados experimentais obtidos.

Outras linhas de pesquisa em aprendizado de máquina fazem críticas à aprendizagem profunda. Em relação ao primeiro problema de seleção \cite{Vapnik2019}, é possível escolher de antemão a estrutura de $h( \cdot , \bm{\theta})$ e são muitas as possibilidades de representações, como funções lineares, polinomiais, árvores de decisão ou redes neurais de várias camadas. Assim, o algoritmo de aprendizagem pode organizar a busca nesse espaço de hipóteses escolhido tirando proveito da sua estrutura. No entanto, para autores do aprendizado estatístico, quando o pesquisador escolhe redes neurais de múltiplas camadas como o subespaço de funções admissíveis, isso não garante, teoricamente, que trará uma boa aproximação da função desejada.

\subsubsection{Por que a otimização no treinamento de ANNs  é suficiente, se não converge para mínimos globais?}\label{sec:otimi2}

A utilização de métodos baseados em gradientes não é infalível e em \cite[pág. 159]{nielsen2018}, o autor discute algumas dificuldades. Na retropropagação, os valores dos gradientes são calculados camada a camada. Uma coisa que pode acontecer é que as camadas aprendam em taxas muito diferentes uma das outras. No problema da dissipação do gradiente, os gradientes vão diminuindo ao longo das camadas de modo que, ao chegar na camada inicial, seus valores estão pequenos demais, o que causa instabilidades.

Outro desafio é que, além das estruturas apresentarem alta dimensionalidade, se a função objetivo é altamente não-linear e não-convexa em relação aos parâmetros, métodos de gradiente descendente provavelmente convergirão para um mínimo local, não para um mínimo global \cite{Adler2021}. Novamente, outras linhas de aprendizado de máquina fazem críticas à aprendizagem profunda. Em relação ao segundo problema de seleção do aprendizado estatístico \cite{Vapnik2019}, não haveria garantia teórica de que utilizando algoritmos iterativos para minimização da função custo será obtida a melhor função dentro do subconjunto, pois se obtém mínimos locais no lugar de mínimos globais.

Autores de aprendizagem profunda entendem que mínimos globais, ou mesmo locais, podem não ser obtidos \cite[págs. 192, 285]{goodfellow2016deep}. Mesmo assim, diminuir razoavelmente a função custo pode ser suficiente para determinadas aplicações \cite[págs. 23-6]{goodfellow2016deep}.

No treinamento de um funcional regularizado da Equação \eqref{eq:ERM4}, os valores da operação realizada pelo regularizador são calculados e somados diretamente na função custo \cite[Seção 7.1]{goodfellow2016deep}, \cite{2017kukacka}. O otimizador não precisa mudar, o código em si é pouco modificado, mas o resultado e seu significado são modificados \cite[pág. 106]{nielsen2018}. Tomando como exemplo o \textit{weigh decay}, regularização com norma $\ell_2$ da Equação \eqref{eq:ERM_RLS}, ela pode ser facilmente implementada pela soma de $\lambda^2 \vert \vert \bm{\theta} \vert \vert^2_2$ na função custo. A forma como ele influência o cálculo do gradiente é explicada em \cite[págs. 78-80]{nielsen2018}. 

A mesma lógica também é aplicada na utilização da norma $\ell_1$ no regularizador \cite[págs. 87-8]{nielsen2018}, que já não é diferenciável em todos os pontos, mas em trabalhos como \cite{Baguer2020} os valores numéricos da variação total foram calculados e somados no funcional, apoiados pela diferenciação automática do Pytorch. A otimização segue da mesma maneira, com o mesmo otimizador do gradiente descendente. As \textit{toolboxes} trazem facilidades na implementação de DNNs e na sua otimização. Por esse mesmo motivo, essa facilidade é alvo de críticas \cite{Cinelli2021}, já que, segundo os autores,  permite o uso generalizado de tais técnicas, sendo possível obter um modelo sem o conhecimento básico de DNNs ou mesmo de cálculo diferencial.

\subsubsection{Quais são as dificuldades provenientes da característica de caixa-preta das ANNs?}\label{Ap:interpretability}

A utilização de modelos preditivos busca automatizar e aumentar a acurácia dos resultados, para tentar ajudar ou mesmo substituir a subjetividade na tomada de decisão. No entanto, seja qual for a tarefa, um modelo deve ser robusto, confiável, compreensível e efetivo \cite{DelosReyes2016}  e ser capaz de explicar o funcionamento do algoritmo faz parte disso, principalmente em áreas sensíveis, como nas áreas da saúde, segurança, privacidade dos dados e tomadas de decisão \cite{Mertz2020, Zhang2020}.  

 Nos últimos anos, algumas preocupações demonstradas foram a tomada de decisão automática, porém opaca, de DNNs. Algoritmos caixa-preta tornam difícil entender  a lógica que levou a determinado resultado \cite[Capítulo 6]{geron2019hands-on},
\cite{Mertz2020}; o \textit{bias} que alguns algoritmos apresentam, como aqueles que reforçam desigualdades e esteriótipos \cite{neudert2020, Mertz2020}; como explicar e como ser robusto à ataques adversariais e ainda se as ANNs são suscetíveis à manipulações maliciosas, ainda pouco conhecidas \cite{Zhang2020}. 

É importante que novas propostas e métodos tragam ideias e \textit{insights} que permitam desenvolver algoritmos que produzam resultados bem definidos, sem deixar de discutir questões de reprodutibilidade, interpretabilidade e explicabilidade.

Reprodutibilidade parte da ideia que um experimento realizado sob as mesmas condições deve ter os mesmos resultados, independente do operador, do instante e do local onde ele é realizado. Em aprendizagem profunda, há autores que apontam que, mesmo partindo dos mesmos conjuntos de dados e utilizando supostamente os mesmos algoritmos, os resultados obtidos são diferentes, o que pode ser causado por erros experimentais, vieses nas publicações e usos impróprios de métodos estatísticos e das técnicas  \cite{Greengard2019}.   Em \cite{Belthangady2019} é discutido como a validação independente de resultados deveria ser natural nas áreas de aprendizado de máquina e aprendizagem profunda, mas a falta de consistência experimental na publicação de diretrizes pode impedir replicação de resultados entre os pesquisadores, bem como raramente é reportada a significância estatística das métricas de performance, por conta do alto custo de tempo e recursos associados à etapa de treinamento. 

A falta de reprodutibilidade pode ser amenizada com a descrição detalhada da metodologia e obtenção dos resultados. Boas práticas incluem a disponibilização dos códigos-fonte, dos pesos dos modelos treinados, dos conjuntos de dados e dos hiperparâmetros que foram utilizados para gerar os resultados  \cite{Belthangady2019, Greengard2019,  watson2021systematic}, além da arquitetura, da função custo e da métrica de performance, do otimizador, descrição das etapas de extração dos dados, pré-processamento e filtragem \cite{watson2021systematic}. É importante que todos os resultados sejam mostrados, mesmo os casos de falha \cite{Belthangady2019} e que haja transparência sobre financiamentos e filiações, para se evitar conflitos de interesses \cite{Greengard2019}.

Conseguir reproduzir os resultados de outros grupos de pesquisa é necessário, mas não resolve totalmente a questão. Muito do sucesso de aprendizagem profunda está relacionado com a sua performance nas mais diversas tarefas, mas se o desejo é utilizá-los em problemas do mundo real para suporte na tomada de decisões, duas outras características desejadas são a interpretabilidade e explicabilidade. Não há uma única definição delas, mas a ideia geral é que se deseja extrair informação da máquina de aprendizagem para se ter mais confiança nos resultados. A explicabilidade está mais relacionada com o racional utilizado na tomada de decisão, enquanto a interpretabilidade envolve dizer qual estrutura no modelo explica o seu funcionamento \cite{Escalante2018}. Saber como um determinado algoritmo chegou na resposta é de extrema importância para que haja confiança nos resultados. 
 
 Soluções baseadas em modelos físicos podem ser comunicadas e ensinadas, mas elas só podem explicar uma pequena variância, enquanto modelos de aprendizado de máquina explicam muita variância, mas são difíceis de serem comunicados \cite{Kording2018}. Especificamente, a utilização de ANNs pode apresentar dificuldades na interpretabilidade de como os resultados são gerados \cite{Belthangady2019}, sem trazer informações e \textit{insights} do problema que está sendo resolvido \cite{LewisD2019}. 
 
Assim, pode-se tentar investigar o funcionamento da caixa-preta \cite{Belthangady2019}, como tentar entender quais características dos dados de entrada são essenciais para predição do resultado; entender quais são as funções das camadas ocultas e separar suas contribuições com estudos de ablação; construir gráficos explanatórias hierárquicos entre as camadas; e propor arquiteturas que sejam interpretáveis pelo próprio \textit{design}. Existe uma área de pesquisa ativa na chamada inteligência artificial explicável que tem foco no desenvolvimento de soluções que tragam \textit{insights} sobre seu comportamento e tomadas de decisão com grande nível de detalhamento \cite{2018Gilpin}.

\subsubsection{Métodos integrados aumentam a interpretabilidade das soluções?}
Ampliando a discussão da Subseção \ref{Ap:interpretability} para métodos integrados, nos últimos anos, muitas das propostas encontradas na literatura trazem explicitamente que a união entre abordagens de modelos físico-matemáticos e aprendizagem profunda ajuda na interpretabilidade e/ou na robustez das soluções:
\begin{itemize}
	\item  Em \cite{He2020} os autores dizem que o método proposto para deconvolução semi-cega, na qual os parâmetros da PSF são estimados utilizando ANNs, aumentam a robustez e generalidade na restauração de imagens;
	\item  Em \cite{Zhang2019plug}, os autores dizem que o método \textit{plug-and-play} com ANNs que eles propõe é altamente interpretável e depende de menos treinamento do que outros métodos com etapas de aprendizagem;
	\item Após dizer que soluções de ponta a ponta são dificilmente interpretáveis, os autores de \cite{Fang2020} dizem que a proposta de obter informação \textit{a priori} utilizando CNNs combinadas com o modelo variacional tradicional é capaz de obter resultados confiáveis, explicáveis e de estado da arte de modo eficiente e mantendo a interpretabilidade dos resultados; 
	\item Em \cite{Monga2021}, os autores consideram que métodos de \textit{deep unrolling} são comparáveis a interpretabilidade de algoritmos iterativos baseados em modelo. Eles dizem que tais algoritmos não apresentam generalização tão alta, mas são mais eficientes e apresentam melhor performance; 
	\item Em \cite{Dardikman20} os autores argumentam que \textit{deep unrolling} permite a inclusão de conhecimento de domínio em métodos \textit{data-driven}, que apresentariam maior interpretabilidade e robustez em relação aos métodos caixa-preta e ainda podem generalizar bem;
	\item Em \cite{Yang2020} são descritas que as soluções propostas, baseadas em ANNs, promovem soluções que possuem significados físicos;
	\item Em \cite{kadkhodaie2021solving}, os autores treinam uma CNN para realização de \textit{blind denoising}. Eles desenvolvem uma forma de obter amostras desse \textit{prior} implícito e de utilizá-las na solução de problemas inversos lineares gerais. Os autores consideram que o algoritmo é interpretável, pois, segundo os autores, parte da relação explícita de Miyasawa entre o mapeamento e o \textit{prior};
	
	\item Em \cite[pág. 4]{Li_2020}, os autores argumentam que sua proposta separa o que é relativo ao ruído e o que é relativo à informação \textit{a priori} das incógnitas, o que permite incluir conhecimento do mecanismo de geração dos dados. Segundo os autores, isso aumenta a interpretabilidade da proposta.
	\end{itemize}
Nos trabalhos citados, essas características de interpretabilidade aparecem como consequência natural das propostas. Porém, para que uma rede neural seja considerada interpretável, é necessário que ela passe por uma avaliação, como realizado em \cite{Margot2021, Meng2022, Molnar2022}. Existem propostas de redes neurais interpretáveis \cite{Du2019}, mas mesmo estas são passíveis de críticas \cite{Zhang2020}. Logo, deve-se verificar se as ferramentas existentes são suficientes ou se são necessárias novas ferramentas para poder afirmar que um método é interpretável. Também é importante que se defina bem o que se entende por explicabilidade e  interpretabilidade. Há trabalhos que relacionam esses conceitos com clareza, completude e parcimônia \cite[Figura 1]{Markus2021}, classificam métodos explicáveis \cite[Tabela 1]{Markus2021} e desenvolvem um  \textit{framework} com recomendações para eles \cite[Figura 2]{Markus2021}. 

Essa discussão leva às perguntas que podem dar base à uma revisão sistemática: 
\begin{itemize}
\item Os conceitos de robustez, interpretabilidade, reprodutibilidade e explicabilidade foram definidos? 
\item Como os autores de trabalhos que combinam ambos os paradigmas abordam tais conceitos em suas propostas? Caso esses conceitos sejam citados, os autores mostram como eles foram avaliados?
\item A integração de problemas inversos, métodos de regularização e aprendizagem profunda resulta, necessariamente, em uma melhor interpretabilidade, explicabilidade e reprodutibilidade das soluções?  Se assim for, como ela acontece e como isso pode ser mostrado?
\end{itemize}

Há também uma discussão sobre limitação do uso de métodos de regularização no contexto de generalização em aprendizagem profunda. Em   \cite{2021Zhang}, os autores questionam ser apenas dois fatores (risco empírico e controle da capacidade) os responsáveis pela generalização. Eles argumentam que utilizar regularização, implícita ou explícita, pode ajudar a generalização da máquina de aprendizagem, mas que a regularização provavelmente não é a sua razão fundamental. Nos seus resultados experimentais, as redes neurais testadas pelos autores apresentaram bons resultados mesmo após remoção dos regularizadores. Assim, pergunta-se:
\begin{itemize}
\item A interpretabilidade, explicabilidade e reprodutibilidade ajuda a entender quais são os fatores para promoção de generalização dentro de aprendizagem profunda?
\end{itemize}

  Responder todas essas perguntas pode permitir um desenvolvimento significativo, tanto pela possibilidade do desenvolvimento de novos métodos de reconstrução quanto por tais métodos poderem trazer maior confiança nos resultados.

\newpage

\section{QUAIS SÃO AS DIFERENTES INTERPRETAÇÕES DE REGULARIZAÇÃO?}\label{sec:polysemy}

\subsection{Introdução}
O conceito de regularização no sentido de Tikhonov foi apresentado nos capítulos iniciais. O Capítulo \ref{sec:illposed} descreveu problemas inversos mal-postos, o Capítulo \ref{sec:variational} descreveu a regularização variacional como uma forma de resolvê-los e o Capítulo \ref{sec:integrated} discutiu como regularização pode diminuir o \textit{overfitting} em aprendizagem profunda, bem como foram apresentadas propostas para unir esses dois pontos de vista. 

Contudo, é possível desenvolver mais ainda o significado da palavra regularização, objetivo deste capítulo. De início, o uso popular da palavra pode trazer \textit{insights} para a discussão. No Michaelis Dicionário Brasileiro da Língua Portuguesa, encontra-se:
\begin{itemize}
		\item \textbf{Regularização}: \textit{Ato ou efeito de regularizar(-se)}\footnote{Retirado de \url{https://michaelis.uol.com.br/palavra/e3wkP/regularização/}.}.
	\item \textbf{Regularizar}: \textit{Tornar regular; voltar à normalidade}\footnote{Retirado de \url{https://michaelis.uol.com.br/palavra/G93dp/regularizar/}.}.
	\item \textbf{Regular}: \textit{Impor ordem ou moderação}; \textit{Seguir ou fazer seguir uma certa orientação ou norteamento}\footnote{Retirado de \url{https://michaelis.uol.com.br/palavra/4b0Bx/regular-2/}.};\textit{Situado entre dois extremos; Que é razoável}\footnote{Retirado de \url{https://michaelis.uol.com.br/palavra/m8Bdn/regular-1/}.}.
\end{itemize}
 Ou seja, busca-se \textit{retornar à normalidade} (em analogia com um problema bem-posto) \textit{seguindo uma orientação}, \textit{impondo uma ordem} (em analogia com restringir o espaço de soluções de acordo com informações \textit{a priori}). 

Regularização sob o ponto de vista de Tikhonov possui um significado próprio \cite{tikhonov1977solutions}, mas também pode assumir diferentes significados dependendo da área em que ela é abordada \cite{Chen2002}. É interessante notar que o desenvolvimento do método de regularização aconteceu em paralelo com a solução de outros problemas. Partindo de contextos diferentes, existem soluções que apresentam semelhanças com a regularização clássica de Tikhonov da Equação \eqref{eq:tiksol} (Subseção \ref{sec:tikhclas}):

\begin{itemize}

\item No Apêndice \ref{Ap:estimador}, discute-se a inversão estatística e quais são as hipóteses para obter uma solução análoga;

\item No Apêndice \ref{App:svd2}, obtém-se uma forma da regularização de Tikhonov a partir da decomposição de valores singulares do operador direto;

\item No Apêndice \ref{AP:Wiener}, discute-se como a regularização clássica de Tikhonov pode ser feita no domínio da frequência e como isso se relaciona com o filtro de Wiener;

\item No Apêndice \ref{Ap:Ridge}, discute-se o a multicolinearidade em problemas de regressão e quais as semelhanças da regressão de Ridge com a regularização clássica de Tikhonov.
\end{itemize}

Nesse sentido, as diferentes propostas compartilham ferramentas matemáticas semelhantes, mas que são justificadas de acordo com seus problemas originais. Isso permite que soluções do tipo da Equação \eqref{eq:tikhonov1} sejam interpretadas de diferentes maneiras dependendo do contexto em que elas se inserem. É razoável considerar que as palavras adquirem significado em função das condições em que são utilizadas. Apesar de tal fato parecer evidente no que se refere ao emprego da linguagem verbal comum, pode-se discutir esse fenômeno na construção do conhecimento científico. 

Em linguagem natural, é comum associar uma palavra com dois ou mais significados que são relacionados entre si, isto é, polissemia \cite{Falkum2015, Vicente2017}. Existem conceitos polissêmicos no vocabulário acadêmico \cite{Skoufaki2021}, como campo \cite{Krapas2008}, calor \cite{Strmdahl2011}, processo \cite{Hyland2007} e solução \cite{Mudraya2006}, para citar alguns. Significados preferenciais podem ocorrer em disciplinas específicas \cite{Hyland2007}, sendo necessário definir o contexto da sua utilização. 

No presente capítulo são revisados diferentes significados de regularização. O objetivo não é desenvolver uma taxonomia, formalmente delimitando-as, mas sim usar vocabulário específico das áreas para descrever regularização para, quando possível, relacioná-las. A definição original é comparada com outros usos da palavra regularização. Algumas dessas interpretações apresentam elevado grau de subjetividade. Mesmo assim, essa diferenciação é importante para deixar claros quais são os pressupostos levados em conta na solução de um problema. Até porque elas acabam sendo utilizadas em conjunto (Por exemplo, ver \cite[pág. 10]{Kaji2019}), mesmo que elas tenham origem em contextos diferentes.

\subsection{O que é regularização no sentido de Tikhonov?}\label{sec:senseof}

A teoria de regularização é a teoria que busca propor e analisar métodos para obter soluções estáveis de problemas mal-postos \cite{bleyer2015novel, Burger2021}. Existem diferentes usos de regularização no sentido de Tikhonov e deve-se diferenciar o que é método de regularização, estratégia de regularização, operador de regularização e funcional de regularização/termo de regularização/regularizador \cite[Seção 4.3]{Bertero2021}. 

Retomando a Equação \eqref{eq:canonical}, Andrey Tikhonov provou em 1943 um lema topológico que dizia que, mesmo em um problema mal-posto, se as soluções $f$ forem restritas a um conjunto compacto\footnote{Uma definição de conjunto compacto pode ser encontrada em \cite[pág. 55]{Shurman2016}} o mapeamento inverso será contínuo e a solução será estável \cite[pág. 29]{tikhonov1977solutions} \cite[pág. 418]{Vapnik2006}. Em outras palavras, isso significa que, com o conhecimento \textit{a priori} de que a solução pertencer a esse conjunto compacto, é possível resolver o problema. Apesar desse lema indicar a possibilidade de obter uma solução estável, naquele momento Tikhonov não descreveu um método para isso \cite[pág. vii] {Morozov1984}. 

Como conta Vapnik \cite[pág. 418]{Vapnik2006}, 20 anos se passaram para que fossem propostas soluções viáveis de problemas mal-postos e muitos autores contribuíram para isso \cite[págs. 60, 203-4]{hansen2010discrete}, \cite[Introdução]{Morozov1984}. No entanto, houve um destaque para o trabalho do próprio Tikhonov, que, em 1963, elaborou o método de regularização \cite[pág. 46]{tikhonov1977solutions}, bem como criou um \textit{framework} com justificativas teóricas para tal. Tikhonov chamou de método de regularização o método de construir soluções estáveis para $\mathcal{A} f = g$, Equação \eqref{eq:canonical}, utilizando um operador de regularização $\mathcal{R}_{\lambda}$ (que será definido na Subseção \ref{sec:abstractmethod}) e um parâmetro de regularização $\lambda$ \cite[pág. 48]{tikhonov1977solutions}. A ideia geral deste método e a relação com outras propostas da época são descritas a seguir. 

\subsection{Historicamente, como que a regularização variacional surgiu?}\label{sec:history}

Métodos de regularização podem ser obtidos a partir de métodos variacionais, na qual o problema inverso é tratado como um problema de otimização. Essa proposta foi apresentada diretamente no Capítulo \ref{sec:variational}, mas aqui há a contextualização histórica.

Antes do método de regularização de Tikhonov, em 1962, V. K. Ivanov propôs uma forma variacional de resolver problemas mal-postos que ficou conhecido como método das quase-soluções \cite{ivanov2002}. Para descrevê-lo, primeiro é necessário definir um funcional $\Omega(f)$ e suas propriedades \cite[págs. 418-9]{Vapnik2006}, tendo como base a notação da Equação \eqref{eq:canonical}:
\begin{itemize}
\item $\Omega(f) \geq 0$ e ele é definido em um subconjunto $M$ do espaço de soluções $F$;
\item O conjunto de elementos $f \in M$ que satisfaz a restrição $\Omega(f)\leq c$ deve ser um subconjunto compacto e convexo de $M$, $\forall c \geq0$;
\item A solução $f_0$ pertence ao domínio da definição de $\Omega(f)$, ou seja, a solução deve pertencer a um conjunto compacto $\Omega(f_0)\leq c_0$, onde $c_0 > 0$ pode ser desconhecido. 
\end{itemize}
Funcionais $\Omega(\cdot)$ que apresentassem essas propriedades eram chamados de \textit{stabilizing functionals} \cite[pág. 51]{tikhonov1977solutions}. Retomando a notação de problemas inversos discretos, a regularização de Ivanov era escrita como um problema de otimização dado por
\begin{equation}
\hat{\mathbf{x}} = \underset{\mathbf{x}}{\arg\min} \vert \vert \mathbf{A}\mathbf{x} - \mathbf{y} \vert \vert^2_2\quad \text{s.t.}\quad \Omega(\mathbf{x}) \leq \bm{\delta}^2_1,
\label{eq:ivanov}
\end{equation}
onde $\bm{\delta}_1$ é uma constante que atua como um limitante superior de $\Omega(\mathbf{x})$. A solução $\mathbf{x}$ obtida a partir dessa restrição do espaço seria uma quase-solução \cite[pág. 19]{baumeister2005topics}, de onde vem o nome do método. 

Em 1963, Tikhonov \cite[pág. 57]{tikhonov1977solutions} utilizou o método dos multiplicadores de Lagrange a partir da formulação com restrições da Equação \eqref{eq:ivanov} e obteve uma formulação sem restrições, conforme 
\begin{equation}
\hat{\mathbf{x}} = \arg\min\limits_{\mathbf{x}} \left[ \vert \vert \mathbf{A} \mathbf{x} - \mathbf{y} \vert \vert^2_2 + \lambda^2 \Omega(\mathbf{x}) \right].
\label{eq:tikhonov_original}
\end{equation}
Dado que o problema fundamental do cálculo de variações é a obtenção de extremos de funcionais \cite[pág. 3]{komzsik2020applied}, máximos ou mínimos, a Equação \eqref{eq:tikhonov_original} obtida a partir de multiplicadores de Lagrange pode ser entendida como um método variacional \cite[Teorema 5.1]{engl1996regularization}, cuja definição e propriedades estão em \cite[págs. 20-9]{Benning2018}, ou regularização variacional.

Nesse contexto, $\Omega(\mathbf{x})$ ficou conhecido como termo de regularização, funcional de regularização \cite[pág. 8]{Benning2018}, ou regularizador \cite[pág. 1129]{Caselle2011}. Ele mede a regularidade da solução \cite[págs. 61, 172]{hansen2010discrete} e sua adição permitiu a obtenção de soluções estáveis em relação a pequenas variações nos dados iniciais, conforme desejado \cite[págs. 22, 47]{tikhonov1977solutions}. 

O resultado de $\Omega(\mathbf{x})$ é sempre positivo ou nulo. Quanto maior o valor deste termo durante a otimização, mais ele é penalizado, o que também depende da escolha de $\lambda$ que dá peso a esse termo. Um exemplo é comparar as soluções suavizadas obtidas com $\Omega(\mathbf{x}) = \vert\vert \mathbf{x} \vert\vert_2^2$ das soluções esparsas obtidas com $\Omega(\mathbf{x}) = \vert\vert \mathbf{x} \vert\vert_1$. Cada escolha de $\Omega(\mathbf{x})$ penaliza as soluções indesejadas de formas diferentes e sua escolha depende da aplicação. Durante a implementação, os problemas de otimização resultantes podem não apresentar soluções de forma fechada, o que também deve ser considerado na escolha de $\Omega(\mathbf{x})$. 

A partir da proposta do método de regularização de Tikhonov, foi definido um problema bem posto\footnote{Há trabalhos como \cite[Capítulo 3]{Dontchev1993} que trazem a definição de um problema mal-posto genérico.} no sentido de Tikhonov \cite[Capítulo 1]{Dontchev1993}, \cite[pág. 4]{Lavrentiev1967}, \cite[págs. 30-1]{tikhonov1977solutions}, também conhecido como bem-posto condicionalmente \cite[pág. 16]{glasko1988inverse}: 
\begin{itemize}
\item Quando se sabe \textit{a priori} que existe uma solução da Equação \eqref{eq:canonical} para uma classe de dados e que ela pertence a um conjunto compacto $M$; 
\item Que nesse conjunto compacto $M$ a solução é única; 
\item Que pequenas perturbações arbitrárias em $g$, que não tirem a solução $f$ do conjunto compacto $M$, correspondem à pequenas variações arbitrárias na solução $f$.
\end{itemize}

Por completude, em 1967, V. A. Morozov estudou uma terceira forma também na forma de otimização com restrições, conforme
\begin{equation}
\hat{\mathbf{x}} = \underset{\mathbf{x}}{\arg\min}\Omega(\mathbf{x}) \quad \text{s.t.} \quad \vert \vert \mathbf{A}\mathbf{x} - \mathbf{y} \vert \vert^2_2 \leq \bm{\delta}^2_2,
\label{eq:morozov}
\end{equation}
onde $\bm{\delta}_2$ é um limitante superior do erro residual. Esse método ficou conhecido como método dos residuais, ou método de discrepância de Morozov e foi bastante estudado no contexto de critérios para a escolha de $\lambda$ na regularização de Tikhonov \cite[págs. X, 32]{Morozov1984}. 

É possível mostrar que, em determinadas condições, é possível escolher o parâmetro de regularização $\lambda$, $\bm{\delta}_1$ e $\bm{\delta}_2$ de modo que os dois problemas de otimização com restrições sejam equivalentes ao método de Tikhonov sem restrições \cite[págs. 2793-4, definição 6]{Chen2002}, \cite[págs. 63-4]{hansen2010discrete}, \cite{Oneto2016}. 

Por outro lado, há casos em que essa equivalência não se verifica como mostrado em \cite{Kaltenbacher2018}, o que mostra que os métodos não são exatamente iguais. Nesse mesmo trabalho, os autores argumentam que, possivelmente, o método de Tikhonov foi mais utilizado no passado por ser um problema de otimização sem restrições, mas que, atualmente, muitas técnicas para problemas de otimização com restrições já foram desenvolvidas e podem ser utilizadas nas formulações de Ivanov e Morozov. 

Por fim, apesar da Equação \eqref{eq:tiksol} ser a forma de regularização de Tikhonov mais conhecida, o caso mais geral da modelagem variacional é dado por 
\begin{equation}
\hat{\mathbf{x}} = \arg\min\limits_{\mathbf{x}} \left[ \mathcal{L}(\mathbf{A}\mathbf{x}, \mathbf{y}_{\bm{\delta}}) + \lambda^2 \Omega(\mathbf{x}) \right],
\label{eq:tikhonovgen}
\end{equation}
onde $\mathcal{L}$ é a função de perda, também chamado de termo de fidelidade, definido para medir uma distância entre $\mathbf{A}\mathbf{x}$ e $\mathbf{y}_{\bm{\delta}}$ \cite{Benning2018}. É usual utilizar a função de perda quadrática, mas existem outras opções como utilizar a norma $\ell_1$ \cite{Bertero2021}. 

\subsection{Qual é a forma mais geral do método de regularização?} \label{sec:abstractmethod}

Em \cite{Chen2002}, os autores argumentaram que existem três princípios para implementar regularização: Navalha de Occam, a descrição de comprimento mínimo e a mínima entropia. Os autores também descreveram como diferentes áreas do conhecimento poderiam estabelecer o \textit{framework} matemático do princípio de regularização: teoria bayesiana, a teoria da informação e a teoria do aprendizado estatístico. Além da forma variacional discutida anteriormente, métodos de regularização também podem ser construídos a partir de diferentes paradigmas, como suavização dos dados, métodos iterativos \cite{Benning2018} ou métodos aprendidos \cite[págs. 3-4]{Burger2021}. Ou seja, regularização pode ser vista a partir de diferentes origens.

De modo mais geral, é possível englobá-los no chamado método de regularização abstrato \cite[Seção 4.1]{Benning2018}, que descreve quais são as propriedades esperadas para um método de regularização qualquer. 

Seja $\bm{\delta}$ o vetor parâmetro que caracteriza o ruído no lado direito $\mathbf{y}_{\bm{\delta}}$ da Equação \eqref{eq:eq7}. Em \cite[pág. 46]{tikhonov1977solutions}, Tikhonov e Arsenin argumentam que se torna natural definir uma solução $\mathbf{x}$ com a ajuda de um operador de regularização $ \mathcal{R}_{\lambda}(\cdot)$, que depende de um parâmetro $\lambda(\bm{\delta})$ escolhido de acordo com $\bm{\delta}$ . Esse operador deve ser capaz de realizar o mapeamento do espaço dos dados para o espaço dos parâmetros, obtendo soluções regularizadas $\mathbf{x} = \mathcal{R}_{\lambda}(\mathbf{y}_{\bm{\delta}})$ que sejam estáveis à pequenas variações \cite[pág. 175]{alvarez2017digital}. Caso o operador $ \mathcal{R}_{\lambda}(\cdot)$ seja linear em relação aos parâmetros $\mathbf{x}$ que se deseja estimar, o método de regularização é dito linear \cite[Definição 4.2]{Benning2018}, \cite[pág. 51]{engl1996regularization}. Este é o caso que será discutido. 

Novamente escrevendo em termos de problemas inversos discretos, as propriedades esperadas para um método de regularização são \cite[pág. 5]{Daubechies2016}, \cite[pág. 50]{engl1996regularization}, \cite[pág. 48] {Mueller2012}:
\begin{itemize}
\item Seja o parâmetro de regularização $\lambda$ positivo e finito. Então 
\begin{equation}
\lim_{\lambda \to 0} \mathcal{R}_{\lambda}(\mathbf{A} \mathbf{x}) = \mathbf{x}_{exato},
\label{eq:limite}
\end{equation}
qualquer que seja $\mathbf{x}_{exato}$ pertencente ao espaço dos modelos. 

A família dos mapas lineares ${R}_{\lambda}$ é chamada de estratégia de regularização. Sabendo que sem regularização ($\lambda = 0$) não é possível resolver o problema, a Equação \eqref{eq:limite} indica que a solução obtida com o método de regularização deve tender à solução exata $\mathbf{x}_{exata}$ na medida que o ruído $\bm{\delta} \rightarrow 0$. Alguns autores identificam ${R}_{\lambda}$ com essa propriedade de esquema linear de regularização e escrevem que, quando $\lambda=0$, $\mathcal{R}_{\lambda = 0}$ seja a inversão ingênua de $\mathbf{A}$, isto é, que $\lim_{\lambda \to 0} \mathcal{R}_{\lambda}(\mathbf{A}) = \mathbf{A}^{-1}$ \cite[pág. 58]{Neto2005};

\item Assumindo que se conhece o nível do ruído $\bm{\delta}$ tal que $ \vert \vert \mathbf{y}_{medido} - \mathbf{y}_{exato} \vert \vert_2 \leq \bm{\delta}$, uma escolha de parâmetro de regularização $\lambda = \lambda(\bm{\delta})$ é admissível se $\lambda(\bm{\delta}) \rightarrow 0$ na medida que $\bm{\delta} \rightarrow 0$. Ou seja, dado que o próprio $\lambda$ é relacionado com $\bm{\delta}$, a partir do momento que $\bm{\delta} \rightarrow 0$, $\lambda(\bm{\delta}) \rightarrow 0$. Em outras palavras, é necessária uma menor regularização quando o ruído é menor. Partindo de um esquema de regularização e escolhendo $\lambda$ dessa forma, alguns autores denominam como estratégia de regularização \cite[pág. 64]{Neto2005}, enquanto outros dizem que essa propriedade define uma escolha admissível de $\lambda$ \cite[pág. 48] {Mueller2012}. Vale notar que alguns autores não assumem convergência para o parâmetro de regularização \cite[pág. 24]{Benning2018}; 
 
\item O erro de reconstrução $\vert \vert \mathcal{R}_{\lambda}(\mathbf{y}_{\bm{\delta}}) - \mathbf{x}_{exato} \vert \vert_2$ deve desaparecer assintoticamente a medida que $\bm{\delta} \rightarrow 0$. 
\end{itemize}

Se o operador $\mathcal{R}_\lambda$ apresentar essas propriedades, ele é um método de regularização convergente para a equação $\mathbf{A} \mathbf{x} = \mathbf{y}$ \cite[pág. 50]{engl1996regularization}, \cite[pág. 48]{Mueller2012}. Nestas propriedades, muito se discute a influência de $\bm{\delta}$ e pouco dos erros $\bm{\epsilon}$ do modelo $\mathbf{A}$. Em muitos casos admite-se que $\mathbf{A}$ é conhecido. No entanto, $\mathbf{A}$ é muitas vezes imperfeito e pode ter o seu próprio erro, como a discretização utilizada ou as hipóteses do próprio modelo, não relacionado com $\bm{\delta}$. 

O objetivo se torna construir operadores $\mathcal{R}_\lambda$ \cite[págs. 50-62]{tikhonov1977solutions} e desenvolver regras para escolhas de $\lambda$ de uma forma ótima \cite{engl1996regularization},com melhores propriedades de estabilidade em respeito à pequenas mudanças nos dados iniciais \cite[pág. 1]{Benning2018}, \cite[pág. 22]{tikhonov1977solutions}. Também há interesse em estimativas de estabilidade e taxas de convergência, a taxa em que a solução regularizada se aproximada da solução exata na medida em que $\bm{\delta} \rightarrow 0$ \cite{Bertero2021, Daubechies2016}. Especificamente, o método variacional é um método convergente de regularização \cite{Benning2018}.

\begin{figure}[htpb]
\centering
\includegraphics[width=0.66\columnwidth]{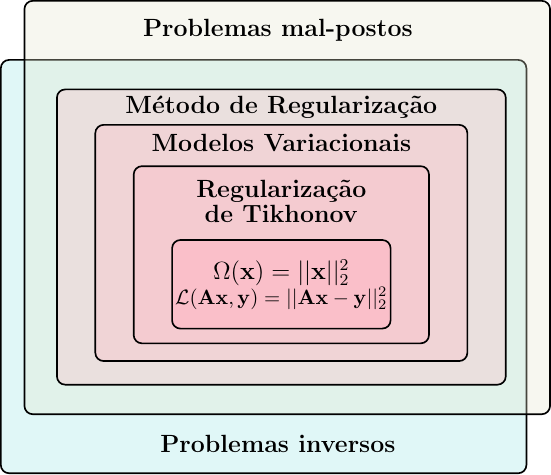} 
\caption[Relação dos diferentes níveis de regularização.]{Relação dos diferentes níveis de regularização. Fonte: Próprio autor.}
\label{fig:01_venn}
\end{figure}

Nesse contexto, ressalta-se que a regularização clássica de Tikhonov da Equação \eqref{eq:tiksol} é apenas um caso possível do método de regularização, como mostra a Figura \ref{fig:01_venn}.

\subsection{Como descrever diferentes interpretações de regularização?} \label{sec:III}

Através dos anos, regularização foi associada a novos significados, nem sempre contraditórios, mas diferentes ao sentido de Tikhonov. A seguir, dez interpretações diferentes de regularização são descritas, destacando-se palavras-chave (em \textit{itálico}) de cada uma e as equações mais relacionadas. Alguns desses significados são entendidos a partir de formas diferentes de implementação da própria regularização de Tikhonov, enquanto outros são soluções novas. Ao mesmo tempo, algumas interpretações são relacionadas com capítulos passados do texto, enquanto outras são desenvolvidas nos apêndices ou nesta própria seção. 

Nota-se que existem intersecções entre elas e palavras-chave poderiam fazer parte de outra interpretação. Com o objetivo mais didático do que uma delimitação formal, foi necessário um grau de arbitrariedade para realizar essa separação.

\subsubsection{Interpretação $1$ (Problemas inversos)}

Esta interpretação é relacionada com as Subseções \eqref{sec:senseof} e \eqref{sec:abstractmethod} do método de regularização abstrato. Se um problema é \textit{mal-posto}, o \textit{método de regularização} busca obter soluções aproximadas, únicas e estáveis, de tal problema, através da sua \textit{substituição por problemas bem-postos vizinhos} \cite{Vapnik2006, Woo2012}. É necessário desenvolver \textit{famílias de operadores de regularização contínuos} $\mathcal{R}_{\lambda}$ \cite{engl1996regularization}, parametrizadas por $\lambda$ \cite{Bertero2021, bleyer2015novel}, de acordo com o método de regularização abstrato dado pela Equação \eqref{eq:limite} e suas propriedades \cite[pág. 48]{Mueller2012}. Tal $\mathcal{R}_{\lambda}$ atua como um \textit{operador inverso aproximado}, ou apenas \textit{inversa regularizada} \cite[Seção 3.4]{Mueller2012}.

\subsubsection{Interpretação $2$ (Método variacional)}

Esta interpretação é relacionada com a Subseção \eqref{sec:history} da regularização variacional. O \textit{lema topológico} demonstrado por Tikhonov, as propriedades esperadas para $\Omega(\mathbf{x})$, junto com o método variacional das Equações \eqref{eq:tikhonov_original} e \eqref{eq:tikhonovgen}, permitem entender a regularização de Tikhonov como uma forma de \textit{restrição do espaço de soluções} de um problema mal-posto. Uma forma implementável de um método de regularização abstrato pode ser obtida com \textit{modelos variacionais}, que consistem de um \textit{termo de fidelidade}, um ou mais \textit{funcionais de regularização} $\Omega(\mathbf{x})$ e seu(s) respectivo(s) \textit{parâmetros de regularização} $\lambda$. 

O termo aditivo $\Omega(\mathbf{x})$ mede a regularidade, a compacidade \cite{Burger2021} ou outras características desejadas da solução $\mathbf{x}$ \cite{Arridge2019, hansen2010discrete}. O regularizador indica como $\mathbf{x}$ deve ser antes do conhecimento de medidas $\mathbf{y}_{\bm{\delta}}$ disponíveis \cite{bleyer2015novel}. Assim, a definição de $\lambda$ faz o balanço entre o ajuste de $\mathbf{A}\mathbf{x}$ aos dados $\mathbf{y}_{\bm{\delta}}$ e a estabilidade da solução $\mathbf{x}$ \cite{aster2019parameter}. Retomando a Seção \ref{sec:abstractmethod}, é possível mostrar que a regularização variacional da Equação \eqref{eq:tikhonov_original} define um operador de regularização \cite[Seção 5]{Benning2018}.

\subsubsection{Interpretação $3$ (Métodos de penalização)}

Se originalmente Tikhonov argumentava pela restrição da solução em conjuntos compactos, a partir do momento em que foi utilizado o método variacional para a solução do problema, outros conceitos puderam ser relacionados. O procedimento de usar \textit{multiplicadores de Lagrange} para obter um problema de \textit{otimização sem restrições} a partir de um problema de \textit{otimização com restrições} de modo geral. Especificamente, ele é utilizado para obter a regularização de Tikhonov, Equação \eqref{eq:tikhonov_original}, a partir da regularização de Ivanov, Equação \eqref{eq:ivanov}, ou da regularização de Morozov, Equação \eqref{eq:morozov}.

Este procedimento advém dos \textit{métodos de penalização} dentro da área de otimização. Nesse contexto, $\Omega(\mathbf{x})$ é também chamado de \textit{termo de penalização}, ou função de penalização \cite[pág. 398]{luenberger2015linear}. O termo de penalização atua como uma restrição suave, no sentido de que seus resultados numéricos servem como medida de quanto as restrições não são respeitadas. Isso é diferente de restrições rígidas, que impedem que o algoritmo siga determinado caminho, por exemplo no caso de positividade que não permite valores negativos \cite[pág. 20]{andreasson2020an}. Em resumo, regularização é um conjunto de abordagens para desencorajar soluções com grande complexidade ou extremas em problemas de otimização \cite[Exemplo 8.3]{Deisenroth2020}.

Muitas vezes o termo de penalização é utilizado de forma intercambiada com o termo de regularização \cite[pág. 263]{Deisenroth2020}, com a ressalva de que apenas o método de regularização surgiu a partir do trabalho de Tikhonov. Essa interpretação também é encontrada em aprendizado de máquina \cite[pág. 70]{cherkassky2007learning}.

Para direcionar a solução \textit{penalizando as soluções indesejadas}, diferentes formas para a função de penalização são encontradas (ver \cite[pág. 404]{luenberger2015linear} para alguns exemplos). No caso da regularização clássica de Tikhonov, ela é definida pela escolha de $\Omega(\mathbf{x}) = \vert \vert \mathbf{x} \vert \vert_2^2$ e pode ser entendida como uma forma de restrição no tamanho de $\mathbf{x}$. A saída desse regularizador é sempre positiva e como $\vert \vert \mathbf{x} \vert \vert_2^2$ é crescente, quanto maior o valor desse termo, mais ele seria penalizado durante a otimização. Ele também pode ser mais ou menos penalizado a partir da escolha de $\lambda$, que tem o papel de dar o peso para o regularizador. Ao final, é necessário resolver o problema de otimização que inclui o regularizador, na qual se espera que apresente melhor convergência do que sem ele \cite{Kaji2019}.

\subsubsection{Interpretação $4$ (Inversão estatística)}

Esta interpretação é relacionada com o Apêndice \ref{Ap:estimador}. Para resolver problemas inversos mal-postos, a teoria da \textit{inversão estatística} trata todas as variáveis como \textit{variáveis aleatórias}. Seja $\mathbf{y}$ a quantidade observável (\textit{realização}) e seja $\pi(\mathbf{x})$ a \textit{densidade de probabilidade} de $\mathbf{x}$. Resolver o problema inverso na abordagem bayesiana significa estimar a densidade de probabilidade \textit{a posteriori} $\pi_{post}(\mathbf{x})$ \cite{kaipio2005statistical} utilizando o \textit{teorema de Bayes} da Equação \eqref{eq:esti14} para relacionar as seguintes variáveis:
\begin{itemize}
\item $\pi_{priori}(\mathbf{x})$ é a densidade \textit{a priori} de $\mathbf{x}$, que codifica informações da solução $\mathbf{x}$ antes do conhecimento de qualquer dado $\mathbf{y}$; 
\item Os dados consistem de valores observáveis $\mathbf{y}$ que apresentam $\pi\left(\mathbf{y} \right)>0$; 
\item $\pi\left(\mathbf{x} \vert \mathbf{y} \right)$ é a densidade de probabilidade condicional de $\mathbf{x}$ em relação a $\mathbf{y}$. 
\end{itemize}

A inversão bayesiana permite a inclusão de \textit{informação a priori}, ou simplesmente \textit{priors}, sobre a solução de forma explícita a partir da definição de $\pi_{priori}$ \cite{kaipio2005statistical}. Enquanto $\pi_{priori}$ é relacionado com o regularizador $\Omega(\mathbf{x})$, a escolha da \textit{função de verossimilhança} $\pi\left(\mathbf{y} \vert \mathbf{x} \right)$ é relacionada ao modelo de ruído (quando $\bm{\delta}$ é aditivo) e resulta em diferentes funções de perda $\mathcal{L}$ \cite{kaipio2005statistical}. 

 Depois de definir essas quantidades, é necessário desenvolver \textit{estimadores} para obter $\pi_{post}(\mathbf{x})$. Um exemplo é o estimador máximo \textit{a posteriori} (MAP) da Equação \eqref{eq:esti15}. Um caso para exemplificar a forma do MAP é considerar que tanto o modelo de \textit{prior} quanto o de ruído são uma forma de ruído branco gaussiano, é possível obter a Equação \eqref{eq:tiksol} a partir da Equação \eqref{eq:esti15}, exceto possivelmente por um fator de escala que pode ser englobado no parâmetro de regularização $\lambda$ \cite{kaipio2005statistical}. 

\subsubsection{Interpretação $5$ (Filtragem espectral)}

Esta interpretação é relacionada com o Apêndice \ref{App:svd2}. Existem duas formas de ver regularização como um processo de \textit{filtragem} \cite{Hansen1998}, \cite[Seção 1.2]{Vogel2002}. A primeira, objeto da presente interpretação, é através da SVD do operador direto $\mathbf{A}$, uma \textit{decomposição matricial} que permite analisar $\mathbf{A}$ e classificá-lo entre \textit{problema deficiente em posto} e \textit{problema mal-posto}, bem como dizer o quanto mal-posto ele é \cite{Hansen1998}. A SVD também ajuda a entender como que a regularização de Tikhonov está atuando na solução. 

Especificamente, pode-se calcular a SVD de $\mathbf{A}$ e substituí-la do lado direito da Equação \eqref{eq:tiksol} conforme descrito no Apêndice \ref{sec:svdtikh}. A solução da Equação \eqref{eq:svdtikh0}, escrita em termos dos $i$ \textit{valores singulares} $\sigma_i$ e \textit{vetores singulares} $\mathbf{u}_i$ e $\mathbf{v}_i$, resulta em um termo $\frac{\sigma_i^2}{\sigma_i^2 + \lambda^2}$, chamados de \textit{fatores de filtro} \cite[pág. 62]{hansen2010discrete}, que amortecem componentes indesejadas da SVD \cite[págs. 106-7]{aster2019parameter}, principalmente relacionado aos vetores singulares de maior frequência. Como a SVD é considerada uma \textit{base espectral}, a Equação \eqref{eq:svdtikh0} é uma forma de \textit{filtragem espectral} \cite[pág. 71]{hansen2006deblurring}, \cite[págs. 53, 177]{hansen2010discrete}, no sentido de modificar os componentes espectrais da solução ingênua \cite{hansen2006deblurring}. 

\subsubsection{Interpretação $6$ (Filtragem linear na frequência)}
Esta interpretação é relacionada com o Apêndice \ref{AP:Wiener}. As regularizações clássica e generalizada de Tikhonov podem ser realizadas no \textit{domínio da frequência} para problemas inversos lineares, visando reduzir amplitudes de componentes de Fourier de alta frequência \cite[pág. 230]{aster2019parameter}. 

Um exemplo é quando tanto o modelo $\mathbf{A}$ quanto a matriz de regularização $\mathbf{L}$ são dados por matrizes circulantes de blocos com blocos circulantes, situação na qual há algoritmos eficientes baseados na transformada rápida de Fourier para realizar a regularização \cite[Algoritmo 5.3.1]{Vogel2002}. Um exemplo prático desse caso é o \textit{deblurring}, uma deconvolução. 

Seja $\mathbf{Y}$ a imagem degradada que está disponível, $\mathbf{X}$ da imagem nítida, $\mathbf{H}$ da PSF do sistema e $\mathbf{N}$ o ruído da imagem. A relação entre elas pode ser escrito diretamente com matrizes conforme $\mathbf{Y} = \mathbf{X} * \mathbf{H} + \mathbf{N}$. Outra forma é escrever a relação entre elas considerando as imagens vetorizadas, o que resulta em um sistema linear do tipo $\mathbf{y} = \mathbf{A}\mathbf{x} + \bm{\delta}$. Em ambos os casos, é possível resolver o problema inverso no domínio da frequência, utilizando respectivamente a Equação \eqref{eq:frequencyvogel} para solução com matrizes e a Equação \eqref{eq:tikhfft1} para a solução com vetores.

Nessas duas soluções, adicionar os valores relativos ao termo de regularização visa impedir que pequenos valores no denominador (próximos de zero) amplifiquem ruídos na inversão. Logo, a regularização de Tikhonov pode ser relacionada a um processo de filtragem \cite{Ribeiro2015} como \textit{filtros inversos regularizados}. A forma da Equação \eqref{eq:frequencyvogel} é semelhante à do \textit{equalizador ideal de Wiener}, cujo desenvolvimento é encontrado  em \cite[págs. 547-8]{press1992numerical}, \cite[págs. 402-3]{Thibaut2005} e \cite[págs. 191-2, 197-8, 425-6]{vaseghi2000advanced}. No próprio livro de Tikhonov e Arsenin esse assunto é abordado \cite[págs. 149-52]{tikhonov1977solutions}.

Não se pode dizer que métodos de regularização sejam equivalentes a métodos de filtragem, nem que a regularização de Tikhonov é equivalente ao \textit{filtro de Wiener} para um caso geral. Muitas hipóteses devem ser feitas sobre o problema e sobre os sinais para que as expressões matemáticas resultantes sejam semelhantes. No entanto, algumas relações podem ser estabelecidas e há diversos trabalhos que provam formalmente que filtragem de Wiener corresponde formalmente à regularização \cite{Anderssen1995} e as condições de equivalência, como \cite{Anderssen1981,Murli1999, Sjberg2012}, de modo que métodos de regularização podem resultar em diferentes a aproximações do filtro de Wiener ideal \cite[págs. 125-6]{Hansen1998}. 

A presente interpretação tem intersecções com outras. No contexto da inversão estatística, também é possível relacionar o filtro de Wiener obtido com um critério de erro médio quadrático mínimo com a estimação MAP quando o modelo de ruído é gaussiano e o \textit{prior} é de também de ruído branco gaussiano, reconhecendo a Equação \eqref{eq:tiksolx} como a solução filtrada de Wiener \cite[pág. 150]{calvetti2007introduction}, \cite{Gribonval2011}, \cite[pág. 79]{kaipio2005statistical}, \cite{Thibaut2005}. 

Apesar de não ser muito usual, há trabalhos que descrevem o filtro de Wiener a partir da SVD como \cite[págs. 38-9]{bai2013acoustic}, \cite[pág. 95]{zhang2010machine} e com a SVD generalizada \cite[pág. 103]{zhang2010machine}. Os autores partem da regularização clássica de Tikhonov, mas chamam os fatores de filtro da Equação \eqref{eq:filterfactor_text} como pesos dos filtros de Wiener. Deve-se lembrar que a SVD pode ser utilizada para problemas inversos lineares gerais, não só para a deconvolução.

\subsubsection{Interpretação $7$ (Análise de regressão)}

Esta interpretação é relacionada com o Apêndice \ref{Ap:Ridge}. O problema de \textit{regressão linear} também apresenta a forma da Equação \eqref{eq:eq1}, mas consiste na busca por um \textit{modelo explanatório} para os dados $\mathbf{y}$, possivelmente ruidosos, através da estimação dos \textit{coeficientes de regressão} $\bm{\beta}$ dada a definição da matriz $\mathbf{A}$, conhecida como \textit{design matrix}, cujas colunas são \textit{regressores}, vetores de predição, variáveis explanatórias ou as variáveis independentes \cite{aster2019parameter}. 

Quando as colunas de $\mathbf{A}$ são \textit{não-ortogonais} umas às outras, existe uma grande chance de que a estimação usando o método dos mínimos quadrados ordinário será insatisfatória, sendo sensível aos erros presentes nos dados. Quando as variáveis independentes são correlacionadas, o problema é conhecido como \textit{multicolinearidade}.
 
Para lidar com a instabilidade da estimação de mínimos quadrados, a \textit{regressão de Ridge} foi proposta \cite{Hoerl1970, Hoerl2020}, conforme Equação \eqref{eq:regression2}. Ela inclui um termo aditivo $\vert\vert \bm{\beta} \vert\vert^2_2$, de modo que tenha forma é idêntica à Equação \eqref{eq:tiksol} da regularização clássica de Tikhonov, mas sem nenhuma menção ao método de regularização no seu trabalho original \cite{Hoerl1970}. Dessa forma, pode-se obter estimativas pontuais mais confiáveis do que através do método dos mínimos quadrados ordinário, mas é introduzido \textit{bias} na solução \cite{Oztrk2000}. Deve-se ressaltar que, como se discute em \cite{Hoerl2020}, o autor explica como tanto a regressão de Ridge quanto a análise de Ridge apresentam similaridades com a solução de Tikhonov, mas tem enfoques diferentes, o que não impede delas serem vistas como análogas em alguns sentidos. 
 
A regressão de Ridge é um estimador no contexto de regressão linear, caracterizada em termos de \textit{bias} e variância. É possível obter uma expressão explícita \cite{Hoerl1970} que mostra que, na medida que $\lambda$ aumenta, o \textit{bias} também aumenta e a variância diminui, um \textit{trade-off entre variância e bias}. Isso permite entender a regularização de Tikhonov como uma forma de controlar o \textit{bias} e a variância da solução \cite[pág. 64]{hansen2010discrete}. 

No caso de regularizadores que usam a norma $\ell_1$, busca-se obter soluções esparsas, o que é importante para seleção de variáveis, já que diversas variáveis se tornam nulas, enquanto as mais relevantes serão não-nulas. A regressão com \textit{least absolute shrinkage and selection operator} (LASSO), \cite{Tibshirani1996} parte da Equação \eqref{eq:norma1} quando $\mathbf{L} = \mathbf{I}$. Já a \textit{regressão generalizada de Lasso} também se baseia na Equação \eqref{eq:norma1}, mas considerando outras formas de $\mathbf{L}$ \cite{Tibshirani2011gen}, uma forma análoga à regularização generalizada de Tikhonov mas que visa promover a esparsidade.

\subsubsection{Interpretação $8$ (Aprendizado de máquina)}

Esta interpretação é relacionada com a Subseção \ref{sec:general}. A área de \textit{aprendizado de máquina} desenvolve sistemas que realizam tarefas sem programação explícita para elas, buscam extrair informações diretamente de conjuntos de dados $\mathbf{x}$ e $\mathbf{y}$ (formando pares ou não). Tarefas incluem \textit{classificação}, \textit{regressão}, \textit{clusterização}, \textit{redução de dimensionalidade}, \textit{detecção de anomalia} e outros, incluindo problemas inversos. 

Para isso, é necessário definir uma \textit{estrutura candidata} $h( \cdot , \bm{\theta})$ que deve ser treinada de acordo com a tarefa para que possa realizar a inferência desejada, como $h( \mathbf{x} , \bm{\theta}) = \mathbf{y}$. Durante a \textit{etapa de treinamento}, os parâmetros $\bm{\theta}$ de $h( \cdot , \bm{\theta})$ são atualizados a partir dos dados. Neste caso, não é sempre necessário desenvolver um operador direto $\mathbf{A}$, pois o mapeamento pode ser realizado diretamente entre conjuntos de dados ou a partir das características extraídas destes. 

Um dos critérios que existem para treinamento é a minimização da Equação \eqref{eq:ERM4}, mas ela é propensa ao \textit{overfitting}. Isso acontece quando $h(\cdot, \bm{\theta})$ se ajusta tão bem aos dados de treinamento que ela não é capaz de ter uma boa performance nos dados de teste, uma baixa generalização. Para suprimir o \textit{overfitting} e melhorar a \textit{generalização} da máquina de aprendizado, é possível utilizar diferentes tipos de regularização. Isso inclui utilizar um termo aditivo na etapa de treinamento conforme Equação \eqref{eq:ERM_RLS} \cite{Deisenroth2020, alvarez2017digital}. Um exemplo é quando $\Omega(\bm{\theta})= \vert \vert\bm{\theta} \vert\vert_2^2$, conhecido como \textit{weight decay} e que é prontamente associado à regularização clássica de Tikhonov. Há trabalhos como \cite{mitchell1997machine, nielsen2018}, que descrevem formas de diminuir \textit{overfitting} sem citar Tikhonov. Outros trabalhos ainda buscam relacionar regularização diretamente com generalização em aprendizado de máquina \cite{devito2005}, ou avaliar tarefas dentro de visão computacional \cite{Poggio1985} ou classificação de padrões \cite{Yee1993} como problemas mal-postos. 

\subsubsection{Interpretação $9$ (Aprendizado estatístico)}

Esta interpretação é uma continuidade das Subseções \ref{sec:vapnik}, \ref{sec:vapnik2} e \ref{sec:otimi2}. Na teoria do aprendizado estatístico, Vapnik e seus colaboradores chegaram a utilizar a regularização de Tikhonov como componente da solução de problemas mal-postos estocásticos \cite[pág. 297]{vapnik1998statistical}. No entanto, eles desenvolveram um novo princípio para problemas preditivos. A minimização do risco estrutural (SRM) \cite[pág. 1952]{Vapnik2019} foi proposta como uma forma mais adequada de controlar a capacidade do conjunto de funções admissíveis \cite[pág. 476]{Vapnik2006}, principalmente quando há poucos dados para treinamento \cite[pág. 219]{vapnik1998statistical}. 

De forma geral, o processo de aprendizagem é controlado e realizado em uma estrutura aninhada \cite{Chen2002} conforme
\begin{equation}
S_1 	\subset S_2 	\subset \dots \subset S_m 	\subset \cdots,
\label{eq:srm}
\end{equation}
onde $S_i$, para $i = 1, 2, ..., m$ são subconjuntos que devem ser definidos. Dependendo da escolha desses subconjuntos, o princípio SRM pode ser universalmente consistente \cite[pág. 1952]{Vapnik2019}. Enquanto a minimização do risco empírico (ERM) busca minimizar o risco a todo custo, a SRM busca pela relação ideal entre a quantidade de dados disponíveis, a capacidade e a qualidade da aproximação dada pelo espaço de hipóteses \cite[pág. 219]{vapnik1998statistical}. 

Na SRM, é imposta uma estrutura no espaço de hipóteses através de um conjunto de subconjuntos aninhados de funções \cite[pág. 221]{vapnik1998statistical}, que são ordenadas e escolhidas a partir da dimensão Vapnik-Chervonenkis (VC) dessas funções, do seu controle de capacidade. Sendo assim, Vapnik argumenta que se utiliza apenas uma informação \textit{a priori} fraca \cite[pág. 702]{vapnik1998statistical}. Diferente de abordagens estatísticas para solução de problemas inversos, a abordagem SRM não necessita informação \textit{a priori} sobre os dados alvo (ou rótulos). 

A partir desse princípio, diversas formas de melhorar a generalização de uma máquina de aprendizagem podem ser deduzidas. Alguns algoritmos obtidos a partir do SRM possuem forma muito parecida com técnicas de regularização, como a adição de um termo de penalização na minimização do risco empírico \cite[pág. 2829]{Chen2002}, \cite[pág. 103]{alvarez2017digital}. Já em \cite[pág. 835]{Vapnik1992}, o autor explica como o \textit{weight decay} pode ser deduzido pela SRM. Assim, em \cite{Chen2002} os autores dizem que diversas técnicas de regularização correspondem ao princípio de aprendizado estrutural, enquanto em \cite[pág. 96-7]{cherkassky2007learning}) os autores dizem que o procedimento construtivo entre os dois é idêntico.

Como consequência, métodos de regularização e SRM acabam sendo relacionados por apresentarem formulações muito parecidas e isso pode causar a impressão que elas seriam equivalentes, como na Subseção \ref{sec:general2}, mas o próprio Vapnik \cite[págs. 421, 477]{Vapnik2006} ou outros autores \cite[págs. 968--9]{Cherkassky2009} deixam claras as diferenças entre os dois.

Em algumas referências, a SRM é relacionada diretamente com a Equação \eqref{eq:ERM_RLS}, sem partir da Equação \eqref{eq:srm} como em \cite{Vapnik1992}, de modo que a regularização é associada diretamente com o \textit{controle de capacidade} (ou \textit{controle de complexidade}) do modelo \cite{Deisenroth2020, alvarez2017digital}. Ou seja, através da \textit{restrição da flexibilidade de uma classe de funções}, a regularização ajudaria a evitar o \textit{overfitting} nos dados de treinamento \cite{alvarez2017digital}. Nestes trabalhos, a distinção entre a SRM e regularização pode ser comprometida caso não se conheça a origem de cada uma. 

Outra forma de relacionar SRM e a expressão de regularização acontece em \textit{support vector machines} (SVM), modelos propostos no aprendizado estatístico que executam o princípio da SRM \cite[pág. 432]{Vapnik2006}. Nesses modelos, há um termo que controla o termo das margens, que muitas vezes são chamados de termo de regularização, de modo que a escolha do parâmetro de regularização possibilita o controle do tamanho da margem dos hiperplanos \cite[Seção 12.2.4]{Deisenroth2020}. Em \cite{Cherkassky2009}, por exemplo, os autores discutem a diferença entre SVM e a regressão de Ridge. Partindo de suas respectivas teorias e arcabouços teóricos, os autores mostram que regularização e o conceito de margem são mecanismos diferentes no controle da complexidade dos modelos.

\subsubsection{Interpretação $10$ (Aprendizagem profunda)}

Esta interpretação está relacionada com a Subseção \ref{sec:new}. A área de \textit{aprendizagem profunda} é uma subárea de aprendizado de máquina que utiliza formas específicas da máquina de aprendizagem, isto é, $h(\mathbf{x}, \bm{\theta})$ são \textit{redes neurais artificiais}. Elas são compostas de \textit{unidades}, que realizam cálculos relativamente simples, e funções de ativação, que introduzem não-linearidades no modelo. Essas unidades são agrupadas em camadas, ligadas umas nas outras, formando uma rede capaz de realizar uma \textit{representação hierárquica dos conceitos}, do mais simples para o mais complexo. Aprendizagem profunda é baseada em redes neurais artificiais com muitas camadas, que podem chegar a milhões de parâmetros do modelo, levando o problema a um \textit{espaço de alta dimensionalidade}. Novamente, uma das formas de tentar melhorar a sua generalização é utilizando métodos de regularização. Em \cite{nielsen2018}, o autor diz que é um fato empírico que redes regularizadas têm melhor performance do que as que não são regularizadas. Além disso, novas definições de regularização foram propostas para a aprendizagem profunda.

Em \cite{goodfellow2016deep}, os autores definem a regularização como qualquer técnica que visa diminuir o \textit{erro de generalização}, mas não o seu \textit{erro de treinamento}. Ou seja, os efeitos da regularização são observados a partir dos resultados da rede neural artificial nos conjuntos de treinamento e de teste. Sendo menos restritiva do que a regularização de Tikhonov, a regularização em aprendizagem profunda pode ser obtida através de formas diversas \cite{goodfellow2016deep, nielsen2018}, entre elas:
\begin{itemize}
\item Através de termos aditivos, como na Equação \eqref{eq:ERM_RLS}; 
\item Pela expansão artificial dos dados de treinamento, ou \textit{data augmentation}; 
\item Pela adição de ruído branco nos dados de treinamento; 
\item Pelo \textit{early stopping}, quando se trunca o número de iterações antes do erro de generalização aumentar; 
\item Pelo \textit{dropout}, quando algumas das unidades são esquecidas (seu peso zerado) durante o treinamento, realizando o treinamento como se a rede fosse diferente); 
\item Pelo otimizador utilizado durante treinamento, como o SGD ou Adam.
\end{itemize}

O conceito de regularização implícita é relevante, pois a maior parte das propostas citadas são implícitas. Quando a regularização é obtida a partir da adição de $\Omega(\bm{\theta})$, como na Equação \eqref{eq:tikhonov_original}, ela é classificada de explícita, mas quando apresenta outras formas, ela é chamada de implícita \cite{Chen2002}. O entendimento de quais fatores são relevantes para melhorar a generalização da ANNs é uma área de pesquisa ativa. 

Em \cite{2017kukacka}, os autores definem regularização como qualquer técnica que ajude a generalizar melhor, isto é, ter melhores resultados no conjunto de teste, sem nenhuma restrição ao conjunto de treinamento como em \cite{goodfellow2016deep}. Isso permite incluir diversos aspectos do processo de aprendizagem, como os dados, a família de modelos selecionada, a função de perda, o termo de regularização em si ou o otimizador utilizado. 

\subsection{Qual é a ocorrência dessas interpretações em livros de referência e artigos científicos?}

Considerando regularização um conceito polissêmico, a Tabela \ref{table:1} apresenta a presença (em lilás) ou ausência (em branco) das $10$ interpretações (linhas) em livros (colunas) de acordo com os seguintes temas:

\begin{itemize}
\item Problemas inversos:  \cite{tarantola2005inverse, hansen2010discrete, Mueller2012, aster2019parameter, Bertero2021}
\item Processamento de sinais \cite{alvarez2017digital} 
\item Aprendizado de máquina e aprendizagem profunda: \cite{Deisenroth2020, goodfellow2016deep, Peng2023, Vandenbussche2023}
\end{itemize}
Os livros são ordenados nas colunas da esquerda para a direita do mais antigo para o mais novo. Em cada uma das interpretações, as palavras-chave são destacadas em negrito. A avaliação de ausência ou presença em cada referência é baseada nessas palavras-chave, em conjunto com as principais equações que as descrevem. Livros específicos da área de problemas inversos são indicados na linha denotada por IP.

\begin{table}[H]
\begin{center}
\caption{Ocorrência das interpretações nos livros de referência.}
\label{table:1}
\begin{tabular}{|c|c | c |c|c|c|c|c|c|c|c|} \hline
\rowcolor[gray]{.9} Interpret. & \cite{tarantola2005inverse} & \cite{hansen2010discrete} & \cite{Mueller2012} & \cite{goodfellow2016deep} & \cite{alvarez2017digital} & \cite{aster2019parameter} & \cite{Deisenroth2020} & \cite{Bertero2021} & \cite{Peng2023} & \cite{Vandenbussche2023}\\ \hline
IP & \cellcolor{blue!25} & \cellcolor{blue!25} & \cellcolor{blue!25} & & & \cellcolor{blue!25} & & \cellcolor{blue!25} & & \\ \hline
1 & & & \cellcolor{blue!25}& & & & &\cellcolor{blue!25} & &\\ 
2 & & \cellcolor{blue!25} &\cellcolor{blue!25} & \cellcolor{blue!25}& \cellcolor{blue!25}& \cellcolor{blue!25}& \cellcolor{blue!25}&\cellcolor{blue!25} & & \\ 
3 & & & & \cellcolor{blue!25}& & & \cellcolor{blue!25} & \cellcolor{blue!25} \cellcolor{blue!25} & \cellcolor{blue!25} & \cellcolor{blue!25}\\ 
4 & \cellcolor{blue!25} & & & \cellcolor{blue!25}& \cellcolor{blue!25}& \cellcolor{blue!25}& \cellcolor{blue!25}&\cellcolor{blue!25} & \cellcolor{blue!25} &\\ 
5 & & \cellcolor{blue!25} & \cellcolor{blue!25} & & \cellcolor{blue!25} & \cellcolor{blue!25}& & \cellcolor{blue!25} & &\\ 
6 & \cellcolor{blue!25} & & & & & \cellcolor{blue!25}& & \cellcolor{blue!25} & &\\ 
7 & \cellcolor{blue!25} & \cellcolor{blue!25} & & \cellcolor{blue!25} & \cellcolor{blue!25} & \cellcolor{blue!25} & \cellcolor{blue!25} & \cellcolor{blue!25} & & \\ 
8 & & & & \cellcolor{blue!25}& \cellcolor{blue!25}& & \cellcolor{blue!25}& & \cellcolor{blue!25} & \cellcolor{blue!25}\\ 
9 & & & & \cellcolor{blue!25} &\cellcolor{blue!25} & &\cellcolor{blue!25} & & &\\ 
10 & & & & \cellcolor{blue!25} & & & & & \cellcolor{blue!25} & \\ \hline
\end{tabular}
\end{center}
\end{table}
{ \vspace{-1cm} \footnotesize IP = Descrição em termos de problemas inversos.}

\vspace{2mm}

Em relação aos livros escolhidos, é possível fazer algumas observações:
\begin{itemize}

\item Apesar das relações entre regularização de Tikhonov e inversão estatística, o autor de \cite{tarantola2005inverse} descreve a solução de problemas inversos sem mencionar métodos de regularização. Isso mostra como a interpretação bayesiana é um método independente de solução, como apontado em \cite{Calvetti2018a}, trazendo também formas eficientes de desenvolver \textit{priors} e quantificar as incertezas das soluções obtidas;
 \item As Equações \eqref{eq:tikhonov1} e \eqref{eq:tiksol}, que descrevem a regularização clássica de Tikhonov, aparecem em praticamente todos esses livros, mas em alguns casos como forma pronta, sem contextualização histórica ou origem. É claro que sua implementação é a mais direta, mas é importante que ela seja vista como apenas um resultado, uma parte da teoria de regularização, assim como representado Figura \ref{fig:01_venn}; 
 \item No caso da regressão linear, há referências como \cite{Bertero2021, hansen2010discrete, tarantola2005inverse} que a descrevem como uma aplicação da regularização clássica de Tikhonov, mas deve-se lembrar que o artigo original da regressão de Ridge não mencionou regularização \cite{Hoerl1970};
\item Apesar de \cite{aster2019parameter, hansen2010discrete} não apresentarem expressões como \textit{penalty method} ou \textit{penalty term}, os autores usam a expressão \textit{penalizes} em diferentes momentos \cite[págs. 187, 190]{hansen2010discrete}, relacionando o regularizador com o conceito de termo de penalização;
 \item A Tabela \ref{table:1} apresenta livros com múltiplas interpretações, o que não significa que elas estejam detalhadas. Isso acontece em \cite{Deisenroth2020, alvarez2017digital}, por exemplo, o que poderia tornar difícil a distinção entre elas sem contextualização histórica; 
 \item Em livros recentes da área de problemas inversos, não é encontrada a interpretação $10$ em termos de erros de generalização e treinamento, que parece estar restrita ao contexto de aprendizagem profunda. Mesmo assim, um livro influente como \cite{goodfellow2016deep}, ou trabalhos com muitas citações como \cite{2017kukacka} tornam essa interpretação relevante para discussão do tópico. Livros que discutem regularização em aprendizado de máquina podem focar mais na interpretação $8$ \cite{Vandenbussche2023} ou também com a $10$ \cite{Peng2023}, dando continuidade para as novas interpretações.
\end{itemize}

Múltiplas interpretações também são verificadas em artigos científicos. Em \cite{Kaji2019} os autores apresentam o propósito de regularização como sendo triplo: 1 - incorporar \textit{priors} para obter saídas de maior qualidade; 2 - suprimir \textit{overfitting} para melhor generalização; 3 - mitigar a característica mal-posta do problema para obter uma melhor convergência da otimização. Eles se relacionam com as interpretações $2$, $4$ e $8$. Assim, o leitor pode entender que elas são intercambiáveis ou simultâneas, mesmo tendo origens diferentes.

\newpage

\section{PROBLEMAS INVERSOS: É POSSÍVEL CAMINHAR EM DIREÇÃO À UMA PRÁTICA INTERDISCIPLINAR?}\label{sec:interdisciplinarity}

\subsection{Introdução}
Problemas inversos específicos foram estudados simultanemente por diferentes áreas, podendo ser descritos a partir de vocabulários próprios. É importante conhecer a nomenclatura para descrever conceitos em áreas diferentes, bem como discutir características de interdisciplinaridade nessa área de pesquisa.  

 Se a construção desse conhecimento for multidisciplinar, será possível extrair perspectivas sobre o assunto a partir de mais de uma disciplina. Caso for interdisciplinar, será possível integrar tais conhecimentos disciplinares  \cite[Capítulo 3]{Greef2017}. Mais do que a classificação em si, a importância da interdisciplinaridade em diminuir as barreiras disciplinares reside nas oportunidades de fomentar novas ideias, beneficiando todas as áreas envolvidas.

\subsection{Existem relações quando diferentes áreas descrevem a solução de um mesmo problema inverso?} 

Seja o exemplo de problema inverso da deconvolução, que busca recuperar os sinais originais tendo apenas o sinal após aquisição por um sistema de medição, ou seja, sob influência da resposta ao impulso do sistema. Nos Capítulos \ref{sec:deblur_forward} e \ref{sec:de_nonblind}, discutiu-se especificamente o \textit{deblurring} para imagens 2D, na qual busca-se reconstruir imagens nítidas a partir de imagens borradas. De acordo com \cite[pág. 216]{Chan2005}, o \textit{deblurring} pode ser visto sob os seguintes pontos de vista:
\begin{itemize}
 \item \textbf{Processamento de sinais clássico}: Inversão de um filtro passa-baixas;
 \item \textbf{Teoria das PDEs lineares hiperbólicas}: Processo \textit{backward} de difusão;
 \item \textbf{Mecânica estatística}: Diminuição da entropia;
 \item \textbf{Análise funcional}: Inversão de operadores compactos.
\end{itemize}
Essa lista ilustra como cada área pode definir um mesmo problema inverso a partir de seus próprios conceitos, trazendo pontos de vista únicos com o mesmo objetivo. 

Logo, diferentes soluções para um mesmo problema inverso podem ser semelhantes, análogas, equivalentes ou mesmo idênticas, mas nem sempre partem das mesmas hipóteses. De fato, é possível obter a regularização clássica de Tikhonov da Equação \eqref{eq:tiksol} a partir de diferentes paradigmas. Na literatura isso se traduz em expressões como dizer que: 
\begin{itemize}
 \item \textit{Priors} são conceitos análogos à regularização \cite{Deisenroth2020}; 
 \item \textit{Priors} fazem o papel de regularizadores \cite{Deisenroth2020}; 
 \item Regularização de Tikhonov é idêntica à regressão de Ridge
 \cite[pág. 133]{aster2019parameter};
 \item Diferentes terminologias da estatística podem ser formuladas no \textit{framework} de Tikhonov, como regressão de Ridge, estimativa pode mínimos quadrados penalizada, estimativa de verossimilhança penalizada, \textit{smoothing splines} e regressão por \textit{averaging kernel regression} \cite{Chen2002}; 
 \item Regularização por variação total é conhecida pelo nome de \textit{Lasso} na literatura estatística \cite[pág. 83]{Mueller2012};
\item \textit{Weight decay} é equivalente à regressão de Ridge , ao mesmo tempo que a regressão de Ridge é uma versão de ordem zero da regularização de Tikhonov \cite[pág. 2820]{Chen2002};
\item A mesma ideia da regularização com norma $\ell_1$ foi utilizada em estatística sob o acrônimo Lasso \cite[pág. 241]{Bertero2021}; 
 \item Uma penalidade de norma $\ell_2$ é usualmente chamada de \textit{weight decay}, mas que em outras comunidades acadêmicas é conhecida como regressão de Ridge ou regularização de Tikhonov \cite{goodfellow2016deep};
 \item Por vezes, técnicas de diferentes áreas são utilizadas em conjunto. Em \cite{Taheri2021}, os autores discutem os limites da regressão de Ridge para ERM em problemas de alta dimensionalidade, mas sem citar SRM, por exemplo.
\end{itemize}

Há compartilhamento de ferramentas matemáticas e computacionais entre diferentes áreas e nem todas partem da solução de problemas mal-postos.  Portanto, não é trivial dizer que haja equivalência entre esses métodos, como discutido no Capítulo \ref{sec:polysemy}. Analogias entre algoritmos na etapa de implementação computacional podem indicar relações e trazer ideias de uma área para a outra, mas expressões como \textit{soluções análogas} podem causar confusão sobre a compatibilidade real entre conceitos de áreas diferentes sob um mesmo ponto de vista (regularização). 

Se esse for o objetivo, é necessário estabelecer essas relações formalmente. Existem trabalhos que reinterpretam suas próprias áreas de pesquisa como problemas inversos. Um exemplo é encontrado em \cite{Gallet2022}, na qual os autores tratam problemas de engenharia estrutural como problemas inversos, apontando como uma das principais vantagens a possibilidade de utilizar as metodologias disponíveis e estabelecidas na área de problemas inversos, trazendo novas possibilidades para problemas antes vistos como intratáveis.

Em problemas inversos e aprendizado de máquina, as relações entre as duas áreas são discutidas em trabalhos como:
\begin{itemize}
\item  \cite{Krkov2005}, que relaciona a habilidade de generalização de ANNs com regularização; 
\item \cite{Mukherjee2006, devito2005}, que relacionam o conceito de consistência de aprendizado de máquina com o conceito de estabilidade em problemas inversos; 
\item \cite{Burger2021}, que reinterpreta problemas de aprendizado de máquina do ponto de vista da teoria de regularização, assim como reinterpreta métodos variacionais sob o ponto de vista da minimização do risco.
\end{itemize}

Em \cite{Chen2002}, os autores trazem diversas facetas de regularização, discutindo as áreas na ciência que permitem estabelecer o princípio da regularização, bem como seu arcabouço teórico, incluindo aplicações e implementações desse conceito \cite[Figura 2]{Chen2002}. Os autores também discutem como falta um princípio universal para a teoria da regularização em aprendizagem de máquina, um princípio que poderia englobar e unificar todos os conceitos apresentados. Com tantas interpretações de áreas diferentes, é possível se perguntar se o princípio universal que os autores de \cite{Chen2002} buscavam estaria mais próximo ou mais longe duas décadas depois dessa pergunta realizada. 

Em \cite[pág. 969]{Cherkassky2009}, os autores falam da importância do consenso sobre os conceitos básicos de algoritmos de aprendizado. No entanto, mesmo que o consenso sobre regularização não exista, é importante que se conheça a origem das interpretações e seus respectivos arcabouços teóricos. 
 Sendo um conceito polissêmico, é importante que um sentido não prevaleça sobre os outros, considerando-se os demais equivocados. Em outras palavras, a falta de diálogo entre as áreas pode privilegiar certas noções de regularização em detrimento de outros. Nem sempre será necessário apresentar todas as possíveis interpretações de regularização em uma única disciplina, mas é importante que o aluno tenha clareza de qual definição está sendo utilizada.

\subsection{Como diferentes áreas visualizam \textit{trade-offs} de termos aditivos?} 
No caso de regularização enquanto adição de termos de penalização no funcional de otimização, há métodos parametrizados por $\lambda$. Na literatura, o \textit{trade-off} devido ao $\lambda$ é descrito de diferentes formas e exemplos em problemas inversos incluem:
\begin{itemize}
 \item Em \cite[págs. 2804-5]{Chen2002}, entre acurácia e estabilidade;
 \item  Em \cite[Equações 4.41-2]{Bertero2021}, entre erro de aproximação e de propagação de ruído;
 \item Em \cite{Braun2017}, entre rejeição ao ruído (robustez) e resolução;
   \item Em \cite[Subseção 4.8]{aster2019parameter}, $\lambda$ influencia nos erros devido aos ruídos e erros devido à regularização na reconstrução.
   \end{itemize}

Esse tipo de avaliação do efeito de $\lambda$ também é feita graficamente entre: 
\begin{itemize}
 \item \textbf{Regressão linear}: \textit{Bias} e variância na regressão de Ridge \cite[Figura 1]{Hoerl1970};
   \item  \textbf{Problemas inversos}: Erro de regularização e de medição \cite[Figura 4.1]{Seo2012}.
   \end{itemize}
   
   Ao mesmo tempo, na literatura são discutidos outros \textit{trade-offs}, fora da área de problemas inversos e não necessariamente relacionados à regularização e $\lambda$. O \textit{trade-off} entre \textit{bias} e variância é analisado em outras áreas, como redes neurais \cite{Geman1992} e \textit{support vector machines} \cite{James2013}. Outras decomposições de erros também existem. Em  \cite{Niyogi1999}, na área de aprendizado de máquina, os autores decompõe o erro total de generalização em função da ordem da classe de hipótese e do número de exemplos. Outros exemplos de gráficos avaliam o efeito de $\lambda$ entre:    
   \begin{itemize}
  \item  \textbf{Problemas inversos} (diferenciação numérica):  Erro de aproximação e o de propagação de ruído a partir do nível de discretização do problema \cite[Figura 1.7]{bleyer2015novel};
  
 \item \textbf{Aprendizado de máquina}: Erro de estimação e o de aproximação na minimização do risco empírico \cite[Figura 4.3]{Mohri2018}; Capacidade do modelo e os erros de treinamento e generalização \cite[Figura 5.3]{goodfellow2016deep};
 \item  \textbf{Aprendizado estatístico}: Complexidade do conjunto de hipóteses, o erro empírico e o termo de penalidade na SRM, a partir do índice dos subconjuntos aninhados que definem o conjunto de hipóteses \cite[Figura 4.4]{Mohri2018}.

\end{itemize}
Enquanto os \textit{trade-offs} são diferentes, as figuras indicadas são praticamente idênticas, o que pode causar a impressão de se tratar dos mesmos conceitos. Porém, cada uma dessas imagens deve ser interpretada por si só, já que as grandezas são definidas no contexto de cada área também porque tais gráficos dependem da aplicação.

\subsection{Regularização é uma teoria ou uma técnica?}

Apesar de Tikhonov ter elaborado a teoria e o método de regularização, as vezes regularização é descrita como um conjunto de técnicas ou procedimentos prontos para uso. Sob o ponto de vista computacional, há técnicas são prontas para uso, como a adição da matriz identidade nas equações normais ou a soma de pesos da ANN na função de perda em um algoritmo de gradiente descendente.

A diferença entre considerá-la teoria ou técnica pode definir a forma em que ela será abordada.  Em livros tradicionais de problemas inversos como \cite[pág. 63]{engl1996regularization}, além do próprio método de regularização, há a expressão de \textit{implementable method}. Já novas propostas que apresentem resultados computacionais promissores nem sempre apresentam prontamente um arcabouço teórico junto. 

 Na literatura, essa classificação é encontrada de diferentes formas:
\begin{itemize}
\item Apesar do impacto de seu trabalho, observa-se que Phillips identificou sua proposta como uma técnica no título de seu artigo \cite{Phillips1962}. Segundo \cite{Morozov1984}, a proposta de Phillips não apresentou muitas justificativas teóricas, de modo que o nome de Tikhonov é o mais associado ao método, mesmo tendo sido publicado depois;

\item Em \cite{Laplante2000}, um dicionário de termos de ciência da computação, engenharia e tecnologia, regularização é definida como o procedimento de adicionar um termo de restrição ao processo de otimização visando efeito de estabilização na solução;
\item  Em \cite[pág. 193]{Sayed2003}, os autores discutem o método de Newton e identificam regularização como a adição de uma matriz identidade $\mathbf{I}$, ponderada por uma constante, para o caso da matriz hessiana do problema ser quase singular. Essa variação é conhecida como método de Levenberg-Marquadt;
\item  Em \cite[pág. 669]{Sayed2003}, os autores discutem uma variação da solução de mínimos quadrados através da incorporação de um termo aditivo de regularização $(\mathbf{x}^T \mathbf{x})^T \Pi (\mathbf{x}^T\mathbf{x})$  para o problema de mínimos quadrados $\vert \vert  \mathbf{A} \mathbf{x} - \mathbf{y}\vert \vert_2^2$, onde $\Pi$ é uma matriz definida positiva. Essa modificação, chamada de mínimos quadrados regularizado,  tem o intuito de adicionar informação \textit{a priori} sobre a solução e para aliviar problemas de quando  $\mathbf{A}$ é mal-condicionada; 
\item No caso de ML, em \cite{mitchell1997machine}, não há menção à regularização. Os autores falam sobre técnicas para tentar resolver o \textit{overfitting}, que incluem o \textit{weight decay}, novamente uma adição relativamente simples sob os parâmetros da máquina de aprendizado;
\item Em aprendizagem profunda, regularização também é descrita em termos de técnicas \cite{2017kukacka} ou de estratégias \cite{goodfellow2016deep} de regularização.
\end{itemize}

\subsection{Qual é o problema essencial do método experimental por Poincaré e qual é a sua relação com problemas inversos?}\label{sec:poincare}

Na edição de 1917 do livro \textit{Ciência e hipóteses} \cite{poincare1917}, traduzido para o inglês em \cite[pág. 131]{poincare2018science}, Poincaré escreveu sobre os problemas das probabilidades das causas (\textit{probabilité des causes}), na qual, ao invés de deduzir os efeitos das causas, se deseja deduzir as causas dos efeitos. Ele diz que estes problemas são os mais interessantes do ponto de vista das aplicações científicas, bem como que é o problema essencial do método experimental \cite[pág. 222]{poincare1917}. 

Muitos problemas de diferentes áreas podem ser considerados \textit{problemas inversos}, seguindo a ideia geral de deduzir as causas a partir dos efeitos. Seguindo essa ideia, Bunge lista os respectivos pares de problemas direto-inverso que são baseados no cotidiano das pessoas \cite[Tabela 2]{Bunge2019}. Em \cite[Seção 3]{Bunge2019}, o autor diferencia explicitamente problemas diretos e inversos utilizando notação lógica. Porém, ele assim o faz não em um contexto estrito de problemas físico-matemáticos, mas sim em um contexto da filosofia da ciência. Assim, em  \cite[Seção 5]{Bunge2019}, ele descreve problemas inversos na área de astronomia, biologia, psicologia, sociologia, historiografia, tecnologia e até teologia. Ainda que Bunge não faça essa relação, em algum nível todos os exemplos dados podem ser relacionados com a discussão de Poincaré.

Por outro lado, é importante entender quais são os seus objetivos e as ferramentas disponíveis em cada área. Especificamente, a área de pesquisa de \textit{inverse problems} visa estudar os problemas inversos que são mal-postos \cite[pág. 3]{Flemming2018} e formas de resolvê-lo, sendo muito associada ao método de regularização, cuja origem se deu na proposta de Tikhonov, mas que segue sendo pesquisado e gerando resultados expressivos \cite{Benning2018}. 

Nesse sentido, a área de \textit{inverse problems} é muito relacionada com determinados problemas físico-matemáticos e formas específicas de solução. Isso é diferente da tentativa de um pai e de uma mãe de entenderem o que um bebê necessita a partir de seu choro, ainda que também esteja na direção consequência $\rightarrow$ causa. 

 O desafio é que, mesmo na área de  \textit{inverse problems}, a distinção entre problemas direto e inverso é intuitiva, porém não definitiva. Retomando a Subseção \ref{sec:forwardp}, as próprias definições de problema direto e inverso permitem alguns comentários adicionais:
\begin{itemize}
\item O nome \textit{problemas inversos} seria resultado da necessidade de inversão de $\mathcal{A}$ \cite[pág. 3]{Flemming2018}, mas esse termo só faz sentido quando o conceito de \textit{problema direto} está definido \cite[pág. 1]{kaipio2005statistical}. Ou seja, dois problemas são inversos um ao outro quando a formulação de um faz parte da solução do outro \cite{Keller1976};
\item Ou seja, sem restrições adicionais nas definições, os problemas direto e inverso poderiam acabar em relações idênticas um com o outro \cite[pág. 4]{Mueller2012}; 
\item Nesse sentido, a definição de problemas diretos e inversos é arbitrária mas, por motivos históricos, o problema direto é aquele que já é estudado a mais tempo e é melhor compreendido \cite{Keller1976};
\item Acrescenta-se que, em \cite[Subseção 2.1]{Bunge2019}, o autor diz que a maioria dos problemas inversos são mais difíceis que o problema direto correspondente. Já em \cite[pág. 4]{Mueller2012}, os autores dizem que, na área de problemas inversos mal-postos, os problemas inversos são mais difíceis. 
\end{itemize}
A diferenciação entre direto e inverso não se dará exclusivamente pela dificuldade na sua solução, até porque há problemas diretos que ainda não foram resolvidos, mas não é difícil imaginar que estimar as causas a partir dos efeitos seja o problema mais difícil. Mesmo assim, para aceitação imediata, a decisão mais fácil parece ser a de classificar problemas inversos da forma mais geral como em \cite{Bunge2019}, ao invés de falar que tal definição pode ser arbitrária como possibilidade descrita em \cite{Keller1976}.  

\subsection{Quais é a ocorrência de regularização na ementa de disciplinas de pós-graduação?}

Na Universidade Federal do ABC\footnote{\url{https://www.ufabc.edu.br/}} (UFABC), uma breve busca realizada em 2024 permitiu observar a ausência de termos como \textit{regularização} e \textit{problemas inversos mal-postos} na ementa das disciplinas dos cursos da área de \textit{Engenharias IV}, que incluem a Engenharia Elétrica, Engenharia Biomédica e Engenharia da Informação\footnote{Definição de acordo com CAPES. Mais informações em \url{https://www.gov.br/capes/pt-br}.} da UFABC. 

A falta de tais palavras-chave nos títulos ou nas ementas não é falta de relação temática. No curso de pós-graduação em Engenharia da Informação\footnote{\url{https://sig.ufabc.edu.br/sigaa/public/programa/portal.jsf?lc=pt_BR\&id=204}} da UFABC, existem diversas disciplinas que poderiam discutir regularização, tais como (em parêntes há o código da disciplina): 
\begin{itemize}
    \item Processamento digital de sinais (INF201); 
    \item TEPS\footnote{Tópicos especiais em processamento de sinais (TEPS)}: Visão computacional (INF209B);
    \item TEPS: Sinais biomédicos (INF209C); 
    \item TEPS: Sinais  de eletroencefalograma  (INF209E);
    \item  TIA\footnote{Tópicos em inteligência artificial (TIA)}: Métodos de engenharia para aprendizado de máquina I (INF317B);
    \item TIA: Aprendizagem profunda (INF317E).
\end{itemize} 
É claro que este é apenas um recorte de uma universidade e dois cursos, com base na vivênvia acadêmica de um dos autores. Uma análise mais aprofundada sobre outras universidades foge do escopo do livro, mas a ausência percebida já possibilita reflexão.

Vale notar que a ausência de conceitos de regularização de Tikhonov foi observada por \cite{Bell1978} na educação de matemática aplicada. Os autores dizem que diversas técnicas apresentadas em \cite{tikhonov1977solutions} seriam acessíveis aos estudantes. Logo, a escolha de um determinado assunto fazer ou não parte da grade também deve ser debatida.

\subsection{Como abordar problemas inversos no ensino?}
Considerando que conceitos científicos vão da área de pesquisa até os diferentes níveis de ensino, é possível apresentar conceitos de problemas inversos mal-postos e regularização de maneiras mais acessíveis para diversas audiências, décadas depois das propostas originais. Para isso, é necessária realizar a transposição didática. Por exemplo, o livro \cite{groetsch1999inverse} de problemas inversos foi proposto para alunos de graduação de acordo com \cite{Yamamoto2003}. Ele traz muitos exemplos de problemas inversos para os dois primeiros anos de graduação, nas áreas pré-calculo, cálculo, equações diferenciais e álgebra linear. 

Trabalhos da literatura já trazem diferentes aspectos e iniciativas sobre ensino de problemas inversos: utilizando interfaces computacionais amigáveis \cite{Crosta1990}; em circuitos elétricos \cite{Curtis2000, Curtis2007}; em cursos de graduação de matemática e matemática aplicada \cite{Liu2003}; no ensino infantil de matemática \cite{Ding2017}; discutindo aspectos filosóficos sobre o ensino de problemas inversos \cite{Kornilov2018}; ensino de problemas inversos em engenharia \cite{Putnik2022}; e em ciência, tecnologia, engenharia e matemática (STEM) \cite{Martinez-Luaces2022}, para citar alguns. Outra possibilidade para tornar mais acessível é a utilização de textos de divulgação científica. Exemplos dos presentes autores estão disponíveis em \cite{ufabcDC5, ufabcDC4, ufabcDC1, ufabcDC2, ufabcDC3, ufabcDC6}.

\subsection{Quais são as implicações da interdisciplinaridade da área de problemas inversos para o ensino?}

Caso se deseje falar sobre regularização, mas não haja um curso específico sobre problemas inversos, seria natural que cada disciplina fizesse o recorte mais adequado. Livros como \cite{aster2019parameter} e \cite{hansen2010discrete} exploram os conceitos partindo da álgebra linear, apresentando menos requisitos do leitor em comparação com do que os livros que partem de análise funcional. Já em uma disciplina mais prática de aprendizagem profunda, é provável que seria suficiente a interpretação $10$ da Subseção \eqref{sec:III}, baseada em \cite{goodfellow2016deep}. 

Conforme \cite{Klein2017}, algumas características de interdisciplinaridade incluem o compartilhamento de componentes metodológicos e ferramentas entre áreas, utilização de conhecimento de uma área para contextualizar problemas de outra área, integração de proposições entre disciplinas e utilização de conceitos entre disciplinas para melhorar resultados de problemas comuns entre elas. Do ponto de vista educacional, reconhecer aspectos desse processo interpretativo pode contribuir para o desenvolvimento de abordagens multidisciplinares e interdisciplinares sobre esses temas, objetivando o desenvolvimento de novas contribuições na área de problemas inversos.

\newpage
\section{CONSIDERAÇÕES FINAIS}\label{sec:conclusion}

 A teoria de regularização de Tikhonov resultou em método de solução de problemas mal-postos de propósito geral. A sua importância histórica e a quantidade de aplicações possíveis justifica a sua revisão.  A escolha do regularizador na Equação \eqref{eq:tiksol} permite que outras características sejam refletidas nas soluções obtidas, como suavidade, transições abruptas ou esparsidade. Essa é uma área ativa de pesquisa. Trabalhos recentes como \cite{Alghamdi2024, Pasha2024, Riis2024} trazem códigos abertos de diferentes algoritmos de inversão utilizando linguagem Python. Isso torna mais acessível o conhecimento da implementação de operadores diretos, matrizes de regularização, otimizadores, entre outros. 

A intenção do livro não foi apenas de ser um catálogo de regularizadores, mas sim de dar base para entender muitas das propostas encontradas na literatura. A definição do regularizador é uma etapa importante, mas a dificuldade pode ser a própria solução do problema de otimização, da minimização do funcional, bem como a comparação entre diferentes algoritmos para o mesmo funcional. Apesar de existirem algoritmos para minimizar o mesmo funcional da Equação \eqref{eq:norma1}, não significa que os mesmos resultados serão obtidos, até porque eles não apresentam as mesmas garantias teóricas \cite{Daubechies2016}. 

A partir do momento que se conhece as suas vantagens e desvantagens de cada proposta, pode-se escolher aquele que é melhor para a aplicação desejada. Para o mesmo funcional há soluções em apenas um passo e há diversos algoritmos iterativos, logo, escolher o mais adequado depende de diversos fatores.  Um algoritmo de inversão deve ser robusto em relação a alguns fatores importantes  \cite{hansen2010discrete}: que o modelo disponível no problema direto não é perfeito na sua representação da realidade, mas sim uma aproximação; para a presença de ruídos nas medidas; e para os casos em que ou não há dados suficientes em comparação com a dimensionalidade do modelo ou há dados em excesso. Outro desafio do método de regularização é a própria disponibilidade de um modelo $\mathbf{A}$, que pode não ter acurácia suficiente ou mesmo não estar disponível \cite[pág. 3]{Arridge2019}. Isso é ainda mais crítico em problemas não-lineares, em que nem sempre existem garantias teóricas para obtenção dos parâmetros ótimos do modelo \cite{Adler2021}. 

Fica claro que já existem muitas propostas e há margem para novas soluções. Mesmo assim, uma pergunta relevante foi feita em \cite{Arnrich2011}: O que podemos esperar realisticamente da solução de um problema mal-posto através da regularização? Neste trabalho, os autores discutem as vantagens e desvantagens de regularização em comparação com outro método de solução para um mesmo problema. É imediato se questionar: quais são os limites da regularização? Em \cite[pág. 32]{engl1996regularization}, os autores falam como não é possível tornar estável um problema inerentemente instável com truques matemáticos. O que regularização faz é recuperar uma informação parcial sobre a solução da forma mais estável possível, mas ainda existe o compromisso entre acurácia e estabiidade. 

Vapnik foi um profundo conhecedor do métodos de regularização e deixou explícito quais eram as suas diferenças em relação à minimização do risco estrutural (SRM), proposto por ele \cite[págs. 419--21, 476--7]{Vapnik2006}: 
\begin{itemize}
\item O método de regularização foi proposto no contexto de problemas mal-postos e busca controlar a suavidade de um conjunto de funções admissíveis, mas requer conhecimento do problema a ser resolvido. De modo geral, não apresenta limitantes garantidos para um número finito de observações. Em problemas de dimensionalidade alta, Vapnik argumenta que regularização de Tikhonov não é suficiente para inferência, além de que os teoremas sobre as quais a regularização de baseia garantem apenas a convergência de uma sequência de soluções; 
\item A SRM foi proposta no contexto de problemas preditivos. Ela controla a diversidade do conjunto de funções admissíveis, sem requerer fortes restrições nesse conjunto, além de apresentar limitantes garantidos para um número finito de observações, controlando a generalização dessa forma. 
\end{itemize}
Em \cite[págs. 968--9]{Cherkassky2009}, os autores reconhecem que o termo regularização as vezes é utilizado para denotar qualquer técnica que permite controle de complexidade. Nessa interpretação tão abrangente, diversas metodologias podem ser consideradas formas de regularização, incluindo a SRM e SVM. Os autores argumentam que tal interpretação abrangente tem pouca substância técnica, já que a importância do controle de complexidade é justificada dentro da Teoria Vapnik–Chervonenkis. Novamente há aspectos de contexto de utilização, vantagens e desvantagens de cada proposta.

Dependendo da referência, observa-se um desvio em maior ou menor grau da definição original de regularização de Tikhonov, o que motiva (ou torna necessário) escrever expressões como \textit{regularização no sentido de Tikhonov} encontrada em \cite{Bertero2021} ou \textit{Regularização (de Tikhonov)} em \cite{Gerth2021}. Dado que o uso da palavra \textit{regularização} não é necessariamente a regularização de Tikhonov, o livro discutiu a importância de que cada trabalho defina a definição de regularização que está sendo utilizada, delimitando seu escopo e tornando mais clara a justificativa de sua utilização.  

\newpage
\clearpage
\phantomsection
\addcontentsline{toc}{section}{REFERÊNCIAS BIBLIOGRÁFICAS}
{\singlespacing
 \renewcommand{\refname}{REFERÊNCIAS BIBLIOGRÁFICAS}
\renewcommand*{\bibfont}{\normalfont\small}
 \printbibliography
}

\newpage
\part*{APÊNDICES}\label{partfinal}
\phantomsection
\addcontentsline{toc}{part}{APÊNDICES}
\vspace{1cm}
\epigraph{«\textit{It often happens that instead of trying to discover an event by means of a more or less imperfect knowledge of the law, the events may be known, and we want to find the law; or that, instead of deducing effects from causes, we wish to deduce the causes from the effects. Now, these problems are classified as probability of causes, and are the most interesting of all from their scientific applications. (...) It may be said that it is the essential problem of the experimental method}.»}{Henri Poincaré \cite{poincare1917, poincare2018science}}

\newpage
\setstretch{1.25}
\begin{appendices}
\normalsize
\section{Notação do Espaço  $\ell _{p}$ das Normas}\label{Ap:normas1}

 A norma é uma função que mapeia números reais ou complexos para números não-negativos e apresenta as propriedades da desigualdade triangular, da homogeneidade absoluta e de ser positiva definida \cite[pág. 68-9]{golub2013matrix}. Quando a última propriedade tal não é respeitada, ou seja, que o resultado possa ser nulo, mesmo com o argumento não-nulo, ela é chamada de seminorma. 

No caso de vetores, seja $p \geq  1$ um número real, $n$ o número de elementos do vetor e $1\leq i \leq n$. A norma $\ell _{p}$ de um vetor $\mathbf{x} =(x_{1},\ldots ,x_{n})$ de números reais é definida por \cite[pág. 69]{golub2013matrix}
\begin{equation}
\begin{aligned}
| | \mathbf{x} ||_{p} & =   \left(\sum _{i=1}^{n}\left | x_{i}\right|^{p}\right)^{1/p}    \\
& =  \left(| x_{1}|^{p}+| x_{2}|^{p}+\dots +| x_{n}|^{p}\right)^{1/p},     
\end{aligned}
\label{eq:lpnorm}
  \end{equation}
  onde $\vert x_i\vert$ é o valor absoluto, ou módulo, do i-ésimo valor de $\mathbf{x}$.  
  
  Nesse caso, a norma pode ser entendida como uma medida de distância em relação à origem. Caso o argumento da norma seja uma diferença entre vetores $| | \mathbf{x} - \mathbf{y} ||_{p}$, a norma representa a distância entre eles. 
  
Seja também o caso em que a Equação \eqref{eq:lpnorm} é elevada à potência $p$, isto é, 
\begin{equation}
\vert \vert \mathbf{x} \vert\vert_{p}^p =    \vert x_{1}\vert^{p}+\vert x_{2}\vert^{p}+\dotsb +\vert x_{n}\vert^{p}.
\label{eq:lpnormp}
  \end{equation}
Partindo das Equações \eqref{eq:lpnorm} e \eqref{eq:lpnormp}, são destacados alguns casos específicos:
\begin{itemize}
 \item Para $p = 1$, tem-se a distância absoluta:
\begin{equation}
\begin{aligned}
\vert \vert \mathbf{x} \vert\vert_{1} & =   \left(\vert{x_{1}\vert}^{1}+{\vert x_{2}\vert}^{1}+\dotsb +{\vert x_{n}\vert}^{1}\right)^{1/1}  \\
     & =    \vert x_{1}\vert + \vert x_{2}\vert + \dotsb +\vert x_{n}\vert.
\end{aligned}
 \end{equation}
 Observa-se que $\vert \vert \mathbf{x} \vert\vert_{1} = \vert \vert \mathbf{x} \vert\vert_{1}^1$, onde o expoente $1$ é usualmente omitido.
 
    \item Para $p = 2$, tem-se a norma euclidiana:
  \begin{equation}
  \begin{aligned} 
\vert \vert \mathbf{x} \vert\vert_{2}& =  \left(\vert x_{1}\vert^{2}+\vert x_{2}\vert^{2}+\dotsb +\vert x_{n}\vert^{2}\right)^{1/2} \\       
& =  \left({x_{1}}^{2}+{x_{2}}^{2}+\dotsb +{x_{n}}^{2}\right)^{1/2}.
    \end{aligned}
\end{equation}
Outras representações e notações são obtidas com o produto interno conforme $\vert \vert \mathbf{x} \vert\vert_{2} = \sqrt{\langle \mathbf{x}, \mathbf{x} \rangle} = \sqrt{\mathbf{x} \cdot \mathbf{x}} = \sqrt{\mathbf{x}^T \mathbf{x}}$. 
 \item O quadrado da norma euclidiana é dado por:
\begin{equation}
\begin{aligned}
\vert \vert \mathbf{x} \vert\vert_{2}^2 & =   \mathbf{x}^T \mathbf{x} \\
& =   {x_{1}}^{2}+{x_{2}}^{2}+\dotsb +{x_{n}}^{2}.
\end{aligned}
\end{equation}
 
 \item O quadrado da norma euclidiana ponderada \cite[pág. 666]{Sayed2003} pode ser denotado:
\begin{equation}
\begin{aligned}
\vert \vert \mathbf{x} \vert\vert_{W}^2  & =   \mathbf{x}^T \mathbf{W} \mathbf{x} \\
 & = \vert \vert \mathbf{W}^{\frac{1}{2}} \mathbf{x} \vert\vert_{2}^2.
\label{eq:weighted}
\end{aligned}
\end{equation}
\item Para $p = \infty$, tem-se a chamada norma infinito, também conhecida como norma do supremo ou norma uniforme, que retorna a maior magnitude entre os elementos de um vetor. Matematicamente isso é representado por 
\begin{equation}
\vert \vert \mathbf{x} \vert\vert_{\infty }=\max \left\{\vert x_{1}\vert,\vert x_{2}\vert,\dotsc ,\vert x_{n}\vert\right\}.  
\end{equation}
\item Para $0 < p<1$, a propriedade da desigualdade triangular não é respeitada, de modo que neste caso não se tem uma norma propriamente dita. Para fins de visualização também será utilizada a Equação \eqref{eq:lpnorm}.
\item Para $p=0$, também não se tem uma norma pela definição, mas há autores que entendem $\vert \vert \mathbf{x} \vert\vert_{0}$ como a contagem de números não-nulos de $\mathbf{x}$ \cite{Donoho2006}.
\end{itemize}

\subsection{Visualização de diferentes normas $\ell_p$}
\begin{figure}[H]
    \centering
    \includegraphics[width = 0.65\linewidth]{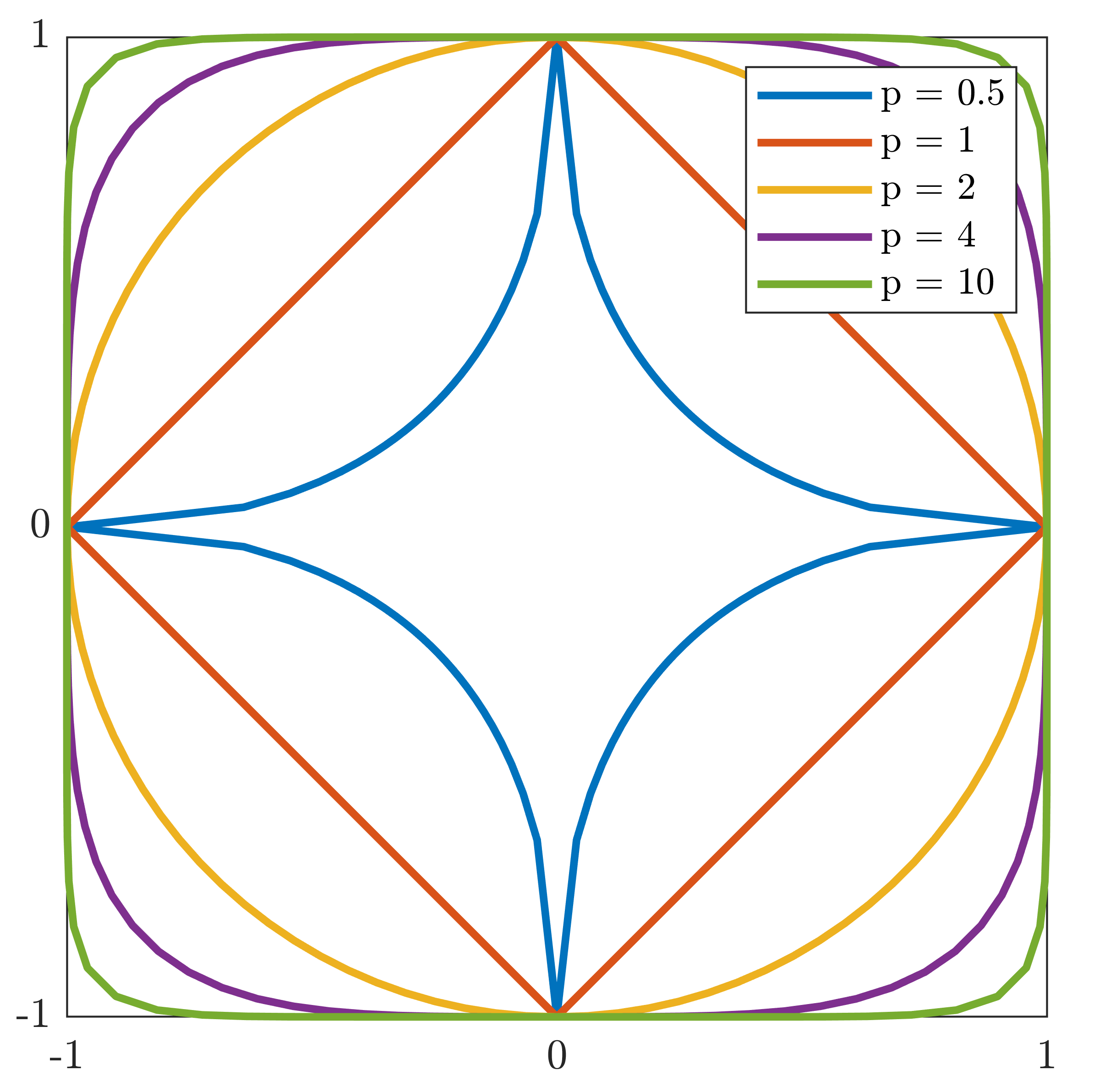}
    \caption[Normas visualizadas no círculo unitário.]{Normas visualizadas no círculo unitário. Fonte: Próprio Autor. }
    \label{fig:my_labelaaa}
   
\end{figure}

Seja o vetor com dois elementos $\mathbf{x} = (x_1, x_2)$ e seja $|| \mathbf{x} ||_{p} = 1$. Na Figura \ref{fig:my_labelaaa} são mostrados os gráficos da Equação \eqref{eq:lpnorm} para diferentes valores de $p$. 

Valores de $p<1$ não definem uma norma, mas é importante a visualização da sua não-convexidade.  Extrapolando essa ideia, a norma para $p = 0$ estaria nos próprios eixos coordenados, que foram omitidos para melhor visualização dos demais gráficos. O outro caso limite é norma infinito, que nessa mesma figura é o quadrado com centro em (0,0) e lado de tamanho = 2, em volta dos demais gráficos. 

\subsubsection{Comparação entre as normas $\ell_2$, $\ell_1$ e $\ell_0$ }
\begin{figure}[H]
     \centering
     \begin{subfigure}[b]{0.45\textwidth}
         \centering
         \includegraphics[width=\textwidth]{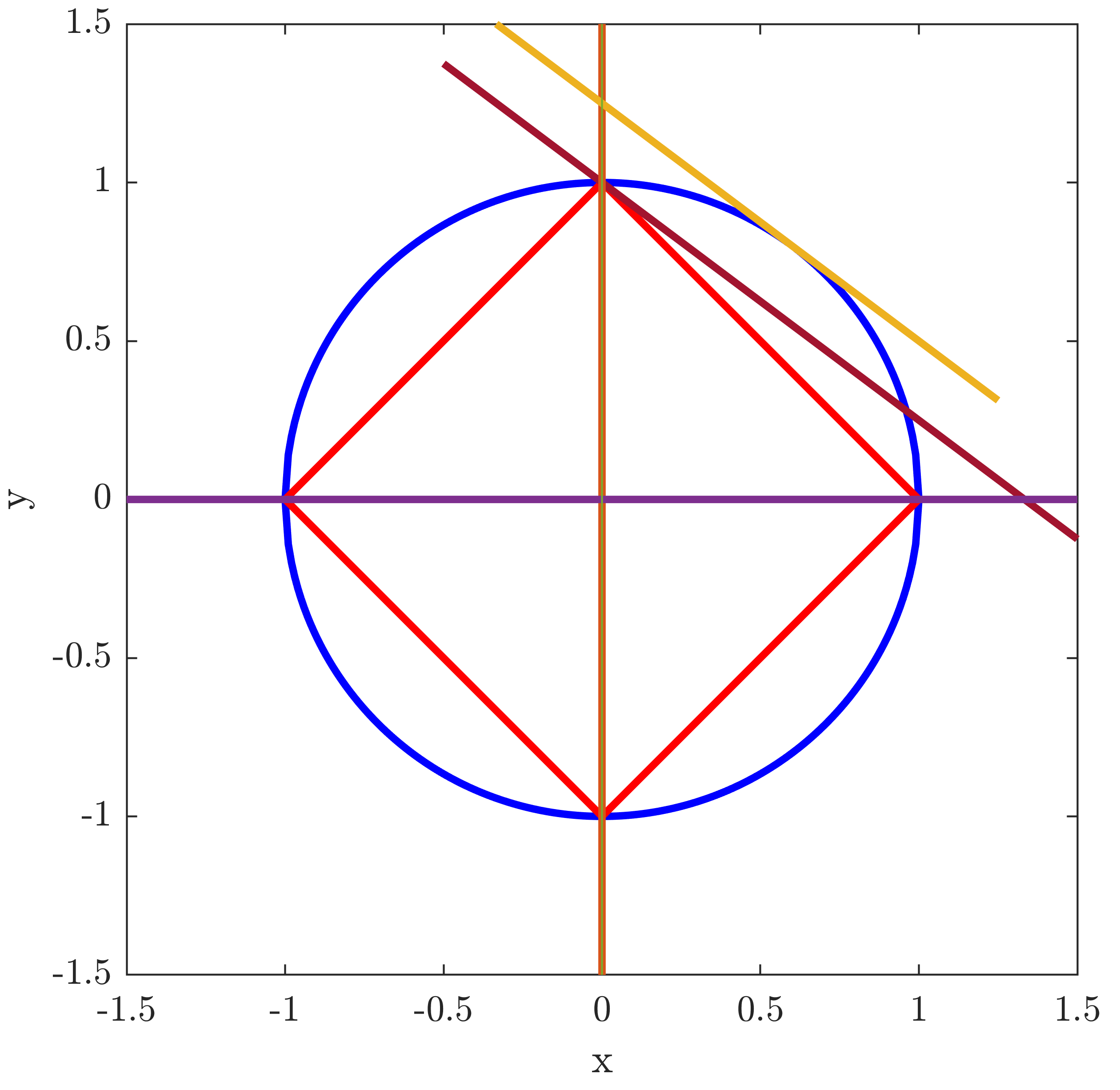}
         \caption{a) Primeira visualização}
         \label{fig:01_021a}
     \end{subfigure}
     \hfill
     \begin{subfigure}[b]{0.45\textwidth}
         \centering
         \includegraphics[width=\textwidth]{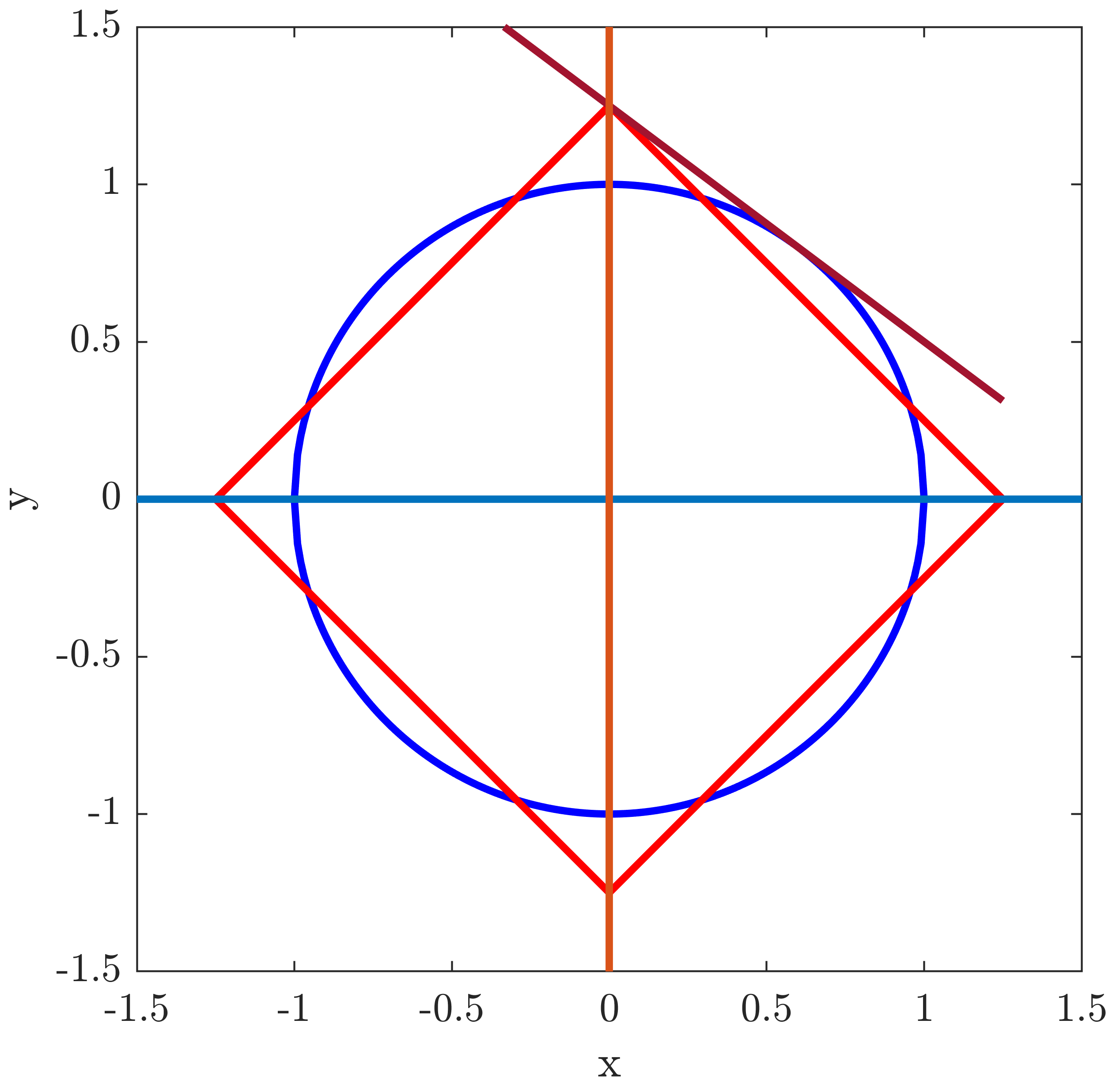}
         \caption{b) Segunda visualização}
         \label{fig:01_021b}
     \end{subfigure}
         \caption[Duas visualizações para soluções com normas $\ell_1$ e $\ell_2$.]{Duas visualizações para soluções com normas $\ell_1$ e $\ell_2$. Fonte: Próprio autor.}
        \label{fig:01_021}
\end{figure}

A Figura \ref{fig:01_021} mostra duas visualizações\footnote{Provenientes do conceito de \textit{ball} \cite[Capítulo 2]{Gopal2020}, nesse caso uma \textit{ball} unitária com norma $\ell_p$.} semelhantes \cite[Subseção 2.10]{Gopal2020}. Elas são importantes para explicar a promoção da esparsidade, bem como utilizada para ilustrar a equivalência entre soluções de normas $\ell_1$ e $\ell_0$, pois, sob certas condições, a norma $\ell_0$ poderia resultar na mesma solução que a norma $\ell_1$ \cite[págs. 30-1]{majumdar2019compressed}. 

Na Figura \ref{fig:01_021a}, baseada em \cite[Figura 2.21]{Gopal2020}:
\begin{itemize}
\item A circunferência azul-escuro é obtida a partir de uma equação de norma $\ell_2$, $\vert \vert \mathbf{x} \vert \vert_2 = 1$ e o quadrado vermelho é obtido a partir de uma equação de norma $\ell_1$, $\vert \vert \mathbf{x} \vert \vert_1 = 1$;
\item Uma solução da minimização da norma $\ell_2$ restringida pela reta amarela $\mathbf{A} \mathbf{x}_1 = \mathbf{y}$ é o ponto onde a reta e a circunferência se tangenciam, com $\mathbf{x}$ e $\mathbf{y}$ não-nulos;
\item Uma solução da minimização da norma $\ell_1$ restringida pela reta roxa $\mathbf{A} \mathbf{x}_2 = \mathbf{y}$ também é o ponto onde elas se interceptam, agora no eixo $\mathbf{y}$, onde um parâmetro é nulo e um é não-nulo, uma solução mais esparsa do que $\mathbf{x}_1$ obtida com a norma $\ell_2$.
\end{itemize}

Na Figura \ref{fig:01_021b}, baseada em \cite[pág. 182]{aster2019parameter}, a ideia é a mesma, mas a norma do quadrado deixa de ser unitária para que a solução esteja contida na mesma reta azul, tangenciando tanto a circunferência quanto o quadrado simultaneamente.

\newpage
\section{Matrizes pseudoinversas} \label{Ap:sobredet}

Neste apêndice, são desenvolvidos os conceitos apresentados na Subseção \ref{sec:excess}. 

\subsection{Sistema sobredeterminado}
Seja um sistema linear de equações do tipo $\mathbf{A}\mathbf{x} = \mathbf{y}$.  Se a matriz $\mathbf{A}$ possuir mais equações do que variáveis (mais linhas do que colunas), um sistema sobredeterminado, não há solução exata e só é possível obter uma solução aproximada. Pode-se buscar, por exemplo, a solução que minimize a soma do quadrado entre os termos da esquerda e da direita do sistema linear \cite[pág. 29]{matrixcookbook}, conforme
\begin{equation}
\hat{\mathbf{x}} = \arg\min\limits_{\mathbf{x}} \left( \vert \vert \mathbf{A}\mathbf{x} - \mathbf{y} \vert \vert^2_2 \right).
    \label{eq:otimizacao12} 
\end{equation}
Pode-se reescrever a expressão anterior como
\begin{equation}
\begin{aligned}
    \vert \vert  \mathbf{A}\mathbf{x} - \mathbf{y}\vert \vert^2_2 & = \left( \mathbf{A}\mathbf{x} - \mathbf{y}\right)^T \left( \mathbf{A}\mathbf{x} - \mathbf{y}\right) \\
    & = \mathbf{y}^T\mathbf{y} - \mathbf{y}^T \mathbf{A}\mathbf{x} - \mathbf{x}^T \mathbf{A}^T \mathbf{y} + \mathbf{x}^T \mathbf{A}^T \mathbf{A} \mathbf{x},
    \label{eq:funcional_app}
      \end{aligned}
\end{equation}
de modo que o seu gradiente em relação a $\mathbf{x}$ é obtido conforme
\begin{equation}
\nabla_\mathbf{x} \left(\vert \vert  \mathbf{A}\mathbf{x} - \mathbf{y}\vert \vert^2_2\right) = - 2 \mathbf{A}^T\mathbf{y} + 2 \mathbf{A}^T \mathbf{A} \mathbf{x}.
    \label{eq:funcional_app2}
\end{equation}
Pela condição de otimalidade de primeira ordem \cite[Equação 4]{Gerth2021}, iguala-se a zero o gradiente em relação a $\mathbf{x}$, obtido na Equação \eqref{eq:funcional_app2}, de acordo com
\begin{equation}
\nabla_\mathbf{x} \left(\vert \vert  \mathbf{A}\mathbf{x} - \mathbf{y}\vert \vert^2_2\right) = \mathbf{0}.
    \label{eq:funcional_app2-2}
\end{equation}
 Em seguida, rearranjando os termos e multiplicando os dois lados por $\frac{1}{2}$ obtém-se
\begin{equation}
\mathbf{A}^T \mathbf{A}\mathbf{x} = \mathbf{A}^T \mathbf{y}.
\end{equation}
Logo, a solução para este problema é dada por
\begin{equation}
\hat{\mathbf{x}}   = \left(\mathbf{A}^T \mathbf{A} \right)^{-1} \mathbf{A}^T \mathbf{y}.
    \label{eq:normalequation_app}
\end{equation}
Em alguns trabalhos como \cite{hansen2010discrete}, a Equação \eqref{eq:otimizacao12} é apresentada apenas com a norma $\ell_2$, sem estar elevada ao quadrado, conforme
\begin{equation}
\vert \vert \mathbf{A}\mathbf{x} - \mathbf{y} \vert \vert_2 = \sqrt{ \sum_{i=1}^m  \vert\left(A x\right)_i - y_i \vert^2 }, 
\label{eq:normalequation67}
\end{equation}
onde $i$ indica a linha do vetor. 

Usualmente evita-se a raiz quadrada para resolver o problema computacionalmente mais simples. Apesar disso, o mínimo obtido pela otimização não se altera, pois a raiz quadrada é uma função monótona crescente.

\subsection{Sistema subdeterminado}\label{subdet} 
Seja o sistema $\mathbf{A}\mathbf{x} = \mathbf{y}$. Se a matriz $\mathbf{A}$ possui mais variáveis do que equações (mais colunas no que linhas), um sistema subdeterminado, e há infinitas soluções. Isso significa que muitas escolhas possíveis para $\mathbf{x}$ podem resultar no mesmo $\mathbf{y}$. Neste caso, uma opção é restringir as soluções cuja norma do vetor $\mathbf{x}$ seja mínima \cite[pág. 29]{matrixcookbook}, conforme
\begin{equation}
\hat{\mathbf{x}} = \underset{\mathbf{x}}{\arg\min}\vert \vert \mathbf{x} \vert \vert_2^2 \quad \text{s.t.} \quad \mathbf{A}\mathbf{x} = \mathbf{y}.
   \label{casosub_sol_app}
\end{equation}
Pode-se transformar este problema de otimização com restrições em um problema de otimização sem restrições utilizando o método dos multiplicadores de Lagrange. Seja
\begin{equation}
\mathcal{M}(\mathbf{x},\lambda) =   \vert \vert \mathbf{x} \vert \vert^2_2  + \lambda \left(\mathbf{y} - \mathbf{A}\mathbf{x} \right).
\end{equation}
A primeira condição de otimalidade é $\nabla_\mathbf{x}  \mathcal{M}(\mathbf{x},\lambda)  = 0$, de modo que
\begin{equation}
\begin{aligned}
\mathbf{0} & =  2\mathbf{x} + \mathbf{A}^T\lambda  \\
 \mathbf{x} & =  -\mathbf{A}^T \frac{\lambda }{2}.
   \label{eq:x}
 \end{aligned}
\end{equation}  
A segunda condição de otimalidade é $ \nabla_{\lambda}  \mathcal{M}(\mathbf{x},\lambda)  = 0$, resultando em
\begin{equation}
   \mathbf{A}\mathbf{x} - \mathbf{y} = \mathbf{0}.
\end{equation}
Substituindo $\mathbf{x}$ obtido na Equação \eqref{eq:x}, obtém-se
\begin{equation}
\begin{aligned}
  \mathbf{0}& =    \mathbf{A}\left( -\mathbf{A}^T  \frac{\lambda }{2}\right) - \mathbf{y}  \\
 \mathbf{y} & =         -\frac{\lambda }{2} \mathbf{A} \mathbf{A}^T  \\   
    \lambda & = -2(\mathbf{A} \mathbf{A}^T)^{-1} \mathbf{y}.   
        \end{aligned}
\end{equation}
Substituindo $\lambda$ de volta na Equação \eqref{eq:x}, a solução é obtida conforme
\begin{equation}
\hat{\mathbf{x}} = \mathbf{A}^T \left(\mathbf{A} \mathbf{A}^T\right)^{-1} \mathbf{y}.
    \label{normalequation4}
\end{equation}

\subsection{Interpolação com matriz pseudoinversa}\label{Ap:pseudo}

\begin{figure}[H]
\centering
	    \includegraphics[width=0.75\textwidth]{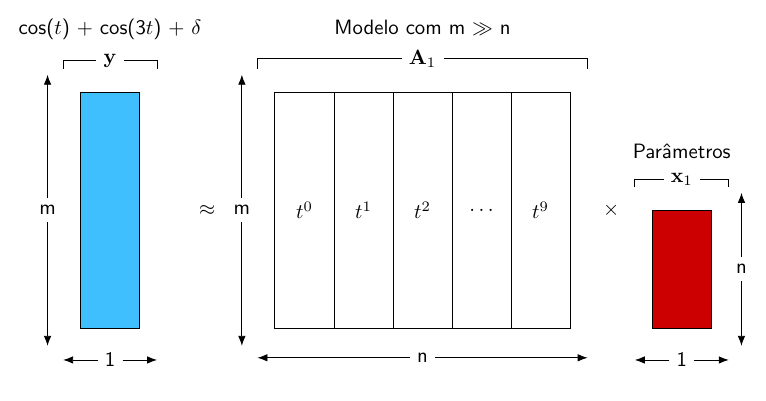}
\caption[Sistema de equações da interpolação com polinômios.]{Sistema de equações da interpolação com polinômios. Fonte: Próprio autor.}
\label{fig:02_100}
\end{figure}

Suponha uma função $y(t) = cos(t) + cos(3t) + \delta$ para $\pi < t < 3\pi$ e $\delta  \sim \mathcal{N}(0, 0.1)$. A partir de medidas discretas $\mathbf{y}$ busca-se uma representação desse sinal em uma base de polinômios dada pela matriz $\mathbf{A}_1$ cujo vetor de parâmetros é $\mathbf{x}_1$, conforme Figura \ref{fig:02_100}.

\begin{figure}[H]
\centering

\begin{subfigure}[b]{0.49\textwidth}
\centering
	    \includegraphics[width=1\textwidth]{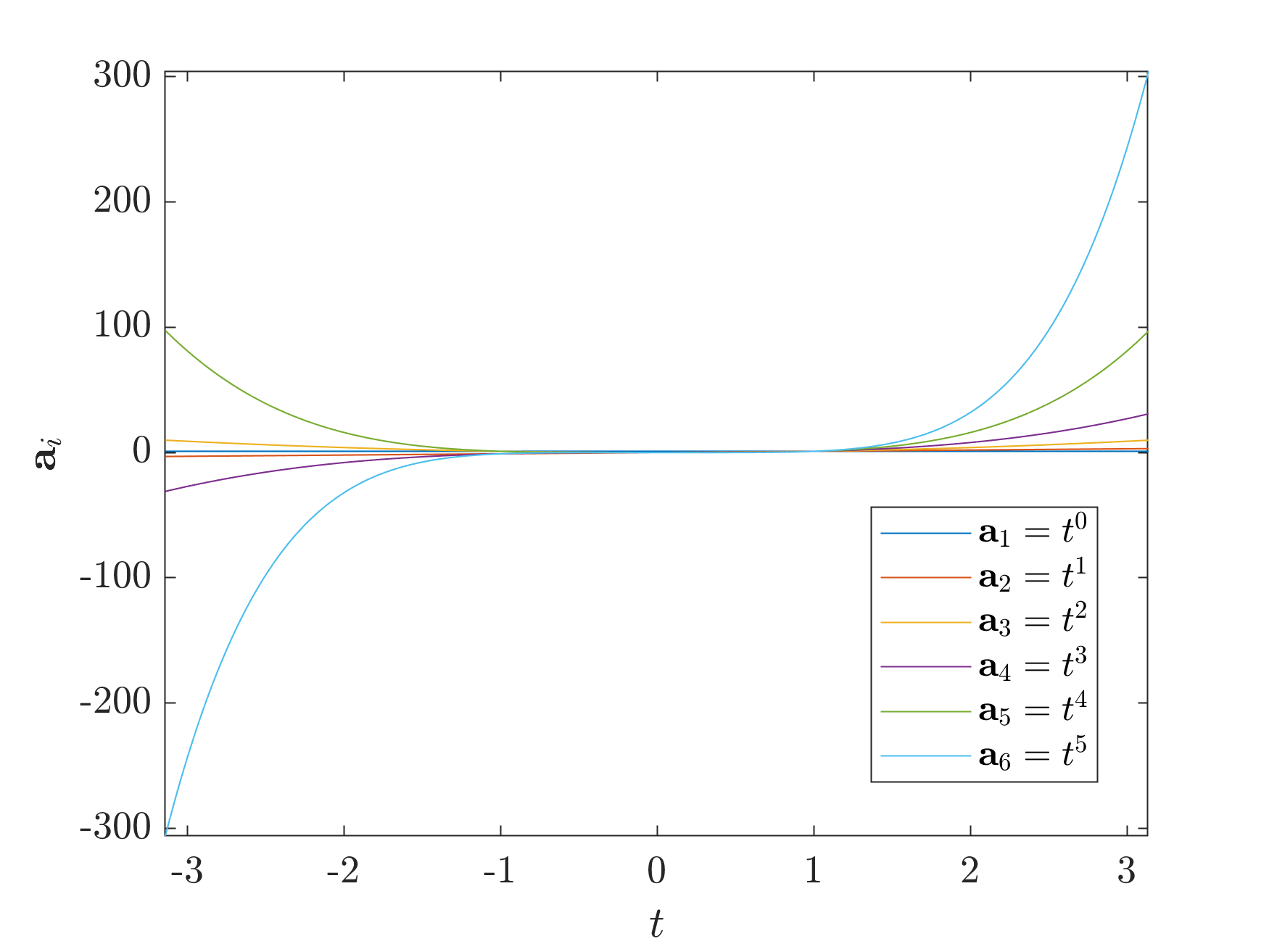}
\caption{Colunas de $\mathbf{A}_1$.}
\label{fig:02_101}
  \end{subfigure}  
\begin{subfigure}[b]{0.49\textwidth}
\centering
	    \includegraphics[width=1\textwidth]{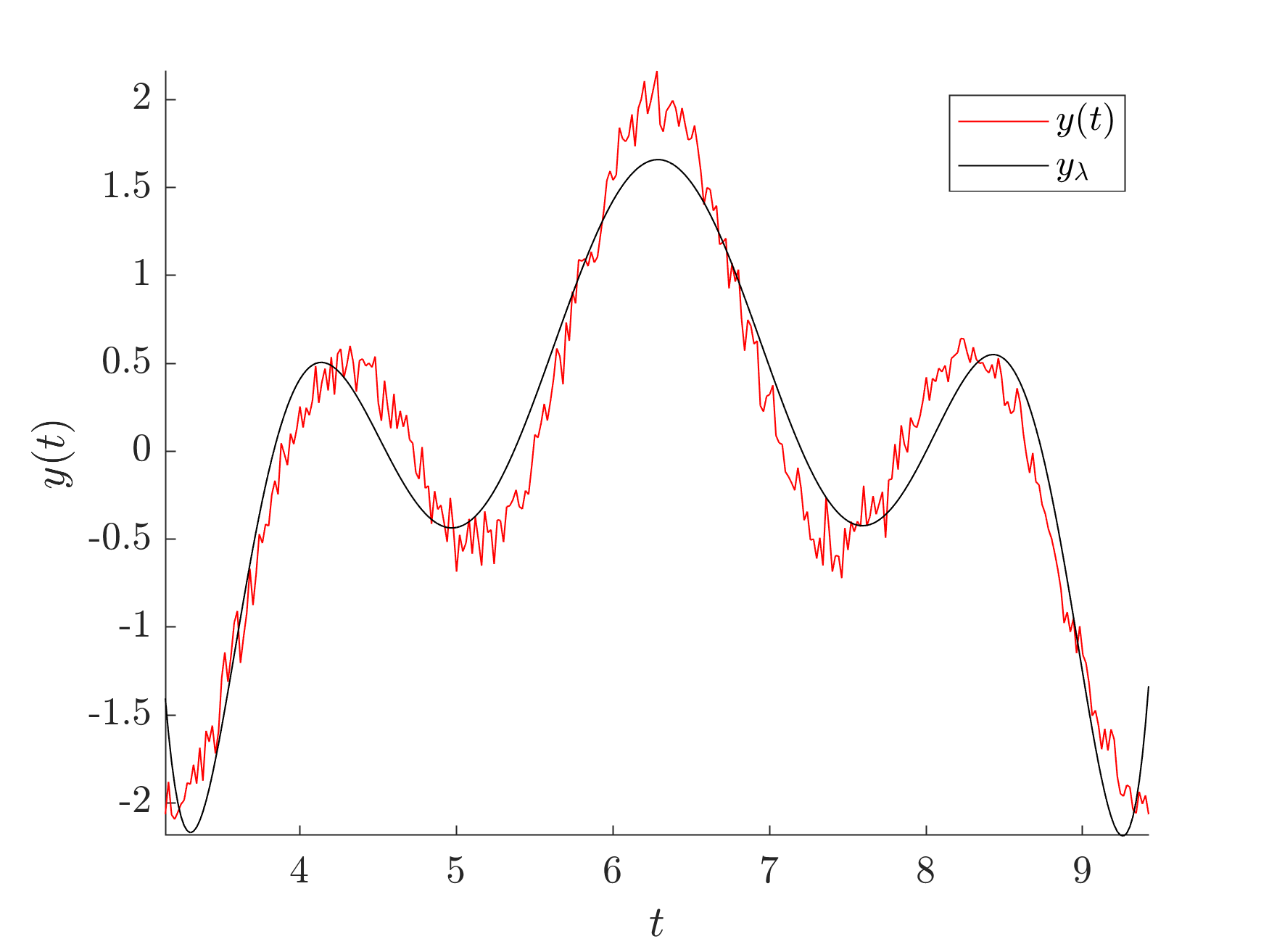}
\caption{Resultado.}
\label{fig:02_102}
      \end{subfigure}  
\caption[Interpolação com polinômios.]{Interpolação com polinômios. Fonte: Próprio autor.}
\label{fig:02_10}
\end{figure}

Cada coluna da matriz $\mathbf{A}_1$ representa a ordem de um polinômio. Para $n= 10$, do grau zero até o polinômio de 9º grau. Ao invés de calcular seus valores no intervalo original, eles foram calculados em torno de zero, no intervalo $-1.7 < t < 1.7$, para evitar valores muito grandes. As primeiras colunas são mostradas na Figura \ref{fig:02_101}.

As colunas da matriz $\mathbf{A}_1$ são independentes, sendo possível realizar a interpolação como um problema de mínimos quadrados conforme Equação \eqref{eq:normalequation}. A saída $\mathbf{y}_{\mathbf{A}1} =  \mathbf{A}_1 \mathbf{x}_1$ é mostrada na Figura \ref{fig:02_102}. A aproximação acompanha a forma geral da curva original, mas não totalmente os pontos onde as direções mudam.

\begin{figure}[H]
\centering
	    \includegraphics[width=0.75\textwidth]{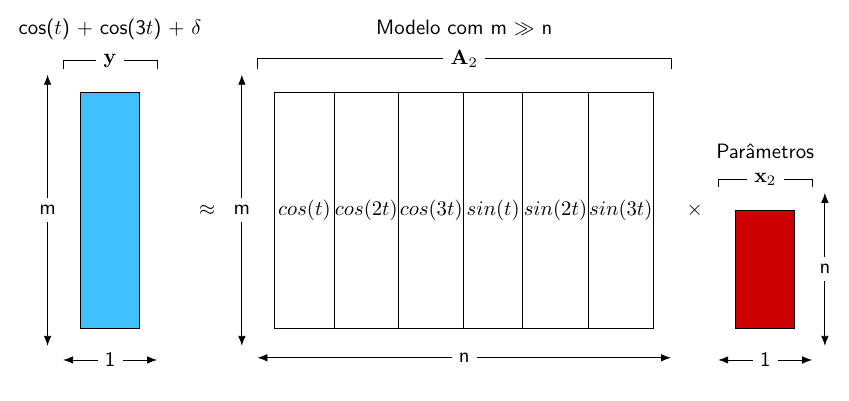}
\caption[Sistema de equações da interpolação com senóides.]{Sistema de equações da interpolação com senóides. Fonte: Próprio autor.}
\label{fig:02_1088}
\end{figure}

\begin{figure}[H]
\centering

\begin{subfigure}[b]{0.49\textwidth}
\centering
	    \includegraphics[width=1\textwidth]{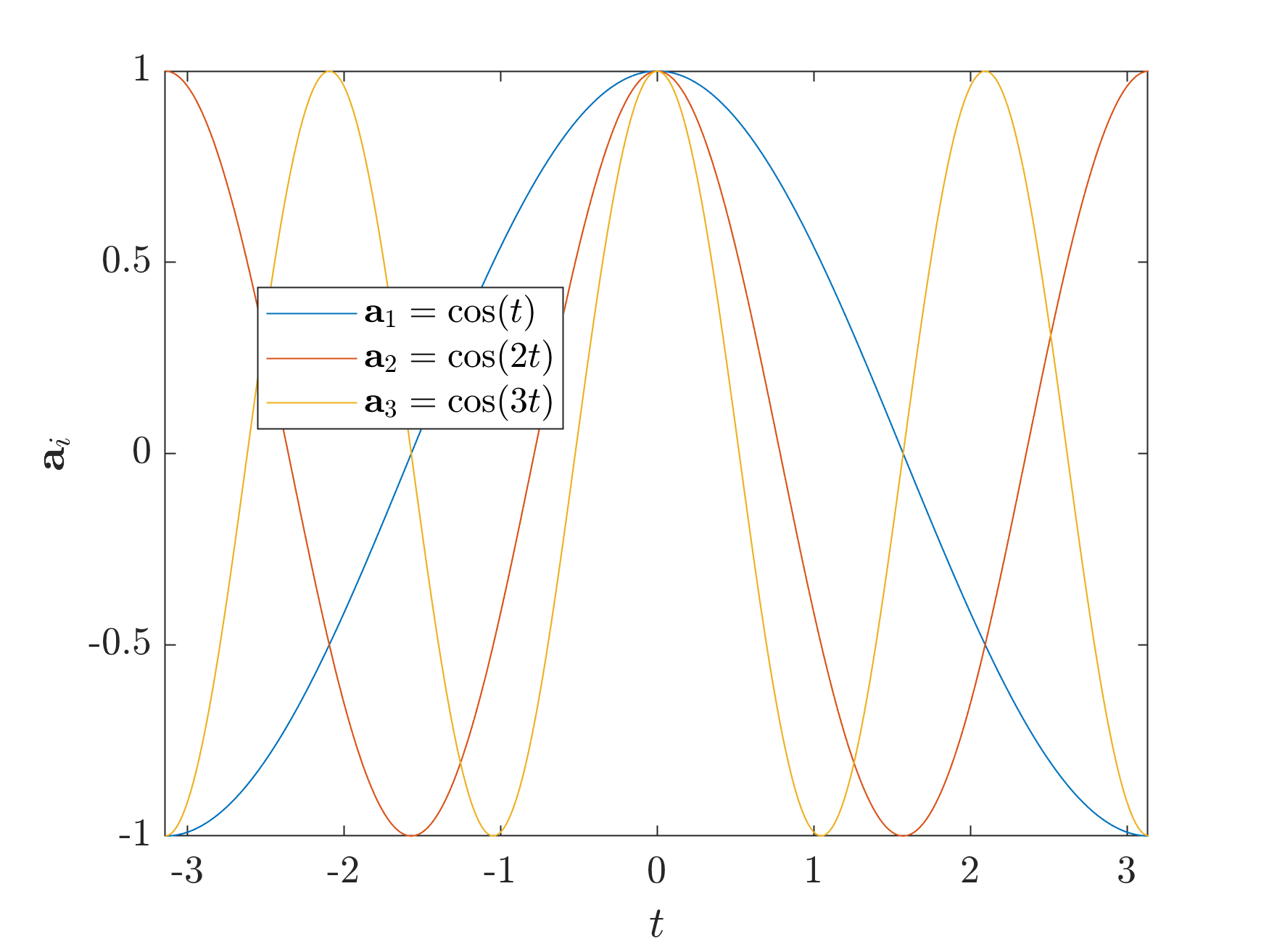}
\caption{Colunas de $\mathbf{A}_2$.}
\label{fig:02_111}
  \end{subfigure}  
\begin{subfigure}[b]{0.49\textwidth}
\centering
	    \includegraphics[width=1\textwidth]{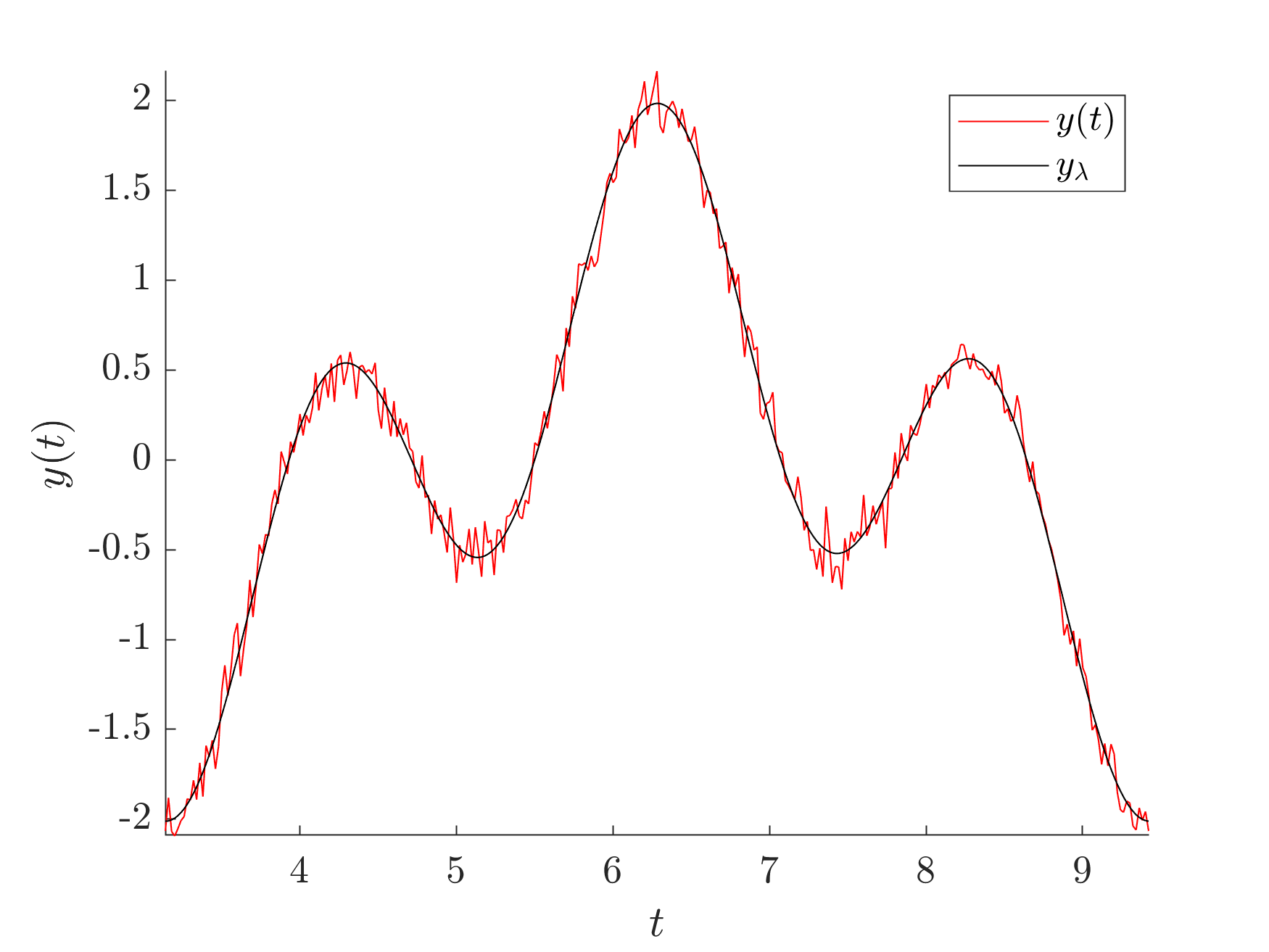}
\caption{Resultado.}
\label{fig:02_112}
      \end{subfigure}  
\caption[Interpolação com senos e cossenos.]{Interpolação com senos e cossenos. Fonte: Próprio autor.}
\label{fig:02_11}
\end{figure}

Na sequência, o mesmo procedimento é realizado, mas através de uma matriz $\mathbf{A}_2$ cujas colunas são dadas por $cos(t)$, $cos(2t)$, $cos(3t)$, $sin(t)$, $sin(2t)$ e $sin(3t)$, ou seja, incluem a resposta certa. Na Figura \ref{fig:02_1088} há o esquema de interpolação e na Figura \ref{fig:02_111} há a visualização das três primeiras colunas dessa matriz. Dessa vez, os valores de suas colunas foram calculados no intervalo $-\pi < t < \pi$, que não é exatamente o mesmo, apesar de também ter comprimento de $2\pi$ como o sinal original.

Calculando-se os parâmetros a partir da Equação 
\begin{equation}
\mathbf{x}_2 = \left(\mathbf{A}_2^T \mathbf{A}_2 \right)^{-1} \mathbf{A}_2^T\mathbf{y},
\end{equation}
o resultado $\mathbf{y}_{\mathbf{A}2} =  \mathbf{A}_2 \mathbf{x}_2$ é mostrado na Figura \ref{fig:02_112}, indicando que essa é uma base melhor de representação para o sinal do que a de polinômios. Agora, o resultado é o esperado. 

Esse exemplo ilustra a importância do operador direto como uma base de representação. Quanto mais correto ele estiver, melhor será a representação do sinal.

\newpage
\section{Inversão bayesiana}\label{Ap:estimador}

O método de regularização de Tikhonov traz soluções determinísticas, trazendo uma estimativa razoável das grandezas de interesse a partir dos dados disponíveis \cite[pág. 2]{kaipio2005statistical}. Em seu livro, Tikhonov já utilizava a norma quadrática como regularizador \cite[págs. 72-3]{tikhonov1977solutions}. Em certas aplicações, a busca por soluções suaves é esperada, mas outras informações disponíveis sobre a solução podem existir. 

Nesse contexto, a teoria de inversão estatística é relevante, pois também busca resolver problemas mal-postos, mas agora como um problema de inferência bayesiana, que exige uma densidade de probabilidade \textit{a priori} sobre a incógnita para eliminar modelos que não fazem sentido, mesmo que eles se ajustem aos dados. 

Essa informação \textit{a priori}\footnote{Vale notar que Tikhonov usava a expressão \textit{informações suplementares} \cite[págs. 9, 148]{tikhonov1977solutions} no contexto de regularização, ou mesmo usava a expressão informação \textit{a priori} quando se considerava as incógnitas e os ruídos como processos estocásticos \cite[pág. 149]{tikhonov1977solutions}, mas sem um \textit{framework} bayesiano.}, também conhecido como \textit{prior}, se refere às informações disponíveis que indiquem como a grandeza de interesse deve ser, mas antes do processo de medição. Isto é, o \textit{prior} deve ser independente dos dados $\mathbf{y}$ observados que serão utilizados na reconstrução \cite[Subseção 3.2]{calvetti2007introduction}, \cite[pág. 5]{Calvetti2018a}, \cite[págs. 51, 113]{kaipio2005statistical}. 

 Entre os \textit{priors} possíveis nesse \textit{framework}, estão aqueles que não são facilmente expressadas em termos quantitativos \cite{Calvetti2018a}, como os atlas anatômicos \cite{Moura2021}. Logo, a abordagem bayesiana também traz novas formas de incorporação de informação \textit{a priori} à solução do problema inverso. Conforme será visto, a escolha do \textit{prior} pode ser relacionado com o regularizador da solução de Tikhonov em uma solução determinística.

A inversão bayesiana também parte de um modelo $\mathbf{A}$ que relaciona os seus parâmetros $\mathbf{x}$ com dados $\mathbf{y}$ que apresentam ruídos $\bm{\delta}$, mas $\mathbf{x}$, $\mathbf{y}$ e $\bm{\delta}$ são considerados realizações de variáveis aleatórias. Modelar o problema considerando $\mathbf{A}$ conhecido significa partir da hipótese da validade do operador direto, mas também é possível incluir incertezas no modelo \cite[pág. 13]{Calvetti2018a}. Em resumo, \cite[pág. 52]{kaipio2005statistical}:
\begin{itemize}
\item Inicia-se pela definição de um modelo de ruído, na forma de uma densidade de probabilidade $\pi_E$. Assumindo que a informação disponível sobre $\mathbf{x}$ antes das medições também pode ser codificada em uma densidade de probabilidade, obtém-se a densidade \textit{a priori}  $\pi_{priori}(\mathbf{x})$; 
\item Definidos o modelo de ruído e o \textit{prior}, as observações $\mathbf{y}$ são relacionadas com as incógnitas $\mathbf{x}$ através de uma função de probabilidade condicional;
\item Por fim, são desenvolvidos estimadores para obter a densidade de probabilidade \textit{a posteriori} $\pi_{post}$. 
\end{itemize}

Nas próximas seções, cada uma dessas etapas será aprofundada. 

\subsection{Modelo de ruído}
Seja uma variável aleatória $Y$ observada através de um processo de medição, onde uma realização $Y = \mathbf{y}$ denotam os dados obtidos. A segunda variável aleatória $X$ é a incógnita que não pode ser observada diretamente, onde uma realização é descrita por $X = \mathbf{x}$. Há ainda um ruído aditivo $E$, da qual $E = \bm{\delta}$ é uma realização da variável aleatória, sendo $X$ e $E$ independentes entre si \cite[pág. 56]{kaipio2005statistical}. Supondo uma relação linear entre $X$ e $Y$, o modelo estocástico resultante é dado por
\begin{equation}
Y = \mathbf{A}X + E, 
\label{eq:esti1}
\end{equation}
onde $\mathbf{A}$ é o operador direto. Isolando-se $E$ na Equação \eqref{eq:esti1}, obtém-se o modelo de ruído
\begin{equation}
E = Y - \mathbf{A}X.
\label{eq:esti2}
\end{equation}
É possível, por exemplo, atribuir uma distribuição gaussiana para a variável aleatória $E$, com média $\mu = 0$, variância $s^2_{\bm{\delta}}$ e covariância $s^2_{\bm{\delta}} \mathbf{I}$ \cite[Exemplos 2 e 5]{kaipio2005statistical}, distribuição que pode ser denotada por $E \sim \mathcal{N}(0,s^2_{\bm{\delta}}\mathbf{I})$.

Para uma única realização de cada variável aleatória, a Equação \ref{eq:esti2} se torna
\begin{equation}
\bm{\delta} = \mathbf{y} - \mathbf{A}\mathbf{x}.
\label{eq:estix}
\end{equation}
 Considerando que as variáveis aleatórias sejam absolutamente contínuas, as suas distribuições de probabilidade podem ser expressas em termos das suas densidades de probabilidade \cite[pág. 50]{kaipio2005statistical}. Associando ao ruído uma densidade de probabilidade $\pi_E(\bm{\delta})$ gaussiana, obtém-se
\begin{equation}
\pi_E(\bm{\delta}) \propto \exp\left( -\frac{1}{2 s^2_{\bm{\delta}} } \vert \vert \bm{\delta}\vert\vert^2 \right).
\label{eq:esti4}
\end{equation}

\subsection{Obtenção da densidade de probabilidade condicional}

Definido o modelo de ruído, é possível obter a densidade de probabilidade condicional. Fixando-se em uma realização $X = \mathbf{x}$, o único fator aleatório em $Y$ é dado por E \cite[pág. 42]{calvetti2007introduction}, \cite[pág. 5]{Calvetti2018a}, \cite[pág. 56]{kaipio2005statistical},  e a densidade condicional de probabilidade se torna
\begin{equation}
\pi_{Y|X}(\mathbf{y}|\mathbf{x}) = \pi_E(\bm{\delta}).
\label{eq:esti5}
\end{equation}
Unindo-se a Equação \eqref{eq:esti4} com a Equação \eqref{eq:esti5} obtém-se
\begin{equation}
\pi_{Y|X}(\mathbf{y}|\mathbf{x}) \propto \exp\left( -\frac{1}{2 s^2_{\bm{\delta}} } \vert \vert \bm{\delta}\vert\vert^2 \right).
\label{eq:esti6}
\end{equation}
Com o modelo de observação linear da Equação \eqref{eq:esti2}, mas no caso de realizações $\mathbf{x}$ e $\mathbf{y}$ como na Equação \eqref{eq:estix}, a densidade condicional de probabilidade fica caracterizada a partir de $\pi_E(\mathbf{A}\mathbf{x}-\mathbf{y})$, isto é, 
\begin{equation}
\pi_{Y|X}(\mathbf{y}|\mathbf{x}) \propto \exp\left( -\frac{1}{2 s^2_{\bm{\delta}} } \vert \vert (\mathbf{A}\mathbf{x}-\mathbf{y})\vert\vert^2 \right),
\label{eq:esti7}
\end{equation}
mas ainda falta um estimador para que a $\mathbf{x}$ seja calculado.

\subsection{Estimador \textit{maximum likelihood}}
Um dos estimadores mais utilizados em estatística mais utilizados é o da máxima verossimilhança (MLE). O seu desenvolvimento matemático para o caso em que os valores observados são realizações independentes de uma mesma variável aleatória é encontrada em \cite[págs. 31-32]{calvetti2007introduction}. Em suma, o MLE é caracterizado por 
\begin{equation}
\hat{\mathbf{x}}_{ML} = \arg\max\limits_{\mathbf{x}} \left[\pi_{Y|X}\left(\mathbf{y} \vert \mathbf{x} \right) \right],
\label{eq:esti8}
\end{equation}
que busca o valor da incógnita $\mathbf{x}$ mais provável de produzir o dado $\mathbf{y}$ \cite[pág. 53]{kaipio2005statistical}. 

O termo $\pi_{Y|X}\left(\mathbf{y} \vert \mathbf{x} \right)$ da Equação \eqref{eq:esti8} é descrito por um produtório da densidade de probabilidade de cada uma das realizações \cite[pág. 31]{calvetti2007introduction}. Assim, quando essas distribuições são gaussianas, elas envolvem funções exponenciais, de modo que pode ser conveniente transformar o lado direito da Equação \eqref{eq:esti8} utilizando a função logaritmo. Como o logaritmo é estritamente crescente, a solução da Equação \eqref{eq:esti8} é equivalente a resolver o problema de minimização dado por
\begin{equation}
\hat{\mathbf{x}}_{ML} = \arg\min\limits_{\mathbf{x}} \left[-\log\left(\pi_{Y|X}\left(\mathbf{y} \vert \mathbf{x} \right) \right) \right],
\label{eq:esti9}
\end{equation}
conhecida como \textit{log-likelihood} \cite[págs. 32, 36]{calvetti2007introduction}.  

Admitindo que o modelo de ruído seja gaussiano e branco e que há um modelo linear que relaciona as variáveis, substituindo-se a Equação \eqref{eq:esti7} na Equação \eqref{eq:esti9}, o problema da máxima verossimilhança se reduz a \cite[págs. 35-7, 42]{calvetti2007introduction}
\begin{equation}
\hat{\mathbf{x}}_{ML} = \arg\min\limits_{\mathbf{x}} \vert \vert \mathbf{A}\mathbf{x} - \mathbf{y}  \vert \vert_2^2, 
\label{eq:esti10} 
\end{equation}
cuja solução é dada pelas equações normais \cite[Seção 9.2.1]{Deisenroth2020}
\begin{equation}
\hat{\mathbf{x}}_{ML} = (\mathbf{A}^T \mathbf{A})^{-1} \mathbf{A}^T \mathbf{y}.
\label{eq:esti11} 
\end{equation}
Logo, o MLE com tal modelo de ruído é equivalente a uma solução não-regularizada e não resolve a instabilidade do problema mal-posto \cite[pág. 28]{aster2019parameter},  \cite[págs. 146-7]{kaipio2005statistical}.

\subsection{Estimador \textit{maximum a posteriori}}

 Como $\mathbf{A}$ é mal-condicionada, são necessárias informações \textit{a priori} sobre $\mathbf{x}$ na forma de uma densidade de probabilidade $\pi_{priori}(\mathbf{x})$, o que era ausente no MLE. O termo $\pi_{priori}(\mathbf{x})$ depende de informações \textit{a priori} de $X$. Especificamente, busca-se $\pi_{priori}$ tal que 
 \begin{equation}
 \pi_{priori}(\mathbf{x}) \gg \pi_{priori}(\mathbf{x}'),
 \end{equation}
  quando $\mathbf{x} \in E$ e $\mathbf{x}' \in U$, onde $E$ é o conjunto dos valores que se pode esperar para realizações da incógnita $X$, enquanto $U$ são os valores de que não se pode esperar para $X$ \cite[pág. 62]{kaipio2005statistical}. A escolha de $\pi_{priori}$ é uma das etapas mais importantes e é possível mostrar que existem alguns casos em há correspondência entre métodos de regularização e o resultado obtido no \textit{framework} bayesiano.

 Pode-se estimar uma densidade \textit{a posteriori} $\pi_{post}(\mathbf{x})$ através do estimador máximo \textit{a posteriori} (MAP) conforme
\begin{equation}
\pi_{post}(\mathbf{x})=\pi_{X|Y}\left(\mathbf{x} \vert \mathbf{y} \right)
\label{eq:esti12}
\end{equation}
\begin{equation}
\hat{\mathbf{x}}_{MAP} = \arg\max\limits_{\mathbf{x}} \pi_{X|Y}\left(\mathbf{x} \vert \mathbf{y} \right).
\label{eq:esti13}
\end{equation}
Os subscritos ${Y|X}$ e ${X|Y}$ serão omitidos daqui para frente por simplicidade na notação.
 
Dada a definição de densidade de probabilidade condicional $\pi\left(\mathbf{x}  \vert  \mathbf{y} \right)$, é possível combinar as informações dos dados $\mathbf{y}$ com a informação \textit{a priori} $\pi_{priori}(\mathbf{x})$ através do teorema de Bayes \cite[pág. 10]{Benning2018}, \cite[pág. 5]{Calvetti2018a}, \cite[Teorema 3.1]{kaipio2005statistical}
\begin{equation}
\pi\left(\mathbf{x} \vert \mathbf{y} \right) = \frac{\pi_{priori}(\mathbf{x})  \pi\left(\mathbf{y} \vert \mathbf{x} \right) }{\pi\left(\mathbf{y} \right) }, 
\label{eq:esti14}
\end{equation}
onde $\pi\left(\mathbf{y} \right) = \int \pi_{priori}(\mathbf{x}) \pi\left(\mathbf{y}\vert \mathbf{x} \right) dx$, termo que pode ser calculado a partir do numerador da própria Equação \eqref{eq:esti14} \cite[pág. 5]{Calvetti2018a}. 

Substituindo a Equação \eqref{eq:esti14} na Equação \eqref{eq:esti13}, 
\begin{equation}
\hat{\mathbf{x}}_{MAP} = \arg\max\limits_{\mathbf{x}} \left( \frac{\pi_{priori}(\mathbf{x})  \pi\left(\mathbf{y} \vert \mathbf{x} \right) }{\pi\left(\mathbf{y} \right) } \right),
\label{eq:esti15}
\end{equation}
dado que o maximizador exista \cite[pág. 53]{kaipio2005statistical}. Em caso afirmativo, maximizar esse funcional é equivalente ao problema de minimização dado por  \cite[pág. 2819-20]{ Chen2002} 
\begin{equation}
\begin{aligned}
\hat{\mathbf{x}}_{MAP} & =   \arg\min\limits_{\mathbf{x}} \left(-\log \left[\frac{\pi_{priori}(\mathbf{x})  \pi\left(\mathbf{y} \vert \mathbf{x} \right) }{\pi\left(\mathbf{y} \right) } \right] \right) \\
& =   \arg\min\limits_{\mathbf{x}} \left[ -\log \pi\left(\mathbf{y} \vert  \mathbf{x}\right) -\log \pi_{priori}(\mathbf{x}) + \log \pi(\mathbf{y}) \right],
\label{eq:esti17}
\end{aligned}
\end{equation}
onde $\log \pi(\mathbf{y})$ resulta em uma constante e, portanto, não afeta o resultado da minimização da Equação \eqref{eq:esti17}, podendo ser desconsiderada \cite[pág. 52]{kaipio2005statistical}.

Observa-se que o problema de otimização inclui uma componente $\pi\left(\mathbf{y} \vert  \mathbf{x}\right)$ igual ao encontrado na Equação \eqref{eq:esti9}, relativo ao termo de fidelidade entre o modelo, seus parâmetros e os dados medidos. Ou seja, partindo das mesmas hipóteses sobre o ruído ser gaussiano e independente de $X$, faz sentido que a expressão $-\log\left(\pi_{Y|X}\left(\mathbf{y} \vert \mathbf{x} \right) \right) $ apareça tanto na Equação \eqref{eq:esti17} quanto na Equação \eqref{eq:esti10}.

\subsubsection{MAP com distribuições normais para ruído e \textit{prior}}\label{sec:normalMAP}
  Além do modelo de ruído ser branco e gaussiano, pode-se considerar que a variável $X$ também possui distribuição gaussiana com média $\mu=0$ e variância $s_{\mathbf{x}}$, definindo o \textit{prior} conforme $X \sim \mathcal{N}(0, s_{\mathbf{x}}^2 \mathbf{I})$. Essa opção resulta na expressão \cite[Seção 12]{Calvetti2018a}
\begin{equation}
\pi_{priori}(\mathbf{x}) \propto \exp\left( -\frac{1}{2 s_{\mathbf{x}}^2} ||\mathbf{x}||^2 \right).
\label{eq:esti19}
\end{equation}
Mesmo se referindo a um modelo de \textit{prior}, e não (necessariamente) ao modelo de ruído aditivo da Equação \eqref{eq:estix}, ele é chamado de \textit{white gaussian noise prior} \cite[pág. 79]{kaipio2005statistical}.
A partir das Equações \eqref{eq:esti7}, (modelo de ruído) e da Equação \eqref{eq:esti19} (\textit{prior}), a expressão da Equação \eqref{eq:esti17} pode ser reescrita como \cite[pág. 77]{kaipio2005statistical}
\begin{equation}
\pi_{priori}(\mathbf{x})  \pi\left(\mathbf{y} \vert \mathbf{x} \right) \propto \exp\left( -\frac{1}{2 s_{\mathbf{x}}^2} ||\mathbf{x}||^2 \right) \exp\left( -\frac{1}{2 s^2_{\bm{\delta}} } \vert \vert (\mathbf{A}\mathbf{x}-\mathbf{y})\vert\vert^2 \right).
\label{eq:esti20_2} 
\end{equation}
O estimador MAP resultante é \cite[pág. 56-7]{calvetti2007introduction}, \cite[Seção 9.2.3]{Deisenroth2020}
\begin{equation}
\begin{aligned}
\hat{\mathbf{x}}_{MAP} & =  \arg\min\limits_{\mathbf{x}} \left( -\log \left[\exp\left( -\frac{1}{2 s_{\mathbf{x}}^2} ||\mathbf{x}||^2 \right) \right] -\log \left[ \exp\left( -\frac{1}{2 s^2_{\bm{\delta}} } \vert \vert \mathbf{A}\mathbf{x}-\mathbf{y}\vert\vert^2 \right) \right] \right) \\
& =   \arg\min\limits_{\mathbf{x}} \left[ \left( \frac{1}{2 s_{\mathbf{x}}^2} ||\mathbf{x}||^2 \right) +\left( \frac{1}{2 s^2_{\bm{\delta}} } \vert \vert \mathbf{A}\mathbf{x}-\mathbf{y}\vert\vert^2 \right) \right] \\
& =   \arg\min\limits_{\mathbf{x}} \frac{1}{2}\left( \vert \vert \mathbf{A}\mathbf{x}-\mathbf{y}\vert\vert^2 + \frac{s^2_{\bm{\delta}}}{s_{\mathbf{x}}^2} ||\mathbf{x}||^2 \right).
\label{eq:esti23}
\end{aligned}
\end{equation}
Definindo $\lambda^2 = s^2_{\bm{\delta}} \backslash s_{\mathbf{x}}^2$, a solução da Equação \eqref{eq:esti23} é 
\begin{equation}
\hat{\mathbf{x}}_{MAP} = \left( \mathbf{A}^T \mathbf{A} + \lambda^2 \mathbf{I} \right)^{-1} \mathbf{A}^T \mathbf{y},
\label{eq:esti24}
\end{equation}
que possui a forma da regularização clássica de Tikhonov da Equação \eqref{eq:tikhonov1}, mas obtida a partir de outras hipóteses \cite[pág. 147]{kaipio2005statistical}. O parâmetro $\lambda^2 $ é entendido a razão entre as variâncias do ruído e do \textit{prior} e denota a confiança que se tem no \textit{prior} \cite{Romano2017}. O \textit{prior} gaussiano dá origem a um regularizador quadrático e a matriz de regularização nesse caso, matriz identidade $\mathbf{I}$, está relacionada com a variância de um modelo de \textit{prior}.  

\subsubsection{MAP com distribuições normais quando a média não é nula e a matriz de covariância não é uma matriz identidade}\label{eq:correcao}
Outro exemplo de modelo de ruído e \textit{prior} considera que a média não é nula e a matriz de covariância não é a matriz identidade, conforme $E \sim \mathcal{N}(\mu_e,s^2_{\bm{\delta}} \mathbf{\Gamma}_{e})$ e $X \sim \mathcal{N}(\mu_{\mathbf{x}},s_{\mathbf{x}}^2\mathbf{\Gamma}_{pr})$. É possível mostrar \cite[pág. 112]{calvetti2007introduction} que a densidade a posteriori é obtida através de
\begin{equation}
\hat{\mathbf{x}}_{MAP} = \arg\min\limits_{\mathbf{x}} \frac{1}{2}\left( \vert \vert \mathbf{L}_e (\mathbf{A}\mathbf{x}-\mathbf{y} - \mu_e)\vert\vert^2 + \lambda^2 \vert \vert \mathbf{L}_{pr}(\mathbf{x} - \mu_{\mathbf{x}}) \vert \vert^2 \right),
\label{eq:esti26}
\end{equation}
onde $\mathbf{\Gamma}_{e}^{-1} = \mathbf{L}_e^T \mathbf{L}_e$ e $\mathbf{\Gamma}_{pr}^{-1} = \mathbf{L}_{pr}^T \mathbf{L}_{pr}$, além das variâncias $s^2_{\bm{\delta}}$ e $s_{\mathbf{x}}^2$ foram englobadas no parâmetro de regularização $\lambda$. 

Ou seja, $\mathbf{\Gamma}_{e}$ torna o termo de fidelidade um termo de mínimos quadrados ponderado \cite[págs. 36-7]{calvetti2007introduction}, considerando também o erro $\mu_e$. Já $\mathbf{\Gamma}_{pr}$ faz com que o termo de regularização se assemelhe àquele da Equação \eqref{eq:eqmult3}, regularização generalizada com valor de referência, o que permite relacionar as distribuições gaussianas com matrizes de regularização na regularização generalizada de Tikhonov.

\subsection{Exemplos de \textit{priors} explícitos e diferença para \textit{priors} implícitos}\label{subsec:implicit}

A solução obtida através do MAP resulta no mesmo problema computacional que métodos de regularização, sendo nesse exemplo análoga à regularização clássica de Tikhonov \cite[pág. 11]{ Calvetti2018a}, \cite[Seção 9.2.4]{Deisenroth2020}, \cite[pág. 79, 146-7]{kaipio2005statistical}, mas trazendo também informações sobre incertezas associada aos seus componentes \cite[pág. 279]{aster2019parameter}. Por conta dessa semelhança, usualmente é feita a referência do termo de penalidade como uma forma de expressar informação \textit{a priori} \cite[pág. 58]{calvetti2007introduction}, \cite[pág. 11]{ Calvetti2018a}, \cite{LiyanMa2013}, mesmo sem referência à métodos bayesianos \cite[pág. 16]{ Calvetti2018a}. 

Retomando a Equação \eqref{eq:tikhonovgeral},  $\pi\left(\mathbf{y} \vert \mathbf{x} \right)$ é relacionado com o termo de fidelidade. No caso de uma densidade de ruído gaussiano branco resulta na sua forma quadrática e partindo de outras distribuições outros termos de fidelidade seriam obtidos. Já $\pi_{priori}(\mathbf{x})$ dá origem ao termo de regularização $\Omega(\mathbf{x})$ e também uma distribuição gaussiana branca resulta na regularização de Tikhonov. Mas, assim como o termo de regularização $\Omega(\mathbf{x})$ pode ter diferentes formas, a densidade da Equação \eqref{eq:esti19} não é a única forma possível e  tanto $ \pi\left(\mathbf{y} \vert \mathbf{x} \right)$ quanto $\pi_{priori}(\mathbf{x})$ podem ser diferentes.

 Considerando a MAP, há \textit{priors} que resultam em termos de regularizações conhecidos (ou, pelo menos, muito semelhantes). Alguns deles são mostrados na Tabela \ref{table:prior} e resumem a relação possível entre a abordagem determinística e a bayesiana para problemas inversos.  Outros \textit{priors} e modelos de ruídos aditivos podem ser vistos em \cite[Tabelas 1 e 2]{Chen2002}, mas  quando o \textit{prior} não é gaussiano, o problema se torna não-linear e são necessários algoritmos iterativos para a sua solução \cite[pág. 15]{ Calvetti2018a}.
 
 Os \textit{priors} da Tabela \ref{table:prior} e das Subseções \ref{sec:normalMAP} e \ref{eq:correcao} são chamados de \textit{priors} explícitos, pois suas densidades de probabilidade são definidas por expressões explícitas (analíticas) \cite[pág. 68]{Arridge2019}, \cite[págs. 58, 62]{kaipio2005statistical}. Por outro lado, \textit{priors} implícitos são aqueles que não são explícitos \cite[pág. 157]{goodfellow2016deep}, de modo que métodos de regularização também podem ser explícitos ou implícitos \cite[pág. 2828]{Chen2002}. 
 
{\centering 
\begin{longtable}{|| l l l || }
\caption{Distribuições de \textit{priors} explícitos}
\label{table:prior}\\ \hline\hline
 \rowcolor{lightyellow} \textbf{\textit{Prior}} $\pi_{prior}(\mathbf{x})$ & \textbf{Ref.} \cite{kaipio2005statistical} & \textbf{Regularização} \\ \hline\hline
\textit{White noise} &Pág. 57& Clássica (Tikhonov) \\
\textit{Smoothness}& Pág. 80& Generalizada (Tikhonov)\\
\textit{Entropy density} & Pág. 64 & Máxima entropia \\
$\ell_1-\textit{norm}$ & Pág. 63 & Esparsa \\
$\ell_p-\textit{norm}$ & Pág. 63 & Esparsa  \\
\textit{Total Variation} & Pág. 68 & Variação total  \\ \hline
\end{longtable}
}

\subsection{Diferenças para a regularização de Tikhonov}

É necessário o cuidado de dizer que abordagens bayesianas podem até chegar em estimadores particulares que coincidem com métodos de regularização clássicos, mas que elas não se limitam a essas soluções\footnote{Por exemplo, comparar as Figuras 1 e 4 de \cite{Calvetti2018a} que trazem a visão geral de ambos.} \cite[pág. XI]{calvetti2007introduction}.  Inclusive, há autores que as consideram mais ricas do que metodologias tradicionais baseadas em regularização \cite{Calvetti2018a} e há autores que argumentam sobre a necessidade da inclusão de informação \textit{a priori} para soluções de problemas quando há infinitas soluções, mas sem citar as expressões de problemas mal-postos ou de regularização \cite{tarantola2005inverse}.

Em abordagens determinísticas, como na regularização de Tikhonov, o resultado é um valor único das incógnitas, além de sempre ser calculado \cite[pág. 112]{kaipio2005statistical}. Na abordagem estatística, o resultado é uma densidade de probabilidade, o que permite estimar o valor mais provável da variável aleatória $X$ ou intervalos de valores das incógnitas com certa probabilidade \cite[pág. 52]{kaipio2005statistical}. Outra diferença é que se pode obter uma densidade \textit{a posteriori} imprópria, o que indicaria que as informações \textit{a priori} junto com os dados não são suficientes para se obter uma estimativa confiável, em contraste com métodos de regularização, da qual sempre se obtém uma solução \cite[pág. 113]{kaipio2005statistical}. 

É também possível produzir diferentes estimativas e avaliar a sua confiabilidade \cite[pág. 2]{kaipio2005statistical}, bem como o cálculo de intervalos de confiança \cite[pág. 52]{kaipio2005statistical}, pois não se busca apenas o valor da variável, mas sim mais informações disponíveis sobre ela \cite[pág. 49]{kaipio2005statistical}.

 \newpage
\section{Interpolação com regularização}\label{Ap:interp}

No Apêndice \ref{Ap:pseudo} foi discutido o problema de interpolação utilizando-se a matriz pseudoinversa para sua solução, buscando comparar como a qualidade do resultado depende da bases de representação utilizadas. Agora, suponha o caso de interpolação cujo modelo não possa ser alterado. O exemplo a seguir permite ilustrar como a regularização pode ajudar a mitigar o \textit{overfitting}, mas também causar o \textit{underfitting}, dependendo da escolha de $\lambda$, relacionando a regularização clássica de Tikhonov com o  do \textit{trade-off} entre \textit{bias} e variância \cite[Exemplo 4.3.1]{alvarez2017digital}.

 Seja $-1.6 < t < 1.6$. O sinal é dado pela função $y(t) = -2 + 2t -t^5$,  representado por amostras que formam o vetor $\mathbf{y}$, cuja norma do vetor de parâmetros $\vert \vert (-2,2,-1) \vert\vert_2 = 3$. Em seguida, o  sinal é corrompido com ruído aditivo gaussiano $ \sim \mathcal{N}(0, 1)$ e o  objetivo é reconstruí-lo com um polinômio de quarta ordem $x_0 t^0+x_1 t^1 + x_2 t^2 + x_3 t^3 + x_4 t^4$, admitindo-se que o modelo não é perfeito. Em uma forma matricial, deseja-se calcular o resultado de $\mathbf{A} \mathbf{x}$, onde cada coluna de $\mathbf{A}$ representa um grau do polinômio e o vetor dos parâmetros é $\mathbf{x} = [x_1, x_2, ..., x_4]$. Mesmo que o modelo $\mathbf{A}$ seja capaz de representar o sinal original, quando há ruídos a solução do problema é prejudicada. Apenas para ilustrar, todas as reconstruções a seguir foram feitas para um mesma mesma realização do ruído. 

\begin{figure}[H]
    \centering
    \includegraphics[width = \textwidth]{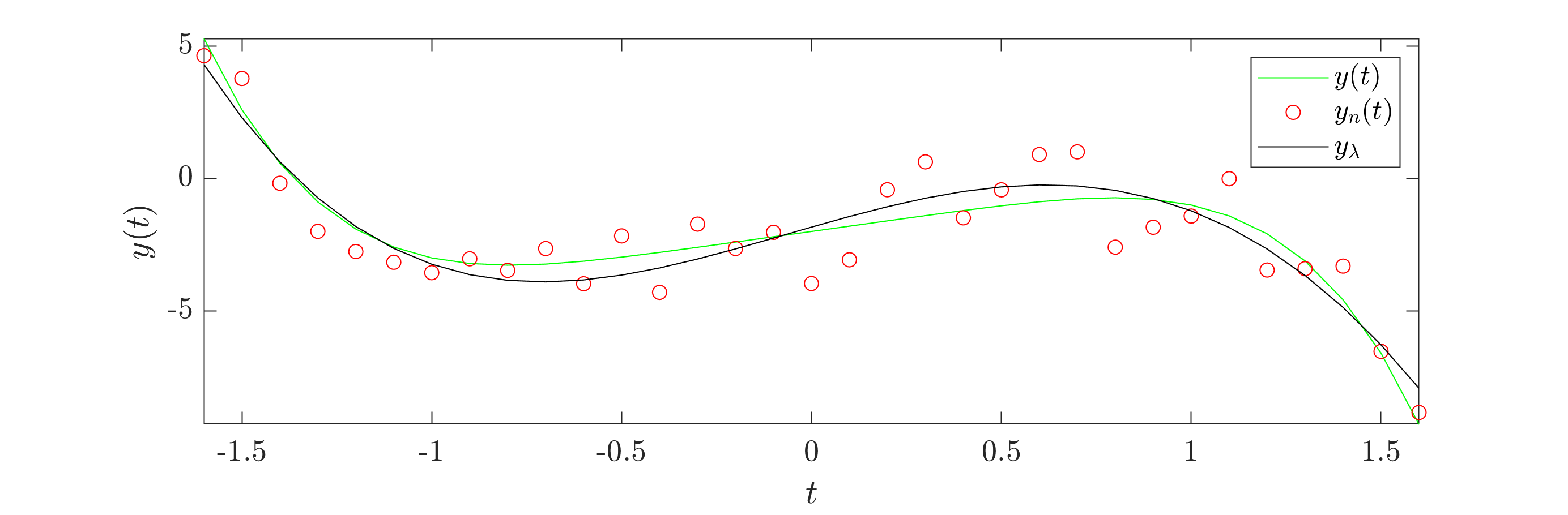}
    \caption[Regressão linear sem regularização.]{Regressão linear sem regularização. Fonte: Próprio autor.}
    \label{fit1}
\end{figure}

Seja a regularização de Tikhonov de ordem zero calculada conforme 
\begin{equation*}
    \mathbf{x}  = \left(\mathbf{A}^T  \mathbf{A}+\lambda \mathbf{I} \right)^{-1} \mathbf{A}^T \mathbf{y}. 
    \label{sinal4}
\end{equation*}
A solução não regularizada ($\lambda = 0$) é vista na Figura \ref{fit1}. Ela apresenta grandes ondulações e a norma euclidiana $\vert \vert \mathbf{x} \vert \vert_2 = 5.50$ é maior a dos que os parâmetros que geraram o sinal.

\begin{figure}[H]
    \centering
    \includegraphics[width = \textwidth]{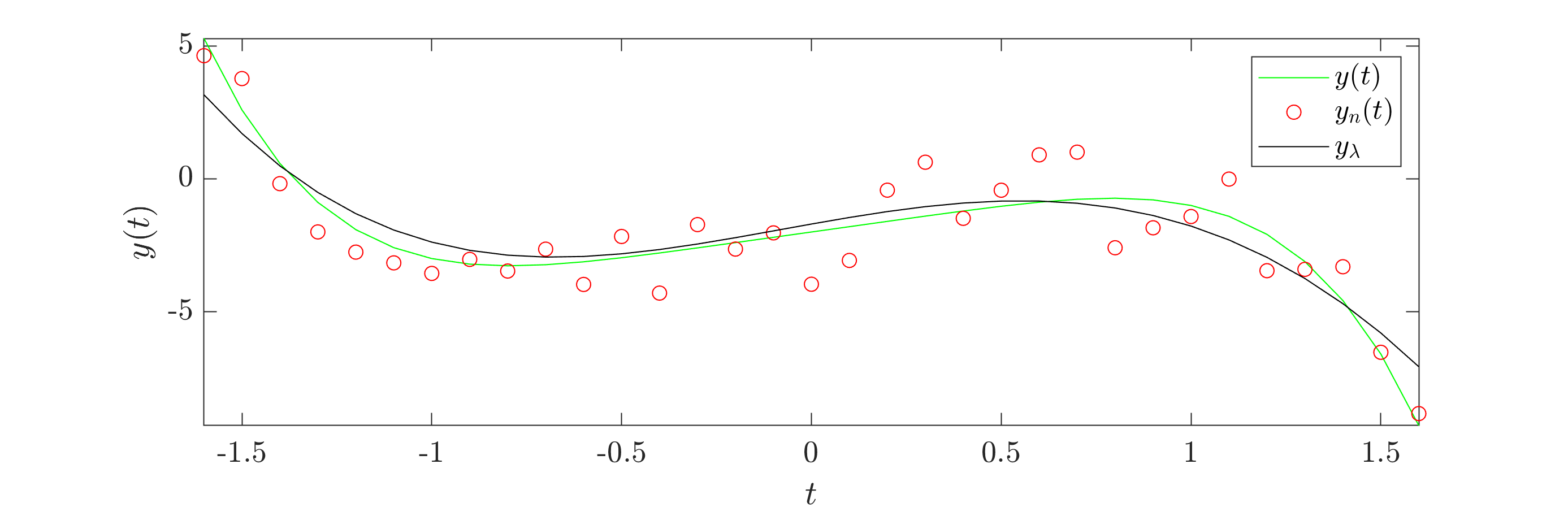}
    \caption[Regressão linear com regularização ($\lambda = 2$).]{Regressão linear com regularização($\lambda = 2$). Fonte: Próprio autor.}
    \label{fit2}
\end{figure}

Na Figura \ref{fit2} é vista uma solução regularizada $\mathbf{x}_{\lambda} = 2$, com $\vert \vert \mathbf{x} \vert \vert_2 \approx 3.84$, valor que é menor do que o obtido com o polinômio original. Observa-se que a curva apresenta menos oscilações por conta dos ruídos, ficando mais próxima do sinal original. 

\begin{figure}[H]
    \centering
    \includegraphics[width = \textwidth]{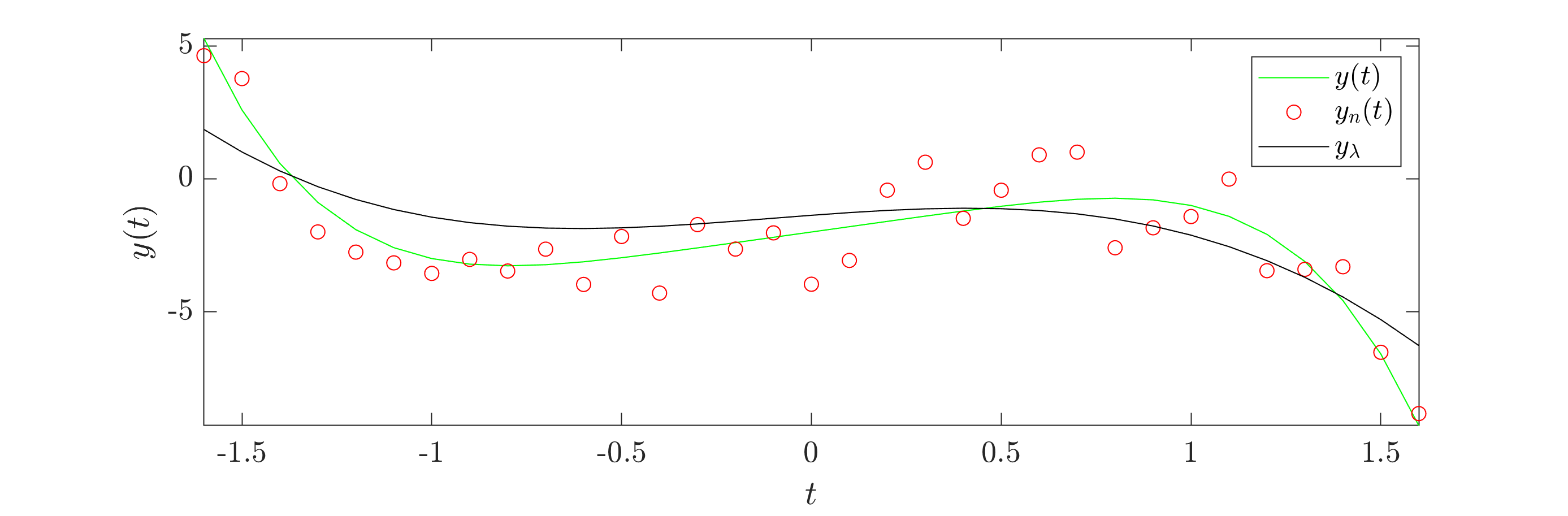}
    \caption[Regressão linear com regularização ($\lambda = 8$).]{Regressão linear com regularização ($\lambda = 8$). Fonte: Próprio autor.}
    \label{fig:fit3}
\end{figure}

Soluções mais regularizadas (com $\lambda$ maior) resultariam em normas ainda menores, o que não significa que as soluções seriam melhores. De fato, elas podem resultar no caso de \textit{underfitting}, ficando mais insensível à alterações nos dados originais. Um exemplo de reconstrução com $\lambda = 8$ é mostrado na Figura \ref{fig:fit3}, com  $\vert \vert \mathbf{x} \vert \vert_2 \approx 2.29$, o menor encontrado até então, mas cuja forma obtida apresenta oscilação menor do que a do sinal original.  

\newpage
\section{\textit{Missing data} com matrizes de regularização derivativas}\label{Ap:missing}

No problema de \textit{missing data}, o objetivo é completar os dados indisponíveis. Em \cite[Figura 8.1]{hansen2010discrete}, o autor compara os resultados utilizando regularização de Tikhonov de ordem zero, de primeira ordem e de segunda ordem. Aqui será desenvolvido um caso análogo, focando em como a amplitude do ruído afeta as soluções regularizadas. 

Seja uma função $y(t) = \sin (t) + \delta$ para $0\leq t \leq 8\pi$. Na sequência, o valor da função é zerado em duas regiões. Seja $\mathbf{A}$ uma matriz identidade e $\mathbf{A_k}$ uma matriz diagonal que contém elementos $a_{ii}$ zerados de acordo com a posição dos dados perdidos. A Equação \eqref{eq:tiksolgen} foi utilizada considerando que $\mathbf{A}$ é conhecida e $\lambda = 0.0001$ na ausência de ruído.

\begin{figure}[H]
\begin{center}
	    \includegraphics[width=0.8\linewidth]{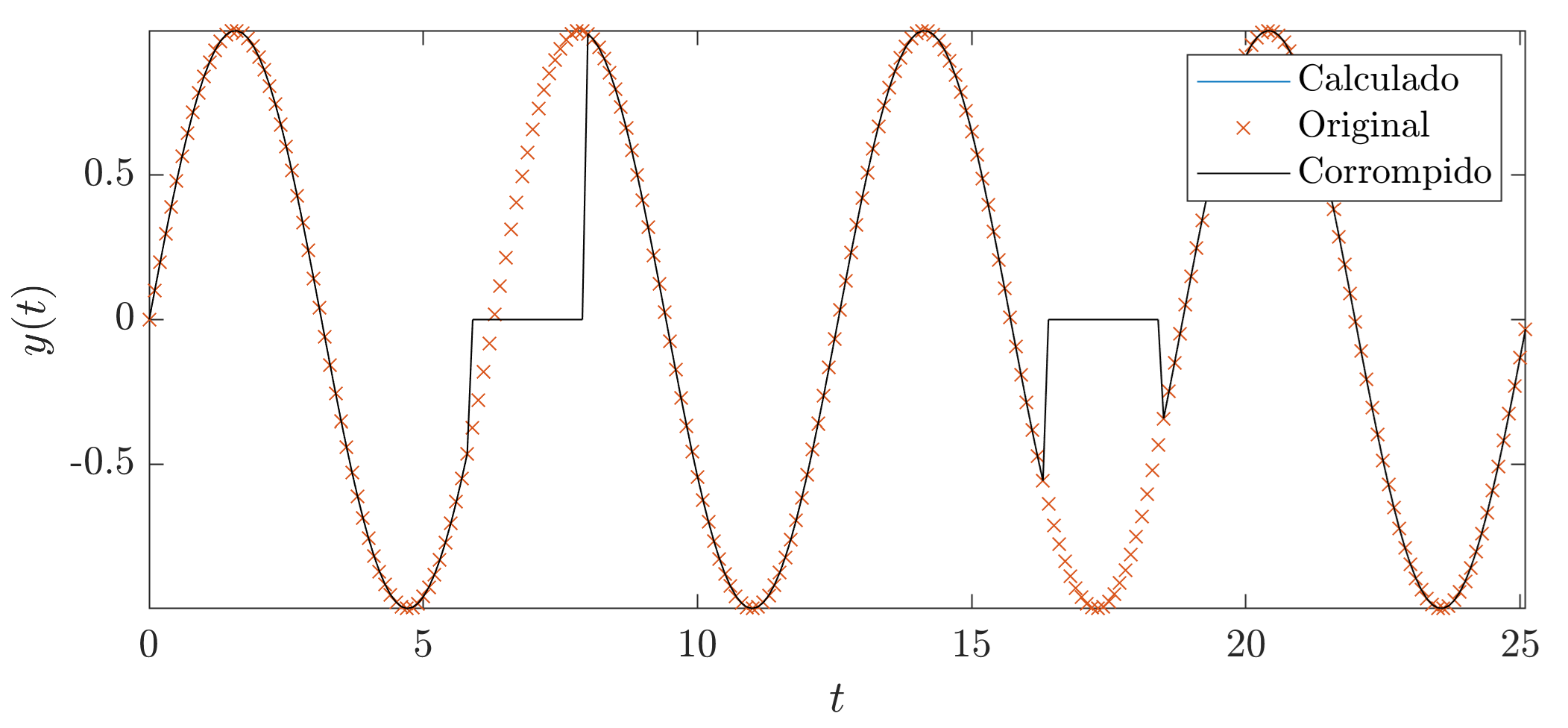}
\caption[Regularização clássica de Tikhonov.]{Regularização clássica de Tikhonov. Fonte: Próprio autor.}
\label{fig:missing2}
  \end{center}
\end{figure}

\begin{figure}[H]
    \begin{center}
	    \includegraphics[width=0.8\linewidth]{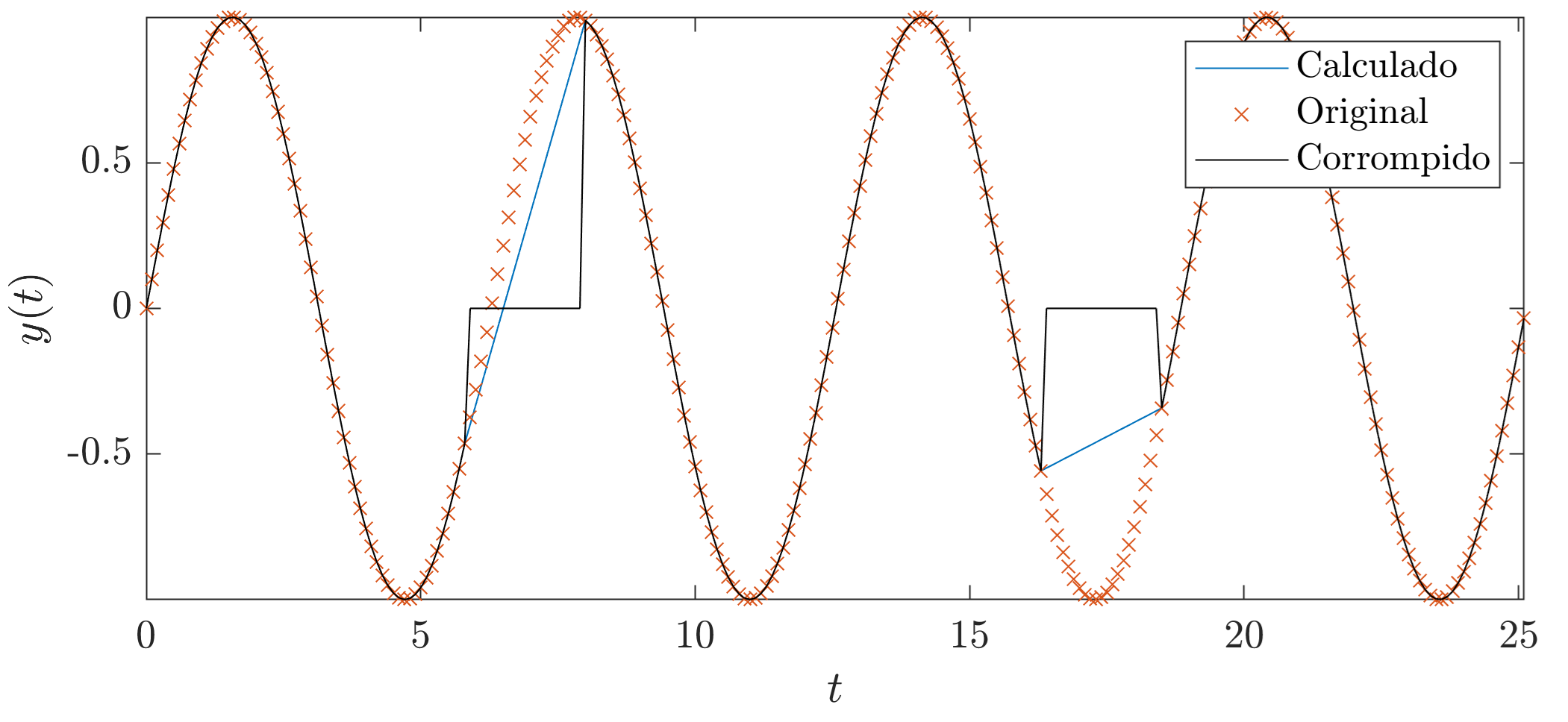}
\caption[Regularização de Tikhonov de primeira ordem.]{Regularização de Tikhonov de primeira ordem. Fonte: Próprio autor.}
\label{fig:missing3}
  \end{center}
\end{figure}

Na Figura \ref{fig:missing2} é mostrada a regularização clássica de Tikhonov, com $\mathbf{L} = \mathbf{I}$. Os valores que faltam não foram recuperados, eles continuaram como valores nulos. 
A Figura \ref{fig:missing3} mostra a regularização de Tikhonov com $\mathbf{L}$ um operador de primeira derivada, Equação \eqref{eq:ld1}. Os valores interpolados lembram uma função linear. 

\begin{figure}[H]
        \begin{center}
	    \includegraphics[width=0.8\linewidth]{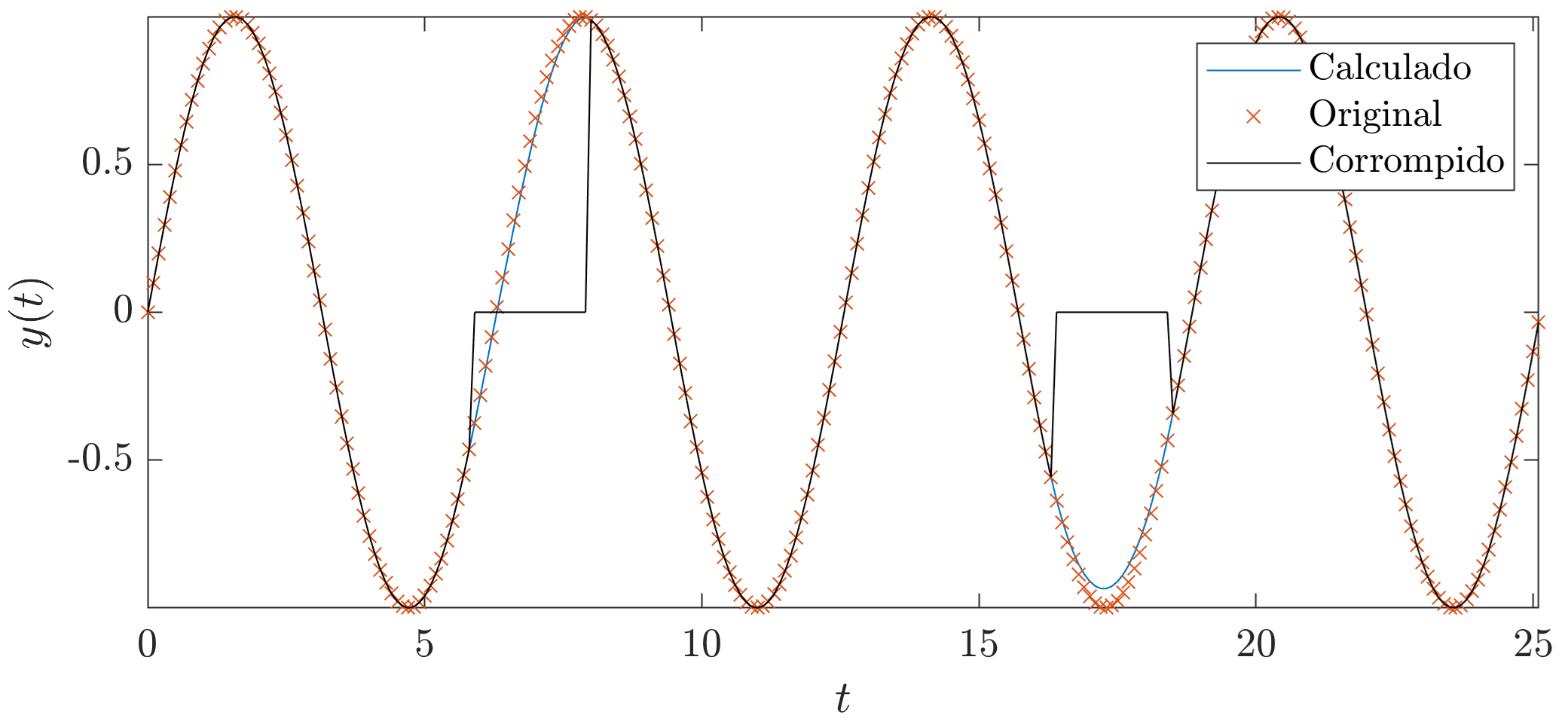}
\caption[Regularização com de Tikhonov de segunda ordem.]{Regularização com de Tikhonov de segunda ordem. Fonte: Próprio autor.}
\label{fig:missing4}
   \end{center}
\end{figure}

 Na Figura \ref{fig:missing4} é mostrada a regularização de Tikhonov, com $\mathbf{L}$ sendo um operador de segunda derivada, Equação \eqref{eq:ld2}. Os valores que faltam foram interpolados com uma função que se assemelha a um polinômio. 

\begin{figure}[H]
        \begin{center}
	    \includegraphics[width=0.9\linewidth]{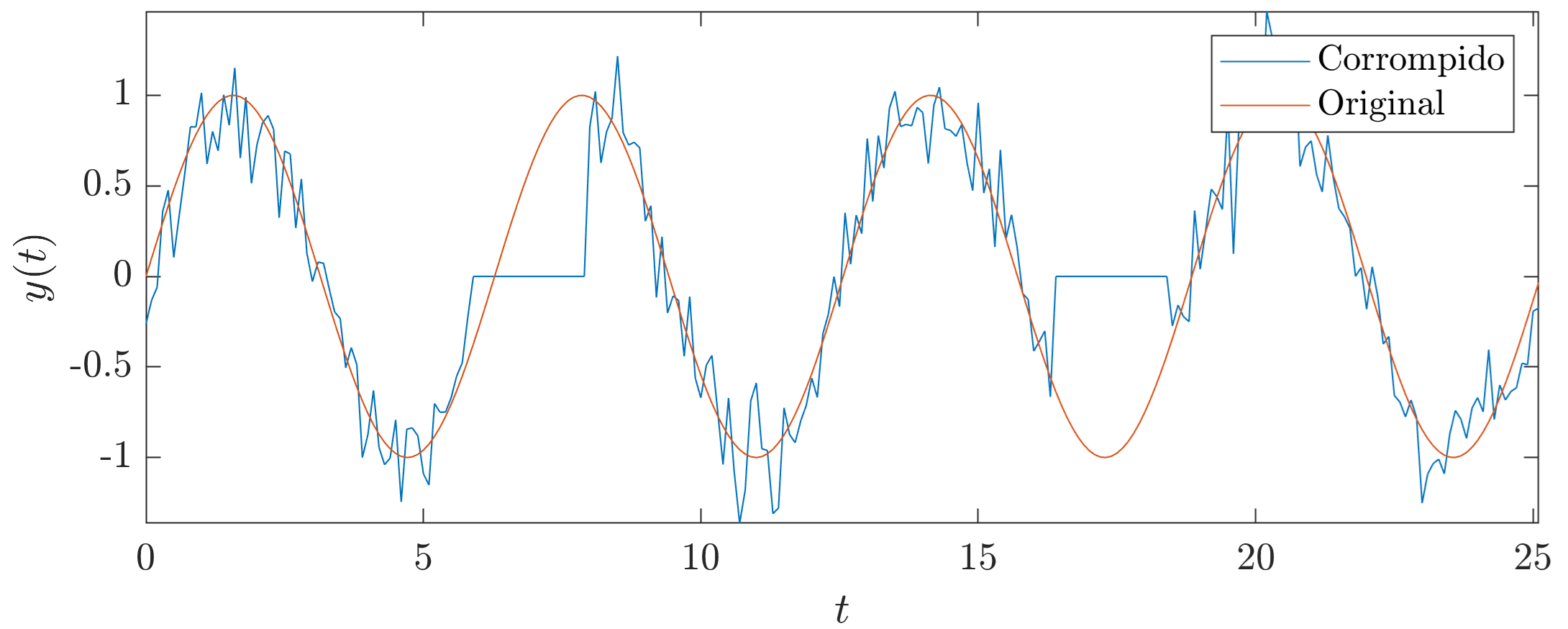}
\caption[Sinal original e sinal com \textit{missing data} e ruídos]{Sinal original e sinal com \textit{missing data} e ruídos. Fonte: Próprio autor.}
\label{fig:missing7}
   \end{center}
\end{figure}

Em problemas do mundo real, os dados apresentam ruído. Dependendo de sua amplitude, é necessário utilizar $\lambda$ com maiores valores. Na Figura \ref{fig:missing7} é mostrado o sinal original e o sinal corrompido, na qual primeiro os dados foram perdidos e depois houve a adição de um ruído gaussiano $\delta  \sim \mathcal{N}(0, 0.2)$, resultando que  mesmo os valores nulos fossem corrompidos.

\begin{figure}[H]
        \begin{center}
	    \includegraphics[width=0.9\linewidth]{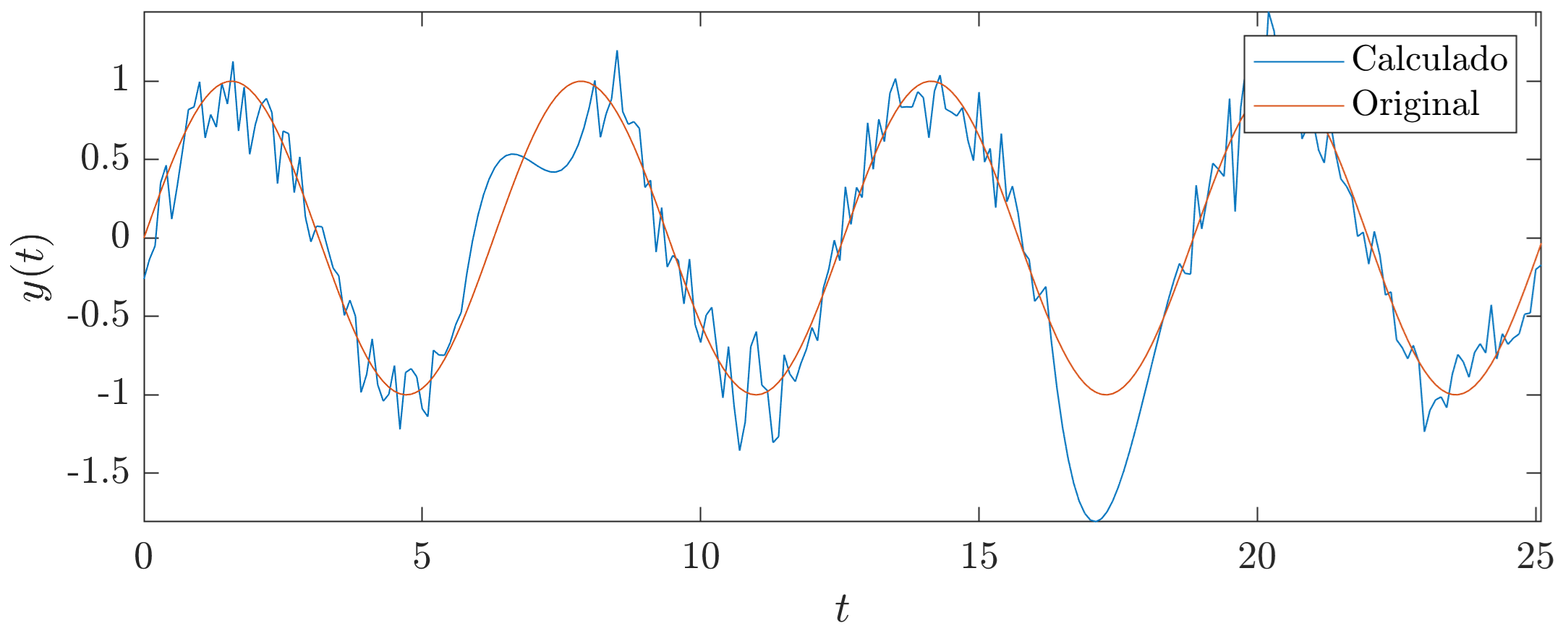}
\caption[Regularização com $\mathbf{L} = \mathbf{L}_{d2}$ e $\lambda = 0.01$.]{Regularização com $\mathbf{L} = \mathbf{L}_{d2}$ e $\lambda = 0.01$. Fonte: Próprio autor.}
\label{fig:missing8}
   \end{center}
\end{figure}

Na Figura \ref{fig:missing8} também é mostrada a regularização de Tikhonov com $\mathbf{L} = \mathbf{L}_{d2}$, só que com um pequeno valor de $\lambda = 0.01$, mostrando uma reconstrução que não representa o sinal esperado. Além disso, a cada realização do ruído, o sinal reconstruído fica diferente, o que mostra que para esse nível de ruído com esse $\lambda$ o algoritmo não é robusto. Quanto maior a ordem desse operador derivativo (como matrizes de regularização de operadores de terceira ou quarta ordem) menor era o seu posto e maior era a sensibilidade ao ruído e, portanto, não foram aqui mostradas.  

\begin{figure}[H]
        \begin{center}
	    \includegraphics[width=0.9\linewidth]{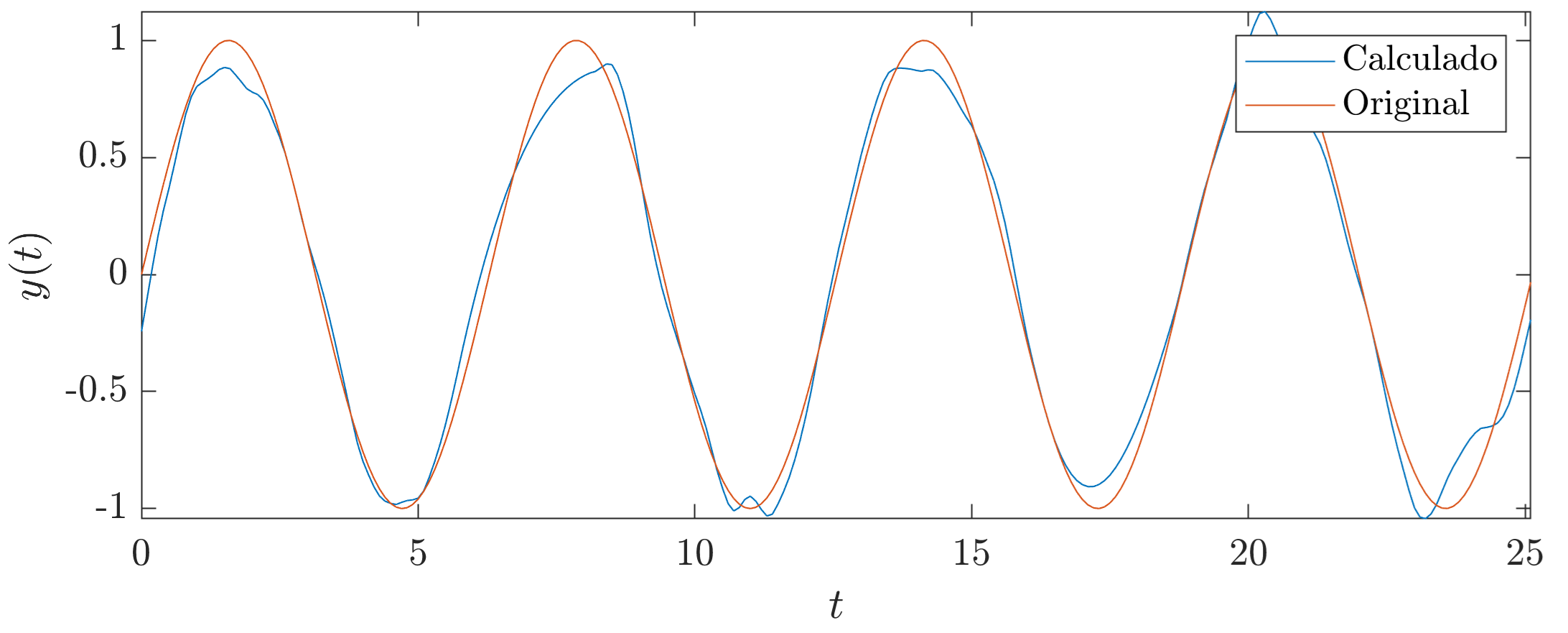}
\caption[Regularização com $\mathbf{L} = \mathbf{L}_{d2}$ e $\lambda = 100$.]{Regularização com $\mathbf{L} = \mathbf{L}_{d2}$ e $\lambda = 100$. Fonte: Próprio autor.}
\label{fig:missing9}
   \end{center}
\end{figure}

Uma solução é tentar aumentar o valor de $\lambda$ para que essa informação \textit{a priori} mais forte compense a presença de ruídos. Na Figura \ref{fig:missing9} é mostrado o caso considerando $\lambda = 10$, em que tanto os pontos perdidos quanto as regiões com ruído foram recuperadas de modo mais adequado, suavizando os dados corrompidos. 

  Outras duas observações podem ser feitas sobre esses exemplos:
  \begin{itemize}
  \item Diferente dos Apêndices \ref{Ap:pseudo} e \ref{Ap:interp}, $\mathbf{x}$ não representa os parâmetros de um modelo para representar $\mathbf{y}$, mas sim é o próprio $\mathbf{y}$ só que sem degradação;
  \item No presente exemplo, é direto de se visualizar que, quando se utiliza matrizes de regularização de primeira ou de segunda derivada, é interessante notar que não é possível inverter $\left( \mathbf{A}^T  \mathbf{A} \right)^{-1}$ nem $\left( \mathbf{L}^T \mathbf{L} \right)^{-1}$, mas que essas duas matrizes singulares permitem a solução através de $\left( \mathbf{A}^T  \mathbf{A} + \lambda_1^2 \mathbf{L}^T \mathbf{L} \right)^{-1} \mathbf{A}^T \mathbf{y}$. 
  \end{itemize}

\newpage
\section{\textit{Solvers} diretos e iterativos de sistemas de equações lineares}\label{sec:class}

 \subsection{Classificação da dimensionalidade do problema inverso}\label{sec:dimens}

No Capítulo \ref{sec:variational}, os conceitos foram apresentados sem discutir o tamanho do sistema resultante e o número de parâmetros que devem ser estimados. É necessário que se estude a performance da implementação computacional dos algoritmos propostos para regularização generalizada de Tikhonov, como o tempo de execução, requisitos de armazenamento e memória de acesso aleatório (RAM), com o objetivo de avaliar seus impactos computacionais.

 Em \cite{luenberger2015linear}, um problema de programação é classificado como um problema de:
 \begin{itemize}
 \item Pequena escala, com unidades de parâmetros; 
 \item Escala intermediária, com dezenas a centenas de variáveis; 
 \item Grande escala, com milhares ou mesmo milhões de variáveis. 
 \end{itemize}
  Esta classificação não é rígida, considerando que a tecnologia e os algoritmos são desenvolvidos possibilitando solução de sistemas cada vez maiores.
 
  No caso de problemas inversos lineares discretos, cada componente $x_i$ de  $\mathbf{x}$ deve ser estimado e depende da solução de um sistema linear de equações do tipo $\mathbf{A}\mathbf{x} = \mathbf{y}$, como por exemplo as Equações \eqref{eq:tikmult},\eqref{eq:irlsiterative1}, \eqref{eq:eqmult2} e \eqref{eq:tiksolx2}. Considerando problemas mal-postos, não se trata mais sobre o problema original $\mathbf{A}\mathbf{x} = \mathbf{y}$, mas sim de um sistema modificado (regularizado) $\tilde{\mathbf{A}}\mathbf{x} = \tilde{\mathbf{y}}$ cuja solução possa trazer informações relevantes sobre $\mathbf{x}$ original.  Logo, são relevantes o número de parâmetros atualizáveis de $\mathbf{x}$ e também o número de componentes não-nulos de $\mathbf{A}$ e  $\mathbf{L}$.
 
 Essa dimensionalidade de alguns problemas inversos pode ser ilustrada pela regularização generalizada de Tikhonov:
\begin{itemize}

\item \textbf{\textit{Deblurring}}: A matriz $\mathbf{A}$ representa a convolução com a PSF. A multiplicação matriz-vetor  $\mathbf{A}\mathbf{x}$ resulta na imagem borrada $\mathbf{y}$. Uma imagem com tamanho de $\left[ n \times n \right]$ \textit{pixels}, quando vetorizada, possui tamanho $\left[ n^2 \times 1\right]$, o que implica que a matriz $\mathbf{A}$ tem tamanho $\left[ n^2 \times n^2\right]$. Se a figura tem tamanho de $[250 \times 250]$ pixels, $\mathbf{A}$ é uma matriz de tamanho $[62500 \times 62500]$. Essa matriz não é cheia. Quanto menores as dimensões da PSF, mais esparsa será $\mathbf{A}$. Em cada caso, deve-se avaliar quando que é possível utilizar funções prontas de convolução para obter $\mathbf{y}$, sem precisar montar $\mathbf{A}$ explicitamente; 
\item  \textbf{Tomografia computadorizada}: Modelo na forma de $\mathbf{A}\mathbf{x} = \mathbf{y}$:
\begin{itemize}
 \item A imagem tomográfica é uma matriz e $ \mathbf{x}$ é essa matriz vetorizada.  A imagem padrão de um equipamento de Tomografia Computadorizada é discretizada no tamanho de $\left[512 \times 512\right]$ \textit{pixels}, de modo que $ \mathbf{x} \in [262144 \times 1]$  valores devem ser estimados. Existem imagens de alta resolução de até $\left[ 2048 \times 2048\right]$ \textit{pixels}, aumentando ainda mais esse número \cite{Hata2018};
 \item O sinograma é uma matriz. O número de linhas é relacionado ao número de ângulos utilizados na aquisição em volta do objeto através da expressão $(2 \times n^o\text{de ângulos} +1)$. O número de colunas é relacionado ao número de feixes/detectores utilizados. Supondo que estes sejam $560$, o tamanho do sinograma resultante é {$ (2 \times n^o\text{ de ângulos} +1) \times 560$}. Assim, $\mathbf{y} \in [((2 \times n^o\text{ de ângulos} +1)\times  560) \times 1]$ seria o sinograma vetorizado;

 \item O número de colunas de $\boldsymbol{A}$ depende da discretização da imagem. Em uma imagem padrão, seria de $512 \times 512 = 262144$ colunas. O número de linhas de $\boldsymbol{A}$ depende do número de amostras do detector de raios-X e dos ângulos disponíveis. Assim, o tamanho de $\boldsymbol{A}$ resultante nesse exemplo seria de $ (2\times n^o\text{ dos ângulos} +1) \times 560 \times 262144$, um problema de grande escala, mas com $\boldsymbol{A}$ sendo esparsa. Nota-se que quanto menor o número de ângulos utilizados na reconstrução, mais subdeterminado se torna o sistema. 

\end{itemize}
\item \textbf{Tomografia por impedância elétrica}: A solução do problema inverso  através do método dos elementos finitos pode ser de pequena, intermediária ou grande escala, a depender da malha utilizada. O seu problema linearizado traz que o modelo $\mathbf{A}$ é o jacobiano do operador direto. As dimensões deste jacobiano dependem da quantidade de eletrodos utilizada na aquisição dos dados (relacionado com o número de linhas) e do número de elementos da malha que serão atualizados (define o número de colunas). Ou seja, mantendo o número de eletrodos, mas utilizando uma malha mais refinada, obtém-se um sistema linear mais subdeterminado. Na literatura, o número de elementos de uma malha pode ter até milhões de elementos no caso 3D \cite{2018Corazza, Moura2021}, influenciando diretamente o tamanho de  $\mathbf{A}$; 
\item  \textbf{Problemas sísmicos}: Com o método das diferenças finitas são de grande escala \cite{Adler2021}.
\end{itemize}

Por tanto, deve-se avaliar se as soluções obtidas são as mais adequadas para o poder computacional disponível. As Equações \eqref{eq:tiksolgen}, \eqref{eq:eqmult3} ou \eqref{eq:eqmult2} podem trazer dificuldades na montagem e inversão explícita de grandes matrizes, lembrando que a matriz de regularização $\mathbf{L}$ teria tamanho da mesma ordem de grandeza de $\mathbf{A}$.

Métodos para solução de sistemas lineares podem ser classificados em métodos diretos e os métodos iterativos, conforme discutido a seguir. Mais do que a classificação entre \textit{solvers} diretos e iterativos, o importante é entender cada um tem suas vantagens e desvantagens e que eles podem trazer  \textit{insights} para novos algoritmos e propostas.

 \subsection{Métodos diretos para solução de sistemas lineares}

 \textit{Solvers} diretos, ou métodos diretos, são aqueles que primeiro fatoram a matriz $\mathbf{A}$ (ou $\mathbf{A}^T$) para depois resolver o problema por sistemas lineares mais simples, sendo necessário que as matrizes $\mathbf{A}$ e $\mathbf{L}$ estejam disponíveis explicitamente \cite[pág. 63]{calvetti2007introduction}. Isso demanda memória para alocação das matrizes, sendo mais adequado para problemas de pequena e média escala. Dependendo da forma de $\mathbf{A}$ e se ela é mal-condicionada ou não, técnicas de decomposição possíveis incluem a decomposição LU, a fatorização de Cholesky, a fatorização QR e a SVD. Em problemas de grande escala, o uso de técnicas de fatoração matricial, como a SVD de $\mathbf{A}$, podem ser inviáveis \cite[pág. 115]{hansen2010discrete}. Mesmo que $\mathbf{A}$ seja esparsa, possivelmente tanto a matriz $\mathbf{U}$ quanto a matriz $\mathbf{V}^T$ serão cheias, trazendo a necessidade de se armazenar matrizes por vezes mais densas que as matrizes originais \cite[pág. 151]{aster2019parameter}.  
 
A solução da Equação \eqref{eq:tiksolgen} traz operações como multiplicações $\mathbf{A}^T \mathbf{A}$ e inversões de matrizes, que são custosos em problemas de grande escala. Por outro lado, observa-se que eles possuem um passo apenas, uma solução fechada, o que é explicado tanto pela presença de termos quadráticos, quanto pelo fato do modelo $\mathbf{A}$ ser linear e do regularizador também depender de uma matriz $\mathbf{L}$ linear com os parâmetros $\mathbf{x}$ (na forma da multiplicação matriz-vetor). Logo, mesmo que existam algoritmos em um só passo para solução problemas inversos lineares, nem sempre eles são adequados. 

\subsection{Métodos iterativos para solução de sistemas lineares}

De modo diferente, algoritmos iterativos geram uma sequência de soluções que, idealmente, convergem para a solução ótima da função custo desejada \cite[pág. 109]{hansen2010discrete}, \cite[Subseção 4.3]{calvetti2007introduction}. Eles são importantes quando a dimensionalidade do problema é muito grande, quando $\mathbf{A}$ não é dada explicitamente ou quando se deseja apenas uma solução aproximada do sistema linear \cite[pág. 67]{calvetti2007introduction}. Nesse contexto, métodos iterativos  podem trazer vantagens na solução dos mesmos funcionais, como menor tempo para reconstrução ou exigir menos poder computacional. 

Há uma grande variedade de algoritmos iterativos, como o CGLS  \cite[págs. 165-6]{aster2019parameter},  o GMRES \cite[págs. 72-7]{calvetti2007introduction}, iterações de Landweber e iterações de Cimmino \cite[Subseção 6.1.1]{hansen2010discrete}. Eles permitem incluir outras formas de restrições na otimização, como restrições rígidas, para que se garanta a representação correta de alguma característica física. As diferenças entre a regularização de Tikhonov e métodos iterativos são ilustradas em \cite[Figura 1]{Calvetti2018a}.  A escolha entre eles depende de características da matriz $\mathbf{A}$, como ela ser quadrada ou retangular, ser esparsa ou cheia e ela apresentar algum padrão na sua estrutura como simetria, para citar alguns. No caso de problemas de mínimos quadrados, diversos métodos são encontrados em \cite{Bjrck1996}.

Na regularização clássica de Tikhonov, Equação \eqref{eq:tiksol}, ao isolar $\mathbf{x}$ é necessária a inversão explícita de $\left( \mathbf{A}^T \mathbf{A} + \lambda^2 \mathbf{I}^T \mathbf{I} \right)^{-1}$. Para evitar isso, é possível utilizar algoritmos iterativos que resolvem sistemas lineares $\mathbf{A}\mathbf{x} = \mathbf{y}$, escrevendo 
\begin{equation}
\begin{aligned}
\left( \mathbf{A}^T \mathbf{A} + \lambda^2 \mathbf{I}^T \mathbf{I} \right) \mathbf{x} & = \mathbf{A}^T \mathbf{y} \\
\tilde{\mathbf{A}} \mathbf{x} & = \tilde{\mathbf{y}},
\label{eq:tiksolalt1}
\end{aligned}
\end{equation}
e então resolver o sistema. Além disso, calcular $\mathbf{A}^T \mathbf{A}$ ou $\mathbf{L}^T \mathbf{L}$ pode ser inviável para grandes matrizes. Assim, em \cite[Subseção 5.2]{Mueller2012}, os autores discutem que a solução dada pela Equação \eqref{eq:eqmult3} pode ser obtida a partir da seguinte forma empilhada 
\begin{equation}
\begin{bmatrix}
\textbf{A} \\
\lambda \textbf{L}\\
\end{bmatrix} \mathbf{x} = \begin{bmatrix}
\mathbf{y} \\
\lambda \mathbf{L} \mathbf{x}^* \\
\end{bmatrix},
\label{eq:stacked2}
\end{equation}
uma formulação que pode ser computacionalmente eficiente para resolver sistemas lineares retangulares. Então, utiliza-se algum algoritmo de solução de sistema de equações lineares \cite[Exemplo 8.1]{calvetti2007introduction} segundo o critério dos mínimos quadrados, sendo equivalente a resolver a Equação \eqref{eq:tiksolalt1}. Nota-se ainda que $\mathbf{A}$ e $\mathbf{L}$ podem ser matrizes retangulares na Equação \eqref{eq:stacked2}, sob a condição de que $\mathbf{L}$ tenha posto linha completo\footnote{Posto da matriz igual ao número de linhas.}. 

\subsection{Semiconvergência e algoritmos iterativos truncados}\label{sec:semic}

Alguns algoritmos iterativos são baseados no conceito da semiconvergência \cite[Figura 6.1]{hansen2010discrete}. Idealmente, eles convergem para a solução exata quando os dados não tem ruídos. Se os dados são ruidosos, o algoritmo parece convergir para a solução correta inicialmente, mas depois diverge, sem apresentar convergência assimptótica para um número infinito de iterações.  Nesse caso, encerrar as iterações antes da convergência assimptótica apresentaria um efeito de regularização \cite[pág. 154]{engl1996regularization}, sendo portanto métodos iterativos truncados \cite[pág. 80]{calvetti2007introduction}, \cite[Subseção 2.4]{kaipio2005statistical}. 

Um método baseado apenas em semiconvergência não apresenta termo de regularização e não se espera que ele apresente convergência assintótica. Logo, o papel do parâmetro de regularização $\lambda$ é substituído pelo número de iterações \cite[pág. 109-10]{hansen2010discrete}, sendo necessário um critério de parada com tal objetivo. Isso pode diminuir o custo computacional da solução ao evitar a obtenção de $\lambda$ por métodos como Curva-L ou GCV, que necessitam que a solução seja calculada para diferentes valores de $\lambda$.

\newpage
\section{Implementação computacional de alguns algoritmos em problemas inversos}\label{sec:otimi1}

 Seja um funcional escrito como a soma de duas partes, conforme
\begin{equation}
\mathcal{R}(\lambda) = \arg\min\limits_{\mathbf{x}} \left[ \mathcal{L} \left(\mathbf{A}, \mathbf{x}, \mathbf{y} \right) +  \Omega(\mathbf{x}, \lambda) \right].
 \label{eq:tikhonovgeralx}
\end{equation}
De início, escolhe-se os termos da Equação \eqref{eq:tikhonovgeralx}  a partir das características esperadas da solução. Por outro lado, a sua utilização depende da capacidade de solução do problema de otimização resultante através de algoritmos numericamente efetivos \cite{Daubechies2016}. O problema de otimização resultante depende das propriedades de $\mathcal{L}(\cdot)$ e $\Omega(\cdot)$: se ambos são lineares em relação à $\mathbf{x}$, se eles convexos e se são diferenciáveis. Por exemplo, se o funcional for convexo e diferenciável, iterações de gradiente descendente podem ser suficientes. Caso algumas dessas características não se verifique, são necessários algoritmos especializados:
\begin{itemize}
    \item Em \cite[Apêndice A]{Arridge2019}, \cite[Seção 3.1]{Boyd2011}, \cite[Seção 4.4]{Parikh2014},  os autores discutem algoritmos para o caso em que tanto $\mathcal{L}(\cdot)$ quanto $\Omega(\cdot)$ são convexos, mas podem não ser diferenciáveis (ou não-suaves);
    \item Em \cite[Seção 9.1]{Benning2018}, os autores discutem algoritmos para quando  $\mathcal{L}(\cdot)$ não é convexo.
    \end{itemize}

Uma das grandes vantagens da otimização convexa é que um mínimo local também é um mínimo global, mas nem sempre o problema é convexo. Um exemplo é na solução de problemas mal-postos, quando o regularizador $\Omega(\mathbf{x})$ apresenta norma $\ell_0$ para explorar a característica de esparsidade da solução, mas essa escolha torna o problema NP-difícil e a otimização não-convexa. 

Outra questão é que existe mais de um algoritmo para minimizar o mesmo funcional, após definição dos termos da Equação \eqref{eq:tikhonovgeralx}. Um exemplo é quando $\Omega(\mathbf{x})$ utiliza norma $\ell_1$, problema convexo para explorar a característica de esparsidade da solução. Nesse caso, algoritmos como ISTA, FISTA, ADMM ou IRLS podem ser utilizados, o que não significa que vão resultar nas mesmas soluções, até porque cada um depende de seus (vários) próprios parâmetros e os algoritmos não apresentam as mesmas garantias teóricas. 

Nesse sentido, a comparação entre otimizadores se torna relevante, seja em termos das suas garantias teóricas ou da sua eficiência computacional. A partir do momento que se conhece as suas vantagens e desvantagens, pode-se escolher aquele que é melhor para a aplicação desejada. Na regularização generalizada de Tikhonov, com normas $\ell_2^2$ em $\Omega(\mathbf{x})$ e $\mathcal{L} \left(\mathbf{A}, \mathbf{x}, \mathbf{y} \right)$, há tanto soluções em apenas um passo como algoritmos iterativos. Entre eles, como escolher o mais adequado?  

Todas essas escolhas são relevantes quando se estudam novas propostas de regularização, incluindo questões de dimensionalidade e dos requisitos computacionais. O ideal seria que existisse um algoritmo de otimização universal, que fosse possível mudar o termo de fidelidade ou o regularizador e ele fosse capaz de resolver de forma adequada. Mesmo o gradiente descendente, amplamente utilizado, requer que a função seja diferenciável. Quando ela não é, adaptações são necessárias. Na prática, cada funcional requer uma estratégia específica. 

  Na Tabela \ref{table:opt-matlab} há exemplos de \textit{toolboxes} para problemas inversos discretos na qual só é necessário fornecer matrizes e vetores para teste de algoritmos de reconstrução, usualmente $\mathbf{A}$, $\mathbf{A}^T$, $\mathbf{x}_0$ e $\mathbf{y}_{\delta}$, além dos parâmetros de reconstrução de cada algoritmo em si. 
 
{\centering 
\begin{longtable}{|| l  c  c|| }
\caption{Exemplos de \textit{toolboxes} para problemas inversos.}
\label{table:opt-matlab}  \\ \hline \hline
 \rowcolor{lightyellow} \textbf{Nome da \textit{toolbox}} & \textbf{Ref.} & \textbf{Ano} \\
 [0.5ex] 
\hline\hline
$\ell_1$-magic\footnote{\url{https://candes.su.domains/software/l1magic/}}  &\cite{l1magic} & 2005 \\
Deblurring images \footnote{\url{http://www.imm.dtu.dk/~pcha/HNO/}}  &\cite{hansen2006deblurring} & 2006 \\
Regularization tools (regtools)\footnote{\url{https://www.mathworks.com/matlabcentral/fileexchange/52-regtools}}  &\cite{Hansen2007} & 2007 \\ 
Linear and nonlinear inverse problems\footnote{\url{https://wiki.helsinki.fi/xwiki/bin/view/mathstatHenkilokunta/Henkil\%C3\%B6t/Siltanen\%2C\%20Samuli/Inverse\%20Problems\%20Book\%20Page/}}  &\cite{Mueller2012} & 2007 \\
IR tools\footnote{\url{https://github.com/jnagy1/IRtools}}  &\cite{Gazzola2018} & 2018 \\
Parameter estimation and inverse problems\footnote{\url{https://github.com/brianborchers/PEIP}}  &\cite{aster2019parameter} & 2019 \\
Compressed sensing for engineers\footnote{Códigos disponíveis no próprio livro}  &\cite{majumdar2019compressed} & 2019 \\ \hline
\end{longtable}
}
Esses algoritmos podem ser entendidos no contexto da álgebra linear, já que na regularização generalizada de Tikhonov, a Equação \eqref{eq:tiksolgen} pode ser reescrita como \begin{equation}
\tilde{\mathbf{A}}^{T} \tilde{\mathbf{A}} \mathbf{x}  = \tilde{\mathbf{A}} \tilde{\mathbf{y}},
\end{equation} 
associado ao problema de mínimos quadrados $\vert \vert \tilde{\mathbf{A}} \mathbf{x} - \tilde{\mathbf{y}} \vert \vert^2_2$, onde os termos foram agrupados conforme $\tilde{\mathbf{A}} = \left[ \mathbf{A} \quad  \lambda \mathbf{L} \right]^T$ e $\tilde{\mathbf{y}} = \left[ \mathbf{y} \quad  \mathbf{0} \right]^T$. Transformações análogas também podem ser feitas para as Equações \eqref{eq:eqmult3} e \eqref{eq:eqmult2}. Isso significa que, neste método, a solução de um sistema de equações lineares mal-condicionado depende da solução de outro sistema de equações lineares, dessa vez melhor condicionado. 

 O presente capítulo discute o método de Newton, a Equação \eqref{eq:eqmult2} da regularização multiparâmetros e o IRLS para aproximar a norma $\ell_1$.

\subsection{Método de Newton na regularização de Tikhonov}\label{Ap:newton}

O método de Newton é um método iterativo de otimização geralmente em problemas não-lineares, cuja forma geral \cite[Subseção 2.1.3]{paivi} é dada por
\begin{equation}
\mathbf{x}_{k+1}=\mathbf{x}_{k}-\alpha_{k}(\nabla^2_\mathbf{x} \mathcal{M}(\mathbf{x}_{k}))^{-1} \nabla_\mathbf{x} \mathcal{M}(\mathbf{x}_{k}),
\label{eq:grad1}
\end{equation}
onde $\nabla^2_\mathbf{x}$ é a matriz hessiana $\alpha_{k}$ é um escalar positivo que define o tamanho do passo, que pode ser fixo ou variável ao longo das iterações $k$. Se $\nabla^2_\mathbf{x} \mathcal{M}(\mathbf{x}_{k})$ é positiva definida e $\mathcal{M}(\mathbf{x}_{k})$ é quadrático, a solução é obtida em apenas um passo \cite[pág. 20]{paivi}. 

Seja a regularização generalizada de Tikhonov para um problema linear  descrita pela Equação \eqref{eq:tikhonov12}, onde $\mathcal{M}(\lambda, \mathbf{x}, \mathbf{y}) = \vert \vert \mathbf{A} \mathbf{x} - \mathbf{y} \vert \vert^2_2 + \lambda^2 \vert \vert \mathbf{L} \mathbf{x} \vert \vert_2^2$. Com termos quadráticos,  $\mathcal{M}(\lambda, \mathbf{x}, \mathbf{y})$  é diferenciável duas vezes em relação à $\mathbf{x}$, obtendo-se
\begin{equation}
\begin{aligned}
\nabla_\mathbf{x} \mathcal{M}(\lambda, \mathbf{x}, \mathbf{y}) & = 2 \left(\mathbf{A}^{T}\mathbf{A}\mathbf{x} - \mathbf{A}^{T}\mathbf{y} + \lambda^2\mathbf{L}^T\mathbf{L}\mathbf{x}\right)\\
\nabla^2_\mathbf{x} \mathcal{M}(\lambda, \mathbf{x}, \mathbf{y}) & = 2 \left(\mathbf{A}^{T}\mathbf{A}  + \lambda^2\mathbf{L}^T\mathbf{L} \right).
\label{eq:grad3}
\end{aligned}
\end{equation}
Considerando $\mathbf{x}_{0} = \mathbf{0}$, $\alpha_k = 1$ e desconsiderando a constante multiplicativa $2$, pois não afeta o resultado final da otimização, o primeiro passo do método de Newton ($k =1$) é dado pela Equação \eqref{eq:gradpasso1}, que é igual à Equação \eqref{eq:tiksolgen} descrita anteriormente.  
\begin{equation}
\begin{aligned}
\mathbf{x}_{1}& = \mathbf{0} - 1 \left(\mathbf{A}^{T}\mathbf{A}  + \lambda^2\mathbf{L}^T\mathbf{L}\right)^{-1}\left(\mathbf{A}^{T}\mathbf{A}\mathbf{0} - \mathbf{A}^{T}\mathbf{y} + \lambda^2\mathbf{L}^T\mathbf{L}\mathbf{0}\right)\\
\mathbf{x}_{1}& = - \left(\mathbf{A}^{T}\mathbf{A}  + \lambda^2\mathbf{L}^T\mathbf{L}\right)^{-1}\left(- \mathbf{A}^{T}\mathbf{y}\right)\\
\mathbf{x}_{1}& = \left(\mathbf{A}^{T}\mathbf{A}  + \lambda^2\mathbf{L}^T\mathbf{L}\right)^{-1}\mathbf{A}^{T}\mathbf{y}.
\label{eq:gradpasso1}
\end{aligned}
\end{equation} 
\subsection{Regularização multiparâmetros de Tikhonov}\label{Ap:multi}

Seja o funcional dado por 
\begin{equation}
\mathcal{M}(\lambda, \mathbf{x}, \mathbf{y}) = \left( \vert \vert \mathbf{A} \mathbf{x} - \mathbf{y} \vert \vert^2_2 
+ \lambda^2_1 \vert \vert \textbf{L}_{1} \left(\mathbf{x} - \mathbf{x}^*_1\right)\vert \vert^2_2
+ \lambda^2_2 \vert \vert \textbf{L}_{2} \left(\mathbf{x} - \mathbf{x}^*_2\right)\vert \vert^2_2 \right).
\label{eq:multiapp2}
\end{equation}
A minimização da Equação \eqref{eq:multiapp2} em relação à $\mathbf{x}$ define a regularização generalizada de Tikhonov com dois regularizadores. Como todos os termos são quadráticos, ela é escrita como um problema de mínimos quadrados linear em relação a $\mathbf{x}$ \cite[pág. 61]{hansen2010discrete}, conforme
\begin{equation}
\hat{\mathbf{x}} = \arg\min\limits_{\mathbf{x}} 
\left|\left|
\begin{pmatrix}
\mathbf{y}\\ 
\lambda_1 \textbf{L}_1 \mathbf{x}_1^*\\ 
\lambda_2 \textbf{L}_2 \mathbf{x}_2^*\\ 
\end{pmatrix} - \begin{pmatrix}
\mathbf{A}\\ 
\lambda_1 \textbf{L}_1\\ 
\lambda_2 \textbf{L}_2\\ 
\end{pmatrix} \mathbf{x} \right| \right|^2_2.
\label{eq:apmult1}
\end{equation} 
Definindo $\tilde{\mathbf{y}} = \begin{pmatrix}
\mathbf{y} ,& 
\lambda_1 \textbf{L}_1 \mathbf{x}_1^*,&
\lambda_2 \textbf{L}_2 \mathbf{x}_2^* 
\end{pmatrix}^T$ e $\tilde{\mathbf{A}} = \begin{pmatrix}
\mathbf{A}, & 
\lambda_1 \textbf{L}_1 ,& 
\lambda_2 \textbf{L}_2 
\end{pmatrix}^T$, reconhece-se um problema de mínimos quadrados dado por
\begin{equation}
\hat{\mathbf{x}} = \arg\min\limits_{\mathbf{x}} \vert\vert \tilde{\mathbf{y}} - \tilde{\mathbf{A}} \mathbf{x} \vert \vert^2_2.
\end{equation}

Prosseguindo da mesma forma que as Equações \eqref{eq:funcional_app}, \eqref{eq:funcional_app2}, \eqref{eq:normalequation_app} e \eqref{eq:otimizacao12}, 
igualando o gradiente a zero, obtém-se as equações normais
\begin{equation}
\tilde{\mathbf{A}}^T \tilde{\mathbf{A}} \mathbf{x} = \tilde{\mathbf{A}}^T \tilde{\mathbf{y}}.
\end{equation}
Substituindo $\tilde{\mathbf{A}}$ e $\tilde{\mathbf{y}}$ de volta, obtém-se
\begin{equation}
\begin{pmatrix}
\textbf{A}^T, \lambda_1 \textbf{L}_1^T, \lambda_2 \textbf{L}_2^T\\ \end{pmatrix}
\begin{pmatrix}
\mathbf{A} \\ 
\lambda_1 \textbf{L}_1 \\ 
\lambda_2 \textbf{L}_2 
\end{pmatrix} \mathbf{x} = \begin{pmatrix}
\textbf{A}^T, \lambda_1 \textbf{L}_1^T, \lambda_2 \textbf{L}_2^T\\ \end{pmatrix}
\begin{pmatrix}
\mathbf{y} \\ 
\lambda_1 \textbf{L}_1 \mathbf{x}_1^* \\
\lambda_2 \textbf{L}_2 \mathbf{x}_2^* 
\end{pmatrix} 
\end{equation}
\begin{equation}
\left( \mathbf{A}^T \mathbf{A} + \lambda_1^2 \textbf{L}_1^T \textbf{L}_1 + \lambda_2^2 \textbf{L}_2^T \textbf{L}_2 \right) \hat{\mathbf{x}} = \left(\mathbf{A}^T \mathbf{y} + \lambda_1^2 \textbf{L}_1^T \textbf{L}_1 \mathbf{x}_1^* + \lambda_2^2 \textbf{L}_2^T \textbf{L}_2 \mathbf{x}_2^*\right),
\label{eq:tikmult}
\end{equation}
que por sua vez pode ser entendida como um sistema linear de equações do tipo $\mathbf{A}\mathbf{x} = \mathbf{y}$ na qual se deseja obter $\mathbf{x}$. É claro que é possível isolar $\mathbf{x}$ para obter a solução fechada da Equação \eqref{eq:eqmult2}, mas dependendo da dimensionalidade do problema utilizar algoritmos iterativos diretamente na Equação \eqref{eq:tikmult} pode trazer vantagens de performance.

\subsection{Considerações sobre a notação utilizada}

Considerando a notação utilizada nesta subseção e no restante do texto, algumas observações podem ser feitas:
\begin{itemize}
\item Nas Equações \eqref{eq:tikhonov1} e \eqref{eq:multiapp2},  os parâmetros de regularização $\lambda_i$ aparecem ao quadrado, de modo que eles aparecem apenas como $\lambda$ quando colocados para dentro da norma quadrática nas Equações \eqref{eq:tikhonov18} E \eqref{eq:apmult1}, respectivamente.  Ressalta-se que essa notação é variada na literatura. Em \cite[Subseção 5.2] {Mueller2012}, denotou-se apenas $\lambda$ na Equação \eqref{eq:tikhonov1}, de modo que na Equação \eqref{eq:tikhonov18} ele teria que aparecer como $\sqrt{\lambda}$. Numericamente não faz diferença, mas optou-se por não utilizar a raiz quadrada;
\item No termo de fidelidade ou no agrupamento de termos em uma norma $\ell_2^2$, a Equação \eqref{eq:funcional_app} mostra que é equivalente escrever $\vert\vert \tilde{\mathbf{y}} - \tilde{\mathbf{A}} \mathbf{x} \vert \vert^2_2$ e $\vert\vert \tilde{\mathbf{A}} \mathbf{x}-\tilde{\mathbf{y}} \vert \vert^2_2$;
\item O cálculo do gradiente em relação a $\mathbf{x}$ da Equação \eqref{eq:apmult1} resulta na constante multiplicativa $= 2$ por conta da norma $\ell_2^2$. Em alguns trabalhos, como \cite[Equação 2.1]{Benning2018}, o funcional $\mathcal{M}(\lambda, \mathbf{x}, \mathbf{y})$ aparece multiplicado por $\frac{1}{2}$, de tal modo que as constantes no numerador e denominador se compensariam. Em outros trabalhos, como \cite[Equação 4.8]{hansen2010discrete}, não há $\frac{1}{2}$. No entanto, o resultado da minimização de $\frac{1}{2} \mathcal{M}(\lambda, \mathbf{x}, \mathbf{y})$ e de  $\mathcal{M}(\lambda, \mathbf{x}, \mathbf{y})$ é o mesmo, a constante multiplicativa não o altera. Assim, optou-se por não incluir $\frac{1}{2}$;
\end{itemize}

\subsection{\textit{Iteratively Reweighted Least Squares}}\label{Ap:IRLS}

\subsubsection{Algoritmo IRLS para norma $\ell_1$ no termo de fidelidade}\label{Ap:IRLSsub1}

Seja o problema de otimização dado por um termo de fidelidade  com norma $\ell_1$ conforme
 \begin{equation}
\hat{\mathbf{x}} = \arg\min\limits_{\mathbf{x}}  \vert \vert \mathbf{A}\mathbf{x} - \mathbf{y} \vert \vert_1. 
\label{eq:IRLS1}
\end{equation}
Essa expressão não é diferenciável em todos os pontos e não há solução fechada para obtenção de $\hat{\mathbf{x}}$. A ideia central do algoritmo \textit{iteratively reweighted least squares} (IRLS) é a de aproximar a solução da Equação \eqref{eq:IRLS1} através de um algoritmo iterativo \cite[págs. 46-7]{aster2019parameter}. A sua utilização no regularizador e para norma $\ell_p$  são análogos e serão mostrados na sequência. Seja $\mathbf{r}$ um vetor residual que representa o argumento da norma,  
\begin{equation}
\mathbf{r}(\mathbf{x}) = \mathbf{A} \mathbf{x} - \mathbf{y}.
\label{eq:IRLS2}
\end{equation}
Dessa forma, o termo de fidelidade com norma $\ell_1$ pode ser reescrito como
\begin{equation}
\vert \vert \mathbf{A} \mathbf{x} - \mathbf{y}\vert \vert_1 = \vert \vert \mathbf{r}(\mathbf{x}) \vert \vert_1
\label{eq:IRLS3}
\end{equation}
\begin{equation}
\vert \vert \mathbf{r}(\mathbf{x}) \vert \vert_1 = \sum_i^m \vert r_i \vert,
\label{eq:IRLS4}
\end{equation}
 onde $r_i$, para $i = 1,\dots,m$ são os componentes do vetor $\mathbf{r}$. De agora em diante o vetor residual é escrito apenas como $\mathbf{r}(\mathbf{x}) = \mathbf{r}$, a menos que se deseje destacar essa relação.

Aplicando a regra da cadeia no lado direito da Equação \eqref{eq:IRLS4} (para considerar tanto o valor absoluto quanto $\mathbf{r}$ que depende de $\mathbf{x}$)  para calcular as derivadas parciais em relação aos componentes do vetor de parâmetros $\mathbf{x}$, é possível mostrar que
\begin{equation}
\frac{\partial \sum_{i=1}^m \vert r_i\vert}{\partial x_j}=  -\sum_{i=1}^m A_{i,j} \sgn(r_i),
\label{eq:IRLS5}
\end{equation}
onde $A_{i,j}$ denota o valor do $(i, j)$-ésimo elemento da matriz $\mathbf{A}$ e $\sgn(\cdot)$ denota a função sinal, que retorna o valor $1$ para entradas positivas e o valor $-1$ para entradas negativas. Considerando que $\sgn(\mathbf{r}_i) = \frac{r_i}{\vert r_i\vert}$, o gradiente do lado direito da Equação \eqref{eq:IRLS3} pode ser calculado e escrito na forma matricial conforme
\begin{equation}
\nabla \vert \vert \mathbf{r} \vert \vert_1 =  - \mathbf{A}^T \mathbf{W}\mathbf{r},
\label{eq:IRLS6}
\end{equation}
onde a matriz de pesos $\mathbf{W}$ é uma matriz diagonal calculada através de
\begin{equation}
W_{i,i} = diag\left( 1\backslash \vert r_i \vert \right).
\label{eq:IRLS7}
\end{equation}
A Equação \eqref{eq:IRLS7}  é válida quando os componentes $r_i \neq 0$, evitando a  divisão por zero. Caso isso não se verifique, é possível definir uma tolerância $\varepsilon > 0$ tal que
\begin{equation}
W_{i,i} = \begin{cases}
    1\backslash \vert r_i \vert,   & \text{para }  \vert r_i \vert \geq \varepsilon \\
    1\backslash \varepsilon, & \text{para } \vert r_i \vert < \varepsilon.
  \end{cases}
  \label{eq:IRLS8}
\end{equation}
Igualando a Equação \eqref{eq:IRLS6} a zero, substituindo $\mathbf{r}$ de volta  e isolando $\mathbf{x}$, obtém-se  
\begin{equation}
\hat{\mathbf{x}}   = \left( \mathbf{A}^T \mathbf{W} \mathbf{A} \right)^{-1}\mathbf{A}^T  \mathbf{W} \mathbf{y},
\label{eq:IRLS9}
\end{equation}
que é a solução de equações normais para
 \begin{equation}
\hat{\mathbf{x}} = \arg\min\limits_{\mathbf{x}}  \vert \vert \sqrt{ \mathbf{W}} (\mathbf{A}\mathbf{x} - \mathbf{y}) \vert \vert_2^2,
\label{eq:IRLS11}
\end{equation} mas agora reponderadas pela matriz $\mathbf{W}$. Assim, no IRLS
 a aproximação da norma $\ell_1$ por um termo quadrático é feita através de
\begin{equation}
\vert \vert \mathbf{A}\mathbf{x} - \mathbf{y} \vert \vert_1 = \vert \vert \sqrt{ \mathbf{W}} (\mathbf{A}\mathbf{x} - \mathbf{y}) \vert \vert_2^2,
\label{eq:IRLS10}
\end{equation}
onde  $\mathbf{W}$ também depende de $\mathbf{x}$, sendo necessária atualizá-laa cada iteração. 

Seja $k = 1, \dots, n$, onde $n$ é o número de iterações. Para enfatizar a característica iterativa do IRLS, reescreve-se a Equação \eqref{eq:IRLS9} como
\begin{equation}
\begin{aligned}
\left( \mathbf{A}^T \mathbf{W}_{k-1} \mathbf{A} \right) \mathbf{x}_{k}  & = \mathbf{A}^T  \mathbf{W}_{k-1} \mathbf{y} \\
\tilde{\mathbf{A}}_{k-1}\mathbf{x}_k & = \tilde{\mathbf{y}}_{k-1},
\end{aligned}
\label{eq:irlsiterative1}
\end{equation}
de modo que após $n$ iterações a aproximação já seja adequada. A inicialização de $\mathbf{W}$ pode ser como matriz identidade \cite[pág. 49]{majumdar2019compressed} ou já de acordo com o residual $\mathbf{r}$  \cite[pág. 197]{aster2019parameter}.

Disso vem o nome do algoritmo, \textit{iteratively reweighted least squares}, pois a Equação \eqref{eq:IRLS11} é um problema de mínimos quadrados e mesmo que o operador direto seja linear, a sua solução deve ser calculada várias vezes, atualizando $\mathbf{W}$ a cada iteração. Na sequência, o caminho para se obter a Equação \eqref{eq:irlsiterative1} será novamente utilizado para aproximar outros termos do funcional.

\subsubsection{Interpretação alternativa do IRLS}
Seja $\mathbf{W}$ calculada pela Equação \eqref{eq:IRLS8}. Conforme  \cite[pág. 173]{Bjrck1996}, outra interpretação da aproximação realizada na Equação \eqref{eq:IRLS10} é possível reescrevendo a Equação \eqref{eq:IRLS4} como
\begin{equation}
\begin{aligned}
\vert \vert \mathbf{r} \vert \vert_1 & = \sum_i^m \frac{\vert r_i\vert^2}{\vert r_i \vert} \quad \\
 & =   \sum_i^m \frac{r_i^2}{\vert r_i \vert}  \\
  & =  \mathbf{r}^T \mathbf{W} \mathbf{r} \\ 
  & =   \vert \vert \sqrt{\mathbf{W}} \mathbf{r} \vert \vert^2_2.
\label{eq:IRLS12}
\end{aligned}
\end{equation}
Ou seja, éoutra forma de se entender a aproximação de termos de norma $\ell_1$ através de termos com norma $\ell_2^2$. 

\subsubsection{IRLS no regularizador} 

Seja o problema linear regularizado dado pelo funcional
\begin{equation}
\hat{\mathbf{x}} = \arg\min\limits_{\mathbf{x}} \left[ \vert \vert \mathbf{A}\mathbf{x} - \mathbf{y} \vert \vert_2^2
+ \lambda^2 \Omega(\mathbf{x}) \right].
\label{eq:IRLS13}
\end{equation}
Seguindo o procedimento da Seção \ref{Ap:IRLSsub1}, é possível obter expressões do IRLS também:
\begin{itemize}
\item Quando $\Omega(\mathbf{x}) = \vert \vert \mathbf{x}  \vert \vert_1$, que remete à regularização clássica de Tikhonov. Pode-se seguir a sequência da Seção \ref{Ap:IRLSsub1} e calcular $\mathbf{W}$ segundo a Equação \eqref{eq:IRLS8}, mas considerando $\mathbf{r} = \mathbf{x}$ ao invés de calcular os pesos da matriz $\mathbf{W}$ em função do residual $\mathbf{r}$ da Equação \eqref{eq:IRLS2}. Nesse caso, o gradiente pode ser escrito na forma matricial
\begin{equation}
\nabla \vert \vert \mathbf{x} \vert \vert_1 =  \mathbf{W}\mathbf{x}.
\label{eq:IRLS14}
\end{equation}
Após cálculo do gradiente em relação a $\mathbf{x}$ da Equação \eqref{eq:IRLS13} como um todo e igualado a zero \cite[págs. 196-7]{aster2019parameter}, uma iteração do IRLS seria
\begin{equation}
\hat{\mathbf{x}}  = \left(2 \mathbf{A}^T \mathbf{A} + \lambda^2 \mathbf{W}\right)^{-1}2\mathbf{A}^T  \mathbf{y},
\label{eq:IRLS15}
\end{equation}
da qual os fatores multiplicativos $=2$ aparecem pois apenas um dos termos do funcional era quadrático, mas é possível reescrever como
\begin{equation}
\begin{aligned}
\hat{\mathbf{x}}  & = \left(2 \left(\mathbf{A}^T \mathbf{A} + \frac{1}{2}\lambda^2 \mathbf{W}\right)  \right)^{-1}2\mathbf{A}^T  \mathbf{y}\\
 & = \left(\mathbf{A}^T \mathbf{A} + \frac{1}{2}\lambda^2 \mathbf{W}\right)^{-1}\mathbf{A}^T  \mathbf{y},
\end{aligned}
\label{eq:IRLS16}
\end{equation}
pois $(c \mathbf{A})^{-1} = c^{-1} \mathbf{A}^{-1}$. Em todo caso, ainda é necessária alguma regra para escolher $\lambda^2$ ou $\lambda^2 \backslash 2 $, mas daqui em diante será escrito apenas como $\lambda^2$.  

A aproximação nesse regularizador (ou em outros) só muda a forma como os pesos de $\mathbf{W}$ são calculados, mas no final 
uma relação análoga à Equação \eqref{eq:IRLS10} é obtida,
 \begin{equation}
\vert \vert \mathbf{x}  \vert \vert_1 = \vert \vert \sqrt{ \mathbf{W}}\mathbf{x} \vert \vert_2^2,
\label{eq:IRLS17}
\end{equation}
o que significa que a Equação \eqref{eq:IRLS16} apresenta a mesma forma da Equação \eqref{eq:tiksolgen} quando $\mathbf{L} = \sqrt{ \mathbf{W}}$.

\item Quando $\Omega(\mathbf{x}) = \vert \vert \mathbf{L}\mathbf{x}  \vert \vert_1$, que remete à regularização generalizada de Tikhonov. Em \cite[págs. 196-7]{aster2019parameter}, o procedimento da Seção \ref{Ap:IRLSsub1} é adotado, mas os valores da matriz $\mathbf{W}$ são calculados considerando $\mathbf{r} = \mathbf{L}\mathbf{x}$, de modo que
\begin{equation}
\vert \vert \mathbf{L}\mathbf{x}  \vert \vert_1 = \vert \vert \sqrt{\mathbf{W}} \mathbf{L} \mathbf{x} \vert \vert_2^2,
\label{eq:IRLS18}
\end{equation}
e como $\left(\sqrt{ \mathbf{W} \mathbf{L}}\right)^T =  \mathbf{L}^T \sqrt{\mathbf{W}}^T$, um passo da solução se torna
 \begin{equation}
\hat{\mathbf{x}}  = \left( \mathbf{A}^T \mathbf{A} + \lambda^2  \mathbf{L}^T\mathbf{W}\mathbf{L}\right)^{-1}\mathbf{A}^T  \mathbf{y}.
\label{eq:IRLS19}
\end{equation}
\item Quando $\Omega(\mathbf{x}) = \vert \vert \mathbf{L}(\mathbf{x}-\mathbf{x}^*) \vert \vert_1$, que remete à regularização generalizada com valor de referência. Deve-se considerar $\mathbf{r} =\mathbf{L}(\mathbf{x}-\mathbf{x}^*)$ de modo que
\begin{equation}
\vert \vert \mathbf{L}(\mathbf{x}-\mathbf{x}^*)  \vert \vert_1 = \vert \vert \sqrt{\mathbf{W}} \mathbf{L} (\mathbf{x}-\mathbf{x}^*) \vert \vert_2^2.
\label{eq:IRLS20}
\end{equation}
Um passo da solução da Equação \eqref{eq:IRLS20}, conforme dedução análoga ao desenvolvido na Subseção \ref{Ap:multi},  é dada por 
 \begin{equation}
\hat{\mathbf{x}} = \left( \mathbf{A}^T \mathbf{A} + \lambda^2  \mathbf{L}^T\mathbf{W}\mathbf{L}\right)^{-1}\left(\mathbf{A}^T  \mathbf{y} + \lambda^2\mathbf{L}^T\mathbf{W}\mathbf{L} \mathbf{x}^* \right).
\label{eq:IRLS21}
\end{equation}
\end{itemize}

Implementar as Equações \eqref{eq:IRLS16}, \eqref{eq:IRLS19} e \eqref{eq:IRLS21} pode não ser eficiente computacionalmente. Uma alternativa é considerar que
\begin{equation}
\hat{\mathbf{x}} = \arg\min\limits_{\mathbf{x}} 
\left|\left|
\begin{pmatrix}
\mathbf{A}\\ 
\lambda \sqrt{\mathbf{W}} \textbf{L}\\ 
\end{pmatrix}\mathbf{x} -\begin{pmatrix}
\mathbf{y}\\ 
\lambda \sqrt{\mathbf{W}}\textbf{L} \mathbf{x}^*\\ 
\end{pmatrix}   \right| \right|^2_2
\label{eq:IRLS22}
\end{equation} 
e resolvê-lo de outra forma, como o método LSQR \cite[pág. 197]{aster2019parameter}.

\subsubsection{IRLS em todos os termos do funcional}
Juntando as abordagens das seções anteriores, é possível a abordagem do IRLS tanto no termo de fidelidade quanto no regularizador 
\begin{equation}
\hat{\mathbf{x}} = \arg\min\limits_{\mathbf{x}} \left[ \vert \vert \mathbf{A}\mathbf{x} - \mathbf{y} \vert \vert_1
+ \lambda^2 \vert \vert \mathbf{L}(\mathbf{x}-\mathbf{x}^*)  \vert \vert_1\right],
\label{eq:IRLS23}
\end{equation}
cuja solução é aproximada por
\begin{equation}
\hat{\mathbf{x}} = \arg\min\limits_{\mathbf{x}} \left[ \vert \vert \sqrt{ \mathbf{W}_1} (\mathbf{A}\mathbf{x} - \mathbf{y}) \vert \vert_2^2 + \lambda^2 \vert \vert  \sqrt{ \mathbf{W}_2} \mathbf{L}(\mathbf{x}-\mathbf{x}^*)  \vert \vert_2^2\right],
\label{eq:IRLS24}
\end{equation}
onde $\mathbf{W}_1$ e $\mathbf{W}_2$ são calculadas separadamente. Um passo da solução da Equação \eqref{eq:IRLS24} é 
\begin{equation}
\hat{\mathbf{x}}   = \left( \mathbf{A}^T \mathbf{W}_1 \mathbf{A} + \lambda^2  \mathbf{L}^T\mathbf{W}_2 \mathbf{L} \right)^{-1}  \left(\mathbf{A}^T  \mathbf{W}_1 \mathbf{y} + \lambda^2\mathbf{L}^T\mathbf{W}_2\mathbf{L} \mathbf{x}^* \right).
\label{eq:IRLS25}
\end{equation}

\subsubsection{IRLS: Aproximação da norma $\ell_p$}
O IRLS  pode aproximar uma norma $\ell_p$ para $p$ qualquer, definida no Apêndice \ref{Ap:normas1}, como um termo quadrático e novamente isto só afeta a forma como os pesos de $\mathbf{W}$ são calculados \cite[pág. 173]{Bjrck1996}, \cite[págs. 48-9]{majumdar2019compressed}.  Seguindo a apresentação de \cite[Apêndice]{Scales1988FastLP}, seja 
\begin{equation}
\vert \vert \mathbf{r}(\mathbf{x}) \vert \vert_p^p = \sum_{i=1}^m \vert r_i\vert^p.
\label{eq:IRLS26}
\end{equation}
Neste caso, quando se deriva essa expressão em relação aos componentes de $\mathbf{x}$ é necessário utilizar a regra da cadeia para uma composição de três funções: o expoente $p$, o valor absoluto e o vetor residual $ \mathbf{r}$ em si que depende de $\mathbf{x}$, isto é
\begin{equation}
\frac{\partial \sum_{i=1}^m \vert r_i\vert^p}{\partial x_j}=  \sum_{i=1}^m  p \vert r_i \vert^{p-1}  \sgn(r_i)   \frac{\partial \sum_{i=1}^m  r_i}{\partial x_j},
\label{eq:IRLS27}
\end{equation}
onde a expressão $\frac{\partial \sum_{i=1}^m  r_i}{\partial x}$ depende da forma de $\mathbf{r}(\mathbf{x})$ adotada. Como $\sgn(\mathbf{r}_i) = \frac{r_i}{\vert r_i\vert}$, 
\begin{equation}
\frac{\partial \sum_{i=1}^m \vert r_i\vert^p}{\partial x_j}=  \sum_{i=1}^m r_i  p \vert r_i \vert^{p-2}   \frac{\partial \sum_{i=1}^m  r_i}{\partial x_j}.
\label{eq:IRLS28}
\end{equation} 
Quando $\mathbf{r}(\mathbf{x}) = \mathbf{x}$, $\frac{\partial \sum_{i=1}^m  x_i}{\partial x_j}$ resulta em uma matriz identidade e 
\begin{equation}
\frac{\partial \sum_{i=1}^m \vert r_i\vert^p}{\partial x_j}=  \sum_{i=1}^m   p \vert x_i \vert^{p-2} x_i.
\label{eq:IRLS29}
\end{equation}
Assim, a Equação \eqref{eq:IRLS29} em notação matricial e igualando a zero se torna
\begin{equation}
\nabla \vert \vert \mathbf{r} \vert \vert_p^p =  \mathbf{W} \mathbf{x}, 
\label{eq:IRLS30}
\end{equation}
\begin{equation}
\mathbf{W} = diag(p \vert x_i \vert^{p-2}),
\label{eq:IRLS31}
\end{equation}
onde se define uma tolerância para não dividir por zero. Da Equação \eqref{eq:IRLS31}:
\begin{itemize}
\item Em \cite{Scales1988FastLP}, a proposta de norma $\ell_p$ considerava $p=0$, mas a Equação \eqref{eq:IRLS29} possui $p$ multiplicando no numerador, anulando-o. Mesmo assim, os autores dizem que essa constante $p$ pode ser ignorada \cite[pág. 332]{Scales1988FastLP} e definem $\mathbf{W} = diag(\vert x_i \vert^{p-2})$ quando utilizaram $p=0$;
\item Quando $p=1$, $\mathbf{W}$ apresenta a mesma forma da Equação \eqref{eq:IRLS8}, mas para serem iguais deve-se definir $\mathbf{r}$ conforme Equação \eqref{eq:IRLS2}, o que não é obrigatório (outros  $\mathbf{r}$ podem ser considerados) ;
\item Quando $p=2$, $\mathbf{W} = 2\mathbf{I}$, mas a constante $=2$ não faria diferença quando a equação fosse igualada a zero. Assim, a solução seria vista como a de mínimos quadrados ordinária.

\end{itemize}

\newpage
\section{Regularização de Tikhonov como filtragem espectral}\label{App:svd2}

Retomando a Subseção \ref{app-svd}, a SVD também permite\footnote{Sejam duas matrizes $\mathbf{A}$ e $\mathbf{B}$ quadradas e não-singulares. Assim, $\left(\mathbf{A} \mathbf{B} \mathbf{A}^T\right)^{-1} = \mathbf{A}^{-T} \mathbf{B}^{-1} \mathbf{A}^{-1}.$} buscar a inversão de $\mathbf{A}$:
\begin{equation}
\begin{aligned}
\mathbf{A}^{-1}  & =   \left(\mathbf{U} \mathbf{\Sigma} \mathbf{V}^T \right)^{-1} \\
& =  \mathbf{V}^{-T} \mathbf{\Sigma}^{-1} \mathbf{U}^{-1} \\
& = \mathbf{V} \mathbf{\Sigma}^{-1} \mathbf{U}^T \\ 
\end{aligned}
\end{equation}
Quando $\mathbf{A}$ é quadrada,  a sua inversa $\mathbf{A}^{-1}$ pode ser escrita como
\begin{equation}
\begin{aligned}
\putunder{\mathbf{A}^{-1}}{m \times m} & = \putunder{\begin{pmatrix}
\vert & & \vert\\
\mathbf{v}_1 & \cdots & \mathbf{v}_n\\
\vert & & \vert\\
\end{pmatrix}}{m \times m}
\putunder{\begin{pmatrix}
\frac{1}{\sigma_1} & & 0\\
& \ddots & \\
0 & & \frac{1}{\sigma_m}\\
\end{pmatrix}}{m \times m}
\putunder{\begin{pmatrix}
\vert & & \vert\\
\mathbf{u}_1 & \cdots & \mathbf{u}_n\\
\vert & & \vert\\
\end{pmatrix}^{T}}{m \times m}.
\label{eq:dvs_inv0}
\end{aligned}
\end{equation}
A SVD de matrizes retangulares resulta em valores singulares nulos. Vetores singulares à direita e à esquerda associados aos valores singulares nulos não são eles próprios nulos e ainda podem não ser únicos \cite[Seção 58]{Hogben2013}, \cite[pág. 49]{Mueller2012}.

\subsection{Solução ingênua}

Se $\mathbf{A}$ é mal-condicionada, o valor numérico da solução pode até ser inadequado. Ainda assim, escreve-se a solução ingênua pela SVD \cite[Equação 2.6.1]{golub2013matrix} conforme
\begin{equation}
\begin{aligned}
\mathbf{x} & =    \left(\mathbf{V} \mathbf{\Sigma}^{-1} \mathbf{U}^{T} \right) \mathbf{y}_{\delta} \\
& =  \putunder{\left(\sum_{i=1}^n \mathbf{v}_i \frac{1}{\sigma_i} \mathbf{u}_i^T\right)}{m \times m} \putunder{\mathbf{y}_{\delta}}{m \times 1} \\
& = \sum^k_{i=1} \frac{\mathbf{u}_i^T \mathbf{y}_{\delta}}{\sigma_i}  \mathbf{v}_i.
\label{eq:dsv11}
\end{aligned}
\end{equation}
No caso de um ruído aditivo, tem-se
\begin{equation}
\mathbf{x} =  \sum_{i=1}^n \frac{\mathbf{u}_i^T \mathbf{y}}{\sigma_i} \mathbf{v}_i + \sum_{i=1}^n \frac{\mathbf{u}_i^T\bm{\delta}}{\sigma_i} \mathbf{v}_i, 
\label{xnaive}
\end{equation}
cujo resultado depende da divisão dos termos do numerador pelos $\sigma_i$, amplificando $\bm{\delta}$ quando há a SVD de $\mathbf{A}$ resulta em pequenos $\sigma_i$ \cite[pág. 30]{hansen2010discrete}. 

\subsection{Regularização clássica de Tikhonov}\label{sec:svdtikh}

A solução $\mathbf{x}_{\lambda}$ da Equação \eqref{eq:tiksol} pode ser reescrita considerando a SVD de $\mathbf{A}$ da Equação \eqref{eq:dvs_inv0}, conforme
\begin{equation}
\begin{aligned}
\mathbf{x}_{\lambda} & = \left( \mathbf{V} \mathbf{\Sigma} \mathbf{U}^T \mathbf{U} \mathbf{\Sigma} \mathbf{V}^T + \lambda^2 \mathbf{I} \right)^{-1} \left(\mathbf{U} \mathbf{\Sigma} \mathbf{V}^T \right)^T \mathbf{y}_{\delta} \\
 & = \left( \mathbf{V} \mathbf{\Sigma}^2 \mathbf{V}^T + \lambda^2 \mathbf{I} \right)^{-1} \left(\mathbf{U} \mathbf{\Sigma} \mathbf{V}^T \right)^T \mathbf{y}_{\delta}.
\end{aligned}
\end{equation}
Considerando que $\mathbf{I} = \mathbf{V} \mathbf{V}^T$ da SVD de $\mathbf{A}$, a solução regularizada se torna \cite[pág. 62]{hansen2010discrete}
\begin{equation}
\begin{aligned}
\mathbf{x}_{\lambda} & =  \left( \mathbf{V} \mathbf{\Sigma}^2 \mathbf{V}^T + \lambda^2 \mathbf{V} \mathbf{V}^T \right)^{-1} \left(\mathbf{U} \mathbf{\Sigma} \mathbf{V}^T \right)^T \mathbf{y}_{\delta} \\
 & =  \left( \mathbf{V}  \left( \mathbf{\Sigma}^2 + \lambda^2 \mathbf{I} \right) \mathbf{V}^T \right)^{-1} \mathbf{V} \mathbf{\Sigma} \mathbf{U}^T \mathbf{y}_{\delta} \\
& =  \mathbf{V}^{-T} \left( \mathbf{\Sigma}^2 + \lambda^2 \mathbf{I} \right)^{-1} \mathbf{V}^{-1} \mathbf{V} \mathbf{\Sigma} \mathbf{U}^T \mathbf{y}_{\delta} \\
& =  \mathbf{V} \left( \mathbf{\Sigma}^2 + \lambda^2 \mathbf{I} \right)^{-1} \mathbf{V}^{T} \mathbf{V} \mathbf{\Sigma} \mathbf{U}^T \mathbf{y}_{\delta} \\
& =\mathbf{V} \left(\mathbf{\Sigma}^2 + \lambda^2\mathbf{I} \right)^{-1} \mathbf{\Sigma} \mathbf{U}^T \mathbf{y}_{\delta}.
\end{aligned} 
\end{equation}
Esta, por sua vez, pode ser escrita em função de vetores e valores singulares, conforme
 \begin{equation}
\begin{aligned}
\hat{\mathbf{x}}_{\lambda} & =  \sum_{i=1}^n \left(\frac{1}{\sigma_i^2 + \lambda^2}\right) \sigma_i \mathbf{u}_i^T \mathbf{y}_{\delta} \mathbf{v}_i \\
& =  \sum_{i=1}^n \left(\frac{1}{\sigma_i^2 + \lambda^2}\right)\left( \frac{\sigma_i}{\sigma_i}\right) \sigma_i \mathbf{u}_i^T \mathbf{y}_{\delta} \mathbf{v}_i \\
& =  \sum_{i=1}^n \left(\frac{\sigma_i^2}{\sigma_i^2 + \lambda^2}\right) \frac{\mathbf{u}_i^T \mathbf{y}_{\delta} }{\sigma_i} \mathbf{v}_i.
\label{eq:svdtikh0}
\end{aligned}
\end{equation}
De modo mais geral, a relação anterior é escrita como 
\begin{equation}
\hat{\mathbf{x}}_{\lambda} =  \sum_{i=1}^n \phi \frac{\mathbf{u}_i^T \mathbf{y}_{\delta} }{\sigma_i} \mathbf{v}_i,
\label{tiksvdsol_repeat05}
\end{equation}
onde $\phi$ são os fatores de filtro \cite[pág. 62]{hansen2010discrete}. 

No caso da regularização clássica de Tikhonov, $\phi$ tem a forma de
\begin{equation}
\phi = \frac{\sigma_i^2}{\sigma_i^2 + \lambda^2} \approx \begin{cases} 
1, & \sigma_i\gg \lambda \\
\sigma_i^2 \backslash \lambda^2, & \sigma_i\ll \lambda.
\end{cases}
\label{eq:filterfactor_text}
\end{equation}
A solução $\hat{\mathbf{x}}_{\lambda}$ da Equação \eqref{tiksvdsol_repeat05} pode ser entendida como uma soma ponderada dos vetores singulares a direita, que formam uma base de representação do sinal:
\begin{itemize}
\item Quanto maior $\lambda$, maior é o peso de $\vert \vert \mathbf{x} \vert \vert_2^2$ e a solução $\hat{\mathbf{x}}_{\lambda}$ será composta de menos coeficientes da SVD (com apenas os maiores $\sigma$), e como seus vetores a direita $\mathbf{v}_i$ apresentam menor frequência, a solução é mais suavizada; 
\item Quanto menor $\lambda$, maior é o peso de $\vert \vert \mathbf{A} \mathbf{x} - \mathbf{y} \vert \vert^2_2$ e mais valores singulares são adicionados à solução $x_{\lambda}$, e seus vetores a direita $\mathbf{v}_i$ associados, incluindo aqueles de maior frequência, e a solução pode ser menos suavizada do que o necessário. 
\end{itemize} 
Uma vez que valores singulares menores são atenuados, e relacionando essa informação com a Figura \ref{fig:01_006}, observa-se que a informação perdida nesse processo de filtragem é relativa às altas frequências da base espectral.

\subsection{Variações sobre a SVD}

Além da regularização clássica de Tikhonov, diversos algoritmos de reconstrução podem ser obtidos a partir da SVD \cite{Hansen2007}, como a regularização a partir do princípio da discrepância de Morozov, e variações como a SVD amortecida, na qual os fatores de filtro decaem mais lentamente do que aqueles da Equação \eqref{eq:filterfactor_text} \cite{Hansen2007}. A escolha eficiente entre os algoritmos depende do problema que se quer resolver e de sua dimensionalidade. Especificamente, pode não ser eficiente o cálculo da SVD de operadores diretos de grande dimensionalidade.

Na regularização generalizada de Tikhonov, é possível realizar a análise da inversão regularizada a partir da decomposição em valores singulares generalizada (GSVD), uma decomposição matricial em conjunto de $\mathbf{A}$ e $\mathbf{L}$ \cite[pág. 104]{aster2019parameter}. Ela permite ver, por exemplo, quando que o ruído é melhor suprimido \cite[pág. 180]{hansen2010discrete}. Em \cite[pág. 103]{zhang2010machine}, os autores obtiveram os fatores de filtro dessa descomposição. Além disso, uma discussão sobre como valores de referência influenciam a solução sob a perspectiva da SVD é encontrada em \cite[págs. 61-3]{Neto2005}.

\newpage
\section{Métodos de projeção}\label{Ap:projection}

Diversos algoritmos iterativos são baseados em métodos de projeção. Esse \textit{framework} traz a possibilidade de levar a solução de um problema inverso para outro espaço quando isso traz vantagens, como ressaltar características desejadas ou em problemas de grande escala. A descrição do método de projeção aqui mostrada foi baseada em \cite[Subseção 6.2]{hansen2010discrete} e é mais computacional. Uma visão mais matemática está disponível em \cite[Subseção 3.3]{engl1996regularization}. Dentro dos métodos de projeção existe ainda a regularização por discretização, cujo efeito da regularização é obtido apenas pela discretização do modelo contínuo, utilizando-se técnicas conhecidas, como colocação, Galerkin ou a aproximação de Ritz \cite[pág. 63]{engl1996regularization}, \cite[Capítulo 3]{Kirsch2011}. No entanto, elas não serão aprofundadas aqui. 

Um método de projeção \cite[págs. 114-6]{hansen2010discrete} busca resolver 
\begin{equation}
\hat{\mathbf{x}} = \arg\min\limits_{\mathbf{x}} \left[ \vert \vert \mathbf{A} \mathbf{x} - \mathbf{y} \vert \vert^2_2 \right] \quad \text{s.t.} \quad \mathbf{x} \in \mathcal{D},
\label{eq:proj1}
\end{equation}
onde $\mathcal{D}$ é o espaço escolhido. Para saber a melhor base é necessário conhecer o sinal de interesse, ao mesmo tempo que o acesso a esse sinal é o problema inverso que se deseja resolver \cite{LewisD2019}. A pergunta que se quer responder é: Qual é a melhor base de representação para uma determinada aplicação? Alguns exemplos são discutidos a seguir. 

\subsection{Subespaços de Krylov}
O subespaço de Krylov $\mathcal{K}_k$ é considerado uma boa escolha para utilização em método de projeção \cite[Subseção 6.3.1]{hansen2010discrete} e é frequentemente utilizado em algoritmos iterativos para problemas de grande escala. Ele é um subespaço que depende tanto do modelo $\mathbf{A}$ quanto dos vetores de medida $\mathbf{y}$ \cite[pág. 119]{hansen2010discrete}, sendo portanto um método adaptativo que constrói uma base específica para aquela aplicação. Ele pode ser formulado como 
\begin{equation}
\hat{\mathbf{x}} = \arg\min\limits_{\mathbf{x}} \left[ \vert \vert \mathbf{A} \mathbf{x} - \mathbf{y} \vert \vert^2_2 \right], \quad \text{s.t.} \quad \mathbf{x} \in \mathcal{K}_k,
\label{eq:cgls}
\end{equation}
onde $\mathcal{K}_k$ é o espaço gerado a partir dos vetores 
\begin{equation}
\mathcal{K}_k = span\{ \mathbf{A} \mathbf{y}, \left(\mathbf{A}^T \mathbf{A} \right)^1\mathbf{A} \mathbf{y}, \left(\mathbf{A}^T \mathbf{A} \right)^2\mathbf{A} \mathbf{y}, \dots, \left(\mathbf{A}^T \mathbf{A} \right)^{k-1}\mathbf{A} \mathbf{y} \},
\label{eq:Krylov}
\end{equation}
buscando levar o problema para um espaço de menor dimensão, mas que seja capaz de carregar informação sobre os dados \cite[pág. 131]{hansen2010discrete}. 

Um exemplo de algoritmo é o CGLS baseado nos subspaços de Krylov \cite[pág. 121]{hansen2010discrete}, que busca resolver as equações normais $\mathbf{A}^T \mathbf{A} \mathbf{x}  = \mathbf{A} \mathbf{y}$ associadas ao problema de mínimos quadrados $\vert \vert \mathbf{A} \mathbf{x} - \mathbf{y} \vert \vert^2_2$. No CGLS, a projeção acontece pela própria definição dos passos do algoritmo iterativo.

\subsection{Métodos de projeção na forma matricial}

Para reescrever a Equação \eqref{eq:proj1} de modo que seja mais fácil de implementá-la computacionalmente \cite[pág. 116]{hansen2010discrete}, seja a equação de síntese \cite[pág. 57]{majumdar2019compressed}, 
\begin{equation}
\mathbf{x} = \mathbf{D} \mathbf{s}_D,
\label{eq:CS1}
\end{equation}
onde $\mathbf{D} = \left( \mathbf{d}_1, \mathbf{d}_2, \dots, \mathbf{d}_k \right) $, as $k$ colunas de $\mathbf{D}$ são vetores da base que geram o espaço $\mathcal{D}$ e $\mathbf{s}_D$ é um vetor contendo os coeficientes de ponderação da base $\mathbf{D}$ de projeção. Enquanto $\mathbf{x}$ é a representação de um sinal no domínio original, $\mathbf{s}_D$ é uma nova representação. Na literatura, $\mathbf{D}$ é chamado de dicionário (ou \textit{frame}) e seus vetores de base são os átomos \cite[pág. 377]{ikeuchi2014computer}. 

Para recuperar o sinal em seu domínio original, a equação de análise traz a relação inversa entre $\mathbf{x}$ e $\mathbf{s}$ conforme
\begin{equation}
\mathbf{s}_D = \mathbf{D}^{-1} \mathbf{x},
\label{eq:CS11}
\end{equation}
dado que $\mathbf{D}^{-1}$ exista, sendo interessante que $\mathbf{D}$ seja ortogonais para que $\mathbf{D}^{-1} = \mathbf{D}^{T}$. 

\begin{figure}[H]
\begin{centering}
\includegraphics[width=0.9\textwidth]{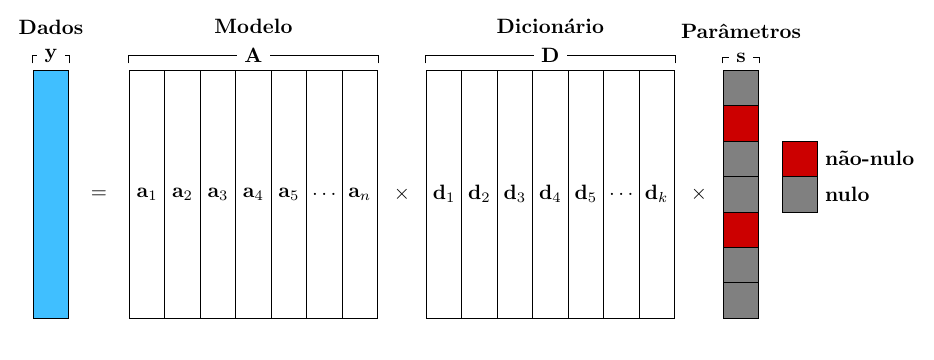}
\caption[Mudança da base de representação de um sinal.]{Mudança da base de representação de um sinal. Fonte: Próprio autor.}
\label{fig:02_042}
\end{centering}
\end{figure}

Em vez de resolver o problema de otimização com restrições da Equação \eqref{eq:proj1}, uma ideia é também tratá-lo como um problema de mínimos quadrados, mas substituindo-se $\mathbf{x}$ conforme Equação \eqref{eq:CS11}, obtendo-se
\begin{equation}
\mathbf{\hat{s}}_D = \arg\min\limits_{s} \left[ \vert \vert \mathbf{A} \mathbf{D} \mathbf{s}_D - \mathbf{y} \vert \vert^2_2 \right],
\label{eq:proj2}
\end{equation}
de modo que a solução desejada seja obtida posteriormente pela Equação \eqref{eq:CS1}.  Vendo uma matriz como uma função, a multiplicação $\mathbf{A} \mathbf{D}$ é a composição de diferentes funções \cite[pág. 153]{interactive}.  Isso é ilustrado na Figura \ref{fig:02_042} e pode ser comparada com a Figura \ref{fig:my_label}.

Há algoritmos que são baseados tanto na forma da Equação \eqref{eq:proj1}  quanto \eqref{eq:proj2}. Nos dois casos, é necessário levar em consideração a base de representação, mesmo que nem sempre seja necessário ou eficiente obter $\mathbf{D}$ explicitamente.

\subsection{Decomposição em valores singulares truncada}

Um problema deficiente em posto apresenta uma queda suave dos valores singulares, seguida de uma queda brusca \cite[pág. 45]{Mueller2012}. Seja a solução ingênua a partir da SVD de $\mathbf{A}$ conforme Equação \eqref{xnaive}. Uma proposta de solução para um problema deficiente em posto é truncar os componentes da SVD antes da queda para conter apenas os maiores valores singulares, pois estes não são dominados pelos ruídos nos primeiros $k$ componentes \cite[pág. 78]{hansen2010discrete}. Retomando e alterando a Equação \eqref{eq:dvs_inv0}, obtém-se 
\begin{equation}
\mathbf{\hat{x}_k} = \sum_{i=1}^k \frac{\mathbf{u}_i^T \mathbf{y}_{\delta} }{\sigma_i} \mathbf{v}_i,
\label{eq:dvs_tsvd}
\end{equation} onde $k$ é o número de componentes da SVD utilizados, usualmente igual ao posto de $\mathbf{A}$.

Este método, chamado de SVD truncada (TSVD),  é intuitivo e fácil de implementar uma vez que a SVD foi calculada, pois os seus fatores de filtro são apenas valores zero e um, dependendo de quantos se deseja incluir. A solução pela TSVD não é iterativa e nem baseada na semiconvergência, sendo importante essa diferenciação de quem nem todo método de projeção resulta em um algoritmo iterativo. Uma desvantagem é a necessidade de se calcular a SVD inteira de $\mathbf{A}$ ou de $k$ valores e vetores singulares. 

Para outro ponto de vista da TSVD, parte-se da Equação \eqref{eq:proj2}, $\mathbf{A}$ é reescrita em termos de seus valores singulares e considera-se $\mathbf{D}$ os primeiros $k$ vetores singulares a direita. Dessa forma, é possível mostrar que a TSVD é um método de projeção  \cite[pág. 116]{hansen2010discrete}.  Comparando os fatores de filtro das Equações \eqref{eq:dvs_tsvd} e \eqref{eq:filterfactor_text}, observa-se a diferença entre a TSVD (apenas projeção) e a regularização clássica de Tikhonov, respectivamente.

Nota-se que, no subespaço da SVD, a matriz $\mathbf{A}$ é decomposta em outras três matrizes, mas o vetor $\mathbf{x}$ continua o mesmo. Como será visto a seguir, a projeção também pode fazer com que o problema seja resolvido em termos de $\mathbf{s}_D$, e $\mathbf{x}$ deve ser recuperado depois.

\subsection{Representação esparsa com bases fixas}

É possível escolher $\mathbf{D}$ de modo que a representação $\mathbf{s}$ seja esparsa nela, inclusive para sinais que originalmente não o eram esparsos na base original. Para isso, é necessária a definição de um sinal compressível, já que na prática os sinais tendem a ser compressíveis, não esparsos \cite{tropp2010}.  A partir da definição de sinal K-esparso, um sinal é compressível quando $\mathbf{s}$ possui alguns poucos coeficientes de grande valor e vários coeficientes pequenos \cite{baraniuk2007}, ou seja, quando $K \gg n$, o que significa que são necessários poucos vetores de base para representá-lo, uma representação mais parcimoniosa. 

O objetivo é obter a representação mais esparsa do sinal. Para isso, não há um dicionário único, pois depende, por exemplo, do número de vetores de base que compõe o dicionário. Seja o dicionário $\mathbf{D}$ de tamanho [$d \times n$]. No caso em que $n < d$, ele é \textit{undercomplete} e no caso em que $n > d$ ele é \textit{overcomplete}. Se o melhor dicionário é aquele consegue a representação mais esparsa, pode-se pensar na busca de um dicionário em que $n \ggg d$, mas isso necessitaria um tempo e requisitos computacionais proibitivos, o que é um \textit{trade-off} com a sua complexidade.

Existem dicionários que são especificados analiticamente, definindo a matriz $\mathbf{D}$ explicitamente e sem necessidade de atualizá-la. Por exemplo, as transformadas discretas de cosseno podem tornar imagens esparsas \cite{aster2019parameter}. Outros modelos podem ser melhores representados pela transformada de Fourier, a transformada Gabor \cite[pág. 139]{majumdar2019compressed}, ou Curvelets \cite{Frikel2013}. A transformada Wavelet também é bastante utilizada dentro de regularização no espaço de Besov \cite[Seção 7.3]{Mueller2012}. 

\subsection{Representação esparsa com bases adaptativas ou aprendidas}\label{sec:baseslearned}

\begin{figure}[H]
\begin{centering}
\includegraphics[width=0.75\textwidth]{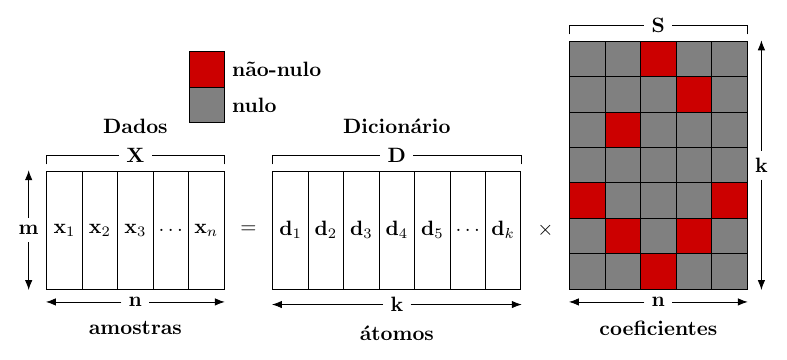}
\caption[Ilustração da aprendizagem de dicionário.]{Ilustração da aprendizagem de dicionário. Fonte: Próprio autor.}
\label{fig:04_2}
\end{centering}
\end{figure}

Pode-se também calcular dicionários $\mathbf{D}$ a partir de exemplos de dados \cite[pág. 54]{Arridge2019}. Na aprendizagem de dicionário esparsa, ilustrada na Figura \ref{fig:04_2}, concatena-se vetores de $\mathbf{x}$ em uma matriz $\mathbf{X} = (\mathbf{x}_1, \mathbf{x}_2, \cdots, \mathbf{x}_p)$, que depois é fatorada em duas outras, uma o dicionário $\mathbf{D}$ e a outra matriz $\mathbf{S} = (\mathbf{s}_1, \mathbf{s}_2, \cdots, \mathbf{s}_p)$ da representação esparsa. A fatoração é tratada como um problema de otimização \cite[pág. 46]{alvarez2017digital} e possibilita que $\mathbf{D}$ seja específica de uma aplicação, trazendo melhores resultados \cite[pág. 139]{majumdar2019compressed}. 

\vspace{-5mm}

Na aprendizagem de dicionário, ainda é importante que os dados utilizados para o cálculo do dicionário sejam independentes dos dados que serão utilizados na reconstrução, para ser uma informação \textit{a priori} do sinal. Mesmo assim, existe a aprendizagem de dicionário simultânea à reconstrução \cite[pág. 54]{Arridge2019}, \cite{LewisD2019}.

Por fim, existe ainda a aprendizagem da transformada, onde se busca uma transformação $\mathbf{D}$ para analisar o vetor $\mathbf{x}$ e produzir os coeficientes $\mathbf{s}$. A aprendizagem de transformada pode ser vista como um problema direto, enquanto a aprendizagem de dicionário um problema inverso \cite[pág. 142]{majumdar2019compressed}. 

\subsection{Amostragem comprimida}\label{Ap:CS}

Um exemplo de problema onde é importante obter uma base de representação esparsa do final é a amostragem comprimida. Em resumo, a amostragem comprimida lida com a solução de sistemas de equações lineares subdeterminados quando se sabe que a solução é esparsa e sua formulação matemática de reconstrução é muito semelhante ao de problemas inversos quando se quer promover a esparsidade do resultado. Uma revisão de suas aplicações pode ser vista em \cite[Figura 19]{Rani2018}.

Suponha que se deseja comprimir um sinal $\mathbf{x}$. Na transformação por codificação, é feita a aquisição de um sinal completo, com todas as $n$ amostras e depois ele é comprimido nesse domínio em um vetor $\mathbf{s}$, através de um dicionário $\mathbf{D}$ onde ele seja esparso. Supondo que $\mathbf{D}$ seja ortonormal, $\mathbf{D}^{-1} = \mathbf{D}^T$, então
\begin{equation}
\mathbf{s} = \mathbf{D}^T \mathbf{x}.
\label{CS2}
\end{equation}
Nessa abordagem, há a necessidade de aquisição de todas as amostras do sinal, do cálculo de todos os coeficientes de $\mathbf{s}$, para depois descartar alguns (ou muitos) desses coeficientes, mas isso não é eficiente \cite{baraniuk2007}. Um dos objetivos na amostragem comprimida é melhorar essa ineficiência através da aquisição direta do sinal comprimido para depois reconstruir o sinal original através de um problema de otimização. 

\begin{figure}[htpb]
\centering
\includegraphics[width = 0.8\textwidth]{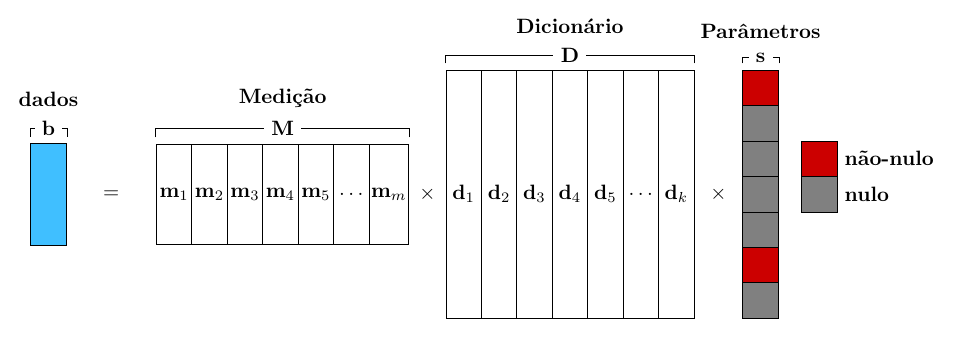}
\caption[Ilustração da amostragem comprimida.]{Ilustração da amostragem comprimida. Fonte: Próprio autor.}
\label{fig:02_046}
\end{figure}

Para isso, pode-se supor um sinal $\mathbf{x}$ que é amostrado por uma matriz de medição $\bm{\Phi}$, aleatória  e retangular, resultando em um vetor $\mathbf{y}$ com muito menos componentes do que o sinal original \cite{baraniuk2007}. Um sinal é dito $K$-esparso se o vetor $\mathbf{s}$ apresenta apenas $K$ valores não-nulos (em outras palavras, $K$ é o número de coeficientes da representação esparsa do sinal). Assim, o sinal $\mathbf{y}$ de tamanho $[ m \times 1]$ é obtido a partir de
\begin{equation}
\mathbf{y} = \mathbf{\Phi} \mathbf{x},
\label{CS3}
\end{equation}
onde $ \mathbf{\Phi}$ é uma matriz de medições de tamanho $[ m \times n]$ não-adaptativa (fixa e independente do sinal $\mathbf{x}$), na qual $K < m \lll n$. Dado que $\mathbf{x} = \mathbf{D}\mathbf{s}$, obtém-se 
\begin{equation}
\mathbf{y} = \mathbf{\Phi} \mathbf{D} \mathbf{s}, 
\label{CS4}
\end{equation}
que leva em conta cada etapa do processo, conforme Figura \ref{fig:02_046}. 

Existem três etapas para soluções em CS:
\begin{enumerate}
\item Definição do dicionário para representação esparsa do sinal \cite{Starck2009}; 

\item Aquisição do sinal sem a necessidade de todas as $n$ amostras que o sinal original teria. Isso é feito através de uma matriz de medições $\mathbf{\Phi}$ que preserve a estrutura do sinal original apesar da redução de dimensionalidade. Foi demonstrado em  \cite{Candes2006} que existem matrizes $\mathbf{\Phi}$ que satisfazem a propriedade de isometria restrita (RIP), porém tanto a construção quanto a verificação dessa propriedade são muito custosas \cite{Oliveri2017}. Uma matriz de fácil construção que satisfaz a RIP é a matriz de medições aleatória \cite{baraniuk2007, Candes2006}, sendo necessária a coleta de amostras aleatórias do sinal. Há diversas estratégias propostas com tal finalidade \cite[Tabela 2]{Rani2018}; 

\item Reconstrução do sinal a partir de apenas $K$ amostras. Ao invés da interpolação do sinal com uma função do tipo $\text{sinc} = \frac{sin(\mathbf{x})}{\mathbf{x}}$, conforme teorema de Nyquist, são necessários outros métodos para a sua solução, mas diversos métodos já foram propostos \cite{majumdar2019compressed}, \cite[Figura 4, Tabela 1]{Qaisar2013}, \cite[Tabela 3]{Rani2018}. Nessa aplicação, o desejado seria utilizar norma $\ell_0$ \cite{Donoho2006} no termo de regularização, mas um dos resultados importantes obtidos foi que quando a matriz $\bm{\Phi}$ respeita a condição RIP, é possível mostrar que resolver o problema com uma norma $\ell_1$ que será equivalente à norma $\ell_0$ \cite{Donoho2006}, \cite[pág. 30]{majumdar2019compressed}. Isso é chamado de relaxação convexa da norma $\ell_0$ e faz sentido porque a norma $\ell_1$ é a função convexa mais próxima da norma $\ell_0$ \cite{tropp2010} e o funcional acaba se tornando semelhante ao da Equação \eqref{eq:norma1}. 
\end{enumerate}

\subsubsection{Reconstrução de imagem nítida com CS}
\begin{figure}[H]
     \centering
     \begin{subfigure}[b]{0.3\textwidth}
         \centering
         \includegraphics[width=0.9\textwidth]{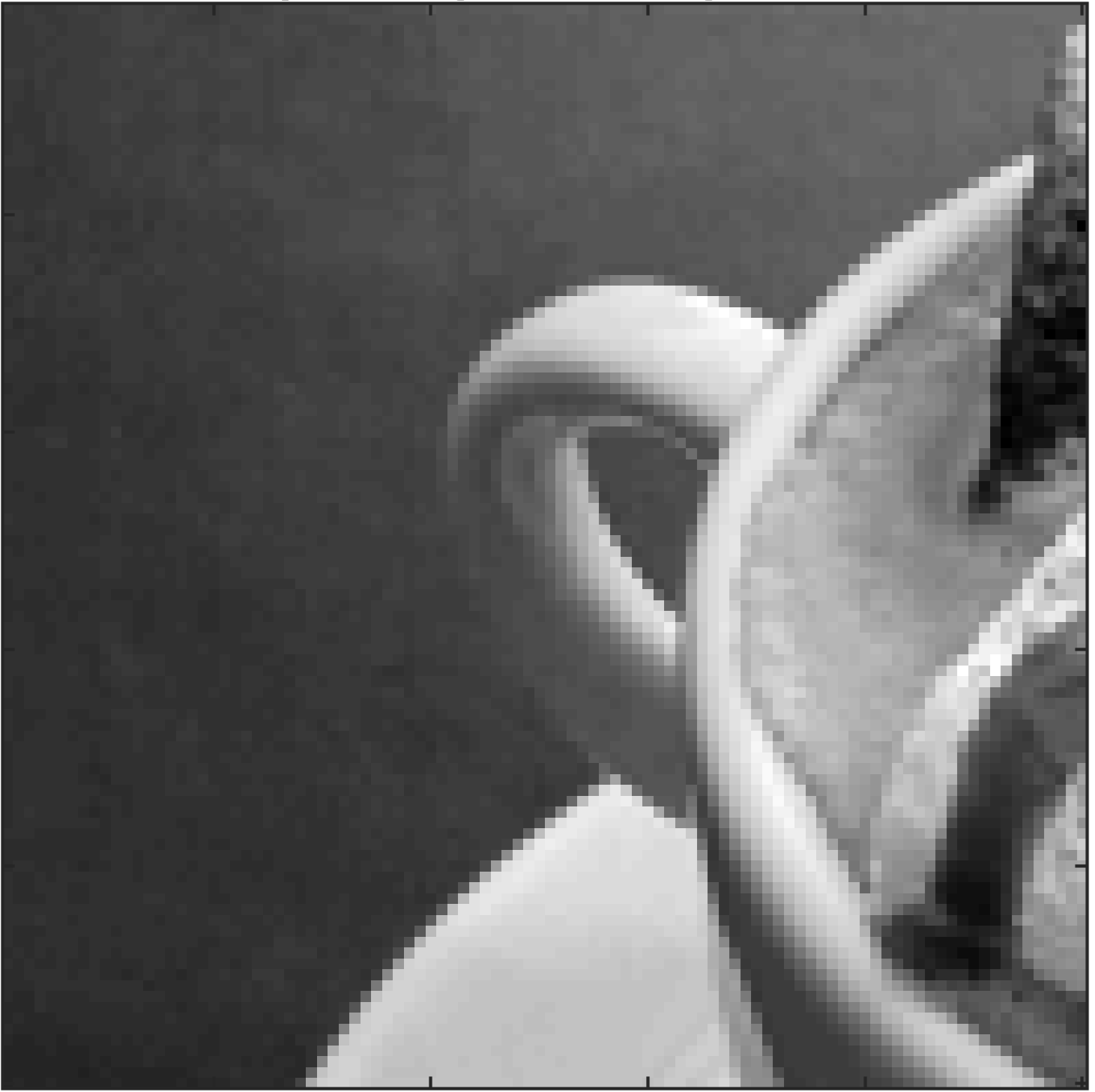}
         \caption{Imagem original}
         \label{fig:02_csa}
     \end{subfigure}
     \hfill
     \begin{subfigure}[b]{0.3\textwidth}
         \centering
                  \includegraphics[width=0.9\textwidth]{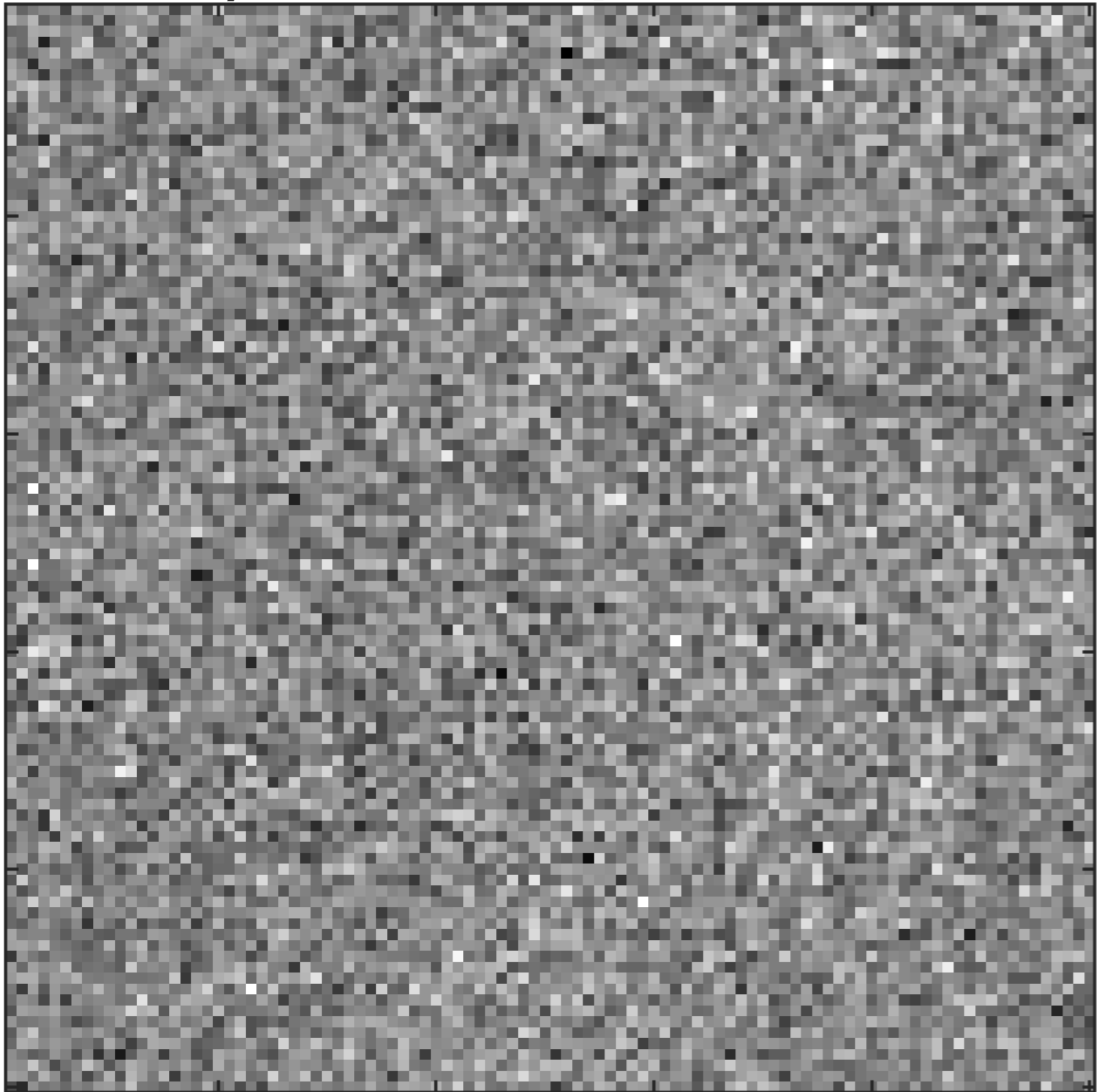}
         \caption{Mínimos quadrados}
         \label{fig:02_csb}
     \end{subfigure}
     \hfill
          \begin{subfigure}[b]{0.3\textwidth}
         \centering
                  \includegraphics[width=0.9\textwidth]{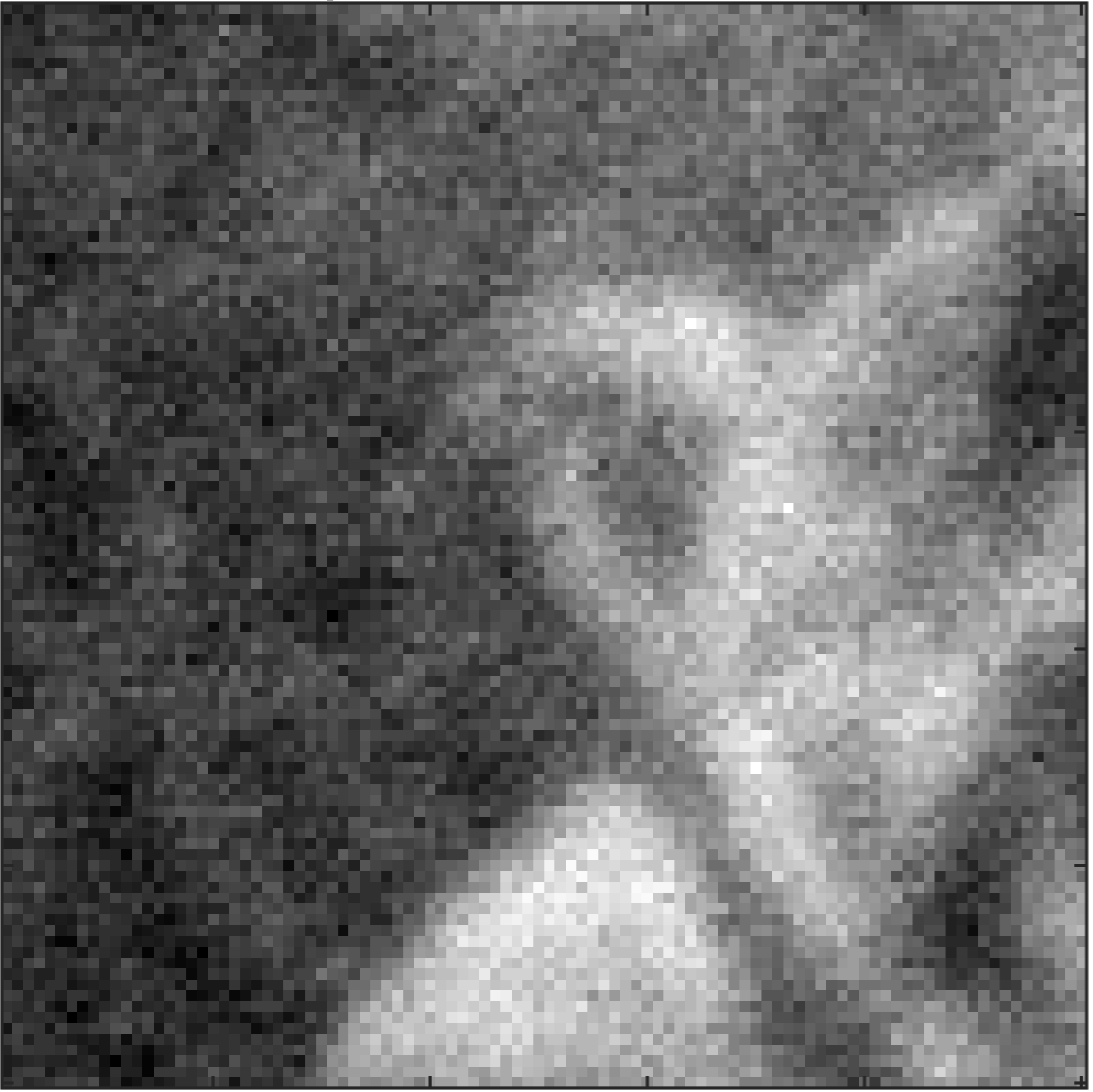}
         \caption{Basis pursuit}
         \label{fig:02_csc}
     \end{subfigure}
\caption[Reconstrução de uma figura subamostrada.]{Reconstrução de uma figura subamostrada. Fonte: Próprio autor.}
\label{fig:02_cs}
\end{figure}

 A partir da notação da Equação \ref{CS4}, seja a imagem nítida de tamanho $[100 \times 100]$ mostrada na Figura \ref{fig:02_csa}. Ela é vetorizada passando a ser um vetor $\mathbf{x}$ de tamanho $[10000 \times 1]$ e suponha que ela tenha uma representação esparsa no domínio da transformada discreta de cosseno (DCT). O objetivo é reconstruir $\mathbf{x}$ a partir de um vetor de medidas $\mathbf{y}$ de tamanho $[1250 \times 1]$, oito vezes menor, obtido a partir da Equação \eqref{CS3} com uma matriz aleatória de medição $\mathbf{\Phi}$  de tamanho $[1250 \times 10000]$. 

A Figura \ref{fig:02_csb} mostra a reconstrução com a utilização de $\mathbf{s} = (\mathbf{\Phi} \mathbf{D})^+\mathbf{y}$ e depois recuperada com $\mathbf{x} = \mathbf{D}^{-1} \mathbf{s}$. A Figura \ref{fig:02_csc} mostra a reconstrução obtida com a relaxação convexa da Equação \eqref{eq:norma0}, também conhecido como \textit{basis pursuit}, conforme
\begin{equation}
\arg\min\limits_{\mathbf{x}} \vert \vert \mathbf{x} \vert \vert_1 \hspace{3mm} \text{ sujeito a }\mathbf{A} \mathbf{x} = \mathbf{y},
\label{eq:normaconvex}
\end{equation}
reescrita com um programa linear e depois utilizando um método de otimização de ponto interior primal-dual \cite{l1magic}.

\subsection{Amostragem comprimida de problemas inversos}
\begin{figure}[H]
\centering
\includegraphics[width = 1\textwidth]{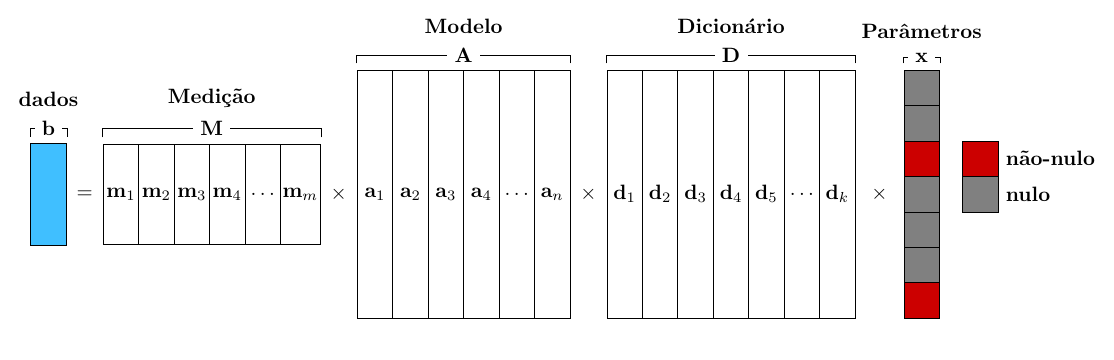}
\caption[Amostragem comprimida de problemas inversos.]{Amostragem comprimida de problemas inversos. Fonte: Próprio autor.}
\label{fig:02_065}
\end{figure}

Pode-se buscar a amostragem comprimida de um problema inverso mal-posto \cite{Herrholz2014, Herrholz_2010}, conforme Figura \ref{fig:02_065}.  Observa-se que há 3 matrizes de transformação, uma relativa à matriz de medições (usualmente aleatória), uma matriz relativa ao modelo do problema inverso e um dicionário que traga a representação esparsa do sinal de interesse. A amostragem comprimida resulta em uma equação de otimização semelhante à encontrada em problemas inversos, mas parte de um contexto diferente.

Nessa união, há trabalhos que discutem o caso de dados incompletos e problemas de grande escala \cite{Herrholz2014}. Há estudos em imagens médicas que discutem desafios como a não-linearidade do problema direto, a falta de uma representação esparsa intrínseca das imagens ou a dificuldade de se usar a informação \textit{a priori} de que a matriz $\mathbf{A}$ deve respeitar a RIP, já que ela vem de um modelo físico do processo \cite{Oliveri2017}, mas se o problema for mal-posto, ele deixa de respeitar a RIP \cite{Herrholz2014, Herrholz_2010}.

\newpage
 \section{\textit{Checklist} de reprodutibilidade para propostas que unem métodos \textit{model-based} e \textit{data-driven}}\label{ap:checklist}

Durante o desenvolvimento de um projeto de pesquisa, inicia-se com uma breve descrição do estado da arte da área de pesquisa e qual é a perspectiva de contribuição científica da proposta. Conforme \cite[págs. 160-3]{hansen2010discrete}, para o \textit{design} e implementação de algoritmos de regularização de problemas lineares mal-postos, ou \textit{solvers}, são necessárias escolhas que incluem a definição do método de regularização, um método de escolha do parâmetro $\lambda$ e sua implementação computacional. Essas escolhas não são únicas. Da própria noção subjetiva de \textit{prior}, pode-se não haver uma escolha sempre melhor do que a outra, mas sim que levam a resultados diferentes. A clareza das escolhas e das hipóteses subjacentes a elas é muito importante para adequar o \textit{solver} ao problema. 

O presente guia tem como objetivo a organização de um \textit{solver} na área de problemas inversos que utilizam técnicas de regularização em conjunto com aprendizagem profunda, ampliando a discussão realizada em \cite[págs. 160-3]{hansen2010discrete}. Sendo importante para o próprio pesquisador e para os demais leitores do trabalho, essas informações são perguntas importantes. Como não existe fórmula pronta para definir uma metodologia, a lista deve ser utilizada e atualizada de acordo com as necessidades de cada um. 

Na implementação computacional, há características desejadas e boas práticas, como a escolha da linguagem de programação adequada, foco em modularidade, clareza do código, mensurar performance e eficiência, como tempo de execução e requisitos de armazenamento \cite[págs. 160-3]{hansen2010discrete}. Este guia aborda detalhes técnicos computacionais, ele apenas parte da organização do próprio autor da proposta visando o leitor.

Finalmente, pode-se argumentar a favor da disponibilização dos códigos implementados, mas apenas isso pode não ser suficiente. Quando se quer estudar e reproduzir de resultados de trabalhos de outros grupos de pesquisa, isso pode envolver versões diferentes dos \textit{softwares} disponíveis, \textit{softwares} pagos e infraestrutura indisponíveis para todos. Além disso, o mesmo código com diferentes bases de dados pode não ter a performance esperada, ou quando há variação nos parâmetros e hiperparâmetros. No limite, um código que gere exatamente as imagens de um artigo seria um caminho, mas dificilmente eles são disponibilizados desta forma. Em todo caso, todo material suplementar ao trabalho que ajude a entendê-lo é bem-vindo. 

\subsection{Checklist de perguntas para reprodutibilidade}
\subsubsection{Solução do problema direto:}
A definição do modelo computacional é necessária tanto para o processo de inversão quanto para geração de dados simulados. A partir da discretização do modelo contínuo, obtém-se uma matriz de transformação linear $\mathbf{A}$ e um vetor de parâmetros $\mathbf{x}$, ou uma transformação não-linear $\mathbf{A}(\mathbf{x})$ dos parâmetros de interesse. 
\begin{itemize}
	\item \textbf{Formulação matemática geral}
		\begin{itemize}
				\item Quais são as informações disponíveis no problema e o que se deseja estimar?  
				\item Quais são as equações físico-matemáticas (modelo contínuo) do problema? 					
				\item Entre quais espaços o operador direto realiza a transformação?
			 	 \item O problema é linear ou não-linear em relação aos parâmetros de interesse? 
				\item Quais são as hipóteses simplificadoras do modelo? 
				\item Quais foram as condições de contorno utilizadas? 
				\item O problema é mal-posto? 
 	 		\end{itemize} 
 	 		\item\textbf{Discretização}
 	 	\begin{itemize} 		 
 		 \item Qual foi o método utilizado para a discretização do operador direto contínuo? 
 		 \item Foi utilizado algum \textit{software} para realizar a discretização? 
 		 \item São necessários arquivos auxiliares, como malhas de elementos finitos, para definição do domínio discreto? Eles foram disponibilizados?
 		  		 \item Há diferenças na discretização do domínio e das condições de contorno?
 		  		 \item O modelo resultante é um problema de grande escala? Qual é o tamanho da maior matriz envolvida?
 \end{itemize} 
 
 \item \textbf{Problemas lineares}:   
		\begin{itemize} 
 \item Qual é o significado da matriz $\mathbf{A}$ (linhas, colunas) e do vetor $\mathbf{x}$?
 \item Sobre essas grandezas, quais são as unidades de medida adequadas?
	 \item A matriz $\mathbf{A}$ é quadrada ou retangular? 
	 \item Qual é o número de condição da matriz $\mathbf{A}$? 
	 \item A matriz $\mathbf{A}$ é inversível?
 \item Há alguma estrutura ou padrão de $\mathbf{A}$ que permita sua implementação de modo eficiente? Exemplos: simetria, esparsidade, circulante, matriz de Toeplitz. 
 \item A visualização das colunas da decomposição em valores singulares (SVD) do problema direto é informativa? 
 \item A partir da visualização da velocidade da queda dos valores singulares de $\mathbf{A}$, qual a classificação do problema inverso? 
 \item Foi gerada a visualização do Gráfico de Picard para verificar a condição de Picard, com e sem ruído? O que o resultado indica?
		\end{itemize}	 
    \item No caso de problemas não-lineares, qual é o significado da matriz $\mathbf{A}(\mathbf{x})$? Existe alguma informação ou padrão que se pode extrair de $\mathbf{A}(\mathbf{x})$ nesse caso?
   \end{itemize}

\subsubsection{Infraestrutura e \textit{softwares}}

\begin{itemize}
\item O poder computacional disponível é suficiente para resolver o problema? Foram necessárias adaptações?
\item Foi utilizado um computador pessoal, uma máquina virtual ou alguma infraestrutura diferente? Quais são suas características?
\item Os \textit{softwares} utilizados são livres ou pagos?
\item Foram utilizadas \textit{toolboxes} adicionais? 
\item É possível utilizá-las quando não há intenção comercial?
\item Em todos os casos, é necessário indicar \textit{copyrights}?
\item Foram disponibilizados os códigos finais utilizados?
\item Nesse caso, há uma lista das versões dos pacotes, \textit{softwares} e \textit{toolboxes} utilizados? 
\end{itemize}

\subsubsection{Dados}
Definida a pergunta, é necessário ter uma base de dados que permita a sua resposta. 
\begin{itemize}
\item \textbf{Disponibilidade}
\begin{itemize}
\item Os dados foram simulados? Nesse caso, como foi evitado o crime de inversão?
\item Os dados serão coletados pelo próprio pesquisador? Como?
\item Os dados foram obtidos de repositórios existentes? 
\item Em caso de dados experimentais, há descrição de como os dados foram coletados e por quais equipamentos?
\item Foi utilizado mais de um repositório para comparação dos resultados? 
\item Quais as licenças ou \textit{copyrights} definidas nos repositórios para os dados? 
\item É possível utilizar os dados quando não há intenção comercial? 
\item Há informações sensíveis nos dados?
\item Há algum código de comitê de ética que deve ser citado?
\end{itemize}

\item \textbf{Manipulação}
\begin{itemize}
\item Há dados em grande quantidade? 
\item A natureza e a quantidade dos dados são adequadas para o poder computacional disponível? 
\item Quais são as extensões dos arquivos? 
\item Os \textit{softwares} necessários para manipular esses dados estão disponíveis? 
\item Há exemplos duplicados, valores ausentes ou \textit{outliers}?
\item Foi necessário extrair, selecionar ou filtrar os dados? Como isso foi realizado?
\item Foi necessário pré-processamento dos dados? Como isso foi realizado?
\end{itemize}

\item \textbf{Caracterização}
\begin{itemize}
\item Os dados foram visualizados? Essa visualização auxilia na interpretação dos fenômenos? 
\item Há gabarito do problema, i.e. dados \textit{ground truth} ou \textit{labels}?
\item Os dados formam pares entrada-saída? É possível utilizar técnicas de aprendizagem supervisionada?
\item Como foi feita a separação em conjunto de treinamento, validação e teste?
\item Os sinais são esparsos? Os sinais são esparsos em alguma base outra base, seja ela fixa ou aprendida?
\item Há ruídos nos dados experimentais? Caso eles sejam conhecidos, qual foi o modelo de ruído considerado? Caso eles sejam desconhecidos, houve tentativa de estimá-los?
\end{itemize}
\end{itemize}

\subsubsection{Definição do problema inverso} 
É necessário definir qual é a forma utilizada para inversão do operador direto. Aqui, o foco será nos métodos variacionais. 

\begin{itemize}
\item Foi utilizado um método de regularização para inversão? Qual sua forma?
\item Esse método é comprovadamente um método de regularização? Ou as conclusões são baseadas nos resultados experimentais?
\item \textbf{Métodos variacionais}:
\begin{itemize}
\item Qual é a função custo utilizada na proposta?
\item Caracterizar o termo de regularização: Ele é linear em relação aos parâmetros? É convexo? É diferenciável? É suave?
\item Caracterizar o funcional resultante como um todo, sob os mesmos termos.
\item O que se esperava com essa função custo? Quais eram as características desejadas na reconstrução ao utilizá-la? 
\end{itemize} 

\item \textbf{Abordagem estatística}:
\begin{itemize}
\item O método utilizado é determinístico ou estatístico? 
\item A estimativa é única ou o resultado é uma variável aleatória?
\item Caso a abordagem utilizada seja estatística, qual foram os modelos de ruído e \textit{prior} utilizados? 
\item É possível gerar visualizações do \textit{prior}?
\item Foram estimados intervalos de confiança dos resultados? 
\item Foram estimadas as incertezas dos resultados?
\end{itemize} 

\item \textbf{Regularização generalizada de Tikhonov}:
\begin{itemize} 
\item O termo de regularização é dado pela regularização generalizada de Tikhonov, com a matriz de regularização $\mathbf{L}$ explícita?
\item Ela foi ``feita à mão'' (\textit{handcrafted}), foi baseada em amostras ou foram utilizadas técnicas de aprendizagem de máquina?
\item Se for uma matriz $\mathbf{L}$ de regularização explícita, como ela foi montada? Há parâmetros dela que devem ser descritos? 
\item É possível utilizá-la indiretamente, como um operador $\mathbf{L} \mathbf{x}$, sem necessidade de sua montagem explícita? Como?
\item O que ela representa? Qual resultado se espera?
\item Além da matriz de regularização, foi utilizado algum valor de referência $\mathbf{x}^*$? O que ele representa? Como ele foi obtido?
\end{itemize} 

\item \textbf{Parâmetro de regularização} $\lambda$
\begin{itemize}
\item Foi necessário um parâmetro de regularização ou foi possível utilizar métodos baseados na semiconvergência?
\item Se sim, como foi escolhido o parâmetro de regularização $\lambda$? Foi utilizado um ou mais critérios objetivos para isso?
\item Há informações sobre o ruído visando calcular $\lambda$?
\item No caso de um parâmetro constante, foi utilizada alguma heurística como critério, como curva-L ou GCV?
\item O parâmetro $\lambda$ é constante ou varia com o número de iterações? 
\end{itemize} 

\item \textbf{Métodos híbridos com regularização e projeção} $\lambda$
\begin{itemize}
\item O método de regularização foi utilizado em conjunto com métodos de projeção? Com qual objetivo?
\item Foi utilizada uma forma de síntese ou de análise?
\item Para qual espaço foi realizada a projeção? Qual vantagem ele traz?
 \end{itemize} 
 \end{itemize} 
 
 \subsubsection{Problema inverso: Otimização}
A partir da definição do funcional, é necessário escolher um otimizador adequado para resolvê-lo, pois não adianta eu propor um método que eu não consiga otimizar. 
\begin{itemize}
\item \textbf{Caracterização geral}
\begin{itemize}
\item Qual é o algoritmo de otimização escolhido?
\item Quais as vantagens e desvantagens do otimizador escolhido?
\item Dadas as características da função custo, há soluções em apenas um passo ou são necessários algoritmos iterativos? 
\item Foram consideradas restrições adicionais?
\item No caso da necessidade de solução de sistema linear de equações, qual foi o método utilizado para sua solução?
\item Foi detectada alguma etapa com maior custo computacional (gargalo)?
\end{itemize}
\item \textbf{Parâmetros do algoritmo de otimização}
\begin{itemize}
\item Qual o tamanho do passo? É variável ou fixo ao longo das iterações?
\item Qual foi o critério de parada utilizado para as iterações? O número de iterações foi fixo ou relativo ao monitoramento de algum erro? Em ambos os casos, qual foi o número de iterações necessárias, em média? 
\item Há outros parâmetros que devem ser definidos? Quais? 
\end{itemize}
\end{itemize}

 \subsubsection{Integração com métodos de aprendizagem profunda}
 Tendo como objetivo unir métodos de regularização e técnicas de aprendizagem profunda, algumas boas práticas em reprodutibilidade são mostradas a seguir. 
 \begin{itemize}
\item \textbf{Definição e treinamento da ANN}
\begin{itemize}
\item Foi utilizada alguma técnica de aprendizagem profunda?
\item A aprendizagem foi supervisionada, não-supervisionada ou outro tipo?
\item Qual foi a arquitetura da ANN utilizada? Qual foi o motivo de sua escolha? 
\item A ANN tem quantos parâmetros treináveis? 
\item Quais são os hiperparâmetros da ANN? Como eles foram definidos? Foi realizado ajuste de hiperparâmetros?
\item Qual foi a função de perda utilizada? Por que ela foi escolhida?
\item Em relação à quantidade de dados disponíveis, a utilização de aprendizagem profunda será propensa ao \textit{overfitting}? 
\item Foi utilizada alguma estratégia de regularização durante o treinamento? 
\item A partir das características do funcional de treinamento da ANN, qual foi o otimizador utilizado? Quais foram seus parâmetros?
\item É possível disponibilizar os pesos obtidos após o treinamento? E a sua implementação como um todo?
\end{itemize}

\item \textbf{Visão geral da proposta}
\begin{itemize}
\item O que se esperada obter ao incluir a ANN na solução do problema?
 \item Por que não usar só modelos com método de regularização ou soluções de ponta a ponta com ANNs? 
 \item Qual é a vantagem esperada do método combinado proposto? 
 \item Qual é a novidade do método combinado proposto? 
\item O método proposto pode ser utilizado em diferentes aplicações? Ou seja, ele é de propósito geral, variando-se apenas a base de dados utilizada no treinamento? Ou ele é específico da aplicação?
\item É possível dizer que a solução que inclui a ANN é interpretável, explicável e reprodutível? Em caso afirmativo, como isso foi avaliado? 
\end{itemize}
\end{itemize}

\subsubsection{Resultados}
Tão importante quanto gerar os resultados é apresentá-los de forma adequada aos leitores.
\begin{itemize}
\item \textbf{Avaliações gerais}
\begin{itemize}
\item Quanto tempo para realizar uma reconstrução da aplicação desejada?
\item Em dados simulados, foi avaliado se o algoritmo é robusto a ruídos? Foram comparados os resultados com e sem ruído, reconstruções para diferentes realizações e níveis de ruído e reconstruções com diferentes tipos de ruídos?
\item Foi realizado um estudo de resolução da proposta?
\item É possível relacionar os resultados com o que se esperava do funcional e \textit{priors}? A performance foi como esperada?
\end{itemize}

\item \textbf{Convergência de algoritmos iterativos}
\begin{itemize}
\item Foi feita uma avaliação formal de convergência do algoritmo proposto?
\item A convergência do algoritmo iterativo (gráficos do valor do funcional em relação ao número de iterações) foi visualizada? 
\item Os resultados obtidos foram avaliados em relação à convergência com respeito ao nível do erro e a taxa de convergência? 
\end{itemize}

\item \textbf{Visualização dos resultados finais}
\begin{itemize}
\item As descrições nos eixos e as unidades de medida estão adequadas?
\item Foram geradas visualizações em domínios diferentes do original?
\item Apresentar casos de falhas, além dos resultados que deram certo
\item Após a reconstrução, foram necessárias etapas de pós-processamento?
\item A partir das imagens reconstruídas, foram feitas análises qualitativas ou quantitativas das imagens, para extração de mais informações?
\end{itemize}

\item \textbf{Performance a partir de métricas}: 
\begin{itemize}
\item Foram utilizadas métricas com referência? Quais?
\item Foram utilizadas métricas sem referência? Quais?
\end{itemize}

\item \textbf{Performance a partir de comparação}: 
\begin{itemize}
\item Foram feitas comparações com métodos de regularização tradicionais? Quais? 
\item Foram feitas comparações com o resultado de ANNs de ponta a ponta? Quais? 
\item Nos dois casos, em que ano os métodos foram propostos? 
\item É possível dizer que eles são métodos de estado da arte para comparação?
\end{itemize}
\end{itemize}

\newpage

\section{Regularização na frequência e  filtragem de Wiener}\label{AP:Wiener}
 
É possível mostrar que sistemas lineares e invariantes no tempo podem  ser expressos como convoluções \cite[págs. 211-2]{aster2019parameter}, de modo que a integral de convolução da Equação \eqref{eq:deco} pode representar diversos problemas inversos lineares. 

A deconvolução é a operação inversa da convolução. Operações de convolução e deconvolução podem ser analisadas no contexto da teoria de Fourier de uma forma efetiva, pois convolução entre duas funções no domínio original se torna uma multiplicação das mesmas na frequência\footnote{Outras propriedades são encontradas em \cite[Tabelas 4.3 e 4.4]{gonzalez2018}}. Com tal objetivo, no caso discreto, pode-se utilizar a transformada discreta de Fourier (DFT) $\mathcal{F}(\cdot)$  e sua respectiva transformação inversa $\mathcal{F}^{-1}(\cdot)$. Outra vantagem da  representação na frequência é que há uma forma computacionalmente eficiente da DFT, conhecida como transformada rápida de Fourier (FFT), e também para sua inversa \cite[pág. 230]{aster2019parameter}. 

\subsection{Inversão ingênua na frequência}\label{app-ingenua}

Para ilustrar o uso do domínio da frequência em problemas inversos, o exemplo do \textit{deblurring} 2D é considerado. Seja $\mathbf{X}_{1}$ a matriz da imagem nítida verdadeira e $\mathbf{X}_{2}$ o resultado do algoritmo de reconstrução. Deseja-se que $\mathbf{X}_{1} \approx \mathbf{X}_{2}$ da melhor maneira possível. No entanto, apenas a imagem degradada está disponível. 

Supondo que exista a PSF verdadeira do sistema $\mathbf{H}_1$ e que o ruído $\mathbf{N}_1$ seja aditivo, o modelo verdadeiro para representar a imagem degradada $\mathbf{Y}_{1}$ é dado por 
 \begin{equation}
\left(\mathbf{X}_{1}*\mathbf{H}_1 + \mathbf{N}_1\right) =  \mathbf{Y}_{1}.
\label{eq:fft000}
\end{equation}
Na prática, parte-se de um modelo disponível, uma PSF $\mathbf{H}_2$ e um modelo de ruído $\mathbf{N}_2$,  para tentar aproximar os dados reais da Equação \eqref{eq:fft000}, conforme
 \begin{equation}
\left(\mathbf{X}_{2}*\mathbf{H}_2  + \mathbf{N}_2\right) \approx \mathbf{Y_1}.
\label{eq:fft001}
\end{equation}
 No domínio da frequência, deseja-se obter $\mathcal{F}(\mathbf{X}_{2})$ a partir de $\mathcal{F}(\mathbf{Y}_{1})$ e de $\mathcal{F}(\mathbf{H}_2)$. Calculando-se a DFT de cada termo da Equação \eqref{eq:fft001} e isolando-se $ \mathcal{F}(\mathbf{X}_2)$, obtém-se
 \begin{equation}
 \begin{aligned}
 \mathcal{F}(\mathbf{X}_2) & =  \left[\mathcal{F}(\mathbf{Y}_{1}) \oslash \mathcal{F}(\mathbf{H}_2)\right] - \left[\mathcal{F}(\mathbf{N}_2) \oslash \mathcal{F}(\mathbf{H}_2)\right]\\
 & = \left[(\mathcal{F}(\mathbf{X}_1) \odot \mathcal{F}(\mathbf{H}_1) + \mathcal{F}(\mathbf{N}_1)) \oslash \mathcal{F}(\mathbf{H}_2)\right] - \left[\mathcal{F}(\mathbf{N}_2) \oslash \mathcal{F}(\mathbf{H}_2)\right] \\
  & = \left[\mathcal{F}(\mathbf{X}_1) \odot \mathcal{F}(\mathbf{H}_1)  \oslash \mathcal{F}(\mathbf{H}_2) \right] +  \left[\mathcal{F}(\mathbf{N}_1) \oslash \mathcal{F}(\mathbf{H}_2)\right] - \left[\mathcal{F}(\mathbf{N}_2) \oslash \mathcal{F}(\mathbf{H}_2)\right],
\label{eq:fft2ssc}
\end{aligned}
\end{equation}
onde $\oslash$ é a divisão ponto a ponto (ou divisão de Hadamard) e $\odot$ é o produto ponto a ponto (ou produto de Hadamard),  seguindo a notação adotada em \cite[págs. 225-7]{aster2019parameter}.

A Equação \eqref{eq:fft2ssc} traz um paralelo com a Equação \eqref{eq:caso1_sol3} da inversão ingênua. A reconstrução é perfeita cometendo-se o crime de inversão com PSF perfeita ($\mathbf{H}_1 = \mathbf{H}_2$) e modelo de ruído perfeito ($\mathbf{N}_1 = \mathbf{N}_2$), ou ruído ausente ($\mathbf{N}_1 = \mathbf{N}_2$ = 0).

No entanto, em dados reais, $\mathbf{H}_1 \neq \mathbf{H}_2$ e $\mathbf{N}_1 \neq 0$,  o que amplifica ruídos quando os valores de $\mathcal{F}(\mathbf{H}_2)$ são próximos de zero. De fato, o próprio modelo de ruído aditivo nem sempre é adequado para representar o problema em questão. Assim, são necessárias outras propostas de solução.

\subsection{Regularização clássica de Tikhonov na frequência}\label{Chap4-wiener}

Problemas inversos lineares usualmente são escritos na forma de um sistema de equações $\mathbf{A}\mathbf{x} = \mathbf{y}$. Ou seja, a convolução poderia ser também representada por uma multiplicação matriz-vetor. Mesmo nesse caso, a regularização clássica de Tikhonov da Equação \eqref{eq:tiksol} pode ser realizada no domínio da frequência \cite[pág. 225]{aster2019parameter}, mas é importante que $\mathbf{A}$ apresente certas propriedades para representação adequada na frequência.  

Quando $\mathbf{A}$ é uma matriz de Toeplitz, como são as matrizes circulantes utilizadas para representar a convolução \cite{Gray2005}, elas são diagonalizáveis pela matriz de Fourier discreta \cite{Murli1999}. Assim, considerando que $\mathbf{A}$ pode ser uma matriz com valores complexos, obtém-se
\begin{equation}
\mathbf{A} = \mathbf{F} \mathbf{D} \mathbf{F}^*, 
\label{eq:tikhfft0}
\end{equation}
onde $\mathbf{D}$ é uma matriz diagonal, $\mathbf{F}$ denota a transformada discreta de Fourier como matriz de transformação, conhecida como matriz DFT \cite{Rao2001}; e $\mathbf{F}^*$ é a matriz transposta do complexo conjugado de $\mathbf{F}$. 

É possível mostrar \cite[Seção 4]{Murli1999}, \cite[pág. 225]{aster2019parameter} que a solução se torna
\begin{equation}
\mathcal{F}(\hat{\mathbf{x}}) =\left(\mathbf{D}^* \odot \mathcal{F}(\mathbf{y})\right) \oslash \left( \mathbf{D}^* \mathbf{D} + \lambda^2 \mathbf{I} \right) ,
\label{eq:tikhfft1}
\end{equation}
de modo que os componentes nulos ou quase nulos de $\mathbf{D}$ não prejudiquem a divisão. Quando $\mathbf{A}$ for uma matriz de Toeplitz com apenas valores reais, utiliza-se a matriz transposta ao invés da transposta conjugada na Equação \eqref{eq:tikhfft1}. 

Conforme Subseção \ref{app-ingenua}, quando o problema é 2D, também é possível escrever o problema direto apenas com matrizes. Na prática, deve-se realizar \textit{zero padding} em $\mathbf{H}$ antes do cálculo da FFT\footnote{Além da razão do \textit{wraparound}, também por uma questão de compatibilidade entre as dimensões das matrizes, já que o \textit{kernel} de convolução costuma ser bem menor do que a imagem em si.}. Então, a relação se torna
\begin{equation}
\mathbf{Y}_{medido} = \left(\mathbf{X} * \mathbf{H}\right) + \mathbf{N}
 \end{equation}
\begin{equation}
\mathcal{F}(\mathbf{Y}) = \mathcal{F}(\mathbf{X}) \odot \mathcal{F}(\mathbf{H}) + \mathcal{F}(\mathbf{N}).
\end{equation}
Seja a notação $\mathcal{F}^*(\mathbf{H})$ referente ao conjugado transposto da matriz $\mathcal{F}(\mathbf{H})$. Nesse caso, a regularização de Tikhonov na frequência seria dada por
\begin{equation}
\mathcal{F}(\mathbf{X}) = \left[\mathcal{F}^*(\mathbf{H}) \odot \mathcal{F}(\mathbf{Y})\right] \oslash \left[ \mathcal{F}^*(\mathbf{H})\odot\mathcal{F}(\mathbf{H}) + \lambda^2  \mathcal{F}(\mathbf{I}) \right].
\label{eq:frequencyvogel}
\end{equation}
Nesse caso, a matriz de regularização é uma matriz identidade $\mathbf{I}$, de modo que $\mathcal{F}(\mathbf{I})$ é uma matriz com todos os valores unitários \cite[págs. 75-6]{Vogel2002}. Na prática, ela atua como uma constante adicionada para evitar a divisão por zero no denominador. 

Em suma, a regularização de Tikhonov na frequência busca reduzir amplitudes de componentes de Fourier de alta frequência \cite[págs. 229-30]{aster2019parameter}. Além da solução clássica, também existem propostas de regularização generalizada de Tikhonov a partir da representação na frequência de $\mathbf{L}$, como \textit{priors} de suavidade \cite{Thibaut2005}, gradientes verticais e horizontais \cite{Schuler2013} ou novas propostas de regularização \cite{Dogariu2021}.

\subsection{Equalizador de Wiener no domínio da frequência}

Na perspectiva da filtragem linear, a reconstrução do sinal desejado é realizada a partir da aplicação de um filtro $\mathbf{H}_{inv}$ ao sinal observado, tal que $\mathbf{Y}*\mathbf{H}_{inv} \approx \mathbf{X}$. Em outras palavras, busca-se um filtro inverso $\mathbf{H}_{inv}$ que anule o efeito de $\mathbf{H}$. Para isso, é necessária a escolha de um critério de otimalidade do filtro e essa escolha é fortemente relacionada com a sua aplicação. 

Uma aplicação é a equalização de canais, que busca a restauração de um sinal distorcido transmitido através de um canal de comunicação. O sinal original $\hat{\mathbf{x}}_W$ recuperado pelo filtro de Wiener é dado por
\begin{equation}
 \hat{\mathbf{x}}_W = \mathbf{h}^{-1}_{W} * \mathbf{y},
\label{eq:tikhfft14}
\end{equation}  
escrevendo $\mathbf{x}$, $\mathbf{y}$ e $\mathbf{h}$ como vetores para cálculo da FFT\footnote{Se é calculada a FFT de uma matriz, cada coluna é tratada de modo independente.} quando necessário.

Um critério para obter o equalizador de Wiener no domínio da frequência é o critério quadrático, na qual se busca minimizar a diferença entre o sinal desejado e o sinal reconstruído considerando $\mathbb{E}(\vert \mathcal{F}(\hat{\mathbf{x}}_W) - \mathcal{F}(\mathbf{x}) \vert^2) $.  Algumas referências que desenvolvem esse resultado são \cite[págs. 547-9]{press1992numerical}, \cite[págs. 402-3]{Thibaut2005} e \cite[págs. 191-2, 197-8, 425-6]{vaseghi2000advanced}. A seguir, as operações matriciais não serão escritas ponto a ponto. Será utilizada uma notação que é mais convencional da área de processamento de sinais. 

Seja a frequência denotada por $f$, o espectro de potência cruzado entre sinal de entrada do canal (que se deseja obter) e a saída do canal (sinal observado) denotado por $P_{XY}(f)$ e o espectro de potência do sinal observado $P_{YY}(f)$. O equalizador de Wiener domínio da frequência passa a ser descrito por 
\begin{equation}
\mathcal{F}(\mathbf{h}^{-1}_{W}) = \frac{P_{XY}(f)}{P_{YY}(f)}.
\label{eq:tikhfft5s}
\end{equation}   
Essa relação pode ser reescrita \cite[págs. 198, 426]{vaseghi2000advanced} em termos do espectro de potência do sinal $P_{XX}(f)$ e do ruído $P_{NN}(f)$, conforme
\begin{equation}
\mathcal{F}(\mathbf{h}^{-1}_{W}) = \frac{P_{XX}(f) \mathcal{F}(\mathbf{h})^*}{P_{XX}(f) \vert \mathcal{F}(\mathbf{h}) \vert^2+P_{NN}(f)}. 
\label{eq:tikhfft5}
\end{equation}      
Após solução da Equação \eqref{eq:tikhfft5}, realiza-se a divisão ponto a ponto na frequência entre o sinal observado $\mathcal{F}(\mathbf{y})$ e o filtro inverso $\mathcal{F}(\mathbf{h}^{-1}_{W})$. Calculando-se a FFT inversa do resultado dessa divisão, $\hat{\mathbf{x}}_W$ é obtido. 

Alguns comentários podem ser feitos sobre isso:
\begin{itemize}
\item A implementação da Equação \eqref{eq:tikhfft5} depende do conhecimento de $P_{XX}(f)$ e $P_{NN}(f)$ é necessário para desenvolvimento do filtro, mas nem sempre elas estão disponíveis \cite{Thibaut2005}, como já identificava o próprio Tikhonov \cite[Subeção 5.1]{tikhonov1977solutions};
 \item Observando as Equações \eqref{eq:tikhfft1} e \eqref{eq:tikhfft5}, se o ruído for branco,  $P_{NN}(f)$ é constante e depende da variância do ruído. Dessa forma, assim como na Equação \eqref{eq:esti23}, $\lambda$ acaba sendo relacionado com informações sobre o ruído;
 \item Existem autores\footnote{ \url{https://blogs.mathworks.com/steve/2007/11/02/image-deblurring-wiener-filter}}$^{,}$\footnote{ \url{https://blogs.mathworks.com/steve/2008/07/21/image-deblurring-using-regularization}} que reconhecem os resultados do filtro de Wiener e da regularização de Tikhonov como semelhantes, mas discutindo que eles partem de pontos de vista diferentes;
 \item A utilização do filtro de Wiener para \textit{deblurring} é discutida em \cite[Seção 5.8]{gonzalez2018}. Em \cite[Seção 5.9]{gonzalez2018} os autores discutem filtros de mínimos quadrados com restrições utilizando \textit{kernel} laplaciano para realizar \textit{deblurring}. Eles não mencionam, mas essa forma também pode ser relacionada à regularização de Tikhonov.
 \end{itemize}

\subsection{Filtragem de Wiener no domínio espacial}

A equalização de canais também pode ser escrita no domínio original. Seja a forma matricial do tipo $\mathbf{A} \mathbf{x} + \bm{\delta} = \mathbf{y}$, onde $\mathbf{y}$ é a saída do canal,  $\mathbf{x}$ é a entrada do canal, $\mathbf{A}$ é a matriz de distorção do canal e $\bm{\delta}$ é um ruído aditivo aleatório, independente de $\mathbf{x}$ \cite[pág. 425]{vaseghi2000advanced}. Agora é necessário considerar que $\mathbf{x}$, $\mathbf{y}$ e $\bm{\delta}$ são realizações de processos estocásticos.

Segundo \cite[págs. 425-6]{vaseghi2000advanced}, para obter o filtro de Wiener, é necessário calcular a matriz de autocorrelação $\mathbf{R}_{yy} = \mathbb{E}(\mathbf{y} \mathbf{y}^t)$ da saída do canal conforme
\begin{equation}
\mathbf{R}_{yy} = \mathbf{A} \mathbf{R}_{xx}\mathbf{A}^T + \mathbf{R}_{nn},
\label{eq:tikhfft11}
\end{equation}  
onde $\mathbb{E}(\cdot)$ é o valor esperado, $\mathbf{R}_{xx}$ é a matriz de autocorrelação da entrada do canal e $\mathbf{R}_{nn}$ é a matriz de autocorrelação do ruído. 

Seja também $\mathbf{r}_{xx} = \mathbb{E}(\mathbf{x} \mathbf{y})$ o vetor de autocorrelação da entrada. O vetor de correlação cruzada entre os sinais de entrada e saída do canal é $\mathbf{r}_{xy} = \mathbf{r}_{xy} = \mathbf{A} \mathbf{r}_{xx}$. Dessa forma,  o equalizador de Wiener $\mathbf{h}^{-1}_{W} = \mathbf{R}^{-1}_{yy} \mathbf{r}_{xy}$ (filtro inverso) pode ser escrito como
\begin{equation}
\mathbf{h}^{-1}_{W} = \left(\mathbf{A} \mathbf{R}_{xx}\mathbf{A}^T + \mathbf{R}_{nn}\right)^{-1} \mathbf{A} \mathbf{r}_{xx},
\label{eq:tikhfft13}
\end{equation}
onde $\mathbf{R}_{xx}$ pondera o termo de fidelidade e $\mathbf{R}_{nn}$ se assemelha a um termo de regularização. Uma vez calculado $\mathbf{h}^{-1}_{W}$, a Equação \eqref{eq:tikhfft14} é utilizada para obter $ \hat{\mathbf{x}}_W$ desejado.

Uma expressão parecida é dada pelo estimador linear para problemas lineares segundo o critério dos mínimos quadrados \cite[Teorema 2.6.1]{Sayed2003}, conforme
\begin{equation}
\hat{\mathbf{x}} = \left(\mathbf{A} \bm{\Gamma}^{-1}_{xx}\mathbf{A}^T + \mathbf{\Gamma}^{-1}_{nn}\right)^{-1} \mathbf{A}^T \mathbf{\Gamma}^{-1}_{xx} \mathbf{y}
\label{eq:tikhfft12}
\end{equation}  
onde $\bm{\Gamma}$ é a matriz de covariância das respectivas quantidades. Essa relação pode ser escrita do modo compacto $\mathbf{A}^+_W \mathbf{y}$ considerando  $\mathbf{A}^+_W = \left(\mathbf{A} \bm{\Gamma}^{-1}_{xx}\mathbf{A}^T + \mathbf{\Gamma}^{-1}_{nn}\right)^{-1} \mathbf{A}^T \mathbf{\Gamma}^{-1}_{xx}$ é uma matriz de reconstrução, permitindo obter o resultado  a partir da multiplicação matriz-vetor. 

A Equação \eqref{eq:tikhfft12} é chamada de mínimos quadrados regularizado e ponderado (ver \cite[Tabela 11.2]{Sayed2003}). Ela é muito semelhante à Equação \eqref{eq:tiksolx}, de modo que esta última também é encontrada como solução filtrada de Wiener \cite[pág. 150]{calvetti2007introduction}, \cite[pág. 79]{kaipio2005statistical}. 

\newpage

\section{Regressão, seleção de variáveis e multicolinearidade}\label{Ap:Ridge}

Seja o problema de regressão linear em estatística, onde se deseja modelar a relação entre os dados $\mathbf{y}$, variável dependente, e as variáveis explicativas (vetores) $\mathbf{a}_1, \mathbf{a}_2, \dots, \mathbf{a}_n$, também conhecidas como regressores ou variáveis independentes, que compõe as colunas da matriz de \textit{design} $\mathbf{A} = (\mathbf{a}_1, \mathbf{a}_2, \dots, \mathbf{a}_n)$. Escrevendo em notação matricial, tem-se 
\begin{equation}
\mathbf{y} = \mathbf{A} \bm{\beta} + \bm{\delta}, 
\label{eq:regression}
\end{equation}
onde $\bm{\beta}$ são pesos para as colunas de $\mathbf{A}$ e ajustados segundo algum critério e  $\bm{\delta}$ é o erro. Ao estimar $\bm{\beta}$ se busca um modelo explicativo para os dados e, no caso da Equação $\eqref{eq:regression}$ o modelo considerado é linear em relação a  $\bm{\beta}$. 

Os dados $\mathbf{y}$ são apenas uma amostra aleatória da população, então quando a amostra é representativa da população e o modelo obtido é adequado, pode-se obter boa generalização estatística para inferir informações sobre populações como um todo. Ou seja, outros dados $\mathbf{y}$ da mesma população, mas que não foram utilizados para cálculo de $\bm{\beta}$, também deveriam ser bem representados no modelo. 

\subsection{Critério dos mínimos quadrados ordinários}

Para estimar $\bm{\beta}$, seja a solução por mínimos quadrados ordinária (OLS) dada por
\begin{equation}
\bm{\hat{\beta}}_{OLS}=\arg\min\limits_{\bm{\beta}} \left[ \vert \vert \mathbf{A}\bm{\beta} - \mathbf{y} \vert\vert^2_2 \right],
\label{eq:ols1}
\end{equation}
cuja solução possui a conhecida forma das equações normais
\begin{equation}
\bm{\hat{\beta}}_{OLS}=  \left(\mathbf{A}^T \mathbf{A} \right)^{-1}\mathbf{A}^T\mathbf{y}, 
\label{eq:ols2}
\end{equation}
que também pode ser vista como análoga a uma solução sem regularização.

\subsection{\textit{Trade-off} entre \textit{bias} e variância na regressão linear}\label{sec:bias}

Para avaliar essa reconstrução $\mathbf{\hat{y}}(\bm{\hat{\beta}}) = \mathbf{A} \bm{\hat{\beta}}$ para dados $\mathbf{y}$ não-observados  há o critério do erro médio quadrático (MSE) dado por
\begin{equation}
\operatorname{MSE}_{OLS} =\mathbb{E}\left[({\mathbf{y}}-\mathbf{\hat{y})}^{2}\right]. 
\label{eq:bias1}
\end{equation}
É possível reescreve-las separando em três partes, deixando explícito que elas dependem de um termo relativo ao \textit{bias}, um termo da diferença entre o resultado obtido e o esperado, um termo relativo à variância, uma forma de se medir a sensibilidade do algoritmo à dispersão e flutuações nos dados de entrada e um termo relativo ao erro irredutível dos dados em si \cite[Capítulo 4]{geron2019hands-on}. Desconsiderando o erro irredutível, obtém-se \cite[Seção 2]{OSullivan1986}
\begin{equation}
\operatorname{MSE}_{OLS} =  \left[\mathbf{y}- \mathbb{E}( \mathbf{\hat{y}}) \right]^{2} + \mathbb{E}\left[\mathbf{\hat{y}} - \mathbb{E}( \mathbf{\hat{y}}) \right]^2, 
\label{eq:bias5}
\end{equation}
mas como $\mathbf{\hat{y}}$ depende de $\bm{\hat{\beta}}$, escreve-se
\begin{equation}
\operatorname{MSE}_{OLS} =  \text{bias}^2(\bm{\beta}) + \text{var}(\bm{\beta}).
\label{eq:bias6}
\end{equation}
A Equação \eqref{eq:bias6} mostra que existe um \textit{trade-off} entre essas duas quantidades, pois aumentar o \textit{bias} resulta em diminuir a variância, e vice-versa \cite[Capítulo 4]{geron2019hands-on}. Cada uma em excesso pode prevenir o algoritmo de funcionar bem além dos dados utilizados para estimação das variáveis explicativas $\bm{\beta}$, isto é, de generalizar\footnote{O conceito de generalização será explorado com mais detalhes na seção de aprendizagem de máquina} bem:
\begin{itemize}

\item Se a sua variância for muito grande, como pode acontecer com modelos com muitos graus de liberdade, o algoritmo poderá sofrer de \textit{overfitting}, ajustando-se tão fielmente aos dados que tanto será difícil para ele generalizar para outros dados quanto até o ruído pode fazer parte desse ajuste;
\item Se o \textit{bias} for muito grande, o modelo pode ser muito simples e algoritmo também não conseguir generalizar bem, sendo incapaz de capturar as relações entre entrada e saída \cite[Capítulo 4]{geron2019hands-on}. 
\end{itemize}

\subsection{Abordando a multicolinearidade}
A Equação \eqref{eq:regression} é semelhante à Equação \eqref{eq:eq7} que descrevia um problema inverso linear, mas enquanto o problema inverso poderia ser mal-posto, a Equação \eqref{eq:regression} pode apresentar o problema da colinearidade ou multicolinearidade. Colinearidade acontece em problemas de regressão quando há uma combinação linear não-trivial (coeficientes não-nulos) entre duas variáveis explicativas que resulta em zero \cite{Stewart1987} e são relacionados com problemas com deficiência de posto \cite[pág. 4]{Hansen1998}. Multicolinearidade acontece quando uma ou mais variáveis independentes do modelo possuindo relação entre si (não sendo, de fato, independentes), o que na prática significa que a matriz de \textit{design} $\mathbf{A}$ é mal-condicionada  \cite[pág. 133]{aster2019parameter}, \cite{Draper2016}. 

Quando discretizados, tanto o problema mal-posto quanto aquele que apresenta multicolinearidade resultam em sistemas mal-condicionados, sendo instáveis para pequenas variações nos dados $\mathbf{y}$, com valores estimados que podem estar longe do valor verdadeiro e que podem afetar a decisão no teste de hipótese \cite{Oztrk2000}. 

Existem muitas propostas de solução para a multicolinearidade e uma delas é similar à solução de problemas mal-postos, através de regularização\footnote{Propostas de solução da multicolinearidade e de preditores redundantes já são tratadas como regularização, como em \url{https://www.mathworks.com/discovery/regularization.html}.} \cite{Oztrk2000}. Além disso, do \textit{trade-off} discutido pode surgir a dúvida se há alguma forma de controlar o de \textit{bias} e a variância da solução e a  contribuição de cada um desses termos pode ser feita através do próprio parâmetro de regularização. 

Na área de estatística, a regularização clássica de Tikhonov com $\mathbf{L} = \mathbf{I}$ e $\mathbf{x}^* = \mathbf{0}$ tem o seu análogo na regressão de Ridge \cite[pág. 101]{Hansen1998}, \cite{Hoerl1970},  cujo funcional e a solução são dadas respectivamente por
\begin{equation}
\bm{\hat{\beta}}_{ridge} = \underset{\bm{\beta}}{\arg\min}\left[ \vert\vert\mathbf{A}\bm{\beta} - \mathbf{y}\vert\vert^2_2 + \lambda^2 \vert\vert \bm{\beta} \vert\vert^2_2 \right],
\label{eq:regression1}
\end{equation}
\begin{equation}
\bm{\hat{\beta}}_{ridge} = \left(\mathbf{A}^T \mathbf{A} + \lambda^2 \mathbf{I}\right) \mathbf{A}^T \mathbf{y}.
\label{eq:regression2}
\end{equation}
Em relação à nomenclatura, a Equação \eqref{eq:ols2} foi chamada de solução por mínimos quadrados ordinária e a Equação \eqref{eq:regression1} de regressão de Ridge. Esta última é, por vezes, encontrada como mínimos quadrados regularizados \cite[pág. 147]{kaipio2005statistical} ou regressão por mínimos quadrados penalizada \cite{Gribonval2011}. No entanto, em alguns casos essa equivalência entre conceitos não seja válida. Por exemplo, em \cite{Gribonval2011} discute-se que a regressão de mínimos quadrados penalizada não é a única interpretação da estimativa por MAP, apesar da formulação semelhante.

\subsection{\textit{Trade-off} entre \textit{bias} e variância de uma solução regularizada}\label{subsec:tradeoff}

A solução de Ridge para o problema de regressão linear ajuda a entender, por analogia, quando hoje se diz que a regularização de Tikhonov controla o \textit{trade-off} entre \textit{bias} e variância pelo parâmetro $\lambda$. Partindo do MSE como critério, a expressão dada pela Equação \eqref{eq:bias1} e, consequentemente da Equação \eqref{eq:bias5}, não é mais válida, pois se referia ao método dos mínimos quadrados ordinário e não está escrita em termos de nenhum parâmetro $\lambda$ para verificar a influência da regularização na solução. 

Deve-se destacar que uma expressão analítica da influência do parâmetro de regularização no trade-off entre \textit{bias} e variância é encontrada no próprio trabalho original da regressão de Ridge \cite{Hoerl1970}, conforme
\begin{equation}
\mathbb{E}\left( (\bm{\hat{\beta}}_{ridge} - \bm{\beta})^T (\bm{\hat{\beta}}_{ridge} - \bm{\beta}) \right) = s^{2} \sum_{i = 1}^{p} \frac{\sigma_{i}^2}{\left( \sigma_{i}^2+\lambda\right)^{2}}+\lambda^{2} \bm{\beta}^{T}\left( \mathbf{A}^{T} \mathbf{A} + \lambda \mathbf{I} \right)^{-2} \bm{\beta},
\label{eq:ridge}
\end{equation}
onde $s^2$ é a variância e $p$ é o posto de $\mathbf{A}$.  Originalmente, o autor escreveu a expressão considerando os autovalores de $\mathbf{A}^{T} \mathbf{A}$. Na Equação \eqref{eq:ridge}, ela foi reescrita considerando que os valores singulares $\sigma_{i}$  não-nulos de $\mathbf{A}$ são iguais às raízes quadradas dos autovalores não-nulos de $\mathbf{A}^{T} \mathbf{A}$, de modo que eles aparecem ao quadrado na expressão.

Na Equação \eqref{eq:ridge}, o termo da esquerda se refere à variância e o termo da direita ao \textit{bias} da solução proposta, o que mostra que um aumento de $\lambda$ resulta em um aumento de \textit{bias} nesse \textit{trade-off}. Além da regressão de Ridge, na regularização de Tikhonov também são discutidas avaliações sobre a diminuição da variância e o aumento do \textit{bias} à medida que $\lambda$ aumenta \cite[págs. 64, 86]{hansen2010discrete}. A partir da geração de gráficos variando-se $\lambda$, pode-se visualizar como esse \textit{trade-off} se traduz na solução,  se é possível observar a tendência de \textit{overfitting} quando não há regularização e \textit{underfitting} quando há muita regularização. A expressão explícita, analítica, da Equação \eqref{eq:ridge} é importante, mas em outros casos como na regularização generalizada de Tikhonov ou de outros regularizadores nem sempre uma expressão analítica está disponível.
 
 A Equação \eqref{eq:ridge} é uma expressão analítica obtida para a regressão linear \cite{Hoerl1970}. Para sua utilização em outros problemas, é necessário notar que:
\begin{itemize}
\item Na prática não é trivial, ou mesmo possível, de utilizá-la para estimar o \textit{bias}, pois a sua expressão depende do conhecimento de  $\bm{\beta}$, que são as variáveis que se deseja estimar. Isso é possível em problemas simulados, o que traz \textit{insights} importantes para o efeito de $\lambda$ na solução;  

\item O gráfico resultante pode ser diferente para modelo, pois essa expressão depende de $\mathbf{A}$ diretamente (\textit{bias}) ou dos valores singulares de $\mathbf{A}$ (variância). Ou seja,  o gráfico também será diferente para cada aplicação. No caso de um modelo de degradação de imagem, o gráfico é diferente para cada nível de degradação, por exemplo;

\item No caso de modelos não-lineares, como redes neurais profundas, tem que se avaliar se é possível obter expressões analíticas para o \textit{trade-off}. Em alguns casos, os gráficos são gerados e visualizados apenas para fins de ilustração do conceito.
 
 \end{itemize}

 \subsection{Exemplo numérico do \textit{trade-off}}\label{subsec:tradeoffblur}
 
\begin{figure}[H]
     \centering
     \begin{subfigure}[b]{0.33\textwidth}
         \centering
         \includegraphics[trim=85 0 85 0,clip,width=0.8\textwidth]{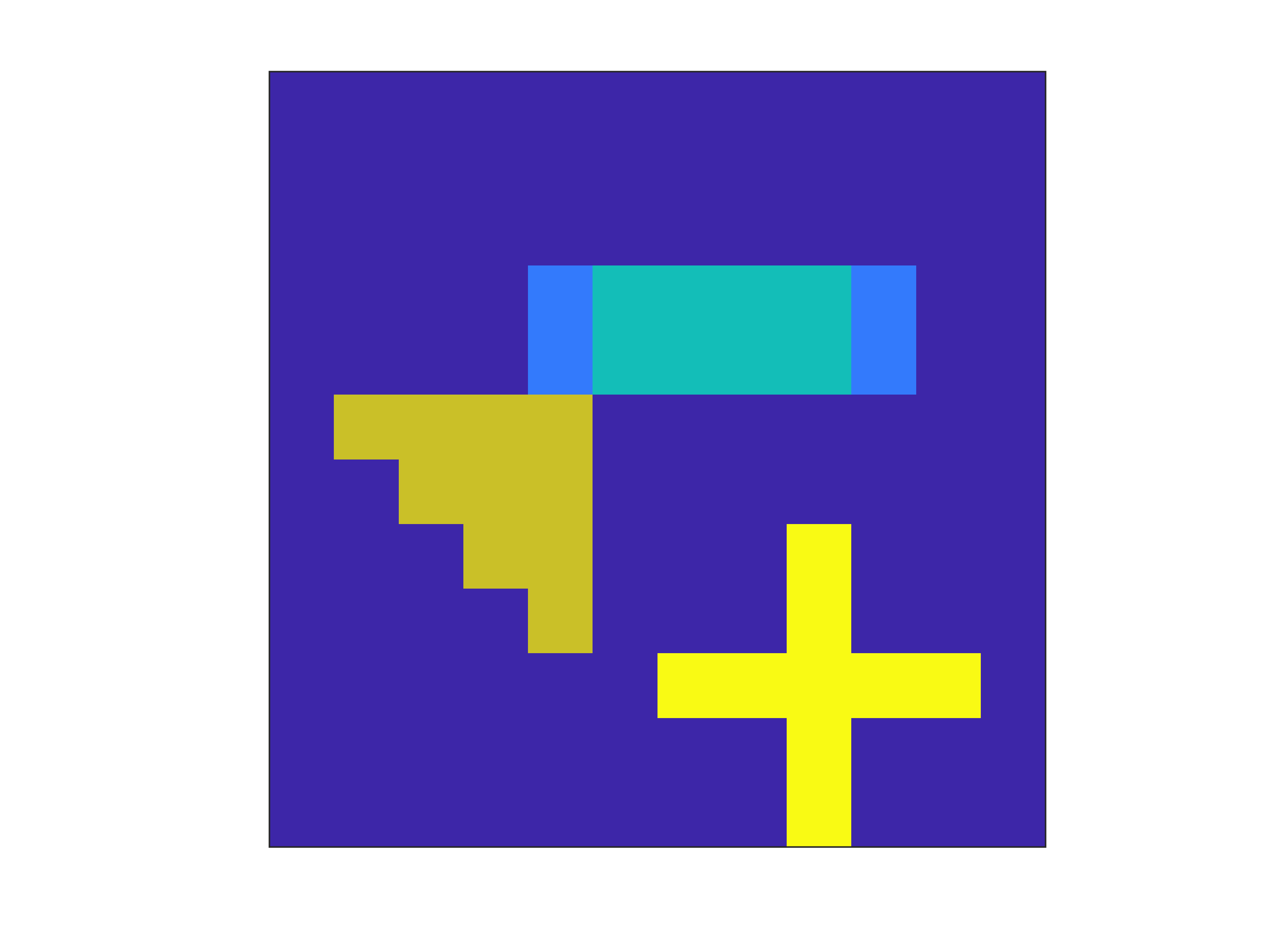}
         \caption{\textit{Ground truth}}
         \label{fig:ridgeblura}
     \end{subfigure}
     \hfill \hspace{-15mm}
     \begin{subfigure}[b]{0.33\textwidth}
         \centering
         \includegraphics[trim=85 0 85 0,clip,width=0.8\textwidth]{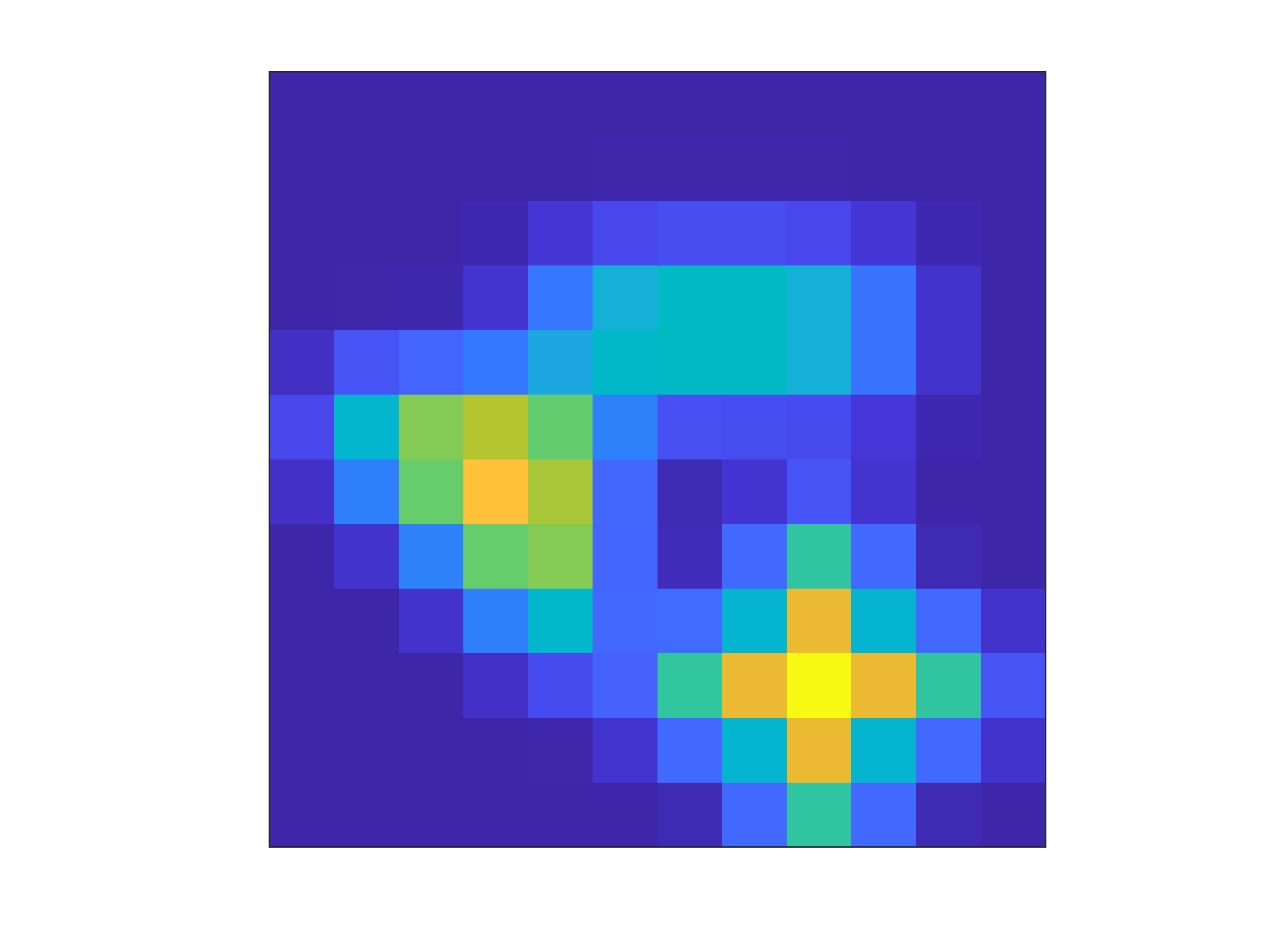}
         \caption{Borrada: $\sigma = 0.7$.}
                  \label{fig:ridgeblurb}
     \end{subfigure}
     \hfill \hspace{-15mm}
     \begin{subfigure}[b]{0.33\textwidth}
         \centering
         \includegraphics[trim=85 0 85 0,clip,width=0.8\textwidth]{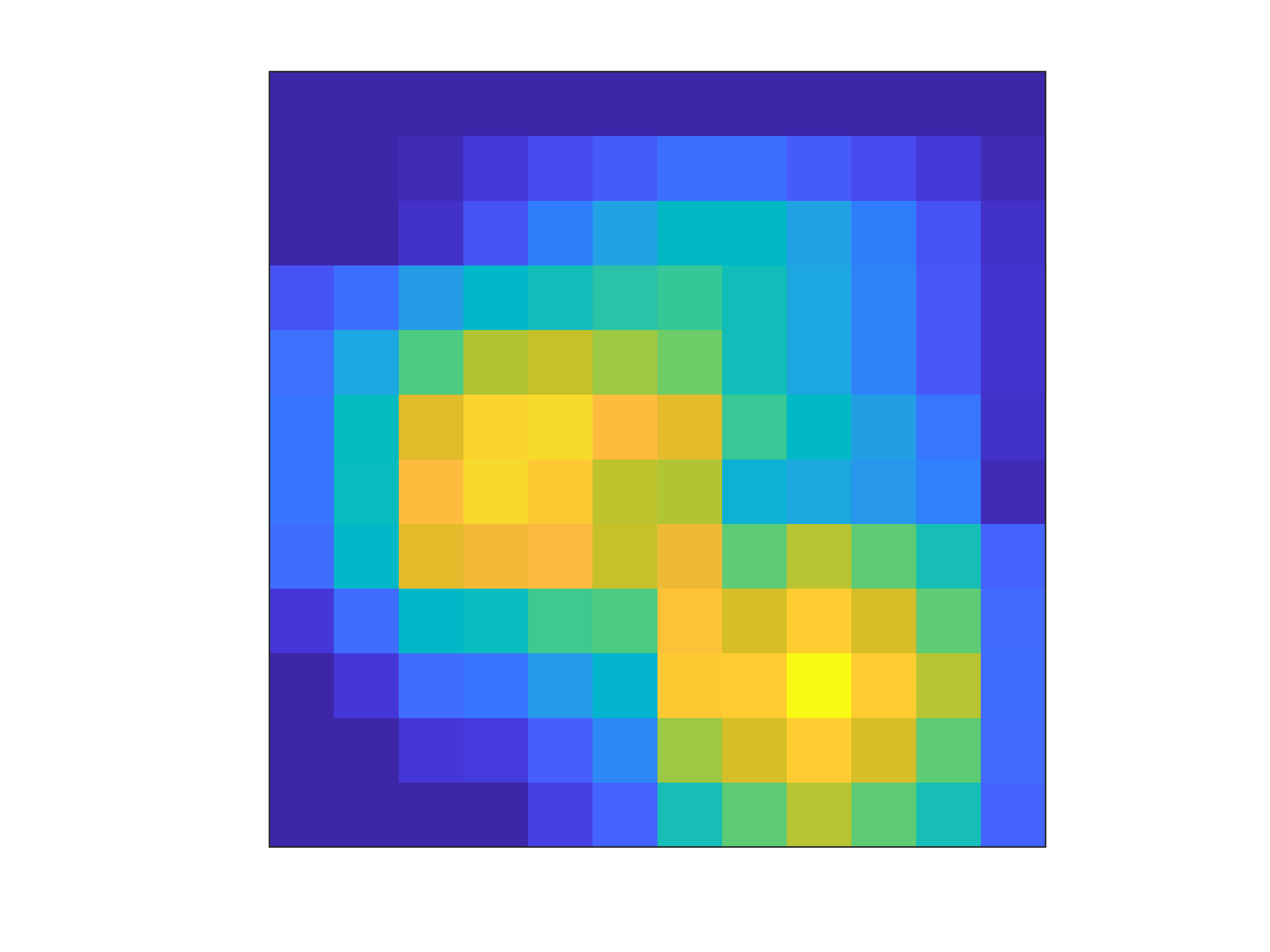}
         \caption{Borrada: $\sigma = 4$.}
                  \label{fig:ridgeblurc}
     \end{subfigure}
          \hfill \hspace{-15mm}    
          \begin{subfigure}[b]{1\textwidth}
         \centering
         \includegraphics[trim=60 0 60 0,clip,width=1\textwidth]{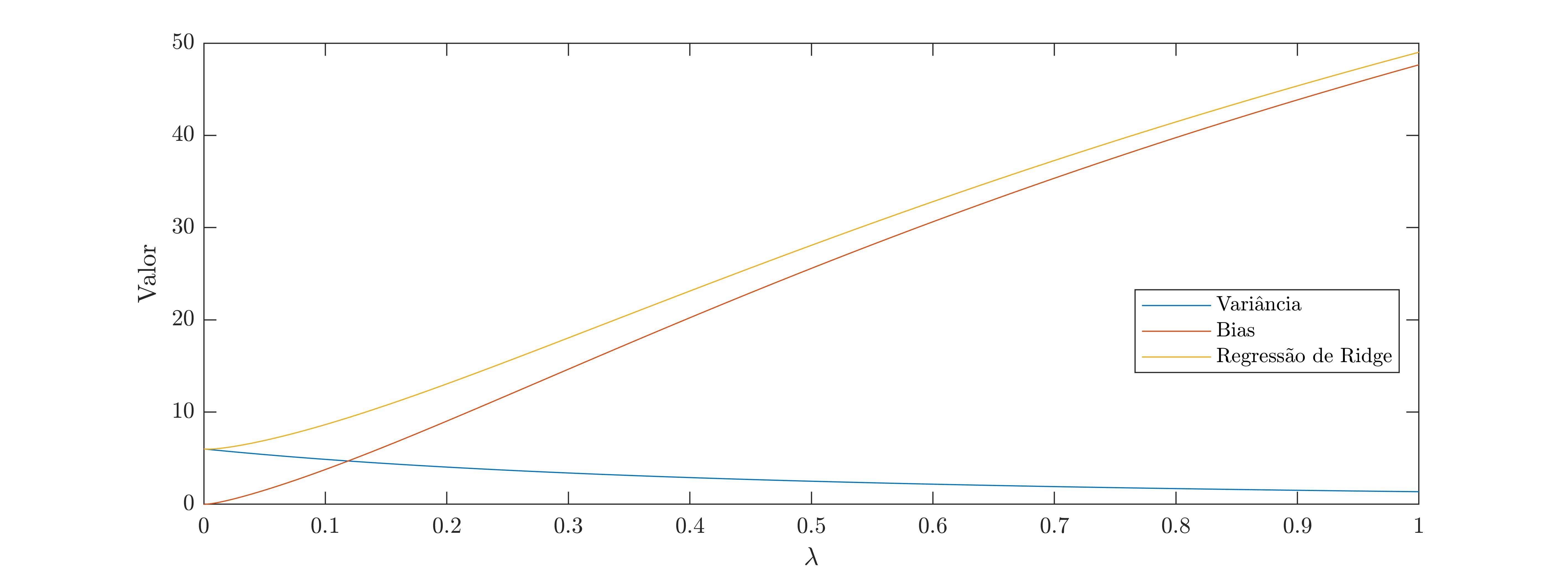}
         \caption{Bias e variância de b)}
                  \label{fig:ridgeblurd}
     \end{subfigure}
               \begin{subfigure}[b]{1\textwidth}
         \centering
         \includegraphics[trim=60 0 60 0,clip,width=1\textwidth]{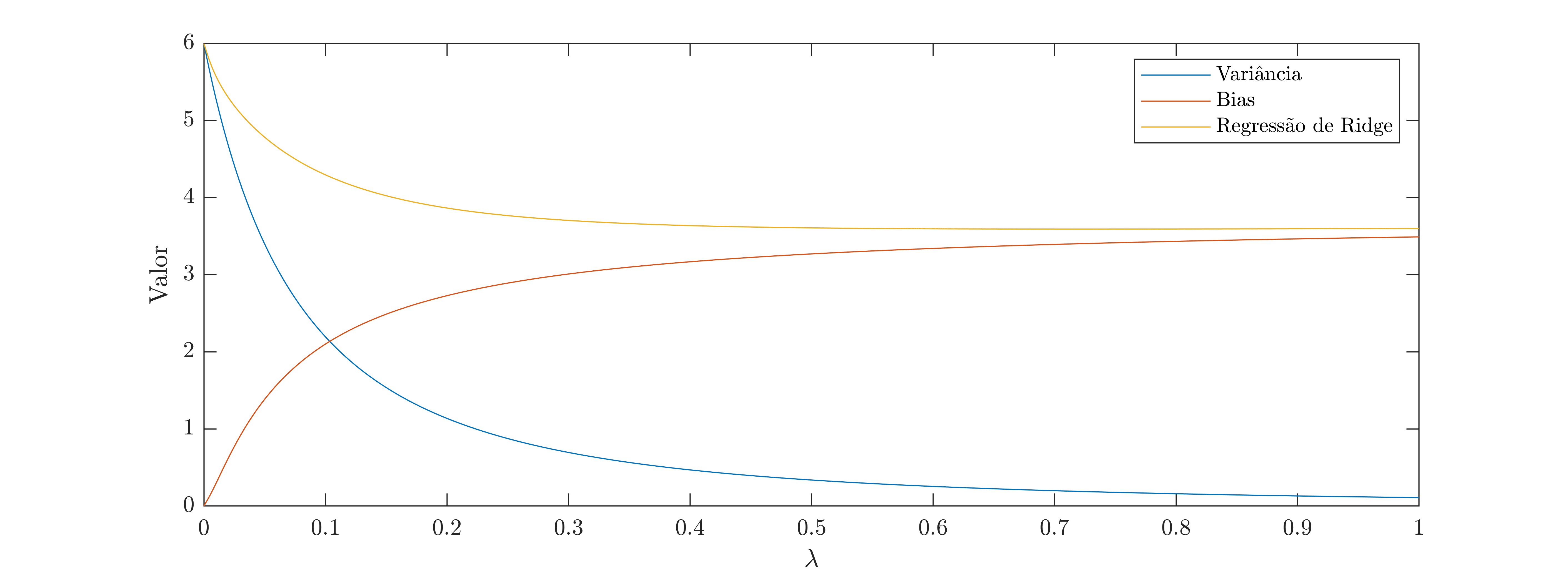}
         \caption{Bias e variância de c)}
                  \label{fig:ridgeblure}
     \end{subfigure}
                         \caption[\textit{Trade-off} entre \textit{bias} e variância em \textit{deblurring}.]{\textit{Trade-off} entre \textit{bias} e variância em \textit{deblurring}. Fonte: Próprio autor. }
        \label{fig:ridgeblur}
\end{figure}

 Para ilustrar a presente discussão, a Figura \ref{fig:ridgeblur} mostra um exemplo numérico da Equação \eqref{eq:ridge} para o caso do \textit{deblurring}. Este exemplo foi baseado na implementação da função \textit{blur} da \textit{toolbox Regtools} \cite{Hansen2007}, na qual variou-se o desvio padrão do \textit{blur} gaussiano para obter dois operadores diretos $\mathbf{A}$ diferentes, um que resulta na Figura \ref{fig:ridgeblurb}, menos borrada, e outro na Figura \ref{fig:ridgeblurc}, mais borrada.  
 
Tendo $\mathbf{A}$ e $\mathbf{x}$ disponíveis, a Equação \eqref{eq:ridge} foi utilizada para obter as Figuras \ref{fig:ridgeblurd} e \ref{fig:ridgeblure}, na qual se observa o efeito de $\lambda$ que decorre da regressão de Ridge. Observa-se que a queda da variância é mais pronunciada quando o \textit{blur} é maior. Neste caso, o gráfico se assemelha mais ao do artigo original da regressão de Ridge em \cite[Figura 1]{Hoerl1970}. 

 \newpage
\subsection{Outras formas de seleção de variáveis}

Na regressão de Ridge, os parâmetros tendem a se se distribuírem ao longo de todo o vetor $\bm{\beta}$, o que não produz um modelo parcimonioso, já que mantém todos os preditores (parâmetros) no modelo \cite{zou2005}. Nesse contexto, um modelo parcimonioso é aquele em que a informação pode ser codificada em poucos coeficientes, mas robustos \cite[pág. 196]{bovik2005handbook}.

Essa não é a única forma de lidar com multicolinearidade e preditores redundantes. A seleção de variáveis é possível aumentando o número de parâmetros nulos do vetor de soluções para identificar as variáveis mais relevantes para descrição da solução, o que anteriormente foi visto como a busca por soluções mais esparsas. Para isso, a norma $\ell_1$ sob os parâmetros é utilizada, o que é conhecido como LASSO em problemas de regressão \cite{Tibshirani2011}, conforme 
\begin{equation}
\bm{\hat{\beta}} = \underset{\bm{\beta}}{\arg\min}\left[ \vert \vert \mathbf{y} - \mathbf{A}\bm{\beta} \vert \vert^2_2+\lambda\vert \vert \bm{\beta} \vert\vert_{1} \right].
\label{eq:lasso}
\end{equation} 
Ela apresenta $\mathbf{L} = \mathbf{I}$, porém com norma $\ell_1$ no regularizador, de modo que $\bm{\beta}$ seja mais esparsa, com mais variáveis nulas e menos variáveis não-nulas. Isso significa que LASSO realiza uma seleção de variáveis mais robusta do que a regressão de Ridge, facilitando a interpretação do modelo. Conforme \cite[pag. 316]{Deisenroth2020}, a LASSO seria equivalente a um modelo de regressão linear com \textit{prior} laplaciano.  

A regressão de LASSO possui limitações, como a tendência de selecionar apenas uma variável entre um grupo de variáveis altamente correlacionadas \cite{zou2005}. Já a rede elástica utiliza uma norma em cada termo de regularização, conforme
\begin{equation}
\bm{\hat{\beta}} = \underset{\bm{\beta}}{\arg\min}\left[ \vert \vert \mathbf{y} - \mathbf{A}\bm{\beta} \vert \vert^2_2+\lambda_{1} \vert\vert \bm{\beta} \vert\vert^{2}_2+\lambda _{2}\vert \vert \bm{\beta} \vert\vert_{1} \right],
\label{eq:elastic_app}
\end{equation} 
o que permite a seleção de variáveis e também permite selecionar grupos de fatores correlacionados \cite{zou2005}.

Por fim, deve-se citar que o LASSO possui outras variações \cite{Tibshirani2011}:
\begin{itemize}
\item Quando o termo aditivo é da forma $\lambda\vert \vert \mathbf{L}\bm{\beta} \vert\vert_{1}$, a proposta é conhecida como LASSO generalizada \cite{Tibshirani2011gen}. Cada uma dentro de seu contexto, mas a diferença entre a regressão de LASSO e da LASSO generalizada é como a diferença entre a regularização clássica e generalizada de Tikhonov, para fazer o paralelo; 
\item A regressão de Bridge é generalização de uma norma $\ell_p$ qualquer \cite{Park2011};
\item Há ainda uma forma com três funções convexas, onde um termo é $\vert\vert\mathbf{x}\vert\vert_1$ e o outro é $\vert\vert\mathbf{L}\mathbf{x}\vert\vert_1$, chamada de LASSO generalizada esparsa  \cite{Ko2019}, que permite fazer uma relação com a regularização multiparâmetros discutida anteriormente.
\end{itemize}

\end{appendices}
   
\end{document}